\documentclass[a4paper,10pt]{book}

\newif\ifcompilefigs
\compilefigstrue

\usepackage{amsmath}
\usepackage{amsfonts}
\usepackage{amssymb}         
\usepackage{amsopn}
\usepackage{amsthm}
\usepackage{thmtools}
\usepackage[nottoc]{tocbibind}
\usepackage{latexsym}
\usepackage{graphicx}        
\usepackage{mathrsfs}		
\usepackage{psfrag}
\usepackage[ruled,algo2e]{algorithm2e}
\usepackage{color}
\usepackage{verbatim}
\usepackage{url}
\usepackage{multirow}
\usepackage{setspace}
\usepackage{booktabs,longtable,rotating}
\usepackage{afterpage}
\usepackage{sidecap}
\usepackage{marvosym}
\usepackage[caption=false]{subfig}
\usepackage{enumerate}
\usepackage[a4paper,
            bindingoffset=0.2in,
            left=1.1in,
            right=1.1in,
            top=1.1in,
            bottom=1.5in,
            footskip=1.5cm,
            headheight=2cm]{geometry}
\usepackage{fancyhdr}
\usepackage{import}
\usepackage{mathbbol}
\usepackage{relsize}
\usepackage{lipsum}
\usepackage{array}
\usepackage{qcircuit}
\usepackage{dashrule}
\usepackage{mathtools}
\usepackage{listings}
\usepackage[usenames,dvipsnames]{xcolor}
\usepackage[breakable, theorems, skins]{tcolorbox}
\usepackage[breaklinks]{hyperref}
\usepackage{breakcites}
\usepackage{mdframed}
\usepackage{bbm}
\usepackage{tikz}
\usetikzlibrary{shapes.geometric}
\usetikzlibrary{calc}
\usetikzlibrary{positioning}
\usetikzlibrary{arrows.meta}
\usepackage[b]{esvect}
\tcbset{enhanced}
\usepackage{makeidx}

\makeindex

\DeclareSymbolFontAlphabet{\mathbb}{AMSb}
\DeclareSymbolFontAlphabet{\mathbbl}{bbold}

\DeclareFontEncoding{FML}{}{}%
\DeclareFontSubstitution{FML}{fncmi}{m}{it}%
\DeclareSymbolFont{fouriernc}{FML}{fncmi}{m}{it}%
\DeclareMathAccent{\fvec}{0}{fouriernc}{"7E}
\DeclareMathAlphabet{\mymathbb}{U}{bbold}{m}{n}

\declaretheorem[name=Theorem,numberwithin=chapter]{theorem}
\declaretheorem[name=Definition,sibling=theorem]{definition}
\declaretheorem[name=Lemma,sibling=theorem]{lemma}
\declaretheorem[name=Corollary,sibling=theorem]{corollary}
\declaretheorem[name=Proposition,sibling=theorem]{proposition}

\declaretheorem[name=Example,sibling=theorem]{example}
\declaretheorem[name=Remark,sibling=theorem]{remark}

\declaretheorem[name=Assumption,numbered=no]{assumption*}
\declaretheorem[name=Postulate]{postulate}
\declaretheorem[name=Postulate,numbered=no]{postulate*}

\newcommand{\floor}[1]{{\left\lfloor{#1}\right\rfloor}}
\newcommand{\ceil}[1]{{\left\lceil{#1}\right\rceil}}
\newcommand{\rint}[1]{{\left\lfloor{#1}\right\rceil}}
\newcommand{\nrm}[1]{{\left\|{#1}\right\|}}
\newcommand{\abs}[1]{{\left|{#1}\right|}}

\newcommand{\N}{\ensuremath{\mathbb{N}}}

\newcommand{\Z}{\ensuremath{\mathbb{Z}}}
\newcommand{\R}{\ensuremath{\mathbb{R}}}
\newcommand{\C}{\ensuremath{\mathbb{C}}}
\newcommand{\di}{\ensuremath{\mathrm{d}}}

\renewcommand{\S}{\ensuremath{\mathbb{S}}}
\newcommand{\zeroes}{\ensuremath{\mathbbl{0}}}

\newcommand{\vh}{\ensuremath{\fvec{h}}}
\newcommand{\vi}{\ensuremath{\fvec{\imath}}}
\newcommand{\vj}{\ensuremath{\fvec{\jmath}}}
\newcommand{\vk}{\ensuremath{\fvec{k}}}
\newcommand{\va}{\ensuremath{\fvec{a}}}
\newcommand{\vp}{\ensuremath{\fvec{p}}}
\newcommand{\vx}{\ensuremath{\fvec{x}}}
\newcommand{\vy}{\ensuremath{\fvec{y}}}

\renewcommand{\v}[1]{\ensuremath{\fvec{#1}}}
\newcommand{\allones}{\ensuremath{\mathbbm{1}}}
\newcommand{\allzeroes}{\ensuremath{\mymathbb{0}}}
\newcommand{\amp}[1]{\ensuremath{\text{amp}\left(#1\right)}}

\newcommand{\polylog}[1]{\ensuremath{\text{polylog}\left(#1\right)}}
\newcommand{\ham}{\ensuremath{\mathcal{H}}}

\newcommand{\book}{{set of lecture notes}}

\DeclareMathOperator{\Tr}{Tr}
\DeclareMathOperator{\sinc}{sinc}

\newcommand{\bigO}[1]{\mathcal{O}\left( #1 \right)}
\newcommand{\bigOt}[1]{\widetilde{\mathcal{O}}\left( #1 \right)}

\newcommand{\trace}[2][]{%
   \ifthenelse{ \equal{#1}{} }
      {\Tr\left(#2\right)}
      {\Tr_{#1}\left(#2\right)}
}

\DeclarePairedDelimiter\bra{\langle}{\rvert}
\DeclarePairedDelimiter\ket{\lvert}{\rangle}
\DeclarePairedDelimiterX\braket[2]{\langle}{\rangle}{#1 \delimsize\vert #2}
\DeclarePairedDelimiterX\ketbra[2]{| }{|}{#1 \delimsize\rangle\!\delimsize\langle #2}
\DeclarePairedDelimiterX\dotp[2]{\langle}{\rangle}{#1, #2}

\SetArgSty{textnormal}

\begin{document}
\begin{titlepage} 
	
  \raggedleft 
	
  \rule{1pt}{\textheight} 
  \hspace{0.05\textwidth} 
  \parbox[b]{0.75\textwidth}{ 
		
    {\Huge\bfseries Quantum algorithms \\[0.5\baselineskip] for optimizers}\\[2\baselineskip] 
    {\large\textit{Lecture notes}}\\[4\baselineskip] 
    {\Large\textsc{giacomo nannicini}} 
		
    \vspace{0.5\textheight} 
    
    {\noindent Developed for course ISE 612 at the University of Southern California}\\[\baselineskip] %
    {\noindent Last update: \today}
  }

\end{titlepage}

This \book{} is a self-contained graduate-level course on quantum
algorithms, with an emphasis on optimization algorithms. It is written
for applied mathematicians and engineers: we do not rely on physics or
physical intuition, but rather, we derive all results in a rigorous
manner starting from first principles and just three postulates. Thus,
knowledge of quantum mechanics or physics is not assumed or
required. A solid background in linear algebra, elementary calculus,
and some knowledge of mathematical optimization are extremely helpful,
if not necessary, to understand the material.

The material contained here started from a set of lecture notes
developed over the years, teaching a ``special topics'' Ph.D.-level
class on quantum algorithms in Industrial \& Systems Engineering
departments: first at Columbia University IEOR in the Fall 2019, then
at Lehigh University ISE in the Spring 2020, finally at the University
of Southern California ISE, first offered in the Fall 2023. The
experience in the classroom had a tremendous impact on the structure
and exposition style of this \book{}. 

This \book{} is intended to be appropriate for graduate students and
researchers who want to learn about quantum algorithms, and who are
particularly interested in mathematical optimization. The choice of
topics is heavily skewed in favor of optimization: among the many
topic areas in quantum computing and quantum algorithms, the material
contained here covers those areas that have already proven
useful for the development of quantum optimization algorithms, or that
seem likely to be useful for that end, because of their strong connections
with classical (i.e., non-quantum) optimization theory or practice. It
is also worth mentioning that the focus of this \book{} is on the
theory of quantum algorithms, i.e., how to design, understand, and
analyze quantum algorithms; we sometimes discuss practical
considerations, but we still assume access to a fully fault-tolerant
quantum computing device, and do not attempt to discuss the (very
interesting, and practically useful) intricacies of running quantum
algorithms on real devices. Throughout the \book{}, we give references
to key results for each of the topics discussed.

The main goal of this \book{} is to equip the reader with the tools
necessary to investigate fundamental questions in quantum optimization
algorithms: Can quantum algorithms be useful for optimization? And if
so, how? What are some of the tasks that quantum computers are good
at, and that can be used for optimization? Although we do not give
definitive answers to these questions, at the end of this course the
reader will be able to form their own informed opinion. Perhaps more
importantly, the reader will have the tools and background to jump
into the beautiful and constantly evolving literature on quantum
algorithms --- especially quantum optimization algorithms --- where
new discoveries are being made at a tremendous rate.

\vspace*{2em}
\paragraph{Acknowledgments.} The author is grateful to the Office of Naval Research for supporting the research behind several chapters of this \book{}, through award \# N000142312585. The author is also grateful to everyone who attended his quantum algorithms classes over the years.

\vspace*{2em}
\paragraph{Copyright notice.} This is a possibly-outdated preprint, using the author's lecture notes, of the book ``Quantum algorithms for optimizers,'' edited by SIAM. All rights to the book's material belong to SIAM. Please refer to the SIAM version of the book.

\thispagestyle{plain}

\tableofcontents

\chapter{Model of computation}
\label{ch:intro}
\thispagestyle{fancy}
Quantum computing is a relatively new area of computing that has the
potential to greatly speed up the solution of certain
problems. Quantum computers work in a fundamentally different way than
classical computers. This \book{} is a course on quantum algorithms, with
a focus on algorithms that may be useful for mathematical
optimization. We begin by introducing the model of computation,
and then proceed to study several quantum algorithms. In the
following, the term ``classical'' is used to mean ``non-quantum'', as
is common in the field.

The quantum computing device is, in abstract terms, similar to a
classical computing device: it has a state, and the state of the
device evolves by applying certain operations. The model of
computation that we consider is the quantum circuit model, which works
as follows:
\begin{enumerate}
\item The quantum computer has a {\em state} that is contained in a
  quantum register and is initialized in a predefined way.
\item The state evolves by applying {\em operations} specified in
  advance in the form of an algorithm.
\item At the end of the computation, some information on the state of
  the quantum register is obtained by means of a special operation,
  called a {\em measurement}.
\end{enumerate}
All terms in italics are the subject of postulates, on which our
exposition builds. Note that this type of computing device is similar
to a Turing machine, except for the presence of a tape. It is possible
to assume the presence of a tape and be more formal in defining a
device that is the quantum equivalent of a Turing machine, but there
is no need to do so for the purposes of this \book{}.  Fundamental
results regarding universal quantum computers (i.e., the quantum
equivalent of a universal Turing machine) are presented in
\cite{deutsch85quantum,yao1993quantum,bernstein1997quantum}.

We use the quantum circuit model throughout this \book{}. This
model of computation closely matches the general-purpose
implementation provided by certain quantum hardware technologies used
by some of the major players in the field. We should note, however,
that the hardware is affected by noise and therefore it does not
provide an exact implementation of the theoretical model. To
understand the effect of noise, we can give the following simple, but
overall quite accurate, intuitive explanation. According to the model
of computation, the state evolves by applying operations, and some
information on the state can be extracted via a measurement; due to
noise, the state may not evolve in the desired way (e.g., applying a
certain operation on the state $s_1$ should yield the state $s_2$, but
we obtain a different state $s_3$ instead), or the information
extracted by a measurement may not be what it is supposed to be (e.g.,
a measurement should produce the output $0$ with probability $p_1$,
but it produces $0$ with a different probability $p_2$ instead).

This \book{} is written for applied mathematicians and engineers, and
aims to be ``physics-free''; for this reason, we do not discuss
further the specifics of existing quantum hardware that follows the
circuit model. However, we note that a different model for quantum
computing exists: the so-called adiabatic model. We provide some notes
and references on the adiabatic model of computation in
Sect.~\ref{sec:adiabaticnotes}, after discussing the adiabatic
theorem. We do not discuss the adiabatic model in detail
for at least two reasons: first, because the adiabatic and the circuit
model are equivalent in theory \cite{aharonov2008adiabatic}; second,
because the circuit model is more commonly used in the literature,
likely due to the fact that it is often easier to analyze.

\section{Basic definitions and notation}
\label{sec:notation}
A formal treatment of quantum algorithms requires familiarity with the
properties of the tensor product. We describe here the necessary
concepts and the notation. It also requires working with the decimal
and the binary representation of integers, discussed subsequently in
this section.

\begin{definition}[Tensor product]
  \label{def:tensor}
  Given two vector spaces $V$ and $W$ over a field $K$ with bases
  $e_1,\dots,e_m$ and $f_1,\dots,f_n$ respectively, the {\em tensor
    product}\index{product!tensor|(}\index{notation!otimes@\ensuremath{\otimes}}\index{tensor product|see{product, tensor}}
  $V \otimes W$ is another vector space (that we call {\em tensor
    product space}) over $K$ of dimension $mn$. The tensor product
  space is equipped with a bilinear operation $\otimes : V \times W
  \to V \otimes W$. The tensor product space $V \otimes W$ has basis
  $e_i \otimes f_j \; \forall i=1,\dots,m, j=1,\dots,n$.
\end{definition}
If the origin vector spaces are complex Euclidean spaces of the form
$\C^n$, and we choose the standard basis (consisting of the
orthonormal vectors that have a $1$ in a single position and 0
elsewhere) in the origin vector spaces, then the bilinear operation of
the tensor product is none other than the Kronecker product. This is
formalized next.
\begin{definition}[Kronecker product]
  \label{def:kronecker}
  Given $A \in \C^{m \times n}, B \in \C^{p \times q}$, the {\em
    Kronecker product}\index{Kronecker product}\index{product!Kronecker} $A \otimes B$ is the
  matrix $D \in \C^{mp \times nq}$ defined as:
  \begin{equation*}
    D := A \otimes B = \begin{pmatrix} a_{11} B & \dots & a_{1n} B\\
      a_{21} B & \dots & a_{2n} B \\
      \vdots & & \vdots \\
      a_{m1} B & \dots & a_{mn} B
      \end{pmatrix}.
  \end{equation*}
  If we choose the standard basis over the vector spaces $\C^{m \times
    n}$ and $\C^{p \times q}$, then the bilinear operation $\otimes$
  of the tensor product $\C^{m \times n} \otimes \C^{p \times q}$
  corresponds to the Kronecker product. This definition naturally
  applies to vectors by taking $n = q = 1$.
\end{definition}
In this \book{} we always work with complex Euclidean spaces of the
form $\C^n$, using the standard basis. With a slight but common abuse
of notation, we therefore use the tensor product to also refer to the
Kronecker product. In this way, we can apply $\otimes$ to vector
spaces, vectors and matrices, and in all cases we obtain precisely the
mathematical object that we want to obtain: the tensor product space,
a vector in the tensor product space, or a linear map on the tensor
product space.

\begin{example}
  \label{ex:tensorproduct}
  We provide an example of the tensor product for normalized vectors,
  which links this concept to probability distributions and 
  hopefully provides an intuition for some of the future
  material. Consider two independent discrete random variables $X$ and
  $Y$ that describe the probability of extracting numbers from two
  urns. The first urn contains the numbers $0$ and $1$, the second urn
  contains the numbers $00, 01, 10, 11$. Assume that the extraction
  mechanism is biased and therefore the outcomes do not have equal
  probability. The outcome probabilities are given below, and for
  convenience we define two vectors containing them:
  \begin{align*}
    x = \begin{pmatrix} \Pr(X = 0) \\ \Pr(X = 1) \end{pmatrix} =
    \begin{pmatrix} 0.25 \\ 0.75 \end{pmatrix} \qquad
    y = \begin{pmatrix} \Pr(Y = 00) \\ \Pr(Y = 01) \\ \Pr(Y = 10) \\ \Pr(Y = 11) \end{pmatrix} = \begin{pmatrix} 0.2 \\ 0.2 \\ 0.2 \\ 0.4 \end{pmatrix}.
  \end{align*}
  Notice that because each vector contains probabilities for all
  possibile respective outcomes, the vectors are normalized so that
  their entries sum up to 1. Then, the joint probabilities for
  simultaneously extracting numbers from the two urns are given by the
  tensor product $x \otimes y$:
  \begin{align*}
    x \otimes y = 
    \begin{pmatrix} 0.25 \\ 0.75 \end{pmatrix} \otimes \begin{pmatrix} 0.2 \\ 0.2 \\ 0.2 \\ 0.4 \end{pmatrix}  =
    \begin{pmatrix} 0.05 \\ 0.05 \\ 0.05 \\ 0.1 \\ 0.15 \\ 0.15 \\ 0.15 \\ 0.3 \end{pmatrix} =
    \begin{pmatrix} \Pr(X = 0) \Pr(Y = 00) \\ \Pr(X = 0) \Pr(Y = 01) \\ \Pr(X = 0) \Pr(Y = 10) \\ \Pr(X = 0) \Pr(Y = 11) \\ \Pr(X = 1) \Pr(Y = 00) \\ \Pr(X = 1) \Pr(Y = 01) \\ \Pr(X = 1) \Pr(Y = 10) \\ \Pr(X = 1) \Pr(Y = 11) \end{pmatrix} =
    \begin{pmatrix} \Pr(X = 0, Y = 00) \\ \Pr(X = 0, Y = 01) \\ \Pr(X = 0, Y = 10) \\ \Pr(X = 0, Y = 11) \\ \Pr(X = 1, Y = 00) \\ \Pr(X = 1, Y = 01) \\ \Pr(X = 1, Y = 10) \\ \Pr(X = 1, Y = 11) \end{pmatrix},
  \end{align*}
  where the last equality is due to the fact that $X$ and $Y$ are
  independent. The vector $x \otimes y$ is also normalized, which is
  easy to verify algebraically.
\end{example}

The next proposition states some properties of the tensor product that
are useful in the rest of this \book{}.
\begin{proposition}
  \label{prop:tensor}
  Let $A, B : \C^{m \times m}, C, D \in \C^{n \times n}$ be linear
  transformations on $\C^m$ and $\C^n$ respectively, $u, v \in \C^m,
  w, x \in \C^n$, and $a, b \in \C$. The tensor product satisfies the
  following properties:
  \begin{enumerate}[(i)]
  \item $(A \otimes C)(B \otimes D) = AB \otimes CD$.
  \item $(A \otimes C)(u \otimes w) = Au \otimes Cw$.
  \item $(u + v)\otimes w = u\otimes w + v\otimes w$.
  \item $u\otimes (w + x) = u\otimes w + u\otimes x$.
  \item $(au) \otimes (bw) = ab (u \otimes w)$.
  \item $(A \otimes C)^\dag = A^\dag \otimes C^\dag$.
  \end{enumerate}
\end{proposition}
Above and in the following, the notation $A^\dag$ denotes the
conjugate transpose\index{notation!dagger@\ensuremath{\dag}} of $A$, which is the matrix defined as follows:
$A^\dag := \bar{A}^{\top}$ ($\bar{A}$ denotes the complex
conjugate). Given a matrix $A$, the notation $A^{\otimes n}$ indicates
the tensor product of $A$ with itself $n$ times, and the same notation
is used for vectors as well as vector spaces $\S$: $$A^{\otimes
  n} := \underbrace{A \otimes A \dots \otimes A}_{n \text{ times}},
\qquad \S^{\otimes n} := \underbrace{\S \otimes \S \dots \otimes
  \S}_{n \text{ times}}.$$\index{product!tensor|)}

The quantum computing literature refers to a Hilbert
space\index{Hilbert space}, typically denoted ${\cal H}$, rather than
a complex Euclidean space $\C^n$. However, the material discussed in
this \book{} does not require any property of Hilbert spaces that is
not already present in complex Euclidean spaces, hence we stick to the
more familiar concept. We reserve the symbol $\ham$ to refer to
Hamiltonians, introduced in Ch.~\ref{ch:hamsim}.

We work extensively with binary strings, using the following
definitions.
\begin{definition}[Binary string]
  \label{def:binaryrep}
  For any integer $q > 0$, we denote by $\vj \in \{0,1\}^q$\index{notation!binarystring@\ensuremath{\vx}} a binary
  string on $q$ digits, where we use the arrow to emphasize that $\vj$
  is a string of binary digits rather than an integer. Given $\vj \in \{0,1\}^q$, we denote its $k$-th digit ($k=1,\dots,q$) by
  $\vj_k$. We use the
  corresponding symbol without the arrow, $j$, to denote the decimal
  number that $\vj$\index{binary!string} corresponds to, i.e., $j = \sum_{k=1}^q \vj_k
  2^{q-k}$. 
\end{definition}
We use the symbol $\v{0}$ to denote the all-zero binary string, and
$\v{1}$\index{notation!allzerostring@\ensuremath{\v{0},\v{1}}} to denote the all-one binary string; the size of these strings
should always be clear from the context. Note that according to
Def.~\ref{def:binaryrep}, we use a little-endian convention for binary
strings, i.e., the first digit is the most significant one (e.g.,
$110$ is the integer $1\cdot 2^2 + 1 \cdot 2^1 + 0 \cdot 2^0 = 6$). We
write $\allzeroes$, $\allones$\index{notation!allzerovec@\ensuremath{\allzeroes,\allones}} to denote the all-zero and all-one
vectors respectively (to distinguish them from the all-zero and
all-one binary strings $\v{0}, \v{1}$), with dimension that is
clear from the context.

An additional piece of notation that we use extensively is the {\em
  bra-ket} notation, used in quantum mechanics. As mentioned earlier,
this \book{} does not rely or touch on quantum physics, however
there is an undeniable advantage in the quantum mechanics notation in that it
puts the most important information in the center of the symbols,
rather than relegate it to a marginal role in the subscript or
superscript. Furthermore, a goal of this \book{} is to equip readers
with the necessary tools to understand quantum computing papers, hence
it is important to familiarize with the bra-ket notation.
\begin{definition}[Bra-ket]
  \label{def:braket}
  Given a complex Euclidean space $\S$ (such as $\S = \C^n$),
  $\ket{\psi} \in \S$ denotes a column vector, and $\bra{\psi} \in
  \S^\dag$\index{notation!braket@\ensuremath{\bra{\psi},\ket{\psi}}}
  denotes a row vector that is the conjugate transpose of
  $\ket{\psi}$, i.e., $\bra{\psi} := \ket{\psi}^\dag$. The vector
  $\ket{\psi}$ is also called a {\em ket}, and the vector $\bra{\psi}$
  is also called a {\em bra}.
\end{definition}
Thus, an expression such as $\braket{\psi}{\phi}$ is an inner
product. (For vectors $x,y$ in the ``usual'', i.e., non-bra-ket notation, we
denote the inner product by $\dotp{x}{y}$\index{notation!dotprod@\ensuremath{\dotp{x}{y}}}.) To remember what is a bra
and what is a ket, it may be helpful to remember that a bra-ket is an
inner product. The complex Euclidean spaces used in this \book{} are
of the form $(\C^2)^{\otimes q}$, where $q$ is a given integer. It is
therefore convenient to give a shorthand notation for the basis
elements of such spaces.
\begin{definition}[Standard basis in bra-ket notation]
  \label{def:standardbasis}
  The standard basis for $\C^2$ is denoted by $\ket{0}
  = \begin{pmatrix} 1 \\ 0 \end{pmatrix}, \ket{1} = \begin{pmatrix}
    0 \\ 1 \end{pmatrix}$\index{basis!standard}. The standard basis for $(\C^2)^{\otimes
    q}$, which has $2^q$ elements, is denoted by $\ket{\vj}, \vj \in
  \{0,1\}^q$.
\end{definition}
\begin{remark}
  \label{rem:implicittensor}
  By convention, given column vectors $\ket{\psi}, \ket{\phi}$, their
  juxtaposition indicates their tensor product, i.e.,
  \begin{equation*}
    \ket{\psi}\ket{\phi} = \ket{\psi} \otimes \ket{\phi}.
  \end{equation*}
  The same convention is used for row vectors. This convention is used
  to shorten expressions whenever it does not create ambiguity, and is
  used extensively in the quantum computing literature.
\end{remark}

According to our notation, for any $q$-digit binary string $\vj \in
\{0,1\}^q$, $\ket{\vj}$ is the $2^q$-dimensional basis vector in
$(\C^2)^{\otimes q}$ corresponding to the binary string $\vj$. Because
we always use the standard basis and the most natural order for its
vectors, it is easy to verify that for $\vj \in \{0,1\}^q$,
$\ket{\vj}$ is the basis vector with a 1 in position $j$ (using zero-based
indices, i.e., 0 corresponds to the first position), and 0
elsewhere.
\begin{example}
  $\ket{110}$ is the 8-dimensional basis vector $(0 \, 0 \, 0 \, 0 \,
  0 \, 0 \, 1 \, 0)^{\top}$, obtained as the tensor product $\ket{1}
  \otimes \ket{1} \otimes \ket{0}$, because the binary string $110$
  corresponds to the number $6$, and $\ket{110}$ has a 1 in position 6
  (if we start counting from 0).
\end{example}

\noindent Whenever useful for clarity, we use a subscript for bras and
kets to denote the dimension of the space that the vector belongs to,
e.g., we write
$\ket{\vj}_q$\index{notation!subscriptdim@\ensuremath{\ket{\cdot}_{q}}}
to emphasize that we are working in a $2^q$ dimensional space (or, in
other words, that the basis elements of the space are associated with
binary strings with $q$ digits). We typically omit the subscript if
the dimension of the space is evident from the context, and we omit it
more often in later parts of the \book{} where such details are less
of a concern, but for now it can be helpful to give rigorous
definitions of the quantities involved in each expression. We provide
a further example of this notation below.
\begin{example}
  Let us write the basis elements of $(\C^2)^{\otimes 2} = \C^2 \otimes \C^2$:
  \begin{align*}
    \ket{00}_2 &= \ket{00} = \ket{0}\ket{0} = \ket{0} \otimes \ket{0} = \begin{pmatrix} 1 \\ 0 \\ 0 \\ 0 \end{pmatrix} &
    \ket{01}_2 &= \ket{01} = \ket{0}\ket{1} = \ket{0} \otimes \ket{1} = \begin{pmatrix} 0 \\ 1 \\ 0 \\ 0 \end{pmatrix}\\
    \ket{10}_2 &= \ket{10} = \ket{1}\ket{0} = \ket{1} \otimes \ket{0} = \begin{pmatrix} 0 \\ 0 \\ 1 \\ 0 \end{pmatrix} &
    \ket{11}_2 &= \ket{11} = \ket{1}\ket{1} = \ket{1} \otimes \ket{1} = \begin{pmatrix} 0 \\ 0 \\ 0 \\ 1 \end{pmatrix}.
  \end{align*}
\end{example}

\noindent In the above example we used the subscript to denote the
dimension of the basis vectors, just to emphasize that $\ket{00}_2$
and $\ket{00}$ are exactly the same. In the remainder of this \book{},
we always write $\ket{01}$ rather than $\ket{01}_2$ because it is
clear that the basis element $\ket{01}$ has two digits and therefore
lives in the space $(\C^2)^{\otimes 2}$.

In the rest of this \book{}, as is frequent in the quantum computing
literature, we use $\vj \in \{0,1\}^q$ or, interchangeably, the
corresponding integer $j$ to index the elements of $2^q$-dimensional
vectors; such an index is well defined because $\{0,1\}^q$ has $2^q$
elements. Thus, whenever we are indexing vectors (or matrices) with
indices that correspond to basis states, we use zero-based indices, as
opposed to the usual one-based indices. For example: if $x \in \R^{2^q}$
we write $\sum_{\vj \in \{0,1\}^q} x_{j}$ to take the sum of its
elements, so the first element of $x$ is indexed by zero; at the same
time, if $x \in \R^q$ (where $q$ is not necessarily a power of 2), we
write $\sum_{j=1}^q x_j$ to take the sum of its elements, so the first
element of $x$ is indexed by one, in the usual manner. This should
always be clear from the context.

To improve clarity when dealing with vectors in $(\C^2)^{\otimes q}$
in bra-ket notation, we always denote basis vectors using spelled-out
binary strings or Roman letters, (e.g., $\ket{01}, \ket{\vj},
\ket{\v{h}}, \ket{\v{x}}, \ket{\v{y}}$ all denote basis vectors),
whereas we use Greek letters to denote vectors that may not be basis
vectors (e.g., $\ket{\psi}, \ket{\phi}$ all denote vectors that may
not be basis vectors). In the same spirit, single-digit binary numbers
are always denoted with Roman letters (e.g., if we need to label a
scalar that is 0 or 1, we denote it by $x$, $y$, or $z$).


We denote by $I_{n \times n}$\index{notation!identity@\ensuremath{I_{n \times n}}} the identity matrix of size $n \times n$. We
generally omit the subscript to refer to the $2 \times 2$ identity
matrix, but sometimes we also omit it if the size of $I$ is clear from
the context, for example if we are using an identity matrix to
``fill'' the unspecified part of an operator on a tensor product space
(e.g., if we are constructing an $n \times n$ operator $A$, and $B$ is
a $2 \times 2$ matrix, then $A = B \otimes I$ implies that the
identity is of size $n/2 \times n/2$.) The reader is not required to
remember these details: experience suggests that the size of the
identity matrix is clear from the context. We use
$\text{diag}(a_1,\dots,a_n)$\index{notation!diag@\ensuremath{\text{diag}(a_1,\dots,a_n)}}
to denote the $n \times n$ diagonal matrix with elements
$a_1,\dots,a_n$ on the diagonal. The
$\ell^p$-norm\index{notation!lpnorm@\ensuremath{\nrm{x}_p}}\index{norm!vector, $\ell^p$-} of a vector $x \in \C^n$ is denoted $\nrm{x}_p :=
\left(\sum_{j=1}^n \abs{x}^p\right)^{1/p}$; if $p$ is omitted then it
is assumed that $p=2$, i.e., the Euclidean norm. The logarithm is
always base $2$, i.e., $\log = \log_2$, unless otherwise specified.

Finally, when discussing efficiency of algorithms we use the
traditional computer science notation $\bigO{\cdot}$, as well as the
perhaps lesser-known $\bigOt{\cdot}$. These are defined below.
\begin{definition}[Big-$\mathcal{O}$ notation]
  \label{def:bigo}
  We write $f(x) = \bigO{g(x)}$\index{notation!bigO@big \ensuremath{\mathcal{O}},\ensuremath{\widetilde{\mathcal{O}}}} if there exist scalars $\ell, \alpha >
  0$, such that $f(x) \leq \alpha g(x) \; \forall x > \ell$.

  We write $f(x) = \bigOt{g(x)}$ if $f(x) = \bigO{g(x)
    \polylog{g(x)}}$, where $\polylog{\cdot}$ denotes a
  polylogarithmic function of the argument. When $\bigOt{\cdot}$ is
  used to express a property (such as the asymptotic running time) of
  an algorithm on a class of instances, we allow the $\polylog{}$ term
  to also depend (still polylogarithmically) on other instance
  parameters that are not explicitly noted in $g(x)$.
\end{definition}
The $\bigOt{\cdot}$ notation is convenient when one does not want to
get bogged down by details: at least from a theoretical standpoint,
polylogarithmic factors are for the most part uninfluential when
determining the asymptotic running time, and keeping track of the
exact expressions can be very cumbersome. Note that to be precise one
should indicate which instance parameters are suppressed by the
$\bigOt{\cdot}$ notation, but we choose not to do it here to avoid
additional notation. In the vast majority of cases, the reader can
rely on the references given in the sections adopting $\bigOt{\cdot}$
notation to track down more precise running time expressions.

\paragraph{Further remarks on our notation.}
\label{sec:remarksnotation}
In this \book{} we use several notational devices that are meant to
enhance clarity, but that are not usually employed in the quantum
computing literature. We list the most important ones here, with the
hope of facilitating the transition from this \book{} to the rest of
the literature.
\begin{itemize}
\item We occasionally use the subscript for bra-ket vectors to
  indicate the dimension of the space, e.g., $\ket{\psi}_q$ for
  $2^q$-dimensional vectors. Typically, the dimension of the space is
  defined elsewhere and/or can be understood from the context. Whenever
  subscripts for kets are used, it is normally to address
  registers. We use capital letter subscripts to address
  registers.
\item We always use the vector arrow, e.g., $\vj$, to indicate binary
  strings. Typically, binary strings are not distinguished from other
  mathematical symbols and are to be identified from the context.
\item We always use Roman letters for basis vectors and
  Greek letters for general, i.e., possibly not basis, vectors. This
  convention is relatively common in the literature, although it is
  adopted with varying degrees of consistency.
\item We use $\v{0}, \v{1}$ to denote the all-zero, all-one binary
  strings. In the literature, these are usually denoted by $0^q$,
  $1^q$ respectively for dimension $q$. (In our notation, the
  dimension is defined elsewhere or denoted by a subscript in the
  ket.)
\end{itemize}

\section{Qubits and quantum states}
\label{sec:qubits}
According to the computational model specified at the beginning of
this chapter, a quantum computing device has a state that is stored in
a quantum register. Qubits are the quantum counterpart of the bits
found in classical computers: a classical computer has registers that
are made up of bits, whereas a quantum computer has quantum registers
that are made up of qubits. It is convenient to assume, without loss
of generality, that there is a single quantum register: one can think
of multiple registers as being placed ``side-by-side'' to form a
single register, and sometimes it is appropriate to address
subregisters of the main single register, as we discuss throughout this
chapter. The state of the quantum register, and therefore of the
quantum computing device, is defined next.
\begin{postulate}
  \label{pos:state}
  The state\index{quantum!state}\index{state!pure} of a $q$-qubit quantum register is a unit vector in
  $\left(\C^2\right)^{\otimes q} = \underbrace{\C^2 \otimes \dots
    \otimes \C^2}_{q \text{ times}}$.
\end{postulate}
\begin{remark}
  A vector $\ket{\psi} \in \C^n$ is a unit vector if $\nrm{\ket{\psi}} =
  \sqrt{\braket{\psi}{\psi}} = 1$.
\end{remark}
\begin{remark}
  Choosing the standard basis for $\C^2$, the state of a single-qubit
  register ($q = 1$) can be represented as $\alpha \ket{0} + \beta
  \ket{1} = \alpha \begin{pmatrix} 1 \\ 0 \end{pmatrix} +
  \beta \begin{pmatrix} 0 \\ 1 \end{pmatrix} = \begin{pmatrix} \alpha
    \\ \beta \end{pmatrix}$ where $\alpha, \beta \in \C$ and
  $|\alpha|^2 + |\beta|^2 = 1$.
\end{remark}
\begin{remark}
  Given the standard basis for $\C^2$ and the definition of the tensor
  product (Def.~\ref{def:tensor}), a basis for
  $\left(\C^2\right)^{\otimes q}$ is given by the following $2^q$
  vectors:
  \begin{align*}
    \vert \underbrace{00 \cdots 00}_{q \text{ digits}} \rangle &=
    \underbrace{\ket{0} \otimes \dots \otimes \ket{0} \otimes
      \ket{0}}_{q \text{ times}} =
    \underbrace{\ket{0}  \dots \ket{0} \ket{0}}_{q \text{ times}} \\
    \vert \underbrace{00 \cdots 01}_{q \text{ digits}} \rangle &=
    \underbrace{\ket{0} \otimes \dots \otimes \ket{0} \otimes
      \ket{1}}_{q \text{ times}} =
    \underbrace{\ket{0} \dots \ket{0} \ket{1}}_{q \text{ times}} \\
    & \vdots \\
    \vert \underbrace{11 \cdots 11}_{q \text{ digits}} \rangle &=
    \underbrace{\ket{1} \otimes \dots \otimes \ket{1} \otimes
      \ket{1}}_{q \text{ times}} =
    \underbrace{\ket{1} \dots \ket{1} \ket{1}}_{q \text{ times}}.
  \end{align*}
  In more compact form, the vectors are denoted by $\ket{\vj}, \vj \in
  \{0,1\}^q$, see Def.~\ref{def:standardbasis}. The state of a
  $q$-qubit quantum register can then be represented as: $\ket{\psi} =
  \sum_{\vj \in \{0,1\}^q} \alpha_{j} \ket{\vj}$, with $\alpha_{j} \in
  \C$ and $\sum_{\vj \in \{0,1\}^q} |\alpha_{j}|^2 = 1$.
\end{remark}
The coefficients of a quantum state (i.e., the elements of the
corresponding vector) are often called \emph{amplitudes}\index{amplitude!definition}. For brevity,
we often write ``state of $q$-qubits'' or ``$q$-qubit state'' to refer
to the state of a $q$-qubit quantum register. This is common in the
literature, where the discussion of qubits is not necessarily limited
to the context of quantum registers. By properties of the tensor
product, as is discussed in the following, sometimes it is appropriate
to refer to the state of just some of the qubits of a quantum
computing device, rather than all of them, and this may still be a
well-defined concept; however, this is not always possible (unlike for
classical computers). We revisit this topic in
Sect.~\ref{sec:entanglement}.

It is important to remark that $\left(\C^2\right)^{\otimes q}$ is a
$2^q$-dimensional space. This is in sharp contrast with the state of
classical bits: given $q$ classical bits, their state is a binary
string in $\{0,1\}^q$, which is a $q$-dimensional space.
\begin{remark}
  Here, to think about the dimension of the space it may be helpful to
  think about how many ``numbers'' are necessary to specify the state
  (formally, the numbers would be the coefficients to express the
  vector in a basis). For a vector in $\left(\C^2\right)^{\otimes q}$
  we need to specify $2^q$ coefficients, whereas for a vector in
  $\{0,1\}^q$, $q$ coefficients suffice.
\end{remark}
Thus, the dimension of the state space of quantum registers grows {\em
  exponentially} in the number of qubits, whereas the dimension of the
state space of classical registers grows {\em linearly} in the number
of bits. Furthermore, to represent a quantum state we need complex
coefficients: the state of a $q$-qubit quantum register is described
by $2^q$ complex coefficients, which is an enormous amount of
information compared to what is necessary to describe a $q$-bit
classical register. However, a consequence of one of our main
postulates (introduced later) is that a quantum state cannot be
accessed directly, therefore even if a description of the quantum
state requires infinite precision in principle, we cannot access such
description as easily as with classical registers. In fact, as it
turns out we cannot extract more than $q$ bits of information out of a
$q$-qubit register! This should be intuitively clear after stating the
effect of quantum measurements in Sect.~\ref{sec:measurement}; for a
formal proof, see \cite{holevo1973bounds}.

\subsection{Basis states and superposition}
\label{sec:superposition}
We continue our study of the space of quantum states by discussing
the concept of superposition.
\begin{definition}[Superposition]
  \label{def:superposition}
  A $q$-qubit quantum state $\ket{\psi}$ is a {\em basis state}\index{state!basis} if
  $\ket{\psi} = \alpha_{k} \ket{\vk}$ for some $\vk \in \{0,1\}^q$ and
  $\alpha_k \in \C, |\alpha_{k}|^2 = 1$. Otherwise, we say that
  $\ket{\psi}$ is a {\em superposition}\index{quantum!superposition}\index{superposition} (of basis states). Similarly,
  we say that $q$ qubits are in a basis state if the state of the
  corresponding register is described by a basis state, and it is a
  superposition otherwise.
\end{definition}
\begin{remark}
  A simpler, more intuitive definition would be to say that a basis
  state is such that $\ket{\psi} = \ket{\vk}$ for some $\vk \in
  \{0,1\}^q$. It is acceptable to use the simpler definition if
  desired: as it turns out, even if the states $\alpha_{k} \ket{\vk}$
  for some $\vk \in \{0,1\}^q$ and $|\alpha_{k}|^2 = 1$ are all
  different in principle, they are equivalent to $\ket{\vk}$ up to the
  multiplication factor $\alpha_{k}$, which is unimportant because
  $\abs{\alpha_k}^2 = 1$, as discussed in Sect.~\ref{sec:measurement}.
\end{remark}

\begin{example}
  Consider two single-qubit registers and their states $\ket{\psi},
  \ket{\phi}$:
  \begin{align*}
    \ket{\psi} &= \alpha_0 \ket{0} + \alpha_1 \ket{1} \\
    \ket{\phi} &= \beta_0 \ket{0} + \beta_1 \ket{1}.
  \end{align*}
  If we put these single-qubit registers side-by-side to form a two-qubit
  register, then the state of the two-qubit register is (recall
  Rem.~\ref{rem:implicittensor}):
  \begin{align*}
    \ket{\psi} \ket{\phi} = \alpha_0 \beta_0 \ket{0}
     \ket{0} + \alpha_0 \beta_1 \ket{0} \ket{1} +
    \alpha_1 \beta_0 \ket{1}  \ket{0} + \alpha_1 \beta_1
    \ket{1}  \ket{1}.
  \end{align*}
  If both $\ket{\psi}$ and $\ket{\phi}$ are basis states, we have
  that either $\alpha_0$ or $\alpha_1$ is zero, and similarly either
  $\beta_0$ or $\beta_1$ is zero, while the nonzero coefficients have
  modulus one. Thus, only one of the coefficients in the expression of
  the state of $\ket{\psi}  \ket{\phi}$ is nonzero, and in fact
  its modulus is one. This implies that if both $\ket{\psi}$ and
  $\ket{\phi}$ are basis states, $\ket{\psi}  \ket{\phi}$
  is a basis state as well. But now assume that $\alpha_0 = \beta_0
  = \alpha_1 = \beta_1 = \frac{1}{\sqrt{2}}$: the qubits $\ket{\psi}$
  and $\ket{\phi}$ are in a superposition. Then the state of
  $\ket{\psi}  \ket{\phi}$ is $\frac{1}{2} \ket{00} +
  \frac{1}{2} \ket{01} + \frac{1}{2} \ket{10} + \frac{1}{2} \ket{11}$,
  which is a superposition as well. Notice that the normalization of
  the coefficients works out, as one can easily check with simple
  algebra: the tensor product of unit vectors is also a unit vector.
\end{example}

The example clearly generalizes to an arbitrary number of
qubits. In fact the following proposition is trivially true:
\begin{proposition}
  A $q$-qubit register, $q > 1$, is in a basis state if and only if
  its state can be expressed as the tensor product of $q$ single-qubit
  registers, each of which is in a basis state.
\end{proposition}
Notice that superposition does not have a classical equivalent: $q$
classical bits are always in a basis state, i.e., a $q$-bit classical
register always contains exactly one of the $2^q$ binary strings
in $\{0,1\}^q$. Indeed, superposition is one of the main features that
differentiate quantum computers from classical computers. Another
important feature is entanglement, discussed next.

\subsection{Product states and entanglement}
\label{sec:entanglement}
We have seen that the state of a $q$-qubit register is a vector in
$\left(\C^2\right)^{\otimes q}$, which is a $2^q$ dimensional
space. Because this is a tensor product of copies of $\C^2$, i.e., the
space in which single-qubit states live, it is natural to ask whether
moving from single qubits to multiple qubits gained us anything beyond
increasing the dimension of the space: are the quantum states
representable on $q$ qubits simply the tensor product of $q$
single-qubit states, or are they a larger set? We can answer this
question by using the definitions and concepts introduced so far. A
$q$-qubit state is a unit vector in $\left(\C^2\right)^{\otimes q}$,
and it can be written as:
\begin{equation*}
  \ket{\psi} = \sum_{\vj \in \{0,1\}^q} \alpha_{j} \ket{\vj}, \qquad
  \sum_{\vj \in \{0,1\}^q} |\alpha_{j}|^2 = 1.
\end{equation*}
Now let us consider the tensor product of $q$ single-qubit states, the
$k$-th of which is given by $\beta_{k,0} \ket{0} + \beta_{k,1}
\ket{1}$, for $k=1,\dots,q$ (the first qubit corresponds to the most
significant bit, according to the little-endian convention). Taking
the tensor product we obtain the vector:
\begin{align*}
  \ket{\phi} &= (\beta_{1,0} \ket{0} + \beta_{1,1} \ket{1}) \otimes
  (\beta_{2,0} \ket{0} + \beta_{2,1} \ket{1}) \otimes \dots \otimes
  (\beta_{q,0} \ket{0} + \beta_{q,1} \ket{1}) \\
  &= \sum_{j_1=0}^{1} \sum_{j_{2}=0}^{1} \cdots
  \sum_{j_q=0}^1 \left(\prod_{k=1}^{q} \beta_{k,j_k}\right) | \underbrace{j_1 j_2
    \dots j_q}_{\substack{\text{taken as a} \\ \text{binary string}}}
  \rangle = \sum_{\vj \in \{0,1\}^q} \left(\prod_{k=1}^{q} \beta_{k,
    \vj_k}\right) \ket{\vj}, \\
  & \text{satisfying } |\beta_{k,0}|^2 + |\beta_{k,1}|^2 = 1
  \;\; \forall k=1,\dots,q.
\end{align*}
It is straightforward to verify (for example, with a proof by induction
on the number of qubits $q$) that the normalization condition for the
coefficients in $\ket{\phi}$ implies the normalization condition for
the coefficients in $\ket{\psi}$, but the converse is not true. That
is, $|\beta_{k,0}|^2 + |\beta_{k,1}|^2 = 1 \; \forall k=1,\dots,q$
implies $\sum_{j_1=0}^{1} \sum_{j_2=0}^{1} \cdots \sum_{j_q=0}^1
\left|\prod_{k=1}^{q} \beta_{k,j_k}\right|^2 = 1$, but not
viceversa. This means that there exist values of the coefficients
$\alpha_{j}$ of $\ket{\psi}$, with $\sum_{\vj \in \{0,1\}^q}
|\alpha_{j}|^2 = 1$, that cannot be obtained as the coefficients of a
state $\ket{\phi}$ by choosing appropriate coefficients $\beta_{k,0},
\beta_{k,1}$ (for $k=1,\dots,q$). This is easily verified with an
example.
\begin{example}
  \label{ex:productstate}
  Consider two single-qubit states:
  \begin{align*}
    \ket{\psi} &= \alpha_0 \ket{0} + \alpha_1 \ket{1} \\
    \ket{\phi} &= \beta_0 \ket{0} + \beta_1 \ket{1}.
  \end{align*}
  Taking the two qubits together in a two-qubit register, the state of
  the two-qubit register is:
  \begin{equation}
    \begin{split}
      \ket{\psi} \ket{\phi} = \alpha_0 \beta_0 \ket{00} + \alpha_0
      \beta_1 \ket{01} + \alpha_1 \beta_0 \ket{10} + \alpha_1 \beta_1
      \ket{11},
    \end{split} \label{eq:productstate}
  \end{equation}
  with the normalization conditions $|\alpha_0|^2 + |\alpha_1|^2 = 1$
  and $|\beta_0|^2 + |\beta_1|^2 = 1$.  The general state of a two-qubit
  register $\ket{\xi}$ is:
  \begin{equation}
  \label{eq:generalstate}
  \ket{\xi} = \gamma_{00} \ket{00} + \gamma_{01} \ket{01} +
  \gamma_{10} \ket{10} + \gamma_{11} \ket{11},
  \end{equation}
  with normalization condition $|\gamma_{00}|^2 + |\gamma_{01}|^2 +
  |\gamma_{10}|^2 + |\gamma_{11}|^2 = 1$. Comparing
  Eq.s~\eqref{eq:productstate} and \eqref{eq:generalstate}, we
  determine that $\ket{\xi}$ is of the form $\ket{\psi} \otimes
  \ket{\phi}$ (i.e., a tensor product of two single-qubit states) if
  and only if it satisfies the relationship:
  \begin{equation}
    \label{eq:gammarel}
    \gamma_{00} \gamma_{11} = \gamma_{01} \gamma_{10}.
  \end{equation}
  Clearly $\ket{\psi} \ket{\phi}$ yields coefficients that satisfy
  this condition. To see the converse, let $\theta_{00}, \theta_{01},
  \theta_{10}, \theta_{11}$ be the phases of $\gamma_{00}, \gamma_{01},
  \gamma_{10}, \gamma_{11}$. Notice that Eq.~\eqref{eq:gammarel} implies:
  \begin{align*}
    |\gamma_{00}|^2 |\gamma_{11}|^2 &= |\gamma_{01}|^2 |\gamma_{10}|^2 \\
    \theta_{00} + \theta_{11} &= \theta_{01} + \theta_{10}.
  \end{align*}
  Using these relationships, we can determine an explicit expression
  for $\alpha_0, \alpha_1, \beta_0, \beta_1$ based on $\gamma_{00},
  \gamma_{01}, \gamma_{10}, \allowbreak \gamma_{11}$; i.e., we show
  that if Eq.~\eqref{eq:gammarel} holds, then $\ket{\xi}$ can be written
  as $\ket{\psi} \otimes \ket{\phi}$. We first define the modulus of
  the coefficients in $\ket{\psi}$, $\ket{\phi}$. We have:
  \begin{align*}
    |\gamma_{00}| &= \sqrt{|\gamma_{00}|^2} = 
    \sqrt{|\gamma_{00}|^2 (|\gamma_{00}|^2 + |\gamma_{01}|^2 + 
      |\gamma_{10}|^2 + |\gamma_{11}|^2)} \\
    &= \sqrt{|\gamma_{00}|^4 + |\gamma_{00}|^2 |\gamma_{01}|^2 + |\gamma_{00}|^2 |\gamma_{10}|^2 + |\gamma_{01}|^2 |\gamma_{10}|^2} \\
    &= \underbrace{\sqrt{|\gamma_{00}|^2 + |\gamma_{01}|^2}}_{|\alpha_0|}\underbrace{\sqrt{|\gamma_{00}|^2 + |\gamma_{10}|^2}}_{|\beta_0|},
  \end{align*}
  and similarly for the other coefficients, we obtain:
  \begin{align*}
    |\gamma_{01}| &= \underbrace{\sqrt{|\gamma_{00}|^2 + |\gamma_{01}|^2}}_{|\alpha_0|}\underbrace{\sqrt{|\gamma_{01}|^2 + |\gamma_{11}|^2}}_{|\beta_1|} \\
    |\gamma_{10}| &= \underbrace{\sqrt{|\gamma_{10}|^2 + |\gamma_{11}|^2}}_{|\alpha_1|}\underbrace{\sqrt{|\gamma_{00}|^2 + |\gamma_{10}|^2}}_{|\beta_0|} \\
    |\gamma_{11}| &= \underbrace{\sqrt{|\gamma_{10}|^2 + |\gamma_{11}|^2}}_{|\alpha_1|}\underbrace{\sqrt{|\gamma_{01}|^2 + |\gamma_{11}|^2}}_{|\beta_1|}.
  \end{align*}
  To fully define the coefficients $\alpha_0, \alpha_1, \beta_0,
  \beta_1$ we must determine their phases. We can assign:
  \begin{align}
    \label{eq:coeffproduct}
    \begin{split}
      \alpha_0 = e^{i\theta_{00}}|\alpha_0|, \qquad \alpha_1 = e^{i\theta_{10}}|\alpha_1|, 
      \qquad
      \beta_0 = |\beta_0|, \qquad \beta_1 = e^{i(\theta_{01}-\theta_{00})}|\beta_1|.
    \end{split}
  \end{align}
  Using the fact that $\theta_{11} = \theta_{01} + \theta_{10} -
  \theta_{00}$, it is now easy to verify that the state $\ket{\xi}$ in
  Eq.~\eqref{eq:generalstate} can be expressed as $\ket{\psi} \otimes
  \ket{\phi}$ in Eq.~\eqref{eq:productstate} with coefficients $\alpha_0,
  \alpha_1, \beta_0, \beta_1$ as given in Eq.~\eqref{eq:coeffproduct}.

  The condition in Eq.~\eqref{eq:gammarel}, to verify if a
  two-qubit state $\ket{\xi}$ can be expressed as a tensor product of
  two single-qubit states, can also be written in matrix form, which
  makes it easier to remember. If we assign the rows of the matrix to
  the first qubit, and the columns to the second qubit, we can arrange
  the coefficients $\gamma$ as follows (notice how the first qubit has
  value 0 in the first row and 1 in the second row; similarly for the
  second qubit and the columns):
  \begin{equation*}
    \begin{pmatrix}
      \gamma_{00} & \gamma_{01} \\
      \gamma_{10} & \gamma_{11}
    \end{pmatrix}.
  \end{equation*}
  Then, $\ket{\xi}$ is a tensor product of two single-qubit states if and
  only if this matrix has rank 1. This is equivalent to
  Eq.~\eqref{eq:gammarel}.
\end{example}

We formalize the concept of expressing a quantum state as a tensor
product of lower-dimensional quantum states as follows.
\begin{definition}[Entangled state]
  \label{def:entanglement}
  A quantum state $\ket{\psi} \in \left(\C^2\right)^{\otimes q}$ is a
  {\em product state}\index{state!product} if it can be expressed as a tensor product
  $\ket{\psi_1} \dots \ket{\psi_q}$ of $q$ single-qubit states. Otherwise,
  it is {\em entangled}\index{state!entangled}\index{entanglement}.
\end{definition}
Notice that a general quantum state $\ket{\psi}$ could be the product
of two or more lower-dimensional quantum states, e.g., $\ket{\psi} =
\ket{\psi_1} \otimes \ket{\psi_2}$, with $\ket{\psi_1}$ and
$\ket{\psi_2}$ being entangled states. In such a situation,
$\ket{\psi}$ exhibits some entanglement, but in some sense it can
still be ``simplified''. Generally, according to the definition above,
we call a quantum state entangled as long as it cannot be fully
decomposed into a tensor product of single-qubit states. In the case of
quantum systems composed of multiple subsystems (rather than just two
subsystems as in the example $\ket{\psi} = \ket{\psi_1} \otimes
\ket{\psi_2}$), the concept of entanglement as discussed in the
literature is not as simple as given in Def.~\ref{def:entanglement}
(and the rank-1 test discussed at the end of Example
\ref{ex:productstate} is not well-defined). However, our simplified
definition works in this \book{} and for most of the literature on
quantum algorithms, therefore we can leave other considerations aside;
we refer to \cite{coffman2000distributed} as an entry point for a
discussion on multipartite entanglement.

\begin{example}
  \label{ex:productentangled}
  Consider the following two-qubit state:
  \begin{equation*}
    \frac{1}{2} \ket{00} + \frac{1}{2} \ket{01} + \frac{1}{2} \ket{10} +
    \frac{1}{2} \ket{11}.
  \end{equation*}
  This is a product state because it is equal to
  $\left(\frac{1}{\sqrt{2}} \ket{0} + \frac{1}{\sqrt{2}}
  \ket{1}\right)\otimes \left(\frac{1}{\sqrt{2}} \ket{0} +
  \frac{1}{\sqrt{2}} \ket{1}\right)$. By contrast, the following
  two-qubit state:
  \begin{equation*}
    \frac{1}{\sqrt{2}} \ket{00} + \frac{1}{\sqrt{2}} \ket{11}
  \end{equation*}
  is an entangled state, because it cannot be expressed as a product
  of two single-qubit states.
\end{example}

\section{Operations on qubits}
\label{sec:operations}
Operations on quantum states must satisfy certain conditions, ensuring
that applying an operation does not break the basic properties of a
quantum state. The required property of operations on quantum states
is stated below, and we treat it as a postulate.
\begin{postulate}
  \label{pos:operations}
  An {\em operation}\index{quantum!gate}\index{quantum!operation}\index{gate!definition} applied by a quantum computer with $q$ qubits,
  also called a {\em gate}, is a unitary matrix in $\C^{2^q \times
    2^q}$.
\end{postulate}
\begin{remark}
  A matrix $U$ is unitary if $U^\dag U = U U^\dag = I$.\index{unitary matrix!definition}
\end{remark}
\begin{remark}
  The term ``unitary,'' used as a noun, is a short form for ``unitary
  matrix.'' This usage is common in the quantum computing literature.
\end{remark}
A well-known property of unitary matrices is that they are
norm-preserving; that is, given a unitary matrix $U$ and a vector $v$,
$\nrm{U v} = \nrm{v}$. Thus, for a $q$-qubit system, the quantum state is
a unit vector $\ket{\psi} \in \C^{2^q}$, a quantum operation is a
matrix $U \in \C^{2^q \times 2^q}$, and the application of $U$ onto
the state $\ket{\psi}$ is the unit vector $U \ket{\psi} \in
\C^{2^q}$. This leads to the following remarks:
\begin{itemize}
\item Quantum operations are {\em linear}.
\item Quantum operations are {\em reversible}.
\end{itemize}
While these properties may initially seem to be extremely restrictive,
it is possible to show that a universal quantum computer is
Turing-complete, implying that it can simulate any Turing-computable
function, given sufficient time; see the notes in
Sect.~\ref{sec:prelimnotes} for several references on a universal
quantum Turing machine\index{Turing machine!quantum}. Out of the two properties indicated above, the
most counterintuitive is perhaps reversibility: the classical notion
of computation does not appear to be reversible, because memory can be
erased and, in the classical Turing machine, symbols can be erased
from the tape. However, \cite{bennett73logical} shows that all
computations (including classical computations) can be made reversible
by means of extra space\index{reversibility}. The general idea to make a function
invertible is to have separate input and output registers: any output
is stored in a different location than the input, so that the input
does not have to be erased. This is a standard trick in quantum
computing that is discussed in Sect.~\ref{sec:uncompute}, but in order
to present it in detail, we first need to introduce some notation for
quantum circuits.

\subsection{Notation for quantum circuits}
\label{sec:circuitnotation}
\begin{figure}[h!]
\leavevmode
\centering
\ifcompilefigs
\Qcircuit @C=1em @R=.7em {
\lstick{\text{qubit } 1} & \multigate{2}{U}  & \qw \\
\lstick{\text{qubit } 2} & \ghost{U}         & \qw \\
\lstick{\text{qubit } 3} & \ghost{U}         & \qw \\
}
\else
\includegraphics{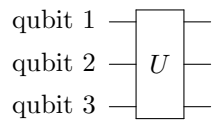}
\fi
\caption{A simple quantum circuit.}
\label{fig:basiccircuit}
\end{figure}

The sequence of operations applied onto the qubits is usually
represented by means of a \emph{quantum circuit}\index{circuit diagram|(}\index{quantum!circuit}. A quantum circuit
indicates which operations are performed on each qubit, or group of
qubits. For a quantum computer with $q$ qubits, we represent $q$ qubit
lines, where the top line indicates qubit $1$ and the rest are given
in increasing order from the top. Operations are represented as gates;
we use the terms ``operation'' and ``gate'' interchangeably. Gates
take qubit lines as input, have the same number of qubit lines as
output, and apply the unitary matrix indicated on the gate to the
quantum state of those qubits. To think about the mathematical
representation coresponding to a circuit, one should imagine that
there is a tensor product symbol between the qubit lines.
Fig.~\ref{fig:basiccircuit} is a simple example, with a gate acting on
all qubits in the circuit.

\begin{figure}[h!]
\leavevmode
\centering
\ifcompilefigs
\Qcircuit @C=1em @R=.7em {
                    & \multigate{2}{A}  & \multigate{2}{B}  & \qw & \\
\lstick{\ket{\psi}} & \ghost{A}         & \ghost{B}         & \qw & \rstick{BA\ket{\psi}}\\
                    & \ghost{A}         & \ghost{B}         & \qw & \\
}
\else
\includegraphics{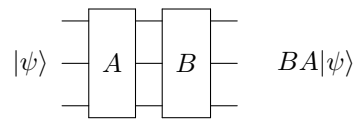}
\fi
\caption{Order of the operations in a quantum circuit.}
\label{fig:circuitorder}
\end{figure}

\noindent Note that circuit diagrams are read from left to right, but
because each gate corresponds to applying a matrix to the quantum
state, the matrices corresponding to the gates should be written from
right to left in the mathematical expression describing the action of
the circuit. For example, in the circuit in
Fig.~\ref{fig:circuitorder},
the outcome of the circuit is the state $BA \ket{\psi}$,
because we start with state $\ket{\psi}$, and we first apply the gate
with unitary matrix $A$, and then $B$.

\begin{figure}[h!]
\leavevmode
\centering
\ifcompilefigs
\Qcircuit @C=1em @R=.5em @!R {
\lstick{\text{qubit } 1} & \qw       & \qw \\
\lstick{\text{qubit } 2} & \qw       & \qw \\
\lstick{\text{qubit } 3} & \gate{U}  & \qw \\
}
\else
\includegraphics{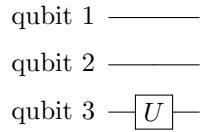}
\fi
\caption{A circuit with a single-qubit gate.}
\label{fig:singlequbit}
\end{figure}

Gates can also be applied to individual qubits. Because a
single qubit is a vector in $\C^2$, a single-qubit gate is a unitary
matrix in $\C^{2 \times 2}$. Consider the same three-qubit device, and
suppose we want to apply the gate only to the third qubit. We would
write it as in Fig.~\ref{fig:singlequbit}.

\begin{figure}[h!]
\leavevmode
\centering
\ifcompilefigs
\Qcircuit @C=1em @R=.7em {
\lstick{\text{qubit } 1} & \gate{I} & \qw \\
\lstick{\text{qubit } 2} & \gate{I}  & \qw \\
\lstick{\text{qubit } 3} & \gate{U}  & \qw \\
}
\hspace{10em}
\Qcircuit @C=1em @R=.7em {
\lstick{\text{qubit } 1} & \multigate{2}{I \otimes I \otimes U} & \qw \\
\lstick{\text{qubit } 2} & \ghost{I \otimes I \otimes U}  & \qw \\
\lstick{\text{qubit } 3} & \ghost{I \otimes I \otimes U}  & \qw \\
}
\else
\includegraphics{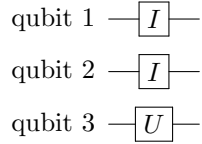}
\fi
\caption{Equivalent representation of the circuit with a single-qubit gate in Fig.~\ref{fig:singlequbit}.}
\label{fig:singlequbitI}
\end{figure}

\noindent From an algebraic point of view, the action of our first
example in Fig.~\ref{fig:basiccircuit} on the quantum state is clear:
the state of the three qubits is mapped onto another three-qubit
state, as $U$ acts on all the qubits. To give a proper mathematical
characterization of the example in Fig.~\ref{fig:singlequbit}, where
$U$ is a single-qubit gate that acts on qubit 3 only, we have to
imagine that an identity gate is applied to all the empty qubit
lines. Therefore, Fig.~\ref{fig:singlequbit} can be thought of as
indicated in Fig.~\ref{fig:singlequbitI}.

\begin{figure}[h!]
\leavevmode
\centering
\ifcompilefigs
\Qcircuit @C=1em @R=.3em @!R {
\lstick{\ket{\psi}} & \qw       & \qw & \rstick{\ket{\psi}} \\
\lstick{\ket{\phi}} & \qw       & \qw & \rstick{\ket{\phi}} \\
\lstick{\ket{\xi}} & \gate{U}  & \qw & \rstick{U\ket{\xi}} \\
}
\else
\includegraphics{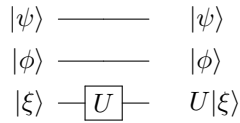}
\fi
\caption{Effect of a single-qubit gate on a product state.}
\label{fig:circuitproduct}
\end{figure}

\noindent This circuit can be interpreted as applying the gate $I
\otimes I \otimes U$ to the three-qubit state. Notice that by
convention the matrix $U$, which is applied to qubit 3, appears in the
rightmost term of the tensor product. This is because qubit 3 is
associated with the least significant digit according to our
little-endian convention, see Def.~\ref{def:binaryrep} and the
subsequent discussion. If we have a product state $\ket{\psi} \otimes
\ket{\phi} \otimes \ket{\xi}$, the effect of $I \otimes I \otimes U$
is as indicated in Fig.~\ref{fig:circuitproduct}.

\begin{figure}[h!]
\leavevmode
\centering
\ifcompilefigs
\Qcircuit @C=1em @R=.3em @!R {
 & \qw  & \qw  \\
 & \qw  & \qw & \rstick{(I \otimes I \otimes U)\ket{\psi}} \\
 & \gate{U}  & \qw  
  \inputgroupv{1}{3}{.8em}{1.5em}{\ket{\psi}}
  {\gategroup{1}{3}{3}{3}{.8em}{\}}} \\
}
\else
\includegraphics{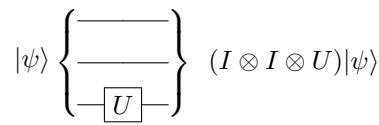}
\fi
\caption{Effect of a single-qubit gate on an entangled state.}
\label{fig:circuitentangled}
\end{figure}

\noindent Indeed, $(I \otimes I \otimes U)(\ket{\psi} \otimes
\ket{\phi} \otimes \ket{\xi}) = \ket{\psi} \otimes \ket{\phi} \otimes
U\ket{\xi}$. If the system is in an entangled state, however, the
action of $(I \otimes I \otimes U)$ cannot be determined in such a
simple way, because the state cannot be factored as a product
state. Thus, for a general entangled input state, the effect of the
circuit is as indicated in Fig.~\ref{fig:circuitentangled}\index{entanglement}.
Notice that this fact is essentially the reason why simulation of
quantum computations on classical computers may take exponential
resources in the worst case: to simulate the effect of even a
single-qubit gate on the entangled state $\ket{\psi}$, we have to
explicitly compute the effect of the $2^q \times 2^q$ matrix $(I
\otimes I \otimes U)$ on the state $\ket{\psi}$. This requires
exponential space with a naive approach (if the matrices and vectors
are stored explicitly), and even with more parsimonious approaches it
may require exponential time (e.g., if we compute elements of the
state vector one at a time). As long as the quantum state is not
entangled computations can be carried out on each qubit independently,
but entanglement requires us to keep track of the full quantum state
in $2^q$-dimensional complex space, leading to large amounts of memory
--- or time --- required.

\subsection{Input-output, and measurement gates}
\label{sec:measurement}
We are almost ready to introduce the last postulate that we need to
formally define the model of computation. To do so, it is useful to
discuss the input-output model for quantum computations. The {\em
  input} of a quantum computation consists of an initial quantum
state, and the description of a quantum circuit.
\begin{remark}
  The quantum state and the quantum circuit must be described in a
  suitable compact way: for a circuit on $q$ qubits, a unitary matrix
  can be of size $2^q \times 2^q$, but for an efficient algorithm we
  require that the circuit contains polynomially-many gates in $q$ and
  each gate has a compact representation. This is discussed further in
  the rest of this chapter.
\end{remark}
By convention, the initial quantum state of the quantum computing
device is assumed to be the all-zero binary string $\ket{\v{0}}$ of
appropriate size (i.e., $\ket{\v{0}}_q$ if we have $q$ qubits in
total), unless otherwise specified. Of course, the circuit can act
on the state and transform it into a more suitable one. Examples of
how this can be done are given in the remainder of this section.

A quantum algorithm consists in the execution of one or more quantum
computations. There are also hybrid algorithms involving classical and
quantum computations. In such situations, the quantum computations can
generally be thought of as subroutines, but this does not change the
principle that each of these quantum computations is described by
an initial quantum state (typically, $\ket{\v{0}}$) and a quantum
circuit.  An important thing to note is that if there is any data that
has to be fed to the algorithm, this data has to be embedded in the
quantum circuit given as part of the input (which may, sometime, have
a significant impact on the number of gates that are necessary to
describe the circuit). This summarizes the input model. But what is
the {\em output} of the quantum computer?

\begin{figure}[h!]
\leavevmode
\centering
\ifcompilefigs
\Qcircuit @C=1em @R=1em {
                    & \qw  & \\
\lstick{\ket{\psi}} & \meter & \\
                    & \qw  & \\
}
\else
\includegraphics{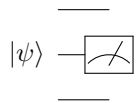}
\fi
\caption{Single-qubit measurement.}
\label{fig:meas}
\end{figure}

So far we characterized properties of quantum states and quantum
gates. Remarkably, the state of a $q$-qubit quantum register is
described by a vector of dimension $2^q$, exponentially larger than
the dimension of the vector required to describe $q$ classical
bits. However, there is a catch: in a classical computer we can simply
read the state of the bits, whereas in a quantum computer we do not
have direct, unrestricted access to the quantum state. Information on
the quantum state is only gathered through a \emph{measurement gate},
depicted in the circuit diagram in Fig.~\ref{fig:meas}.\index{circuit diagram|)} We now
formally define the effect of a single-bit measurement gate.
\begin{postulate}
  \label{pos:meas}
  Information on the state of a quantum computing device can only be
  obtained through a {\em measurement}\index{measurement!definition}.  Given a $q$-qubit quantum
  state $\ket{\psi} = \sum_{\vj \in \{0,1\}^q} \alpha_{j} \ket{\vj}$,
  a {\em measurement gate} on qubit $k$ outputs a sample from a random
  variable $\mathcal{Q}_k$ with sample space $\{0,1\}$ and the
  following distribution:
  \begin{align*}
    \Pr(\mathcal{Q}_k = 0) = \sum_{\vj \in \{0,1\}^q: \vj_k =
      0} |\alpha_{j}|^2, \\
    \Pr(\mathcal{Q}_k = 1) = \sum_{\vj \in
    \{0,1\}^q: \vj_k = 1} |\alpha_{j}|^2.    
  \end{align*}
  Let $x \in \{0,1\}$ be the observed value. After the measurement,
  the quantum state becomes: $$\sum_{\substack{\vj \in
      \{0,1\}^q:\\ \vj_k = x}} \frac{\alpha_{j}}{\sqrt{\sum_{\v{\ell} :
        \v{\ell}_k = x} |\alpha_{\ell}|^2}} \ket{\vj}.$$ The original quantum
  state is no longer recoverable.
\end{postulate}
\begin{remark}
  The state of the quantum computer after a measurement collapses to a
  linear combination of only those basis states that are consistent
  with the outcome of the measurement, i.e., basis states $\ket{\vj}$
  with $\vj_k = x$. The coefficients $\alpha_{j}$ for such basis
  states are normalized to yield a unit vector.
\end{remark}

The rule for single-qubit measurements leads to a very simple and
natural expression for the probability of observing a given binary
string when measuring all the qubits.
\begin{proposition}
  \label{prop:meas}
  Given a $q$-qubit quantum state $\ket{\psi} = \sum_{\vj \in \{0,1\}^q}
  \alpha_{j} \ket{\vj}$, applying a measurement gate to the $q$ qubits
  in any order yields $\vj$ with probability $|\alpha_{j}|^2$, for $\vj
  \in \{0,1\}^q$.
\end{proposition}
\begin{proof}
  We need to show that the probability of observing $\vj$ after $q$
  single-qubit measurements is equal to $|\alpha_{j}|^2$. We can do
  this by induction on $q$. The case $q=1$ is trivial. We now show how
  to go from $q-1$ to $q$. As in Post.~\ref{pos:meas}, we write
  $\Pr(\mathcal{Q}_k = x)$ to denote the probability that the
  measurement of qubit $k$ yields $x \in \{0,1\}$. If it is important
  to indicate the quantum state on which the measurement is performed,
  we use the notation $\displaystyle \Pr_{\ket{\psi}}(\mathcal{Q}_k =
  x)$ to indicate that the measurement is applied to state
  $\ket{\psi}$.
  
  Suppose we apply a measurement to all qubits in an abitrary order,
  and the qubit in position $h$ is the first to be measured. (The
  order of the remaining measurements does not matter for the proof,
  because after the first measurement we rely on the inductive
  hypothesis). The probability of obtaining the outcome $\vj$ is:
  \begin{align*}
    \Pr_{\ket{\psi}}\left(\mathcal{Q}_1 = \vj_{1}, \dots, \mathcal{Q}_q =
    \vj_{q} \right) = \\ \Pr_{\ket{\psi}}\left(\mathcal{Q}_1 =
    \vj_{1}, \dots, \mathcal{Q}_{h-1} = \vj_{h-1}, \mathcal{Q}_{h+1} = \vj_{h+1}, \dots,
    \mathcal{Q}_q = \vj_{q} | \mathcal{Q}_h = \vj_{h}\right) \Pr_{\ket{\psi}}
    \left(\mathcal{Q}_h = \vj_{h}\right) = \\
    \Pr_{\ket{\phi}}\left(\mathcal{Q}_1 =
    \vj_{1}, \dots, \mathcal{Q}_{h-1} = \vj_{h-1}, \mathcal{Q}_{h+1} = \vj_{h+1}, \dots,
    \mathcal{Q}_q = \vj_{q} \right) \Pr_{\ket{\psi}}
    \left(\mathcal{Q}_h = \vj_{h}\right),
  \end{align*}
  where $\ket{\phi}$ is the state obtained from $\ket{\psi}$ after
  measuring the qubit in position $h$ and observing
  $\vj_h$. By Post.~\ref{pos:meas}, we have:
  \begin{equation*}
    \ket{\phi} = \sum_{\substack{\vk \in \{0,1\}^q :\\ \vk_{h} = \vj_{h}}}
    \frac{\alpha_{k}}{\sqrt{\sum_{\v{\ell} \in \{0,1\}^q : \v{\ell}_{h} =
          \vj_{h}} |\alpha_{\ell}|^2}} \ket{\vk} =: \sum_{\substack{\vk \in
      \{0,1\}^q :\\ \vk_{h} = \vj_{h}}} \beta_{k} \ket{\vk},
  \end{equation*}
  and the coefficients $\beta_{k}$, defined as above, are only
  defined for $\vk \in \{0,1\}^q : \vk_{h} = \vj_{h}$.  By Post.~\ref{pos:meas}, applying a single-qubit measurement, we also have:
  \begin{equation*}
    \Pr_{\ket{\psi}}\left(\mathcal{Q}_h = \vj_{h}\right) = \sum_{\vk
      \in \{0,1\}^q: \vk_{h} = \vj_{h}} |\alpha_{k}|^2.
  \end{equation*}
  By the induction hypothesis:
  \begin{align*}
    \Pr_{\ket{\phi}}\left(\mathcal{Q}_1 =
    \vj_{1}, \dots, \mathcal{Q}_{h-1} = \vj_{h-1}, \mathcal{Q}_{h+1} = \vj_{h+1}, \dots,
    \mathcal{Q}_q = \vj_{q} \right) = |\beta_{j}|^2,
  \end{align*}
  because: $\ket{\phi}$ is the state after measuring the qubit in
  position $h$ and obtaining $\vj_h$ as the outcome, therefore it only
  contains basis states $\vk$ with $\vk_h = \vj_h$; and the induction
  hypothesis imposes that the probability of observing the entire
  binary string $\vj$ (for qubits other than qubit $h$, because qubit
  $h$ was already measured, i.e., value $\vj_{\ell}$ in position
  $\ell$, $\ell \neq h$) is simply $\abs{\beta_{j}}^2$.  Remembering
  that $\beta_{k} = \alpha_{k}/\left( \sqrt{\sum_{\v{\ell} \in
      \{0,1\}^q : \v{\ell}_{h} = \vj_{h}} |\alpha_{\ell}|^2}\right)$,
  we finally obtain:
  \begin{equation*}
    \Pr_{\ket{\psi}}\left(\mathcal{Q}_1 = \vj_{1}, \dots, \mathcal{Q}_q =
    \vj_{q} \right) =  
    \frac{|\alpha_{j}|^2}{\displaystyle
      \sum_{\v{\ell} \in \{0,1\}^q : \v{\ell}_{h} = \vj_{h}}
      |\alpha_{\ell}|^2} \left(\sum_{\substack{\vk \in \{0,1\}^q : \\
        \vk_{h} = \vj_{h}}} |\alpha_{k}|^2\right) = |\alpha_{j}|^2. 
  \end{equation*}
\end{proof}

\begin{figure}[h!]
\leavevmode
\centering
\ifcompilefigs
\Qcircuit @C=1em @R=0.7em {
                    & \multigate{2}{\metersymb}        & \\
\lstick{\ket{\psi}} & \ghost{\metersymb}         & \\
                    & \ghost{\metersymb}         & \\
}
\hspace{10em}
\Qcircuit @C=1em @R=0.7em {
                    & \meter         & \\
\lstick{\ket{\psi}} & \meter         & \\
                    & \meter         & \\
}
\else
\includegraphics{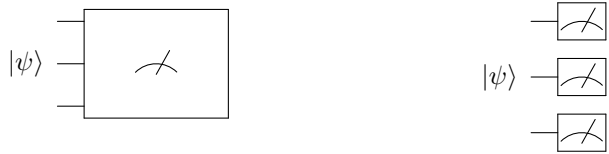}
\fi
\caption{Multiple-qubit measurement.}
\label{fig:multimeas}
\end{figure}

\noindent Proposition \ref{prop:meas} above shows that the two
circuits in Fig.~\ref{fig:multimeas} are equivalent.
In other words, the single-qubit measurement gate as defined
in Post.~\ref{pos:meas} is sufficient to measure any number of qubits
in the most natural way, i.e., the measurement outcome $\vj$ on the
$q$ qubits occurs with probability that is exactly equal to
$\abs{\alpha_{j}}^2$. Notice that with this simple rule, it is easy to
compute the probability of obtaining a given string on a given subset
of the qubits: we just need to add up the modulus squared of the
coefficients for all those basis states that contain the desired
string in the desired position.

\begin{example}
  \label{ex:prodvsent}
  Consider again the following two-qubit state:
  \begin{equation*}
    \alpha_{00} \ket{00} + \alpha_{01} \ket{01} + \alpha_{10} \ket{10}
    + \alpha_{11} \ket{11} = \frac{1}{2} \ket{00} + \frac{1}{2}
    \ket{01} + \frac{1}{2} \ket{10} + \frac{1}{2} \ket{11}.
  \end{equation*}
  We noted in Ex.~\ref{ex:productentangled} that this is a product
  state. As usual, let qubit 1 the first qubit (i.e., the one
  corresponding to the first digit in the two-digit binary strings),
  and let qubit 2 be the second qubit (i.e., the one corresponding to
  the second digit in the two-digit binary strings). Then:
  \begin{align*}
    \Pr(\mathcal{Q}_1 = 0) &= |\alpha_{00}|^2 + |\alpha_{01}|^2 = \left(\frac{1}{2}\right)^2 + \left(\frac{1}{2}\right)^2 = \frac{1}{2} \\
    \Pr(\mathcal{Q}_1 = 1) &= |\alpha_{10}|^2 + |\alpha_{11}|^2 = \left(\frac{1}{2}\right)^2 + \left(\frac{1}{2}\right)^2 = \frac{1}{2} \\
    \Pr(\mathcal{Q}_2 = 0) &= |\alpha_{00}|^2 + |\alpha_{10}|^2 = \left(\frac{1}{2}\right)^2 + \left(\frac{1}{2}\right)^2 = \frac{1}{2} \\
    \Pr(\mathcal{Q}_2 = 1) &= |\alpha_{01}|^2 + |\alpha_{11}|^2 = \left(\frac{1}{2}\right)^2 + \left(\frac{1}{2}\right)^2 = \frac{1}{2}.
  \end{align*}
  Suppose we measure qubit 2 and we obtain 1 as the outcome of the
  measurement. Then the state of the two-qubit system collapses to:
  \begin{equation*}
    \frac{1}{\sqrt{2}} \ket{01} + \frac{1}{\sqrt{2}} \ket{11}.
  \end{equation*}
  The measurement outcome distribution for qubit 1 for this new state is:
  \begin{equation*}
    \Pr(\mathcal{Q}_1 = 0) = \frac{1}{2} \qquad \Pr(\mathcal{Q}_1 = 1) = \frac{1}{2}.
  \end{equation*}
  Hence, the probability of observing 0 or 1 when measuring qubit 1
  did not change after the measurement.

  Consider now the following entangled two-qubit state:
  \begin{equation*}
    \beta_{00} \ket{00} + \beta_{11} \ket{11} = \frac{1}{\sqrt{2}}
    \ket{00} + \frac{1}{\sqrt{2}} \ket{11}.
  \end{equation*}
  Doing the calculations, we still have:
  \begin{align*}
    \Pr(\mathcal{Q}_1 = 0) &= |\beta_{00}|^2 = \frac{1}{2} 
    & \Pr(\mathcal{Q}_1 = 1) &= |\beta_{11}|^2 = \frac{1}{2} \\
    \Pr(\mathcal{Q}_2 = 0) &= |\beta_{00}|^2 = \frac{1}{2} & 
    \Pr(\mathcal{Q}_2 = 1) &= |\beta_{11}|^2 = \frac{1}{2}.
  \end{align*}
  As before, suppose we measure qubit 2 and we obtain 1 as the outcome
  of the measurement. Then the state of the two-qubit system collapses
  to:
  \begin{equation*}
    \ket{11}.
  \end{equation*}
  If we measure qubit 1 from this state, we obtain:
  \begin{equation*}
    \Pr(\mathcal{Q}_1 = 0) = 0 \qquad \Pr(\mathcal{Q}_1 = 1) = 1.
  \end{equation*}
  The situation is now very different: the probability distribution of
  $\mathcal{Q}_1$ has changed after measuring qubit 2 (obtaining a
  sample from $\mathcal{Q}_2$). This is exactly the concept of
  entanglement\index{entanglement}: when two or more qubits are
  entangled, they affect each other, and measuring one qubit changes
  the probability distribution characterizing the measurement outcomes
  for the other qubits.
\end{example}

\noindent The example above can be seen in terms of conditional
probabilities: if, for all $x, y \in \{0,1\}$, we have
$\Pr(\mathcal{Q}_1 = x) = \Pr(\mathcal{Q}_1 = x | \mathcal{Q}_2 = y)$
(i.e., the random variables $\mathcal{Q}_1, \mathcal{Q}_2$ are
independent), then the two qubits are not entangled (product state),
whereas if $\Pr(\mathcal{Q}_1 = x) \neq \Pr(\mathcal{Q}_1 = x |
\mathcal{Q}_2 = y)$ for some $x, y$, there is entanglement. Indeed,
recall that taking the tensor product of two vectors containing
outcome probabilities for independent random variables yields the
joint probability distribution. Quantum state vectors do not contain
outcome probabilities, but the modulus squared of the components of
the state vector corresponds to a probability. Furthermore, for any
two complex numbers $\alpha, \beta \in \C$ we have $|\alpha\beta|^2 =
|\alpha|^2|\beta|^2$, so the operation to compute probabilities from
state coefficients is distributive with respect to multiplication.  A
product state is a tensor product of smaller-dimensional state
vectors, hence it leads to outcome probabilities that are simply the
product of the outcome probabilities corresponding to measuring each
of the qubits independently. Conversely, an entangled state is not a
product state, and the random variables associated with measuring each
of the qubits are no longer independent.
\begin{remark}
  Despite the above discussion, it would be wrong to think of the
  quantum state as a probability distribution: the quantum state {\em
    induces} a probability distribution by taking the modulus squared
  of its entries, but it is not a probability distribution! Indeed,
  the coefficients in a quantum state are complex numbers unrestricted
  in sign, while probabilities are nonnegative real numbers.
  Furthermore, just as there is an infinite set of complex numbers
  that have the same modulus (i.e., the set $\{a \in \C: |a| = v\}$
  for some real number $v > 0$ is infinite), there is an infinite
  number of quantum state vectors in $(\C^2)^{\otimes q}$ that yield
  the same distribution. After applying the same sequence of
  operations to two states that induce the same probability
  distribution, we may or may not obtain quantum states that induce
  the same outcome distribution: this is shown in the next two
  examples.
\end{remark}
\begin{example}
  \label{ex:globalphase}
  Suppose we have two $q$-qubit quantum states $\ket{\psi},
  \ket{\phi}$ satisfying $\ket{\psi} = e^{i\theta} \ket{\phi}$ for
  some $\theta \in \R$. Now consider the application of some unitary
  matrix $U$ onto $\ket{\psi}$ and $\ket{\phi}$, followed by a
  measurement of all the qubits. Define:
  \begin{equation*}
    U \ket{\phi} := \sum_{\vj \in \{0,1\}^q} \alpha_{j} \ket{\vj}
  \end{equation*}
  for some (normalized) coefficients $\alpha_{j}$, which implies:
  \begin{equation*}
    U \ket{\psi} = Ue^{i\theta} \ket{\phi} = \sum_{\vj \in \{0,1\}^q}
    e^{i\theta} \alpha_{j} \ket{\vj}.
  \end{equation*}
  This means that for a given $\vk$:
  \begin{equation*}
    \Pr_{U\ket{\phi}} (\mathcal{Q}_1 = \vk_1, \dots, \mathcal{Q}_q = \vk_q) = |\alpha_{k}|^2,
    \qquad
    \Pr_{U\ket{\psi}} (\mathcal{Q}_1 = \vk_1, \dots, \mathcal{Q}_q = \vk_q) = |e^{i\theta} \alpha_{k}|^2 = |\alpha_{k}|^2,
  \end{equation*}
  so the probability of obtaining $\vk$ as the outcome of a
  measurement is the same for both $U\ket{\psi}$ and
  $U\ket{\phi}$. Because this is true after applying an arbitrary
  unitary $U$, it is also true after applying a whole circuit, which
  is just a sequence of unitaries. Hence, if the vectors $\ket{\psi},
  \ket{\phi}$ satisfy the relationship $\ket{\psi} = e^{i\theta}
  \ket{\phi}$, they induce the same outcome distribution, no matter
  what (unitary) operations we apply to them. The factor $e^{i\theta}$
  is usually called {\em global phase}\index{phase!global}\index{global phase|see{phase, global}} and can safely be ignored in the
  calculations.
\end{example}
\begin{example}
  \label{ex:probdistdanger}
  Consider the following two single-qubit states:
  \begin{equation*}
    \ket{\psi} = \frac{1}{\sqrt{2}} \ket{0} + \frac{1}{\sqrt{2}} \ket{1} \qquad
    \ket{\phi} = \frac{1}{\sqrt{2}} \ket{0} - \frac{1}{\sqrt{2}} \ket{1}.
  \end{equation*}
  Both induce the same probability distribution on the measurement
  outcomes:
  \begin{align*}
    \Pr_{\ket{\psi}}(\mathcal{Q}_1 = 0) = \frac{1}{2} \qquad \Pr_{\ket{\psi}}(\mathcal{Q}_1 = 1) = \frac{1}{2}\phantom{.} \\
    \Pr_{\ket{\phi}}(\mathcal{Q}_1 = 0) = \frac{1}{2} \qquad \Pr_{\ket{\phi}}(\mathcal{Q}_1 = 1) = \frac{1}{2}.
  \end{align*}
  But $\ket{\psi}$ and $\ket{\phi}$ are very different states! If we
  apply a certain unitary matrix to both (this gate is called Hadamard
  gate, see Sect.~\ref{sec:basicops}), we obtain very
  different results -- orthogonal vectors, in fact:
  \begin{align*}
    \frac{1}{\sqrt{2}} \begin{pmatrix} 1 & 1 \\ 1 & -1 \end{pmatrix} \ket{\psi} &= \frac{1}{2} \begin{pmatrix} 1 & 1 \\ 1 & -1 \end{pmatrix} \begin{pmatrix} 1 \\ 1 \end{pmatrix} = \frac{1}{2} \begin{pmatrix} 2 \\ 0 \end{pmatrix} = \ket{0} \\
    \frac{1}{\sqrt{2}} \begin{pmatrix} 1 & 1 \\ 1 & -1 \end{pmatrix} \ket{\phi} &= \frac{1}{2} \begin{pmatrix} 1 & 1 \\ 1 & -1 \end{pmatrix} \begin{pmatrix} 1 \\ -1 \end{pmatrix} = \frac{1}{2} \begin{pmatrix} 0 \\ 2 \end{pmatrix} = \ket{1}.
  \end{align*}
  This illustrates the dangers of thinking about the quantum state as
  a probability distribution, rather than a set of complex
  coefficients that induce a probability distribution.
\end{example}

\subsection{The no-cloning principle}
Because measurement destroys the quantum state, it is natural to look
for a way to create a copy of a quantum state. If a clone could be
created, it would be possible to perform measurements on the clone, so
that the original state would not be destroyed. Furthermore, cloning
would allow us to take several measurements of the same set of qubits
without having to repeat the circuit that creates the quantum state.
However, it turns out that cloning\index{cloning} is impossible: this is a direct
consequence of the properties of quantum gates, in particular the fact
that gates are unitary matrices.
\begin{theorem}[No-cloning principle]
  There does not exist a unitary matrix that maps $\ket{\psi}_q
  \ket{\v{0}}_q$ to $\ket{\psi}_q \ket{\psi}_q$ for
  an arbitrary quantum state on $q$ qubits $\ket{\psi}$.
\end{theorem}

\begin{proof}
  Suppose there exists such a unitary $U$. Then for any two quantum
  states $\ket{\psi}, \ket{\phi}$ on $q$ qubits, we have (all registers in this proof are $q$ qubits each):
  \begin{align*}
    U(\ket{\psi} \ket{\v{0}}) &= \ket{\psi} \ket{\psi} \\
    U(\ket{\phi} \ket{\v{0}}) &= \ket{\phi} \ket{\phi}.
  \end{align*}
  Using these equalities, and remembering that $U^{\dag}U = I$, we can
  write:
  \begin{align*}
    \braket{\phi}{\psi} &= \braket{\phi}{\psi}\underbrace{\braket{\v{0}}{\v{0}}}_{= 1} = 
    \braket{\phi}{\psi} \otimes \left(\braket{\v{0}}{\v{0}}\right) =
    (\bra{\phi} \otimes \bra{\v{0}})(\ket{\psi} \otimes \ket{\v{0}}) 
    \\
    &= (\bra{\phi} \otimes \bra{\v{0}})U^{\dag}U(\ket{\psi} \otimes \ket{\v{0}})  = (\bra{\phi} \otimes \bra{\phi})(\ket{\psi} \otimes \ket{\psi}) =
    \braket{\phi}{\psi}^2.    
  \end{align*}
  But $\braket{\phi}{\psi} = \braket{\phi}{\psi}^2$ is only true if
  $\braket{\phi}{\psi}$ is equal to 0 or to 1, contradicting the fact
  that $\ket{\phi}, \ket{\psi}$ are arbitrary quantum states.
\end{proof}

\noindent The above theorem shows that we cannot copy an arbitrary
quantum state. We remark that the proof does not rule out the
possibility of constructing a unitary matrix that copies a specific
quantum state. In other words, if we know what quantum state we want
to copy, one could construct a unitary matrix to do that; but it is
impossible to construct a single unitary matrix to copy all possible
states.  This establishes that we cannot ``cheat'' the destructive
effect of a measurement by simply cloning the state before the
measurement. Hence, whenever we run a circuit that produces an output
quantum state, in general we can reproduce the output quantum state
only by repeating all the steps of the circuit. So if we think of the
circuit as implementing an algorithm, followed by a measurement to get
a sample from a certain probability distribution, in general we need
to run the algorithm again from scratch to get another sample.

\subsection{Basic operations and universality}
\label{sec:basicops}
Quantum computation does not allow the user to apply every possible
unitary matrix in the circuit that they want to run (i.e., their
code), just as classical computations do not allow the user to apply
every possible classical function. Rather, the user is limited to
gates (unitary matrices) that are efficiently specifiable and
implementable, just as classically one can only write efficient
programs by specifying a polynomial-size sequence of basic operations
on bits. The specification of a unitary matrix must be done by
combining gates out of a basic set, which can be thought of as the
instruction set of the quantum computer. We now discuss what
these basic gates are, and how they can be combined to form other
operations.

We use the following two definitions of operations on binary strings;
these are frequently used in this and subsequent chapters.
\begin{definition}[Bitwise XOR]
  \label{def:xor}
  For any integer $q > 0$ and binary strings $\vj, \vk \in \{0,1\}^q$, we
  denote by $\vj \oplus \vk$\index{notation!oplus@\ensuremath{\oplus}} the bitwise modulo-2 addition of $q$-digit
  strings (bitwise XOR), defined as:
  \begin{equation*}
    \vj \oplus \vk := \vh, \text{ with } \vh \in \{0,1\}^q \text{ and } 
    \v{h}_p = \begin{cases} 0 & \text{if } \vj_p = \vk_p \\
      1 & \text{otherwise} \end{cases} \text{ for all } p=1,\dots,q.
  \end{equation*}
\end{definition}
\begin{definition}[Bitwise dot product]
  \label{def:bullet}
  For any integer $q > 0$ and binary strings $\vj, \vk \in \{0,1\}^q$,
  we denote by $\vj \bullet \vk$\index{notation!bullet@\ensuremath{\bullet}} the bitwise dot product of $q$-digit
  strings, defined as:
  \begin{equation*}
    \vj \bullet \vk := \sum_{h = 1}^{q} \vj_h \vk_h.
  \end{equation*}
\end{definition}
We also use this (common) definition of matrix norm, which we state
for completeness. It is usually referred to as the operator norm
induced by the Euclidean norm. 
\begin{definition}[Matrix norm]
  \label{def:opnorm}
  For a given matrix $A$, we denote $\nrm{A} = \sup_{x : \nrm{x} = 1}
  \nrm{Ax}$\index{norm!matrix}\index{norm!operator}.
\end{definition}

\paragraph{Single-qubit gates.}The first operations that we discuss are the {\em Pauli gates}.
\begin{definition}[Pauli gates]
  \label{def:pauli}
  The four Pauli gates\index{Pauli gates}\index{gate!Pauli} are the following single-qubit gates:
  \begin{align*}
    I &= \begin{pmatrix} 1 & 0 \\ 0 & 1 \end{pmatrix} & X &= \begin{pmatrix} 0 & 1 \\ 1 & 0 \end{pmatrix} \\
    Y &= \begin{pmatrix} 0 & -i \\ i & 0 \end{pmatrix} & Z &= \begin{pmatrix} 1 & 0 \\ 0 & -1 \end{pmatrix}. 
  \end{align*}
\end{definition}
\begin{proposition}
  The Pauli gates form a basis for $\C^{2 \times 2}$, they are
  Hermitian, and they satisfy the relationship $XYZ = iI$.
\end{proposition}
The proof is left as an exercise. The $X$ gate is the equivalent of the
NOT gate in classical computers, as it implements a bit (rather,
qubit) flip, changing from $\ket{0}$ to $\ket{1}$ and vice versa:
\begin{equation*}
  X \ket{0} = \ket{1} \qquad X \ket{1} = \ket{0}.
\end{equation*}
The $Z$ gate is also called a phase flip gate: it leaves $\ket{0}$
unchanged, and maps $\ket{1}$ to $-\ket{1}$.
\begin{equation*}
  Z \ket{0} = \ket{0} \qquad Z \ket{1} = -\ket{1}.
\end{equation*}

\begin{figure}[h!]
\leavevmode
\centering
\ifcompilefigs
\Qcircuit @C=1em @R=0.8em {
\lstick{\ket{0}} & \gate{H}  & \qw & \rstick{\frac{1}{\sqrt{2}}\left(\ket{0} + \ket{1}\right)}\\
\lstick{\ket{0}} & \gate{H}  & \qw & \rstick{\frac{1}{\sqrt{2}}\left(\ket{0} + \ket{1}\right)}\\
\lstick{\ket{0}} & \gate{H}  & \qw & \rstick{\frac{1}{\sqrt{2}}\left(\ket{0} + \ket{1}\right)}\\
}
\hspace{15em}
\Qcircuit @C=1em @R=0.8em {
\lstick{\ket{0}} & \multigate{2}{H^{\otimes 3}}  & \qw & \rstick{\frac{1}{\sqrt{2}}\left(\ket{0} + \ket{1}\right)}\\
\lstick{\ket{0}} & \ghost{H^{\otimes 3}}  & \qw & \rstick{\frac{1}{\sqrt{2}}\left(\ket{0} + \ket{1}\right)}\\
\lstick{\ket{0}} & \ghost{H^{\otimes 3}}  & \qw & \rstick{\frac{1}{\sqrt{2}}\left(\ket{0} + \ket{1}\right)}\\
}
\else
\includegraphics{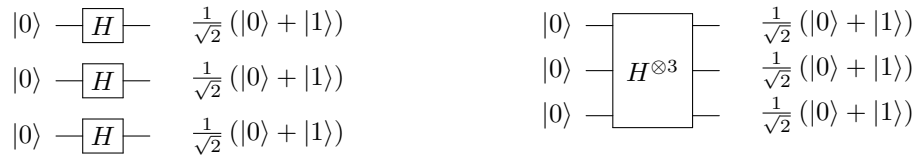}
\fi
\caption{Two representations for multiple Hadamard gates.}
\label{fig:hadamard}
\end{figure}

A single-qubit gate that is used in many quantum algorithms is the
so-called Hadamard\index{Hadamard gate}\index{gate!Hadamard} gate:
\begin{equation*}
  H = \frac{1}{\sqrt{2}} \begin{pmatrix} 1 & 1 \\ 1 & -1 \end{pmatrix}.
\end{equation*}
The action of $H$ is as follows:
\begin{equation*}
  H \ket{0} = \frac{1}{\sqrt{2}}\left(\ket{0} + \ket{1}\right) \qquad
  H \ket{1} = \frac{1}{\sqrt{2}}\left(\ket{0} - \ket{1}\right).
\end{equation*}
In the rest of this \book{} we sometimes need an algebraic
expression for the action of Hadamard gates on basis states. The
effect of $H$ on a single-qubit basis state $\ket{x}$, $x \in
\{0,1\}$, can be summarized as follows:
\begin{equation*}
  H \ket{x} = \frac{1}{\sqrt{2}}(\ket{0} + (-1)^x\ket{1}) =
  \frac{1}{\sqrt{2}} \sum_{k=0}^1 (-1)^{k x} \ket{k}.
\end{equation*}
This is consistent with the previous expression for $H$. Using our
notation, we can define the effect of $H^{\otimes q}$ on a $q$-qubit
basis state $\ket{\v{x}}_q$ as:
\begin{equation}
  \label{eq:hadamard}
  \begin{split}
  H^{\otimes q} \ket{\v{x}}_q &= \frac{1}{\sqrt{2^q}} \sum_{k_{1}=0}^1 \cdots \sum_{k_{q}=0}^1 (-1)^{\sum_{h=1}^{q} k_h \v{x}_{h}} \ket{k_{1}} \otimes \cdots \otimes \ket{k_q} \\
  &= \frac{1}{\sqrt{2^q}} \sum_{\vk \in \{0,1\}^q} (-1)^{\vk \bullet \v{x}} \ket{\vk},
  \end{split}
\end{equation}
where $\bullet$ is the bitwise dot product, see
Def.~\ref{def:bullet}. When considering multiple Hadamard gates in
parallel, it is sometimes useful to rely on the following relationship, that
can be easily verified using the definition:
\begin{equation*}
  H^{\otimes q} = \frac{1}{\sqrt{2}} \begin{pmatrix} H^{\otimes q-1} & H^{\otimes q-1} \\ H^{\otimes q-1} & -H^{\otimes(q-1)} \end{pmatrix}.
\end{equation*}
The next proposition shows one of the
reasons why the Hadamard gate is frequently employed in many quantum
algorithms.
\begin{proposition}
  Given a $q$-qubit quantum register initially in the state
  $\ket{\v{0}}_q$, applying the Hadamard gate to all qubits, or
  equivalently the matrix $H^{\otimes q}$, yields the uniform
  superposition of basis states $\frac{1}{\sqrt{2^q}} \sum_{\vj \in
    \{0,1\}^q} \ket{\vj}$.
\end{proposition}
\begin{proof}
  We have:
  \begin{equation*}
    H^{\otimes q} \ket{\v{0}}_q = H^{\otimes q} \ket{0}^{\otimes q} =
    \left(H\ket{0}\right)^{\otimes q} = \left(\frac{1}{\sqrt{2}}\ket{0}
    + \frac{1}{\sqrt{2}}\ket{1}\right)^{\otimes q} =
    \frac{1}{\sqrt{2^q}} \sum_{\vj \in \{0,1\}^q} \ket{\vj}. 
  \end{equation*}
\end{proof}
\begin{remark}
  The uniform superposition of the $2^q$ basis states on $q$ qubits
  can be obtained from the initial state $\ket{\v{0}}_q$ by
  applying $q$ gates only.
\end{remark}
The multiple Hadamard can be represented by one of the equivalent
circuits given in Fig.~\ref{fig:hadamard}.
Several quantum algorithms start by setting the state of the quantum
device to a uniform superposition, and then apply further operations
which, by linearity, are simultaneously applied to all the possible
binary strings. This is a remarkable advantage of quantum computing
over classical computing.

Readers with advanced knowledge of theoretical computer science might
be wondering how this compares to classical probabilistic computation, i.e.,
probabilistic Turing machines\index{Turing machine!probabilistic}, a well-known concept in
computational complexity theory. A probabilistic Turing machine is
initialized with a set of random bits that take an unknown value and
influence the state transition. The state is described by a
probability distribution over all the possible states, because we do
not know the value of the random bits with which the machine is
initialized. When a state transition occurs, to update the description
of the state we need to apply the transition to all states that appear
with positive probability. In this sense, operations in a
probabilistic Turing machine can be thought of as being simultaneously
applied to many (possibly all) binary strings. However, a
probabilistic Turing machine admits a more compact description of the
state: if we know the random bits with which the machine is
initialized, then the state becomes deterministically known. Hence,
for a given value of the random bits, the state of the probabilistic
Turing machine can be described in linear space, and operations map
one state into another state. On the other hand, it is not known how
to obtain such a compact description for a quantum computer: there is
no equivalent for the random bits, and a characterization of the state
truly requires an exponential number of complex coefficients. In fact,
it is believed that quantum computers are more powerful than
probabilistic Turing machines, although there is no formal proof.

To conclude our discussion on single-qubit gates, we note that all
single-qubit can be represented by the following parameterized
matrix that describes all unitary matrices (up to a global phase
factor):
\begin{equation*}
  U(\theta, \phi, \lambda) = 
    \begin{pmatrix}
      e^{-i(\phi + \lambda)/2} \cos (\theta/2) &  -e^{-i(\phi - \lambda)/2} \sin (\theta/2) \\
      e^{i(\phi - \lambda)/2} \sin (\theta/2) & e^{i(\phi + \lambda)/2} \cos (\theta/2)
    \end{pmatrix}
\end{equation*}
All single-qubit gates can be obtained by an appropriate choice of
parameters $\theta, \phi, \lambda$. This parameterized expression is
used in some of the existing programming languages for quantum
circuits.

\begin{figure}[h!]
\leavevmode
\centering
\ifcompilefigs
\Qcircuit @C=1em @R=0.7em {
  & \ctrl{1}  & \qw & \\
  & \targ  & \qw & 
}
\else
\includegraphics{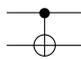}
\fi
\caption{The $\text{CX}_{12}$, or controlled NOT, gate with control
  qubit 1 and target qubit 2.}
\label{fig:cnot}
\end{figure}

\paragraph{Two-qubit gates.} Another fundamental gate is the C$X$ gate, also called ``controlled
NOT'' or ``CNOT''\index{gate!CNOT} (because the $X$ gate acts as a NOT). The C$X$ gate is
a two-qubit gate that has a control bit and a target bit, and acts as
follows: if the control bit is $\ket{0}$, nothing happens, whereas if
the control bit is $\ket{1}$, the target bit is bit-flipped (i.e., the
same effect as the $X$ gate). The corresponding circuit diagram is
given in Fig.~\ref{fig:cnot}.

\noindent The matrix description of the gate with control qubit 1 and
target qubit 2 is as follows:
\begin{equation*}
  \text{C}X_{12} = 
  \begin{pmatrix}
    1 & 0 & 0 & 0 \\ 0 & 1 & 0 & 0 \\ 0 & 0 & 0 & 1 \\ 0 & 0 & 1 & 0
  \end{pmatrix}.
\end{equation*}
Thus, the effect of C$X$ is:
\begin{align*}
  \text{C}X_{12} \ket{00} &= \ket{00} & \text{C}X_{12} \ket{01} &= \ket{01} \\
  \text{C}X_{12} \ket{10} &= \ket{11} & \text{C}X_{12} \ket{11} &= \ket{10}.
\end{align*}
This leads to the following formula for the effect of C$X$:
\begin{equation*}
  \text{C}X_{12} \ket{x} \ket{y} = \ket{x} \ket{y\oplus x}, \qquad \forall x,y \in \{0,1\}.
\end{equation*}
The C$X$ gate can create and destroy entanglement, as shown in
the next example.
\begin{example}
  Consider the circuit in Fig.~\ref{fig:cxentangled}.
  \begin{figure}[h!]
    \leavevmode
    \centering
    \ifcompilefigs
    \Qcircuit @C=1em @R=0.7em {
     \lstick{\ket{0}} & \qw &\gate{H}  & \ctrl{1}  & \qw & \\
     \lstick{\ket{0}} & \qw &  \qw     & \targ     & \qw & 
    }
    \else
    \includegraphics{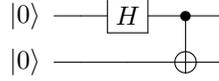}
    \fi
    \caption{A circuit that produces an entangled state.}
    \label{fig:cxentangled}
  \end{figure}
  
  The circuit yields the following state:
  \begin{align*}
    \text{C}X_{12} (H \otimes I) \ket{00} = \text{C}X_{12} \left(\frac{1}{\sqrt{2}}(\ket{0} + \ket{1}) \ket{0} \right) = \frac{1}{\sqrt{2}} \left(\ket{00} + \ket{11}\right).
  \end{align*}
  As we have seen in Ex.~\ref{ex:prodvsent}, this is an entangled
  state. In this case it is also easy to break entanglement: just
  apply C$X$ one more time, which reverses the last operation and
  brings us back to the state $\frac{1}{\sqrt{2}}(\ket{0} +
  \ket{1}) \ket{0}$.
\end{example}

\noindent An interesting feature of the C$X$ gate is that it can be
used to swap two qubits. A swap\index{gate!SWAP} between two qubits
$j$ and $k$ is defined as the operation that maps a quantum state to
a new quantum state in which every basis state has its $j$-th and
$k$-th digit permuted; this corresponds to the following matrix:
\begin{equation*}
  \text{SWAP} = \begin{pmatrix}
    1 & 0 & 0 & 0 \\ 0 & 0 & 1 & 0 \\ 0 & 1 & 0 & 0 \\ 0 & 0 & 0 & 1
  \end{pmatrix}.
\end{equation*}
If two qubits are in a product state $\ket{\psi}_1 \otimes
\ket{\phi}_1$, then $\text{SWAP}(\ket{\psi}_1 \otimes \ket{\phi}_1) =
\ket{\phi} \otimes \ket{\psi}$. The circuit diagram for the
$\text{SWAP}$ gate is given in Fig.~\ref{fig:swap}.
\begin{figure}[h!]
\leavevmode
\centering
\ifcompilefigs
\Qcircuit @C=1em @R=2em {
  & \qswap  & \qw & \\
  & \qswap \qwx  & \qw & 
}
\else
\includegraphics{figures/swap.pdf}
\fi
\caption{The $\text{SWAP}$ gate.}
\label{fig:swap}
\end{figure}
Considering that C$X$, like all quantum gates, is a linear map, it may
sound surprising that it can implement a swap. However, the SWAP gate
can indeed be constructed out of C$X$ gates using the circuit depicted
in Fig.~\ref{fig:swapcx}.
\begin{figure}[h!]
\leavevmode
\centering
\ifcompilefigs
\Qcircuit @C=1em @R=0.7em {
  & \ctrl{1}  & \targ      & \ctrl{1} & \qw & \\
  & \targ     & \ctrl{-1}  & \targ    & \qw & 
}
\else
\includegraphics{figures/swapcx.pdf}
\fi
\caption{Implementation of $\text{SWAP}$ using C$X$ gates.}
\label{fig:swapcx}
\end{figure}
\begin{proposition}
  The circuit in Fig.~\ref{fig:swapcx}, constructed with three C$X$
  gates, swaps qubits 1 and 2.
\end{proposition}
\begin{proof}
  By linearity, it suffices to show that the circuit above maps
  $\ket{00} \to \ket{00}, \ket{01} \to \ket{10}, \ket{10} \to
  \ket{01}$, and $\ket{11} \to \ket{11}$. We have:
  \begin{align*}
    \text{C}X_{12} \text{C}X_{21} \text{C}X_{12} \ket{00} = \text{C}X_{12} \text{C}X_{21} \ket{00} 
    = \text{C}X_{12} \ket{00} = \ket{00}. \\
    \text{C}X_{12} \text{C}X_{21} \text{C}X_{12} \ket{01} = \text{C}X_{2}1 \text{C}X_{21} \ket{01} 
    = \text{C}X_{12} \ket{11} = \ket{10}. \\
    \text{C}X_{12} \text{C}X_{21} \text{C}X_{12} \ket{10} = \text{C}X_{12} \text{C}X_{21} \ket{11} 
    = \text{C}X_{12} \ket{01} = \ket{01}. \\
    \text{C}X_{12} \text{C}X_{21} \text{C}X_{12} \ket{11} = \text{C}X_{12} \text{C}X_{21} \ket{10} 
    = \text{C}X_{12} \ket{10} = \ket{11}.
  \end{align*}
  Therefore, the SWAP circuit maps:
  \begin{equation*}
  \alpha_{00} \ket{00} + \alpha_{01} \ket{01} + \alpha_{10} \ket{10} +
  \alpha_{11} \ket{11} \to  \alpha_{00} \ket{00} + \alpha_{01} \ket{10} +
  \alpha_{10} \ket{01} + \alpha_{11} \ket{11}. 
  \end{equation*}
\end{proof}

\noindent The SWAP circuit is particularly important for practical
reasons: sometimes, in the current generation of quantum computing
hardware, two-qubit gates can only be applied among certain pairs of
qubits. For example, when employing one of the most prevalent quantum
hardware technologies (superconducting qubits, see
e.g.~\cite{devoret2013superconducting,castelvecchi2017leap}),
two-qubit gates can only be applied to qubits that are physically
adjacent on a chip.  Thanks to the SWAP, as long as the graph
representing the qubit adjacency in the hardware device is a connected
graph, two-qubit gates can be applied to any pair of qubits: if the
qubits are not directly connected by an edge on the graph (e.g.,
physically located next to each other on the chip), we just need to
SWAP one of them as many times as is necessary to bring it to a
location adjacent to the other qubit. In this way, we can assume that
each qubit can interact with all other qubits from a theoretical point
of view, even if from a practical perspective this may require extra
SWAP gates.

\paragraph{Multiple-qubit gates.} Defining and constructing every possible unitary matrix of every size seems like an impossible endeavor: clearly there must be some methodology by which general multiple-qubit gate can be implemented in a finite way without requiring a brand-new physical construction, otherwise implementing all possible gates would be a never-ending engineering task. Indeed, it can be shown that a set of gates consisting of (some) single-qubit gates plus C$X$ is sufficient to construct any unitary matrix with
arbitrary precision. This is the concept of {\em universality}.\index{gate!universality}\index{approximation!gate|see{gate, approximation}}
\begin{definition}[Universal set of gates]
  \label{def:universal}
  A unitary matrix $V$ is an $\epsilon$-approximation of a unitary
  matrix $U$ if $\nrm{U - V} = \sup_{x : \nrm{x} = 1} \nrm{ (U - V)x }
  \le \epsilon$. A finite set of gates that can be used to construct
  an $\epsilon$-approximation of any unitary matrix, for any $\epsilon
  > 0$ and on any given number of qubits, is called a {\em universal}\index{universal set}
  set of gates.
\end{definition}
To build a universal set of gates, the first step is to show how to
construct arbitrary single-qubit gates from a finite set of basic
gates, then we use arbitrary single-qubit gates (together with some
entangling gate) to build multiple-qubit gates. A possible universal
set of gates is described in the next result.
\begin{theorem}[Solovay-Kitaev theorem; \cite{kitaev97quantum,nielsen02quantum}]
  \label{thm:sk}
  Let $U \in \C^{2 \times 2}$ be an arbitrary unitary matrix. Then
  there exists a sequence of gates of length $\bigO{\log^c
    \frac{1}{\epsilon}}$, where $c$ is a constant, that yields an
  $\epsilon$-approximation\index{gate!approximation} of $U$ and consists only of $H$, $T
  = \begin{pmatrix} 1 & 0 \\ 0 & e^{i\frac{\pi}{4}} \end{pmatrix}$ and
  C$X$ gates.
\end{theorem}
The theorem implies that just two single-qubit gates together with
C$X$ allow us to build any single-qubit gate with arbitrary
precision. We discuss the value of the constant $c$ in the notes in
Sect.~\ref{sec:prelimnotes}. The crucial observation is that the
length of the sequence is polylogarithmic in the precision, so we can
obtain high-precision approximations with a relatively small gate
count. To go from single-qubit gates to general $q$-qubit gates, one
needs at most $\bigO{q^2 4^q}$ basic gates (i.e., the gates of Theorem
\ref{thm:sk}); intuitively, this is because each gate on $q$ qubits
has $2^q \times 2^q = 4^q$ elements, and it takes $q^2$ basic gates to
``fill'' an arbitrary element of a large matrix --- for a detailed
discussion of how this can be done, see
\cite[Ch.~4]{nielsen02quantum}. This implies that the set of gates
consisting of just $H, T$ and C$X$ is universal: with a very small set
of basic gates, we can construct any unitary matrix in any dimension
to high precision, although this may require many operations. As
mentioned above, this fact is extremely important for practical
reasons: when constructing a quantum computer, it is sufficient to
focus on a small number of gates (e.g., some single-qubit gates and
C$X$), and all other gates can be constructed from these. Although
many existing hardware platforms offer the possibility of applying
arbitrary single-qubit gates (i.e., parameterized with continuous
parameters) in a seemingly native way, the sufficiency of a small,
finite set of gates assumes tremendous practical importance when
considering the necessity of fault tolerance. Without going into
details (consistent with the stated goal of this \book{} to be
``physics-free'', and ignore the specifics of quantum hardware), fault
tolerance\index{fault tolerance} refers to the ability to correct physical errors that occur
in the course of a quantum computation; such errors are bound to
happen in practice. Thanks to the above discussion, it is sufficient
to provide a fault-tolerant implementation only for gates forming a
universal set: all remaining gates can be constructed from those. On
the other hand, and still remaining at a high level, implementing a
family of gates with continuous parameters in a fault-tolerant manner
would be impossible.

From now on, we ignore any issue related to physical errors, and
assume that the gates in the chosen universal set can be implemented
exactly, i.e., in a fault-tolerant manner. Still, with
Thm.~\ref{thm:sk} (and its generalization to unitaries of arbitrary
dimension) we only construct an approximation of the target unitary,
so one may wonder how the errors due to this approximation accumulate
throughout the computation. Note that such errors are different from
physical errors in the universal set of gates: even if the basic gates
are implemented exactly, which is what we assume, we have errors
arising from the finite approximation of arbitrary unitaries. We study
this aspect in Sect.~\ref{sec:vardistance}.

\begin{figure}[h!]
\leavevmode
\centering
\ifcompilefigs
\Qcircuit @C=1em @R=0.7em @!R {
  & \ctrl{2}  & \qw & \\
  & \ctrl{1}  & \qw & \\
  & \targ  & \qw & 
}
\else
\includegraphics{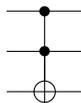}
\fi
\caption{The CC$X$, or doubly-controlled NOT, gate.}
\label{fig:ccnot}
\end{figure}

We conclude our discussion on basic operations with a quantum circuit
for the logic AND gate. We already know that the $X$ gate performs the
logic NOT: having access to the AND gate guarantees that we can
construct any Boolean circuit --- because we already stated that quantum
computers are Turing-complete, being able to perform Boolean logic is
of course implied.
\begin{remark}
  \label{rem:classicalfun}
  With quantum gates that perform the logic AND and NOT operations, we
  can simulate \emph{any} classical Boolean circuit with a quantum
  circuit, possibly using a polynomial amount of additional resources
  (space, i.e., qubits, or time, i.e., gates). In fact, we are about
  to see that the quantum equivalent of the AND gate has separate
  input and output registers, so it requires one extra qubit to store
  the output of the computation. This implies that any function $f:
  \{0,1\}^m \to \{0,1\}^n$ that can be classically computed using $T$
  logic gates (i.e., in time $T$) can be computed by a quantum circuit
  using $\bigO{T}$ quantum gates and additional qubits. See also
  Sec.~\ref{sec:uncompute} for a discussion on how to make
  computations reversible.
\end{remark}
The quantum version of the AND gate is the CC$X$ (doubly-controlled
NOT)\index{gate!CCNOT} gate, that acts on three qubits: it has two control qubits, and
it flips (bit flip, i.e., as the $X$ gate) the third qubit if and only
if both control qubits are $\ket{1}$. The gate is depicted in
Fig.~\ref{fig:ccnot}.
The action of CC$X$ can be described as: $\ket{x} \otimes \ket{y}
\otimes \ket{z} \to \ket{x} \otimes \ket{y} \otimes \ket{z
  \oplus (x \cdot y)}$, where $x, y, z \in \{0,1\}$. Notice that if
$z = 0$, CC$X$ indeed outputs the logical AND between $x$ and $y$
because $0 \oplus (x \cdot y) = x \wedge y$.

\begin{figure}[h!]
  \leavevmode
  \centering
  \ifcompilefigs
  \Qcircuit @C=1em @R=0.7em @!R {
    & \qw      & \qw      & \qw        & \ctrl{2} & \qw      & \qw      & \qw        & \ctrl{2} & \qw        & \ctrl{1} & \qw        & \ctrl{1} & \gate{T} & \qw      & \qw & \\
    & \qw      & \ctrl{1} & \qw        & \qw      & \qw      & \ctrl{1} & \qw        & \qw      & \gate{T^{\dag}} & \targ    & \gate{T^{\dag}} & \targ    & \gate{T} & \gate{T} & \qw &\\
    & \gate{H} & \targ    & \gate{T^{\dag}} & \targ    & \gate{T} & \targ    & \gate{T^{\dag}} & \targ    & \gate{T}   & \gate{H} & \qw        & \qw      & \qw      & \qw      & \qw
  }
  \else
  \includegraphics{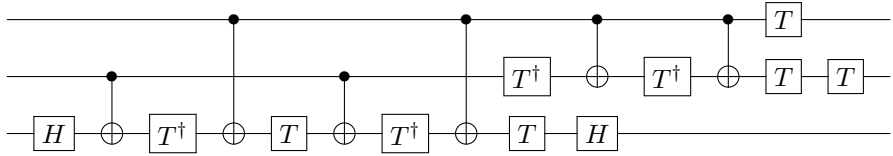}
  \fi
  \caption{Decomposition of CC$X$ in terms of the universal set of gates of Theorem \ref{thm:sk}.}
  \label{fig:ccnotbasic}
\end{figure}

Following our earlier discussion, CC$X$ can be constructed using only
the basic gates indicated in Theorem \ref{thm:sk}. For this, we can
use the circuit in Fig.~\ref{fig:ccnotbasic}, see
\cite{nielsen02quantum}. In this circuit we also use the conjugate
transpose $T^{\dag}$ of the $T$ gate, but it is easy to see that if we
really want to stick to the gates $H, T$, C$X$ only, $T^{\dag}$ can be
constructed from $T$ because $e^{-i\frac{\pi}{4}} =
e^{i\frac{7\pi}{4}}$.  Verifying correctness of the construction in
Fig.~\ref{fig:ccnotbasic} requires a few calculations, that we leave
as an exercise. One way is to carry out the matrix multiplications;
another way, probably more manageable if doing calculations by hand,
is to use linearity and look at the effect of the circuit on each of
the $2^3$ possible basis states. We show only part of the calculations
here. Suppose the circuit is applied to the basis state $\ket{11x}$
with $x \in \{0,1\}$. After performing several simplifications ($T$
and $T^{\dag}$ cancel out, the $T$ gate has no effect on a qubit in
state $\ket{0}$, and we can transform the C$X$s on the third qubit
line into $X$ gates because we already know that the first and second
qubit are in state $\ket{1}$), we find out that the circuit maps:
  \begin{equation*}
    \ket{1} \otimes \ket{1} \otimes \ket{x} \to (T\ket{1}) \otimes
    (T\ket{1}) \otimes (H T X T^{\dag} X T X T^{\dag} X H \ket{x}).
  \end{equation*}
  Doing the calculations, we see that:
  \begin{align*}
    H T X T^{\dag} X T X T^{\dag} X H = \begin{pmatrix} 0 & -i \\ -i & 0 \end{pmatrix},
  \end{align*}
  so that the mapping reads:
  \begin{align*}
    \ket{1} \otimes \ket{1} \otimes \ket{1} \to &(T\ket{1}) \otimes
    (T\ket{1}) \otimes (H T X T^{\dag} X T X T^{\dag} X H \ket{1}) = \\
    & (e^{i\frac{\pi}{4}} \ket{1}) \otimes (e^{i\frac{\pi}{4}} \ket{1}) \otimes (-i \ket{0}) = \ket{1} \otimes \ket{1} \otimes \ket{0} \\
    \ket{1} \otimes \ket{1} \otimes \ket{0} \to &(e^{i\frac{\pi}{4}}\ket{1})
    \otimes (e^{i\frac{\pi}{4}}\ket{1}) \otimes (-i \ket{1}) =
    \ket{1} \otimes \ket{1} \otimes \ket{1}.
  \end{align*}  
  In general, coming up with these constructions requires a good deal
  of experience, or a piece of code implementing the algorithms
  referenced in Sect.~\ref{sec:prelimnotes} to approximate any unitary
  with a universal set of gates.

\subsection{Can we solve NP-hard problems?}
\label{sec:solvenphard}
It is important to remark that even if we can easily create a uniform
superposition of all basis states, the rules of measurement imply that
using this easily-obtained superposition does not allow us to
immediately solve NP-hard\index{complexity!class NP} problems such as, for example, SAT (the
satisfiability problem). Indeed, suppose we have a quantum circuit
$U_f$ that computes the truth value of a SAT formula on $q$ boolean
variables; in other words, a unitary $U_f : \ket{\vj}_q \ket{0} \to
\ket{\vj}_q \ket{f(\vj)}$, where $f(\vj)$ is 1 if the binary string
$\vj$ satisfies the formula, and 0 if not.
\begin{remark}
  \label{rem:unspecifiedaction}
  The definition of $U_f$ is somewhat imprecise, because we only
  defined it for certain basis states rather than for all basis
  states; for the sake of exposition we ignore this, and come back to
  this topic in Sect.~\ref{sec:uncompute}.
\end{remark}
We might be tempted to apply $H^{\otimes q}$ to the initial state
$\ket{\v{0}}_q$ to create the uniform superposition
$\frac{1}{\sqrt{2^q}} \sum_{\vj \in \{0,1\}^q} \ket{\vj}$, append an
additional register with a single qubit initialized in the state
$\ket{0}$, apply $U_f$ to this superposition (which evaluates the
truth assignment of all possible binary strings, using the second
register to store the output), and then perform a measurement on all
$q+1$ qubits. But measuring the state:
\begin{equation*}
  U_f \left(\frac{1}{\sqrt{2^q}}
\sum_{\vj \in \{0,1\}^q} \ket{\vj} \ket{0}\right) = \frac{1}{\sqrt{2^q}}
\sum_{\vj \in \{0,1\}^q} \ket{\vj} \ket{f(\vj)}
\end{equation*}
returns a binary string that satisfies the formula if and only if
the last qubit has value 1 after the measurement, and this happens
with a probability that depends on the number of binary assignments
that satisfy the formula. If the SAT problem at hand is solved by
exactly $m$ assignments out of $2^n$ possible assignments, then the
probability of finding the solution after one measurement is
$\frac{m}{2^n}$: we have done nothing better than randomly sampling
a binary string and hoping that it satisfies the SAT formula. Clearly,
this is not a very smart algorithm. In fact, in general solving NP-hard
problems (such as SAT) in polynomial time is not believed to be
possible with quantum computers: most researchers believe that the
complexity class BQP\index{complexity!class BQP}, which is the class of problems solvable in
polynomial time by a quantum computer with bounded (and small) error
probability (see Def.~\ref{def:bqp}), does not contain the class
NP. Of course, one cannot hope to prove this unconditionally, because
showing NP $\not\subseteq$ BQP would resolve the famous P vs NP
problem. Nevertheless, it is strongly believed that NP $\not\subseteq$
BQP, due to the lower bound on black-box search of
\cite{bennett1997strengths}, and the inability of quantum computing
researchers to develop an efficient quantum algorithm for SAT (and not
for lack of trying).

Even if we cannot solve all difficult problems in polynomial time
using a quantum computer, we discuss in the next chapters some
examples of quantum algorithms that are faster than any known
classical algorithm.

\subsection{Dealing with errors}
\label{sec:vardistance}
In Sect.~\ref{sec:basicops} we observed that we can construct an
approximation of an arbitrary unitary matrix to some precision
$\epsilon$ in an efficient manner, i.e., using a number of elementary
gates (from some universal set) that scales as $\bigO{\log^c
  \frac{1}{\epsilon}}$. This is a positive result, because such a
scaling indicates that we can approximate the unitary with high
precision with a small increase in the required resources. However, in
principle we still need to concern ourselves with the total error of a
circuit that is composed of several unitaries, all of which may be
only an approximation of the ideal unitaries that we wanted to apply in
the first place. In this section we show that in fact we do not need
to be too concerned about this fact: the total error of a circuit is
at most the sum of the errors of the individual gates, therefore if we
want to approximate a circuit $U$ with $m$ gates up to precision
$\epsilon$, it suffices to approximate each gate to precision
$\epsilon/m$.\index{gate!approximation} In light of the logarithmic
error scaling of Thm.~\ref{thm:sk}, from a theoretical perspective
this is a fully satisfactory answer: if we have a quantum circuit that
solves a problem in a polynomial number of ``ideal'' gates, $m$ is a
polynomial of the input size, therefore compiling these ideal gates to
a universal set with error scaling $\bigO{\log \frac{m}{\epsilon}}$
only adds a small (i.e., polylogarithmic) number of gates. Even if $m$
is not polynomial in the input size, e.g., if we are implementing an
exponential-time algorithm, the number of extra gates remains very
small compared to the total complexity of the circuit.
\begin{proposition}
  \label{prop:unitaryerror}
  Let $U_1 U_2 \dots U_T$, $U_1' U_2' \dots U_T'$ be two sequences of
  unitaries of the same length.\index{unitary matrix!approximation} Then
  \begin{equation*}
    \nrm{U_1 U_2 \dots U_T - U_1' U_2' \dots U_T'} \le \sum_{j=1}^T \nrm{U_j - U_j'}.
  \end{equation*}
\end{proposition}
\begin{proof}
  By induction on the length $T$. When $T=1$ it is obvious. For larger
  $T$, we have:
  \begin{align*}
    &\phantom{=}\;\; \nrm{U_1 U_2 \dots U_T - U_1' U_2' \dots U_T'} \\
    &= \nrm{U_1 U_2 \dots U_{T-1} U_T - U_1' U_2' \dots U_{T-1}' U_T + U_1' U_2' \dots U_{T-1}' U_T - U_1' U_2' \dots U_{T-1}' U_T'} \\
    &= \nrm{(U_1 U_2 \dots U_{T-1} - U_1' U_2' \dots U'_{T-1}) U_T + U_1' U_2' \dots U_{T-1}' (U_T - U_T')} \\
    &\le \nrm{U_1 U_2 \dots U_{T-1} - U_1' U_2' \dots U'_{T-1}}\nrm{U_T} + \nrm{U_1' U_2' \dots U_{T-1}'} \nrm{U_T - U_T'} \\
    &\le \sum_{j=1}^{T-1} \nrm{U_j - U_j'} + \nrm{U_T - U_T'},
  \end{align*}
  where we used the induction hypothesis for the terms with
  $j=1,\dots,T-1$, the triangle inequality, Cauchy-Schwarz and the
  fact that unitary matrices (and their products) have unit operator
  norm.
\end{proof}

Throughout this discussion there is an implicit assumption that the
approximation metric of Def.~\ref{def:universal}, i.e., the
operator norm of the difference between a target unitary and its
approximation, is the right metric to use. We now show that this is
indeed the case, in the sense that a circuit $V$ that approximates a
target circuit $U$ up to some operator norm distance $\epsilon$ yields
almost the same output. To do so, we show that two quantum states with
Euclidean distance at most $\epsilon$ yield measurement outcome
distributions with total variation distance at most $\epsilon$.
\begin{definition}[Total variation distance]
  \label{def:tvd}
  Given two discrete probability distributions $P$ and $Q$ with the
  same sample space $\Omega = \{1,\dots,n\}$, let $p,q$ be the
  $n$-dimensional vectors with entries corresponding to the
  probability of $j=1,\dots,n$ according to $P, Q$ respectively.  The
  \emph{total variation distance}\index{total variation distance|see{variation distance, total}}\index{variation distance, total} between $P$ and $Q$ is
  \begin{equation*}
    d_{\text{TV}}(P, Q) := \frac{1}{2} \sum_{j=1}^n |p_j - q_j|.
  \end{equation*}
\end{definition}
\begin{remark}
  It is not difficult to show that the total variation distance
  between two probability distributions, as defined in
  Def.~\ref{def:tvd}, is also the maximum difference of the
  probability that these two distributions can assign to any event. In
  other words, $d_{\text{TV}}(P, Q) = \sup_{S \subseteq \{1,\dots,n\}}
  |\Pr_P(S) - \Pr_Q(S)|$.
\end{remark}
\begin{proposition}
  \label{prop:euclideantotvd}
  Let $\ket{\psi} = \sum_{\vj \in \{0,1\}^q} \alpha_{j} \ket{\vj},
  \ket{\phi} = \sum_{\vj \in \{0,1\}^q} \beta_{j} \ket{\vj}$ be two
  quantum states on $q$ qubits. Let $P, Q$ be the discrete probability
  distributions over $\{0,1\}^q$ induced, respectively, by
  $\ket{\psi}, \ket{\phi}$ when performing a measurement of all
  qubits. Suppose $\nrm{\ket{\psi} - \ket{\phi}} \le \epsilon$.\index{state!distance} Then
  $d_{\text{TV}}(P,Q) \le \epsilon$.
\end{proposition}
\begin{proof}
  Let us define vectors $a, b \in \R^{2^q}$ with entries
  $|\alpha_{j}|, |\beta_{j}|$ respectively. Furthermore, define the
  vectors $u, v \in \R^{2^q}$ with entries $u_j = |a_j + b_j|, v_j =
  |a_j - b_j|$. By Def.~\ref{def:tvd} we can write:
  \begin{align*}
    d_{\text{TV}}(P,Q) &= \frac{1}{2} \sum_{j} |a_j^2 - b_j^2| = \frac{1}{2} \sum_{j} |(a_j + b_j)(a_j - b_j)| \le \frac{1}{2} \sum_{j} |a_j + b_j||a_j - b_j| \\
    &= \frac{1}{2} u^{\top} v \le \frac{1}{2} \nrm{u} \nrm{v}.
  \end{align*}
  Let us analize $\nrm{u}, \nrm{v}$. For $\nrm{u}$, recalling that $\nrm{a}
  = \nrm{b} = 1$, we have:
  \begin{align*}
    \nrm{u}^2 &= \sum_{j} |a_j + b_j|^2 = \sum_{j}(a_j^2 + b_j^2 +
    2a_jb_j) = \nrm{a}^2 + \nrm{b}^2 + 2 a^{\top}b \\
    & \le \nrm{a}^2 + \nrm{b}^2 +
    2 \nrm{a}\nrm{b} \le 4.
  \end{align*}
  For $\nrm{v}$, we have:
  \begin{align*}
    \nrm{v}^2 &= \sum_{j} |a_j - b_j|^2 = \sum_{j}(a_j^2 + b_j^2 -
    2a_jb_j) = \sum_{j}(|\alpha_j|^2 + |\beta_j|^2 - 2 |\alpha_j||\beta_j|) \\
    &\le \sum_{j} (|\alpha_j|^2 + |\beta_j|^2 - 2 \Re(\alpha_j^{\dag} \beta_j)) = \nrm{ \ket{\psi} - \ket{\phi} }^2 \le \epsilon^2,
  \end{align*}
  where we used the fact that $|\alpha_j||\beta_j| \ge
  |\alpha_j^{\dag} \beta_j| \ge \Re(\alpha_j^{\dag} \beta_j)$. Putting
  everything together, we find:
  \begin{equation*}
    d_{\text{TV}}(P,Q) \le \frac{1}{2} \nrm{u}\nrm{v} \le \epsilon. 
  \end{equation*}
\end{proof}

\noindent Prop.~\ref{prop:euclideantotvd} tells us that if two quantum
states are close to each other in Euclidean norm, then any measurement
on the two states yields similarly-distributed outcomes. Thus, suppose
our goal is to prepare some state $\ket{\psi}$, encoding the answer to
some problem using an algorithm that is successful with probability
$1-\delta$. Suppose also that we can only prepare $\ket{\phi}$
instead, with the property that $\nrm{\ket{\psi} - \ket{\phi}} \le
\epsilon$; for example, this may happen because we do not know how to
implement the unitary that prepares $\ket{\psi}$ exactly, but we can
find an $\epsilon$-approximation of it. Eventually, to obtain the
answer to the problem from $\ket{\psi}$ we have to perform a
measurement; thanks to Prop.~\ref{prop:euclideantotvd}, we can perform
the measurement on $\ket{\phi}$ instead, knowing that we get the
correct answer with probability at least $1 - \delta - \epsilon$.

\subsection{Implicit measurement, reversibility, and uncomputation}
\label{sec:uncompute}
As a direct consequence of the laws of measurement
(Post.~\ref{pos:meas}), we can introduce the following general
principle of quantum computing that is often helpful when thinking
about what happens to a quantum state after measurement, see
\cite{nielsen02quantum} for further discussion.
\begin{proposition}[Principle of implicit measurement]
  \label{prop:implicitmeas}
  Any qubits that are not measured at the end of a quantum circuit may
  be assumed to be measured, and the corresponding information is
  discarded.\index{measurement!implicit}
\end{proposition}
The principle is quite intuitive if one thinks about the necessity of
quantum states to be consistent with the observations after a
measurement. Indeed, suppose we have a multiple-qubit state, and we
plan to apply a measurement on one specific qubit. For consistency,
the distribution of the measurement outcomes on that qubit before we
perform any measurement is the same as if we were instead planning to
measure all qubits at the same time, rather than just one. Thus, if we
apply a measurement to only some of the qubits, and there remain some
other qubits on which we never perform a measurement, we could also
assume that those other qubits have been measured as well, but the
corresponding measurement outcomes were discarded. (Assuming that some
qubits have been measured, and the corresponding outcomes are
discarded without looking at them, does not give us any additional
information and therefore does not modify the distribution of
measurement outcomes.)

This raises an issue concerning any information that might be stored
into working registers\index{register!working}: if we have a register that is used as working
space for some computation, and is subsequently discarded (i.e.,
ignored), we can assume that, at the end of the circuit, a measurement
is applied onto the working register as well. Moreover, the contents
of the working register may very well be entangled with other
registers, so we cannot reuse the working register: due to
entanglement, any additional operation on the working register risks
affecting the ``main'' (other) registers. This is easily clarified
with an example.
\begin{example}
  \label{ex:workingreg}
  Suppose we have a function $f$ that takes as
  input a binary string, and outputs a binary string, i.e., $f :
  \{0,1\}^m \to \{0,1\}^n$, where $m$ is not necessarily equal to
  $n$. Every operation on a quantum computer has to be reversible, so
  we must find an appropriate form of this function that can be
  represented as a valid operation for a quantum computer. The
  conventional way of constructing such a function is with a unitary
  that implements the following map:
  \begin{equation*}
    U_f \ket{\vx}_m \ket{\vy}_n = \ket{\vx}_m \ket{\vy \oplus f(\vx)}_n.
  \end{equation*}
  This map is defined for all input states (because it is defined for
  all input basis states), it allows us to read the value of $f(\vx)$
  (if we apply it when the second register contains $\v{0}$: $U_f
  \ket{\vx} \ket{\v{0}} = \ket{\vx} \ket{\v{0} \oplus f(\vx)} =
  \ket{\vx} \ket{f(\vx)}$), and it is reversible:
  \begin{equation*}
    U_f U_f (\ket{\v{x}} \ket{\v{y}}) = U_f (\ket{\v{x}}  \ket{\v{y}
      \oplus f(\v{x})}) = \ket{\v{x}}  \ket{\v{y} \oplus {f(\v{x})} \oplus f(\v{x})} =
    \ket{\v{x}}  \ket{\v{y}},
  \end{equation*}
  i.e., applying the circuit $U_f$ twice goes back to the initial
  state. (This way of specifying the function, whereby we apply
  $\oplus$ in the second register to write the output in a reversible
  way, lets us clarify the confusion regarding the unspecified action
  of $U_f$ in Rem.~\ref{rem:unspecifiedaction}.)

  Now assume that the computation carried out by $U_f$ requires some
  additional working space, as is often the case for all but the
  simplest functions (e.g., recall that the quantum version of the
  logic AND gate already requires a separate output qubit, see
  Sect.~\ref{sec:basicops}; so if we construct a circuit that contains
  many AND gates, we may use several auxiliary qubits to store
  intermediate results of AND operations). W.l.o.g.\ we can assume
  that there is a third register, containing $q$ bits and typically
  initialized in the all-zero basis state, used as working space. The
  mapping then becomes:
  \begin{equation*}
    U_f \ket{\vx}_m \ket{\vy}_n \ket{\v{0}}_q = \ket{\vx}_m \ket{\vy \oplus f(\vx)}_n \ket{g(\vx)}_q,
  \end{equation*}
  where $g(\vx)$ is some function of the input $\vx$ that outputs a
  binary string representing the final state of the working space. (We
  only define the output of this function when the working register
  contains $\v{0}$; if it does not, the circuit computes \emph{some}
  function of the input registers, but we do not need to characterize
  it because we plan to apply $U_f$ only if the working register
  contains $\v{0}$.) If we apply this map with $\v{y} = \v{0}$, we
  compute $f(\vx)$:
  \begin{equation*}
    U_f \ket{\vx} \ket{\v{0}} \ket{\v{0}} = \ket{\vx} \ket{f(\vx)} \ket{g(\vx)}.
  \end{equation*}
  The last register still contains $g(\vx)$, which is uninfluential,
  but it is still there, and it depends on $\vx$. Therefore, if we had
  a superposition over different values of $\vx$, all three registers
  --- including the last one --- would be entangled. If we apply a
  measurement onto the second register, and observe $f(\vx)$, the
  implicit measurement principle tells us that the last register also
  collapses to $g(\vx)$. Worse, we cannot reuse the last register as
  working register for additional function evaluations, because it
  still contains $g(\vx)$.
\end{example}

\begin{figure}[h!]
\leavevmode
\centering
\ifcompilefigs
\Qcircuit @C=1em @R=0.7em {
\lstick{\ket{\v{x}}} & \qw & {/^m} \qw & \qw & \multigate{2}{U_f} & \qw & \qw      & \qw & \multigate{2}{U_f^{\dag}} & \qw & \qw & \rstick{\ket{\v{x}}} \\
\lstick{\ket{\v{0}}} & \qw & {/^n} \qw & \qw & \ghost{U_f}        & \qw & \ctrl{2} & \qw & \ghost{U_f^{\dag}}        & \qw & \qw & \rstick{\ket{\v{0}}} \\
\lstick{\ket{\v{0}}} & \qw & {/^q} \qw & \qw & \ghost{U_f}        & \qw & \qw      & \qw & \ghost{U_f^{\dag}}        & \qw & \qw & \rstick{\ket{\v{0}}} \\
\lstick{\ket{\v{y}}} & \qw & {/^n} \qw & \qw & \qw                & \qw & \targ    & \qw & \qw                & \qw & \qw & \rstick{\ket{\v{y} \oplus f(\v{x})}}
}
\else
\includegraphics{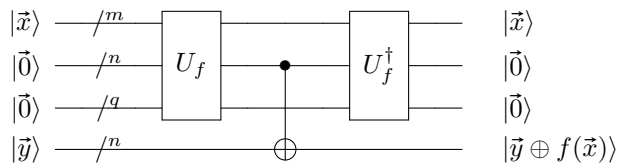}
\fi
\caption{A circuit implementing $U_f$ with an uncomputation step.}
\label{fig:uncompute}
\end{figure}

\noindent The fact that we cannot reuse the working register would
seem to imply that every function application needs its own working
register, so that a large number of qubits is needed even for
relatively simple calculations. In fact, we can avoid this issue, as
well as the issue of a working register entangled with the other
registers, by using a technique called \emph{uncomputation}.
\begin{definition}[Uncomputation]
  Let $U_f$ be a unitary that implements some Boolean function $f$
  using a working register, relying on the assumption that the working
  register is initialized to the all-zero binary string. To
  \emph{uncompute}\index{uncomputation} the function means to apply a sequence of
  operations to reset the state of the working register to the
  all-zero binary string.
\end{definition}
To uncompute a function, we introduce an auxiliary register with the
same size as the output register. Thus, in total, we have four
registers, which we order as follows: input, auxiliary, working, and
output register. The auxiliary and working registers are initialized
with the all-zero basis state. We first apply $U_f$ onto the input,
auxiliary, and working register; this writes the output of $U_f$, say,
$f(\vx)$, onto the auxiliary register. We then perform bitwise C$X$
from the auxiliary register onto the output register, to ``copy''
$f(\vx)$ into the output register using bitwise modulo-2
addition. Finally, we apply $U_f^{\dag}$ onto the input, auxiliary, and
working register, erasing the last two registers and resetting them to
the all-zero binary string (because $U_f \ket{\vx} \ket{\v{0}} \ket{\v{0}} = \ket{\vx} \ket{f(\vx)} \ket{g(\vx)}$, we have $U_f^{\dag} \ket{\vx} \ket{f(\vx)} \ket{g(\vx)} = \ket{\vx} \ket{\v{0}} \ket{\v{0}}$). The corresponding circuit is shown in
Fig.~\ref{fig:uncompute}.
Because the working and auxiliary registers are reset to $\v{0}$ at
the end of the circuit, they are no longer entangled with the rest and
can be reused for other purposes, thereby saving working space. We
could also SWAP the auxiliary and output register if we want the
output in the second register, obtaining the same order of the
registers as discussed in Ex.~\ref{ex:workingreg}.

\section{Mixed states and purifications}
\label{sec:mixedstate}
Our discussion so far has been based on Post.~\ref{pos:state}: the
state of a $q$-qubit quantum computer is a unit vector in
$(\C^{2})^{\otimes q}$. This is correct, but there are situations
where such a formalism is not the best approach to describe the state
of a quantum register, or to determine its evolution. For example,
consider the situation of a system with several qubits, but starting
from a certain point of the computation we only apply operations onto
some of the qubits, and discard (i.e., ignore) the rest. How should we
characterize the state of a system when some of the qubits have been
discarded?

\begin{example}
  \label{ex:qubitsubset}
  Suppose we are in the state:
  \begin{equation*}
    \frac{1}{2} \ket{00} + \frac{1}{2} \ket{01} + \frac{1}{\sqrt{2}}
    \ket{11},
  \end{equation*}
  but we only have access to the first qubit; this might be the case
  if the second qubit is physically distant (i.e., the above state was
  constructed over a quantum network), or simply because from now on
  we only want to perform computation on the first qubit line. The
  principle of implicit measurement tells us that ignoring the second
  qubit is equivalent to performing a measurement and discarding the
  information that we obtained. The measurement outcomes for both
  qubits are:
  \begin{align*}
    \Pr(00) = \frac{1}{4} \qquad \Pr(01) = \frac{1}{4} \qquad \Pr(11) = \frac{1}{2}.
  \end{align*}
  If the outcome of the measurement on the second qubit is 0, the
  first qubit is in state $\ket{0}$ with certainty. If the outcome of
  the measurement on the second qubit is 1, the first qubit might be
  in state $\ket{0}$ or $\ket{1}$. Because the outcome of the
  measurement on the second qubit is assumed to be discarded (by
  principle of implicit measurement), we do not know how to
  characterize the state of the first qubit within the formalism for
  quantum states used so far: there is no single-qubit state (i.e., a
  unit vector in $\C^2$) that would accurately describe the state of
  the first qubit. We could find a quantum state that matches the
  probability distribution of the measurement outcomes, but we know
  from Ex.~\ref{ex:probdistdanger} that this is not sufficient to
  correctly characterize a quantum state.
\end{example}

As seen in the above example, our formalism to represent the state of
a quantum register does not work very well when we want to consider
only a subset of the qubits of a larger system. There is another
formalism to express the state of a quantum computer: it is the
language of \emph{mixed states}, as opposed to the \emph{pure states}
that we have studied so far. Mixed states generalize pure states, and
they are better able to deal with situations such as the one in
Ex.~\ref{ex:qubitsubset}, at the price of more cumbersome notation and
calculations.
\begin{definition}[Pure state]
  \label{def:purestate}
  A \emph{pure state}\index{state!pure} $\ket{\psi}$ on $q$ qubits is a unit vector in
  $(\C^{2})^{\otimes q}$, i.e., the state of a $q$-qubit quantum
  register.
\end{definition}
\begin{definition}[Mixed state]
  \label{def:mixedstate}
  An \emph{ensemble of pure states} on $q$ qubits is a collection
  $\{p_j, \ket{\psi_j}\}_{j=1,\dots,m}$ of pure states $\ket{\psi_j}$
  and corresponding probabilities $p_j$. A $q$-qubit quantum register
  that is in state $\ket{\psi_j}$ with probability $p_j$ for
  $j=1,\dots,m$ is said to be in a \emph{mixed
  state}\index{quantum!state}\index{state!mixed}, described by the
  \emph{density matrix}\index{matrix!density}\index{density matrix!definition} corresponding to the ensemble, defined as the
  matrix $\rho := \sum_{j=1}^{m} p_j \ketbra{\psi_j}{\psi_j}.$
\end{definition}
\begin{remark}
  Recall our notation: $\ket{\psi_j}$ is a column vector,
  $\bra{\psi_j}$ is a row vector, so $\ketbra{\psi_j}{\psi_j}$ is the
  matrix that performs the orthogonal projection onto $\ket{\psi_j}$.
\end{remark}
\begin{remark}
  Density matrices are also called \emph{density
  operators}. Technically the matrix is the representation of the
  operator once we choose a basis, but because we always use the
  standard orthonormal basis, in the context of this \book{} these terms
  are fully interchangeable. We use density matrix in the following.
\end{remark}
It is relatively straightforward to rewrite
Post.s~\ref{pos:operations} and \ref{pos:meas} to work with mixed
states. It is sufficient to write the expression for the application
of an operation to each state $\ket{\psi_j}$ in the ensemble, and
define the resulting collection of states as the new ensemble, with
the same probability distribution as before. This yields the
following.
\begin{itemize}
\item The application of a $q$-qubit gate $U$ to the register in state
  $\rho = \sum_{j=1}^{m} p_j \ketbra{\psi_j}{\psi_j}$ evolves the
  system to the state: $U \rho U^{\dag} = \sum_{j=1}^{m} p_j
  U\ketbra{\psi_j}{\psi_j}U^{\dag}$.
\item Define the matrices:
  \begin{align*}
    M^{(0)}_k = \underbrace{I_{2 \times 2} \otimes I_{2 \times 2} \otimes \cdots \otimes \overbrace{\ketbra{0}{0}}^{k\text{-th position}} \otimes \cdots \otimes I_{2 \times 2} \otimes I_{2 \times 2}}_{q \text{ times}} \\
    M^{(1)}_k = \underbrace{I_{2 \times 2} \otimes I_{2 \times 2} \otimes \cdots \otimes \overbrace{\ketbra{1}{1}}^{k\text{-th position}} \otimes \cdots \otimes I_{2 \times 2} \otimes I_{2 \times 2}}_{q \text{ times}} .
  \end{align*}
  A measurement\index{measurement!definition} gate on qubit $k$ yields a sample from a random variable $\mathcal{Q}_k$ with sample space $\{0,1\}$, $\Pr(\mathcal{Q}_k = x) = \trace{M^{(x)}_k\rho}$ for $x \in \{0,1\}$, and the state after the measurement becomes:
  \begin{equation*}
    \frac{M^{(x)}_k\rho (M^{(x)}_k)^{\dag}}{\trace{M^{(x)}_k\rho}} = \frac{M^{(x)}_k\rho M^{(x)}_k}{\trace{M^{(x)}_k\rho}}.
  \end{equation*}
\end{itemize}
\begin{remark}
  \label{rem:projectivemeas}
  The derivation of the expression for the state after a measurement
  highlights the fact that the effect of this type of measurement is
  captured by a projection matrix: $M^{(x)}_k$ projects a pure state
  onto the components that are consistent with the measurement.
  Indeed, this is called a \emph{projective measurement}.\index{measurement!projective} 
\end{remark}
\begin{remark}
  \label{rem:densitymatmeas}
  If we apply measurements to all qubits, by repeating the procedure
  described above for single-qubit measurements we see that the
  probability of obtaining outcome $\ket{\v{k}}$, $\vk \in \{0,1\}^q$,
  is $\trace{\ketbra{\vk}{\vk} \rho} = \bra{\vk} \rho \ket{\vk}$ (due
  to the cyclic property of the trace), and if we observe $\ket{\vk}$,
  the state after measurement becomes $\ketbra{\vk}{\vk}$. This is
  consistent with the pure state formalism, because each state in the
  ensemble collapses to $\ket{\vk}$, therefore now the quantum
  register is in the state $\ket{\vk}$ with certainty.
\end{remark}
There are more general types of measurements than the projective
measurements described above. The most general expression for a
measurement operator $M_k$ is that we observe the corresponding
outcome with probability $\trace{M_k\rho M_k^{\dag}}$, and the state
after the measurement becomes:
\begin{equation*}
  \frac{M_k\rho M_k^{\dag}}{\trace{M_k\rho M_k^{\dag}}},
\end{equation*}
see \cite[Sect.~2.4.1]{nielsen02quantum}. However, we never use the
general case in this \book{}: the projective measurement onto the
states $\ket{0}$ or $\ket{1}$ (usually called \emph{measurement in the
computational basis})\index{basis!computational} suffices. It is easy
to see that for measurements in the computational basis, the general
formulas for the outcome probabilities and for the state after the
measurement reduce to the simplified ones given before
Rem.~\ref{rem:projectivemeas}, due to the fact that the matrices
$M_k^{(x)}$ are Hermitian projections, and using the cyclic property
of the trace.

\subsection{Properties of density matrices}
Density matrices are precisely characterized by two properties: they
have unit trace, and they are positive semidefinite. We use the
standard notation for positive (semi)definite matrices: $\rho \succeq
0$ indicates that $\rho$ is positive semidefinite, $\rho \succeq
\rho'$ indicates that $\rho - \rho'$ is positive semidefinite, and
similar notation with $\succ$ instead of $\succeq$ to indicate
positive definiteness.
\begin{theorem}[Characterization of density matrices]
  The matrix $\rho$ is a density matrix associated with some ensemble
  of pure states $\{p_j, \allowbreak \ket{\psi_j}\}_{j=1,\dots,m}$ (for some
  unknown $m$) if and only if it satisfies the following two
  properties: $(i)$ it has unit trace, and $(ii)$ it is positive
  semidefinite.\index{density matrix!characterization}
\end{theorem}
\begin{proof}
  First let us suppose $\rho$ is a density matrix associated with the
  ensemble of pure states $\{p_j, \ket{\psi_j}\}_{j=1,\dots,m}$. Then,
  using the cyclic property of the trace:
  \begin{align*}
    \trace{\rho} = \trace{\sum_{j=1}^m p_j \ketbra{\psi_j}{\psi_j}} = \sum_{j=1}^m p_j \trace{\ketbra{\psi_j}{\psi_j}} = \sum_{j=1}^m p_j = 1,
  \end{align*}
  and
  \begin{align*}
    \bra{\phi} \rho \ket{\phi} = \sum_{j=1}^m p_j \braket{\phi}{\psi_j}\braket{\psi_j}{\phi} = \sum_{j=1}^m p_j |\braket{\phi}{\psi_j}|^2 \ge 0
  \end{align*}
  for every vector $\ket{\phi}$ (even unnormalized ones).

  Then, let us suppose $\rho$ has unit trace and $\rho \succeq 0$. We
  want to show it corresponds to some ensemble of pure states. Because
  $\rho$ is a Hermitian positive semidefinite matrix, it admits a
  spectral decomposition with an orthonormal eigenbasis, and its
  eigenvalues are real and nonnegative. So
  \begin{equation*}
    \rho = \sum_{j=1}^m p_j \ketbra{\psi_j}{\psi_j}
  \end{equation*}
  for some $m \le$ number of columns of $\rho$,
  values $p_j$, and vectors $\ket{\psi_j}$. Because $\trace{\rho} = 1$
  we also have $\sum_{j=1}^m p_j = 1$, therefore $\rho$ describes an
  ensemble of pure states.
\end{proof}
\begin{remark}
  There could be multiple ensembles that correspond to the same
  density matrix, i.e., the spectral decomposition may not be
  unique. For example, suppose we have a unitary transformation $U$,
  and we define $\sqrt{q_i} \ket{\phi_i} = \sum_{j} U_{ij} \sqrt{p_j}
  \ket{\psi_j}$. Then:
  \begin{align*}
    \sum_{i} q_i \ketbra{\phi_i}{\phi_i} &= \sum_{i} (\sum_{j} U_{ij}
    \sqrt{p_j} \ket{\psi_j}) (\sum_{j} \bra{\psi_j} \sqrt{p_j} (U_{ij})^{\dag}) = \sum_{i,j,k} U_{ij} (U_{ik})^{\dag}
    \sqrt{p_jp_k} \ketbra{\psi_j}{\psi_k}  \\
    &= \sum_{j,k} \sum_i \left(U_{ij} (U_{ik})^{\dag}\right)
    \sqrt{p_jp_k} \ketbra{\psi_j}{\psi_k} = \sum_{j,k} \sum_i \left(U_{ij} U_{ki}^{\dag}\right)
    \sqrt{p_jp_k} \ketbra{\psi_j}{\psi_k} \\
    &= \sum_{j} p_j \ketbra{\psi_j}{\psi_j},
  \end{align*}
  where the fourth equality follows by definition of conjugate
  transpose of a matrix (notice that the indices $i, k$ get swapped),
  and the last equality is due to the fact that $U$ is unitary matrix
  so the term in parentheses is $1$ if $j=k$, and $0$ otherwise.
  This shows that the ensembles $\{p_j, \ket{\psi_j}\}_{j}$ and
  $\{q_i, \ket{\phi_i}\}_{i}$ are represented by the same density
  matrix. In fact, it is possible to show that this type of
  transformation between two ensembles is not only a sufficient
  condition to have the same density matrix, but also necessary, see
  \cite[Thm.~2.6]{nielsen02quantum}.
\end{remark}

\subsection{Reduced density matrix}
\label{sec:reduceddensmat}
Arguably one of the greatest advantages of the density matrix
formalism is the fact that it allows a rigorous treatment of the
situation discussed in Ex.~\ref{ex:qubitsubset}: we have a quantum
register of a certain size, but we want to study the state of only a
subset of the qubits, and continue the computation on those qubits
while disregarding the rest. Naturally we could look at the evolution
of the pure state of the entire register, but sometimes this is not
feasible, or it is mathematically cumbersome; even when it is
possible, it still faces the issue that in the pure state formalism,
we can no longer describe the state of only the qubits that we are
interested in. However, by implicit measurement it is easy to see that
the state of a subset of qubits in the register is described by an
ensemble of pure states, where the probability distribution is
obtained by looking at the distribution of the measurement outcomes of
the other qubits (i.e., the ones that are disregarded). The density
matrix formalism provides an abstraction and computational rules for
this concept.

Formally, suppose we have a quantum register $AB$ whose state is
described by the density matrix $\rho^{(AB)}$, and we split the
register into two distinct quantum registers $A$ and $B$. We now
define the \emph{reduced density matrix} obtained by tracing out one
of the registers, and we claim that this describes the state of only
one of the registers, in some sense that is formally specified in
Prop.~\ref{prop:reddensmatmeas}.
\begin{definition}[Partial trace]
  \label{def:partialtrace}
  Let $AB$ be a quantum register composed of two registers $A,B$ with
  $m_a, m_b$ qubits respectively. The \emph{partial trace}\index{trace!partial}\index{partial trace} over
  register $B$ is the operation $\Tr_B$ defined as follows:
  \begin{enumerate}[(i)]
  \item for any $\vj, \vk \in \{0,1\}^{m_a}, \v{h}, \v{\ell} \in
    \{0,1\}^{m_b}$, we have:
    \begin{equation*}
      \trace[B]{\ketbra{\vj}{\vk} \otimes \ketbra{\v{h}}{\v{\ell}}} := \ketbra{\vj}{\vk} \trace{\ketbra{\v{h}}{\v{\ell}}} = \ketbra{\vj}{\vk} \braket{\v{\ell}}{\v{h}};
    \end{equation*}
  \item $\Tr_B$ is linear.
  \end{enumerate}
\end{definition}
This is a proper definition, because $\Tr_B$ is linear, and we are
defining its effect on every basis element for the space of
density matrices over $AB$. Computing the partial trace over register
$B$ is often called \emph{tracing out} register $B$. In the setting of
Def.~\ref{def:partialtrace}, let $\rho^{(AB)}$ be the density matrix
describing the state of register $AB$. An alternative definition of
the partial trace, that might appear more intuitive to some readers,
is given by the following expression:
\begin{equation*}
  \trace[B]{\rho^{(AB)}} := \sum_{\vj \in \{0,1\}^{m_b}} (I^{\otimes m_a} \otimes \bra{\vj}) \rho^{(AB)} (I^{\otimes m_a} \otimes \ket{\vj}),
\end{equation*}
where $I^{\otimes m_a}$ is the identity matrix of size $2^{m_a} \times
2^{m_a}$, i.e., the appropriate size for register $A$. We can of
course give similar definitions swapping the role of registers $A$
and $B$, and obtain the partial trace over register $A$, which is the
following operation:
\begin{equation*}
  \trace[A]{\rho^{(AB)}} := \sum_{\vj \in \{0,1\}^{m_a}} (\bra{\vj} \otimes I^{\otimes m_b}) \rho^{(AB)} (\ket{\vj} \otimes I^{\otimes m_b}),
\end{equation*}
or, in the style of Def.~\ref{def:partialtrace}, it is a linear
operation with $\trace[A]{\ketbra{\vj}{\vk} \otimes
  \ketbra{\v{h}}{\v{\ell}}} := \trace{\ketbra{\vj}{\vk}}
\ketbra{\v{h}}{\v{\ell}}.$
\begin{remark}
  The original and alternative definitions are equivalent because the
  elements $\ketbra{\vj}{\vk} \otimes \ketbra{\v{h}}{\v{\ell}}$
  constitute a basis for the space of density matrices $\rho^{(AB)}$,
  and if we express $\rho^{(AB)}$ in this basis and apply the linear
  operator $\Tr_B$, we obtain:
  \begin{align*}
    \sum_{\vj, \vk, \v{h}, \v{\ell}} \trace[B]{\rho^{(AB)}_{\ketbra{\vj}{\vk} \otimes \ketbra{\v{h}}{\v{\ell}}} \ketbra{\vj}{\vk} \otimes \ketbra{\v{h}}{\v{\ell}}} =
    \sum_{\vj, \vk, \v{h}} \rho^{(AB)}_{\ketbra{\vj}{\vk} \otimes \ketbra{\v{h}}{\v{h}}} \ketbra{\vj}{\vk} \braket{\v{h}}{\v{h}} = \sum_{\vj, \vk, \v{h}} \rho^{(AB)}_{\ketbra{\vj}{\vk} \otimes \ketbra{\v{h}}{\v{h}}} \ketbra{\vj}{\vk},
  \end{align*}
  where we denoted by $\rho^{(AB)}_{\ketbra{\vj}{\vk} \otimes
    \ketbra{\v{h}}{\v{\ell}}}$ the element of $\rho^{(AB)}$ in the
  position corresponding to the nonzero element of the subscript
  matrix (one can think of the subscript as a ``mask'' to identify the
  correct element). Because $\ketbra{\vj}{\vk}$ is a basis for
  register $A$ and we keep all the corresponding elements for register
  $A$, we are effectively ``acting as the identity'' on the first
  register, but we only consider the elements of $\rho^{(AB)}$ that
  characterize the quantum state if we were to perform a measurement
  on the second register, collapsing to some basis state
  $\ket{\v{h}}$; then, we sum over all $\ket{\v{h}}$, by principle of
  implicit measurement.
\end{remark}
\begin{definition}[Reduced density matrix]
  \label{def:reddensitymat}
  In the setting of Def.~\ref{def:partialtrace}, let $\rho^{(AB)}$ be
  the density matrix describing the state of register $AB$. The
  \emph{reduced density matrix}\index{density matrix!reduced} for register $A$ is $\rho^{(A)} :=
  \trace[B]{\rho^{(AB)}}$, and similarly, the reduced density matrix
  for register $B$ is $\rho^{(B)} := \trace[A]{\rho^{(AB)}}$.
\end{definition}
A reduced density matrix characterizes the state of a subsystem (i.e.,
subregister) of the entire register, as can be seen in the following
examples.
\begin{example}
  Consider the state:
  \begin{equation*}
    \frac{1}{\sqrt{2}} (\ket{00} + \ket{11}).
  \end{equation*}
  The corresponding density matrix is:
  \begin{align*}
    \frac{1}{2} (\ket{00} + \ket{11})(\bra{00} + \bra{11}) =
    \frac{1}{2}(\ketbra{00}{00} + \ketbra{00}{11} + \ketbra{11}{00} + \ketbra{11}{11}) 
    = \begin{pmatrix} \frac{1}{2} & 0 & 0 & \frac{1}{2} \\ 0 & 0 & 0 & 0 \\ 0 &
      0 & 0 & 0 \\ \frac{1}{2} & 0 & 0 & \frac{1}{2} \end{pmatrix} =
    \rho^{(AB)}.
  \end{align*}
  Let us denote the first qubit as register $A$ and the second qubit
  as register $B$.  If we now want to drop the second qubit and
  consider only the first, its state is represented by the following
  reduced density matrix:
  \begin{align*}
    \trace[B]{\frac{1}{2}(\ketbra{00}{00} + \ketbra{00}{11} + \ketbra{11}{00} + \ketbra{11}{11})} &=
    \frac{1}{2} (\ketbra{0}{0} \braket{0}{0} + \ketbra{0}{1}\braket{0}{1} + \ketbra{1}{0}\braket{1}{0} + \ketbra{1}{1}\braket{1}{1}) \\
    &= \frac{1}{2}(\ketbra{0}{0} + \ketbra{1}{1}).
  \end{align*}
  Using the alternative definition, we equivalently obtain:
  \begin{align*}
    \rho^{(A)} &= \trace[B]{\rho^{(AB)}} = \sum_{j=0,1} (I \otimes \bra{j} ) \rho^{(AB)} (I \otimes \ket{j}) \\
    &= \begin{pmatrix} 1 & 0 & 0 & 0 \\ 0 & 0 & 1 & 0 \end{pmatrix}
    \begin{pmatrix} \frac{1}{2} & 0 & 0 & \frac{1}{2} \\ 0 & 0 & 0 & 0 \\ 0 &
      0 & 0 & 0 \\ \frac{1}{2} & 0 & 0 & \frac{1}{2} \end{pmatrix}
    \begin{pmatrix} 1 & 0 \\ 0 & 0 \\ 0 & 1 \\ 0 & 0 \end{pmatrix} +
    \begin{pmatrix} 0 & 1 & 0 & 0 \\ 0 & 0 & 0 & 1 \end{pmatrix}
    \begin{pmatrix} \frac{1}{2} & 0 & 0 & \frac{1}{2} \\ 0 & 0 & 0 & 0 \\ 0 &
      0 & 0 & 0 \\ \frac{1}{2} & 0 & 0 & \frac{1}{2} \end{pmatrix}
    \begin{pmatrix} 0 & 0 \\ 1 & 0 \\ 0 & 0 \\ 0 & 1 \end{pmatrix} \\
    &= \begin{pmatrix} \frac{1}{2} & 0 \\ 0 & 0 \end{pmatrix}
    + \begin{pmatrix} 0 & 0 \\ 0 & \frac{1}{2} \end{pmatrix}
    = \begin{pmatrix} \frac{1}{2} & 0 \\ 0 & \frac{1}{2} \end{pmatrix} =
    \frac{1}{2}(\ketbra{0}{0} + \ketbra{1}{1}).
  \end{align*}
  Intuitively this makes sense: from the initial state, if we ignore
  the second qubit we still have a system that is $\ket{0}$ or
  $\ket{1}$ with probability $0.5$ each, which is what we see from the
  reduced density matrix. We do not have additional information on the
  first qubit because, even though it is entangled with the second
  qubit and it might be considered measured by principle of implicit
  measurement, we discard the outcome of the measurement on the second
  qubit.
\end{example}
\begin{example}
  Let us study Ex.~\ref{ex:qubitsubset} using the formalism of
  reduced density matrices. Recall that in that example, we are
  considering the pure state:
  \begin{equation*}
    \frac{1}{2} \ket{00} + \frac{1}{2} \ket{01} + \frac{1}{\sqrt{2}}
    \ket{11},
  \end{equation*}
  and we want to analyze what happens if we want to describe the state
  of the first qubit only. The density matrix for the entire system
  is:
  \begin{align*}
    \rho^{(AB)} =& \left(\frac{1}{2} \ket{00} + \frac{1}{2} \ket{01} + \frac{1}{\sqrt{2}}
    \ket{11}\right)\left(\frac{1}{2} \bra{00} + \frac{1}{2} \bra{01} + \frac{1}{\sqrt{2}}
    \bra{11}\right) \\
    =& \;\frac{1}{4} \ketbra{00}{00} + \frac{1}{4} \ketbra{00}{01} + \frac{1}{2\sqrt{2}} \ketbra{00}{11} + \frac{1}{4} \ketbra{01}{00} + \frac{1}{4} \ketbra{01}{01} + \\ & \;\frac{1}{2\sqrt{2}}\ketbra{01}{11} + \frac{1}{2\sqrt{2}} \ketbra{11}{00} + \frac{1}{2\sqrt{2}} \ketbra{11}{01} + \frac{1}{2}\ketbra{11}{11}.
  \end{align*}
  Calling $B$ the register with the second qubit, and tracing it out, we obtain:
  \begin{align*}
    \trace[B]{\rho^{(AB)}} = \frac{1}{2} \ketbra{0}{0} + \frac{1}{2\sqrt{2}} \ketbra{0}{1} + \frac{1}{2\sqrt{2}} \ketbra{1}{0} + \frac{1}{2} \ketbra{1}{1}.
  \end{align*}
  Recalling Ex.~\ref{ex:qubitsubset}, let us consider the ensemble of
  pure states where a qubit is in state $\ket{0}$ with probability
  $1/4$, and is in state $\frac{1}{\sqrt{3}} \ket{0} +
  \sqrt{\frac{2}{3}} \ket{1}$ with probability $3/4$; this is a
  natural description of the state of the first qubit, if we apply the
  principle of implicit measurement and look at what happens if we
  observe the second qubit to be $\ket{0}$ or $\ket{1}$. (The
  probabilities $1/4$, $3/4$ are the eigenvalues obtained from an
  eigendecomposition of $\trace[B]{\rho^{(AB)}}$, and the pure states
  in the ensemble are the corresponding eigenvectors; here we
  constructed them from intuition, looking for probabilities and pure
  states that yield the probability distribution that we expect to
  see.) The density matrix corresponding to this ensemble is:
  \begin{align*}
    \frac{1}{4} \ketbra{0}{0} + \frac{3}{4}\left(\frac{1}{\sqrt{3}} \ket{0} +
  \sqrt{\frac{2}{3}} \ket{1}\right)\left(\frac{1}{\sqrt{3}} \bra{0} +
  \sqrt{\frac{2}{3}} \bra{1}\right) = \frac{1}{2} \ketbra{0}{0} + \frac{1}{2\sqrt{2}} \ketbra{0}{1} + \frac{1}{2\sqrt{2}} \ketbra{1}{0} + \frac{1}{2} \ketbra{1}{1},
  \end{align*}
  which is precisely $\trace[B]{\rho^{(AB)}}$. Thus, the chosen
  ensemble yields the reduced density matrix obtained above after
  tracing out register $B$, and is, in fact, a natural expression for
  the state of the first qubit. In general, the eigendecomposition of
  a reduced density matrix is not unique: not only there could be
  eigenvalues with algebraic multiplicity greater than one, but when
  we consider global phases, there are infinite eigenvectors that are
  indistinguishable in the sense that they yield the same probability
  distribution of measurement outcomes. So, there are are multiple
  ensembles of pure states that equivalently describe the state of the
  same system.
\end{example}

\noindent We can now precisely state in what sense density matrices
correctly characterize the state of a register after discarding (or
ignoring) some other registers: they lead to the correct probability
distribution of the measurement outcomes. We state this result using
the reduced density matrix for register $A$, but clearly we can obtain
a symmetric result for register $B$.
\begin{proposition}
  \label{prop:reddensmatmeas}
  Let register $AB$ be in a state characterized by the density matrix
  $\rho^{(AB)}$, and let $\rho^{(A)} = \trace[B]{\rho^{(AB)}}$. Then
  $\rho^{(A)}$ correctly characterizes the probabilities of the
  measurement outcomes for register $A$, when discarding register $B$.
\end{proposition}
\begin{proof}
  We need to show that the probability of observing outcome
  $\vj$ from a measurement on register $A$ is the same whether we
  compute it from $\rho^{(A)}$ or starting from the original mixed
  state $\rho^{(AB)}$.

  Using the entire system, the probability of observing outcome $\vj$
  is (recall Rem.~\ref{rem:densitymatmeas}):
  \begin{equation*}
    \sum_{\vk \in \{0,1\}^{m_b}} \bra{\vj}\bra{\vk} \rho^{(AB)}
    \ket{\vj}\ket{\vk},
  \end{equation*}
  because it is equal to the probability of observing any string
  starting with $\vj$ if we apply a measurement on all qubits. Using
  the reduced density matrix, the probability of observing outcome
  $\vj$ in the first register is:
  \begin{align*}
    \bra{\vj} \rho^{(A)}
    \ket{\vj} = \bra{\vj} \left(\sum_{\vk \in \{0,1\}^{m_b}} (I^{\otimes m_a} \otimes \bra{\vk}) \rho^{(AB)} (I^{\otimes m_a} \otimes \ket{\vk})\right) \ket{\vj} = \sum_{\vk \in \{0,1\}^{m_b}} \bra{\vj} \bra{\vk} \rho^{(AB)} \ket{\vj} \ket{\vk},
  \end{align*}
  so we obtain the same probability as above.
\end{proof}

\subsection{Purifications}
\label{sec:purifications}
An important concept in the study of density matrices is the idea of a
purification; this is also crucial in several quantum algorithms for
semidefinite optimization, and in particular in Ch.~\ref{ch:mmwu} we
exploit purifications to construct mixed states corresponding to
candidate solutions for semidefinite optimization problems. Our
previous discussion shows that a mixed state is described by a density
matrix. We show next that given a density matrix $\rho$, we can
construct a \emph{pure state} on two registers such that tracing out
one of the registers yields a mixed state corresponding to $\rho$.
\begin{theorem}[Every density matrix admits a purification]
  \label{thm:purification}
  Let $d = 2^q$. Let $\rho \in \C^{d \times d}$ be a given density
  matrix. Then, there exists a pure state $\ket{\phi}$ over two
  registers $A$, $B$ such that $A$ has $q$ qubits and tracing out $B$
  yields a mixed state described by $\rho$ in register $A$. Moreover,
  it is possible to choose register $B$ so that it has at most $q$
  qubits.
\end{theorem}
\begin{proof}
  Let $\rho = \sum_{j=0}^{d-1} \lambda_j \ketbra{\psi_j}{\psi_j}$ be
  an eigendecomposition of $\rho$, which always exists because $\rho$
  is a Hermitian positive semidefinite matrix. Note that $\lambda_j
  \in \R$ and they are nonnegative. Furthermore, we can assume that
  there are $d$ eigenvalues without loss of generality: if there are
  fewer we can simply add some zero eigenvalues, and clearly there
  cannot be more because the rank of $\rho$ is at most $d$. Let
  $\ket{\psi_j} := \sum_{\vk} \alpha^{(j)}_{k} \ket{\vk}$ for some
  vector of coefficients $\alpha^{(j)}$. Consider the pure state
  \begin{equation*}
    \ket{\phi} = \sum_{\vj \in \{0,1\}^q} \sqrt{\lambda_j}
    \ket{\psi_j} \ket{\vj}
  \end{equation*}
  over $2q$ qubits (each register has $q$ qubits). Tracing out the
  second register, which we call register $B$, yields:
  \begingroup
  \allowdisplaybreaks
  \begin{align*}
    \trace[B]{\ketbra{\phi}{\phi}} &= \trace[B]{\sum_{\vj, \vk} \sqrt{\lambda_j \lambda_k} \ket{\psi_j} \ketbra{\vj}{\psi_k} \bra{\vk}}\\
    &= \trace[B]{\sum_{\vj, \vk} \sqrt{\lambda_j \lambda_k} (\sum_{\v{h}} \alpha^{(j)}_{h} \ket{\v{h}}) \ket{\vj} (\sum_{\v{\ell}} (\alpha^{(k)}_{\ell})^{\dag} \bra{\v{\ell}}) \bra{\vk}} \\
    &= \sum_{\vj, \vk} \sqrt{\lambda_j \lambda_k} (\sum_{\v{h}} \alpha^{(j)}_{h} \ket{\v{h}})  (\sum_{\v{\ell}} (\alpha^{(k)}_{\ell})^{\dag} \bra{\v{\ell}}) \trace{\ketbra{\vj}{\vk}} \\
    &= \sum_{\vj, \vk} \sqrt{\lambda_j \lambda_k} \ketbra{\psi_j}{\psi_k} \braket{\vk}{\vj} = \sum_{j=0}^{d-1} \lambda_j \ketbra{\psi_j}{\psi_j} = \rho,
  \end{align*}
  \endgroup
  concluding the proof. (In the third equality, we applied the
  definition of $\Tr_{B}$.)
\end{proof}

\noindent Essentially, the second register is used to construct the
ensemble of pure states on the first register by assigning the correct
probability to each state of the ensemble. This leads to the concept
of a \emph{purification}. We use purifications in
Ch.s~\ref{ch:blockenc} and \ref{ch:mmwu}.
\begin{definition}[Purification]
  \label{def:purification}
  Given a density matrix $\rho$ describing the state of register $A$,
  a \emph{purification}\index{density matrix!purification}\index{purification} of $\rho$ is a pure state over two registers
  $A,B$ such that tracing out register $B$ yields $\rho$.
\end{definition}
\noindent The register $B$ that is traced out is typically called
\emph{purifying register}.\index{register!purifying}

We conclude this section with another result that is often useful in
the study of composite systems (i.e., registers with subregisters),
and that plays a crucial role in some classical simulation algorithms
for quantum circuits, e.g.,
\cite{markov2018quantum,nanni49qubitsjournal}. The result can be seen
as a restatement of the singular value decomposition, but in quantum
information theory it is referred to as \emph{Schmidt decomposition}.\index{Schmidt decomposition}\index{decomposition!Schmidt}
\begin{theorem}[Schmidt decomposition]
  \label{thm:schmidt}
  Let $\ket{\psi}$ be a pure state for a composite register $AB$. Then,
  there exist orthonormal states $\ket{\phi^A_j}$ for register $A$,
  and $\ket{\phi^B_j}$ for register $B$, such that $\ket{\psi} =
  \sum_{j} \lambda_j \ket{\phi^A_j}\ket{\phi^B_j}$, where $\lambda_j$
  are nonnegative reals such that $\sum_j \lambda_j^2 = 1$.
\end{theorem}
\begin{proof}
  Let $\ket{\psi} = \sum_{\vj, \vk} \alpha_{jk}
  \ket{\vj}\ket{\vk}$. Arrange the coefficients $\alpha_{jk}$ into
  a matrix $M$ where $j$ indexes the rows and $k$ indexes the columns,
  i.e., $M_{jk} = \alpha_{jk}$. The matrix $M$ admits a singular
  value decomposition: $M = U \Sigma V^{\dag}$, and in particular
  $M_{jk} = \sum_{h} U_{jh} \sigma_h V^{\dag}_{hk}$ where $\sigma_h$ is
  the $h$-th diagonal element of $\Sigma$. Then:
  \begin{align*}
    \ket{\psi} &= \sum_{\vj, \vk} M_{jk} \ket{\vj}\ket{\vk}
    = \sum_{\vj, \vk} \left(\sum_{h} U_{jh} \sigma_h V^{\dag}_{hk}\right) \ket{\vj}\ket{\vk} \\
    &= \sum_{h} \sigma_h \left(\sum_{\vj} U_{jh} \ket{\vj}\right) \left(\sum_{\vk} V^{\dag}_{hk} \ket{\vk}\right) = \sum_{h} \sigma_h \ket{\phi^A_h} \ket{\phi^B_h},
  \end{align*}
  where we defined $\ket{\phi^A_h} := \sum_{\vj} U_{jh} \ket{\vj}$ and
  $\ket{\phi^B_h} := \sum_{\vk} V^{\dag}_{hk} \ket{\vk}$. Note that
  these are indeed orthonormal vectors, because $U, V$ are unitary and
  the vectors $\ket{\phi^A_h}, \ket{\phi^B_h}$ are simply the columns
  of $U, V$. This yields the desired decomposition up to
  relabeling. Finally, note that the $\sigma_h$ are real because they
  are the singular values, and $1 = \nrm{\ket{\psi}} = \sum_{h}
  \sigma_h^2$, because all the cross terms in the expression for
  $\nrm{\ket{\psi}}$ cancel out due to orthonormality of the vectors
  $\ket{\phi^A_h}, \ket{\phi^B_h}$.
\end{proof}

\section{Notes and further reading}
\label{sec:prelimnotes}
On the topic of Turing machines\index{Turing machine}, quantum Turing machines, and their
equivalence, fundamental references are
\cite{deutsch85quantum,bennett73logical,bernstein1997quantum}. \cite{deutsch85quantum}
shows the existence of a universal quantum Turing machine, that can
simulate any quantum Turing machine; the simulation overhead is only
polynomial, as shown in \cite{bernstein1997quantum}. Results from
\cite{bennett73logical} imply that polynomial-time quantum Turing
machines can simulate polynomial-time deterministic and probabilistic
Turing machines.

Regarding upper bounds on the length of the sequence of gates from a
universal set that is necessary to construct an arbitrary single-qubit
gate\index{gate!approximation}\index{gate!universality}\index{universal set}, which we discussed in Sect.~\ref{sec:basicops} and more
specifically in Thm.~\ref{thm:sk}, \cite{dawson05solovay} gives a
detailed proof with $c \approx 3.98$, and in general,
Solovay-Kitaev-type algorithms yield $c = 3 + \delta$ for $\delta > 0$
\cite{kitaev2002classical}. Lower values are possible: for general
gate sets, \cite{kuperberg2023breaking} reduces $c$ to $1.44 +
\delta$, and for special sets of gates, even $c = 1$ is possible
\cite{selinger12efficient,kliuchnikov16practical}. The exact values do
not matter much for the high-level exposition in this \book{}: it is
sufficient to know that a polylogarithmic number of gates suffices. It
can, however, be very important for practical implementations.

We conclude these notes by giving references to additional reading
material that discusses the fundamentals of quantum computing and
quantum algorithms. The most celebrated reference is
\cite{nielsen02quantum}, a comprehensive treatment of quantum
computing, including error correction and quantum algorithms. Due to
its sheer size and scope, the book is often used as a reference, and
may not be the most suitable instrument (or the fastest way) for an
applied mathematician who wants to learn about quantum algorithms from
scratch. It does, however, contain a rigorous treatment of many
important topics, and is especially valuable for readers with a
background in physics. \cite{rieffel07quantum} is another extensive
treatment of quantum computing, with a sizeable discussion of quantum
algorithms, and it uses a language that may be more familiar for
applied mathematicians. \cite{kaye07introduction} is a concise but
rigorous introduction to quantum computing, quantum algorithms and
quantum error correction. Being more recent than
\cite{nielsen02quantum}, it has the added benefit of covering certain
topics in quantum algorithms using a modern, and possibly clearer,
approach.

In addition to the three well-known books above, there are excellent
sets of lecture notes by prominent scientists available on the arXiv
or directly on their author's website. We mention three in
particular. \cite{childs2017lecture} is an advanced treatment of
multiple topics in quantum algorithms, including quantum algorithms
for algebraic problems (e.g., the hidden subgroup problem, see
Sect.~\ref{sec:qftnotes}) and quantum walks. This set of lecture notes
is written with a computer science perspective. Another set of
lectures notes with a computer science perspective is
\cite{dewolf2019quantum}; the style in \cite{dewolf2019quantum} is
more informal (although it is precise), and as a result, it may be a
more accessible starting point for some
readers. \cite{dewolf2019quantum} covers a vast number of topics,
including some that have gained steam quite recently. Finally,
\cite{lin2022lecture} is a set of lecture notes on quantum algorithms
for scientific computation, and it includes an in-depth discussion of
quantum phase estimation and operations on matrices (which is also the
subject of our Ch.~\ref{ch:blockenc}) via block-encodings and the
quantum singular value transformation.

\chapter{Early examples of quantum algorithms}
\label{ch:earlyalg}
\thispagestyle{fancy}
In this chapter we explore some of the principles of quantum algorithm
design, together with a discussion of some of the (chronologically)
first algorithms providing evidence of a quantum speedup of some
type. These algorithms are not directly useful for optimization, but
they serve the purpose of familiarizing the reader with, and building
intuition on, the analysis of quantum algorithms; for this purpose, it
is generally helpful to start from simple algorithms.

\section{Phase kickback}
\label{sec:phasekickback}
As discussed in Sect.~\ref{sec:basicops}, the C$X$ gate may create
entangled states. Recall that the C$X$ gate takes a control qubit and
a target qubit. However, one should not make the mistake of thinking
of C$X$ as acting on the target qubit only. When C$X$ is applied onto
a basis state it is natural to think of the control qubit as being
left alone while acting on the target qubit, but overall, the effect
of C$X$ (like any other two-qubit gate) is dependent on the state on
which it is applied, and one cannot think of each qubit in
isolation. To see this, we show an example of a controlled gate
sandwiched between some single-qubit gates, and overall this makes it
look like the target qubit is acting on the control (instead of the
converse), if we follow the same intuitive interpretation.
\begin{example}
  \label{ex:swapctrltarg}
  Consider this operation on a two-qubit state:
  \begin{equation*}
    H^{\otimes 2} \text{C}X_{12} H^{\otimes 2}.
  \end{equation*}
  We claim that this is the same as $\text{C}X_{21}$ (note the swapped
  role of control and target qubit). Indeed:
  \begin{align*}
    \frac{1}{2}
    \begin{pmatrix}
      1 & 1 & 1 & 1 \\
      1 & -1 & 1 & -1 \\
      1 & 1 & -1 & -1 \\
      1 & -1 & -1 & 1 
    \end{pmatrix}
    \begin{pmatrix}
      1 & 0 & 0 & 0 \\
      0 & 1 & 0 & 0 \\
      0 & 0 & 0 & 1 \\
      0 & 0 & 1 & 0 
    \end{pmatrix}
    \frac{1}{2}
    \begin{pmatrix}
      1 & 1 & 1 & 1 \\
      1 & -1 & 1 & -1 \\
      1 & 1 & -1 & -1 \\
      1 & -1 & -1 & 1 
    \end{pmatrix}
    =
    \begin{pmatrix}
      1 & 0 & 0 & 0 \\
      0 & 0 & 0 & 1 \\
      0 & 0 & 1 & 0 \\
      0 & 1 & 0 & 0 
  \end{pmatrix},
  \end{align*}
  and the last matrix swaps the second and third row, i.e., it maps
  $\ket{01} \to \ket{11}$ and $\ket{11} \to \ket{01}$.
\end{example}

\begin{figure}[h!]
\leavevmode
\centering
\ifcompilefigs
\Qcircuit @C=1em @R=0.7em {
 & \gate{H}  & \ctrl{1} & \gate{H} & \qw & \\
 & \gate{H}  & \targ    & \gate{H} & \qw & 
}
\hspace{10em}
\Qcircuit @C=1em @R=1.5em {
 & \targ     & \qw & \\
 & \ctrl{-1} & \qw & 
}
\else
\includegraphics{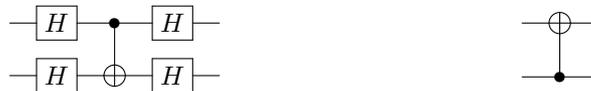}
\fi
\caption{Interchanging the control and target qubit of C$X$.}
\label{fig:cxswap}
\end{figure}

\noindent In circuit form, Ex.~\ref{ex:swapctrltarg} implies that the
circuits in Fig.~\ref{fig:cxswap} are equivalent.
The left part of Fig.~\ref{fig:cxswap} is essentially an indication of
a basis change\index{basis!change}: instead of expressing each qubit
in the standard orthonormal basis, via Hadamard gates, we are
expressing it in a different basis (with basis elements $H\ket{0}$ and
$H\ket{1}$). We know from linear algebra that we can express a linear
transformation in a different basis by premultiplying and
postmultiplying by a matrix containing the new basis as its columns,
or its inverse, depending on the direction of the transformation. So
if $A$ is expressed in the basis ${\cal B}$, and $U$ maps each element
of ${\cal B}$ to some other basis ${\cal B}'$, we have that $UAU^{-1}$
is the expression of the linear transformation associated with $A$ in
terms of the basis ${\cal B}'$. Going back to
Ex.~\ref{ex:swapctrltarg}, this means that the operation C$X_{21}$ is
the same as the operation C$X_{12}$ in a different basis.
\begin{example}
  \label{ex:swapctrltarg2}
  Instead of the standard orthonormal basis, consider the basis
  $H\ket{0} = \frac{1}{\sqrt{2}}(\ket{0} + \ket{1})$, $H\ket{1} =
  \frac{1}{\sqrt{2}}(\ket{0} - \ket{1})$ for each qubit. (This is
  often called the \emph{Hadamard basis}\index{basis!Hadamard}, for
  obvious reasons.) Given a vector in the Hadamard basis, we can
  express it in the standard orthonormal basis by multiplying by
  $H^{-1} = H^{\dag} = H$. In a tensor product space, the basis change
  operation is applied identically on each vector space taking part in
  the tensor product. So, for example, given the two-qubit state
  $\ket{01}$ in the Hadamard basis (for each qubit), its expression in
  the standard orthonormal basis is:
  \begin{equation*}
    (H^{\dag} \otimes H^{\dag}) \ket{01} = \frac{1}{\sqrt{2}}(\ket{0} + \ket{1}) \otimes \frac{1}{\sqrt{2}}(\ket{0} - \ket{1}) = \frac{1}{2}(\ket{00} - \ket{01} + \ket{10} - \ket{11}).
  \end{equation*}
  If we apply C$X_{12}$ in the standard orthonormal basis, we obtain:
  \begin{equation*}
    \frac{1}{2}(\ket{00} - \ket{01} - \ket{10} + \ket{11}) = 
    \frac{1}{\sqrt{2}}(\ket{0} - \ket{1}) \otimes \frac{1}{\sqrt{2}}(\ket{0} - \ket{1})
  \end{equation*}
  Transforming from the standard orthonormal basis back to the
  Hadamard basis, the expression for this state is:
  \begin{equation*}
    (H \otimes H) \frac{1}{\sqrt{2}}(\ket{0} - \ket{1}) \otimes
    \frac{1}{\sqrt{2}}(\ket{0} - \ket{1}) = \ket{11} = \text{C}X_{21}
    \ket{01}.
  \end{equation*}
  We have recovered the equivalence shown in Fig.~\ref{fig:cxswap}.
\end{example}

As stated in Ch.~\ref{ch:intro}, we always work with the standard
orthonormal basis, so we do not explicitly use the Hadamard basis in
the rest of this \book{}. But the concept exemplified in
Ex.~\ref{ex:swapctrltarg2} stands: a controlled gate in a certain
basis may look like a completely different operation in another basis,
even one with control and target qubit exchanged. Unitary matrices
encode a basis change between orthonormal bases, so for general
quantum states, we can never assume that the control qubit of a
controlled operation does not get affected by it. Indeed, depending on
the state on which a gate is applied, a C$X$ gate can be interpreted
as having an effect on the control qubit, rather than the target
qubit. We can generalize this idea further, and exploit it for
computation, so that applying a C$X$ has some quantifiable effect on
the control qubit. In this section we develop a technique that
encodes information on the value of certain types of functions as the
phase of the control qubit (or, in general, of some basis states).\index{phase!kickback|(} The
technique relies on properties of the \emph{eigenstate}
$\frac{1}{\sqrt{2}}(\ket{0} - \ket{1})$ of the $X$ gate.
\begin{definition}[Eigenstate]
  \label{def:eigenstate}
  Given a unitary $U \in \C^{2^q \times 2^q}$, we say that the
  $q$-qubit state $\ket{\psi}$ is an {\em eigenstate}\index{eigenstate} of $U$ if the
  $2^q$-dimensional vector corresponding to $\ket{\psi}$ is an
  eigenvector of $U$, i.e., $U\ket{\psi} = e^{i \theta} \ket{\psi}$ for
  some $\theta$.
\end{definition}
\noindent In other words, ``eigenstate'' simply means that the quantum
state is an eigenvector of a given operator.
\begin{remark}
  All eigenvalues of a unitary matrix have modulus 1, so they can be
  written as $e^{i \theta}$ for some $\theta \in [0, 2\pi)$.
\end{remark}
Notice that:
\begin{equation*}
  \begin{pmatrix}
    0 & 1 \\ 1 & 0
  \end{pmatrix}
  \frac{1}{\sqrt{2}}(\ket{0} - \ket{1}) = \frac{1}{\sqrt{2}}(\ket{1} - \ket{0}) = -\frac{1}{\sqrt{2}}(\ket{0} - \ket{1})
\end{equation*}
and:
\begin{equation*}
  \begin{pmatrix}
    1 & 0 \\ 0 & 1
  \end{pmatrix}
  \frac{1}{\sqrt{2}}(\ket{0} - \ket{1}) = \frac{1}{\sqrt{2}}(\ket{0} - \ket{1}),
\end{equation*}
i.e., $\frac{1}{\sqrt{2}}(\ket{0} - \ket{1})$ is an eigenstate with
eigenvalue $-1$ of $X$, and (trivially) it is an eigenstate with
eigenvalue $+1$ of the identity gate $I$. The C$X$ gate applies $X$ to
target qubit if the control is $1$, and applies $I$ to the target
qubit if the control is $0$. Thus, if the target qubit is in the state
$\frac{1}{\sqrt{2}}(\ket{0} - \ket{1})$, depending on the value of the
control qubit we ``obtain'' a different eigenvalue, i.e., multiply the
quantum state by a different scalar. It is easy to check that we can
write:
\begin{equation*}
  \text{C}X\left(\ket{x} \otimes \frac{1}{\sqrt{2}}(\ket{0} -
  \ket{1})\right) = (-1)^{x} \left(\ket{x} \otimes
  \frac{1}{\sqrt{2}}(\ket{0} - \ket{1})\right).
\end{equation*}
With this operation, some information on $x$ becomes encoded in the
coefficient of the quantum state: if $x = 0$ nothing happens, but if
$x = 1$ the entire quantum state gets sign-flipped.

This effect can be applied even more in general. Let us study a
two-qubit operation $U_f$ that implements the map $\ket{x} \ket{y} \to
\ket{x} \ket{y \oplus f(x)}$, where, as usual, $x,y \in \{0,1\}$ and
we also assume that $f(x) \in \{0,1\}$. (Recall the definition of
$\oplus$ given in Def.~\ref{def:xor}.) As we discussed in
Sect.~\ref{sec:uncompute}, this particular form of the function, with
a second register used to contain the output by means of $\oplus$, is
typical of the quantum computing world. Let us apply the two-qubit
operation $U_f$ on a target qubit that is prepared in the state
$\frac{1}{\sqrt{2}}(\ket{0} - \ket{1}) = H \ket{1}$. We have:
\begin{equation*}
  U_f \left(\ket{x} \otimes \frac{1}{\sqrt{2}}(\ket{0} -
  \ket{1})\right) = \ket{x} \otimes \frac{1}{\sqrt{2}}(\ket{0 \oplus
    f(x)} - \ket{1 \oplus f(x)}).
\end{equation*}
If $f(x) = 0$, this has no effect on the second qubit. If $f(x) = 1$,
this bit-flips the second qubit (i.e., $\ket{0}$ becomes $\ket{1}$ and
$\ket{1}$ becomes $\ket{0}$), which has the overall effect of changing
the sign of the second qubit. Thus, we can write the effect of $U_f$
as:
\begin{equation}
  \label{eq:ufeigenvalue}
   U_f \left(\ket{x} \otimes \frac{1}{\sqrt{2}}(\ket{0} -
   \ket{1})\right) = (-1)^{f(x)} \left(\ket{x} \otimes
   \frac{1}{\sqrt{2}}(\ket{0} - \ket{1})\right).
\end{equation}
The C$X$ gate can be obtained from Eq.~\ref{eq:ufeigenvalue} with $f(x) = x$.
If the control qubit is in a general state $\alpha_0 \ket{0} +
\alpha_1 \ket{1}$ rather than a basis state $\ket{x}$, we have:
\begin{equation*}
   U_f \left((\alpha_0 \ket{0} + \alpha_1 \ket{1}) \otimes
   \frac{1}{\sqrt{2}}(\ket{0} - \ket{1})\right) = \left((-1)^{f(0)} \alpha_0
   \ket{0} + (-1)^{f(1)} \alpha_1 \ket{1}\right) \otimes
   \frac{1}{\sqrt{2}}(\ket{0} - \ket{1}).
\end{equation*}

If the second qubit is prepared in the state
$\frac{1}{\sqrt{2}}(\ket{0} - \ket{1})$, applying $U_f$ yields the
situation that is depicted in Fig.~\ref{fig:ufeigenvalue}.
\begin{figure}[h!]
\leavevmode
\centering
\ifcompilefigs
\Qcircuit @C=1em @R=0.7em {
 \lstick{\ket{x}}  & \multigate{1}{U_f}              & \qw & \rstick{\ket{x}} \\
 \lstick{\frac{1}{\sqrt{2}}(\ket{0} - \ket{1})}  & \ghost{U_f} & \qw & \rstick{\frac{(-1)^{f(x)}}{\sqrt{2}}(\ket{0} - \ket{1})} \\
}
\else
\includegraphics{figures/ufeigenvalue.pdf}
\fi
\caption{Application of $U_f$ when the second qubit is prepared with
  an eigenstate of $X$.}
\label{fig:ufeigenvalue}
\end{figure}
By properties of the tensor product, we can interpret the
multiplicative factor $(-1)^{f(x)}$ as being applied to the first
qubit, rather than the second one, writing the mapping as
\begin{equation*}
 \ket{x} \otimes \frac{1}{\sqrt{2}}(\ket{0} - \ket{1}) \to
 (-1)^{f(x)}\ket{x} \otimes \frac{1}{\sqrt{2}}(\ket{0} - \ket{1}).
\end{equation*}
The relative phase
that is --- in principle --- applied to the second qubit is now
``kicked back'' to the first qubit, by virtue of the fact that
$\frac{1}{\sqrt{2}}(\ket{0} - \ket{1})$ is an eigenstate of the
addition $\oplus f(x)$ that $U_f$ applies to the second qubit. The net
effect is to flip the sign of a basis state $\ket{x}$ such that $f(x)
= 1$. This technique, called \emph{phase kickback}\index{phase!kickback|)}, is at the heart of
several quantum algorithms discussed in this \book{}.

\section{The first quantum algorithm: Deutsch's algorithm}
\label{sec:deutsch}
We discuss Deutsch's algorithm\index{Deutsch's algorithm|(}\index{algorithm!Deutsch's|(} \cite{deutsch85quantum} as a direct
application of phase kickback, and as a way to introduce the idea of
quantum interference that is exploited in Simon's algorithm as
well, in Sect.~\ref{sec:simon}. Historically, Deutsch's algorithm was
the first to show a quantum speedup over classical algorithms for the
same problem; it also has the tremendous benefit of being simple to
understand.

For Deutsch's algorithm, we are given access to a function $f :
\{0,1\} \to \{0,1\}$, and the goal is to find $f(0) \oplus f(1)$ by
querying the function the smallest number of times.
\begin{remark}
  \label{rem:complexity}
  For the first two algorithms discussed in this \book{} (i.e.,
  Deutsch's and Simon's), as well as some of the subsequent
  algorithms, the complexity of the algorithm is determined only in
  terms of the number of calls to a function $f$ given as part of the
  input. Considerations on what the function $f$ actually implements,
  and how many operations are performed inside of $f$, or between the
  calls to $f$, are not part of how we determine this type of
  complexity. This model is known as {\em query
    complexity}\index{complexity!query}\index{oracle!complexity|see{complexity, query}}\index{complexity!oracle|see{gate}}, because --- as the name implies --- it defines the
  complexity of an algorithm as the number of queries to a given
  function (in this case, $f$). Query complexity is used as a model to
  answer important theoretical questions. There are many quantum
  algorithms that yield speedups under the query complexity model;
  some others, e.g., Shor's algorithm, are faster than (known)
  classical algorithms under the more traditional computational
  complexity model, according to which complexity is measured by the
  number of basic operations --- usually called \emph{gate
  complexity}\index{complexity!gate}\index{complexity!time|see{gate}}
  because it counts the number of (elementary) gates. In fact, even
  for cases where we are interested in the query complexity, we may
  separately discuss the number of gates applied between calls to $f$
  to give a more precise characterization of the gate complexity as
  well. The gate complexity is equivalently called \emph{time
  complexity}.
\end{remark}
Classically, solving the problem described above requires two queries
to $f$: if we query both $f(0)$ and $f(1)$ we can easily compute $f(0)
\oplus f(1)$, and if we query only one of the two we cannot determine
the correct answer with certainty. Surprisingly, we can solve the
problem with only one quantum query using an appropriate quantum
circuit. We assume that $f$ is given in the form of a quantum oracle
$U_f : \ket{x} \ket{y} \to \ket{x} \ket{y \oplus f(x)}$, as we have
seen before.

\begin{figure}[h!]
\leavevmode
\centering
\ifcompilefigs
\Qcircuit @C=1em @R=0.7em {
 \lstick{\ket{0}}  & \qw      & \gate{H} & \multigate{1}{U_f} & \gate{H} & \qw & \meter \\
 \lstick{\ket{0}}  & \gate{X} & \gate{H} & \ghost{U_f}        & \qw & \qw & \\
}
\else
\includegraphics{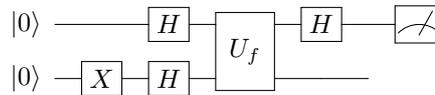}
\fi
\caption{Circuit to solve Deutsch's algorithm.}
\label{fig:deutsch}
\end{figure}
Deutsch's algorithm works by applying the circuit depicted in
Fig.~\ref{fig:deutsch}. Let us study the evolution of the quantum
state. The final quantum state before the measurement is:
\begin{equation*}
  (H \otimes I) U_f (H \otimes H) (I \otimes X) (\ket{0} \otimes \ket{0}).
\end{equation*}
We have:
\begin{align*}
  (I \otimes X) (\ket{0} \ket{0}) &= \ket{0} \ket{1} \\
  (H \otimes H) (\ket{0} \ket{1}) &= \frac{\ket{0} + \ket{1}}{\sqrt{2}} \otimes \frac{\ket{0} - \ket{1}}{\sqrt{2}} \\ 
  U_f \left(\frac{\ket{0} + \ket{1}}{\sqrt{2}} \otimes \frac{\ket{0} - \ket{1}}{\sqrt{2}}\right) &= \frac{(-1)^{f(0)}}{\sqrt{2}} \ket{0} \otimes \frac{\ket{0} - \ket{1}}{\sqrt{2}} + \frac{(-1)^{f(1)}}{\sqrt{2}} \ket{1} \otimes \frac{\ket{0} - \ket{1}}{\sqrt{2}} \\
  &= (-1)^{f(0)} \left(\frac{\ket{0} + (-1)^{f(0) \oplus f(1)} \ket{1}}{\sqrt{2}} + \right) \otimes \frac{\ket{0} - \ket{1}}{\sqrt{2}}.
\end{align*}
In the third equation above we applied phase kickback\index{phase!kickback},
Eq.~\eqref{eq:ufeigenvalue}, and in the last line we simply collected
the term $(-1)^{f(0)}$. Finally, we apply $(H \otimes I)$ to this
state, and doing the calculations we obtain:
\begin{equation*}
  (-1)^{f(0)} \left(\frac{(1 + (-1)^{f(0) \oplus f(1)}) \ket{0} + (1 - (-1)^{f(0) \oplus f(1)}) \ket{1}}{2}\right) \otimes \frac{\ket{0} - \ket{1}}{\sqrt{2}}.
\end{equation*}
The outcome of the measurement operation then depends on the value of
$f(0) \oplus f(1)$. We can ignore the global phase factor
$(-1)^{f(0)}$: it is irrelevant when we take the modulus squared to
look at the measurement outcome probabilities, see
Ex.~\ref{ex:globalphase}. We are measuring the first qubit, and the
state is in a product state, so we only need to look at the first
qubit. If $f(0) \oplus f(1) = 0$, then the coefficient for $\ket{0}$
is $(1 + (-1)^0)/2 = 1$, implying that we have probability 1 of
observing $0$ as the measurement outcome. If, on the other hand, $f(0)
\oplus f(1) = 1$, the coefficient for $\ket{0}$ is $(1 + (-1)^1)/2 =
0$ and the coefficient for $\ket{1} = 1$, implying that we have
probability 1 of obtaining 1 as the measurement outcome. Thus, with
this measurement we can determine with probability 1 the value of
$f(0) \oplus f(1)$. Notice that Fig.~\ref{fig:deutsch} contains a
single application of $U_f$, as opposed to the two function
evaluations required classically: a quantum speedup, in the query
complexity model!\index{Deutsch's algorithm|)}\index{algorithm!Deutsch's|)}

\section{Quantum interference and period finding:
  Simon's algorithm}
\label{sec:simon}
In the final part of this chapter we describe a quantum algorithm,
known as Simon's algorithm \cite{simon97power}\index{Simon's algorithm|(}\index{algorithm!Simon's|(}, that gives an expected
exponential speedup with respect to classical algorithms. Although
Simon's algorithm has not been directly helpful for quantum
optimization algorithms (at least so far), we discuss it because it has
many interesting features from an educational perspective: namely, it
uses both classical and quantum computation, and it yields an
exponential speedup.

Admittedly, the problem that Simon's algorithm solves is not very
useful (just as Deutsch's algorithm), but the ideas shown here give us
further intuition of what quantum computing can do. In fact, this
algorithm was an inspiration for the well-known and groundbreaking
work of Shor on integer factorization \cite{shor97polynomial}: a large
part of Shor's algorithm relies on the solution of a period finding
problem, and Simon's algorithm solves a simplified problem of the same
flavor. Shor's algorithm is, however, much more involved than Simon's
algorithm, and a full treatment requires several number-theoretical
results that are beyond the scope of this \book{}. Thus, we focus
on Simon's algorithm; some notes on Shor's algorithm are given in
Sect.~\ref{sec:earlyalgnotes}.

\begin{figure}[h!]
\leavevmode
\centering
\ifcompilefigs
\Qcircuit @C=1em @R=0.7em {
\lstick{\ket{\v{x}}} & \qw & {/^n} \qw & \qw & \multigate{1}{U_f} & \qw & {/^n} \qw & \qw & \rstick{\ket{\v{x}}} \\
\lstick{\ket{\v{y}}} & \qw & {/^n} \qw & \qw & \ghost{U_f}        & \qw & {/^n} \qw & \qw & \rstick{\ket{\v{y} \oplus f(\v{x})}}
}
\else
\includegraphics{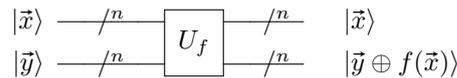}
\fi
\caption{The circuit implementing $U_f$ for Simon's problem, with
  basis states $\v{x}, \v{y} \in \{0,1\}^n$.}
\label{fig:simonuf}
\end{figure}

For Simon's algorithm, we are given access to a function $f :
\{0,1\}^n \to \{0,1\}^n$ with the property that $f(\v{x}) = f(\v{z})$
if and only if $\v{x} = \v{z} \oplus \v{a}$, for some unknown $\v{a}
\in \{0,1\}^n$. We do not know anything else about the function, and
the goal is to find $\v{a}$ by querying the function the smallest
number of times, again using a query complexity model.  Notice that if
$\v{a} = \v{0}$ then the function is one-to-one, whereas if $\v{a}
\neq \v{0}$ the function is two-to-one, because for every $\v{x}$,
there is exactly another number in domain for which the function has
the same value. The function $f$ is assumed to be given as a quantum
circuit on $q = 2n$ qubits, via the unitary $U_f$ depicted in
Fig.~\ref{fig:simonuf}, therefore we are allowed to query the function in
superposition (this gives us an advantage over classical computation,
but remember that a classical circuit to compute $f$ can be turned
into a quantum circuit to compute $U_f$, see
Rem.~\ref{rem:classicalfun}). Recall that by linearity, to describe
the effect of $U_f$ it is enough to describe its behavior on all basis
states, so Fig.~\ref{fig:simonuf} fully specifies the unitary.

\subsection{Classical algorithm}
\label{sec:simonclassical}
Because we do not know anything about the binary string $\v{a}$, the
best we can do to recover its value is to feed inputs to the function
$f$, and try to extract information from the output. The string
$\v{a}$ is determined once we find two distinct inputs $\v{x}, \v{z}$
such that $f(\v{x}) = f(\v{z})$, because then $\v{x} = \v{z} \oplus
\v{a}$ which implies $\v{x} \oplus \v{z} = \v{a}$. So we are
successful once $\v{a}$ belongs to the set $\{\v{x} \oplus \v{z}
\text{ for all pairs of distinct tested inputs } \v{x}, \v{z}\}$.


Suppose we have evaluated $m$ distinct input values and we did not
find a match. Then $\v{a} \neq \v{x} \oplus \v{z}$ for all $\v{x},
\v{z}$ previously evaluated, therefore we have eliminated at most
$m(m-1)/2$ possible values of $\v{a}$ from the search space. (Fewer
values may have been eliminated if we test inputs equal to $\v{x}
\oplus \v{y} \oplus \v{z}$ for any three input values $\v{x}, \v{y},
\v{z}$ already tested. In fact, if we test $\v{w}$ such that $\v{w} =
\v{x} \oplus \v{y} \oplus \v{z}$, we have that $\v{w} \oplus \v{z} =
\v{x} \oplus \v{y}$, therefore the value $\v{w} \oplus \v{z}$ had
already been eliminated from the list of possible values of $\v{a}$.)
In the worst case all possible values for $\v{a}$ have equal
probability, so this approach is successful with probability
$\frac{m(m-1)}{2^{n+1}}$, i.e., the number of values of $\v{a}$ that
have already been tested divided by the probability that any one of
them is the correct answer.  Because $m(m-1)/2$ is small compared to
$2^n$, the probability of success of this algorithm (i.e.,
$\frac{m(m-1)}{2^{n+1}}$) is very small until we have evaluated a
number of inputs that is in the order of $2^n$. In particular, to
guarantee a probability of success of at least $\delta$, we need
$m(m-1) \ge \delta 2^{n+1}$, which implies that $m =
\bigO{\sqrt{\delta 2^n}}$. Hence, for any positive constant $\delta$,
the number of required iterations is exponential: $\bigO{2^{n/2}}$.
After evaluating $\frac{1 +\sqrt{2^{n+3}+1}}{2} = \bigO{2^{n/2}}$
distinct input values satisfying the condition outlined above for
non-matching triplets (to obtain this number, we found the smallest
value of $m$ such that $m(m-1) \ge 2^{n+1}$), we are guaranteed that a
matching pair has been found, or we can safely determine that $\v{a} =
\v{0}$.

\subsection{Quantum algorithm}
\label{sec:simonquantum}
\begin{figure}[h!]
\leavevmode
\centering
\ifcompilefigs
\Qcircuit @C=1em @R=0.7em {
\lstick{\ket{\v{0}}} & \qw & {/^n} \qw & \gate{H^{\otimes n}} & \qw & {/^n} \qw & \multigate{1}{U_f} & \qw & {/^n} \qw & \gate{H^{\otimes n}} & \qw & {/^n} \qw & \qw & \meter \\
\lstick{\ket{\v{0}}} & \qw & {/^n} \qw & \qw                  & \qw & \qw       & \ghost{U_f}        & \qw & {/^n} \qw & \qw & \qw & \qw & \qw & \qw \\
}
\else
\includegraphics{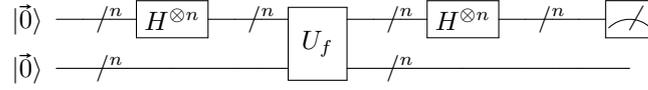}
\fi
\caption{Quantum circuit used in Simon's algorithm.}
\label{fig:simoncircuit}
\end{figure}

Using a quantum computer, we can determine $\v{a}$ much faster. The idea,
first described in \cite{simon97power}, is to apply the circuit in
Fig.~\ref{fig:simoncircuit}.

\noindent From an algebraic point of view, the state produced by this
circuit just before the measurement is described by the following
expression
\begin{equation*}
  (H^{\otimes n} \otimes I^{\otimes n})U_f(H^{\otimes n} \otimes
  I^{\otimes n}) (\ket{\v{0}}_n \otimes \ket{\v{0}}_n).
\end{equation*}
We now analyze the output of the quantum circuit, by looking at the
quantum states at intermediate steps of the circuit. Define the following three states, used to track the evolution of the quantum state through the end of the circuit:
\begin{align*}
  \ket{\psi} &=   (H^{\otimes n} \otimes
  I^{\otimes n}) (\ket{\v{0}} \ket{\v{0}}) \\
  \ket{\phi} &=   U_f(H^{\otimes n} \otimes
  I^{\otimes n}) (\ket{\v{0}} \ket{\v{0}}) \\
  \ket{\chi} &=   (H^{\otimes n} \otimes I^{\otimes n})U_f(H^{\otimes n} 
  \otimes I^{\otimes n}) (\ket{\v{0}} \ket{\v{0}}).
\end{align*}
For $\ket{\psi}$, we know that $H^{\otimes n}$ applied onto
$\ket{\v{0}}_n$ creates a uniform superposition of $\ket{\vj}, \vj \in
\{0,1\}^n$. Therefore we have:
\begin{equation*}
  \ket{\psi} = (H^{\otimes n} \otimes I^{\otimes n}) (\ket{\v{0}}
  \ket{\v{0}}) = \frac{1}{\sqrt{2^n}} \sum_{\vj \in \{0,1\}^n}
  \ket{\vj} \ket{\v{0}}.
\end{equation*}
By linearity, applying $U_f$ onto this state yields:
\begin{equation*}
  \ket{\phi} = U_f \ket{\psi} = \frac{1}{\sqrt{2^n}} \sum_{\vj \in
    \{0,1\}^n} \ket{\vj} \ket{\v{0} \oplus f(\vj)} =
  \frac{1}{\sqrt{2^n}} \sum_{\vj \in \{0,1\}^n} \ket{\vj} 
    \ket{f(\vj)}.
\end{equation*}
We now need to analyze the effect of applying further Hadamard gates
on the top lines of the circuit, i.e., the first register. Using the
algebraic expression for the Hadamard gate found in
Eq.~\eqref{eq:hadamard}, the next step in the circuit yields the
state:
\begin{align}
  \ket{\chi} &= (H^{\otimes n} \otimes I^{\otimes n}) \ket{\phi}
  = (H^{\otimes n} \otimes I^{\otimes n})
  \frac{1}{\sqrt{2^n}} \sum_{\vj \in \{0,1\}^n} \ket{\vj} 
  \ket{f(\vj)} \notag \\
  &= \frac{1}{2^n} \sum_{\vj \in \{0,1\}^n} \sum_{\vk \in \{0,1\}^n} 
  (-1)^{\vk \bullet \vj} \ket{\vk} 
  \ket{f(\vj)}. \label{eq:simonbasecoeff}
\end{align}
When we make a measurement on the top $n$ qubit lines of $\ket{\chi}$
(i.e., the first $n$-qubit register, containing qubits 1 through $n$),
we obtain any given binary string $\vk$ with probability equal to the
sum of the modulus squared of the coefficient of the states $\ket{\vk}
\ket{f(\vj)}$, for all $\vj$. This is a direct consequence of the
principle of implicit measurement (Prop.~\ref{prop:implicitmeas}): if
we only measure the first register, and discard the second, we can
assume that the measurement is applied to the second register as well;
thus, we observe $\vk$ in the first register if measurement of the
entire state yields any string that starts with $\vk$, hence the sum
over all the possibilities in the second register.

It is easy to verify that for fixed $\vk$, the probability of
observing $\vk$ in the first measurement (i.e., the sum of the modulus
squared of the coefficient of the states $\ket{\vk} \ket{f(\vj)}$, for
all $\vj$) is equal to:
\begin{equation*}
  \nrm{\frac{1}{2^n} \sum_{\vj \in \{0,1\}^n} (-1)^{\vk \bullet \vj}
    \ket{f(\vj)}}^2.
\end{equation*}
A simple formal argument for why this is the case can be derived by
using the density matrix formalism. The density matrix corresponding
to the pure state in Eq.~\eqref{eq:simonbasecoeff} is:
\begin{equation*}
  \rho =  \left(\frac{1}{2^n}\sum_{\vj \in \{0,1\}^n} \sum_{\vx \in \{0,1\}^n} 
  (-1)^{\vx \bullet \vj} \ket{\vx} 
  \ket{f(\vj)}\right) \left(\frac{1}{2^n}\sum_{\vh \in \{0,1\}^n} \sum_{\vy \in \{0,1\}^n} 
  (-1)^{\vy \bullet \vh} \bra{\vy} 
  \bra{f(\vh)}\right).
\end{equation*}
Computing the reduced density matrix by tracing out the second
register (see Sect.~\ref{sec:reduceddensmat}), and then applying a
measurement on all qubits in the first register (recall
Rem.~\ref{rem:densitymatmeas}), the probability of observing $\vk$ in
the first register is:
\begin{align*}
  \trace{ \left(\ketbra{\vk}{\vk} \otimes I^{\otimes n}\right) \rho} &=
  \trace{ \left(\bra{\vk} \otimes I^{\otimes n}\right) \rho \left(\ket{\vk} \otimes I^{\otimes n}\right)}  \\
  &= \trace{ \left(\frac{1}{2^n}\sum_{\vj \in \{0,1\}^n} 
  (-1)^{\vk \bullet \vj} 
  \ket{f(\vj)}\right) \left(\frac{1}{2^n}\sum_{\vh \in \{0,1\}^n} 
  (-1)^{\vk \bullet \vh}  \bra{f(\vh)} \right)} \\
  &= \nrm{\frac{1}{2^n} \sum_{\vj \in \{0,1\}^n} (-1)^{\vk \bullet \vj} \ket{f(\vj)}}^2,
\end{align*}
so we have the following relationship:
\begin{equation*}
  \Pr(\vk) = \nrm{\frac{1}{2^n} \sum_{\vj \in \{0,1\}^n} (-1)^{\vk \bullet \vj} \ket{f(\vj)}}^2,
\end{equation*}
where we denote by $\Pr(\vk)$ the probability of observing string
$\vk$ after applying a measurement to the first register. Now we
analyze the expression for $\Pr(\vk)$. First, we deal with the case
$\va = \v{0}$, which is easier to analyze. In this case the function
$f$ is one-to-one, so the summation $\sum_{\vj \in \{0,1\}^n}
(-1)^{\vk \bullet \vj} \ket{f(\vj)}$ is over every basis vector, and:
\begin{equation*}
  \nrm{\frac{1}{2^n} \sum_{\vj \in \{0,1\}^n} (-1)^{\vk \bullet \vj}
    \ket{f(\vj)}}^2 = \frac{1}{2^n}.
\end{equation*}
This means that we have probability $\frac{1}{2^n}$ to observe a given
binary string $\vk$, i.e., each measurement gives an $n$-digit binary
string uniformly at random. Let us now analyze the case $\va \neq
\v{0}$, which is a bit more involved but also more
interesting. Assuming $\va \neq \v{0}$, the function $f$ is
two-to-one: $f(\vj) = f(\vj \oplus \va)$. So $\ket{\vk} \ket{f(\vj)} =
\ket{\vk} \ket{f(\vj \oplus \va)}$, which means that there are only
$2^n/2 = 2^{n-1}$ nonzero entries in the vector $\frac{1}{2^n}
\sum_{\vj \in \{0,1\}^n} (-1)^{\vk \bullet \vj} \ket{f(\vj)}$. Let $R$
be a set of cardinality $2^{n-1}$ with the following property: $R \cup
\{\vj \oplus \va : \vj \in R\} = \{0,1\}^n$. In other words, for every
$\vj \in \{0,1\}^n$, $R$ contains either $\vj$ or $\vj \oplus \va$,
but not both --- it does not matter which one of these two we choose,
as long as we pick only one. (For readers familiar with the concept of
quotient sets, $R$ picks a representative from the quotient set
$\{0,1\}^n/\sim$ where $\sim$ is the equivalence relationship defined
as: $\v{x} \sim \v{y}$ if and only if $\v{x} = \v{y} \oplus \va$.)
\begin{example}
  Suppose $n=3$ and $\va = 101$. Then the following holds:
  \begin{align*}
    f(000) &= f(101) \\
    f(001) &= f(100) \\
    f(010) &= f(111) \\
    f(011) &= f(110). 
  \end{align*}
  In this example, the set $R$ contains four ($=2^{n-1}$) elements,
  chosen as follows: for every row of the above set of equations, we
  either pick the binary string on the l.h.s., or the one on the
  r.h.s. It does not matter which ones we choose.
\end{example}

\noindent For each $\vk$, the string $\vk$ appears in the first
register precisely in the $2^{n-1}$ basis states $\ket{\vk}
\ket{f(\vj)}$ for $\vj \in R$. For each $\vj \in R$, the coefficient
of the basis state $\ket{\vk} \ket{f(\vj)}$ is the sum of the
coefficients in Eq.~\eqref{eq:simonbasecoeff} for $\ket{\vk}
\ket{f(\vj)}$ and $\ket{\vk} \ket{f(\vj \oplus \va)}$ (because the
content of the second register is the same, so these basis states
coincide and the corresponding coefficients must be added together);
that is, it is equal to:
\begin{align*}
  \frac{(-1)^{\vk \bullet \vj} + (-1)^{\vk \bullet (\vj \oplus \va)}}{2^n} &= \frac{(-1)^{\vk \bullet \vj} + (-1)^{\vk \bullet \vj}(-1)^{\vk \bullet \va}}{2^n} \\
  &=  \frac{(-1)^{\vk \bullet \vj}\left(1 + (-1)^{\vk \bullet \va}\right)}{2^n}.
\end{align*}
Therefore the probability of obtaining the binary string $\vk$ after
applying a measurement to all qubits in the first register is:
\begin{align*}
  \sum_{\vj \in R} 
  \left(\frac{(-1)^{\vk \bullet \vj}\left(1 + (-1)^{\vk \bullet \va}\right)}{2^n}\right)^2 =
  2^{n-1}\left(\frac{\left(1 + (-1)^{\vk \bullet \va}\right)}{2^n}\right)^2 =
  \begin{cases} \frac{1}{2^{n-1}} & \text{if } \vk \bullet \va \equiv 0 \mod 2 \\
    0 & \text{if } \vk \bullet \va \equiv 1 \mod 2, \end{cases}
\end{align*}
where the multiplication factor $2^{n-1}$ comes from the fact that
$|R| = 2^{n-1}$. Thus, the only binary strings that have positive
probability to be observed are those strings $\vk$ for which $\vk
\bullet \va \equiv 0 \mod 2$. The remaining strings are never
observed: by carefully applying quantum operations we have reduced
their state coefficients to zero, a phenomenon known as {\em
  destructive interference}. Notice that unless $\vk = \v{0}$, then
there is a nonempty set of bits for which the modulo-2 sum of the
corresponding coefficients of $\va$ must vanish. Thus, excluding
``unlucky'' or edge cases (see Sect.~\ref{sec:simonanalysis}), we can
express one of the bits of $\va$ as a modulo-2 sum of the others, and
we eliminate approximately half of the possible values from the search
space for $\va$.

Our discussion shows that with a single quantum query to $U_f$, in the
case $\va \neq \v{0}$ with high probability we learn very valuable
information about $\va$, and we can approximately halve the search
space for $\va$. In the case $\va = \v{0}$, we instead obtain a
binary string uniformly at random. It now remains to fully specify, in
a more precise manner, how this information can be used.

\subsection{Full description and analysis}
\label{sec:simonanalysis}
The quantum algorithm described in the previous section yields
information on $\va$, but it does not output $\va$ directly. To
recover $\va$, further calculations have to be performed. This is a
situation that can be fairly common in quantum algorithms: a quantum
computation measures some properties of the desired answer; then,
classical computations are used to analyze these properties and obtain
the desired answer. Thus, even if the quantum algorithm does not
explicitly output the desired answer, it allows us to get closer to
our goal.

In the specific case of the problem discussed here in
Sect.~\ref{sec:simon}, the quantum computation allows us to learn
$\vk$ such that $\vk \bullet \va \equiv 0 \mod 2$: we already
discussed why this is the case when $\va \neq \v{0}$, and this is also
the case when $\va = \v{0}$ because then trivially $\vk \bullet \va =
0$. Because all $\vk$ with this property have the same probability of
being output by the measurement, we obtain a uniformly random sample
from the set $\{\vk \in \{0,1\}^n: \vk \bullet \va \equiv 0 \mod
2\}$. We embed the equation $\vk \bullet \va \equiv 0 \mod 2$ into a
solution algorithm as follows: we initialize the set of equations $E$
to the empty set; then, while the system of equations $E$ has multiple
nonzero solutions, we apply the circuit described in
Sect.~\ref{sec:simonquantum} (Fig.~\ref{fig:simoncircuit}) to obtain
$\vk$, and add the equation $\vk \bullet \va \equiv 0 \mod 2$ to
$E$. Notice that $\va = \v{0}$ is always a solution of the homogeneous
system $E$, but we are interested in determining if there are some
nonzero solutions as well. In other words, we want to determine if the
null space of the matrix of the linear system contains any nonzero
vector. The termination condition for the algorithm (i.e., the system
of equations $E$ does not have multiple nonzero solutions) triggers in
one of two possible situations: either the system has a uniquely
determined nonzero solution $\va \neq \v{0}$, or the only possible
solution is $\va = \v{0}$; in both cases, the solution must be the
hidden string $\va$. Because there are $n$ unknowns and we are dealing
with a homogeneous system, to identify which of the two situations
happens we need $E$ to contain $n-1$ linearly independent vectors
$\vk$, where independence is intended in the set $\mathbb{F}_2^n$.
\begin{remark}
  $\mathbb{F}_2$, also denoted $\text{GF}(2)$, is the finite field
  with two elements: 0 and 1, where addition and multiplication are
  performed modulo 2. $\mathbb{F}_2^n$ is essentially the set of
  $n$-digit binary strings, combined using XOR and AND.
\end{remark}
In total, $n-1$ linearly independent vectors are sufficient because
the space of all vectors perpendicular to $\va$ modulo 2 has dimension
$n-1$. At every iteration we obtain a random $\vk$ for which $\vk
\bullet \va \equiv 0 \mod 2$; thus, we need to analyze how many
iterations are needed to obtain $n-1$ linearly independent such vectors
with high probability.

In continuous space, uniform random sampling of vectors yields
linearly independent vectors with probability 1. In this case we are
considering linear independence among vectors that have coefficients 0
or 1, and independence is in terms of the modulo-2 sum, so the
argument is less clear; however, we show that even in
$\mathbb{F}_2^n$, the probability of obtaining $n-1$ linearly
independent vectors after sampling $n + t$ times is very large, and in
particular it is bounded below by $1 - \frac{1}{2^{t+1}}$. The proof
technique is taken from \cite[Apx.~G]{mermin07quantum}.
\begin{proposition}
  \label{prop:simonindlb}
  Let $\vk^{(1)},\dots,\vk^{(n+t)}$ be vectors in $\mathbb{F}_2^{n}$,
  i.e., $n$-dimensional vectors with coefficients $\{0,1\}$ and
  addition and multiplication modulo 2. Suppose all vectors are drawn
  uniformly at random from the subspace of dimension $n-1$ of all
  vectors orthogonal to some vector $\va$. Then, the probability
  that the set $\{\vk^{(1)},\dots,\vk^{(n+t)}\}$ contains $n-1$
  linearly independent vectors (where independence is in
  $\mathbb{F}_2^{n}$) is at least $1 - \frac{1}{2^{t+1}}$.
\end{proposition}
\begin{proof}
  Consider the $(n+t) \times n$ matrix obtained by stacking the
  vectors as rows:
  \begin{equation*}
    M = 
    \begin{pmatrix}
      \rule[.6ex]{3.5em}{0.4pt} \vk^{(1)} \rule[.6ex]{3.5em}{0.4pt}\\
      \rule[.6ex]{3.5em}{0.4pt} \vk^{(2)} \rule[.6ex]{3.5em}{0.4pt} \\
      \vdots \\
      \rule[.6ex]{3.5em}{0.4pt} \vk^{(n+t)} \rule[.6ex]{3.5em}{0.4pt}
    \end{pmatrix}.
  \end{equation*}
  If this matrix has rank $n-1$, there are $n-1$ linearly independent
  vectors among $\vk^{(1)},\dots,\vk^{(n+t)}$. By assumption, the
  vectors are drawn uniformly at random from the subspace of vectors
  orthogonal to $\va$. Therefore, in a basis of this
  $(n-1)$-dimensional subspace, the vectors have coefficients that are
  drawn uniformly at random from $\{0,1\}$. Thus, if we express each
  row of $M$ in such a basis, we obtain a matrix $M'$ of size $(n+t)
  \times (n-1)$, with coefficients that are drawn uniformly at random
  from $\{0,1\}$, and with the property that if its rank is $(n-1)$,
  there are $n-1$ linearly independent vectors among
  $\vk^{(1)},\dots,\vk^{(n+t)}$.

  Let $B$ be a set of linearly independent columns of $M'$, to be
  constructed, and initialized to be the empty set.  Pick a column of
  $M'$ uniformly at random. This column is an $(n+t)$-dimensional
  vector, therefore it is nonzero with probability
  $1-\frac{1}{2^{n+t}}$. If so, add it to $B$. Now for
  $h=2,\dots,n-1$, repeat the following steps as long as each step
  successfully adds a column to $B$. Pick a random column of $M'$ that
  was not picked in previous iterations. If this column is linearly
  independent of the columns in $B$, add it to $B$. The probability of
  this happening is at least $1-\frac{1}{2^{n+t+1-h}}$: at iteration
  $h$, $|B| = h-1$, so the subspace of vectors spanned by $B$ contains
  at most $2^{h-1}$ vectors (only the coefficients $\{0,1\}$ are
  allowed in each linear combination), implying that the probability
  of sampling a random vector and obtaining a vector in the subspace
  spanned by $B$ is at most $\frac{2^{h-1}}{2^{n+t}} =
  \frac{1}{2^{n+t+1-h}}$. The probability that this algorithm
  terminates at $h=n-1$ with $|B| = n-1$ (i.e., after finding $n-1$
  linearly independent columns) is therefore at least:
  \begin{equation*}
    p = \prod_{j=1}^{n-1} \left( 1 - \frac{1}{2^{n+t+1-j}} \right).
  \end{equation*}
  We show below that $\prod_{j=1}^{n-1} \left( 1 -
  \frac{1}{2^{n+t+1-j}} \right) \ge 1 - \sum_{j=1}^{n-1}
  \frac{1}{2^{n+t+1-j}}$, which allows us to lower bound $p$ and prove
  the main result:
  \begin{equation*}
    p = \prod_{j=1}^{n-1} \left( 1 - \frac{1}{2^{n+t+1-j}} \right) \ge 1 - \sum_{j=1}^{n-1} \frac{1}{2^{n+t+1-j}} = 1 - \sum_{j=0}^{n-2} \frac{1}{2^{t+2+j}} \ge 1 - \frac{1}{2^{t+1}}.
  \end{equation*}
  It remains to show that $\prod_{j=1}^{n-1} \left( 1 -
  \frac{1}{2^{n+t+1-j}} \right) \ge 1 - \sum_{j=1}^{n-1}
  \frac{1}{2^{n+t+1-j}}$. We prove a more general statement: given
  scalars $0 \le c_1,\dots,c_n \le 1$, $\prod_{j=1}^{n} (1 - c_j) \ge
  1 - \sum_{j=1}^n c_j$. The proof is by induction on $n$. For $n=1$
  the inequality is trivially satisfied at equality. For the induction
  step (from $n-1$ to $n$), we have:
  \begin{align*}
    \prod_{j=1}^{n} (1 - c_j) &= (1-c_n) \prod_{j=1}^{n-1} (1 - c_j) \ge (1-c_n) \left(1 - \sum_{j=1}^{n-1} c_j\right)\\
    &= \left(1 - \sum_{j=1}^{n} c_j\right) + c_n \sum_{j=1}^{n-1} c_j \ge 1 - \sum_{j=1}^{n} c_j,
  \end{align*}
  where the first inequality used the induction hypothesis and the
  fact that $1-c_n \ge 0$, and the last inequality used the fact that
  all $c_j$ are nonnegative.
\end{proof}

\noindent The lower bound of Prop.~\ref{prop:simonindlb} does not
depend on $n$: if we pick, e.g., $t=9$, the probability of success it
at least $99.9\%$. Hence, with overwhelming probability after slightly
more than $n$ executions of the quantum circuit, and therefore
$\bigO{n}$ queries to the function $f$, we determine the solution to
the problem with a classical computation that can be performed in
polynomial time (i.e., $\bigO{n^2}$ to determine a solution to the
system of linear equations modulo 2). We remark that once the unique
nonzero $\va$ is determined, we can easily verify that it is the
sought binary string by querying the function $f$. On the other hand,
if $\va = \v{0}$, the algorithm detects that this is the case because
at some point the system of linear equations $E$ has $\va = \v{0}$ as
the only possible solution. Compare the $\bigO{n}$ queries of this
approach with the $\bigO{2^{n/2}}$ queries that are required by a
classical algorithm, and we have shown an exponential speedup in the
query complexity.\index{Simon's algorithm|)}\index{algorithm!Simon's|)}

This algorithm shows a typical feature of many quantum algorithms:
oftentimes, there is a classical computation to complement the quantum
computation. For example, the classical computation could be used to
verify, with certainty, that the correct solution to the problem has
indeed been found. In this case, the verification is carried out by
checking whether the system of equations has a unique nonzero
solution. Indeed, quantum algorithms are probabilistic algorithm, and
we can only try to increase the probability that the correct answer is
returned; only in rare cases the solution can be obtained with
probability 1, see e.g.~\cite{brassard2002quantum}. For this reason,
it is desirable to have a way to deterministically verify
correctness. This may require a classical computation. In other words,
sometimes the quantum algorithm is applied to a problem for which it
is difficult to classically compute the solution, but once the
solution (or some information about it) is obtained, it is easy to
classically verify that we have the right answer. Note that this may not 
be possible in general, because the complexity class BQP
(Def.~\ref{def:bqp}) is not known or believed to be contained in NP
(recall that NP is the class of problems that admit efficient
classical verification). Some of the quantum algorithms presented in
this \book{} admit simple classical verification.

\section{Notes and further reading}
\label{sec:earlyalgnotes}
The Deutsch-Jozsa algorithm \cite{deutsch1992rapid} generalizes
Deutsch's algorithm. The Bernstein-Vazirani algorithm is another
quantum algorithm developed in the early days of the field
\cite{bernstein1997quantum}, and it is based on a modified version of
the Deutsch-Jozsa construction. \cite{bernstein1997quantum}
additionally lays the mathematical foundations for computational
complexity theory of quantum algorithms. Two other notable and
groundbreaking examples of early work on quantum algorithms are Shor's
prime factorization algorithm \cite{shor97polynomial} and Grover's
search algorithm \cite{grover96fast}. An ample discussion of Grover's
algorithm is given in Ch.~\ref{ch:ampamp}. We do not discuss Shor's
algorithm, although one of its most important building blocks, the
quantum Fourier transform, is the subject of Ch.~\ref{ch:qft}. Some
notes on the relationship between the quantum Fourier transform and
Shor's algorithm are given therein, Sect.~\ref{sec:qftnotes}. In fact,
Simon's algorithm is a specific instance of the hidden subgroup
problem, discussed in the notes for Ch.~\ref{ch:qft}.

On the topic of classical verification of quantum computation, we
mention that it is an active topic of research to design verification
protocols for generic quantum computations, see, e.g.,
\cite{broadbent2009universal,aharonov2017interactive,reichardt2013classical,mahadev2018classical}. In
particular \cite{mahadev2018classical} proposes a scheme that allows a
classical computer to verify the output of a quantum computation, with
an interactive protocol in which the classical computer uses the
quantum computer to run some quantum computations and report the
results of measurements.

\chapter{Quantum Fourier transform and phase estimation}
\label{ch:qft}
\thispagestyle{fancy}
In this chapter we present two fundamental building blocks for quantum
algorithms: the quantum Fourier transform, and one of its direct
applications known as phase estimation. Both building blocks are be
used extensively in the rest of this \book{}.

\section{Quantum Fourier transform}
\label{sec:qft}
The definition of the quantum Fourier transform\index{quantum!Fourier transform|(}\index{Fourier transform|(} originates directly
from the classical discrete Fourier transform (DFT). The DFT finds
numerous applications in science and engineering, and it is so crucial
in many areas that the fast Fourier transform algorithm --- a
classical algorithm to compute the DFT --- is considered one of the
most important algorithms of the 20th century. Given $x \in \C^{2^n}$,
its DFT is defined as the vector $y \in \C^{2^n}$ with components:
\begin{equation}
  \label{eq:dft}
  y_j = \sum_{k=0}^{2^n-1} x_k e^{2\pi i jk/2^n} \quad \forall
  j=0,\dots,2^n-1.
\end{equation}
Our goal in this section is to discuss a quantum algorithm to compute
the DFT, or at least something similar to it. To do so, it is
convenient to look at the value of the DFT when applied onto the
$k$-th standard orthonormal basis vector; i.e., suppose the input
vector $x$ coincides with $\ket{\vk}$. Then the output vector $y$ has
components:
\begin{equation*}
  y_j= e^{2\pi i jk/2^n} \quad \forall j=0,\dots,2^n-1,
\end{equation*}
which in the bra-ket notation can be written as $y = \sum_{\vj \in
  \{0,1\}^n} e^{2\pi i jk/2^n} \ket{\vj}$. The vector $y$, as written,
is not normalized, so it is not a quantum state. In a very natural
way, we then define the quantum Fourier transform as follows.
\begin{definition}[Quantum Fourier transform]
  \label{def:qft}
  The \emph{quantum Fourier transform} (QFT) on $n$ qubits is the operation $Q_n$ that implements the following map:
  \begin{equation}
    \label{eq:qftdef}
    Q_n \ket{\vk} = \frac{1}{\sqrt{2^n}} \sum_{\vj \in \{0,1\}^n}
    e^{2\pi i jk/2^n} \ket{\vj} \qquad \forall \vk \in \{0,1\}^n.
  \end{equation}
\end{definition}
With this definition, given a quantum state of the form $\sum_{\vk \in
  \{0,1\}^n} x_{k} \ket{\vk}$, the $j$-th component of the quantum
state $\ket{\psi}$ obtained by applying the QFT to $\sum_{\vk \in
  \{0,1\}^n} x_{k} \ket{\vk}$, i.e., the coefficient of the $j$-th
basis state in $\ket{\psi}$, is given by:
\begin{equation*}
  \braket{\vj}{\psi} = \bra{\vj}\left(\frac{1}{\sqrt{2^n}} \sum_{\vk
    \in \{0,1\}^n} x_{k} \sum_{\vh \in \{0,1\}^n} e^{2\pi i hk/2^n}
  \ket{\vh}\right) = \frac{1}{\sqrt{2^n}} \sum_{\vk \in \{0,1\}^n}
  x_{k} e^{2\pi i jk/2^n}.
\end{equation*}
This is consistent with the classical definition in
Eq.~\eqref{eq:dft}: the only difference is the normalization factor,
which is necessary to ensure that the QFT can be implemented as a
unitary (because it has to output a unit vector when applied onto a unit
vector).

Define $\omega_{n} = e^{2 \pi i / 2^n}$. Then the matrix $Q_n$ that
implements the $n$-qubit QFT has elements:
\begin{equation*}
  (Q_n)_{j k} = \frac{1}{\sqrt{2^n}} \omega_{n}^{jk} \quad \forall
  \vj, \vk \in \{0,1\}^n.
\end{equation*}
In matrix form, this yields:
\begin{equation*}
  Q_n = 
  \frac{1}{\sqrt{2^n}}
  \begin{pmatrix}
    1 & 1 & 1 & \dots & 1 \\
    1 & \omega_{n} & \omega_{n}^{2} & \dots & \omega_{n}^{2^n-1} \\
    1 & \omega_{n}^2 & \omega_{n}^{4} & \dots & \omega_{n}^{2(2^n-1)} \\
    \vdots & & & \ddots & \vdots \\
    1 & \omega_{n}^{(2^n-1)} & \omega_{n}^{2(2^n-1)} & \dots & \omega_{n}^{(2^n-1)(2^n-1)} \\
  \end{pmatrix} =
    \frac{1}{\sqrt{2^n}}
  \begin{pmatrix}
    1 & 1 & 1 & \dots & 1 \\
    1 & \omega_{n} & \omega_{n}^{2} & \dots & \omega_{n}^{-1} \\
    1 & \omega_{n}^{2} & \omega_{n}^{4} & \dots & \omega_{n}^{-2} \\
    \vdots & & & \ddots & \vdots \\
    1 & \omega_{n}^{-1} & \omega_{n}^{-2} & \dots & \omega_{n}^{} \\
  \end{pmatrix},
\end{equation*}
using the fact that $\omega_{n}^{2^n} = 1$. The conjugate transpose of
this matrix is:
\begin{equation*}
  Q_n^{\dag} =     \frac{1}{\sqrt{2^n}}
  \begin{pmatrix}
    1 & 1 & 1 & \dots & 1 \\
    1 & \omega_{n}^{-1} & \omega_{n}^{-2} & \dots & \omega_{n}^{} \\
    1 & \omega_{n}^{-2} & \omega_{n}^{-4} & \dots & \omega_{n}^{2} \\
    \vdots & & & \ddots & \vdots \\
    1 & \omega_{n}^{} & \omega_{n}^{2} & \dots & \omega_{n}^{-1} \\
  \end{pmatrix}.
\end{equation*}
If $Q_n$ is to be implemented as a quantum algorithm, it has to be a
unitary matrix. We can verify that it is by showing $Q_n^{\dag} Q_n =
I^{\otimes n}$. We have:
\begin{equation*}
  (Q_n^\dag Q_n)_{j k} = \sum_{\v{\ell} \in \{0,1\}^n} (Q_n^\dag)_{j \ell} (Q_n)_{\ell k} = \frac{1}{2^n} \sum_{\v{\ell} \in \{0,1\}^n} \omega_{n}^{-j\ell} \omega_{n}^{\ell k} = \frac{1}{2^n} \sum_{\v{\ell} \in \{0,1\}^n} \omega_{n}^{\ell(k-j)}.
\end{equation*}
This last expression is $1$ if $j = k$ (because all terms in the
summation are equal to 1, and there are $2^n$ of them), and it is
equal to 0 otherwise, because of the formula for a geometric series
and using again the fact that $\omega_{n}^{2^n} = 1$:
\begin{equation*}
  \frac{1}{2^n} \sum_{\v{\ell} \in \{0,1\}^n}
  \left(\omega_{n}^{k-j}\right)^{\ell} = \frac{1}{2^n} \sum_{\ell=0}^{2^n-1}
  \left(\omega_{n}^{k-j}\right)^{\ell} = \frac{1}{2^n} \frac{1 -
    \omega_{n}^{2^n(k-j)}}{1 - \omega_{n}^{k-j}} = 0.
\end{equation*}
Thus, $(Q_n^\dag Q_n)_{j k} = 1$ if $j = k$ and 0 otherwise, implying
that $(Q_n^\dag Q_n)_{j k} = I^{\otimes n}$, i.e., it is the identity
matrix of size $2^n \times 2^n$. This confirms that $Q_n$ is unitary,
so there may exist an efficient quantum circuit that implements it. We
describe such a circuit in the next section.

\subsection{A useful way of expressing the QFT}
To construct a circuit that implements the QFT and therefore the
matrix $Q_n$, we first show that the image of a basis state after
applying the QFT is a product state, implying that it can be
decomposed as a tensor product of smaller-dimensional quantum
states. The expression for the decomposition leads to a circuit
construction. We make use of the following fact.
\begin{remark}
  The exponential $e^{2\pi i j / 2^n}$ can always be expressed in
  terms of $j \mod 2^n$: $e^{2\pi i}$ is equal to 1, therefore any
  integer multiple of $2\pi i$ in the exponent can be neglected.
\end{remark}
To express the QFT as a tensor product, we write the definition of
$Q_n \ket{\vk}$, and then split the corresponding sum into basis
states ending with $0$ and basis states ending with $1$:
\begin{align*}
  Q_n \ket{\vk} &= \frac{1}{\sqrt{2^n}} \sum_{\vj \in \{0,1\}^n} e^{2\pi
    i jk/2^n} \ket{\vj} \\
  &= \frac{1}{\sqrt{2^n}} \sum_{\vj \in \{0,1\}^{n-1}} e^{2\pi
    i (2j) k/2^n} \ket{\vj0} + \frac{1}{\sqrt{2^n}} \sum_{\vj \in \{0,1\}^{n-1}} e^{2\pi
    i (2j+1) k/2^n} \ket{\vj1} \\
  &= \frac{1}{\sqrt{2^n}} \sum_{\vj \in \{0,1\}^{n-1}} e^{2\pi
    i j k/2^{n-1}} \ket{\vj}\otimes\ket{0} + \frac{e^{2\pi i k / 2^n}}{\sqrt{2^n}} \sum_{\vj \in \{0,1\}^{n-1}} e^{2\pi
    i j k/2^{n-1}} \ket{\vj}\otimes\ket{1} \\
  &= \left(\frac{1}{\sqrt{2^{n-1}}} \sum_{\vj \in \{0,1\}^{n-1}} e^{2\pi
    i j k/2^{n-1}} \ket{\vj} \right) \otimes \frac{1}{\sqrt{2}} \left(\ket{0} + e^{2\pi i k / 2^n} \ket{1} \right) \\
  &= \left(Q_{n-1} \ket{\vk_2\vk_3\dots\vk_n}\right) \otimes \frac{1}{\sqrt{2}} \left(\ket{0} + e^{2\pi i k / 2^n} \ket{1} \right).
\end{align*}
In the expression above, $Q_{n-1}$ is the $2^{n-1} \times 2^{n-1}$
unitary representing the QFT on $n-1$ qubits (consistent with
Def.~\ref{def:qft}), and $\ket{\vk_2\vk_2\dots\vk_{n}}$ is the basis
state corresponding to dropping the first (most significant) digit of
$\vk$. For the last equality, we used the fact that:
\begin{equation*}
  \frac{1}{\sqrt{2^{n-1}}} \sum_{\vj \in \{0,1\}^{n-1}} e^{2\pi
    i j k/2^{n-1}} \ket{\vj} = Q_{n-1} \ket{\vk_2\vk_3\dots\vk_n}
\end{equation*}
because the value of the summation does not depend on $\vk_1$, the
first digit of $\vk$ (if $\vk_1 = 0$ we obtain $Q_{n-1}
\ket{\vk_2\vk_3\dots\vk_n}$ directly, if $\vk_1 = 1$ --- corresponding
to adding $2^{n-1}$ to the value of the integer $k$ --- we would
simply add $2\pi i j$ to the exponent, which has no effect on the
entire expression because $e^{2 \pi i j} = 1$ for integer $j$). Thus,
we have expressed the $n$-qubit transformation $Q_n \ket{\vk}$
recursively in terms of the $(n-1)$-qubit transformation $Q_{n-1}$
acting on the last $n-1$ bits of $\ket{\vk}$. We can further simplify
and better understand this expression using a little additional
notation, which is the natural extension of the notation for decimal
fractions to binary strings.
\begin{definition}[Binary fraction]
  \label{def:dot}
  For any integer $q > 0$ and binary string $\vj \in \{0,1\}^q$, we
  denote by $0.\vj$\index{notation!binary fraction@\ensuremath{0.\vj}}
  the fractional number defined as:
  \begin{equation*}
    0.\vj := \sum_{k=1}^q \frac{\vj_k}{2^k} = \frac{j}{2^{q}}.
  \end{equation*}
\end{definition}
\begin{example}
  With this notation, $0.011 = 0\cdot \frac{1}{2} + 1 \cdot \frac{1}{4} + 1 \cdot \frac{1}{8} = \frac{3}{8}$.
\end{example}

\noindent Substituting all terms of the recursion from $n$ down to
$1$, and simplifying integer powers of $e^{2\pi i}$, we then obtain
the following equation, that uses different ``shifts'' of the binary
point (separating the integer from the fractional part) or the
notation for binary fractions introduced in Def.~\ref{def:dot}:
\begin{align}
  Q \ket{\vk} &= \frac{1}{\sqrt{2}} \left(\ket{0} + e^{2\pi i k / 2} \ket{1} \right) \otimes \frac{1}{\sqrt{2}} \left(\ket{0} + e^{2\pi i k / 4} \ket{1} \right) \otimes \notag \\
  &\phantom{=} \frac{1}{\sqrt{2}} \left(\ket{0} + e^{2\pi i k / 8} \ket{1} \right) \otimes \dots \otimes \frac{1}{\sqrt{2}} \left(\ket{0} + e^{2\pi i k / 2^n} \ket{1} \right) \label{eq:qftexpr} \\
  &= \frac{1}{\sqrt{2}} \left(\ket{0} + e^{2\pi i 0.\vk_n} \ket{1} \right) \otimes \frac{1}{\sqrt{2}} \left(\ket{0} + e^{2\pi i 0.\vk_{n-1}\vk_n} \ket{1} \right) \otimes \notag\\
  &\phantom{=} \frac{1}{\sqrt{2}} \left(\ket{0} + e^{2\pi i 0.\vk_{n-2}\vk_{n-1}\vk_n} \ket{1} \right) \otimes \dots \otimes \frac{1}{\sqrt{2}} \left(\ket{0} + e^{2\pi i 0.\vk} \ket{1} \right). \notag
\end{align}
Eq.~\eqref{eq:qftexpr} shows that the QFT maps $\ket{\vk}$ to a
product state, enabling the recursive definition, and gives a precise
description of the state of each qubit in the product state.

\subsection{Implementation of the QFT}
\label{sec:qftimplementation}
\begin{figure}[h!]
\leavevmode
\centering
\ifcompilefigs
\Qcircuit @C=1em @R=1.2em {
  \lstick{\ket{\vk_1}} & \gate{H}  & \gate{P(\frac{\pi}{2})} & \qw & \dots &  & \gate{P(\frac{\pi}{2^{n-2}})} & \gate{P(\frac{\pi}{2^{n-1}})} & \qw & \rstick{\frac{1}{\sqrt{2}} \left(\ket{0} + e^{2\pi i 0.\vk}
\ket{1} \right)}\\
  \lstick{\ket{\vk_2}} & \qw       & \ctrl{-1}                & \qw & \dots &  & \qw                            & \qw                        & \qw & \rstick{\ket{\vk_2}}\\
  \vdots \\
  \\
  \lstick{\ket{\vk_{n-1}}} & \qw       & \qw                      & \qw & \dots &  & \ctrl{-4}                      & \qw                        & \qw & \rstick{\ket{\vk_{n-1}}} \\
  \lstick{\ket{\vk_{n}}} & \qw       & \qw                      & \qw & \dots &  & \qw                            & \ctrl{-5}                  & \qw & \rstick{\ket{\vk_{n}}} \\      
}
\else
\includegraphics{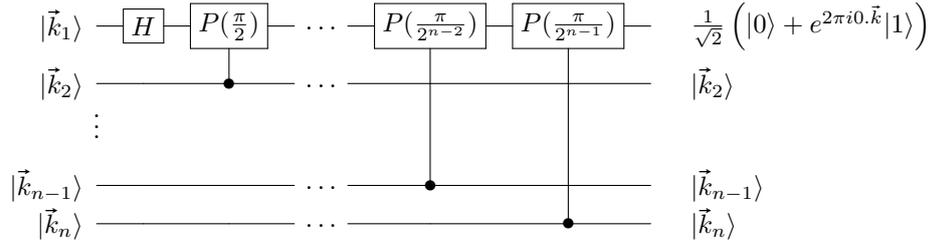}
\fi
\caption{Implementation of one qubit of the QFT.}
\label{fig:qftpart1}
\end{figure}

We have established that the QFT of a basis state is a product state,
and this helps in the construction of a circuit that implements it. We
now describe this circuit, one qubit at a time. We first discuss the
construction of the least-significant qubit $\frac{1}{\sqrt{2}}
\left(\ket{0} + e^{2\pi i 0.\vk} \ket{1} \right)$; the others
follow similarly. Note that the expression for the least-significant
qubit can be rewritten in the following way, by definition of $0.\vk$:
\begin{align*}
  \frac{1}{\sqrt{2}} \left(\ket{0} + e^{2\pi i
  0.\vk} \ket{1} \right) = \frac{1}{\sqrt{2}} \left(\ket{0} + e^{2\pi i
  0.\vk_1} e^{2\pi i 0.0\vk_2} \cdots e^{2\pi i 0.00\dots\vk_n} \ket{1} \right)
\end{align*}
Each exponential $e^{2\pi i 0.\vk_1}, e^{2\pi i 0.0\vk_2}, \dots$
applies a phase shift of a certain magnitude to $\ket{1}$ if the qubit
corresponding to $\vk_1, \vk_2, \dots$ is $\ket{1}$, and acts as the
identity otherwise. Therefore, these operations can be implemented as
controlled phase shifts. Let us define the following gate.
\begin{definition}[Phase shift gate]
  \label{def:phaseshiftgate}
  The {\em phase shift gate}\index{gate!phase shift} $P(\theta)$ is
  defined as the matrix:
  \begin{equation*}
    P(\theta) :=
    \begin{pmatrix}
      1 & 0 \\ 0 & e^{i \theta}
    \end{pmatrix}.
  \end{equation*}
\end{definition}
Notice that the $Z$ gate is a particular case of the phase shift gate,
setting $\theta = \pi$; see also Def.~\ref{def:rz} and the surrounding
discussion. The effect of the phase shift gate is to apply a phase to
$\ket{1}$, while $\ket{0}$ is untouched. Because we need to apply the
phase shift only if certain values of $\v{k}$ are 1, we use a
controlled version of the $P(\theta)$ gate, easy to obtain from a
basic set of operations, see Sect.~\ref{sec:basicops}; thus, we can
efficiently implement the unitary that constructs $\frac{1}{\sqrt{2}}
\left(\ket{0} + e^{2\pi i 0.\vk} \ket{1} \right)$. The corresponding
circuit is given in Fig.~\ref{fig:qftpart1}. The desired qubit state
is found in the first (topmost) qubit.

\begin{figure}[htb!]
\leavevmode
\centering
\ifcompilefigs
\Qcircuit @C=0.8em @R=0.7em {
  & \gate{H}  & \gate{P(\frac{\pi}{2})} & \qw & \dots &  & \gate{P(\frac{\pi}{2^{n-2}})} & \gate{P(\frac{\pi}{2^{n-1}})} & \qw      & \qw & \dots &  & \qw                            & \qw                        & \qw      & \qw                      & \qw      & \qswap      & \qw         & \qw & \dots & \\
  & \qw       & \ctrl{-1}                & \qw & \dots &  & \qw                            & \qw                        & \gate{H} & \qw & \dots &  & \gate{P(\frac{\pi}{2^{n-3}})} & \gate{P(\frac{\pi}{2^{n-2}})} & \qw      & \qw                      & \qw      & \qw \qwx    & \qswap      & \qw & \dots &  \\
  \vdots &    &                          &     &       &  &                                &                            &          &     &       &  &                                &                            &          &                          &          & \qwx        & \qwx        & \vdots      &  \\
  &           &                          &     &       &  &                                &                            &          &     &       &  &                                &                            &          &                          &          & \qwx        & \qwx        &     &       &  \\
  & \qw       & \qw                      & \qw & \dots &  & \ctrl{-4}                      & \qw                        & \qw      & \qw & \dots &  & \ctrl{-3}                      & \qw                        & \gate{H} & \gate{P(\frac{\pi}{2})} & \qw      & \qw \qwx    & \qswap \qwx & \qw & \dots &   \\
  & \qw       & \qw                      & \qw & \dots &  & \qw                            & \ctrl{-5}                  & \qw      & \qw & \dots &  & \qw                            & \ctrl{-4}                  & \qw      & \ctrl{-1}                & \gate{H} & \qswap \qwx & \qw         & \qw & \dots &  \\      
}
\else
\includegraphics{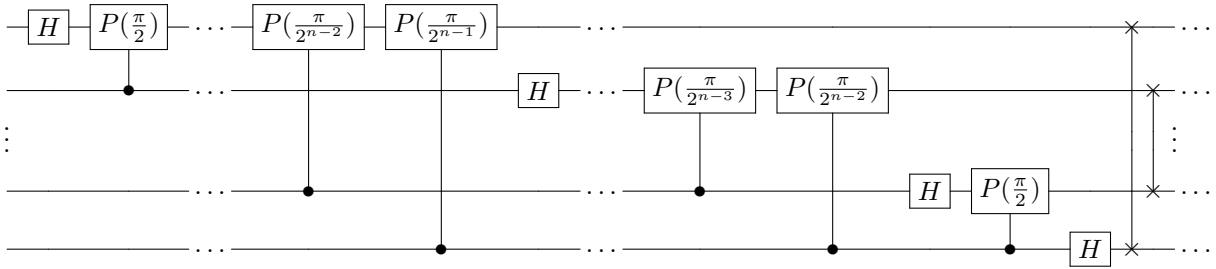}
\fi
\caption{Implementation of the QFT.}
\label{fig:qft}
\end{figure}

The circuit first applies the Hadamard gate $H$ on the top qubit;
we claim that this is equivalent to the transformation:
\begin{equation*}
  \ket{\vk_1} \to \frac{1}{\sqrt{2}} \left(\ket{0} + e^{2\pi i
  0.\vk_1} \ket{1} \right).
\end{equation*}
Indeed, this is exactly what the Hadamard does, because $e^{2\pi i
  0.\vk_1} = (-1)^{\vk_1}$, see Eq.~\eqref{eq:hadamard}. Next, it is
obvious that each of the subsequent controlled gates applies one of
the phase factors $e^{2\pi i 0.0\vk_2}, \cdots, e^{2\pi i
  0.00\dots\vk_n}$ to the $\ket{1}$ state in the first qubit, because
these are given by the $P(\theta)$ gates controlled by qubits
$2,\dots,n$. Hence, the circuit in Fig.~\ref{fig:qftpart1} constructs
the rightmost qubit of the QFT applied to $\ket{\vk}$, i.e., the
single-qubit state $\frac{1}{\sqrt{2}} \left(\ket{0} + e^{2\pi i
  0.\vk} \ket{1} \right)$, mapping the first qubit of the input (the
one containing $\vk_1$) to the last qubit of the expression of the
QFT.  We can now proceed by induction, because this qubit is in a
product state with the remaining qubits, see
Eq.~\eqref{eq:qftexpr}. Thus, we recursively apply the same circuit,
with its size reduced by one each time, to the qubit lines containing
$\vk_2\cdots \vk_n$, yielding the full QFT implementation. Note that
we no longer have access to $\vk_1$ after applying
Fig.~\ref{fig:qftpart1}, but this is not an issue: from
Eq.~\eqref{eq:qftexpr}, we can see that $\vk_1$ only appears in the
rightmost qubit on the r.h.s.\ of the equation. Hence, we do not need
$\vk_1$ after the the rightmost qubit of the QFT (top qubit in
Fig.~\ref{fig:qftpart1}) is computed.
\begin{remark}
  Because each application of the circuit in Fig.~\ref{fig:qftpart1}
  outputs the \emph{last} qubit of the desired output in the
  \emph{first} position of the output lines, at the end of the
  computation we should swap all qubits to restore the initial
  order. Alternatively, we do not need to swap as long as we keep
  track of the position of each qubit and consequently adjust all
  subsequent operations in the circuit, including measurements. (If
  the QFT is the last operation of a circuit, followed by a
  measurement of all qubits, we can simply rearrange the classical
  bits after measurement.)
\end{remark}
Putting it all together, we obtain the circuit depicted in
Fig.~\ref{fig:qft}. This circuit contains $\bigO{n^2}$ one and
two-qubit gates. Rather than use oracle complexity to determine the
runtime of an algorithm --- which is useful for
information-theoretical purposes --- for this algorithm there is no
natural query concept (i.e., no function is being queried), and it
makes more sense to use gate complexity, in which we assess the
performance of an algorithm by looking at how many elementary gates it
uses, see Rem.~\ref{rem:complexity}. We consider all single-qubit and
two-qubit gates as elementary, because they can all be constructed
with a constant number of basic gates (if the precision is fixed),
i.e., gates from a minimal, universal set of gates. (Technically, to
keep the same accuracy of the entire implementation as $n$ grows, we
need to increase the precision with which each gate is approximated
with basic gates, but remember that the cost for doing so is only
polylogaritmically-large in the precision, see Thm.~\ref{thm:sk}.)
Thus, the QFT uses a number of basic gates polynomial (in fact,
quadratic) in $n$; this is an exponential improvement over the
classical fast Fourier transform, which uses $\bigO{n2^n}$ basic
operations and therefore time.

\begin{example}
  We show a full example of the QFT circuit on three qubits. It is
  given in Fig.~\ref{fig:qftex}.
  \begin{figure}[h!]
    \leavevmode
    \centering
    \ifcompilefigs
    \Qcircuit @C=1em @R=0.7em {
      & \gate{H}  & \gate{P(\frac{\pi}{2})} & \gate{P(\frac{\pi}{4})} & \qw & \qw & \qw & \qswap & \qw \\
      & \qw & \ctrl{-1} & \qw & \gate{H}  & \gate{P(\frac{\pi}{2})} & \qw & \qw \qwx & \qw \\
      & \qw & \qw & \ctrl{-2} & \qw & \ctrl{-1} & \gate{H} & \qswap \qwx & \qw \\
    }
    \else
    \includegraphics{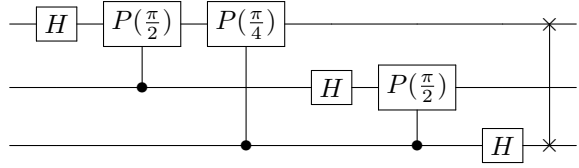}
    \fi
    \caption{Example of the QFT on three qubits.}
    \label{fig:qftex}
  \end{figure}
  The gates $P(\frac{\pi}{2})$ and $P(\frac{\pi}{4})$ are commonly
  called $S$ and $T$, respectively. We have seen the $T$ gate in
  Thm.~\ref{thm:sk}. Using these new names and substituting the SWAP
  in terms of C$X$, we obtain the circuit in Fig.~\ref{fig:qftexfull}.
  \begin{figure}[h!]
    \leavevmode
    \centering
    \ifcompilefigs
    \Qcircuit @C=1em @R=0.7em {
      & \gate{H}  & \gate{S} & \gate{T} & \qw & \qw & \qw & \ctrl{2} & \targ & \ctrl{2} & \qw \\
      & \qw & \ctrl{-1} & \qw & \gate{H}  & \gate{S} & \qw & \qw & \qw &\qw & \qw \\
      & \qw & \qw & \ctrl{-2} & \qw & \ctrl{-1} & \gate{H} & \targ & \ctrl{-2} & \targ & \qw \\
    }
    \else
    \includegraphics{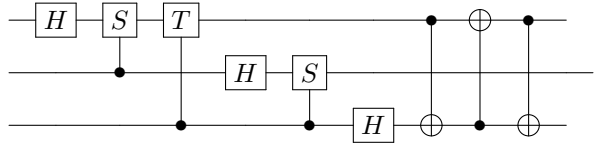}
    \fi
    \caption{Example of the QFT on three qubits, with the SWAP gate
      decomposed into basic operations.}
    \label{fig:qftexfull}
  \end{figure}
  As an exercise, perhaps aided by computer code, we could carry out
  the calculations to compute the unitary matrix corresponding to this
  circuit, and verify that it implements the matrix $Q_{3}$.
\end{example}
\index{quantum!Fourier transform|)}\index{Fourier transform|)}

\section{Phase estimation}
\label{sec:qpe}
The QFT is a crucial building block of many useful quantum
subroutines, and its application leads almost directly to another
crucial building block for quantum algorithms, called \emph{phase
estimation} (quantum phase estimation, QPE). Under an appropriate
input model, phase estimation gives an exponential speedup with
respect to classical algorithms for the same problem.

The purpose of phase estimation is to determine the eigenvalue of a
given eigenstate of a given unitary. Let $U$ be a unitary on $n$
qubits that we can efficiently implement. Let $\ket{\psi}$ be an
eigenstate (Def.~\ref{def:eigenstate}) of $U$. Because $U$ is unitary, its
eigenvalues have modulus one. Hence, we can write:
\begin{equation*}
  U \ket{\psi} = e^{2\pi i \varphi} \ket{\psi},
\end{equation*}
where $\varphi \in [0,1)$.  Phase estimation\index{quantum!phase estimation|(}\index{algorithm!phase estimation|(}\index{phase!estimation|(} solves the following
  problem: given $\epsilon > 0$, quantum circuits for a controlled
  version of $U^{2^k}$ for any $k \le \ceil{\log \frac{1}{\epsilon}}$,
  and a circuit to construct the state $\ket{\psi}$ such that $U
  \ket{\psi} = e^{2\pi i \varphi} \ket{\psi}$, determine
  $\tilde{\varphi}$ such that $\min\{|\varphi - \tilde{\varphi}|, 1
  -|\varphi -\tilde{\varphi}|\} \le \epsilon$.
\begin{remark}
  The distance between $\varphi$ and $\tilde{\varphi}$ is intended
  with period $1$, i.e., $0.99$ is close to $0.01$: all angles in the
  exponential can be interpreted modulo $2\pi$, and the angle $2\pi
  (0.99)$ is close to the angle $2\pi (0.01)$. This is why we take
  $\min\{|\varphi - \tilde{\varphi}|, 1 -|\varphi
  -\tilde{\varphi}|\}$: if $\varphi = 0.99$ and $\tilde{\varphi} =
  0.01$, the expression $\min\{|\varphi - \tilde{\varphi}|, 1
  -|\varphi -\tilde{\varphi}|\}$ evaluates to 0.02.
  \end{remark}
Note that classically, $\varphi$ can be computed by carrying out
the multiplication $U \ket{\psi}$, but this takes time
$\bigO{4^{n}}$ in general because $U$ is a $2^n \times 2^n$ matrix.

\subsection{Main idea for quantum phase estimation}
\label{sec:qpeidea}
Let $m = \ceil{\log \frac{1}{\epsilon}}$: if we obtain a
representation $\tilde{\varphi}$ of $\varphi$ with $m$ correct binary
digits, then $\tilde{\varphi}$ is no more than $\epsilon$ away from
$\varphi$. For now, we make the simplifying assumption that $\varphi$
is exactly representable on $m$ bits. More formally, we assume that
there exists $\vp \in \{0,1\}^m$ such that $\varphi = p/2^m =
0.\vp$. Thus, to obtain a representation of $\varphi$ with $m$ correct
digits we need to output $\vp$. (If the assumption is not verified,
i.e., $\varphi$ requires more than $m$ digits to be written in binary,
then we would like to output the closest representation of $\varphi$
on $m$ bits; we discuss this case more precisely in
Sect.~\ref{sec:qpegeneral}.)

\begin{figure}[h!]
  \leavevmode
  \centering
  \ifcompilefigs
  \Qcircuit @C=1em @R=0.7em {
    & \qw & \gate{H}  & \qw & \qw & \qw & \dots & & \ctrl{4} & \qw & \qw \\
    & \vdots &        &     &     &     &       & &          &     &     \\
    & \qw & \gate{H}  & \qw & \ctrl{2} & \qw & \dots & & \qw & \qw & \qw \\
    & \qw & \gate{H}  & \ctrl{1} & \qw & \qw & \dots & & \qw & \qw & \qw \\
    \lstick{\ket{\psi}} & {/^n} \qw & \qw & \gate{U^{2^0}} & \gate{U^{2^1}} & \qw & \dots & & \gate{U^{2^{m-1}}} & \qw & \qw
    \inputgroupv{1}{4}{.8em}{2.2em}{\ket{\v{0}}_m} \\
  }
  \else
  \includegraphics{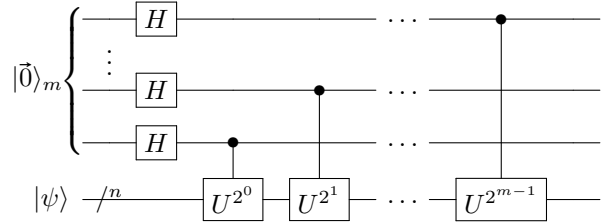}
  \fi
  \caption{State preparation for the quantum phase estimation.}
  \label{fig:qpestateprep}
\end{figure}

We exploit the fact that quantum computation is reversible: we would
like to produce the state $\ket{\vp}$, because it encodes the desired
answer, so we study how to obtain a state that can be transformed into
$\ket{\vp}$. In particular, we study the Fourier state obtained from
$\ket{\vp}$. The image of the basis state $\ket{\vp}$ under the QFT on
$m$ qubits is:
\begin{equation*}
  Q_m \ket{\vp} = \frac{1}{\sqrt{2^m}} \sum_{\vj \in \{0,1\}^m}
  e^{2\pi i j p/2^m} \ket{\vj} = \frac{1}{\sqrt{2^m}} \sum_{\vj \in \{0,1\}^m}
  e^{2\pi i j 0.\vp} \ket{\vj}.
\end{equation*}
Using the definition of $0.\vp$, relying on the same argument that we
used in Eq.~\eqref{eq:qftexpr}, this expression can be rewritten as:
\begin{equation}
  \begin{split}
  \label{eq:qpeinteger}
  Q_m \ket{\vp} &= \frac{1}{\sqrt{2}} \left(\ket{0} + e^{2\pi i 0.\vp_m} \ket{1} \right) \otimes \frac{1}{\sqrt{2}} \left(\ket{0} + e^{2\pi i 0.\vp_{m-1}\vp_m} \ket{1} \right) \otimes \cdots \otimes \frac{1}{\sqrt{2}} \left(\ket{0} + e^{2\pi i 0.\vp} \ket{1} \right) \\
  &= \frac{1}{\sqrt{2}} \left(\ket{0} + e^{2\pi i p/2} \ket{1} \right) \otimes \frac{1}{\sqrt{2}} \left(\ket{0} + e^{2\pi i p/4} \ket{1} \right) \otimes \cdots \otimes \frac{1}{\sqrt{2}} \left(\ket{0} + e^{2\pi i p/2^m} \ket{1} \right) .
  \end{split}
\end{equation}
Note that $e^{2\pi i 0.\vp_m} = e^{2\pi i p/2} = e^{2\pi i
  2^{m-1}\varphi}$, because any integer multiple of $2\pi i$ in the
exponent cancels out, and similarly, each qubit can be expressed using
$e^{2 \pi i \varphi}$ with $\varphi$ multiplied by some power of
2. This gives the following equivalent expression for the QFT applied
to $\ket{\vp}$:
\begin{equation}
  \label{eq:qpetarget}
  Q_m \ket{\vp} = \frac{1}{\sqrt{2}} \left(\ket{0} + e^{2\pi i 2^{m-1} \varphi} \ket{1} \right) \otimes \frac{1}{\sqrt{2}} \left(\ket{0} + e^{2\pi i 2^{m-2} \varphi} \ket{1} \right) \otimes \cdots \otimes \frac{1}{\sqrt{2}} \left(\ket{0} + e^{2\pi i 2^0 \varphi} \ket{1} \right).
\end{equation}
Then, if we could construct this state, the inverse QFT $Q_m^{\dag}$
would recover $\vp$, allowing us to determine $\varphi$. To prepare
the state in Eq.~\eqref{eq:qpetarget}, we must be able to construct a
quantum state that has several phase factors of the form $e^{2 \pi i
  2^k \varphi}$. We can obtain any such phase factor with phase
kickback and repeated applications of $U$, exploiting the fact that
$\ket{\psi}$ is an eigenstate with eigenvalue $e^{2 \pi i
  \varphi}$. Indeed, we have:
\begin{equation*}
  U^{2^k} \ket{\psi} = \left(e^{2 \pi i \varphi}\right)^{2^k} \ket{\psi} = e^{2 \pi i 2^k \varphi} \ket{\psi}.
\end{equation*}
Thus, we start with the state $\frac{1}{\sqrt{2^m}} \left(\ket{0} +
\ket{1}\right)^{\otimes m} \otimes \ket{\psi}$, that can be
constructed using $m$ Hadamard gates applied to $\ket{\v{0}}$ and the
circuit to prepare $\ket{\psi}$, which is given as input by
assumption. Next, we apply a controlled version of $U$ to the qubit
lines corresponding to $\ket{\psi}$, controlled by the $m$-th qubit;
we follow up with an application of $U^{2^{1}}$ to the qubit lines
corresponding to $\ket{\psi}$, controlled by the $(m-1)$-th qubit; and
so on, applying the unitary $U^{2^j}$ controlled by qubit $m-j$ for
$j=0,\dots,m-1$. This leads to the circuit given in
Fig.~\ref{fig:qpestateprep}.

\begin{figure}[h!]
  \leavevmode
  \centering
  \ifcompilefigs
  \Qcircuit @C=1em @R=0.7em {
    & \qw & \gate{H}  & \qw & \qw & \qw & \dots & & \ctrl{4} & \qw & \multigate{3}{Q_{m}^{\dag}} & \qw \\
    &\vdots&        &     &     &     &       & &          &     &     &  \\
    & \qw & \gate{H}  & \qw & \ctrl{2} & \qw & \dots & & \qw & \qw & \ghost{Q_{m}^{\dag}} & \qw & \rstick{\begin{array}{l}\ket{\vp}\\\phantom{\_}\end{array}} \\
    & \qw & \gate{H}  & \ctrl{1} & \qw & \qw & \dots & & \qw & \qw & \ghost{Q_{m}^{\dag}} & \qw \\
    \lstick{\ket{\psi}} & {/^n} \qw & \qw & \gate{U^{2^0}} & \gate{U^{2^1}} & \qw & \dots & & \gate{U^{2^{m-1}}} & \qw & {/^n} \qw & \qw & \rstick{\ket{\psi}}
    \inputgroupv{1}{4}{.8em}{2.2em}{\ket{\v{0}}_m}
    \gategroup{1}{12}{4}{12}{1.75em}{\}} \\
  }
  \else
  \includegraphics{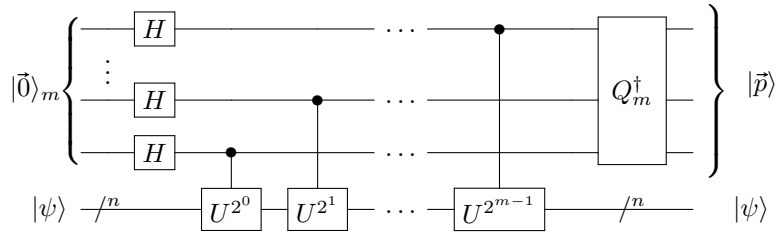}
  \fi
  \caption{Quantum phase estimation circuit on $m$ qubits.}
  \label{fig:qpefull}
\end{figure}

To understand Fig.~\ref{fig:qpestateprep}, we examine the effect of
applying a Hadamard on qubit $m-j$, followed by controlled-$U^{2^j}$,
where qubit $m-j$ is the control, and $U^{2^j}$ acts on $\ket{\psi}$
when the control is active (i.e., $\ket{1}$). We have:
\begin{align*}
  \text{C}U^{2^j} (H \otimes I) \ket{0} \otimes \ket{\psi} = \text{C}U^{2^j} \left( \frac{1}{\sqrt{2}}(\ket{0} + \ket{1}) \otimes \ket{\psi} \right) = \frac{1}{\sqrt{2}} \left(\ket{0} \otimes \ket{\psi} + \ket{1} \otimes e^{2\pi i 2^j \varphi} \ket{\psi} \right).
\end{align*}
This is exactly the state that we want to construct for qubit $m-j$ in
Eq.~\eqref{eq:qpetarget}. It is also a product state,
because it can be expressed as the tensor product:
\begin{equation*}
  \frac{1}{\sqrt{2}} \left(\ket{0}  + e^{2\pi i 2^j \varphi} \ket{1}\right) \otimes \ket{\psi}
\end{equation*}
By induction, starting from qubit $m$ down to $1$, it is easy to prove
that the circuit in Fig.~\ref{fig:qpestateprep} leaves the first $m$
qubit lines in a product state, and produces
Eq.~\eqref{eq:qpetarget}. Thus, we can construct the full quantum
phase estimation circuit as given in Fig.~\ref{fig:qpefull}.
The correctness of this construction is ensured by the above
discussion: the first part of the circuit, i.e., the circuit in
Fig.~\ref{fig:qpestateprep}, outputs the state $\frac{1}{\sqrt{2}}
\left(\ket{0} + e^{2\pi i 2^{m-1} \varphi} \ket{1} \right) \otimes
\frac{1}{\sqrt{2}} \left(\ket{0} + e^{2\pi i 2^{m-2} \varphi} \ket{1}
\right) \otimes \cdots \otimes \frac{1}{\sqrt{2}} \left(\ket{0} +
e^{2\pi i 2^0 \varphi} \ket{1} \right) \otimes \ket{\psi}$ as in
Eq.~\eqref{eq:qpetarget}. Then the inverse QFT, which is exactly the
inverse of the transformation in Eq.~\eqref{eq:qpeinteger}, produces
the state $\ket{\vp}$ in the top $m$ output lines.

\subsection{General phase estimation algorithm}
\label{sec:qpegeneral}
The previous section shows an implementation of the QPE under the
assumption that $\varphi$ is exactly representable on $m$ bits, but
this is a restrictive assumption that may not hold in practice. More
importantly, a priori we have no way of verifying whether the
assumption holds unless we know $\varphi$ in the first place, which
would defeat the purpose of the algorithm. Fortunately it is easy to
relax this assumption. Recall that the QFT is a continuous
transformation, because it is a linear map on a finite-dimensional
vector space. The forward transformation of the QFT is given in
Eq.~\eqref{eq:qpeinteger}, mapping $\ket{\vp}$ to its image state. By
continuity, if we apply the inverse QFT to a state that is close to
$Q_m \ket{\vp} = \frac{1}{\sqrt{2}} \left(\ket{0} + e^{2\pi i 0.\vp_m}
\ket{1} \right) \otimes \frac{1}{\sqrt{2}} \left(\ket{0} + e^{2\pi i
  0.\vp_{m-1}\vp_m} \ket{1} \right) \otimes \cdots \otimes
\frac{1}{\sqrt{2}} \left(\ket{0} + e^{2\pi i 0.\vp} \ket{1} \right)$,
we obtain a state that is close to $\ket{\vp}$. In particular, while
$2^m \varphi$ may not be an integer, it is close to some
$m$-bit-representable integer, hence applying the inverse QFT yields
the binary string corresponding to that integer with high
probability. This result is formalized next, by stating that if we
want to obtain an accurate $q$-digit representation of the phase
$\varphi$, we need to run the quantum phase estimation using slightly
more than $q$ qubits; this suffices to ensure that the first $q$
digits of the output are correct with good probability.
\begin{theorem}[Phase estimation; based on Sect.~5.2 in \cite{nielsen02quantum}]
  \label{thm:qpe}
  When applying phase estimation, let $\vp$ be the output of the
  procedure (corresponding to estimating the eigenvalue as $e^{2 \pi i
    0.\vp}$) when applied to an eigenstate with eigenvalue $e^{2 \pi i
    \varphi}$, i.e., with phase $2 \pi i \varphi$. If we use $q +
  \ceil{\log(2 + \frac{1}{2\delta})}$ qubits of precision, i.e.,
  execute the circuit in Fig.~\ref{fig:qpefull} setting $m = q +
  \ceil{\log(2 + \frac{1}{2\delta})}$, then the first $q$ bits of
  $\vp$ are accurate with probability at least $1 - \delta$, i.e.,
  $\Pr(\min\{|\varphi - 0.\vp|, 1 - |\varphi - \vp|\} < 2^{-q}) >
  1 - \delta$.
\end{theorem}
The proof is technical, so we skip it; a detailed proof for the above
statement can be found in \cite{nielsen02quantum}. However, there is a
precise characterization of the probability distribution of phase
estimation that is sometimes very useful in the analysis of certain
quantum algorithms, and it is worth mentioning. For the most intuitive
version of this result (including the edge case where the phase is
representable exactly with the number of qubits used), it is helpful
to rely on the $\sinc$ function, used in signal processing and defined
below.
\begin{definition}[Normalized sinc function]
  \label{def:sinc}
  The \emph{normalized sinc function} is defined as $\sinc(x) :=
  \frac{\sin(\pi x)}{\pi x}$ for $x \neq 0$, and $\sinc(x) := 1 =
  \lim_{y \to 0} \frac{\sin(\pi y)}{\pi y}$ for $x = 0$.
\end{definition}
\begin{proposition}[Lem.~7.1.2 in \cite{kaye07introduction}]
  \label{prop:qpedistr}
  Suppose the phase estimation circuit (Fig.~\ref{fig:qpefull}) is
  applied to an eigenstate $\ket{\psi}$ with phase $\varphi$.
  Let $X$ be the random variable describing the measurement outcomes
  of the output register (top $m$ qubit lines). Then $X$ satisfies:
  \begin{equation*}
    \Pr(X = \vk) = \frac{\sinc^2 \left(2^m(\varphi - 0.\vk)\right)}{\sinc^2 (\varphi - 0.\vk)}.
  \end{equation*}
\end{proposition}
For a proof of this result, see
\cite[Sect.~7.1.1]{kaye07introduction},
\cite[Lem.~10]{brassard2002quantum}. Using Prop.~\ref{prop:qpedistr},
one can prove a somewhat simpler (and easier to remember) version of
Thm.~\ref{thm:qpe} to characterize the probability of success of phase
estimation.
\begin{theorem}[Phase estimation, alternative version; Thm.~7.1.5 in \cite{kaye07introduction}]
  \label{thm:qpesimple}
  Let $m$ be fixed, and suppose we apply quantum phase estimation to
  an eigenstate $\ket{\psi}$ with eigenvalue $e^{2\pi i
    \varphi}$. Suppose further that $\varphi$ satisfies $\frac{k}{2^m}
  \le \varphi \le \frac{k+1}{2^m}$ for $k \in \{0,\dots,2^m-1\}$. Then
  the phase estimation circuit (Fig.~\ref{fig:qpefull}) outputs $k$ or
  $k+1$ (expressed in binary) as measurement outcomes with probability
  at least $\frac{8}{\pi^2} \approx 0.81$, corresponding to estimating
  the eigenvalue as $e^{2 \pi i k/2^m}$ or $e^{2 \pi i (k+1)/2^m}$.
\end{theorem}
Note that in Thm.~\ref{thm:qpesimple}, the integers $k$ and $k+1$ are
the two closest representations of the exact phase $\varphi$ using $m$
digits of precision (the denominator $2^m$ is fixed, we are aiming to
determine a fraction), and they both lead to error $\le 2^{-m}$. The
result states that we obtain one of these two integers with high
probability, $\approx 0.81$. The bound for the probability of
obtaining the closest integer, rather than one of the two closest
integers, is $\frac{4}{\pi^2} \approx 0.41$, see
\cite{kaye07introduction} for details.

\paragraph{Gate complexity.} The gate complexity of the QPE algorithm is $\bigO{m^2}$, which is the gate complexity of the inverse QFT,
plus the cost of applying all the controlled unitaries $U^{2^j}$ for
$j=0,\dots,m-1$. If each gate is compiled down to basic gates from a
universal set, the total gate complexity increases slightly to
$\bigOt{n^2}$ for arbitrarily high (but constant) precision.
\begin{remark}
  \label{rem:qpeuexp}
  Constructing $U^{2^j}$ may not be easy: a trivial construction that
  applies $U$ repeatedly a total of $2^j$ times incurs an exponential
  cost in $m$, in general. Sometimes this is acceptable: usually we
  want to estimate the phase with error at most $\epsilon$, and we
  choose $m = \bigO{\log \frac{1}{\epsilon}}$, so $2^m =
  \bigO{\frac{1}{\epsilon}}$. In other words, even if the number of
  calls to $U$ (query complexity) depends exponentially on the number
  of qubits, the number of qubits is usually polylogarithmic in one of
  the input parameters, i.e., the desired precision. This yields a
  number of applications of $U$ that is polynomial in $\epsilon$:
  $\bigO{\frac{1}{\epsilon} \log \frac{1}{\delta}}$ for probability of
  failure at most $\delta$. Other times the exponential cost in $m$
  may not be acceptable. Unfortunately, in general we cannot ``fast
  forward'' the implementation of $U^{2^j}$ using a polynomial number
  operations. However, for some specific matrices $U$ it may be
  possible to construct $U^{2^j}$ more efficiently, avoiding
  exponential costs. One situation where this is known to be the case
  is the modular exponentiation function used in Shor's algorithm
  \cite{shor97polynomial}\index{algorithm!Shor's}\index{Shor's algorithm}, in which $U$ applies the function $f(x) = a^x \mod
  2^n$. Because this is the power function, in can be implemented
  efficiently using the repeated squaring algorithm, i.e., computing
  $a^2, a^4, a^8, \dots$ simply by squaring the result each time. This
  implies that constructing $U^{2^j}$ is much less expensive than
  expected. Another situation where this occurs is discussed in
  Sect.~\ref{sec:jordanlinear}, in the context of the gradient
  algorithm; there, we exploit a binary representation of a certain
  function and phase kickback to efficiently construct $U^{2^j}$.
\end{remark}

\paragraph{Inexact eigenstates.} We conclude our study of the QPE by analyzing the effect of applying
QPE to a state that is not an eigenstate of $U$. Let $\ket{\psi_j},
j=0,\dots,2^{n-1}$ be an orthonormal eigenbasis for $U$ with
corresponding eigenvalues $e^{2\pi i \varphi_j}$. Then there exist
coefficients $\alpha_j$ such that:
\begin{equation*}
  \ket{\psi} = \sum_{j=0}^{2^n-1} \alpha_j \ket{\psi_j},
\end{equation*}
with $\sum_{j=0}^{2^n-1} |\alpha_j|^2 = 1$ due to the normalization
condition. Because QPE maps $\ket{\v{0}}_m \ket{\psi_j} \to
\ket{\vp^{(j)}} \ket{\psi_j}$ with probability at least
$1-\delta$, where $\vp^{(j)}$ is some integer representation of the
(potentially fractional) phase $\varphi_j$, by linearity it maps:
\begin{equation*}
  \ket{\v{0}}_m \otimes \left(\sum_{j=0}^{2^n-1} \alpha_j \ket{\psi_j}\right) \longrightarrow \sum_{j=0}^{2^n-1} \alpha_j \left(\ket{\vp^{(j)}} \otimes \ket{\psi_j}\right)
\end{equation*}
with probability at least $1-\delta$. Hence, if we perform a
measurement of the first register, with probability $1-\delta$ we are
able to obtain the phase of one of the eigenvalues as the output of
the circuit; which eigenvalue is produced depends on the overlap
$|\braket{\psi_j}{\psi}| = |\alpha_j|$. If we want to obtain the phase
of a specific eigenvalue $\varphi_j$, we must be able to produce a
state with large $|\alpha_j|^2$. We formalize this as follows.
\begin{proposition}
  Suppose we want to estimate the eigenvalue $e^{2 \pi i \varphi^*}$
  of the eigenstate $\ket{\psi^*}$ of a unitary $U$, but we only have
  the ability to prepare a state $\ket{\xi}$ that may not coincide
  with $\ket{\psi^*}$. Then, applying phase estimation with precision
  $\epsilon$ and probability of success $\ge 1-\delta$, gives us a
  binary description $\v{p}^*$ of $\varphi^*$ with precision
  $\epsilon$ with probability at least
  $(1-\delta)|\braket{\xi}{\psi^*}|^2$.
\end{proposition}
\begin{proof}
  Using an eigenbasis of $U$ to express $\ket{\xi}$, just as in the
  discussion before the proposition statement, we have:
  \begin{equation*}
    \ket{\xi} = \sum_{j=0}^{2^n-1} \alpha_j \ket{\psi_j}.
  \end{equation*}
  Suppose the desired eigenstate is $\ket{\psi^*} = \ket{\psi_h}$. By
  linearity, the action of the QPE is:
  \begin{align*}
    \text{QPE}\left(\ket{\v{0}} \otimes \ket{\xi}\right) = \text{QPE}\left(\ket{\v{0}} \otimes \sum_{j=0}^{2^n-1} \alpha_j \ket{\psi_j}\right) =
    \sum_{j=0}^{2^n-1} \alpha_j \left(\ket{\vp^{(j)}} \otimes \ket{\psi_j}\right),
  \end{align*}
  and this operation is successful with probability at least
  $1-\delta$. It follows that the probability of obtaining
  $\v{p}^{(h)} = \v{p}^*$ is at least $|\alpha_h|^2 (1-\delta)$. Then,
  note that $\alpha_h = \braket{\xi}{\psi^*}$. This concludes the
  proof.
\end{proof}

\section{Iterative phase estimation}
\label{sec:iterativeqpe}
Rather than performing phase estimation in a single pass, i.e.,
obtaining at the same time the entire bit description of the phase
with a given accuracy, it is possible to break the procedure down to
simpler steps, obtaining one bit of the phase at a time. This line of
work was initially developed by Kitaev, and a detailed description can
be found in \cite{kitaev2002classical}. Iterative phase estimation
allows splitting the phase estimation circuit into several smaller
circuits, which are more likely to be executable by a quantum
computing device with limited capabilities. Although the query
complexity (i.e., number of calls to the unitary whose eigenvalue is
being estimated) for the iterative phase estimation algorithm is not
better than the standard algorithm described in Sect.~\ref{sec:qpe},
its analysis is instructive for at least two reasons. First, it gives
an avenue to obtain a different tradeoff regarding the requirement of
computational resources, i.e., number of qubits and number of
gates. Second, it uses an idea that we have not seen so far in this
\book{}, and that has proven very successful in a variety of
situations: start by obtaining a coarse approximation of the answer,
and then, based on that approximation, define a new problem that
iteratively improves over the current estimate. In a different context
and with different benefits and costs, this idea can be powerful also
for linear algebra and optimization; we discuss some approaches based
on a related scheme in Sect.s~\ref{sec:hhlir} and
\ref{sec:blockencnotes}.\index{iterative refinement}

Let us formally state the goal of the algorithm described in this
section. It is exactly the same as in Sect.~\ref{sec:qpe}: given
$\epsilon > 0$, quantum circuits for a controlled version of $U^{2^k}$
for any $k \le \ceil{\log \frac{1}{\epsilon}}$, and a circuit to
construct the state $\ket{\psi}$ such that $U \ket{\psi} = e^{2\pi i
  \varphi} \ket{\psi}$, determine a binary string $\vp$ with the
property that $\Pr(\min\{|\varphi - 0.\vp|, 1 - |\varphi -
0.\vp|\} \le \epsilon) > 1 - \delta$ for a given probability
$\delta$. Let $q$ be the number of correct binary digits of $\vp$ that
we aim to obtain, similarly to the notation used in the statement of
Thm.~\ref{thm:qpe}. By choosing $q = \ceil{\log \frac{1}{\epsilon}}$,
we ensure that obtaining $q$ accurate digits yields error $< 2^{-q}
\le \epsilon$. Thus, we aim to obtain an estimate of $\varphi$ with
$q$ correct digits. (Note that in Sect.~\ref{sec:qpe} we run QPE using
an $m$-qubit register, but not all of those bits may be accurate in
the output, see Thm.s~\ref{thm:qpe} and \ref{thm:qpesimple}.)

\subsection{Algorithm for constant precision}
\label{sec:constantprec}
\begin{figure}[h!]
  \leavevmode
  \centering
  \ifcompilefigs
  \Qcircuit @C=1em @R=0.7em {
    \lstick{\ket{0}}    & \gate{H}  & \gate{P(\theta)} & \ctrl{1}   & \gate{H} & \meter \\
    \lstick{\ket{\psi}} & {/^n} \qw & \qw               & \gate{U^{2^k}} & \qw      & \qw \\
  }
  \else
  \includegraphics{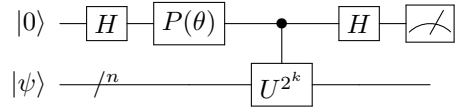}
  \fi
  \caption{Iterative phase estimation circuit.}
  \label{fig:ipecircuit}
\end{figure}

The basic circuit executed by the algorithm is the one indicated in
Fig.~\ref{fig:ipecircuit}, which has two parameters: the integer $k$,
and the angle $\theta$.

The state at the end of this circuit is given by:
\begin{equation*}
  \left( \frac{1 + e^{2 \pi i 2^k \varphi + \theta}}{2} \ket{0} +
  \frac{1 - e^{2 \pi i 2^k \varphi + \theta}}{2} \ket{1} \right) \otimes \ket{\psi},
\end{equation*}
as can be easily verified by noticing that the controlled-$U^k$ gate
introduces a phase kickback of $e^{2 \pi i 2^k \varphi}$, and the
$P(\theta)$ gate adds an extra phase $\theta$ (see
Def.~\ref{def:phaseshiftgate}). With a slight abuse of notation, in
this section we use the notation of conditional probabilities to
indicate the value of the parameter $\theta$, e.g., we write
$\Pr(\mathcal{Q}_1 = 0 | \theta = 0)$ to denote the probability of
measuring $0$ on the first qubit given that the value of $\theta$ is 0
($\theta$ is an algorithmic parameter, not a random variable). With relatively straightforward calculations, it can be verified that the
measurement in the circuit in Fig.~\ref{fig:ipecircuit} yields the
following outcome probabilities:
\begingroup
\allowdisplaybreaks
\begin{align*}
  \Pr(\mathcal{Q}_1 = 0 | \theta = 0) &=  \left|\frac{1 + e^{2 \pi i 2^k \varphi}}{2} \right|^2 = \left|\frac{1 + \cos 2\pi 2^k \varphi  + i \sin 2 \pi 2^k \varphi}{2} \right|^2 \\
  &= \frac{1 + \cos^2 2\pi 2^k \varphi + 2 \cos 2\pi 2^k \varphi + \sin^2 2 \pi 2^k \varphi}{4}  \\
  &= \frac{2 + 2 \cos 2\pi 2^k \varphi}{4} = \frac{1 + \cos 2\pi 2^k \varphi}{2}, \\
  \Pr(\mathcal{Q}_1 = 1 | \theta = \frac{\pi}{2}) &=  \left|\frac{1 - e^{2 \pi i (2^k \varphi + \frac{1}{4})}}{2} \right|^2 = \left|\frac{1 - i\cos 2\pi 2^k \varphi  + \sin 2 \pi 2^k \varphi}{2} \right|^2 \\
  &= \frac{1 + \cos^2 2\pi 2^k \varphi + 2 \sin 2\pi 2^k \varphi + \sin^2 2 \pi 2^k \varphi}{4}  \\
  &= \frac{2 + 2 \sin 2\pi 2^k \varphi}{4} = \frac{1 + \sin 2\pi 2^k \varphi}{2}.
\end{align*}
\endgroup
Using $\theta = 0$ and performing the observation multiple times, we
can obtain an estimate of $\cos 2 \pi (2^k \varphi) = 2
\Pr(\mathcal{Q}_1 = 0 | \theta = 0) - 1$ with a prescribed level of
confidence; the number of samples is discussed subsequently in this
section. Notice that estimating $\cos 2 \pi (2^k \varphi)$ does not
give us full knowledge of $2^k \varphi$, due to the symmetry of the
cosine. To fully estimate $2^k \varphi$ we should also determine
information on the sine of the angle. This is straightforward to do,
because using $\theta = \frac{\pi}{2}$, as indicated above, allows us
to determine $\sin 2 \pi (2^k \varphi) = 2 \Pr(\mathcal{Q}_1 = 1 |
\theta = \frac{\pi}{2}) - 1$.

Because we can estimate $\cos 2 \pi (2^k \varphi), \sin 2 \pi (2^k
\varphi)$ with the circuit in Fig.~\ref{fig:ipecircuit}, we can
estimate $2^k \varphi$, thereby solving the goal of phase estimation
at least partially.  We now examine the question of how many samples
from the circuit are necessary to estimate $\cos 2 \pi (2^k \varphi),
\sin 2 \pi (2^k \varphi)$ up to a certain precision: this tells us how
many calls to $U^{2^k}$ are necessary, which we want to know to assess
the query complexity of this phase estimation algorithm. Recall that
the values are estimated by observing frequencies of a certain
outcome, i.e., the cosine is estimated from $\Pr(\mathcal{Q}_1 = 0 |
\theta = 0)$, while the sine is estimated from $\Pr(\mathcal{Q}_1 = 1
| \theta = \frac{\pi}{2})$. Let us discuss the estimation of the
cosine, as the analysis for the sine is essentially the same. Given
$t$ samples from measuring qubit 1, the observed frequency of
$\mathcal{Q}_1 = 0$ can be expressed as $\frac{1}{t} \sum_{j=1}^{t}
X_j$, where $X_j$ are independent Bernoulli trials with probability of
success equal to $p^* = \frac{1 + \cos 2\pi 2^k \varphi}{2}$ (success
is defined as the outcome that we are interested in; in this case,
$\mathcal{Q}_1 = 0$). An error estimate on $\left|\frac{1}{t}
\sum_{j=1}^{t} X_j - p^*\right|$ translates into an error estimate on
$\cos 2 \pi 2^k \varphi$ using the formula $p^* = \frac{1 + \cos 2\pi
  2^k \varphi}{2}$. The Chernoff bound tells us that:
\begin{equation*}
  \Pr\left(\left|\frac{1}{t} \sum_{j=1}^{t} X_j - p^*\right| \ge
  \Delta \right) \le 2 e^{-2 \Delta^2 t}.
\end{equation*}
This implies that for any fixed $\Delta$, in order to reduce the
probability of error below a certain threshold $\delta_c$ we need a
number of trials $t = \bigO{\log \frac{1}{\delta_c}}$.
\begin{remark}
  For additional clarity, it may be worth emphasizing the role of the
  different error parameters here. We have the maximum difference
  $\Delta$ between the true value $p^*$ and its estimate $\frac{1}{t}
  \sum_{j=1}^{t} X_j$, and we have the maximum probability $\delta_c$
  that the estimate fails to satisfy the difference upper bound
  $\Delta$. We discovered that, to ensure that
  $\Pr\left(\left|\frac{1}{t} \sum_{j=1}^{t} X_j - p^*\right| \ge
  \Delta \right) \le \delta_c$, it suffices to choose the number of
  samples $t$ in the order of $\log \frac{1}{\delta_c}$. Thus, if the
  precision $\Delta$ for the cosine estimation is fixed (as is the
  case in the iterative phase estimation algorithm, described below),
  the number of samples scales \emph{logarithmically} in the
  reciprocal of the error probability.
\end{remark}
The final piece needed to determine the accuracy of the estimate for
$\varphi$ is to combine estimates for $\cos 2\pi 2^k \varphi$ and
$\sin 2\pi 2^k \varphi$ into a single estimate for $2^k\varphi$, and
assess its distance from the true value. There are many possible ways
to do so; the next result gives an error bound on one of the most
straightforward ways, combining sine and cosine estimates to form a
``box'' around the correct angle.
\begin{proposition}
  For any $0 \le \eta \le \frac{\pi}{2}$ and $\varphi \in [0,2\pi]$,
  estimates $\tilde{c}, \tilde{s}$ for cosine and sine of $\varphi$ with
  errors $|\tilde{c} - \cos \varphi| \le \frac{\sin{\eta}}{\sqrt{2}}$,
  $|\tilde{s} - \sin\varphi| \le \frac{\sin{\eta}}{\sqrt{2}}$ yield a value 
  $\tilde{\varphi}$ such that $|\tilde{\varphi} - \varphi| \le \eta$.
\end{proposition}
A proof is given in \cite{van2020practical}. The result tells us that
if we want an error of $\eta$ for the angle estimate
$\tilde{\varphi}$, we need to choose $\Delta \le \sin{\eta}/\sqrt{2}$
as the maximum error for the sine and cosine estimates. So if $\eta$
is fixed, $\Delta$ is also fixed. In summary, as long as we want to
estimate $2^k \varphi$ up to constant precision, it is sufficient to
take $\bigO{\log \frac{1}{\delta_c}}$ samples from the circuit in
Fig.~\ref{fig:ipecircuit}, where $\delta_c$ is the maximum probability
of failure of the algorithm.

\subsection{Iterative algorithm}
\label{sec:qpeiterative}
We can now describe an iterative phase estimation algorithm that uses
the single-qubit constant-precision phase estimation of
Sect.~\ref{sec:constantprec} as a subroutine.

Recall that we aim to obtain $\vp$ such that $\Pr(\min\{|\varphi
- 0.\vp|, 1 - |\varphi - 0.\vp|\} \le 2^{-q}) > 1 - \delta$. Let $h = q
- 2$ and let $0.\vp = 0.\vp_1 \vp_2 \dots \vp_{h+2}$. The algorithm
estimates the digits of $\vp$ starting from the least significant
digit, $\vp_{h+2}$, and down to $\vp_{1}$. It can be described as
follows.
\begin{itemize}
\item Initialization: use the single-qubit estimation procedure of
  Sect.~\ref{sec:constantprec} with $k = h-1$, maximum error
  probability $\delta_c = \delta/h$, and estimation error at most
  $\Delta = \frac{1}{16}$; round the result to the closest multiple of
  $\frac{1}{8}$. Because we use $k = h-1$, we are estimating the phase
  of the eigenvalue $e^{2\pi i 2^{h-1} \varphi}$, and because integer
  multiples of $2\pi$ can be ignored, this yields the estimate
  $0.\vp_h \vp_{h+1} \vp_{h+2}$ of the last three digits of $\vp$. The
  maximum approximation error at this step is $< \frac{1}{8}$: an
  error of at most $\frac{1}{16}$ comes from the estimation of $2^{h-1}
  \varphi$ (because $\Delta = \frac{1}{16}$), and an additional error
  of at most $\frac{1}{16}$ comes from rounding the estimate to the
  closest multiple of $\frac{1}{8}$ (because we are using three digits
  after the binary point).
\item Iteration step, for $j=h-1,\dots,1$:
  \begin{itemize}
  \item Use the single-qubit estimation procedure of
    Sect.~\ref{sec:constantprec} with $k = j-1$, maximum error
    probability $\delta_c = \delta/h$, and estimation error at most
    $\Delta = \frac{1}{16}$, obtaining an estimate $\omega_j$ of the
    angle $2^k \varphi$.
  \item Set:
    \begin{equation*}
      \vp_j = \begin{cases} 0 & \text{if } |0.0\vp_{j+1}\vp_{j+2} - \omega_j| < \frac{1}{4} \\
        1 & \text{if } |0.1\vp_{j+1}\vp_{j+2} - \omega_j| < \frac{1}{4}. \end{cases}
    \end{equation*}
    One of these two conditions is always satisfied because otherwise
    the digits $\vp_{j+1}, \vp_{j+2}$ would have to be incorrect, but
    we show in Prop.~\ref{prop:iterativeqpeerr} below that both are
    correct.
  \end{itemize}
\end{itemize}
This iterative procedure yields an approximation with precision
$2^{-q} = 2^{-(h+2)}$, as shown below.
\begin{proposition}
  \label{prop:iterativeqpeerr}
  At each step $j=h,\dots,1$, if the angle $\omega_j$ is an
  approximation of $2^{j-1}\varphi$ with error at most $\frac{1}{16}$,
  then the approximation computed by the above algorithm satisfies:
  \begin{equation*} 
    \left|0.\vp_{j}\dots\vp_{h+2} - 2^{j-1}\varphi\right| < 2^{-(h+3-j)}.
  \end{equation*}
\end{proposition}
\begin{proof}
  We show it by induction for $j=h,\dots,1$. The base step $j=h$ is
  obvious from the Initialization step of the algorithm. Now suppose
  we are at step $j$. By the induction hypothesis the digits
  $\vp_{j+1}, \vp_{j+2}$ are correct (i.e., they correspond to the
  closest binary representation of $\varphi$),
  because:
  \begin{equation*}
    \left|0.\vp_{j+1}\vp_{j+2}\dots\vp_{h+2} - 2^{j}\varphi\right| <
    2^{-(h+2-j)},
  \end{equation*}
  and there are only $h+2-j$ digits in total in the string, so all of
  them have to be correct. Because $\vp_{j+1}, \vp_{j+2}$ are correct,
  and $\omega_j$ is an approximation of $2^{j-1}\varphi$ with error at
  most $\frac{1}{16}$, the estimate for $\vp_{j}$ must be correct as
  well, otherwise $|0.\vp_{j}\vp_{j+1}\vp_{j+2} - \omega_j| \ge
  \frac{1}{4}$. This implies that the total error is less than
  $2^{-\ell}$ where $\ell$ is the number of digits, i.e.,
  \begin{equation*}
    \left|0.\vp_{j}\dots\vp_{h+2} - 2^{j-1}\varphi\right| < 2^{-(h+3-j)}. 
  \end{equation*}
\end{proof}

\noindent Prop.~\ref{prop:iterativeqpeerr} shows that the algorithm
returns the desired binary string under the assumption that each
intermediate angle $\omega_j$ is estimated correctly. Because each of
these estimations has probability at most $\delta/h$ to fail, and
there are $h$ such estimations in total (equal to the number of steps
of the algorithm), by the union bound the probability that at least
one estimation fails is at most $h\frac{\delta}{h} = \delta$, so the
entire algorithm is successful (i.e., no estimation fails) with
probability at least $1-\delta$. The total number of samples required
by the algorithm is $\bigO{q \log \frac{q}{\delta}} =
\bigO{\frac{1}{\epsilon} \log \frac{1}{\epsilon \delta}}$, and the
gate complexity for each step is that of implementing controlled-$U^k$
(where the largest $k$ is at most $q-3 = \bigO{\log
  \frac{1}{\epsilon}}$) plus three basic gates. A different tradeoff
as compared to the full phase estimation is thus realized: we execute
several smaller circuits, rather than a large circuit, at the cost of
performing significantly more measurements.\index{quantum!phase estimation|)}\index{algorithm!phase estimation|)}\index{phase!estimation|)}
  
\section{Notes and further reading}
\label{sec:qftnotes}
The QFT is one of the main components of Shor's celebrated quantum
algorithm for prime factorization
\cite{shor97polynomial}.\index{algorithm!Shor's}\index{Shor's algorithm} Shor's prime factorization algorithm uses several results
from number theory, combined with a quantum algorithm for the solution
of the discrete logarithm problem: given a prime number $p$, the
multiplicative group of integers modulo $p$ consisting of the numbers
$\{1,\dots,p-1\}$, and a generator $g$ of the group, the
\emph{discrete logarithm} of $x$ in the group, denoted $\log_g x$ is
the smallest nonnegative integer $a$ such that $x^a = g$. Here,
multiplication is the group operation, i.e., it is intended modulo
$p$. So, for example, for $p = 7$ and the group $G = \{1,\dots,6\}$,
the number $3$ is a generator of the group, and $\log_3 6 = 3$ because
$3^3 \mod 7 = 6$. Shor's work showed that quantum computers can be
used to solve the discrete logarithm problem faster than classical
computers. In fact, the discrete logarithm is a special case of the
hidden subgroup problem \cite{jozsa2001quantum}\index{hidden subgroup problem}, which can be solved efficiently by quantum computers for
certain types of groups. Abelian groups are discussed in
\cite{simon97power,shor97polynomial} and admit efficient quantum
algorithms for the hidden subgroup problem. Non-Abelian groups do not
enjoy the same positive results in general \cite{grigni2001quantum}. A
subexponential-time algorithm for the case of the dihedral group is
discussed in
\cite{kuperberg2005subexponential,kuperberg2011another}. See also
\cite{van2006quantum} for a generalization to the hidden coset
problem.

Several papers discuss efficient methods to implement the QFT\index{quantum!Fourier transform}\index{Fourier transform}, and how
to improve the gate count or depth of the corresponding circuit; see,
e.g., \cite{cleve2000fast} for a low-depth implementation, or
\cite{nam2020approximate} for an approximate QFT implementation on $n$
qubits with only $\bigO{n \log n}$ gates of a certain important class.

The phase estimation algorithm\index{quantum!phase estimation}\index{algorithm!phase estimation}\index{phase!estimation} of Sect.~\ref{sec:qpe} is a fundamental
subroutine in many quantum algorithms, and is used multiple times
throughout this \book{}. One of the limitations of quantum phase
estimation is the fact that the estimate produced by the algorithm
after measurement can be biased (recall that the possible outcomes are
discrete and not necessarily symmetric around the exact real value,
see the distribution in Prop.~\ref{prop:qpedistr}): this can sometimes
interfere with desirable statistical properties. For discussions on
how to remove the bias from the estimator, see
\cite{cornelissen2023sublinear,linden2022average,lu2022unbiased,nannitomography}. One
of the simplest techniques to control the bias, discussed in
\cite{nannitomography}, is to apply --- before estimation --- a random
phase shift to the eigenvalue being estimated, and subtract the same
shift after the estimation procedure. If the phase shift is chosen
uniformly at random, this reduces the bias. Some care needs to be
taken because the phase shift eventually needs to be discretized to
obtain a physically-realizable implementation, and the discretization
may introduce a bias, but such bias can be shown to be exponentially
small.

\chapter{Amplitude amplification and estimation}
\label{ch:ampamp}
\thispagestyle{fancy}
In Ch.s~\ref{ch:earlyalg} and \ref{ch:qft} we described several
quantum algorithms that are exponentially faster than classical
algorithms for the same problem, under some measure of
complexity. Unfortunately, the algorithms of Ch.~\ref{ch:earlyalg}
apply to rather contrived problems that do not have known direct
applications. We now describe a quantum algorithm that gives only a
polynomial --- more specifically, quadratic --- speedup with respect
to classical algorithms, but it applies to a large class of problems
and finds many useful applications, both directly and as an
algorithmic design tool. The algorithm is known as Grover's search
\cite{grover96fast}, and its generalization is known as amplitude
amplification \cite{brassard2002quantum}. Amplitude amplification is
widely used as part of many quantum algorithms. It also serves as the
basis for amplitude estimation, a way of estimating probabilities
quadratically faster than with classical Monte Carlo, and for quantum
minimum finding. All these topics are discussed in this chapter.

\section{Grover's algorithm for black-box search}
\label{sec:grover}
The problem solved by Grover's algorithm\index{algorithm!Grover's|(}\index{Grover's algorithm!base algorithm|(} is usually described as
\emph{black-box} (or \emph{unstructured}) search: we are given a
circuit that computes an unknown function of a binary string, and we
want to determine for which value of the input the function gives
output 1. In other words, we are trying to find a binary string that
satisfies a given property; the property is tested by a circuit that
outputs 1 to ``mark'' any string that satisfies the property. For now,
we assume that there is a single binary string that satisfies the
property. The original paper \cite{grover96fast} describes this as
looking for a certain element in a database. The algorithm can be
applied whenever we are searching for a specific element in a set, we
have a way of testing if an element is the desired element (in fact,
this test must be implementable as a quantum subroutine --- see
below), and we do not have enough information to do anything smarter
than a brute force search, i.e., testing all elements in the set.

The basic idea of the algorithm is to start with the uniform
superposition of all basis states, and iteratively increase the
magnitude of the coefficients of basis states that correspond to
binary strings for which the unknown function gives output
1. Crucially, this can be done even without knowing in advance which
basis states have their coefficients increased, using phase kickback:
we discuss this in Sect.~\ref{sec:groverquantum}.

We need to introduce some notation to formally state the problem
solved by Grover's algorithm. Let $f : \{0,1\}^n \to \{0,1\}$, and
assume that there exists a unique $\v{\ell} \in \{0,1\}^n :
f(\v{\ell}) = 1$, i.e., there is a unique element in the domain of the
function that yields output 1. We call this the \emph{marked
element}. We want to determine $\v{\ell}$. The function $f$ is assumed
to be encoded by a unitary as follows:
\begin{equation*}
  U_f : \ket{\vj} \ket{y} \to \ket{\vj} \ket{y \oplus f(\vj)}.
\end{equation*}
Note that this is an $(n+1)$-qubit unitary. As usual, we are allowed
to query the function in superposition.
\begin{remark}
  Grover's search can also be applied to the case in which there are
  multiple input values that yield output 1, and we want to retrieve
  any of them: we discuss this in Sect.~\ref{sec:groverallmarked}.
\end{remark}

\subsection{Classical algorithm}
\label{sec:groverclassical}
Given the problem definition, classical search has query complexity
$\bigO{2^n}$. Indeed, any deterministic classical algorithm may need
to explore all $2^n$ possible input values before finding $\v{\ell}$:
given any deterministic classical algorithm, there exists a
permutation $\pi$ of $\{0,1\}^n$ that represents the longest execution
path (i.e., sequence of values at which $f$ is queried) of such
algorithm. Then, if $\v{\ell} = \pi(\v{1})$ (i.e., it is the last
element queried by the algorithm in its longest execution path) the
algorithm takes $2^n$ queries to determine the answer.

At the same time, a randomized algorithm with a constant positive
probability to determine $\v{\ell}$ uses $\bigO{2^n}$ function
calls. This can be verified as follows. Suppose we apply a randomized
algorithm that tries one untested binary string uniformly at random at
each iteration. The expected number of function calls that this
algorithm performs until we determine $\v{\ell}$ is given by:
\begin{equation*}
  \sum_{k=1}^{2^n} k \Pr(\v{\ell} \text{ is found at the } k\text{-th function evaluation}).
\end{equation*}
We can expand this by noticing that the probability that $\v{\ell}$ is
found at the $k$-th evaluation is the product of the probability that
$\v{\ell}$ is selected at the $k$-the iteration, and the probability
that $\v{\ell}$ is not found for the first $k-1$ evaluations. This is
equal to:
\begin{align*}
  \sum_{k=1}^{2^n} k \frac{1}{2^n-(k-1)} \prod_{j=1}^{k-1}
  \frac{2^n-j}{2^n-(j-1)} &= \sum_{k=1}^{2^n} k \frac{1}{2^n-(k-1)}
  \frac{2^n-(k-1)}{2^n} \\
  &= \sum_{k=1}^{2^n} \frac{k}{2^n} = \frac{2^n(2^n+1)}{2} \frac{1}{2^n} =
  \frac{2^n+1}{2}.
\end{align*}
Hence, such an algorithm needs approximately $2^{n-1}$ function
calls. By Yao's principle (the worst case expected cost of a
randomized algorithm is no better than the cost of the best
deterministic algorithm against the worst probability distribution),
no randomized algorithm can do better than the above. Thus, the query
complexity of classical algorithms for this problem is $\bigO{2^n}$,
and it cannot be significantly improved.

\subsection{Grover's search: algorithm description}
\label{sec:groverquantum}
The query complexity can be improved with Grover's quantum
algorithm. The algorithm uses $q = n+1$ qubits, which is equal to the
number of qubits of the unitary $U_f$.

The outline of the algorithm is as follows. The algorithm starts with
the uniform superposition of all basis states on $n$ qubits. The last
qubit (i.e., the one in position $n+1$) is used as an auxiliary qubit,
and it is initialized to $H\ket{1}$; this specific state is helpful
for phase kickback, see Sect.~\ref{sec:phasekickback}. Then, these
operations are repeated several times:
\begin{enumerate}[(i)]
\item Flip the sign of the vectors for which $U_f$ gives output 1.
\item Invert all the coefficients of the quantum state around the
  average coefficient; we explain the precise mapping
  implemented by this operation in Sect.~\ref{sec:inversionavg}.
\end{enumerate}
A full cycle of the two operations above increases the coefficient of
$\ket{\v{\ell}} \otimes \frac{1}{\sqrt{2}}(\ket{0} - \ket{1})$, and
after a certain number of cycles (to be specified later), the
coefficient of the state $\ket{\v{\ell}} \otimes
\frac{1}{\sqrt{2}}(\ket{0} - \ket{1})$ is large enough that it can be
obtained from a measurement with probability close to 1. This
phenomenon is known as {\em amplitude amplification}, see
Sect.~\ref{sec:ampamp}.

\begin{figure}[t!b]
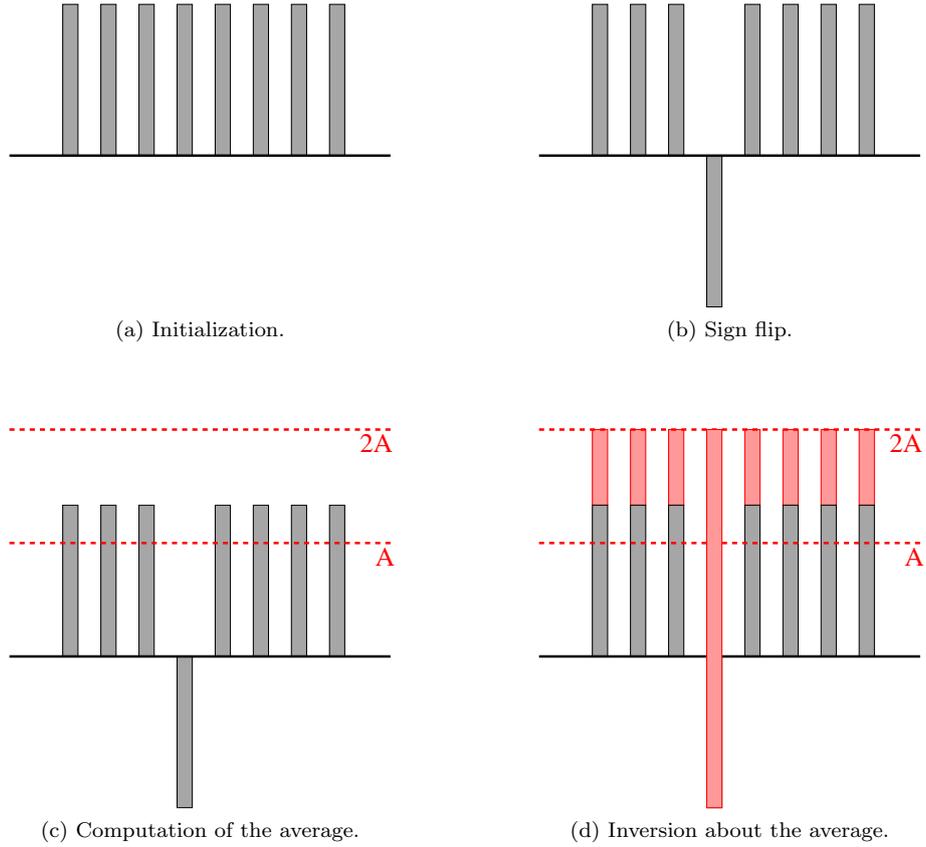

  \centering
  \subfloat[Initialization.]{%
    \resizebox{0.4\textwidth}{!}{%
      \ifcompilefigs
      \begin{tikzpicture}
        \draw[very thick] (0,0) -- (19, 0);
        \draw[draw=black,fill=cyan] (2,0) rectangle ++(1,6);
        \draw[draw=black,fill=cyan] (4,0) rectangle ++(1,6);
        \draw[draw=black,fill=cyan] (6,0) rectangle ++(1,6);
        \draw[draw=black,fill=cyan] (8,0) rectangle ++(1,6);
        \draw[draw=none] (8,0) rectangle ++(1,-6);
        \draw[draw=black,fill=cyan] (10,0) rectangle ++(1,6);
        \draw[draw=black,fill=cyan] (12,0) rectangle ++(1,6);
        \draw[draw=black,fill=cyan] (14,0) rectangle ++(1,6);
        \draw[draw=black,fill=cyan] (16,0) rectangle ++(1,6);
        \draw[very thick,dashed,draw=none] (0, 9) -- (19, 9);
        \draw[very thick,dashed,draw=none] (0, 4.5) -- (19, 4.5);
      \end{tikzpicture}
      \else
      \includegraphics{figures/gr1.pdf}
      \fi
    }
    \label{fig:gr1}
  }\hspace{3em}
  \subfloat[Sign flip.]{%
    \resizebox{0.4\textwidth}{!}{%
      \ifcompilefigs
      \begin{tikzpicture}
        \draw[very thick] (0,0) -- (19, 0);
        \draw[draw=black,fill=cyan] (2,0) rectangle ++(1,6);
        \draw[draw=black,fill=cyan] (4,0) rectangle ++(1,6);
        \draw[draw=black,fill=cyan] (6,0) rectangle ++(1,6);
        \draw[draw=black,fill=cyan] (8,0) rectangle ++(1,-6);
        \draw[draw=black,fill=cyan] (10,0) rectangle ++(1,6);
        \draw[draw=black,fill=cyan] (12,0) rectangle ++(1,6);
        \draw[draw=black,fill=cyan] (14,0) rectangle ++(1,6);
        \draw[draw=black,fill=cyan] (16,0) rectangle ++(1,6);
        \draw[very thick,dashed,draw=none] (0, 9) -- (19, 9);
        \draw[very thick,dashed,draw=none] (0, 4.5) -- (19, 4.5);
      \end{tikzpicture}
      \else
      \includegraphics{figures/gr2.pdf}
      \fi
    }
    \label{fig:gr2}
  }\\[2em]
  \subfloat[Computation of the average.]{%
    \resizebox{0.4\textwidth}{!}{%
      \ifcompilefigs
      \begin{tikzpicture}
        \draw[very thick] (0,0) -- (19, 0);
        \draw[draw=black,fill=cyan] (2,0) rectangle ++(1,6);
        \draw[draw=black,fill=cyan] (4,0) rectangle ++(1,6);
        \draw[draw=black,fill=cyan] (6,0) rectangle ++(1,6);
        \draw[draw=black,fill=cyan] (8,0) rectangle ++(1,-6);
        \draw[draw=black,fill=cyan] (10,0) rectangle ++(1,6);
        \draw[draw=black,fill=cyan] (12,0) rectangle ++(1,6);
        \draw[draw=black,fill=cyan] (14,0) rectangle ++(1,6);
        \draw[draw=black,fill=cyan] (16,0) rectangle ++(1,6);
        \draw[very thick,dashed,draw=red] (0, 9) -- (19, 9);
        \draw[very thick,dashed,draw=red] (0, 4.5) -- (19, 4.5);
        \node at (18,8.5) {\Huge \color{red}{2A}};
        \node at (18,4) {\Huge \color{red}{A}};
      \end{tikzpicture}
      \else
      \includegraphics{figures/gr3.pdf}
      \fi
    }
    \label{fig:gr3}}
  \hspace{3em}
  \subfloat[Inversion about the average.]{%
    \resizebox{0.4\textwidth}{!}{%
      \ifcompilefigs
      \begin{tikzpicture}
        \draw[very thick] (0,0) -- (19, 0);
        \draw[draw=black!10,fill=cyan!10] (2,0) rectangle ++(1,6);
        \draw[draw=black!10,fill=cyan!10] (4,0) rectangle ++(1,6);
        \draw[draw=black!10,fill=cyan!10] (6,0) rectangle ++(1,6);
        \draw[draw=black!10,fill=cyan!10] (8,0) rectangle ++(1,-6);
        \draw[draw=black!10,fill=cyan!10] (10,0) rectangle ++(1,6);
        \draw[draw=black!10,fill=cyan!10] (12,0) rectangle ++(1,6);
        \draw[draw=black!10,fill=cyan!10] (14,0) rectangle ++(1,6);
        \draw[draw=black!10,fill=cyan!10] (16,0) rectangle ++(1,6);
        \draw[very thick] (0,0) -- (8, 0);
        \draw[very thick] (9,0) -- (19, 0);
        \draw[draw=black,fill=cyan] (2,9) rectangle ++(1,-3);
        \draw[draw=black,fill=cyan] (4,9) rectangle ++(1,-3);
        \draw[draw=black,fill=cyan] (6,9) rectangle ++(1,-3);
        \draw[draw=black,fill=cyan] (8,9) rectangle ++(1,-15);
        \draw[draw=black,fill=cyan] (10,9) rectangle ++(1,-3);
        \draw[draw=black,fill=cyan] (12,9) rectangle ++(1,-3);
        \draw[draw=black,fill=cyan] (14,9) rectangle ++(1,-3);
        \draw[draw=black,fill=cyan] (16,9) rectangle ++(1,-3);
        \draw[very thick,dashed,draw=red] (0, 9) -- (19, 9);
        \draw[very thick,dashed,draw=red] (0, 4.5) -- (19, 4.5);
        \node at (18,8.5) {\Huge \color{red}{2A}};
        \node at (18,4) {\Huge \color{red}{A}};
      \end{tikzpicture}
      \else
      \includegraphics{figures/gr4.pdf}
      \fi
    }
    \label{fig:gr4}}
  \caption{Sketch of Grover's algorithm. The bars represent the coefficients of the basis states.}
  \label{fig:grover}
\end{figure}

A sketch of the ideas for the algorithm is depicted in
Fig.~\ref{fig:grover}: we have eight basis states, and suppose the
fourth basis state is the target basis state $\ket{\v{\ell}}$. The
representation is purely meant to convey intuition on the behavior of
the coefficients in the quantum state, and does not geometrically
represent the vectors encoding the quantum state. In
Fig.~\ref{fig:gr1}, all basis states have the same coefficient. In
Fig.~\ref{fig:gr2}, the coefficient of the target basis state has its
sign flipped. In Fig.~\ref{fig:gr3}, we can see that the average value
of the coefficients is slightly below the coefficient for the
undesired states. Taking twice the average and subtracting each
coefficient now yields the new bars in Fig.~\ref{fig:gr4},
where the target basis state $\ket{\v{\ell}}$ has a coefficient with
much larger value than the rest, leading to a higher probability of
being output as the outcome of a measurement. Of course, we need to
show that these steps can be implemented with unitary matrices that
can be constructed with a polynomial number of basic gates.

Next, we describe each step in more detail. 

\paragraph{Initialization.}
The algorithm is initialized by applying the operation $H^{\otimes
  (n+1)}(I^{\otimes n}\otimes X)$ onto the state
$\ket{\v{0}}_{n+1}$. We can express the quantum state as follows:
\begin{align*}
  (I^{\otimes n} \otimes X) \ket{\v{0}}_{n+1} &= \ket{\v{0}}_n \ket{1} \\
  H^{\otimes (n+1)} (I^{\otimes n} \otimes X) \ket{\v{0}}_{n+1} &= 
  \sum_{\vj \in \{0,1\}^n} \frac{1}{\sqrt{2^n}}
  \ket{\vj} \otimes \frac{(\ket{0} - \ket{1})}{\sqrt{2}} =
  \sum_{\vj \in \{0,1\}^n} \alpha_{j}
  \ket{\vj} \otimes \frac{(\ket{0} - \ket{1})}{\sqrt{2}} = \ket{\psi},
\end{align*}
where $\alpha_{j} = \frac{1}{\sqrt{2^n}}$ and we labeled the quantum
state after the initialization phase as $\ket{\psi}$. We evolve
$\ket{\psi}$ through the rest of the algorithm. Note that the initial
coefficients $\alpha_{j}$ of the state $\ket{\psi}$ are real
numbers. Because all the other steps of the algorithm map real numbers
to real numbers (as is discussed later), we only need to consider real
numbers through the course of the algorithm.

\paragraph{Sign flip: step (i).}
To flip the sign of the target state $\ket{\v{\ell}} \otimes
\frac{1}{\sqrt{2}}(\ket{0} - \ket{1})$, we apply $U_f$ to
$\ket{\psi}$. This is just an application of phase kickback, because we
are applying a function of the form $\ket{\vj} \ket{y} \to \ket{\vj}
\ket{y \oplus f(\vj)}$ after preparing the last qubit in the
eigenstate $\frac{1}{\sqrt{2}}(\ket{0}-\ket{1})$ of modulo-2 addition
$y \oplus f(\vj)$. Indeed, using the expressions derived in
Sect.~\ref{sec:phasekickback}, we have:
{\allowdisplaybreaks
\begin{align*}
  U_f \ket{\psi} &= U_f\left(\sum_{\vj \in \{0,1\}^n} 
  \alpha_{j} \ket{\vj} \otimes \frac{1}{\sqrt{2}} (\ket{0} - \ket{1})\right) \\
  &= \sum_{\vj \in \{0,1\}^n} 
  (-1)^{f(\vj)} \alpha_{j} \ket{\vj} \otimes \frac{1}{\sqrt{2}} (\ket{0} - \ket{1}) \\
  &= \left(-\alpha_{\v{\ell}} \ket{\v{\ell}}
   + \sum_{\substack{\vj \in \{0,1\}^n \\ \vj \neq \v{\ell}}}
  \alpha_{j} \ket{\vj}\right) \otimes \frac{1}{\sqrt{2}} (\ket{0} - \ket{1}).
\end{align*}
}
As the expression above suggests, we can always think of the last
qubit as being in the state $\frac{1}{\sqrt{2}}(\ket{0} - \ket{1})$
and unentangled from the rest of the qubits, with the sign flip
affecting only the first $n$ qubits. Therefore, the state that we
obtain by applying $U_f$ to $\ket{\psi}$ is the same as $\ket{\psi}$
except that the sign of $\ket{\v{\ell}} \otimes
\frac{1}{\sqrt{2}}(\ket{0}-\ket{1})$ has been flipped.

\paragraph{Inversion about the average: step (ii).}
\label{sec:inversionavg}
To perform the inversion about the average, we apply the following
operation:
\begin{equation*}
  \sum_{\vj \in \{0,1\}^n} \alpha_{j} \ket{\vj} \to \sum_{\vj \in
    \{0,1\}^n} \left(2\left(\sum_{\vk \in \{0,1\}^n} \frac{\alpha_{k}}{2^n}\right) - \alpha_{j}\right) \ket{\vj},
\end{equation*}
where $\sum_{\vk \in \{0,1\}^n} \frac{\alpha_{k}}{2^n}$ is the
average, and therefore we are taking twice the average and subtracting
each coefficient from it. It is not clear yet that this is a unitary
operation, but it follows from our discussion below. This mapping
is realized by the following matrix:
\begin{equation*}
  W := \begin{pmatrix} \frac{2}{2^n} - 1 & \frac{2}{2^n} & \dots & \frac{2}{2^n} \\
    \frac{2}{2^n} & \frac{2}{2^n} -1 & \dots & \frac{2}{2^n} \\
    \vdots & \vdots & \ddots & \vdots \\
    \frac{2}{2^n} & \frac{2}{2^n} & \dots & \frac{2}{2^n} - 1
    \end{pmatrix} =
  \begin{pmatrix}
    \frac{2}{2^n} & \frac{2}{2^n} & \dots & \frac{2}{2^n} \\
    \frac{2}{2^n} & \frac{2}{2^n} & \dots & \frac{2}{2^n} \\
    \vdots & \vdots  & \ddots & \vdots \\
    \frac{2}{2^n} & \frac{2}{2^n} & \dots & \frac{2}{2^n}
  \end{pmatrix}
  - I^{\otimes n},
\end{equation*}
where in each element of the matrix, the denominator $\frac{1}{2^n}$
computes the average coefficient, the numerator $2$ of the fraction
takes twice the average, and finally we subtract the identity to
subtract each individual coefficient from twice the average. From the
definition of the Hadamard gate in Eq.~\eqref{eq:hadamard}, we can see
that the entry of $H^{\otimes n}$ in position $j,k$ is
$\left(H^{\otimes n}\right)_{jk} = \frac{1}{\sqrt{2^n}}(-1)^{\vj
  \bullet \vk}$.  If we let:
\begin{equation*}
  M := \begin{pmatrix} 2 & 0 & \dots & 0 \\ 0 & 0 & \dots & 0 \\
    \vdots & & \ddots & \vdots \\
    0 & 0 & \dots & 0 \end{pmatrix} = 2 \ketbra{\v{0}}{\v{0}} \in \R^{2^n \times 2^n},
\end{equation*}
then we can write $(H^{\otimes n}M H^{\otimes n})_{jk} =
\left(H^{\otimes n}\right)_{j0}M_{00}\left(H^{\otimes
  n}\right)_{0k} = \frac{2}{2^n}$, because $M_{jk} = 0$ for $j \neq
0$ or $k \neq 0$. Therefore, using the fact that $H^{\otimes n}H^{\otimes
  n} = I^{\otimes n}$, we have:
\begin{equation}
  \label{eq:Tdecomp}
  \begin{split}
    W &= H^{\otimes n}MH^{\otimes n} - I^{\otimes n} = H^{\otimes n}(M - I^{\otimes n})H^{\otimes n} \\
    &= H^{\otimes n} \underbrace{\text{diag}(\underbrace{1,-1,\dots,-1}_{2^n \text{ elements}})}_{=: F} H^{\otimes n} = H^{\otimes n} F H^{\otimes n}.
  \end{split}
\end{equation}
The expression in Eq.~\eqref{eq:Tdecomp}, besides providing a
decomposition for $W$, also shows that $W$ is unitary, because it is a
product of unitary matrices ($H^{\otimes n}$ is a tensor product of
unitary matrices, $F$ is diagonal with ones on the diagonal hence it
is unitary). We must find a way to construct the matrix $F =
\text{diag}(1,-1,\dots,-1)$. This is discussed below. For now, we
summarize our analysis of the inversion about the average by
concluding that it can be performed by applying $W = H^{\otimes n} F
H^{\otimes n}$ to the $n$ qubits of interest (i.e., the input lines of
$U_f$ --- all qubits except the output qubit of $U_f$, which we use
for the sign flip of step (i) via phase kickback).

\begin{figure}[h!]
  \leavevmode
  \centering
  \ifcompilefigs
  \Qcircuit @C=1em @R=.7em {
    & \qw      & \qw      & \ctrl{1}    & \qw & \qw      & \qw \\
    & \qw      & \qw      & \gate{Z}    & \qw & \qw & \qw \\
  }
  \hspace{10em}
  \Qcircuit @C=1em @R=.7em {
    & \qw      & \qw      & \ctrl{1} & \qw & \qw      & \qw \\
    & \gate{H} & \qw      & \targ    & \qw & \gate{H} & \qw \\
  }
  \else
  \includegraphics{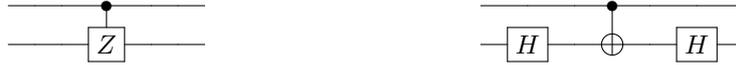}
  \fi
  \caption{Controlled Z gate on two qubits: two possible representations.}
  \label{fig:cz}
\end{figure}

\paragraph{Constructing the matrix $F$.}
\label{sec:matrixd}
We give a sketch of the idea of how to construct $F =
\text{diag}(1,-1,\dots,-1)$. Notice that the effect of this quantum
operation is to flip the sign of the coefficient of every basis state
except $\ket{\v{0}}_n$. We are going to implement $-F$ rather than
$F$.
\begin{remark}
  As discussed in Ex.~\ref{ex:globalphase}, a global phase factor in
  a gate has no effect on the outcome of the computation, as it gets
  canceled out during measurement. The matrices $F$ and $-F$ are equal
  up to a global phase factor of $-1$, hence they implement the same
  operation.
\end{remark}
The matrix $-F$ flips the sign of $\ket{\v{0}}$ and leaves other basis
state untouched. Instead of flipping the sign of $\ket{\v{0}}$, let us
start by seeing how to flip the sign of $\ket{\v{1}}$ while leaving
all other coefficients untouched. Let C${}^{n-1}Z$ be the gate that
applies $Z$ to qubit $n$ if qubits $1,\dots,n-1$ are $\ket{1}$, and
does nothing otherwise. This is similar to the C$X$ gate, except that
it has multiple controls, and it applies a $Z$ gate rather than an $X$
(i.e., $X$) gate when the control qubits are $\ket{1}$. It is called a
``multiply-controlled $Z$''. C${}^{n-1}Z$ in the case of two qubits
($n=2$) is given by the following matrix:
\begin{equation*}
  \text{C}Z = 
  \begin{pmatrix}
    1 & 0 & 0 & 0 \\
    0 & 1 & 0 & 0 \\
    0 & 0 & 1 & 0 \\
    0 & 0 & 0 & -1 \\
  \end{pmatrix}. 
\end{equation*}
Notice that in the two-qubit case ($n = 2$), the two circuits depicted
in Fig.~\ref{fig:cz} are equivalent: carrying out the matrix
multiplications confirms that the circuit on the right in
Fig.~\ref{fig:cz} implements exactly the C$Z$ matrix as defined
above. Thus, the controlled $Z$ gate can be easily realized with a
C$X$ and two Hadamard gates. 

\begin{figure}[h!]
  \leavevmode
  \centering
  \ifcompilefigs
  \Qcircuit @C=1em @R=.7em {
    & \gate{X} & \ctrl{1} & \gate{X} & \qw \\
    & \gate{X} & \ctrl{1} & \gate{X} & \qw \\
    & \gate{X} & \ctrl{3} & \gate{X} & \qw \\
    & \vdots   &          & \vdots   & \\
    &  \\
    & \gate{X} & \gate{Z} & \gate{X} & \qw \\
  }
  \else
  \includegraphics{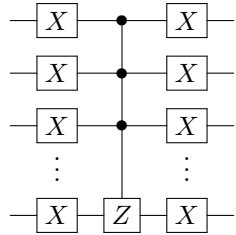}
  \fi
  \caption{Quantum circuit implementing the $F$ operation (up to a global phase factor) used in Grover's algorithm.}
  \label{fig:grovercircuit}
\end{figure}

If we have access to the C${}^{n-1}Z$ gate, we can write:
\begin{equation*}
  -F = X^{\otimes n} (\text{C}^{n-1}Z) X^{\otimes n},
\end{equation*}
because, as can be easily verified, this sequence of operations flips
the sign of the coefficient of a basis state if and only if all qubits
have value $\ket{0}$ in the basis state. In circuit form, these
operations can be represented as depicted in
Fig.~\ref{fig:grovercircuit}.

\begin{figure}[h!]
  \leavevmode
  \centering
  \ifcompilefigs
  \Qcircuit @C=1em @R=1em @!R {
    & \ctrl{4} & \qw      & \qw       & \qw      & \ctrl{4} & \qw \\ 
    & \ctrl{3} & \qw      & \qw       & \qw      & \ctrl{3} & \qw \\
    & \qw      & \qw      & \ctrl{1}  & \qw      & \qw      & \qw \\
    & \qw      & \gate{H} & \targ     & \gate{H} & \qw      & \qw \\
    \lstick{\ket{0}} & \targ    & \qw      & \ctrl{-1} & \qw      & \targ    & \qw & \rstick{\ket{0}}
    \gategroup{3}{3}{5}{5}{2.2em}{--}
  }
  \else
  \includegraphics{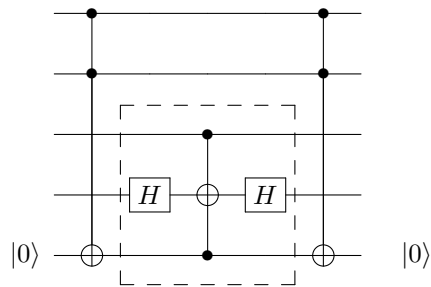}
  \fi
  \caption{Decomposition of $\text{C}^{n-1}Z$ for $n=4$. The fifth
    (bottom) qubit is initialized to $\ket{0}$ and is used as
    working space. This implements $\text{C}^3Z$ for the top four
    qubits.}
    \label{fig:czdecomposition}
\end{figure}

Of course, one has to construct the operation $\text{C}^{n-1}Z$. There
are several ways to do so. The construction that is simplest to
understand is perhaps the one suggested in \cite{barenco95elementary},
according to which we use a $\text{C}^{n-2}X$ gate and a controlled Z
gate. The $\text{C}^{n-2}X$ is actually easy to implement with some
auxiliary qubits. We show this scheme in
Fig.~\ref{fig:czdecomposition} with an example for $n=4$ qubits, but
clearly it can be generalized to an arbitrary number of qubits. We
first implement a $\text{C}^{n-2}X$ gate, with an auxiliary qubit
(which is initialized to $\ket{0}$, as one can see from the bottom
qubit in Fig.~\ref{fig:czdecomposition}) as the target of the
$\text{C}^{n-2}X$. We then implement a $\text{CC}Z$ gate using a CC$X$
and two Hadamard gates on the target qubit; the reader can easily
verify that this implements a doubly-controlled $Z$, using the
identity $HXH = Z$ and carrying out the calculations (in the large
unitary matrix for $\text{CC}Z$, the gate being controlled appears in
the bottom right, just as in C$X$).  Summarizing, this yields a
decomposition of $\text{C}^{n-1}Z$ with a number of gates and
auxiliary qubits that is linear in $n$. It is possible to forsake the
initialization of the auxiliary qubit, see \cite{barenco95elementary}
for details. To conclude, the construction of $F$ (more precisely,
$-F$), and therefore of the whole circuit implementing step (ii) of
Grover's search, can be done using $\bigO{n}$ gates and auxiliary
qubits.

\subsection{Determining the number of iterations}
\label{sec:grovernumiter}
A single iteration of Grover's search consists of steps (i) and (ii)
described in Sect.~\ref{sec:groverquantum}. It is paramount to
determine how many iterations should be performed, so that the
coefficient of the desired basis state $\ket{\v{\ell}} \otimes
(\ket{0} - \ket{1})$ is as large as possible (in modulus), and the
binary string $\v{\ell}$ is the outcome of a measurement with high
probability. In this section we study how the amplitude of the target
basis state changes through the iterations, and determine the optimal
number of iterations.

Because the last, auxiliary qubit is always in the state
$\frac{1}{\sqrt{2}}(\ket{0} - \ket{1})$ and unentangled with the rest,
we can ignore it in this section.  Let
\begin{equation*}
  \ket{\psi_G} := \ket{\v{\ell}}, \qquad \ket{\psi_B} :=
  \left(\sum_{\substack{\vj \in \{0,1\}^n \\ \vj \neq \v{\ell}}}
  \frac{1}{\sqrt{2^n-1}}\ket{\vj} \right)
\end{equation*}
be the ``good'' and ``bad'' quantum states, respectively. These are
valid quantum states, as they have unit norm. (One should think of
them as a desirable and undesirable part of the state, i.e., some part
that we wish to have as the outcome of a computation, and one that we
do not wish to have. The quantum computing literature usually labels
these states ``good'' and ``bad''.) We claim that after iteration $k$
of the algorithm, the quantum state can be expressed as:
\begin{equation*}
 \ket{\psi_k} = d_k \ket{\psi_G} + u_k \ket{\psi_B},
\end{equation*}
with $\abs{d_k}^2 + \abs{u_k}^2 = 1$. We show this by
induction. Initially, we pick $d_0 = \frac{1}{\sqrt{2^n}}$ and $u_0 =
\sqrt{\frac{2^n - 1}{2^n}}$; this yields the initial state
$\ket{\psi}$ with uniform coefficients equal to
$\frac{1}{\sqrt{2^n}}$. Thus, the claim is true for $k=0$. Notice that
to obtain $u_0$ from the value of an individual coefficient in
$\ket{\psi}$ we multiplied by $\sqrt{2^n - 1}$ for normalization,
because there are $2^n-1$ basis states in $\ket{\psi_B}$. We now need
to show the induction step: assuming $\ket{\psi_{k}} = d_{k}
\ket{\psi_G} + u_{k} \ket{\psi_B}$ with coefficients $d_{k}, u_{k}$
satisfying $\abs{d_{k}}^2 + \abs{u_{k}}^2 = 1$, we must show
$\ket{\psi_{k+1}} = d_{k+1} \ket{\psi_G} + u_{k+1} \ket{\psi_B}$ with
$\abs{d_{k+1}}^2 + \abs{u_{k+1}}^2 = 1$.

At step (i) of the algorithm, the algorithm flips the sign of the
coefficient in front of $\ket{\psi_G}$; formally, it applies the
mapping $d_k \ket{\psi_G} + u_k \ket{\psi_B} \to -d_k \ket{\psi_G} +
u_k \ket{\psi_B}$.

At step (ii), the algorithm maps $\alpha_h \to 2A_k - \alpha_h$ for
each coefficient $\alpha_h$, where $A_k$ is the average
coefficient at iteration $k$. Therefore:
\begin{align*}
  -\alpha_{\ell} &\to 2A_k + \alpha_{\ell} \\
  \alpha_{h} &\to 2A_k - \alpha_{h} \quad \forall h \neq \ell.
  \end{align*}
To compute $A_k$, we need to determine the value of each individual
coefficient. The coefficient for $\ket{\v{\ell}}$ is clearly $d_k$,
as there is only one such state. On the other hand, there are
$2^n-1$ states with coefficient $u_k$, so the value of the
coefficient for each of the states $\ket{\vj}, \vj \neq \v{\ell}$ is
$\frac{u_k}{\sqrt{2^n -1}}$ (the square root is due to
normalization, see above). The average coefficient at iteration $k$
is therefore:
\begin{equation*}
  A_k = \frac{(2^n-1)\frac{1}{\sqrt{2^n - 1}} u_k - d_k}{2^n} = \frac{\sqrt{2^n - 1} u_k - d_k}{2^n}.
\end{equation*}
Again, to obtain $u_k$ from one of the coefficients $\alpha_{h}$ we
multiply by $\sqrt{2^n -1}$, so the mapping of steps (i) and (ii) can
be summarized as:
\begin{align*}
  d_k \ket{\psi_G} + u_k \ket{\psi_B} &\to -d_k \ket{\psi_G} +
u_k \ket{\psi_B} & \text{(step (i))} \\
  -d_k\ket{\psi_G} + u_k \ket{\psi_B} &\to (2A_k + d_k)\ket{\psi_G} +
  \sqrt{2^n -1}(2A_k - \frac{u_k}{\sqrt{2^n -1}}) \ket{\psi_B} & \text{(step (ii))} \\
  &= d_{k+1} \ket{\psi_G} + u_{k+1} \ket{\psi_B},
\end{align*}
where we defined:
\begin{align*}
  d_{k+1} &:= 2A_k + d_k \\
  u_{k+1} &:= 2A_k\sqrt{2^n - 1} - u_k.
\end{align*}
With simple algebraic calculations we can verify that $\abs{d_{k+1}}^2
+ \abs{u_{k+1}}^2 = 1$. This finishes the proof of the induction step,
and therefore of the entire claim.

Now we analyze the coefficients $d_{k+1}, u_{k+1}$ more
closely. Performing the substitution of $A_k$, we obtain:
\begin{align*}
  d_{k+1} &= 2\frac{\sqrt{2^n - 1} u_k - d_k}{2^n} + d_k = \left(1 - \frac{1}{2^{n-1}}\right)d_k + \frac{2\sqrt{2^n - 1}}{2^n} u_k \\
  u_{k+1} &= 2\frac{\sqrt{2^n - 1} u_k - d_k}{2^n}\sqrt{2^n - 1} - u_k = -\frac{2\sqrt{2^n - 1}}{2^n} d_k + \left(1-\frac{1}{2^{n-1}}\right)u_k.
\end{align*}
This transformation is exactly a clockwise rotation of the vector
$\begin{pmatrix} d_k \\ u_k \end{pmatrix}$ by a certain angle
$2\theta$, because it has the form: $$\begin{pmatrix} \cos 2\theta &
  \sin 2\theta \\ -\sin 2\theta & \cos
  2\theta \end{pmatrix} \begin{pmatrix} d_k \\ u_k \end{pmatrix}$$ and
it satisfies the relationship $\sin^2 2\theta + \cos^2 2\theta =
1$. (The reason why we call this angle $2\theta$, rather than simply
$\theta$, is discussed in Sect.~\ref{sec:geomgrover}; this choice
also makes our exposition more consistent with the literature.) The
angle $\theta$ must satisfy:
\begin{equation}
  \label{eq:groversintheta}
  \sin 2\theta = \frac{2\sqrt{2^n - 1}}{2^n}.
\end{equation}
Note that because this value of the sine is very small for large
$n$, we can use the approximation $\sin x \approx x$ (when $x$
is close to 0) to write:
\begin{equation}
  \label{eq:grovertheta}
  \theta \approx \frac{\sqrt{2^n - 1}}{2^n} \approx \frac{1}{\sqrt{2^n}}.
\end{equation}

Summarizing, the above analysis shows that $\ket{\psi_k}$ belongs to
the plane spanned by $\ket{\psi_G}$ and $\ket{\psi_B}$, and at each
iteration we perform a rotation of $\ket{\psi_k}$ in that plane by an
angle $2\theta$. Thus, after $k$ iterations the coefficients $d_k,
u_k$ satisfy the following equation:
\begin{equation*}
  \begin{pmatrix} d_k \\ u_k \end{pmatrix} = \begin{pmatrix} \cos 2\theta & \sin 2\theta \\ - \sin 2\theta & \cos 2\theta \end{pmatrix}^k \begin{pmatrix} d_0 \\ u_0 \end{pmatrix},
\end{equation*}
and because this is an application of a rotation $k$ times, the above
equation can be rewritten as:
\begin{align*}
  d_{k} &= \cos 2 k \theta \, d_0 + \sin 2 k \theta \, u_0 \\
  u_{k} &= -\sin 2 k \theta \, d_0 + \cos 2 k \theta \, u_0.
\end{align*}
In order to maximize the probability of obtaining $\ket{\psi_G}$ after
a measurement, we recall that $|u_0| \gg |d_0|$ and $|u_0|$ is close
to 1 for large $n$, so we need to set $\sin 2 k \theta$ as large as
possible. Thus, the best choice is to pick $2 k \theta =
\frac{\pi}{2}$, which yields the largest value of $|d_k|$. Using
Eq.~\eqref{eq:grovertheta}, and noting that the number of iterations
has to be integer, the optimal number of iterations of Grover's search
algorithm is:
\begin{equation}
  \label{eq:groveroptiter}
  k = \rint{\frac{\pi}{4 \theta}} \approx \rint{\frac{2^{n} \pi}{4\sqrt{2^n - 1}}} \approx
  \frac{\pi}{4}\sqrt{2^n} = \bigO{\sqrt{2^n}},
\end{equation}
where we write $\rint{\cdot}$ to denote the rounding to the nearest
integer. After this many iterations, we have a probability close to 1
of measuring $\ket{\psi_G}$ and obtaining the sought state
$\ket{\v{\ell}}$. In each iteration we apply $U_f$ once (for the sign
flip step). Comparing this with a classical algorithm, that performs
$\bigO{2^n}$ queries to the oracle $f$, we obtained a quadratic
speedup.
\begin{remark}
  If we perform more iterations of Grover's algorithm than the optimal
  number, the probability of measuring the desired state actually goes
  down, and reduces our chances of success. Therefore, it is important
  to choose the right number of iterations, see the discussion in
  Sect.~\ref{sec:ampamp} for ways to avoid this issue.
\end{remark}
Of course, the approximation for $\theta$ given in
Eq.~\eqref{eq:grovertheta} is only valid for large $n$: for smaller
$n$, it is better to compute the optimal number of iterations deriving
$\theta$ from Eq.~\eqref{eq:groversintheta}. We conclude this section
by noting that in case there are multiple input values on which $f$
has value 1, we should amend the above analysis adjusting the values
for $d_0$ and $u_0$, but the main steps remain the same: we discuss
this in Sect.~\ref{sec:ampamp}. Another situation that can be analyzed
is that in which it is not known in advance for how many input strings
the function $f$ outputs 1, i.e., we do not know the number of marked
elements; this is discussed in Sect.~\ref{sec:groverunknown}.

\subsection{A geometric interpretation of the algorithm}
\label{sec:geomgrover}
To continue our study of Grover's algorithm, and help us transition to
a more general version of it, it is helpful to provide a geometric
view of the algorithm's effect. We are also going to restate the
algorithm's input and objective in a slightly more general form.

Suppose that we want to approximately construct a certain $n$-qubit
state $\ket{\psi_G}$, having access to the following two circuits:
\begin{enumerate}[(i)]
\item A circuit $S$ acting on $n$ qubits that has the following
  action:
  \begin{equation*}
    S \ket{\v{0}}_n := \ket{\psi_0} = \sin \theta \ket{\psi_G} + \cos \theta \ket{\psi_B},
  \end{equation*}
  where $\braket{\psi_G}{\psi_B} = 0$ (i.e., the states are
  orthogonal) and $\theta \in (0, \frac{\pi}{2})$.
\item A circuit $R$ acting on $n$ qubits that has the following
  action:
  \begin{equation*}
    R(\alpha_G \ket{\psi_G} + \alpha_B \ket{\psi_B}) = -\alpha_G
    \ket{\psi_G} + \alpha_B \ket{\psi_B}
  \end{equation*}
  for any $\alpha_G, \alpha_B \in [-1,1]: |\alpha_G|^2 + |\alpha_B|^2
  = 1$.
\end{enumerate}
This is a generalization of our previous discussion of Grover's
algorithm: if we take $S = H^{\otimes n}$, and we let $R$ be the
``sign flip'' unitary (that can be constructed with $U_f$ and phase
kickback), we obtain exactly the statement of Grover's problem with
$\ket{\psi_G}, \ket{\psi_B}$ defined as in
Sect.~\ref{sec:grovernumiter}.  Let us call the operator $G =
SFS^{\dag}R$ a \emph{Grover iteration}, and let us look at the effect
of the Grover iteration on the plane spanned by $\ket{\psi_G}$ and
$\ket{\psi_B}$.
\begin{remark}
  \label{rem:groverreflect}
  The operation $S F S^{\dag}$ implements a reflection through
  $\ket{\psi_0} = S\ket{\v{0}}$, because:
  \begin{equation*}
    S F S^{\dag} = S (2\ketbra{\v{0}}{\v{0}} - I^{\otimes n}) S^{\dag} =
    2 S \ketbra{\v{0}}{\v{0}} S^{\dag} - S S^{\dag} = 2 \ketbra{\psi_0}{\psi_0} - I^{\otimes n},
  \end{equation*}
  which is exactly the desired reflection. ($\ketbra{\psi_0}{\psi_0}$
  is the orthogonal projection onto $\ket{\psi_0}$; if we take twice
  the projection and subtract the identity, we reflect through
  $\ket{\psi_0}$.) We have shown in Sect.~\ref{sec:inversionavg} that
  we know how to construct the matrix $F = \text{diag}(1,-1,\dots,-1)
  = M - I^{\otimes n} = 2\ketbra{\v{0}}{\v{0}} - I^{\otimes n}$. Then,
  the operation to reflect through $\ket{\psi_0}$ can be efficiently
  implemented using $F$ and $S, S^{\dag}$.
\end{remark}
\begin{figure}[t!b]
  \centering
  \subfloat[Initial position.]{%
    \centering
    \ifcompilefigs
    \begin{tikzpicture}[>=Stealth]
      \draw[draw=none] (0,0) -- (-3,0);
      \draw[thick,->] (0,0) -- (2,0) node[anchor=west] {$\ket{\psi_B}$};
      \draw[thick,->] (0,0) -- (0,2) node[anchor=south] {$\ket{\psi_G}$};
      \draw[thick,->] (0,0) coordinate (O) -- (20:2) coordinate (oc) 
      node[anchor=west] {$\ket{\psi_0}$};
      \draw[black] (0:1) arc(0:20:1) node[midway,right] {$\theta$};
      \draw[red, thick] (O) circle (2 cm);
    \end{tikzpicture}
    \else
    \includegraphics{figures/groverplane1.pdf}
    \fi
    \label{fig:groverplane1}
  }\hspace{3em}
  \subfloat[Application of $R$ (reflection through $\ket{\psi_B}$).]{%
    \centering
    \ifcompilefigs
    \begin{tikzpicture}[>=Stealth]
      \draw[draw=none] (0,0) -- (-3,0);
      \draw[thick,->] (0,0) -- (2,0) node[anchor=west] {$\ket{\psi_B}$};
      \draw[thick,->] (0,0) -- (0,2) node[anchor=south] {$\ket{\psi_G}$};
      \draw[thick,->] (0,0) coordinate (O) -- (20:2) coordinate (oc) 
      node[anchor=west] {$\ket{\psi_0}$};
      \draw[thick,->,dashed] (0,0) coordinate (O) -- (-20:2) coordinate (oc) 
      node[anchor=west] {$R\ket{\psi_0}$};
      \draw[black] (0:1) arc(0:-20:1) node[midway,right] {$\theta$};
      \draw[red, thick] (O) circle (2 cm);
    \end{tikzpicture}
    \else
    \includegraphics{figures/groverplane2.pdf}
    \fi
    \label{fig:groverplane2}
  }\\[2em]
  \subfloat[Reflection through $\ket{\psi_0}$.]{%
    \centering
    \ifcompilefigs
    \begin{tikzpicture}[>=Stealth]
      \draw[draw=none] (0,0) -- (-3,0);
      \draw[thick,->] (0,0) -- (2,0) node[anchor=west] {$\ket{\psi_B}$};
      \draw[thick,->] (0,0) -- (0,2) node[anchor=south] {$\ket{\psi_G}$};
      \draw[thick,->] (0,0) coordinate (O) -- (20:2)
      coordinate (oc) node[anchor=west] {$\ket{\psi_0}$};
      \draw[thick,->,dashed] (0,0) coordinate (O) -- (-20:2)
      coordinate (oc) node[anchor=west] {$R\ket{\psi_0}$};
      \draw[thick,->] (0,0) coordinate (O) -- (60:2)
      coordinate (oc) node[anchor=south west] {$SFS^{\dag}R\ket{\psi_0}$};
      \draw[thick,->,dotted,blue] (-20:2)
      coordinate (oc) -- (60:2) coordinate (oc) {};
      \draw[black] (0:1) arc(0:20:1) node[midway,right] {$\theta$};
      \draw[black] (0:0.8) arc(0:60:0.8) node[near end,right] {$3\theta$};
      \draw[red, thick] (O) circle (2 cm);
    \end{tikzpicture}
    \else
    \includegraphics{figures/groverplane3.pdf}
    \fi
    \label{fig:groverplane3}
  }\hspace{3em}
  \subfloat[Quantum state $\ket{\psi_1}$ after one iteration.]{%
    \centering
    \ifcompilefigs
    \begin{tikzpicture}[>=Stealth]
      \draw[draw=none] (0,0) -- (-3,0);
      \draw[thick,->] (0,0) -- (2,0) node[anchor=west] {$\ket{\psi_B}$};
      \draw[thick,->] (0,0) -- (0,2) node[anchor=south] {$\ket{\psi_G}$};
      \draw[thick,->] (0,0) coordinate (O) -- (20:2)
      coordinate (oc) node[anchor=west] {$\ket{\psi_0}$};
      \draw[thick,->] (0,0) coordinate (O) -- (60:2)
      coordinate (oc) node[anchor=south west] {$\ket{\psi_1}$};
      \draw[black] (0:1) arc(0:20:1) node[midway,right] {$\theta$};
      \draw[black] (0:0.8) arc(0:60:0.8) node[near end,right] {$3\theta$};
      \draw[red, thick] (O) circle (2 cm);
    \end{tikzpicture}
    \else
    \includegraphics{figures/groverplane4.pdf}
    \fi
    \label{fig:groverplane4}
  }
  \caption{Sketch of Grover's algorithm on the plane spanned by $\ket{\psi_G}$ and $\ket{\psi_B}$.}
  \label{fig:groverplane}
\end{figure}
We can assume that the mutual relationship between the states is as
given in Fig.~\ref{fig:groverplane1}. An application of the operator
$R$ reflects $\ket{\psi_0}$ through $\ket{\psi_B}$, obtaining the
dashed arrow in Fig.~\ref{fig:groverplane2}. Then, an application of
$SFS^{\dag}$ reflects $R\ket{\psi_0}$ through $\ket{\psi_0}$, see
Rem.~\ref{rem:groverreflect}; this is depicted in
Fig.~\ref{fig:groverplane3}. Thus, because $\theta$ is the angle between
$\ket{\psi_0}$ and $\ket{\psi_B}$, reflecting $R\ket{\psi_0}$ through
$\ket{\psi_0}$ rotates $\ket{\psi_0}$ closer to $\ket{\psi_G}$ by an
angle of $2\theta$; this is shown in Fig.~\ref{fig:groverplane3}. At
this point we have performed one full iteration of Grover's algorithm:
the quantum state is denoted $\ket{\psi_1}$ in
Fig.~\ref{fig:groverplane4}, and it is closer to $\ket{\psi_G}$ by an
angle $2\theta$, compared to the initial state $\ket{\psi_0}$.

These operations (reflection through $\ket{\psi_B}$, reflection through
$\ket{\psi_0}$) can be repeated multiple times until we obtain
$\ket{\psi_k}$ that is close to $\ket{\psi_G}$. The initial angle
between $\ket{\psi_0}$ and $\ket{\psi_G}$ is $\frac{\pi}{2} - \theta$,
hence the number of iterations is:
\begin{equation*}
  \left\lfloor \frac{\frac{\pi}{2} - \theta}{2\theta} \right\rceil =
  \left\lfloor \frac{\pi}{4\theta} - \frac{1}{2} \right\rceil =
  \floor{\frac{\pi}{4\theta}},
\end{equation*}
exactly as derived in Eq.~\eqref{eq:groveroptiter}. Using
Eq.s~\eqref{eq:groversintheta} and \eqref{eq:grovertheta} we obtain
once again the optimal number of iterations previously shown.
\begin{remark}
  The angle $\theta$ in this section is defined by the action of $S$,
  because $S \ket{\v{0}} = \sin \theta \ket{\psi_G} + \cos \theta
  \ket{\psi_B}$. For Grover's algorithm, where $S = H^{\otimes n}$, we
  obtain exactly the same angle as in Eq.~\eqref{eq:groversintheta}:
  using the double angle formula, we have
  \begin{equation*}
    \frac{2\sqrt{2^n - 1}}{2^n} = \sin 2\theta = 2 \sin \theta \cos \theta = 2 \underbrace{\frac{1}{\sqrt{2^n}}}_{\sin \theta} \underbrace{\sqrt{\frac{2^n - 1}{2^n}}}_{\cos \theta}.
  \end{equation*}
\end{remark}
Knowing that a Grover iteration performs a rotation, we summarize the
effect of applying $k$ Grover iterations as:
\begin{equation*}
  G^k \ket{\psi_0} = \sin ((2k + 1) \theta) \ket{\psi_G} +
  \cos ((2k + 1) \theta) \ket{\psi_B}.
\end{equation*}
\index{algorithm!Grover's|)}\index{Grover's algorithm!base algorithm|)}

\section{Amplitude amplification}
\label{sec:ampamp}
The geometric interpretation of Grover's algorithm given in
Sect.~\ref{sec:geomgrover}, which is more general than the original
Grover's algorithm, leads to a technique known as \emph{amplitude
amplification}\index{algorithm!amplitude amplification|(}\index{amplitude!amplification|(}, first introduced in \cite{brassard2002quantum}. The
algorithm discussed in Sect.~\ref{sec:geomgrover} takes as input a
circuit $S$ preparing a superposition of a ``good'' state
$\ket{\psi_G}$ and a ``bad'' state $\ket{\psi_B}$, and a circuit $R$
that flips the sign of $\ket{\psi_G}$; its goal is to find a state
with large overlap with $\ket{\psi_G}$. We have seen that in Grover's
algorithm, $R$ is constructed with phase kickback: the function $U_f$
marks the basis states in $\ket{\psi_G}$ by performing modulo-2
addition on an auxiliary qubit, and if the auxiliary qubit is prepared in
the state $H\ket{1}$ (which is an eigenvector of addition modulo $2$,
with eigenvalue $-1$) this applies a sign flip. Note that to implement
$R$ in this way, the only requirement is that we are able to recognize
(``mark'') the basis states in $\ket{\psi_G}$. Amplitude amplification
is precisely the algorithm that we described with a geometric
interpretation in Sect.~\ref{sec:geomgrover}, to prepare
$\ket{\psi_G}$ given a unitary that marks it.

Using the geometric intuition described in Sect.~\ref{sec:geomgrover},
we found that the optimal number of iterations is
$\frac{\pi}{4\theta}$, where $\theta$ is the angle such that $S
\ket{\v{0}} = \sin \theta \ket{\psi_G} + \cos \theta
\ket{\psi_B}$. Relying once again on the approximation $\sin \theta
\approx \theta$ for small angles, and calling $p = \sin^2 \theta$, we
obtain the following result, which follows directly from our analysis
in Sect.~\ref{sec:grover} and more specifically
Sect.~\ref{sec:geomgrover}.
\begin{theorem}[Amplitude amplification; \cite{brassard2002quantum}]
  \label{thm:ampamp}
  Let $S$ be an $n$-qubit unitary such that $S\ket{\v0} = \sqrt{p}
  \ket{\psi_G} + \sqrt{1-p} \ket{\psi_B}$, where for some $M \subset
  \{0,1\}^n$, we have $\ket{\psi_G} = \frac{1}{\sqrt{p}} \sum_{j \in
    M} \alpha_j \ket{\vj}$, $\ket{\psi_B} =
  \frac{1}{\sqrt{1-p}}\sum_{j \not\in M} \alpha_j \ket{\vj}$ and $p =
  \sum_{j \in M} |\alpha_j|^2$. Let $R$ be a unitary that maps
  $\ket{\psi_G} \to -\ket{\psi_G}$, $\ket{\psi_B} \to
  \ket{\psi_B}$. The amplitude amplification algorithm produces a
  quantum state such that its overlap with $\ket{\psi_G}$ is at least
  $2/3$ using $\bigO{\frac{1}{\sqrt{p}}}$ applications of $S$ and $R$,
  and additional gates.
\end{theorem}
This result can potentially be used to boost the probability of
success of any randomized algorithm with the property that success can
be recognized. The most typical case is the one in which we have some
``flag qubits'' that indicate success of the algorithm: these can be
obtained by running a verification procedure, or sometimes they are
obtained as a byproduct of the algorithm itself. Then we define
$\ket{\psi_G}$ as the superposition of all basis states in which the
flag qubits indicate success, and $\ket{\psi_B}$ as its orthogonal
complement. If the original randomized algorithm is successful
with probability $p$, amplitude amplification increases the
probability of success to some number very close to one using
$\bigO{\frac{1}{\sqrt{p}}}$ applications of the circuit that
implements the algorithm, whereas classical repetition of the
algorithm until success would take $\bigO{\frac{1}{p}}$ executions of
the circuit in expectation.
\begin{remark}
  The statement of Thm.~\ref{thm:ampamp} only guarantees probability
  of success $2/3$, but this can easily be boosted to a number
  arbitrarily close to one by doing a few repetitions of the
  algorithm: using standard arguments, with $k$ repetitions the
  probability that all of them fail is at most $1/3^k$, so for
  probability $1-\delta$ of obtaining at least one success, it
  suffices to pick $k = \bigO{\log \frac{1}{\delta}}$.
\end{remark}
\begin{example}
  \label{ex:groverampamp}
  Let us consider Grover's problem again: we want to find a value
  $\v{\ell} \in \{0,1\}^n$ such that $f(\v{\ell}) = 1$. We can
  determine this using a simple randomized algorithm: sample a binary
  string $\vj$ uniformly at random, and evaluate $f$ until we find
  $f(\vj) = 1$. The probability of success of a single sample is $p =
  1/2^n$. A quantum circuit implementation of this algorithm can be
  obtained by applying a layer of Hadamard gates $H^{\otimes n}$ onto
  the initial state $\ket{\v{0}}$, applying the unitary $U_f$ that
  evaluates $f$ on the uniform superposition, and following with a
  measurement of all qubits. If we obtain a string with $f(\vj) = 1$,
  the algorithm has succeeded. Repeating this procedure until success
  results in $\bigO{1/p} = \bigO{2^n}$ applications of the quantum
  circuit in expectation. With amplitude amplification, the
  probability of success is close to one after only $\bigO{1/\sqrt{p}}
  = \bigO{\sqrt{2^n}}$ applications of $H^{\otimes n}$ and $U_f$,
  where $U_f$ allows us to implement the reflection circuit $R$ via
  phase kickback; so we only need that many applications of $U_f$
  in expectation.
\end{example}

\noindent Thm.~\ref{thm:ampamp} can be restated for the (simpler) case
of quantum search using a binary marking oracle, just as in Grover's
algorithm, generalizing the argument discussed in
Ex.~\ref{ex:groverampamp}.
\begin{corollary}
  \label{cor:ampampsearch}
  Let $U_f$ be a quantum (binary) oracle implementing a Boolean
  function $f : \{0,1\}^n \to \{0,1\}$, and let $M = \{\vj \in
  \{0,1\}^n : f(\vj) = 1\}$ be the set of marked elements. Then we can
  determine an element of $M$ with $\bigO{\sqrt{\frac{2^n}{|M|}}}$
  applications of $U_f$, if $|M|$ is known.
\end{corollary}
When $|M|$ is unknown we can achieve the same expected running time
$\bigO{\sqrt{\frac{2^n}{|M|}}}$ with a randomized algorithm discussed
in Sect.~\ref{sec:groverunknown}: we postpone its description because
the analysis is simpler after developing a few additional tools.

One potential issue of amplitude amplification (and thus Grover's
algorithm) is that we need to have a reasonable estimate of the value
of $p$ before executing the algorithm. Indeed, if we have no such
estimate, we cannot compute the right number of iterations $k$ for the
algorithm. As a result, the overlap between the target state
$\ket{\psi_G}$ and the state produced by the algorithm may be too
small. Intuitively, this is easy to see: the overlap is expressed by
the function $\sin ((2k+1) \theta)$, and if $k$ is too small or too
large we may obtain a small value for the sine.

To overcome this issue, several approaches are possible. A simple one
is to use the \emph{amplitude estimation} algorithm to get an estimate
of $p$ \cite{brassard2002quantum}, see Sect.~\ref{sec:ampest}. This
incurs an extra cost, but asymptotically we still obtain a quadratic
speedup over classical algorithms. In fact, this algorithm can be
executed in a different way that avoids explicit estimation: we
discuss a version of it in the context of quantum search, in
Sect.~\ref{sec:groverunknown}. Other approaches are discussed in the
notes in Sect.~\ref{sec:ampampnotes}.

\subsection{Obtaining all marked states}
\label{sec:groverallmarked}
Suppose we have a known number $t = |M|$ of marked states in total
(the set $M$ is defined as in Cor.~\ref{cor:ampampsearch}), and we
want to find them all. Note that this is a direct generalization of
the black-box search problem of Sect.~\ref{sec:grover}\index{algorithm!Grover's}\index{Grover's algorithm!obtaining all solutions}\index{amplitude!amplification!obtaining all solutions}\index{quantum!search, all solutions}. The most
natural approach is to use amplitude amplification to construct a
state with a large overlap with $\ket{\psi_G}$, measure in the
computational basis, and obtain $\vj \in M$ with high
probability. Then we can ``unmark'' the string $\vj$, in the following
way: construct a lookup-table circuit that for a given $\vj$, checks
whether $\vj$ is a previously observed marked element (i.e., $f(\vj) =
1$), and if so it returns 0, otherwise it returns $f(\vj)$. In other
words, we implement the following function:
\begin{equation*}
  f_E'(\vj) = \begin{cases}
    0 & \text{if } \vj \in E \\
    f(\vj) & \text{otherwise},
    \end{cases}
\end{equation*}
where $E \subset M$ is the set of previously observed marked
elements. Implementing this function is easy: one call to $f_E'$ can
be implemented with at most one call to $f$ and $\bigO{|E|}$
additional gates.

We then apply the scheme suggested earlier: initialize $E \leftarrow
\emptyset$; apply amplitude amplification to the function $f_E'$ with
known number of marked elements $t - |E|$ to determine a new element
in $M$; repeat until all elements in $M$ are found. Using
Corollary~\ref{cor:ampampsearch}, the number of calls to $f_E'$ (and
hence $f$) before we find all elements can be upper bounded as follows
(this is an order-of-magnitude upper bound, i.e., we ignore possible
constants in front of the expression):
\begin{align*}
  \sum_{k=0}^{t-1} \sqrt{\frac{2^n}{t-k}} &= \sqrt{2^n} \sum_{k=1}^{t} \frac{1}{\sqrt{k}} \le \sqrt{2^n} \int_{1}^{t} \frac{1}{\sqrt{x}} \di x = \sqrt{2^n} \left(\left.2 x^{1/2}\right|_{1}^{t} \right) = \sqrt{2^{n+1}} (\sqrt{t} - 1) \\
  &= \bigO{\sqrt{t2^n}}.
\end{align*}
This implies the following.
\begin{corollary}
  Let $M$ be the set of marked elements, and let $|M|$ be known. Then
  we can determine all elements in $M$ using $\bigO{\sqrt{|M|2^n}}$
  applications of $f$. 
\end{corollary}

\subsection{Oblivious amplitude amplification}
\label{sec:obliviousampamp}
We discussed how to amplify the relative weight of certain quantum
states in a superposition, which allows us to increase the probability
of success of quantum algorithms in a very general way. To do so, so
far we required access to a circuit to prepare an initial state that
we can reflect through. In this section we show that amplitude
amplification can be applied to select a ``useful'' part of the state,
obtained by applying a unitary to the initial state, even if we do not
know the initial state itself and hence cannot reflect through
it. This is called \emph{oblivious} amplitude amplification
\cite{berry2014exponential}\index{amplitude!amplification!oblivious}. We
use this technique in Sect.~\ref{sec:lcu}.

The setup for oblivious amplitude amplification is the
following. Suppose we apply some unitary $U$ to an initial state
$\ket{\psi}$ on $n$ qubits, and this unitary produces a superposition
of a ``good state'' that we want to obtain with some large
probability, and a ``bad state''.  If we have a way of identifying the
good state, for example if we know that all good states are marked by
one or more flag qubits, it would seem that we can apply amplitude
amplification to increase the probability of observing the good state
up to the desired level. Formally, suppose the unitary $U$ has the
following action:
\begin{equation*}
  U \ket{0} \ket{\psi} = \sin \theta \ket{0} V \ket{\psi} + \cos \theta \ket{1} \ket{\phi},
\end{equation*}
where $V\ket{\psi}$ is the good state, i.e., the state that we are
interested in, and $\ket{\phi}$ is the bad state, which is allowed to
depend on $\ket{\psi}$ in the above expression. The first qubit serves
as a flag qubit to mark the good state. To apply amplitude
amplification, as we have seen in Sect.~\ref{sec:geomgrover}, we need
a way of reflecting through $\ket{\psi_G} = \ket{0} V
\ket{\psi}$. This is easy to do if we have a circuit to construct
$\ket{\psi}$ from $\ket{\v{0}}$ and we are willing to execute this
circuit repeatedly: in this case, we can apply the amplitude
amplification algorithm as discussed in the preceding sections (the
circuit $S$ of Thm.~\ref{thm:ampamp} is then given by the circuit that
constructs $\ket{\psi}$ from $\ket{\v{0}}$, followed by $U$). However,
suppose that we do not have a circuit to construct $\ket{\psi}$, or we
have the circuit but we choose not to use it more than once because it
requires a large amount of computational resources. Standard amplitude
amplification fails because we do not have the circuit $S$ of
Thm.~\ref{thm:ampamp}, i.e., a circuit that prepares the initial state
starting from $\ket{\v{0}}$. As it turns out we can still apply
amplitude amplification, as we show next.

The first step in studying amplitude amplification in this setting is
to identify a two-dimensional subspace in which we can do reflection,
and such that the Grover operator never leaves that subspace. In the
basic version of Grover search, that was the subspace spanned by
$\ket{\psi_G}$ and $\ket{\psi_B}$. Here we define it slightly
differently: the two fundamental states are $\ket{0}\ket{\psi}$ and
$\ket{1}\ket{\phi}$, where, as mentioned earlier, the first qubit is
used to identify the good subspace.
\begin{lemma}[Based on Lem.~3.7 in \cite{berry2014exponential}]
  \label{lem:2dsubspace}
  Let $U, V$ be unitaries on $n+1$ and $n$ qubits respectively, and let $\theta \in (0, \pi/2)$. Suppose that for any $n$-qubit state $\ket{\psi}$, we have
  \begin{equation*}
    U \ket{0} \ket{\psi} = \sin \theta \ket{0} V \ket{\psi} + \cos \theta \ket{1} \ket{\phi},
  \end{equation*}
  where $\ket{\phi}$ may depend on $\ket{\psi}$. Then the state $\ket{\Psi^{\perp}}$, defined as:
  \begin{equation*}
    \ket{\Psi^{\perp}} = U^{\dag}\left(\cos \theta \ket{0} V \ket{\psi} - \sin \theta \ket{1} \ket{\phi} \right),
  \end{equation*}
  is orthogonal to $\ket{\Psi} = \ket{0} \ket{\psi}$ and has no
  support on the basis states that have $\ket{0}$ as their first
  qubit, i.e., $(\ketbra{0}{0} \otimes I^{\otimes n}) \ket{\Psi^{\perp}} = 0$.
\end{lemma}
\begin{proof}
  For the first part, we need to show that $\braket{\Psi}{\Psi^{\perp}}
  = 0$. We have:
  \begin{align*}
    \braket{\Psi}{\Psi^{\perp}} &= (\bra{0} \bra{\psi}) U^{\dag}\left(\cos \theta \ket{0} V \ket{\psi} - \sin \theta \ket{1} \ket{\phi} \right) \\
    &= \left(\sin \theta \bra{0}  \bra{\psi} V^{\dag} + \cos \theta \bra{1} \bra{\phi}\right)\left(\cos \theta \ket{0} V \ket{\psi} - \sin \theta \ket{1} \ket{\phi} \right) \\
    &= \sin \theta \cos \theta - \cos \theta \sin \theta = 0.
  \end{align*}

  For the second part, we want to show that:
  \begin{equation*}
    (\ketbra{0}{0} \otimes I^{\otimes n}) \ket{\Psi^{\perp}} =
    (\ketbra{0}{0} \otimes I^{\otimes n}) U^{\dag}\left(\cos \theta
    \ket{0} V \ket{\psi} - \sin \theta \ket{1} \ket{\phi} \right) = 0.
  \end{equation*}
  To show this, we first make a couple of observations. We want to
  study $(\ketbra{0}{0} \otimes I^{\otimes n}) U^{\dag}$, so let us
  study $(\bra{0} \otimes I^{\otimes n}) U^{\dag}$ to begin with. From
  the definition of $U\ket{0}\ket{\psi}$, with algebraic manipulations
  we see that $\ket{0} V \ket{\psi} = \frac{1}{\sin
    \theta}(\ketbra{0}{0} \otimes I^{\otimes n}) U \ket{0}
  \ket{\psi}$. Thus, we have:
  \begin{equation}
    \label{eq:lemmaqdef}
    \begin{split}
    (\bra{0} \otimes I^{\otimes n}) U^{\dag} \ket{0} V \ket{\psi} &= \frac{1}{\sin \theta} (\bra{0} \otimes I^{\otimes n}) U^{\dag} (\ketbra{0}{0} \otimes I^{\otimes n}) U \ket{0} \ket{\psi}\\
      &= \frac{1}{\sin \theta} \underbrace{(\bra{0} \otimes I^{\otimes n}) U^{\dag} (\ketbra{0}{0} \otimes I^{\otimes n}) U (\ket{0} \otimes I^{\otimes n})}_{=: Q} \ket{\psi}.
    \end{split}
  \end{equation}
  The operator $Q$ defined above satisfies:
  \begin{align*}
    \bra{\psi} Q \ket{\psi} &= \nrm{(\ketbra{0}{0} \otimes I^{\otimes n}) U \ket{0}\ket{\psi}}^2 = \nrm{(\ketbra{0}{0} \otimes I^{\otimes n})(\sin \theta \ket{0} V \ket{\psi} + \cos \theta \ket{1} \ket{\phi})}^2 \\
    &= \nrm{\sin \theta \ket{0} V \ket{\psi}}^2 = \sin^2 \theta.
  \end{align*}
  Because this holds for any $\ket{\psi}$, it holds for a basis of
  eigenvectors of $Q$, so we can assume that $Q = \sin^2 \theta I^{\otimes n}$ by
  working in the corresponding basis. Further, note that:
  \begin{align*}
    U(\sin \theta \ket{0}\ket{\psi} + \cos \theta \ket{\Psi^{\perp}}) & = U(\sin \theta \ket{0}\ket{\psi} + \cos \theta U^{\dag}\left(\cos \theta \ket{0} V \ket{\psi} - \sin \theta \ket{1} \ket{\phi} \right)  \\
    &= \sin^2 \theta \ket{0}V \ket{\psi} + \sin \theta \cos \theta \ket{1}\ket{\phi} +  \cos^2 \theta \ket{0} V \ket{\psi} - \cos \theta \sin \theta \ket{1} \ket{\phi} \\
    &= \ket{0}V \ket{\psi}.
  \end{align*}
  Thus, using Eq.~\eqref{eq:lemmaqdef}:
  \begin{align*}
    \sin^2 \theta \ket{\psi} &= Q \ket{\psi} = \sin\theta (\bra{0} \otimes I^{\otimes n}) U^{\dag} \ket{0} V \ket{\psi} = \sin\theta (\bra{0} \otimes I^{\otimes n}) (\sin \theta \ket{0}\ket{\psi} + \cos \theta \ket{\Psi^{\perp}}) \\
    &= \sin^2 \theta \ket{\psi} + \sin \theta \cos \theta (\bra{0} \otimes I^{\otimes n}) \ket{\Psi^{\perp}},
  \end{align*}
  which implies $\sin \theta \cos \theta (\bra{0} \otimes I^{\otimes n})
  \ket{\Psi^{\perp}} = 0$ and hence $(\bra{0} \otimes I^{\otimes n})
  \ket{\Psi^{\perp}} = 0$, because $\sin \theta \cos \theta \neq 0$
  due to $\theta \in (0, \pi/2)$. It follows that $(\ketbra{0}{0} \otimes I)
  \ket{\Psi^{\perp}} = 0$.
\end{proof}

\noindent We use Lem.~\ref{lem:2dsubspace} to show that the evolution
of the state when using the Grover operator remains in a
two-dimensional subspace spanned by $\ket{0}\ket{\psi}$ and
$\ket{1}\ket{\phi}$.
\begin{theorem}[Oblivious amplitude amplification; Lem.~3.6 in \cite{berry2014exponential}]
  \label{thm:obliviousampamp}
  Let $U, V$ be unitaries on $n+1$ and $n$ qubits respectively, and let $\theta \in (0, \pi/2)$. Suppose that for any $n$-qubit state $\ket{\psi}$, we have
  \begin{equation*}
    U \ket{0} \ket{\psi} = \sin \theta \ket{0} V \ket{\psi} + \cos \theta \ket{1} \ket{\phi},
  \end{equation*}
  where $\ket{\psi}$ may depend on $\ket{\psi}$. Let $R =
  2\ketbra{0}{0} \otimes I^{\otimes n} - I^{\otimes (n+1)}$ and $G =
  -UR^{\dag}U^{\dag}R$. Then for any integer $k > 0$ we have:
  \begin{equation*}
    G^{k} U \ket{0}\ket{\psi} = \sin ((2k + 1)\theta) \ket{0} V \ket{\psi} + \cos ((2k + 1)\theta) \ket{1} \ket{\phi}.
  \end{equation*}
\end{theorem}
\begin{proof}
  Let $\ket{\Phi} = \ket{0}V\ket{\psi}, \ket{\Phi^{\perp}} =
  \ket{1}\ket{\phi}$ and let $\ket{\Psi}, \ket{\Psi^{\perp}}$ be
  defined as in Lem.~\ref{lem:2dsubspace}. Then:
  \begin{equation}
    \label{eq:upsiperp}
    \begin{split}
    U \ket{\Psi} &= \sin \theta \ket{\Phi} + \cos \theta \ket{\Phi^{\perp}}  \\
    U \ket{\Psi^{\perp}} &= \cos \theta \ket{\Phi} - \sin \theta \ket{\Phi^{\perp}},
    \end{split}
  \end{equation}
  where the last equation is by definition of
  $\ket{\Psi^{\perp}}$. Adding these two equations with coefficients
  $(\sin \theta, \cos \theta)$ and $(\cos \theta, -\sin \theta)$, and
  then multiplying through by $U^{\dag}$, yields:
  \begin{equation}
    \label{eq:uphiperp}
    \begin{split}
    U^{\dag} \ket{\Phi} &= \sin \theta \ket{\Psi} + \cos \theta \ket{\Psi^{\perp}}\\
    U^{\dag} \ket{\Phi^{\perp}} &= \cos \theta \ket{\Psi} - \sin \theta \ket{\Psi^{\perp}}.
    \end{split}
  \end{equation}
  Then, noting that $R \ket{\Phi} = \ket{\Phi}$ ($R$ is a reflection
  through the states that have $\ket{0}$ as their first qubit, and
  $\ket{\Phi}$ is a superposition of states satisfying that property),
  by relying on basic geometry and Lem.~\ref{lem:2dsubspace} we
  can study the effect of $G$ on $\ket{\Phi}$: \begingroup
  \allowdisplaybreaks
  \begin{align*}
    G \ket{\Phi} &= -UR^{\dag}U^{\dag}R \ket{\Phi} \\
    &= -UR^{\dag}(\sin \theta \ket{\Psi} + \cos \theta \ket{\Psi^{\perp}}) \\
    &= -U( \sin \theta \ket{\Psi} - \cos \theta \ket{\Psi^{\perp}}) \\
    &= (\cos^2 \theta - \sin^2 \theta)\ket{\Phi} -2 \cos\theta \sin\theta \ket{\Phi^{\perp}} \\
    &= \cos 2\theta \ket{\Phi} - \sin 2\theta \ket{\Phi^{\perp}},
  \end{align*}
  \endgroup
  where for the second equality we used Eq.~\eqref{eq:uphiperp}, and
  for the fourth equality we used Eq.~\eqref{eq:upsiperp}. With very
  similar calculations ($\ket{\Phi^{\perp}}$ is orthogonal to
  $\ket{\Phi}$, hence $R$ acts as a sign flip), using Eq.s~\eqref{eq:upsiperp} and \eqref{eq:uphiperp} we find:
  \begingroup
  \allowdisplaybreaks
  \begin{align*}
    G \ket{\Phi^{\perp}} &= -UR^{\dag}U^{\dag}R \ket{\Phi^{\perp}} = UR^{\dag}(\cos \theta \ket{\Psi} - \sin \theta \ket{\Psi^{\perp}}) \\
    &= U( \cos \theta \ket{\Psi} + \sin \theta \ket{\Psi^{\perp}}) \\
    &= 2 \cos\theta \sin\theta \ket{\Phi} + (\cos^2 \theta - \sin^2 \theta) \ket{\Phi^{\perp}} \\
    &= \sin 2\theta \ket{\Phi} + \cos 2\theta \ket{\Phi^{\perp}},
  \end{align*}
  \endgroup
  thereby showing that $G$ acts as a rotation by $2\theta$ in the
  subspace spanned by $\ket{\Phi} = \ket{0}\ket{\psi}$ and
  $\ket{\Phi^{\perp}} = \ket{1}\ket{\phi}$, from which the desired
  result follows.
\end{proof}

\noindent Thus, we have shown that we can perform amplitude
amplification even with just a copy of $\ket{\psi}$, i.e., without
repeatedly using a circuit to prepare it: applying
Thm.~\ref{thm:obliviousampamp} we choose $k$ to maximize the
probability of obtaining $V\ket{\psi}$, and the asymptotic scaling is
the same as with standard amplitude amplification. All that is
necessary to apply Thm.~\ref{thm:obliviousampamp} is the unitary $U$
that performs the desired operation $V$ on $\ket{\psi}$, and a way to
recognize (and reflect through) the subspace where the term
$V\ket{\psi}$ appears. In particular, note that for simplicity we
considered the case of a single flag qubit, but it is straightforward
to extend the analysis to the case with multiple flag qubits, i.e.,
\begin{equation*}
  U \ket{\v{0}}_q \ket{\psi}_n = \sin \theta \ket{\v{0}}_q V \ket{\psi}_n + \cos \theta \ket{\phi}_{n+q}
\end{equation*}
where $\ket{\phi}$ is a state that has no support on $\ket{\v{0}}_q$
(formally, $\ket{\v{0}}_q \bra{\v{0}}_q \otimes I) \ket{\phi} =
0$). For example, one could simply use a unitary that checks whether
all the flag qubits are set correctly, and performs a controlled
operation to reduce to the case of a single flag qubit. A formal
analysis is available in \cite{berry2014exponential}.\index{algorithm!amplitude amplification|)}\index{amplitude!amplification|)}

\section{Amplitude estimation}
\label{sec:ampest}
Amplitude estimation\index{algorithm!amplitude estimation|(}\index{amplitude!estimation|(} uses the amplitude amplification framework to
estimate the magnitude of an amplitude. In the context of Grover's
problem, it leads to the following algorithmic building block: rather
than identifying a marked binary string $\v{\ell}$ in $\{0,1\}^n$,
i.e., a string that can be recognized by a function, we can {\it
  count} the total number of marked strings. Trivially, counting the
number of solutions also answers the question of existence of a
solution. Amplitude estimation has many other applications, besides
counting the number of solutions; some of them are discussed in
subsequent sections. The technique was introduced in
\cite{brassard2002quantum}.

The problem solved by amplitude estimation can be phrased as
follows. Similarly to the discussion in Sect.~\ref{sec:geomgrover},
the input of the algorithm is:
\begin{enumerate}[(i)]
\item A circuit $S$ acting on $n$ qubits that prepares the state:
  \begin{equation*}
    S \ket{\v{0}}_n := \ket{\psi_0} = \sin \theta \ket{\psi_G} + \cos
    \theta \ket{\psi_B},
  \end{equation*}
  with $ \braket{\psi_G}{\psi_B} = 0$ and $\theta \in [0,
  \frac{\pi}{2}]$.
\item A controlled circuit C$R$ acting on $n$ qubits, plus one control
  qubit, that has the following action:
  \begin{equation*}
    \text{C}R\left( \ket{x}_1 \otimes (\alpha_G \ket{\psi_G} + \alpha_B
    \ket{\psi_B})\right) = \ket{x} \otimes \left((-1)^x \alpha_G
    \ket{\psi_G} + \alpha_B \ket{\psi_B}\right),
  \end{equation*}
  for any $\alpha_G, \alpha_B \in [-1,1]: \abs{\alpha_G}^2 +
  \abs{\alpha_B}^2 = 1$ and $x \in \{0,1\}$. Note that C$R$ is the
  controlled version of the reflection operator $R$ in
  Sect.~\ref{sec:geomgrover}.
\item A precision parameter $\epsilon > 0$.
\item A maximum failure probability $\delta > 0$.
\end{enumerate}
The goal is to determine $\tilde{\theta}$ such that $\abs{\theta -
  \tilde{\theta}} \le \epsilon$ with a probability of success at least
$1 - \delta$. Note that estimating $\theta$ is equivalent to
estimating $\braket{\psi_0}{\psi_G} = \sin \theta$, up to some
(constant) conversion factor to translate between the angle $\theta$
and its sine.
\begin{remark}
  \label{rem:counting}
  It is easily established that the problem stated in this form allows
  counting the number of marked items, e.g., solutions of a
  satisfiability problem if the function $f$ computes the value of a
  given Boolean formula for a given truth assignment of the Boolean
  variables. Indeed, suppose we have access to $f : \{0,1\}^n \to
  \{0,1\}$ and we want to determine $|M|$ where $M := \{\vj \in
  \{0,1\}^n : f(\vj) = 1\}$; the function $f$ identifies the marked
  binary strings. Using phase kickback, as discussed in
  Sect.~\ref{sec:phasekickback}, the circuit $U_f$ can be used to
  implement $R$ using an additional qubit set in the state $(\ket{0} -
  \ket{1})/\sqrt{2}$; then we can take $S = H^{\otimes n}$ so that
  $\ket{\psi_G} = \frac{1}{\sqrt{|M|}} \sum_{\vj \in M} \ket{\vj},
  \ket{\psi_B} = \frac{1}{\sqrt{2^n - |M|}} \sum_{\vj \in \{0,1\}^n
    \setminus M} \ket{\vj}$. These two states are orthogonal, and:
  \begin{equation*}
    S \ket{\v{0}} = H^{\otimes n} \ket{\v{0}} = \sqrt{\frac{|M|}{2^n}}
    \ket{\psi_G} + \sqrt{\frac{2^n - |M|}{2^n}} \ket{\psi_B}.
  \end{equation*}
  According to our definition, $\sin \theta = \sqrt{\frac{|M|}{2^n}}$,
  which implies $|M| = 2^n \sin^2 \theta$, so that estimating $\theta$
  (or, equivalently, $\sin \theta$) allows us to recover an estimate
  on the number of marked items.
\end{remark}

\subsection{Solution strategy}
To solve this problem we rely on properties of the Grover
iteration operator $G = S F S^{\dag} R$.
\begin{remark}
  In Sect.~\ref{s:qaecircuit} we clarify why we assume that $R$, i.e.,
  the reflection through $\ket{\psi_G}$, is given in the form of a
  controlled operator. Note that Grover search does not require the
  controlled version of $R$. Because we have a controlled version of
  $R$, we can of course also apply $R$, so it is legitimate to study
  the operator $G$.
\end{remark}
\begin{proposition}
  \label{prop:grovereig}
  The states $\ket{\phi_+} = \frac{1}{\sqrt{2}}(\ket{\psi_G} + i
  \ket{\psi_B}), \ket{\phi_-} = \frac{1}{\sqrt{2}}(\ket{\psi_G} - i
  \ket{\psi_B})$ are orthogonal eigenstates of $S F S^{\dag} R$ with
  eigenvalues $e^{2 i \theta}, e^{-2 i \theta}$ respectively.
\end{proposition}
\begin{proof}
  To check that they are eigenstates and find the corresponding
  eigenvalue, we carry out the matrix-vector multiplication. We
  use the fact that $(2\ketbra{\psi_0}{\psi_0} - I^{\otimes n})$ is a
  reflection through $\ket{\psi_0} = \sin \theta \ket{\psi_G} + \cos
  \theta \ket{\psi_B}$. To see the effect of such a reflection on
  $-\ket{\psi_G}$ and on $\ket{\psi_B}$, we can rely on
  Fig.~\ref{fig:grovereig}, together with simple geometry: reflecting
  $-\ket{\psi_G}$ through $\ket{\psi_0}$ yields $\cos 2\theta
  \ket{\psi_G} - \sin 2\theta \ket{\psi_B}$ (see
  Fig.~\ref{fig:grovereig1}), while reflecting $\ket{\psi_B}$ through
  $\ket{\psi_0}$ yields $\sin 2\theta \ket{\psi_G} + \cos 2\theta
  \ket{\psi_B}$ (see Fig.~\ref{fig:grovereig2}). Then we have:
  \begin{align*}
    S F S^{\dag} R \ket{\phi_+} &= S F S^{\dag}
    \frac{1}{\sqrt{2}}(-\ket{\psi_G} + i \ket{\psi_B}) =
    (2\ketbra{\psi_0}{\psi_0} - I^{\otimes n})
    \frac{1}{\sqrt{2}}(-\ket{\psi_G} + i \ket{\psi_B}) \\
    &= \frac{1}{\sqrt{2}} \left(\cos 2\theta \ket{\psi_G} - \sin 2\theta \ket{\psi_B} + i\left(\sin 2\theta \ket{\psi_G} + \cos 2\theta \ket{\psi_B}\right)\right) \\
    &= \frac{1}{\sqrt{2}} \left(e^{2 i \theta} \ket{\psi_G} + i e^{2 i \theta} \ket{\psi_B}\right) = e^{2 i \theta} \ket{\phi_+},
  \end{align*}
  which shows that $\ket{\phi_+}$ is an eigenstate with eigenvalue
  $e^{2 i \theta}$.
  \begin{figure}[t!b]
    \centering
    \subfloat[Reflection of $-\ket{\psi_G}$ through $\ket{\psi_0}$.]{%
      \centering
      \ifcompilefigs
      \begin{tikzpicture}[>=Stealth,scale=0.78]
        \draw[draw=none] (0,0) -- (-3,0);
        \draw[thick,->] (0,0) -- (2,0) node[anchor=west] {$\ket{\psi_B}$};
        \draw[thick,->] (0,0) -- (0,2) node[anchor=south] {$\ket{\psi_G}$};
        \draw[thick,->,dashed] (0,0) -- (0,-2) node[anchor=north] {$\ket{-\psi_G}$};
        \draw[thick,->] (0,0) coordinate (O) -- (20:2) coordinate (oc) 
        node[anchor=west] {$\ket{\psi_0}$};
        \draw[thick,->] (0,0) coordinate (O) -- (130:2) coordinate (oc) 
        node[anchor=east] {$-(2\ketbra{\psi_0}{\psi_0} - I^{\otimes
            n})\ket{\psi_G}$};
        \draw[thick,->,dotted,blue] (0,-2)
        coordinate (oc) -- (130:2) coordinate (oc) {};
        \draw[black] (0:1) arc(0:20:1) node[midway,right] {$\theta$};
        \draw[black] (90:1) arc(90:130:1) node[midway,above] {$2\theta$};
        \draw[red, thick] (O) circle (2 cm);
      \end{tikzpicture}
      \else
      \includegraphics[scale=0.85]{figures/grovereig1.pdf}
      \fi
      \label{fig:grovereig1}
    }\hspace{0.01em}
    \subfloat[Reflection of $\ket{\psi_B}$ through $\ket{\psi_0}$.]{%
      \centering
      \ifcompilefigs
      \begin{tikzpicture}[>=Stealth,scale=0.78]
        \draw[draw=none] (0,0) -- (-3,0);
        \draw[thick,->] (0,0) -- (2,0) node[anchor=west] {$\ket{\psi_B}$};
        \draw[thick,->] (0,0) -- (0,2) node[anchor=south] {$\ket{\psi_G}$};
        \draw[draw=none,thick,->,dashed] (0,0) -- (0,-2) node[anchor=north] {$\phantom{\ket{-\psi_G}}$};
        \draw[thick,->] (0,0) coordinate (O) -- (20:2) coordinate (oc) 
        node[anchor=west] {$\ket{\psi_0}$};
        \draw[thick,->] (0,0) coordinate (O) -- (40:2) coordinate (oc) 
        node[anchor=west] {$(2\ketbra{\psi_0}{\psi_0} - I^{\otimes
            n})\ket{\psi_B}$};
        \draw[thick,->,dotted,blue] (2,0)
        coordinate (oc) -- (40:2) coordinate (oc) {};
        \draw[black] (0:1) arc(0:20:1) node[midway,right] {$\theta$};
        \draw[black] (20:1) arc(20:40:1) node[near end,right] {$\theta$};
        \draw[red, thick] (O) circle (2 cm);
      \end{tikzpicture}
      \else
      \includegraphics[scale=0.85]{figures/grovereig2.pdf}
      \fi
      \label{fig:grovereig2}
    }
    \caption{Sketch of the two-dimensional plane spanned by $\ket{\psi_G}$ and $\ket{\psi_B}$, to understand the effect of reflections through $\ket{\psi_0}$.}
    \label{fig:grovereig}
\end{figure}

  The calculations to find the eigenvalue corresponding to
  $\ket{\phi_-}$ are very similar:
  \begin{align*}
    S F S^{\dag} R \ket{\phi_-} &= S F S^{\dag}
    \frac{1}{\sqrt{2}}(-\ket{\psi_G} - i \ket{\psi_B}) =
    (2\ketbra{\psi_0}{\psi_0} - I^{\otimes n})
    \frac{-1}{\sqrt{2}}(\ket{\psi_G} + i \ket{\psi_B}) \\
    &= \frac{-1}{\sqrt{2}} \left(-\cos 2\theta \ket{\psi_G} + \sin 2\theta \ket{\psi_B} + i\left(\sin 2\theta \ket{\psi_G} + \cos 2\theta \ket{\psi_B}\right)\right) \\
    &= \frac{1}{\sqrt{2}} \left(-e^{-2 i \theta} \ket{\psi_G} - i e^{-2 i \theta} \ket{\psi_B}\right) = e^{-2 i \theta} \ket{\phi_-}.
  \end{align*}
  Finally, orthogonality can be checked by computing the inner product
  of the two eigenstates, and verifying that it is zero.
\end{proof}

\noindent Given the result in Prop.~\ref{prop:grovereig}, a strategy
to compute $\theta$ becomes apparent, via the quantum phase estimation
algorithm: we can apply phase estimation to the Grover operator $G$,
with the goal of estimating the eigenvalue of one of the two
eigenstates $\ket{\phi_+}, \ket{\phi_-}$. However, to apply phase
estimation two ingredients are needed: we must be able to prepare one
of the two eigenstates $\ket{\phi_+}, \ket{\phi_-}$, and we must be
able to implement the controlled operators $G^{2^k}$ for integer $k$.

Let us turn our attention to the first issue, namely, preparing one of
the two eigenstates $\ket{\phi_+}, \ket{\phi_-}$. Recall that by
assumption we only know how to prepare $S \ket{\v{0}} = \sin \theta
\ket{\psi_G} + \cos \theta \ket{\psi_B}$. We show that this state is a
linear combination of the two eigenstates above. Indeed, we have:
\begin{align*}
  S \ket{\v{0}} &= \sin \theta \ket{\psi_G} + \cos \theta \ket{\psi_B} =
  \frac{e^{i\theta} - e^{-i\theta}}{2i} \ket{\psi_G} +
  \frac{e^{i\theta} + e^{-i\theta}}{2} \ket{\psi_B} \\
  &= \frac{i}{2}\left((-e^{i\theta} + e^{-i\theta}) \ket{\psi_G} - i (e^{i\theta} + e^{-i\theta}) \ket{\psi_B} \right) \\
  &= \frac{i}{2}\left(-e^{i\theta} (\ket{\psi_G} + i\ket{\psi_B}) + e^{-i\theta} (\ket{\psi_G} - i\ket{\psi_B})\right) \\
  &= \frac{i}{\sqrt{2}}\left(-e^{i\theta} \ket{\phi_+} + e^{-i\theta} (\ket{\phi_-} \right).
\end{align*}
Thus, $S \ket{\v{0}}$ is a linear combination (with complex
coefficients) of the two eigenstates, and the corresponding
coefficients have equal weight in the superposition, i.e.,
$\left|\frac{-ie^{i\theta}}{\sqrt{2}}\right|^2 =
\left|\frac{ie^{-i\theta}}{\sqrt{2}}\right|^2 = \frac{1}{2}$. Because
the eigenstates $\ket{\phi_+}, \ket{\phi_-}$ are orthogonal
(Prop.~\ref{prop:grovereig}), this decomposition of $S \ket{\v{0}}$ in
terms of eigenstates of the Grover operator is unique (it corresponds
to the decomposition of $S \ket{\v{0}}$ in terms of an eigenbasis of
the operator). It follows that if we apply phase estimation to $S
\ket{\v{0}}$, we obtain with equal probability a binary representation
of the phase of the eigenvalue corresponding to either of the
eigenstates $\ket{\phi_+}, \ket{\phi_-}$, namely, $2\theta$ or
$-2\theta$.
\begin{remark}
  More precisely, we obtain a finite-precision representation of
  $\frac{\theta}{\pi}$ or $-\frac{\theta}{\pi}$, because phase
  estimation assumes that the phase of the eigenvalue is of the form
  $2\pi \theta$.
\end{remark}
However, which eigenvalue is obtained does not matter, because by
assumption $\theta \in [0, \frac{\pi}{2}]$, hence we can determine
with certainty if we obtained $\frac{\theta}{\pi}$ or
$-\frac{\theta}{\pi}$ by simply looking at whether we obtained $\theta
\le \frac{1}{2}$ (in which case we must have collapsed onto the
eigenstate with eigenvalue $2\theta$) or $\theta \ge \frac{1}{2}$ (in
which case we have collapsed onto the eigenstate with eigenvalue
$-2\theta$). Thus, phase estimation leads to an estimate of $\theta$,
no matter which of the two eigenvalues we collapse to after
measurement. It remains to determine how to construct the controlled
operators $G^{2^k}$ for integer $k$.

\subsection{Implementation of the amplitude estimation circuit}
\label{s:qaecircuit}
\begin{figure}[h!]
  \leavevmode
  \centering
  \ifcompilefigs
  \Qcircuit @C=1em @R=.7em {
    \lstick{\ket{x}}    & \qw       & \ctrl{1} & \qw           & \ctrl{1} & \qw      & \qw & \rstick{\ket{x}} \\
    \lstick{\ket{\psi}} & \qw {/^n} & \gate{R} & \gate{S^\dag} & \gate{F} & \gate{S} & \qw & \rstick{(SFS^{\dag}R)^{x} \ket{\psi}} \\
  }
  \else
  \includegraphics{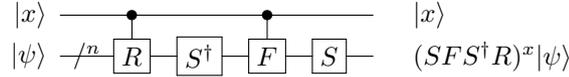}
  \fi
  \caption{Controlled version of the Grover operator, with $x \in \{0,1\}$.}
  \label{fig:contrgrover}
\end{figure}

We now come to the central question of implementing the algorithm
described above, so as to estimate its resource requirements. To do
so, we first study how to implement the controlled $(S F S^{\dag} R)$
operator; then, implementing powers of this operator can be done by
simply concatenating multiple copies of the controlled operator.
\begin{remark}
  Implementing a controlled version of $G^{2^k}$ by chaining $2^k$
  copies of controlled-$G$ requires resources (i.e., number of gates)
  that grow with $2^k$, and may therefore be large if $k$ is
  large. For our purposes, $k$ is as large as $m$, where $m$ is
  the number of bits of the phase estimation, which in turn determines
  the precision of our estimate for $\theta$. In general we cannot do
  better than this, because we are not assuming much structure on
  $S$. However, it is possible that for a specific problem at hand, a
  more efficient implementation of this operator exists, leading to
  smaller resource requirements, see also the discussion in
  Rem.~\ref{rem:qpeuexp}.
\end{remark}
By assumption we are given access to C$R$, the controlled version of
$R$. A crucial observation is the fact that to obtain
$\text{C}G=\text{C}(S F S^{\dag} R)$, it is sufficient to implement
C$F$. Indeed, consider the circuit in Fig.~\ref{fig:contrgrover}.
In this circuit, when the control qubit $\ket{x}$ is $\ket{1}$ the
full Grover operator $(S F S^{\dag} R)$ is applied to the second
register; if, on the other hand, $\ket{x} = \ket{0}$, the
transformation acts as the identity on the second register, because $S
S^{\dag} = I^{\otimes n}$. Thus, we only need to determine how to
implement C$F$.

\begin{figure}[h!]
  \leavevmode
  \ifcompilefigs
  \hspace*{14em}
  \Qcircuit @C=1em @R=.7em {
    \lstick{\text{Control} \; \ket{x}}& \gate{Z} & \ctrl{1} & \qw      & \qw \\
    & \gate{X} & \ctrl{1} & \gate{X} & \qw \\
    & \gate{X} & \ctrl{1} & \gate{X} & \qw \\
    & \gate{X} & \ctrl{3} & \gate{X} & \qw & \rstick{\raisebox{2.7em}{$\left(\ketbra{0}{0}\otimes I^{\otimes n} + \ketbra{1}{1} \otimes
    \left(2\ketbra{\v{0}}{\v{0}} - I^{\otimes n}\right)\right)\ket{x}\ket{\psi}$}} \\
    & \vdots   &          & \vdots   & \\
    &  \\
    & \gate{X} & \gate{Z} & \gate{X} & \qw 
    \inputgroupv{2}{7}{.8em}{3.68em}{\ket{\psi}}
    {\gategroup{1}{5}{7}{5}{.8em}{\}}}
    \\
  }
  \else
  \includegraphics{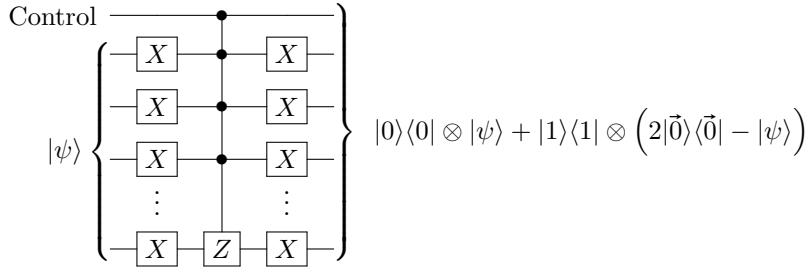}
  \fi
  \caption{Quantum circuit implementing the controlled-$F$ operation
    (up to a global phase factor) used in the amplitude estimation
    algorithm.}
  \label{fig:contrF}
\end{figure}

The operator $-F$ is implemented by the circuit in
Fig.~\ref{fig:grovercircuit}, with a global phase factor $-1$, see
Sect.~\ref{sec:matrixd}. Because $-F$ is already a controlled
operation with multiple controls, to obtain C$F$ we simply add one
more control and one $Z$ gate to get rid of the $-1$ (otherwise we
would implement C$(-F)$, which is not the same as $-\text{C}F$, see
Rem.~\ref{rem:relativephasecontrol}). We obtain the circuit in
Fig.~\ref{fig:contrF}. This can be easily decomposed in terms of CC$X$
gates with some auxiliary qubits.

\subsection{Summary and resource requirements}
\label{sec:amplestsummary}
We summarize the phase estimation algorithm: the circuits $S$ and
C$R$ are given as input (see Sect.~\ref{sec:ampest} for a definition),
together with parameters $\epsilon, \delta > 0$; the goal is to
determine $\tilde{\theta}$ such that $|\theta - \tilde{\theta}| \le
\epsilon$ with a probability of success at least $1 - \delta$.

Let $m = \ceil{\log \frac{\pi}{\epsilon}} + 2$; by Thm.~\ref{thm:qpe},
with this number of qubits for phase estimation we obtain $\theta$ to
precision $\frac{\epsilon}{\pi}$ with probability at least $3/4$. In
this setting, it is convenient to pick a constant probability of
success for phase estimation, and then repeat the algorithm a few
times to boost the probability of obtaining the correct answer.
\begin{remark}
  The term $\frac{\pi}{\epsilon}$, rather than $\frac{1}{\epsilon}$,
  is due to the fact that phase estimation applied to eigenvalues
  $e^{\pm 2 i \theta}$ outputs $\pm \frac{\theta}{\pi}$ rather than
  $\pm \theta$, so we need to increase the precision slightly.
\end{remark}
The algorithm works as follows:
\begin{itemize}
\item Initialize the state as $\ket{\v{0}}_{m} \ket{\v{0}}_{n}$.
\item Apply $S$ to the last $n$ qubits (second register) to obtain
  $\ket{\v{0}}_m \otimes (\sin \theta \ket{\psi_G} + \cos \theta
  \ket{\psi_B})$.
\item Run the quantum phase estimation algorithm applied to the
  operator $G = S F S^\dag R$, using the first $m$ qubits (first
  register) to store the phase, and the bottom $n$ qubits to store the
  ``eigenstate'' $\sin \theta \ket{\psi_G} + \cos \theta
  \ket{\psi_B}$. (This is in fact a linear combination of
  eigenstates.)
\item Let $\v{b}$ be the $m$-digit binary string obtained as output of
  phase estimation by measuring the first $m$ qubits. If $\v{b}_1 =
  1$, i.e., $0.\v{b} > \frac{1}{2}$, return $\tilde{\theta} = \pi (1 -
  0.\v{b})$; otherwise, return $\tilde{\theta} = \pi 0.\v{b}$.
\end{itemize}
\begin{figure}[h!]
  \leavevmode
  \centering
  \ifcompilefigs
  \Qcircuit @C=1em @R=0.7em {
    & \qw & \gate{H}  & \qw & \qw & \qw & \dots & & \ctrl{4} & \qw & \multigate{3}{Q_{m}^{\dag}} & \qw \\
    &\vdots&        &     &     &     &       & &          &     &     &  \\
    & \qw & \gate{H}  & \qw & \ctrl{2} & \qw & \dots & & \qw & \qw & \ghost{Q_{m}^{\dag}} & \qw \\
    & \qw & \gate{H}  & \ctrl{1} & \qw & \qw & \dots & & \qw & \qw & \ghost{Q_{m}^{\dag}} & \qw \\
    \lstick{\ket{\v{0}}} & {/^n} \qw & \gate{S} & \gate{(S F S^\dag R)^{2^0}} & \gate{(S F S^\dag R)^{2^1}} & \qw & \dots & & \gate{(S F S^\dag R)^{2^{m-1}}} & \qw & {/^n} \qw & \qw 
    \inputgroupv{1}{4}{.8em}{2.2em}{\ket{\v{0}}_m} \\
  }
  \else
  \includegraphics{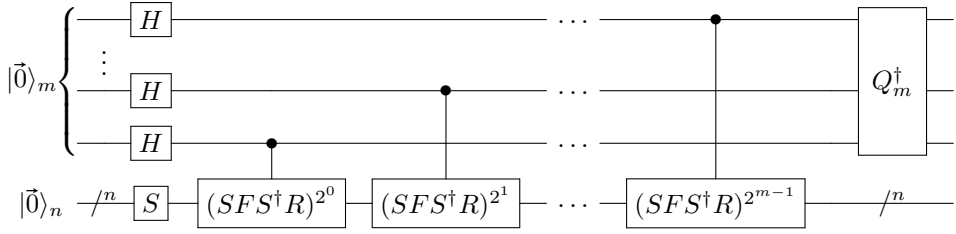}
  \fi
  \caption{Amplitude estimation circuit with $m$ bits of precision.}
  \label{fig:ampest}
\end{figure}

This algorithm uses $n + m$ qubits, plus all qubits necessary for
auxiliary space, for example for the implementation of the C$F$
operation, as well as the implementation of the black-box circuits $S$
and C$R$. The gate complexity is $\bigO{2^m (n + T_S + T_{\text{C}R})
  + m^2}$, where $T_S$ is the gate complexity of $S$, and
$T_{\text{C}R}$ is the gate complexity of C$R$. This is because phase
estimation uses $\bigO{2^m}$ applications of the Grover operator, and
each application involves one call to $S$, one call to $S^\dag$, one
call to C$R$, and one application of C$F$, which uses $\bigO{n}$ gates
if implemented as in Fig.~\ref{fig:contrF}. The final $\bigO{m^2}$
gates are for the inverse quantum Fourier transform. Considering our
choice of $m = \ceil{\log \frac{\pi}{\epsilon}} + 2$, the complexity
amounts to $\bigO{\frac{1}{\epsilon}}$ applications of $S$ and C$R$,
and $\bigO{\frac{1}{\epsilon} \log^2 \frac{1}{\epsilon}}$ additional
gates: this yields the correct answer with probability 3/4. To boost
the probability of success to $1-\delta$ we can repeat the algorithm a
few times and output the majority answer, with $\bigO{\log
  \frac{1}{\delta}}$ repetitions of the algorithm with constant
success probability.

\subsection{Amplitude estimation for counting and probability estimation}
\label{sec:counting}
We discuss an application of amplitude estimation to count marked
items, as introduced in Rem.~\ref{rem:counting}. In this setting,
suppose we have access to $f : \{0,1\}^n \to \{0,1\}$ and we want to
determine $|M|$ where $M := \{\vj \in \{0,1\}^n : f(\vj) = 1\}$; the
function $f$ identifies the marked binary strings. We use $S =
H^{\otimes n}$ and set $\ket{\psi_G} = \frac{1}{\sqrt{|M|}} \sum_{\vj
  \in M} \ket{\vj}, \ket{\psi_B} = \frac{1}{\sqrt{2^n - |M|}}
\sum_{\vj \in \{0,1\}^n \setminus M} \ket{\vj}$. As remarked in
Rem.~\ref{rem:counting}, $\sin \theta = \sqrt{\frac{|M|}{2^n}}$, which
implies $|M| = 2^n \sin^2 \theta$. We first give a bound on the
distance between $\sin^2 \tilde{\theta}$ and $\sin^2 \theta$ based on
the distance between the angles.
\begin{proposition}[Lem.~7 in \cite{brassard2002quantum}]
  \label{prop:abound}
  Let $a = \sin^2 \theta, \tilde{a} = \sin^2 \tilde{\theta}$ with $0
  \le \theta, \tilde{\theta} \le 2\pi$. Then
  \begin{equation*}
    |\theta - \tilde{\theta}| \le \epsilon \implies |a - \tilde{a}| \le 2\epsilon \sqrt{a(1-a)} + \epsilon^2.
  \end{equation*}
\end{proposition}
\begin{proof}
  Using trigonometric identities, we have:
  \begin{align*}
    \sin^2(\theta + \epsilon) - \sin^2 \theta &= (\sin\theta \cos \epsilon + \sin \epsilon \cos \theta)^2 - \sin^2 \theta \\
    &= \sin^2 \theta \cos^2 \epsilon + \cos^2\theta \sin^2 \epsilon + 2 \sin\theta \sin\epsilon \cos\theta \cos\epsilon - \sin^2 \theta.    
  \end{align*}
  We rewrite this, using $\cos^2 \epsilon = 1 - \sin^2
  \epsilon$, $2 \sin\epsilon \cos\epsilon = \sin
  2\epsilon$, $\sin \theta = \sqrt{a}$, $\cos\theta =\sqrt{1 - a}$,
  and obtain:
  \begin{equation*}
    \sin^2(\theta + \epsilon) - \sin^2 \theta = \sqrt{a(1-a)} \sin
    2\epsilon + (1 - 2a) \sin^2 \epsilon.
  \end{equation*}
  Using similar transformations, we obtain:
  \begin{equation*}
    \sin^2 \theta - \sin^2(\theta - \epsilon)= \sqrt{a(1-a)} \sin
    2\epsilon + (2a - 1) \sin^2 \epsilon.
  \end{equation*}
  Finally, using the fact that $\sin x \le x \; \forall x \ge 0$, and
  $|2a - 1| \le 1$, we have:
  \begin{equation*}
    |a - \tilde{a}| \le \max\{\sin^2(\theta + \epsilon) - \sin^2 \theta,
    \sin^2 \theta - \sin^2 (\theta + \epsilon)\} \le 2\epsilon \sqrt{a(1-a)} + \epsilon^2. 
  \end{equation*}
\end{proof}

\noindent Based on this, we can already determine if $\theta = 0$ or
not.
\begin{example}
  Suppose our goal is to determine if $|M| = 0$ or not. We choose
  $\delta = 1/4$ and $\epsilon = 2^{-(n/2+2)}$; this tells us that we
  should use $m = \ceil{\log(\frac{2^{n/2 + 5}}{\pi})} =
  \bigO{\frac{n}{2}}$ qubits of precision for phase estimation. When
  $|M| = 0$, $\theta = 0$, $a = 0$ and Prop.~\ref{prop:abound} tells
  us that we have:
  \begin{equation*}
    |a - \tilde{a}| \le \frac{1}{2^{2\ceil{\frac{n}{2} + 2}}} = \frac{1}{2^4 \cdot 2^n}.
  \end{equation*}
  Hence, from our estimate $\tilde{\theta}$ we compute $\tilde{a}$,
  and when rounding this value to the closest integer (because the
  number of solutions has to be an integer) we obtain $0$ with
  probability at least $3/4$.

  Now assume $|M| = 1$. Then by Prop.~\ref{prop:abound} have:
  \begin{equation*}
    |a - \tilde{a}| \le 2 \frac{\sqrt{\frac{1}{2^n}(1-\frac{1}{2^n})}}{2^{\ceil{\frac{n}{2} + 2}}} + \frac{1}{2^{2\ceil{\frac{n}{2} + 2}}} < \frac{1}{4} \frac{\sqrt{(1 - \frac{1}{2^n})}}{2^n} + \frac{1}{2^4 \cdot 2^n} < \frac{1}{2 \cdot 2^n},
  \end{equation*}
  so the number of solutions, rounded to the nearest integer, is a
  value $\neq 0$ with probability at least $3/4$. If $|M| > 1$, the
  probability to obtain $0$ is even lower. It is therefore easy to
  distinguish the two cases $|M| = 0$ and $|M| \neq 0$. Because $m =
  \bigO{\frac{n}{2}}$, the query complexity of this algorithm (number
  of applications of the unitary $U_f$ implementing the marking
  function $f$) is $\bigO{\sqrt{2^n}}$.
\end{example}

\noindent Using a similar analysis to the one employed in the
derivation of Thm.~\ref{thm:qpesimple} for phase estimation with $m$
qubits, \cite{brassard2002quantum} also shows the following (which is
the direct analog of Thm.~\ref{thm:qpesimple} in the context of
amplitude estimation).
\begin{proposition}[Thm.~12 in \cite{brassard2002quantum}]
  \label{prop:brassardbound}
  Suppose that we apply the amplitude estimation algorithm using $m$
  qubits for phase estimation. Define $a = \sin^2 \theta, \tilde{a} =
  \sin^2 \tilde{\theta}$ with $0 \le \theta, \tilde{\theta} \le
  2\pi$. Then the algorithm produces an estimate $\tilde{a}$ such
  that:
  \begin{equation}
    \label{eq:brassardbound}
    |a - \tilde{a}| \le 2 \pi \frac{\sqrt{a(1-a)}}{2^{m}} + \frac{\pi^2}{2^{2m}}
  \end{equation}
  with probability at least $\frac{8}{\pi^2}$.
\end{proposition}
This implies the following simplified statement of the complexity of
amplitude estimation.
\begin{corollary}
  \label{cor:brassardbound}
  Suppose we want to estimate the probability $\sin^2 \theta$ of
  obtaining $\ket{\psi_G}$ from a measurement of $S\ket{\v{0}}$. To
  obtain an estimate with absolute error at most $\epsilon$ using
  amplitude estimation, it is sufficient to use $m = \bigO{\log
    \frac{1}{\epsilon}}$ qubits for phase estimation, leading to
  $\bigO{\frac{1}{\epsilon}}$ queries to the input unitaries $S$ and
  C$R$ to carry out the algorithm.
\end{corollary}
\begin{remark}
  \label{rem:ampestunbiased}
  Cor.~\ref{cor:brassardbound} provides a quadratic speedup over
  classical estimation via the empirical average of a number of
  samples (see Sect.~\ref{sec:montecarlo}), but it does not discuss
  the bias of the estimator for $a$, which unfortunately can be
  poor. However, with more advanced techniques amplitude estimation
  can be made unbiased
  \cite{cornelissen2023sublinear,rall2023amplitude}, or rather, it can
  be modified to efficiently reduce the bias (i.e., the cost of the
  reduction of the bias is merely polylogarithmic in the reduction
  factor), see the notes in Sect.~\ref{sec:ampampnotes}.
\end{remark}



\subsection{Application to Monte Carlo simulation}
\label{sec:montecarlo}
Amplitude estimation finds a direct application in Monte Carlo\index{simulation!Monte Carlo|(}
simulation. The crucial observation is that the precision of the
estimate grows linearly with the number of samples, rather than with
the square root of the number of samples. Indeed, this can be observed
in Prop.~\ref{prop:brassardbound}, by looking at the error estimates:
obtaining $a = \sin^2 \theta$ with precision $\epsilon$ requires
$\bigO{\log \frac{1}{\epsilon}}$ qubits to store the outcome of phase
estimation, and therefore $\bigO{\frac{1}{\epsilon}}$ calls to the
unitary $S$ preparing $\sin \theta \ket{\psi_G} + \cos \theta
\ket{\psi_B}$. The linear scaling is stated explicitly in
Cor.~\ref{cor:brassardbound}. This is better scaling than in classical
Monte Carlo techniques, where the number of samples that one needs to
obtain from a random variable grows quadratically with the precision;
i.e., we generally need $\bigO{\frac{1}{\epsilon^2}}$ samples, and
therefore calls to a function constructing a sample from the desired
probability distribution, to obtain an estimate with precision
$\epsilon$. We formalize this next, also explaining why the comparison
between calls to $S$ (for the quantum case) and classical samples
makes sense, at least from some point of view --- see
Rem.~\ref{rem:classicalqsample}.

Suppose we are given a discrete random variable $X$ with sample space
$\Omega = \{0,1\}^n$ and $\Pr(X = \vj) = p_{j}$. Let $P$ be the
unitary that maps:
\begin{equation}
  \label{eq:qsampleampest}
  P \ket{\v{0}}_n = \sum_{\vj \in \{0,1\}^n} \sqrt{p_{j}} \ket{\vj}.
\end{equation}
Such a unitary can be implemented with $\bigO{2^n}$ basic gates in
general, see Sect.~\ref{sec:ampencalg}; more efficient
implementations, possibly with complexity polynomial in $n$, exist
for distributions with certain properties, see the notes in
Sect.~\ref{sec:ampampnotes}. Suppose further that our goal is to
compute the expected value of a function $f : \{0,\frac{1}{2^n},\dots,\frac{2^n-1}{2^n}\} \to [0,1]$,
which we assume to be given as the following quantum oracle on $n+1$
qubits:
\begin{equation*}
  U_f (\ket{\vj}_n \otimes \ket{0}) = \ket{\vj} \otimes \left(\sqrt{1 - f(0.\vj)} \ket{0} + \sqrt{f(0.\vj)} \ket{1}\right).
\end{equation*}
Because $f$ is defined over fractions, the binary string $\vj$
corresponds to the argument $0.\vj$ of the function (i.e., $\vj$
represents the number $j/2^n$): this yields a natural mapping that
simplifies subsequent calculations.  Note that if we have a binary
oracle for $f$, we can implement the above transformation as a
controlled rotation on the last qubit (i.e., we first use the binary
oracle to output a binary description of $f(0.\vj)$, then we apply
rotations controlled on the individual bits of $f(0.\vj)$ to rotate
the last qubit by the corresponding amount). Thus, we can apply the
amplitude estimation algorithm onto the state:
\begin{equation*}
  U_f (P \ket{\v{0}}_n \otimes \ket{0}) = \sum_{\vj \in \{0,1\}^n} \sqrt{1 - f(0.\vj)} \sqrt{p_{j}} \ket{\vj} \ket{0}  + \sum_{\vj \in \{0,1\}^n} \sqrt{f(0.\vj)} \sqrt{p_{j}} \ket{\vj} \ket{1},
\end{equation*}
aiming to estimate the amplitude of the part of the state that has
$\ket{1}$ in the last qubit, i.e., $\sum_{\vj \in \{0,1\}^n}
\sqrt{f(0.\vj)} \sqrt{p_{j}} \ket{\vj} \ket{1}$.
\begin{remark}
  \label{rem:classicalqsample}
  In this setting, one application of the unitary $P$ is equivalent to
  one classical sample in the following sense: if we prepare the state
  $P\ket{\v{0}}$ and then apply a measurement to all qubits, we obtain
  the string $\vj$ with probability $p_j$. This is exactly what we
  would obtain from a classical sample from the discrete probability
  distribution encoded by the vector $p$. Thus, we can simulate one
  classical sample by applying the unitary $P$ once on a quantum
  computer. This implies that an application of $P$ is more powerful
  than the construction of one classical sample: using $P$ once we can
  simulate a classical sample, but the converse may not be true. The
  state produced in Eq.~\ref{eq:qsampleampest} is often called a
  \emph{qsample}\index{qsample}\index{quantum!sample} in the literature, because it is the quantum
  generalization of a classical sample.
\end{remark}
To solve the problem stated above using amplitude estimation, we let:
\begin{align*}
  \ket{\psi_G} &:= \frac{1}{\sqrt{\sum_{\vj \in \{0,1\}^n} f(0.\vj) p_{j}}} \sum_{\vj \in \{0,1\}^n} \sqrt{f(0.\vj)} \sqrt{p_{j}} \ket{\vj} \ket{1} \\
  \ket{\psi_B} &:= \frac{1}{\sqrt{\sum_{\vj \in \{0,1\}^n} (1 - f(0.\vj)) p_{j}}} \sum_{\vj \in \{0,1\}^n} \sqrt{1 - f(0.\vj)} \sqrt{p_{j}} \ket{\vj} \ket{0},
\end{align*}
and these are orthogonal states because there is no overlap in the
last qubit. The state preparation circuit $S$ is set to be $U_f P$,
and we have:
\begin{equation}
  \label{eq:ampeststate}
  S \ket{\v{0}}_{n+1} = U_f P \ket{\v{0}} = \sqrt{\sum_{\vj \in \{0,1\}^n} f(0.\vj) p_{j}} \ket{\psi_G} + \sqrt{\sum_{\vj \in \{0,1\}^n} (1 - f(0.\vj)) p_{j}} \ket{\psi_B}.
\end{equation}
With these definitions, we have $\sin^2 \theta = \sum_{\vj \in
  \{0,1\}^n} f(0.\vj) p_{j} = \mathbb{E}[f(X)]$. Thus, estimating
$\theta$ and therefore $\sin^2 \theta$ leads to an estimation of
$\mathbb{E}[f(X)]$. If we are interested in the expected value of $X$,
we can, for example, choose $f(0.\vj) = \frac{j}{2^n-1}$ (recall that
$j \in \{0,\dots,2^n-1\}$, so this normalizes the output), which leads
to estimating $\mathbb{E}[\frac{X}{2^n-1}]$ and hence
$\mathbb{E}[X]$. If we are interested in higher-order moments of $X$,
we could choose $f(0.\vj) = (\frac{j}{2^n-1})^2$ to estimate
$\mathbb{E}[X^2]$, and so on.

Let us discuss the sample complexity of estimating $\mathbb{E}[f(X)]$.
Computing $\mathbb{E}[f(X)]$ with classical Monte Carlo yields a
standard deviation of the estimator that scales with the square root
of the number of samples, by central limit theorem; thus, to obtain an
estimate with error $\pm\epsilon$ with high probability, classically
we use $\bigO{\frac{1}{\epsilon^2}}$ samples and therefore we perform
that many calls to an algorithm that collects the samples (in the
quantum setting, the complexity of that algorithm is the same as the
gate complexity of circuit $P$), followed by evaluation of $f$ on each
sample. On the other hand, by Prop.~\ref{prop:brassardbound}, quantum
amplitude estimation has query complexity $\bigO{\frac{1}{\epsilon}}$,
i.e., it performs that many calls to $U_f$ and $P$. Indeed, it is
sufficient to use $q = \bigO{\log \frac{1}{\epsilon}}$ qubits to store
the output of the phase estimation, leading to $\bigO{2^q} =
\bigO{\frac{1}{\epsilon}}$ applications of $U_f$ and $P$. The
discussion for different choices of the function $f$ is similar.
\begin{remark}
  The statement on the asymptotic behavior in terms of the number of
  calls to $f$ is accurate, but potentially misleading: in order to
  apply the quantum algorithm we need access to $P$ and to a quantum
  oracle for $f$, i.e., to $U_f$, whereas classically we just need
  sampling access to $f$. In other words, classically we only need to
  be able to draw samples from $f(X)$, whereas in the quantum
  algorithm described above we must have access to a circuit that
  prepares the natural quantum encoding of the distribution of $X$
  (i.e., the unitary of Eq.~\eqref{eq:qsampleampest}), and we must be
  able to implement $f$ as a quantum circuit. In theory this is not an
  issue, because any function that can be classically computed can
  also be simulated with a quantum circuit with at most polynomial
  overhead; however, in practice this requires knowing an explicit
  algorithm (that can then be translated into a Boolean circuit) to
  compute $f$. Certain types of function may not be computable by a
  Boolean circuit. For example, if we are trying to estimate how many
  people like the color green more than the color red, a classical
  sampling algorithm may involve asking people on the street what they
  prefer and recording the answer, but there is no quantum oracle that
  can perform the same task.
\end{remark}

We end this section with an example, but we first need to define a
certain gate.
\begin{definition}[$Y$ rotation gate]
  \label{def:yrot}
  The gate $R_Y(\theta)$\index{gate!Y rotation} is defined as the matrix:
  \begin{equation*}
      R_Y(\theta) := \begin{pmatrix} \cos \theta/2 & -\sin \theta/2
        \\ \sin \theta/2 & \cos \theta/2 \end{pmatrix}.
  \end{equation*}
\end{definition}
The factor $\frac{1}{2}$ in the angles appearing in $R_Y(\theta)$ may
look confusing, but this is the convention, and it comes from an
interpretation of this gate as a rotation in a certain geometric
representation of the space of single-qubit quantum states. We discuss
other gates of this form in Ch.~\ref{ch:adiabatic}. (In the language
of matrix exponentials, $R_Y(2\theta) = e^{-i \theta Y}$; for readers
not familiar with this notation yet, this remark can be ignored until
the corresponding discussion in Ch.~\ref{ch:adiabatic}.)

\begin{example}
  Let us look at a toy amplitude estimation example, inspired by
  \cite{woerner2019quantum}. (For this toy problem, all calculations
  could be easily done by hand.) Suppose we are trying to determine
  the expected value of a quantity that takes the value $V_{h}$ with
  probability $p$, and the value $V_{\ell}$ with probability
  $1-p$. This can correspond to a number of situations, e.g.,
  determining the value of an asset whenever there are two possible
  outcomes for a random event that determines its value. This example
  can also be generalized to multiple possible outcomes, but the
  construction of the circuit becomes considerably more involved and
  it would complicate this example.  The expected value that we want
  to estimate is thus:
  \begin{equation*}
    V = (1-p) V_{\ell} + p V_h.
  \end{equation*}
  Let us renormalize so that $V_{\ell} = 0, V_h = 1$. Note that this
  renormalization does not affect the final outcome: if we can
  estimate the expected value in the rescaled range $[0, 1]$, we can
  transform it back to the original range $[V_{\ell}, V_h]$ with a
  linear transformation; the absolute error attained by the estimation
  procedure would have to be rescaled accordingly.

  Because there are two possible scenarios, uncertainty can be
  represented with a single qubit. Furthermore, the amplitude
  coefficients $\sqrt{f(0.\vj)}, \sqrt{1-f(0.\vj)}$ in the state in
  Eq.~\eqref{eq:ampeststate}, onto which amplitude estimation is
  applied, are either 0 or 1, because we normalized the value $f(0) =
  V_{\ell} = 0, f(1) = V_h = 1$. Hence, the target state is:
  \begin{align*}
    S \ket{00} &= \sqrt{1-p} \ket{0} \otimes (\sqrt{1-f(0)} \ket{0} + \sqrt{f(0)} \ket{1}) + \sqrt{p} \ket{1} \otimes (\sqrt{1-f(1)} \ket{0} + \sqrt{f(1)} \ket{1}) \\
    &= \sqrt{1-p} \ket{00} + \sqrt{p} \ket{11}.
  \end{align*}
  We can prepare this state by applying the circuit given in
  Fig.~\ref{fig:stateprepex} onto the state $\ket{00}$. In this
  circuit we use the $Y$ rotation $R_Y$, as defined in
  Def.~\ref{def:yrot}, with angle $\gamma$ set to $\gamma = 2
  \sin^{-1} \sqrt{p}$ to obtain the correct amplitudes.
  \begin{figure}[h!]
    \leavevmode
    \centering
    \ifcompilefigs
    \Qcircuit @C=1em @R=.7em {
       & \gate{R_Y(\gamma)} & \ctrl{1} & \qw \\
       & \qw                & \targ    & \qw \\
    }
    \else
    \includegraphics{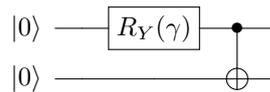}
    \fi
    \caption{State preparation circuit $S$ for the amplitude estimation example.}
    \label{fig:stateprepex}
  \end{figure}
  
  We then need to implement the controlled reflection circuits: C$R$
  (reflection through $\ket{\psi_B}$) and C$F$ (reflection through
  $\ket{\v{0}}$). In this case, we can implement both circuits with a
  doubly-controlled-$Z$ gate. Indeed, because $\ket{\psi_G} = \ket{11}$,
  a C$Z$ gate applies a sign-flip to $\ket{11}$ and implements C$R$;
  and the reflection through $\ket{\v{0}}$, given by
  $2\ketbra{\v{0}}{\v{0}} - I^{\otimes 2}$, can be constructed in the
  way discussed in Sect.~\ref{sec:inversionavg}, i.e., with
  multiply-controlled $Z$ and some $X$ gates. Thus, an application of
  the controlled Grover operator $\text{C}G = (I \otimes S) \text{C}F
  (I \otimes S^{\dag}) \text{C}R$ is given by the circuit in
  Fig.~\ref{fig:groveropex}.
  \begin{figure}[h!]
    \leavevmode
    \centering
    \ifcompilefigs
    \Qcircuit @C=1em @R=.7em {
      \lstick{\text{Control}} & \ctrl{1} & \qw      & \qw                 & \gate{Z} & \ctrl{1} & \qw      & \qw                & \qw      & \qw \\
                              & \ctrl{1} & \ctrl{1} & \gate{R_Y(-\gamma)} & \gate{X} & \ctrl{1} & \gate{X} & \gate{R_Y(\gamma)} & \ctrl{1} & \qw \\
                              & \gate{Z} & \targ    & \qw                 & \gate{X} & \gate{Z} & \gate{X} & \qw                & \targ    & \qw 
      \gategroup{1}{2}{3}{2}{.8em}{.}
      \gategroup{2}{3}{3}{4}{.8em}{.}
      \gategroup{1}{5}{3}{7}{.8em}{.}
      \gategroup{2}{8}{3}{9}{.8em}{.}\\      
    }
    \else
    \includegraphics{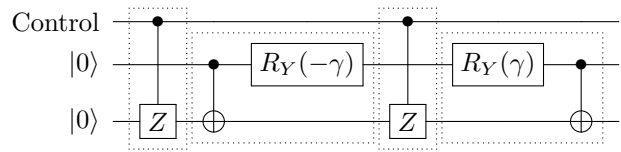}
    \fi
    \caption{Controlled Grover operator $S F S^{\dag} R$. The four boxes represent the circuits C$R$, $S^{\dag}$, C$F$, $S$ (from the left to the right, in this order).}
    \label{fig:groveropex}
  \end{figure}

  We can now construct the full quantum amplitude estimation
  algorithm. If we use three qubits to store the outcome of the
  estimation, the full circuit is given in Fig.~\ref{fig:ampestex}.
  \begin{figure}[h!]
    \leavevmode
    \centering
    \ifcompilefigs
    \Qcircuit @C=1em @R=.7em {
      \lstick{\ket{0}} & \gate{H}           & \qw      & \qw              & \qw                & \ctrl{3}           & \qswap & \qw      & \qw                       & \qw      & \gate{P(-\frac{\pi}{4})} & \gate{P(-\frac{\pi}{2})} & \gate{H} & \meter \\
      \lstick{\ket{0}} & \gate{H}           & \qw      & \qw              & \ctrl{2}           & \qw                &\qw \qwx& \qw      & \gate{P(-\frac{\pi}{2})} & \gate{H} & \qw                       & \ctrl{-1}                 & \qw      & \meter \\
      \lstick{\ket{0}} & \gate{H}           & \qw      & \ctrl{1}         & \qw                & \qw                &\qswap\qwx&  \gate{H} & \ctrl{-1}                & \qw       & \ctrl{-2}                 & \qw                       & \qw      & \meter \\
      \lstick{\ket{0}} & \gate{R_Y(\gamma)} & \ctrl{1} & \multigate{1}{G} & \multigate{1}{G^2} & \multigate{1}{G^4} & \qw    & \qw      & \qw                      & \qw      & \qw    & \qw               & \qw                       &\qw       & \\
      \lstick{\ket{0}} & \qw                & \targ    & \ghost{G}        & \ghost{G^2}        & \ghost{G^4}        & \qw    & \qw      & \qw                      & \qw      & \qw    & \qw               & \qw                       &\qw       &  }
    \else
    \includegraphics{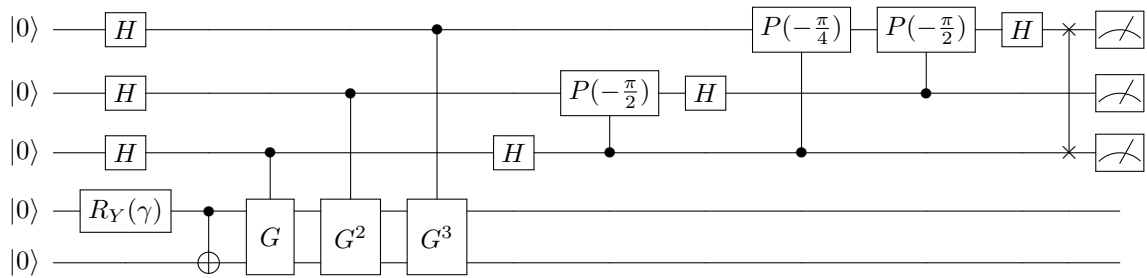}
    \fi
    \caption{Quantum amplitude estimation circuit on two qubits.}
    \label{fig:ampestex}
  \end{figure}
  Running some simulations, for $p = 0.75$ with 2048 repetitions, we
  find that the the distribution of the measurement outcomes is as
  given in Tab.~\ref{tab:ampestex}.
  \begin{table}[htb!]
    \centering
    \begin{tabular}{|l|r|r|}
      \hline
      $\v{b}$ & $\sin^2 \tilde{\theta}$ & Count \\
      \hline
      000 & 0.00000 & 38 \\
      001 & 0.14645 & 44 \\
      010 & 0.50000 & 177 \\
      011 & 0.85355 & 687 \\
      100 & 1.00000 & 104 \\
      101 & 0.85355 & 764 \\
      110 & 0.50000 & 186 \\
      111 & 0.14645 & 48 \\ 
      \hline      
    \end{tabular}
    \caption{Measurement outcomes of the amplitude estimation example;
      we use the notation of Sect.~\ref{sec:amplestsummary}.}
    \label{tab:ampestex}
  \end{table}
  Recall that $\sin^2 \tilde{\theta}$ is the estimate of the
  probability of observing $\ket{\psi_G}$, and therefore it should be
  close to $p$. We find that the most likely outcome corresponds to
  $\approx 0.85$, which is fairly close to the original value $p =
  0.75$. Notice that with only 3 qubits of precision, the
  discretization of the possible output values is coarse, and $0.85$
  is the closest value to the correct answer given this level of
  granularity. \index{algorithm!amplitude estimation|)}\index{amplitude!estimation|)}\index{simulation!Monte Carlo|)}
\end{example}

\subsection{Searching when the number of marked items is not known}
\label{sec:groverunknown}
We now revisit Grover's algorithm and amplitude amplification\index{algorithm!Grover's|(}\index{Grover's algorithm!without knowing \#solutions|(}\index{amplitude!amplification!without knowing \#solutions|(}\index{algorithm!amplitude amplification|(}\index{quantum!search|(} to show
how one can search in time $\bigO{\sqrt{2^n/|M|}}$ even when $|M| =
|\{\v{\ell} \in \{0,1\}^n : f(\v{\ell}) = 1\}|$ is not known. To
achieve this result, we rely on amplitude estimation. In the setting
of Sect.~\ref{sec:ampamp}, assume $\theta \in (0,
\frac{\pi}{2})$. Consider the amplitude estimation circuit in
Fig.~\ref{fig:ampest}; we have seen that the bottom $n$ qubit lines
(second register) are in a superposition of two eigenstates
$\ket{\phi_+}, \ket{\phi_-}$ of the Grover operator. Then we can express the output of the
amplitude estimation circuit in the following way:
\begin{equation}
  \label{eq:ampestfinalstate}
  \frac{i}{\sqrt{2}}\left(-e^{i\theta} \ket{\vartheta_+} \ket{\phi_+} + e^{-i\theta} (\ket{\vartheta_-} \ket{\phi_-} \right) = \frac{1}{\sqrt{2}}\left(\ket{\vartheta_+} \ket{\phi_+} - e^{-i2\theta} (\ket{\vartheta_-}  \ket{\phi_-} \right),
\end{equation}
up to a global phase factor (we can ignore the $i$ in front and
multiply everything by $-e^{-i\theta}$ to obtain the expression on the
r.h.s.), where $\ket{\vartheta_+}, \ket{\vartheta_-}$ are the
$m$-qubit normalized quantum states produced by the amplitude
estimation circuit, and which we analyze in the following. We
claim that as the number $m$ of qubits used for the phase estimation
increases, the two states $\ket{\vartheta_+}, \ket{\vartheta_-}$ get
more and more orthogonal.
\begin{lemma}
  \label{lem:ampestfinalstate}
  For the quantum state in Eq.~\eqref{eq:ampestfinalstate}, with $m$
  qubits for the first register (containing $\ket{\vartheta_+},
  \ket{\vartheta_-}$), it holds that
  $|\braket{\vartheta_-}{\vartheta_+}| = \bigO{\frac{1}{2^m \theta}}$.
\end{lemma}
Before giving a formal proof, we provide an intuitive
argument. Suppose that the angle $\theta$ is exactly representable on
$m$ bits: then $\ket{\vartheta_+}, \ket{\vartheta_-}$ are basis states
corresponding to the binary representation of $\theta, -\theta$, and
they are orthogonal for $\theta \neq 0$ because $\theta \neq -\theta$. However in general
$\theta$ is not representable on $m$ bits (in fact, in this section we
have not specified how we plan to choose $m$ yet). In this case
$\ket{\vartheta_+}, \ket{\vartheta_-}$ are superpositions with
amplitudes that concentrate on some $m$-bit representation of the
angles $\theta, -\theta$, but their overlap may not be zero. If $m$ is
chosen large enough, however, almost all of the weight in the
superposition is on the basis states corresponding to the binary
representation of $\theta, -\theta$, hence $\ket{\vartheta_+},
\ket{\vartheta_-}$ are almost orthogonal. This is the main idea; we
now proceed with the proof.
\begin{proof}
  \allowdisplaybreaks
  Let us write down analytical expressions for $\ket{\vartheta_+}$ and
  $\ket{\vartheta_-}$. For $\ket{\vartheta_+}$ we can assume that the
  bottom $n$ qubit lines in Fig.~\ref{fig:ampest} contain
  $\ket{\phi_+}$, and similarly for $\ket{\vartheta_-}$ we can assume
  that they contain $\ket{\phi_-}$. The circuit in
  Fig.~\ref{fig:ampest} first creates a uniform superposition over $m$
  qubits, then applies phase kickback, finally applies the inverse
  QFT. Carrying out the calculations, and using index $\vk$ to sum
  over the initial uniform superposition, and index $\vj$ for the
  superposition created by the inverse QFT, we obtain:
  \begin{align*}
    \ket{\vartheta_+} = \frac{1}{\sqrt{2^m}} \sum_{\vk \in \{0,1\}^m}  e^{2 i \theta k} \sum_{\vj \in \{0,1\}^m} \frac{1}{\sqrt{2^m}} e^{-2 \pi i j k/2^m} \ket{\vj} = \frac{1}{2^m} \sum_{\vj, \vk \in \{0,1\}^m}  e^{2 i (\theta - \pi j/2^m) k} \ket{\vj} \\
    \ket{\vartheta_-} = \frac{1}{\sqrt{2^m}} \sum_{\vk \in \{0,1\}^m}  e^{-2 i \theta k} \sum_{\vj \in \{0,1\}^m} \frac{1}{\sqrt{2^m}} e^{-2 \pi i j k/2^m} \ket{\vj} = \frac{1}{2^m} \sum_{\vj, \vk \in \{0,1\}^m}  e^{2 i (-\theta - \pi j/2^m) k} \ket{\vj}.
  \end{align*}
  Now we take the inner product of the two states:
  \begin{align}
    \braket{\vartheta_-}{\vartheta_+} &= \frac{1}{4^m} \sum_{\vj \in \{0,1\}^m} \sum_{\vk \in \{0,1\}^m} e^{2 i (\theta - \pi j/2^m) k}  \sum_{\v{\ell} \in \{0,1\}^m}  e^{2 i (\theta +\pi j/2^m) \ell} \notag \\
    &= \frac{1}{4^m} \sum_{\vj \in \{0,1\}^m} \sum_{\vk, \v{\ell} \in \{0,1\}^m} e^{2 i \theta(k+\ell)} e^{2\pi i j/2^m (\ell-k)} \notag \\
    &= \frac{1}{4^m} \sum_{\vk, \v{\ell} \in \{0,1\}^m} \left[ e^{2 i \theta(k+\ell)} \sum_{\vj \in \{0,1\}^m}  e^{2\pi i j/2^m (\ell-k)} \right]. \label{eq:varthetainnerprod}
  \end{align}
  Using the formula for a geometric series, when $\ell \neq k$ we have:
  \begin{equation*}
    \sum_{\vj \in \{0,1\}^m}  e^{2\pi i j/2^m (\ell-k)} = \sum_{j=0}^{2^m-1}  e^{2\pi i j/2^m (\ell-k)} = \frac{1-e^{2\pi i (\ell-k)}}{1-e^{2\pi i (\ell-k)/2^m}} = 0,
  \end{equation*}
  where the last equality is due to the fact that for $\ell \neq k$, $e^{2\pi i (\ell-k)} = 1$. On the other hand, if $\ell = k$ we have:
  \begin{equation*}
    \sum_{\vj \in \{0,1\}^m}  e^{2\pi i j/2^m (\ell-k)} = 2^m,
  \end{equation*}
  because each term of the summation is $1$. Back to
  Eq.~\ref{eq:varthetainnerprod}: using what we just discussed, the
  only nonzero terms in the summation satisfy $\ell = k$, and
  simplifying using the expression for the case $\ell = k$ we have:
  \begin{align*}
    \braket{\vartheta_-}{\vartheta_+} &= \frac{1}{2^m} \sum_{\vk \in \{0,1\}^m} e^{4 i \theta k} = \frac{1}{2^m} \frac{1 - e^{4 i \theta 2^m}}{1 - e^{4 i \theta}}.
  \end{align*}
  Now taking the modulus, we obtain:
  \begin{align*}
    |\braket{\vartheta_-}{\vartheta_+}| &\le \frac{1}{2^m} \frac{2}{|1 - e^{4 i \theta}|} = \frac{1}{2^m} \frac{2}{\sqrt{(1 - \cos 4 \theta)^2 + \sin^2 4 \theta}} = \frac{1}{2^m} \sqrt{\frac{2}{1 - \cos 4 \theta}} = \frac{1}{2^{m} \sin 2\theta} \\
    &= \bigO{\frac{1}{2^m \theta}}.
  \end{align*}
  For the third equality, we used the half-angle identity $\sin
  (\alpha/2) = \pm\sqrt{(1-\cos\alpha)/2}.$
\end{proof}

\noindent This shows that increasing $m$ makes the $\ket{\vartheta_+},
\ket{\vartheta_-}$ more and more orthogonal. In particular, whenever
$2^m \gg \frac{1}{\theta}$, the two states are essentially
orthogonal. Suppose, for the sake of analysis, that the states
$\ket{\vartheta_+}, \ket{\vartheta_-}$ are indeed orthogonal. To
compute the probability of observing $\ket{\psi_G}$ when performing a
measurement on the second register, we switch to the density matrix
formalism. The density matrix associated with the entire system
described in Eq.~\eqref{eq:ampestfinalstate} is:
\begin{align*}
  \frac{1}{2}\left(\ketbra{\vartheta_+}{\vartheta_+} \otimes \ketbra{\phi_+}{\phi_+} 
  -e^{i2\theta}\ketbra{\vartheta_+}{\vartheta_-} \otimes \ketbra{\phi_+}{\phi_-}
  -e^{-i2\theta}\ketbra{\vartheta_-}{\vartheta_+} \otimes \ketbra{\phi_-}{\phi_+}
  +\ketbra{\vartheta_-}{\vartheta_-} \otimes \ketbra{\phi_-}{\phi_-}
  \right),
\end{align*}
and if we trace out the first register, using
$\braket{\vartheta_-}{\vartheta_+} = 0$ and then using the definition
of $\ket{\phi_+}, \ket{\phi_-}$, we obtain:
\begin{align*}
  \frac{1}{2}\left(\ketbra{\phi_+}{\phi_+} + \ketbra{\phi_-}{\phi_-}
  \right) = \frac{1}{2} \left(\ketbra{\psi_G}{\psi_G} -
  \ketbra{\psi_B}{\psi_B}\right).
\end{align*}
From this equation, we see that the probablity of observing
$\ket{\psi_G}$ (rather, one of the basis states constituting
$\ket{\psi_G}$) when performing a measurement on the second register
is $\frac{1}{2}$.

In this analysis, we used the orthogonality of $\ket{\vartheta_+}$ and
$\ket{\vartheta_-}$ to simplify the expression of the state when
tracing out the first register; otherwise, the expression still
contains some cross-terms. If $\ket{\vartheta_+}, \ket{\vartheta_-}$
are not orthogonal, then the probability of observing a basis state
from $\ket{\psi_G}$ is at least $\frac{1}{2} - \bigO{\frac{1}{2^m
    \theta}}$, because in the worst case the probability decreases by
$|\braket{\vartheta_-}{\vartheta_+}|$, and
Lem.~\ref{lem:ampestfinalstate} bounds this value. Hence, repeating
this circuit twice, we obtain a marked item (i.e., a binary string
corresponding to a basis state in $\ket{\psi_G}$, and hence in the set
of all solutions $M$) with probability at least $\frac{3}{4} -
\bigO{\frac{1}{2^m \theta}}$. Relying on this idea,
Alg.~\ref{alg:quantumsearch} determines a marked item by increasing
$m$ in an iterative fashion.
\begin{algorithm2e}[htb]
  \SetAlgoLined
  \LinesNumbered
\KwIn{Unitary $U_f$ to evaluate $f : \{0,1\}^n \to \{0,1\}$.} 
\KwOut{Index $\v{\ell}$ such that $f(\v{\ell}) = 1$, or ``no solution'' if no such $\v{\ell}$ exists.}
\textbf{Initialize}: As in Rem.~\ref{rem:counting}, let $S = H^{\otimes n}$, and let $R$ be the reflection unitary mapping $\ket{\vj} \to (-1)^{f(\vj)}\ket{\vj}$ constructed using $U_f$.\\
Set $m \leftarrow 1$.\\
\While{$m < n$}{
  \For{$i=1,2$}{
    Apply the amplitude estimation circuit (Fig.~\ref{fig:ampest})
    with $m$ qubits of precision. \\
    Measure the second register to obtain a string $\vj \in
    \{0,1\}^n$. \\
    \If{$f(\vj) = 1$}{
      \Return $\vj$.
    }    
  }
  Set $m \leftarrow m+1$.
}
\For{$\vj \in \{0,1\}^n$}{
  \If{$f(\vj) = 1$}{
    \Return $\vj$.
  }
} 
\Return ``no solution''.
\caption{Quantum search algorithm (without knowing the number of marked items).}
\label{alg:quantumsearch}
\end{algorithm2e}
If a marked item is not found before reaching $m = n$, then the
algorithm resorts to full enumeration of the binary strings. 
The intuition is that as soon as $2^m > \theta$, it only takes a few
iterations of the ``while'' loop to have a high probability of
success, and each loop execution uses $\bigO{\frac{1}{\theta}}$
applications of $f$, i.e., of the Grover operator.
\begin{theorem}[Quantum search; \cite{boyer1998tight,brassard2002quantum}]
  If $\theta > 0$, Alg.~\ref{alg:quantumsearch} returns a value $\vj$
  such that $f(\vj) = 1$. The expected number of applications of the
  circuit $U_f$ implementing $f$ is $\bigO{\frac{1}{\theta}}$. If
  $\theta = 0$, the algorithm returns ``no solution'' and uses
  $\bigO{2^n}$ queries to $f$.
\end{theorem}
\begin{proof}
  We give a sketch of the proof. Let $m_0$ be chosen so that
  $1/(2^{m_0} \sin 2\theta) < 1/10$; hence, $m_0 = \bigO{\log
    \frac{1}{\theta}}$. Let us consider the number of applications of
  $U_f$ that the algorithm performs when $m \le m_0$. Because for each
  value of $m$ we use $\bigO{2^m}$ applications of $U_f$, this number
  is:
  \begin{equation*}
    \bigO{ \sum_{k=1}^{m_0} 2^k } = \bigO{ 2^{m_0} } =
    \bigO{\frac{1}{\theta}}.
  \end{equation*}
  Now let us consider the expected number of applications of $U_f$
  that the algorithm performs when $m > m_0$. By our choice of $m_0$,
  each iteration is successful with probability at least $3/4 -
  \bigO{\frac{1}{2^m \theta}} \ge 3/5$. Following a geometric
  distribution where each trial has success probability $p$, the
  probability that we obtain the first success at iteration $k$ (and
  not before) is $p(1-p)^{k-1}$. Thus, the expected number of
  applications is:
  \begin{align*}
    \sum_{k=1}^{n-m_0} \frac{3}{5}\left(1 - \frac{3}{5}\right)^{k-1} 2^{m_0+k} &=
    \sum_{k=1}^{n-m_0} \frac{3}{5}\left(\frac{2}{5}\right)^{-1}\left(\frac{2}{5}\right)^{k} 2^{m_0+k} =
    \frac{3}{2} 2^{m_0} \sum_{k=1}^{n-m_0} \left(\frac{4}{5}\right)^{k} \\
    &= \bigO{2^{m_0}} = \bigO{\frac{1}{\theta}}.
  \end{align*}
  By adding the worst-case number of iterations when $m \le m_0$ and
  the expected number of iterations when $m > m_0$, we obtain the
  bound $\bigO{\frac{1}{\theta}}$ on the total expected number of
  iterations.
\end{proof}

\noindent More details can be found in
\cite{boyer1998tight,brassard2002quantum}.
\begin{corollary}
  \label{cor:ampampsearchunknown}
  Let $U_f$ be a quantum (binary) oracle implementing a Boolean
  function $f : \{0,1\}^n \to \{0,1\}$, and let $M = \{\vj \in
  \{0,1\}^n : f(\vj) = 1\}$ be the set of marked
  elements. Alg.~\ref{alg:quantumsearch} is a randomized quantum
  algorithm that does not require knowledge of $|M|$, and that
  determines an element of $M$ with $\bigO{\sqrt{\frac{2^n}{|M|}}}$
  applications of $U_f$ in expectation.
\end{corollary}

\begin{figure}[h!]
  \leavevmode
  \centering
  \ifcompilefigs
  \Qcircuit @C=1em @R=0.7em {
    & \qw & \gate{H}  & \qw & \qw & \qw & \dots & & \ctrl{4} & \qw & \multigate{3}{Q_{m}^{\dag}} & \qw \\
    &\vdots&        &     &     &     &       & &          &     &     &  \\
    & \qw & \gate{H}  & \qw & \ctrl{2} & \qw & \dots & & \qw & \qw & \ghost{Q_{m}^{\dag}} & \qw \\
    & \qw & \gate{H}  & \ctrl{1} & \qw & \qw & \dots & & \qw & \qw & \ghost{Q_{m}^{\dag}} & \qw \\
    \lstick{\ket{\v{0}}} & {/^n} \qw & \gate{S} & \gate{(S F S^\dag R)^{2^0}} & \gate{(S F S^\dag R)^{2^1}} & \qw & \dots & & \gate{(S F S^\dag R)^{2^{m-1}}} & \qw & {/^n} \qw & \meter
    \inputgroupv{1}{4}{.8em}{2.2em}{\ket{\v{0}}_m} \\
  }
  \else
  \includegraphics{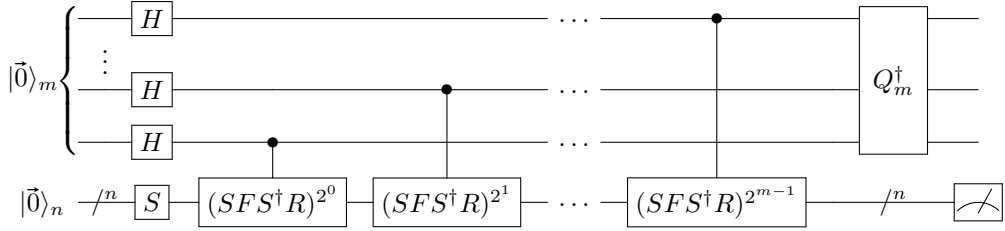}
  \fi
  \caption{Amplitude estimation circuit with $m$ bits of precision, in the context of searching when the number of marked items is unknown.}
  \label{fig:ampestsolcount}
\end{figure}

We can simplify this algorithm with the following observation. Note
that the circuit used by Alg.~\ref{alg:quantumsearch} is the same as
in Fig.~\ref{fig:ampest}, with a variable qubit count $m$, and
measurement gates added to the second register: we thus obtain the
circuit given in Fig.~\ref{fig:ampestsolcount}.
An interesting feature of this circuit is the fact that we only
measure the second register, which contains $\vj$ with $f(\vj) = 1$ as
the answer if the algorithm is successful. Furthermore, recall that
the argument for the correctness of Alg.~\ref{alg:quantumsearch} is
based on the near-orthogonality of $\ket{\vartheta_+},
\ket{\vartheta_-}$, but the same near-orthogonality relationship also
holds before we apply the inverse QFT: trivially,
$|\braket{\vartheta_-}{\vartheta_+}| = |\bra{\vartheta_-}Q_m^{\dag}
Q_m\ket{\vartheta_+}|$. So even if we remove the inverse QFT block,
the probability of observing a basis state from $\ket{\psi_G}$ is
still at least $\frac{1}{2} - \bigO{\frac{1}{2^m \theta}}$. Now
consider the circuit in Fig.~\ref{fig:ampestsolcount} without
$Q^{\dag}_m$. By the principle of implicit measurement
(Prop.~\ref{prop:implicitmeas}), because the first register is discarded
without looking at it, we can assume it is also measured at the end of
the circuit. Because the first register contains a uniform superposition
of all basis states on $m$ qubits, applying a measurement yields a
uniformly random integer in the set $\{0,\dots,2^m-1\}$. That integer
represents the number of times that the Grover operator was applied to
the second register. Thus, we can replicate the effect of the circuit
without the inverse QFT (which still yields the correct result) with
the following, simpler algorithm.
\begin{enumerate}
\item Set $m=1$.
\item Pick a random $y \in \{0,\dots,2^m-1\}$. Repeat twice:
  \begin{itemize}
  \item Compute $(S F S^\dag R)^{y}\ket{\v{0}}$, and apply a
    measurement to obtain a string $\ket{\vj}$. If $f(\vj) = 1$,
    output $\vj$ and stop.
  \end{itemize}
\item If $2^m < 2^n$, increase $m \leftarrow m+1$ and go back to step
  (2). If $2^m > 2^n$, then do full enumeration of the $2^n$ binary
  strings; if $\vj : f(\vj) = 1$ is found, return $\vj$, otherwise
  return ``no solution''.
\end{enumerate}
Historically, this simplified version of the algorithm was the first
to be proposed in \cite{brassard2002quantum}, rather than the approach
based on the amplitude estimation circuit in
Alg.~\ref{alg:quantumsearch}. Because the two approaches are equivalent,
the query complexity is the same.\index{algorithm!Grover's|)}\index{Grover's algorithm!without knowing \#solutions|)}\index{amplitude!amplification!without knowing \#solutions|)}\index{algorithm!amplitude amplification|)}\index{quantum!search|)}


\section{Quantum minimum finding}
\label{sec:quantummin}
Grover's algorithm solves the problem of finding an element satisfying
some easy-to-check property in a set, using a subroutine that
checks if the property holds for a given element. The generalization
of Grover search to the case where the number of marked items is
unknown can be turned into an algorithm for finding the minimum\index{quantum!search|(}\index{minimum finding|(} of an
unstructured function. That algorithm is the subject of this section.

\subsection{Base algorithm}
\label{sec:quantumminbase}
Let $f : \{0,1\}^n \to \Z$, and assume a circuit $U_f$ to evaluate $f$
is given in the usual form:
\begin{equation*}
  U_f: \ket{\vj} \ket{\vk} \to \ket{\vj} \ket{\vk \oplus \vv{f(\vj)}}.
\end{equation*}
Suppose we want to find the minimum of $f$, but we do not know
anything about the function: it may be completely unstructured. Then,
classically we may have to scan all elements in the set $\{0,1\}^n$.
We can do better than that with a quantum computer. The idea is to
``guess'' the value $\gamma$ of the minimum, and then perform repeated
binary search, searching for some $\vj$ such that $f(\vj) \le \gamma$
using amplitude amplification. If such $\vj$ exists, we can update
$\gamma$. If no such $\vj$ exists, then we have found the minimum
(provided we know some value $\v{\ell}$ such that $f(\v{\ell}) =
\gamma$). We can turn this idea into an algorithm, described in
Alg.~\ref{alg:fmin}. In this algorithm we use the notation
$\mathbb{I}(\text{event})$ to denote the indicator function\index{indicator function}\index{notation!indicator@\ensuremath{\mathbb{I}(\cdot)}} of a
certain event; more specifically, we write $\mathbb{I}(f(\vj) <
f(\v{\ell}))$ to denote the function that returns $1$ if $f(\vj) <
f(\v{\ell})$, and $0$ otherwise.
\begingroup
\SetArgSty{textnormal}
\begin{algorithm2e}[htb]
  \SetAlgoLined
  \LinesNumbered
\KwIn{Unitary $U_f$ to evaluate $f$, total number of evaluations $T$.} 
\KwOut{Index $\v{\ell}$ such that $f(\v{\ell}) \le f(\vj)$ for all $\vj$ (if $T$ is chosen appropriately).}
\textbf{Initialize}: Randomly choose $\v{\ell} \in \{0,1\}^n$.\\
\While{the number of evaluations of $U_f$ does not exceed $T$}{
Construct marking unitary $U_m: \ket{\vj}\ket{y} \to \ket{\vj}\ket{y \oplus \mathbb{I}(f(\vj) < f(\v{\ell}))}$.\\
Apply the search algorithm in Alg.~\ref{alg:quantumsearch} using the marking unitary $U_m$.\\
Let $\vk$ be the index returned by the search algorithm. If $\vk : f(\vk) < f(\v{\ell})$, set $\v{\ell} \leftarrow \vk$.
}
\Return $\v{\ell}$.
\caption{Quantum minimum finding algorithm.}
\label{alg:fmin}
\end{algorithm2e}
\endgroup

\begin{theorem}[Quantum minimum finding; \cite{durr1996quantum}]
  \label{thm:fmin}
  Let $f : \{0,1\}^n \to \Z$, let $U_f$ be a circuit that evaluates
  $f$ in binary, and let $\delta > 0$. Using Alg.~\ref{alg:fmin}, we
  can determine the global minimum of $f$ with probability at least
  $1-\delta$ using $\bigO{\sqrt{2^n} \log \frac{1}{\delta}}$
  applications of $U_f$ in total, and $\bigOt{\sqrt{2^n}}$ additional
  gates.
\end{theorem}
\begin{proof}
  We call \emph{rank} of an element of $\{0,1\}^n$, denoted
  $\text{rank}(\vj)$, its position in the set $\{0,1\}^n$ ordered by
  nondecreasing value of $f$; our goal is to show that we can
  determine the element of rank 1, i.e., the global minimum.

  The proof consists of three steps. We first consider the case where
  Alg.~\ref{alg:fmin} is executed with $T = \infty$, which we call the
  \emph{infinite-time} algorithm, and analyze the probability that the
  index of the rank-$r$ element is returned. Then we use that
  probability to compute the expected running time of the
  infinite-time algorithm before it returns the element of rank
  1. Finally, we apply Markov's inequality and show the desired
  result.

  The main loop of Alg.~\ref{alg:fmin} consists of running the search
  algorithm of Cor.~\ref{cor:ampampsearchunknown}
  (Alg.~\ref{alg:quantumsearch}) for a fixed $\v{\ell}$; therefore,
  the search algorithm is executed to determine an element of the set
  of marked items $M := \{\vj \in \{0,1\}^n : f(\vj) <
  f(\v{\ell})\}$. Denote $|M| = t$. Let us call $p(r, t)$ the
  probability that the element of rank $r$ is obtained as $\vk$ on
  line 5 when running the infinite-time algorithm with a set of marked
  items of size $t$. We claim that $p(r, t) = 1/r$ if $r \le t$, and
  $p(r, t) = 0$ otherwise. The case $r > t$ is obvious because no
  element of rank $r$ exists. Then for each fixed $r$ we perform
  induction on $t = r, r+1, \dots, 2^n$. If $t=r$, we have $p(r, r) =
  1/r$ because the search algorithm returns an element of $M$ chosen
  uniformly at random, so we have $1/|M|=1/r$ chance of observing the
  element of rank $r$. For any $t > r$, we can express $p(r, t)$ as
  the sum of two terms: the probability that the index $\vk$ on line 5
  is the element of rank $r$, and the probability that $\vk$ it is not
  the element of rank $r$ but it becomes so in a subsequent iteration,
  i.e.,
  \begin{align*}
    p(r, t) = &\Pr(\text{rank}(\vk) = r)  + \\
    &\sum_{\substack{s=1\\s \neq r}}^{t} \Pr(\text{element of rank } r \text{ is chosen subsequently}|\text{rank}(\vk) = s) \Pr(\text{rank}(\vk)= s).
  \end{align*}
  Because the index $\vk$ is chosen uniformly at random from
  $M$, $\Pr(\text{rank}(\vk) = s) = 1/|M| = 1/t$ for any
  $t$. Furthermore, by definition $\Pr(\text{element of rank } r
  \text{ is chosen subsequently}|\text{rank}(\vk) = s) = p(r,
  s-1)$. Using the induction hypothesis, we know $p(r, s-1) = 0$ if $s
  \le r$, and $p(r, s-1) = 1/r$ if $r < s \le t-1$. Thus:
  \begin{align*}
    p(r, t) &= \frac{1}{t}  + \sum_{s=r+1}^{t} p(r, s-1) \frac{1}{t} 
    = \frac{1}{t}  + \frac{1}{t} \sum_{s=r+1}^{t} \frac{1}{r}
    = \frac{1}{t}  + \frac{1}{t}  \frac{t-r}{r} = \frac{1}{r}.
  \end{align*}

  We now turn to computing the expected running time (in terms of
  number of calls to $U_f$) of the infinite-time algorithm before
  $\v{\ell}$ contains the index of the global minimum. Note that once
  $\text{rank}(\v{\ell}) = 1$, the index $\v{\ell}$ no longer
  changes during the course of the algorithm. By
  Cor.~\ref{cor:ampampsearchunknown}, the number of applications of
  the marking unitary $U_m$ to find the index of a marked item among
  $2^n$ items, where $t$ items are marked, is $\bigO{\sqrt{2^n/t}}$ in
  expectation. Let $c$ be the constant of the $\bigO{\cdot}$
  expression, i.e., Alg.~\ref{alg:quantumsearch} uses $\le c
  \sqrt{2^n/t}$ applications of the marking unitary.  (To be more
  concrete, \cite{boyer1998tight} gives a slightly different quantum
  search algorithm for which $c = \frac{9}{2} \sqrt{2^n/t}$; so we
  can take $c = \frac{9}{2}$ below.) Then we upper bound the total
  expected running time in the following way:
  \begin{equation*}
    \sum_{r=1}^{2^n} \Pr(\text{rank}(\v{\ell})= r \text{ at some iteration}) (\text{Expected running time to find a better element than rank } r).
  \end{equation*}
  This is an upper bound because the search at line 4 only looks for
  better elements, therefore $\text{rank}(\v{\ell})=r$ can only occur
  once in the course of the algorithm. We can expand this expression
  as follows: when $r=1$ we are done with a single application of
  $U_m$, and otherwise, the formula evaluates to at most:
  \begin{align}
    \label{eq:fminiterbound}
    \begin{split}
    \sum_{r=2}^{2^n} p(r, 2^n) c \sqrt{\frac{2^n}{r-1}} &=
    c \sqrt{2^n} \sum_{r=2}^{2^n} \frac{1}{r} \frac{1}{\sqrt{r-1}}
    = c \sqrt{2^n} \left(\frac{1}{2} +  \sum_{r=2}^{2^n-1} \frac{1}{r+1} \frac{1}{\sqrt{r}}\right) \\
    &\le c \sqrt{2^n} \left(\frac{1}{2} +  \sum_{r=2}^{2^n-1} r^{-3/2}\right) \le c \sqrt{2^n} \left(\frac{1}{2} +  \int_{r=1}^{2^n-1} r^{-3/2}\right)\\
    &\le c \sqrt{2^n} \left(\frac{1}{2} +  \left(\left. -2r^{-1/2}\right|_1^{2^n-1} \right)\right) \le c \sqrt{2^n} \left(\frac{1}{2} + 2\right) \le 3c \sqrt{2^n}.
    \end{split}
  \end{align}
  Each application of the marking unitary $U_m$ can be implemented
  with a single call to $U_f$ plus some binary arithmetic
  operations. Thus, the expected number of calls of the infinite-time
  algorithm before $\v{\ell}$ contains the index of the global minimum
  is $\bigO{\sqrt{2^n}}$. The number of gates also follows from
  Cor.~\ref{cor:ampampsearchunknown} and
  Sect.~\ref{sec:groverunknown}.

  From this bound on the expected number of iterations of the
  infinite-time algorithm to obtain the global minimum, we can finish
  the proof using standard tools.  Let $X$ be the random variable
  corresponding to the number of applications of $U_f$ before
  Alg.~\ref{alg:fmin} finds the global minimum. Let $\bar{t} = 3c
  \sqrt{2^n} \ge \mathbb{E}[X]$. By Markov's inequality, if we run
  Alg.~\ref{alg:fmin} with $T = 3 \bar{t}$, the probability that we do
  not find the minimum is at most:
  \begin{equation*}
    \Pr(X \ge 3 \bar{t}) \le \frac{\mathbb{E}[X]}{3 \bar{t}} \le
    \frac{1}{3}.
  \end{equation*}
  We execute Alg.~\ref{alg:fmin} $k$ times in total, setting $T = 3
  \bar{t}$ each time, and take the index $\v{b}$ of the best element
  returned among the $k$ executions as the global minimum. The
  probability that $\v{b}$ is not the global minimum is at most
  $(1/3)^k$. Setting $k = \ceil{\log_3 \frac{1}{\delta}} = \bigO{\log
    \frac{1}{\delta}}$ ensures the success of the algorithm with
  probability at least $1-\delta$, and concludes the proof.
\end{proof}

\subsection{Function evaluations with errors}
\label{sec:quantumminerrors}
We can now address the more general case where we want to find an
approximate minimizer of a function that cannot be evaluated
exactly. In the previous section $f$ could be evaluated exactly and in
binary. But we may not always be so lucky: for example, it may be the
case that evaluating $f$ requires acting on quantities obtained from
an amplitude/probability estimation procedure, and such an estimation
incurs a probability of error. It may be helpful to distinguish the
type of errors that are easy to deal with, and those that require a
more careful treatment. We proceed in increasing order of
difficulty. Here we only discuss cases where positive results are
possible; in other settings quantum speedups are negated by noise, see
the notes in Sect.~\ref{sec:ampampnotes} at the end of this chapter,
therefore it is important to pay attention to the details of the error
model.

\paragraph{Deterministic noisy evaluation.}
In this case $f$ is evaluated through a unitary $U_f$ that
acts as:
\begin{equation*}
  U_f: \ket{\vj} \ket{\vk} \to \ket{\vj} \ket{\vk \oplus \vv{f(\vj) +
      \epsilon_j}},
\end{equation*}
where $\epsilon_j$ is some error that may depend on $\vj$. An example
where such a situation may occur is when $f$ is a trigonometric
function evaluated on a binary string: the exact value of $f(\vj)$ may
not be computable in finite precision, so $U_f$ computes a
finite-precision approximation whose error depends on $\vj$. Let
$\epsilon_{\max}$ be an upper bound for all errors: $|\epsilon_j| \le
\epsilon_{\max}| \; \forall j$. Under this assumption, in each
execution of the ``while'' loop in Alg.~\ref{alg:fmin} the marking
unitary $U_m$ is always consistent: each state $\ket{\vj}$ is
entangled with a specific binary string $\ket{\vv{f(\vj) +
    \epsilon_j}}$, and the state is marked or not marked depending if
$f(\vj) + \epsilon_j \le f(\v{\ell})$. We might miss the true global
minimum of the function because of the error terms $\epsilon_j$, but
it is straightforward to conclude that Alg.~\ref{alg:fmin} with high
probability determines a value that is at most $2 \epsilon_{\max}$
away: the proof of Thm.~\ref{thm:fmin} applies directly.

\paragraph{Nondeterministic evaluation with exogenous randomness.}
In this case $f$ is evaluated through a unitary $U_f$ that
acts as:
\begin{equation*}
  U_f: \ket{\vj} \ket{\vk} \to \ket{\vj} \ket{\vk \oplus \vv{f(\vj) +
      \epsilon_j}},
\end{equation*}
where $\epsilon_j$ is an exogenous random variable. Conceptually, we
can think of this situation as having a separate ``random seed''
register, and the values of $\epsilon_j$ are determined once the
random seed is fixed, but we do not know the value of the random seed
a priori. An example where such a situation may occur is when $U_f$
performs Monte Carlo estimation of a difficult-to-compute function
(e.g., a complicated integral): the output of the computation can be
different in every execution, and depends on a random seed. We can
think of this as ``classical randomness.'' Applying
Alg.~\ref{alg:fmin} directly may not work here: in different
executions of the ``while'' loop the samples from the random variables
$\epsilon_j$ may be different, so it is theoretically possible that
the marking unitary $U_m$ accepts more values $\vj$ than we
anticipated (recall that in principle we only want to mark $\vj :
f(\vj) < f(\v{\ell})$). The proof of Thm.~\ref{thm:fmin} explicitly
relies on the assumption that if the element of rank $r$ is the
incumbent, we can find a better element with expected running time
$\bigO{\sqrt{\frac{2^n}{r-1}}}$; the exogenous randomness model might
violate that assumption, because the marking unitary is affected by
randomness that is not directly under our control (i.e., it depends on
an external random seed) and therefore it may mark (i.e., accept) more
than $r-1$ binary strings.

It is however not difficult to recover the same running time as
Thm.~\ref{thm:fmin} with some slightly weaker guarantees. There are
multiple ways to do so. If we have some information on the
distribution of the errors $\epsilon_j$, we can exploit it to our
advantage. A weaker but simpler approach can be employed if we know
that $|\epsilon_j| \le \epsilon_{\max} \; \forall j$. We change the
marking unitary on line 3 of the ``while'' loop of Alg.~\ref{alg:fmin}
to:
\begin{equation*}
  U_m: \ket{\vj}\ket{y} \to \ket{\vj}\ket{y \oplus \mathbb{I}(f(\vj) < f(\v{\ell}) - 2 \epsilon_{\max})}
\end{equation*}
and the acceptance criterion on line 5 to $f(\vk) < f(\v{\ell}) - 2
\epsilon_{\max}$. Because $|\epsilon_j| \le \epsilon_{\max}$, these
modifications ensure that we only accept elements with a function
value guaranteed to be better than the function value of the incumbent
$\v{\ell}$. Thus, in the proof of Thm.~\ref{thm:fmin} we can rely on
$p(r, t) = 1/r$ being an upper bound on the true probability of
selecting the element of rank $r$. Furthermore,
Eq.~\eqref{eq:fminiterbound} is still a valid upper bound, because the
more stringent acceptance criterion $f(\vk) < f(\v{\ell}) - 2
\epsilon_{\max}$ may only result in skipping some of the terms of the
summation, thereby decreasing the value of the upper
bound. Formally, we can upper bound the expected running time as:
\begin{align*}
  \sum_{\substack{r:\text{rank}(\v{\ell}) = r\\\text{at some}\\\text{iteration}}} \Pr(\text{rank}(\v{\ell}) = r \text{ at some iteration}) c \sqrt{\frac{2^n}{r-1}} &\le
  \sum_{r=2}^{2^n} \Pr(\text{rank}(\v{\ell}) = r \text{ at some iteration}) c \sqrt{\frac{2^n}{r-1}} \\
  &\le
  \sum_{r=2}^{2^n} p(r, 2^n) c \sqrt{\frac{2^n}{r-1}},
\end{align*}
and the chain of inequalities continues in
Eq.~\eqref{eq:fminiterbound}. This shows that, similarly to
Thm.~\ref{thm:fmin}, with $\bigO{\sqrt{2^n} \log \frac{1}{\delta}}$ we
can determine $\v{\ell}$ such that $f(\v{\ell}) \le f_{\min} + 2
\epsilon_{\max}$, where $f_{\min}$ is the value of the global minimum
of $f$.

\paragraph{Nondeterministic evaluation with endogenous randomness.}
In this case $f$ is evaluated through a unitary $U_f$ that
acts as:
\begin{equation*}
  U_f: \ket{\vj} \ket{\v{0}} \to \ket{\vj} \left( \sqrt{p_j} \ket{\vv{f(\vj)}} + \sum_{\vk \neq \vv{f(\vj)}} \alpha_{j,k} \ket{\vk} \right)
\end{equation*}
where $\sum_{\vk \neq \vv{f(\vj)}} |\alpha_{j,k}|^2 = 1-p_j$. The
interpretation of this model is the following: the unitary $U_f$ uses
a register, initialized in the state $\ket{\v{0}}$, to output the
correct value of $f$ with some probability $p_j$, and with the
complementary probability it outputs some other value. An example
where such a situation may occur is mentioned at
the beginning of this section, where evaluating $f$ requires some
amplitude estimation or phase estimation procedure: in that case the
output of the final QFT is a superposition of basis states, one of
which leads to the ``correct'' function value; all other basis states
lead to a potentially erroneous computation, and they may also appear
with a nonzero but potentially negligible amplitude.

A more detailed example should help clarify the difficulties
encountered in this setting.
\begin{example}
  \label{ex:fminerror}
  Consider the case in which we have a state $\sum_{\vj \{0,1\}^n}
  \alpha_j \ket{\vj}$, we know that all $\alpha_j$ are real and
  nonnegative up to some global phase, and we want to return the index
  $j$ such that $|\alpha_j|$ is minimized. One way to do so is to
  use amplitude estimation followed by quantum minimum finding. For
  concreteness, let $\ket{\psi} = 0.6 \ket{0} + 0.8 \ket{1}$. Below we
  perform several approximations and simplifications for the sake of
  the example: an actual implementation could yield very different
  results. The setup is the following: we use three registers, the
  first one to store the index $j$ over which we search (in this
  case, $j \in \{0,1\}$, so it is in fact a single-digit binary
  string), the second to store $\ket{\psi}$, and the third to contain
  the output of amplitude estimation.

  Suppose we use 4 qubits for the third register. Recall that for an
  amplitude $\sin \theta$, the phase estimation procedure at the end
  of amplitude estimation outputs $\pm \theta/\pi$; the ``ideal''
  4-digit output of amplitude estimation for each of the two
  amplitudes $0.6, 0.8$ is then:
  \begin{align*}
    \ket{0011} = 0.1875 \text{ in decimal } & (\sin 0.1875 \pi = 0.5557 \approx 0.6) \\
    \ket{0101} = 0.3125 \text{ in decimal } & (\sin 0.3125 \pi = 0.8314 \approx 0.8).
  \end{align*}
  We consider these the ``correct'' function values $f(0), f(1)$.
  For the sake of this example, assume that amplitude estimation
  outputs the ideal number with probability $0.81 = (0.9)^2$, and the
  ideal number $\pm 1/16$ with probability $0.095 \approx (0.308)^2$
  each. (In reality the output distribution might have quite a
  different shape than what we assumed.) Then the amplitude estimation
  circuit (where, conditioned on some value $\ket{j}$ in the first
  single-qubit register, we estimate the amplitude of $\ket{j}$ in
  the second single-qubit register, and write the answer in the third
  register) would perform the following mapping:
  \begin{align*}
    \ket{0} (0.6 \ket{0} + 0.8 \ket{1}) \ket{0000} &\to \ket{0}(0.6 \ket{0} + 0.8 \ket{1})(0.308 \ket{0010} + 0.9 \ket{0011} + 0.308 \ket{0100}) \\
    \ket{1} (0.6 \ket{0} + 0.8 \ket{1}) \ket{0000} &\to \ket{1}(0.6 \ket{0} + 0.8 \ket{1})(0.308 \ket{0100} + 0.9 \ket{0101} + 0.308 \ket{0110}).
  \end{align*}

  Let us now apply Alg.~\ref{alg:fmin}, where in the initialization
  phase we randomly choose $\v{\ell} = 1$. The search on line 4 of the
  ``while'' loop tries to determine the index of an element with
  function value better than $\v{\ell}$. An issue immediately arises:
  what is the function value associated with the incumbent? For the
  incumbent $\v{\ell} = 1$, the function value is the superposition
  $(0.308 \ket{0100} + 0.9 \ket{0101} + 0.308 \ket{0110})$. And how do
  we compare the function value $f(1)$ with the function value $f(0)$?
  The output of amplitude estimation for index $j = 0$ is not
  deterministically strictly smaller than the output of amplitude
  estimation for index $j = 1$! Assume that we implement the test
  $f(0) < f(1)$ by using three register: one to store the output of
  the procedure that attempts to compute $f(0)$, one for the procedure
  that attempts to compute $f(1)$, and one to store the outcome of the
  comparison. The output of amplitude estimation for the index $j =
  0$, is $(0.308 \ket{0010} + 0.9 \ket{0011} + 0.308 \ket{0100})$. For
  the index $j = 1$, it is $(0.308 \ket{0100} + 0.9 \ket{0101} + 0.308
  \ket{0110})$. If we compare the binary values stored in the
  superpositions for the two indices, by doing pairwise comparisons,
  we see that with probability $(0.308)^2 \cdot (0.308)^2 \approx
  0.009$ the comparison yields ``$f(0)$ is larger than or equal to
  $f(1)$'', whereas, for correctness of the algorithm, we wanted
  ``$f(0)$ is strictly smaller than $f(1)$''.
\end{example}

Ex.~\ref{ex:fminerror} describes a situation where it is not
immediately apparent how to execute Alg.~\ref{alg:fmin}: to proceed
with the algorithm we must be able to compare two function values and
determine which one is smaller, but if the function values are
produced in a superposition, the comparison is not deterministic. The
result of the comparison depends on the outcome of a measurement of
the registers involved: this is what we mean by ``endogenous
randomness'' of the evaluation. (Sometimes this setting is called
``relational oracles'' in the literature, see
\cite{apeldoorn2020convex}.)\index{oracle!relational}

Without loss of generality, we can simplify the exposition by
considering the following related search problem: given a function $f
: \{0,1\}^n \to \{0,1\}$, we want to determine some $\v{\ell} \in
\{0,1\}^n : f(\v{\ell}) = 1$, while having access only to a
nondeterministic version $\tilde{f}$ of $f$ satisfying the following:
\begin{equation*}
  \text{if } f(\vj) = 1 \text{ then } \Pr(\tilde{f}(\vj) = 1) \ge 9/10, \qquad \text{if } f(\vj) = 0 \text{ then } \Pr(\tilde{f}(\vj) = 0) \ge 9/10.
\end{equation*}
In other words, we cannot access $f$ directly, but we have access to a
``noisy'' function that outputs the correct function value at least
90\% of the time.\index{oracle!bounded error}
\begin{remark}
  The threshold value 9/10 is chosen arbitrarily: as long as it is $>
  1/2$, we can always boost it it with a few repetitions.
\end{remark}
This setting is a direct generalization of the quantum search problem:
now the function $f$ is not correct all the time, but we have a
bound on the failure probability; the corresponding problem is usually
called \emph{search with bounded error probability}
\cite{hoyer2003quantum}. Clearly, if we can solve search with bounded
error probability, then we can generalize Alg.~\ref{alg:fmin} to the
same setting, in which function values are computed correctly with
bounded error probability: the crucial component of
Alg.~\ref{alg:fmin} is precisely the application of quantum search on
line 4.  In the setting of nondeterministic evaluation with endogenous
randomness, the marking unitary can be considered as the noisy
function $\tilde{f}$ that outputs the correct value with reasonably
large (e.g., $> 9/10$) probability.

The quantum search algorithm in Alg.~\ref{alg:quantumsearch} does not
directly work for the bounded error case, because errors could
accumulate too quickly. There is a very simple approach to design a
search algorithm that works in this setting, at the expense of
additional (but polynomial) query complexity. Let $m$ be the number of
queries to (exact) $f$ that the quantum search algorithm would have to
apply to find a solution $\v{\ell}$; by
Cor.~\ref{cor:ampampsearchunknown}, $m =
\bigO{\sqrt{\frac{2^n}{|M|}}}$. Suppose the failure probability of
$\tilde{f}$ could be reduced to:
\begin{equation*}
  \text{if } f(\vj) = 1 \text{ then } \Pr(\tilde{f}(\vj) = 1) \ge 1 - \frac{1}{100 m}, \qquad \text{if } f(\vj) = 0 \text{ then } \Pr(\tilde{f}(\vj) = 0) \ge 1 - \frac{1}{100 m},
\end{equation*}
and apply quantum search (Alg.~\ref{alg:quantumsearch}) using
$\tilde{f}$. Because each call to $\tilde{f}$ differs from a call $f$
with probability only $\frac{1}{100 m}$, and errors in quantum
computation accumulate linearly (see Sect.~\ref{sec:vardistance} and
in particular Prop.~\ref{prop:unitaryerror}), we can bound the
difference between the quantum state produced by the search algorithm
using $\tilde{f}$ and the quantum state produced by the search
algorithm using $f$ as:
\begin{equation*}
  (\text{number of calls to } \tilde{f})(\text{failure probability of } \tilde{f}) \le m \frac{1}{100 m} \le \frac{1}{100}
\end{equation*}
in the Euclidean norm. This implies (Prop.~\ref{prop:euclideantotvd})
that the quantum search algorithm using $\tilde{f}$ succeeds with
probability at most $\frac{1}{100}$ worse than the success probability
of quantum search using $f$. Thus, if we can reduce the failure
probability of $\tilde{f}$ to $\frac{c}{100 \sqrt{2^n}}$, where $c$ is
the constant in the $\bigO{\cdot}$ for quantum search, applying
quantum search substituting $\tilde{f}$ for $f$ obtains the correct
answer with high probability.

To boost the probability of success, using some extra queries to
$\tilde{f}$, we can do the following: we construct a function
$\tilde{f}_{\text{maj}}$ that makes $k=\bigO{n}$ queries to
$\tilde{f}$, stores the output in separate working registers, takes
the majority vote of the outputs, and finally uncomputes the working
registers. More formally, the unitary $U_{\tilde{f}_{\text{maj}}}$,
when applied onto the basis state $\ket{\vj}$ and several fresh
registers, implements the following steps:
\begin{align*}
  U_{\tilde{f}_{\text{maj}}} \ket{\vj} \underbrace{\ket{0} \dots \ket{0}}_{k \text{ times}} \ket{0} &\to \ket{\vj} \ket{\tilde{f}_1(\vj)} \dots \ket{\tilde{f}_k(\vj)} \ket{0} \\
  &\to \ket{\vj} \ket{\tilde{f}_1(\vj)} \dots \ket{\tilde{f}_k(\vj)} \ket{\text{MAJ}(\tilde{f}_1(\vj),\dots,\tilde{f}_k(\vj))} \\
  &\xrightarrow{\text{uncompute}} \ket{\vj} \underbrace{\ket{0} \dots \ket{0}}_{k \text{ times}} \ket{\text{MAJ}(\tilde{f}_1(\vj),\dots,\tilde{f}_k(\vj))},
\end{align*}
where $\tilde{f}_1(\vj),\dots,\tilde{f}_k(\vj)$ denote $k$ different
evaluations of $\tilde{f}$, and the function $\text{MAJ}$ takes the
majority vote. The uncomputation process cleans up the working
registers to a large degree, see Sect.~\ref{sec:uncompute} for a
discussion on uncomputation.
\begin{remark}
  In fact, the uncomputation may not be perfect: the issue stems from
  the fact that after computing
  $\tilde{f}_1(\vj),\dots,\tilde{f}_k(\vj)$ and
  $\text{MAJ}(\tilde{f}_1(\vj),\dots,\tilde{f}_k(\vj))$, we want to
  copy $\text{MAJ}(\tilde{f}_1(\vj),\dots,\tilde{f}_k(\vj))$ to a safe
  register before uncomputation. However, the register containing
  $\text{MAJ}(\tilde{f}_1(\vj),\dots,\tilde{f}_k(\vj))$ in general
  could be a superposition of $\ket{0}$ and $\ket{1}$ rather than
  containing only the correct value. This happens because with a small
  probability, not enough values among
  $\tilde{f}_1(\vj),\dots,\tilde{f}_k(\vj)$ agree with the correct
  result. In this case, carefully writing down the effect of
  uncomputation shows that with small probability we do not uncompute
  the working registers in a ``clean'' way. Fortunately, the state
  that we obtain after the imperfect uncomputation is still very close
  to the perfect uncomputation state: its overlap with the ideal state
  is the same as the probability of success for computing
  $\text{MAJ}(\tilde{f}_1(\vj),\dots,\tilde{f}_k(\vj))$, which is
  shown in the following to be very large. Thus, we can neglect this
  issue for ease of exposition, and it only affects the success
  probability of the algorithm by a very small amount.
\end{remark}
Next, we show that the function $\tilde{f}_{\text{maj}}$ implemented
in this way has very low failure probability --- below the necessary
threshold. We analyze the failure probability of
$\tilde{f}_{\text{maj}}$ using standard arguments: let $k$ be the
number of queries to $\tilde{f}$ for the majority
vote. $\tilde{f}_{\text{maj}}$ outputs the correct answer if at least
$k/2$ queries to $\tilde{f}$ give the correct answer. Let
$X_1,\dots,X_k$ be Bernoulli random variables that take value $1$ if
the corresponding query to $\tilde{f}$ gives the correct answer, an
event that happens with probability at least $9/10$ by assumption. Let
$X = \sum_{j=1}^k X_j$. Using the multiplicative Chernoff bound, the
probability that fewer than $k/2$ queries to $\tilde{f}$ give the
correct answer can be bounded above as follows:
\begin{align*}
  \Pr\left( X \le \frac{k}{2}\right) &= \Pr\left( X \le \left(1 - \frac{4}{9}\right) \frac{9k}{10}\right) = \Pr\left( X \le \left(1 - \frac{4}{9}\right) \mathbb{E}[X] \right) \le e^{- \frac{16}{162} \mathbb{E}[X]}.
\end{align*}
We want $e^{- \frac{16}{162} \mathbb{E}[X]} \le e^{- \frac{16}{162}
  \frac{9}{10}k} \le \frac{c}{100\sqrt{2^n}}$, and taking the natural
logarithm on both sides, we find that $k = \bigO{n}$ is sufficient to
make this inequality hold. Summarizing, if we are willing to perform
$\bigO{n}$ queries to $\tilde{f}$ to simulate one almost-exact query
to $f$, we can employ Alg.~\ref{alg:fmin} using the almost-exact query
and no further modifications; this brings the total query complexity
of the algorithm to $\bigO{n\sqrt{2^n} \log \frac{1}{\delta}}$.

With some ingenuity it is possible to reduce the query complexity to
$\bigO{\sqrt{2^n} \log \frac{1}{\delta}}$, as in
Thm.~\ref{thm:fmin}. We describe an idea introduced in
\cite{hoyer2003quantum} in the context of quantum search: as remarked
earlier, if we can perform quantum search the extension to quantum
minimum finding is straightforward. We execute quantum search by
interleaving iterations of amplitude amplification and error
reduction. We start with one iteration of amplitude amplification step
to amplify all quantum states $\ket{\vj}$ such that $\tilde{f}(\vj) =
1$: this includes states for which $f(\vj) = 1$, but also ``false
positives'', i.e., branches of the computation where $f(\vj) = 0$ and
$\tilde{f}$ outputs an incorrect value. Then we run an error reduction
step: for all $\vj : \tilde{f}(\vj) = 1$ in the previous step we
perform $k$ evaluations of $\tilde{f}$, take a majority vote, and use
the outcome of the majority vote to reduce the probability of
observing a false positive to $\bigO{2^{-k}}$. At this point we go
back to the amplitude amplification step and iterate. Note that as we
do so, we need to add new registers as working registers to store the
outcome of the majority votes. The details of this idea can be found
in \cite{hoyer2003quantum}, with a detailed proof showing that when
$f$ has bounded error probability, the asymptotic complexity of
quantum search stays the same, although the algorithm gets more
involved and the constants in $\bigO{\cdot}$ notation get worse.\index{quantum!search|)}\index{minimum finding|)}

\section{Notes and further reading}
\label{sec:ampampnotes}
Even before Grover presented his algorithm for unstructured quantum
search with a quadratic speedup over classical algorithms, it was
known that at a quadratic speedup is optimal for unstructured search,
i.e., relative to an oracle that identifies the optimal solution
\cite{bennett1997strengths}.

Amplitude amplification is a fundamental component of most of the
optimization algorithms discussed in subsequent chapters, if only as a
way to boost the probability of success of the algorithms. Among the
direct applications of Grover's unstructured search algorithm to
optimization, one of the most notable is the acceleration of the
solution of certain types of dynamic programming problems,\index{dynamic programming} discussed
in \cite{ambainis2019quantum}. The main feature of these dynamic
programs is that they are defined by a recursion across subsets: to
determine the optimal decision over a set of given cardinality, one
must loop over all of its subsets, potentially with some cardinality
constraints. \cite{ambainis2019quantum} initializes the dynamic
programming recursion with some classical computation, then uses
Grover's algorithm to fill out the rest of the dynamic programming
table by looping over all the possible subsets. The classical running
time $\bigOt{2^n}$ for doing so gets reduced to an exponential with a
smaller base. Notably, for the Bellman-Held-Karp dynamic programming
formulation of the traveling salesman problem,
\cite{ambainis2019quantum} reduces the classical $\bigOt{2^n}$ running
time to quantum $\bigOt{1.728^n}$ running
time. \cite{grange2023quantum} uses this framework to give quantum
speedups for (exponential-time) single-machine job scheduling problem
solved by dynamic programming across subsets. It is important to
remark that this line of work requires QRAM (quantum RAM, see
Sect.~\ref{sec:qram}) to achieve a quantum speedup, as the values used
to initialize the dynamic programming table, on which the recursion is
built, are assumed to be available via a constant-time oracle in
superposition, and this can be done with QRAM.

As highlighted multiple times throughout the chapter, one of the
potential issues of amplitude amplification is that if the number of
iterations exceeds the optimal number, the probability of success of
the algorithm starts decreasing. In addition to the ideas of
estimating the probability of success, or using the quantum search
algorithm without knowing the probability of success of
Sect.~\ref{sec:groverunknown}, another possibility is to use
fixed-point quantum search: this avoids the problem of choosing too
large $k$ altogether \cite{yoder2014fixed}. The main idea of
fixed-point quantum search is to implement a polynomial function of
the amplitudes of the target state $\ket{\psi_G}$ with the property
that even when $k$ increases past the optimal value, these amplitudes
oscillate between values that are still sufficiently large. Hence, we
never go back to amplitudes that are too small: after reaching a large
enough probability of observing $\ket{\psi_G}$ when applying a
measurement, further iterations may increase this probability
slightly, but we have a lower bound ensuring that the probability does
not get too small.

There is a version of amplitude amplification that is tailored for
algorithms with multiple branches, each of which has different time
complexity. This version is called \emph{variable-time} amplitude
amplification. We use it in Sect.~\ref{sec:hhlimprovements}, but do
not give all details as we only need it in that specific section. A
general treatment can be found in \cite{ambainis2010variable}.

In Sect.~\ref{sec:qftnotes} we mentioned that phase estimation yields
a biased estimator, and that in some contexts this is undesirable. The
same considerations apply to amplitude estimation, because in turns it
relies on phase estimation (see \cite{suzuki2020amplitude} for a
version of amplitude estimation that employs a maximum likelihood
estimator rather than phase estimation). Unbiased amplitude estimation
is discussed in \cite{cornelissen2023sublinear,rall2023amplitude}. It
is an important technique in quantum algorithms for the estimation of
partition functions, a task that can be used to approximately count
combinatorial objects such as matchings or independent sets in a graph
\cite{cornelissen2023sublinear,harrow2020adaptive}. Work on the
estimation of partition function has also led to \emph{nondestructive}
amplitude estimation, i.e., a technique to apply quantum amplitude
estimation on a state while restoring a copy of the state after
measurement --- as opposed to the standard amplitude estimation
described in this chapter, where the final measurement would collapse
the quantum state irreversibly.

In general, the preparation of a qsample\index{qsample} (i.e., a quantum state
encoding a probability distribution) such as the one in
Eq.~\eqref{eq:qsampleampest} has gate complexity $\bigO{2^n}$, i.e.,
linear in the size of the vector encoding the probability
distribution. We can improve on this worst-case complexity with
additional assumptions. Two such assumptions are common in the
literature. The first assumption is that we have access to QRAM, see
Sect.~\ref{sec:qram}. Because QRAM implements some operations faster
than the standard circuit model, one has to be careful that a
potential speedup obtained by encoding probability distributions in
quantum states using QRAM is due to some algorithmic quantum
advantage, rather than to QRAM only. The second assumption is that the
probability distribution being encoded is efficiently integrable, as
defined in \cite{grover2002creating}. Note that this is a strong
assumption, as it implies that we can integrate the probability
density function between arbitrary endpoints, which usually means we
know it analytically and often leads to efficient classical sampling
as well.

The topic of quantum search, or quantum minimum finding,\index{quantum!search}\index{minimum finding} in the
presence of errors has produced both positive and negative results,
and these depend on the error model. We discussed several positive
results in Sect.~\ref{sec:quantumminerrors}, in particular for those
error models that appear to be more directly relevant for
fault-tolerant computation (e.g., errors due to oracles that rely on
bounded-error subroutines). If the errors are due to hardware noise,
results can be markedly more negative. \cite{regev2008impossibility}
shows that if the oracle $U_f$ is faulty, i.e., it applies identity
instead with some constant probability, then no quantum speedup can be
achieved. An analogous result for continuous-time quantum queries
(rather than the discrete-time queries discussed in this chapter) is
shown in \cite{temme2014runtime}. Note that if there are no marked
elements, then the computation is not affected by noise, because $U_f$
would be the identity map anyways. With a different form of noise
(depolarizing noise), that turns the state register of Grover search
into a uniform superposition with probability $p$, at least
$\Omega(p2^n)$ queries are necessary \cite{vrana2014fault}. A tight
characterization of the complexity of quantum search in the presence
of depolarizing noise, as well as additional types of noise, is given
in \cite{rosmanis2023quantum}. These noise models are inspired by
commonly used models for faulty hardware.

\chapter{Quantum gradient algorithm and vector input/output}
\label{ch:quantumgradient}
\thispagestyle{fancy}
In this chapter we discuss a quantum algorithm to estimate the
gradient of a multivariate function, given access to a quantum circuit
to compute the function. The core of the algorithm is a
multidimensional version of phase estimation. After introducing the
quantum gradient algorithm, we discuss two of its applications: an
algorithm to extract the classical description of the vector
describing a quantum state, and an algorithm to separate a point from
a convex set when we only have an indirect description of the set, via
an oracle that determines if a point belongs to it or not. Whereas
algorithms discussed in previous chapters always output a scalar, in
this chapter they output vectors. We also present an algorithm to
create the natural quantum encoding of a given classical vector, which
is used as a subroutine in a variety of quantum algorithms, sometimes
to encode part of the input data.

\section{The quantum gradient algorithm}
\label{sec:quantumgradient}
In this section we show how, using phase estimation, we can
simultaneously compute multiple components of the
gradient\index{gradient computation|(}\index{algorithm!gradient|(} of
a function using an evaluation circuit for the function. This approach
is, in some sense, the quantum analog of numerical computation of the
gradient using finite differences in the classical world. The
algorithm based on this idea, first introduced in
\cite{jordan2005fast}, is often called ``Jordan's gradient algorithm''
in the literature.

The problem that this algorithm solves is defined as follows. We are
given oracle access to a function $f : \R^d \to \R$, and we want to
output its gradient at the origin. The choice of the origin is
w.l.o.g., as one can always translate the function. Furthermore, we
only evaluate the function in a certain hyperbox (i.e., $[0, 1]^d$ in
the simplest version of the algorithm), rather than over all of
$\R^d$, so we could restrict its domain. We do not have an analytical
description of $f$, so the gradient must be estimated
numerically. Without further knowledge on the structure of the
function, a classical algorithm that outputs the gradient up to some
level of precision takes $\Omega(d)$ function evaluations: otherwise,
there is some direction in $\R^d$ on which we do not have enough
information to compute the gradient. (The computer science notation
$\Omega(d)$\index{notation!omega@\ensuremath{\Omega(\cdot)}} means
``at least $d$ asymptotically, up to a multiplicative factor,''
similarly to how $\bigO{d}$ means ``at most $d$ asymptotically, up to a
multiplicative factor.'') A simple and natural algorithm to
perform gradient estimation with a classical computer is to evaluate
the objective function along the coordinate axes, with some small
step size $\eta > 0$, and output the gradient estimate obtained doing
finite differences:
\begin{equation*}
  \frac{\partial f}{\partial x_j}(\zeroes) \approx \frac{f(\zeroes + \eta e_j) - f(\zeroes)}{\eta} \text{ for all } j=1,\dots,d,
\end{equation*}
where $e_j$ is the $j$-th orthonormal basis vector. Using this
formula, we estimate the gradient with $d+1$ function evaluations. (A
different formula for a finite-difference approximation is given in
Def.~\ref{def:finitediff}; that formula evaluates $f$ at $2d$ points.)
We can do better with a quantum algorithm. We divide our discussion on
this topic based on properties of the function $f$, and how it is
specified. Note that in this chapter we generally use $d$ to denote
the dimension of the vector space in which a generic vector $x$ lives
(e.g., the argument of the function $f$, or the vector that we want to
encode in the amplitudes of a quantum state), because we reserve $n$
to denote the number of qubits in certain sections.

\subsection{Linear functions with a binary oracle}
\label{sec:jordanlinear}
We first present the quantum algorithm in the setting studied by Jordan
\cite{jordan2005fast}. In this setting, we have access to a binary
oracle for the function $f$, which we define below after introducing
some notation.
\begin{definition}[Addition modulo the largest representable integer]
  \label{def:boxplus}
  For any given integer $q >0$ and integers $j, k \in \{0,\dots,2^q-1\}$, we
  define $j \boxplus k := (j+k \mod 2^q)$.\index{notation!boxplus@\ensuremath{\boxplus}} In other words, $\boxplus$
  of two integers representable on $q$ digits is the addition modulo
  $2^q$.

  We extend the definition to binary strings: given $\vj, \vk \in
  \{0,1\}^q$, $\vj \boxplus \vk := \v{h}$ where $h = (j+k \mod 2^q)$.
\end{definition}
\noindent The function $f$ is then given to us via the following oracle:
\begin{equation*}
  U_f \ket{\vv{x_1}} \ket{\vv{x_2}} \cdots  \ket{\vv{x_d}}  \ket{\v{y}} = \ket{\vv{x_1}} \ket{\vv{x_2}} \cdots  \ket{\vv{x_d}}  \ket{\v{y} \boxplus \vv{f(x_1,\dots,x_d)}},
\end{equation*}
where each register is represented on $q$ qubits. We call this a
\emph{binary oracle}.\index{binary!oracle} We make a further simplifying assumptions in
this section: that the function $f$ is linear or approximately linear.

\paragraph{Linear function.} The simplest case to analyze is when
the function is exactly (as opposed to approximately) linear, i.e.,
$f(x_1,\dots,x_d) = \dotp{g}{x} + b$ for some vector $g \in \R^d$ and
$b \in \R$. (Recall that $\dotp{g}{x}$ denotes an inner product; here,
since the vectors lie in real Euclidean space, $\dotp{g}{x} = g^{\top}
x$.)  
\begin{remark}
  The precise assumption is that we know that the function is
  linear, but we do not know what its gradient (i.e., the vector $g$)
  is. We could estimate it with a classical algorithm using $d+1$
  function evaluations, and we seek to do better than that.
\end{remark}
\noindent Thus, $\nabla f (\zeroes) = g$ and the answer that we seek
is precisely the vector $g$. For the sake of building intuition, we
also assume that all numbers (i.e., the components of $g$, as well as
$f(x)$ for all $x$) are representable on $q$ bits. This allows us to
ignore many important but cumbersome issues. We discuss how to relax
the assumption and give a more precise treatment beginning in the
second part of this subsection, and continuing in
Sect.~\ref{sec:jordanpoly}.

Assume that $f : [0,1]^d \to \R_+$ (one could translate and rescale
the function to make this assumption hold, if the function is
defined over a compact set and is bounded). Because each argument is
represented on $q$ bits, this creates a grid of points with integer numerator, say
$0, \frac{1}{2^q}, \frac{2}{2^q}, \dots, \frac{2^q-1}{2^q}$ along each
axis. Let us study the following state:
\begin{equation}
  \label{eq:jordanstate}
  \ket{\psi} = \frac{1}{\sqrt{2^{dq}}} \sum_{\v{x} \in \{0,1\}^{dq}} e^{2 \pi i (\dotp{g}{x} + b)/2^q} \ket{\vv{x_1}}\ket{\vv{x_2}}\dots\ket{\vv{x_d}},
\end{equation}
where $x$ is obtained from the $dq$-dimensional binary string $\v{x}$
by reshaping it as a $d$-dimensional vector with $q$-digit entries
(that is: the first $q$ digits indicate the first component of $x$,
the next $q$ indicate the second component, and so on). It is not too
difficult to see that the state $\ket{\psi}$ is a tensor product of
Fourier states: $e^{2 \pi i b/2^q}$ is a constant that can be
collected and taken out of the summation; then we use the fact that
$\dotp{g}{x} = g_1 x_1 + \dots g_d x_d$ to write:
\begin{align*}
  \ket{\psi} &= \frac{1}{\sqrt{2^{dq}}} \sum_{\v{x} \in \{0,1\}^{dq}} e^{2 \pi i b/2^q} e^{2 \pi i (g_1 x_1/2^q)} \ket{\vv{x_1}} e^{2 \pi i (g_2 x_2/2^q)}\ket{\vv{x_2}}\dots e^{2 \pi i (g_d x_d/2^q)} \ket{\vv{x_d}} \\
  &= e^{2 \pi i b/2^q} \left(\frac{1}{\sqrt{2^{q}}} \sum_{\vv{x_1} \in \{0,1\}^q} e^{2 \pi i (g_1 x_1/2^q)} \ket{\vv{x_1}}\right) \otimes \left(\frac{1}{\sqrt{2^{q}}} \sum_{\vv{x_2} \in \{0,1\}^q} e^{2 \pi i (g_2 x_2/2^q)} \ket{\vv{x_2}}\right) \otimes \dots\\
  &\dots \otimes \left(\frac{1}{\sqrt{2^{q}}} \sum_{\vv{x_d} \in \{0,1\}^q} e^{2 \pi i (g_d x_d/2^q)} \ket{\vv{x_d}}\right).
\end{align*}
The term $e^{2 \pi i b/2^q}$ is a global phase that can be ignored. In
the remaining $q$-qubit registers we have precisely the Fourier states
corresponding to the scalars $g_1,\dots,g_d$. Thus, if each component
$g_j$ of $g$ is exactly representable on $q$ bits, applying the
inverse QFT to each $q$-qubit register recovers a binary description
of $g_1,\dots,g_d$:
\begingroup
\allowdisplaybreaks
\begin{align*}
  Q_{q}^{\dag} \left(\frac{1}{\sqrt{2^{q}}} \sum_{\vv{x_1} \in \{0,1\}^q} e^{2 \pi i (g_1 x_1/2^q)} \ket{\vv{x_1}}\right) &= \ket{\vv{g_1}} \\
  Q_{q}^{\dag} \left(\frac{1}{\sqrt{2^{q}}} \sum_{\vv{x_2} \in \{0,1\}^q} e^{2 \pi i (g_2 x_2/2^q)} \ket{\vv{x_2}}\right) &= \ket{\vv{g_2}} \\
  \vdots \\
  Q_{q}^{\dag} \left(\frac{1}{\sqrt{2^{q}}} \sum_{\vv{x_d} \in \{0,1\}^q} e^{2 \pi i (g_d x_d/2^q)} \ket{\vv{x_d}}\right) &= \ket{\vv{g_d}}.
\end{align*}
\endgroup
It follows that if we could create the state $\ket{\psi}$ as in
Eq.~\eqref{eq:jordanstate}, then the application of the $q$-qubit
inverse QFT for $d$ times would output the correct answer to the
problem of computing the gradient.  As it turns out, creating
$\ket{\psi}$ is relatively straightforward given our assumptions, by using
phase kickback and exploiting the fact that $Q_{q} \ket{\v{1}}$ is an
eigenstate of modular addition.

\begin{figure}[h!]
  \leavevmode
  \centering
  \ifcompilefigs
  \Qcircuit @C=1em @R=0.7em {
    \lstick{\ket{\v{0}}} & {/^q} \qw & \gate{H^{\otimes q}}  & \qw & \multigate{4}{U_f} & \qw & \gate{Q^{\dag}_q} & \qw & \meter & \rstick{\vv{g_1} }\\
    \lstick{\ket{\v{0}}} & {/^q} \qw & \gate{H^{\otimes q}}  & \qw & \ghost{U_f}        & \qw & \gate{Q^{\dag}_q} & \qw & \meter & \rstick{\vv{g_2} }\\
    \lstick{\ket{\v{0}}} & {/^q} \qw & \gate{H^{\otimes q}}  & \qw & \ghost{U_f}        & \qw & \gate{Q^{\dag}_q} & \qw & \meter & \rstick{\vv{g_3} }\\
    \lstick{\ket{\v{0}}} & {/^q} \qw & \gate{H^{\otimes q}}  & \qw & \ghost{U_f}        & \qw & \gate{Q^{\dag}_q} & \qw & \meter & \rstick{\vv{g_4} }\\
    \lstick{\ket{\v{1}}} & {/^q} \qw & \gate{Q_q}          & \qw &  \ghost{U_f}        & \qw & \qw              & \qw\\
  }
  \else
  \includegraphics{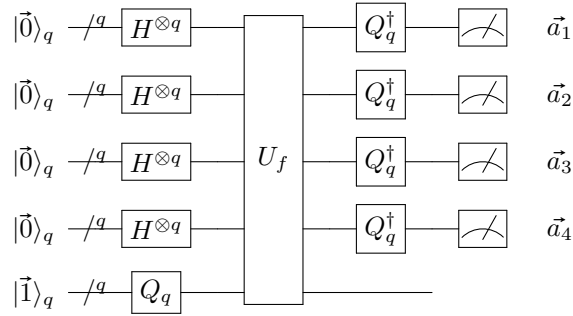}
  \fi
  \caption{Circuit for the quantum gradient algorithm for a
    four-dimensional function ($d = 4$, the gradient has four
    components $g_1, g_2, g_3, g_4$).}
  \label{fig:gradient}
\end{figure}

Recall that $U_f$ acts on $d+1$ registers, each of which is on
$q$-qubits: $d$ registers for the input arguments of the function $f$,
one register for the output. We initialize the algorithm with the
following state composed of $d+1$ $q$-qubit registers:
\begin{equation*}
  \ket{\v{0}}_q \otimes \ket{\v{0}}_q \otimes \dots \otimes \ket{\v{0}}_q \otimes \ket{\v{1}}_q.
\end{equation*}
Then we apply $H^{\otimes q}$ to each of the first $d$ registers, and
the QFT to the last register. We obtain:
\begin{align*}
  H^{\otimes q} \ket{\v{0}} \otimes H^{\otimes q} \ket{\v{0}} \otimes \dots H^{\otimes q} \ket{\v{0}} \otimes Q_{q} \ket{\v{1}} = \\
  \frac{1}{\sqrt{2^{dq}}} \sum_{x \in \{0,1\}^{dq}} \ket{\vv{x_1}} \ket{\vv{x_2}} \dots\ket{\vv{x_d}} \otimes \frac{1}{\sqrt{2^q}} \sum_{\vj \in \{0,1\}^q} e^{-2 \pi i j/2^q} \ket{\vj}.
\end{align*}
(The expression for the last register can be obtained by noticing that
the $q$-digit all-ones string corresponds to the integer $2^q-1$, and
$2^q$ can be simplified and eliminated from the expression.) Now we
apply $U_f$ to perform modular addition of $f(x_1,\dots,x_d)$ to the
last register, obtaining:
\begin{align}
  U_f \left(\frac{1}{\sqrt{2^{dq}}} \sum_{\v{x} \in \{0,1\}^{dq}} \ket{\vv{x_1}} \ket{\vv{x_2}} \dots\ket{\vv{x_d}} \otimes \frac{1}{\sqrt{2^q}} \sum_{\vj \in \{0,1\}^q} e^{-2 \pi i j/2^q} \ket{\vj}\right) = \notag \\
  \frac{1}{\sqrt{2^{dq}}} \sum_{\v{x} \in \{0,1\}^{dq}} \left( \ket{\vv{x_1}} \ket{\vv{x_2}} \dots\ket{\vv{x_d}} \otimes \sum_{\vj \in \{0,1\}^q}\frac{e^{-2 \pi i j/2^q}}{\sqrt{2^q}} \ket{\vj \boxplus \vv{f(x_1,\dots,x_d)}}\right). \label{eq:jordanfinalstate}
\end{align}
For every fixed $(\vv{x_1},\dots,\vv{x_d}) \in \{0,1\}^{dq}$, relabeling $k = j \boxplus f(x_1,\dots,x_d)$ and looking at the second summation in the last part of Eq.~\ref{eq:jordanfinalstate}, we have:
\begin{align*}
  \frac{1}{\sqrt{2^q}}\sum_{\vj \in \{0,1\}^q} e^{-2 \pi i j/2^q} \ket{\vj \boxplus \vv{f(x_1,\dots,x_d)}} &=
  \frac{1}{\sqrt{2^q}}\sum_{\vk \in \{0,1\}^q} e^{-2 \pi i (k - f(x_1,\dots,x_d))/2^q} \ket{\vk} \\
  &= e^{2\pi i f(x_1,\dots,x_d)/2^q} \underbrace{\frac{1}{\sqrt{2^q}} \sum_{\vk \in \{0,1\}^q} e^{-2 \pi i k/2^q} \ket{\vk}}_{= Q_q \ket{\v{1}}}.
\end{align*}
In the above expression, the first equality is due to the fact that we
are still summing over all possible binary strings, because modular
addition applies a constant shift to the index of the sum, together
with the fact that $j = (k - f(x_1,\dots,x_d) \mod 2^q)$; the second
equality holds because the term $f(x_1,\dots,x_d)$ in the exponent
does not depend on the summation index, hence it can be taken out of
the sum. Thus, only the phase factor depends on $x$, whereas the
Fourier state $Q_q \ket{\v{1}}$ is untouched an unentangled with the
rest. It follows that Eq.~\ref{eq:jordanfinalstate} can be rewritten
as a tensor product of a summation over $\v{x}$, and $Q_q
\ket{\v{1}}$:
\begin{align*}
  \frac{1}{\sqrt{2^{dq}}} \sum_{\v{x} \in \{0,1\}^{dq}} e^{2 \pi i f(x_1,\dots,x_d)/2^q} \ket{\vv{x_1}} \ket{\vv{x_2}} \dots\ket{\vv{x_d}} \otimes  \frac{1}{\sqrt{2^q}} \sum_{\vj \in \{0,1\}^q} e^{-2 \pi i j/2^q} \ket{\vj}.
\end{align*}
The first $dq$ qubits in this expression are precisely the state
$\ket{\psi}$ in Eq.~\ref{eq:jordanstate}, showing that we have
constructed the desired state from which the inverse QFT recovers the
correct answer. To construct this state via phase kickback, we have
used Hadamards everywhere in the first $d$ registers, QFT in the last
register, and one application of $U_f$. The full circuit implementing
the algorithm described in this section is given in
Fig.~\ref{fig:gradient}.

We have shown the following.
\begin{proposition}
  \label{prop:jordanlinear}
  Let $U_f$ be given as a binary oracle that performs modular addition
  in the last register. Assume $f$ is linear, and the value $f(x)$ as
  well as each component of the gradient of $f$ is exactly
  representable on $q$ bits. Then the circuit in
  Fig.~\ref{fig:gradient} recovers the gradient of $f$ with a single
  application of $U_f$.
\end{proposition}
To highlight the connection between the circuit in
Fig.~\ref{fig:gradient} and phase estimation, consider the case with
$d=1$. Then, we have a single inverse QFT block, similar to the phase
estimation circuit in Fig.~\ref{fig:qpefull}. The last register is
prepared in an eigenstate of modular addition with phase $2\pi i
0.\vv{g_1}$ (the vector $g$ has a single component with $d=1$), and
the application of $U_f$ prepares the Fourier state. We encourage the
reader to verify that the application of $U_f$ has the same effect as
applying the controlled-$U^{2^k}$ operations in the phase estimation
circuit: if the binary string $\vv{x_1}$ is the number $2^{q-j}$,
i.e., the integer whose binary representation has $1$ in position $j$
and $0$ elsewhere, the product $g_1 x_1/2^q$ is equivalent to shifting
the binary fractional point of $0.\vv{g_1}$, so we kick back the phase
$2 \pi i 2^{q-j} 0.\vv{g_1}$, precisely as in the quantum phase
estimation circuit.
\begin{remark}
  The gradient algorithm discussed in this section is a
  multidimensional version of phase estimation\index{phase!estimation}\index{algorithm!phase estimation}. In standard phase
  estimation, we use phase kickback to construct the Fourier state
  corresponding to the sought phase --- a scalar. In multidimensional
  phase estimation, we use phase kickback to construct the tensor
  product of multiple Fourier states, each corresponding to one
  component of a vector. Rather than a single inverse QFT at the end
  of the circuit, we apply multiple inverse QFT
  blocks. Multidimensional phase estimation is a powerful technique to
  output vectors with a quantum algorithm, when it is applicable.
\end{remark}
Prop.~\ref{prop:jordanlinear} has two severe limitations: it assumes
that $f$ is linear and calculated exactly, and furthermore, it assumes
that $f(x)$ and $g$ are exactly representable on $q$ bits. To broaden
the applicability of the gradient algorithm we need to relax these
assumptions.

\paragraph{Approximately linear function.} If $f$ is exactly linear, it does not matter where it is evaluated: with phase kickback we construct the Fourier states corresponding to the components of the gradient $g$ at the origin, without any error.
If $f$ is not exactly linear, it seems more reasonable to evaluate it
at a grid centered at the origin, rather than a grid in the hypercube
$[0,1]^d$: Taylor's theorem tells us that, if we get further away from
the origin, there may be a larger discrepancy between the value of the
closest linear approximation to $f$ constructed at the origin (i.e., its
first-order Taylor approximation) and the true value of $f$. A grid
symmetric around the origin is also helpful to borrow tools from
classical numerical analysis to study the error of finite-difference
formulas for gradient approximation, e.g., the central difference in
the next section (see Def.~\ref{def:centraldiff}). For this discussion
we assume $f: [-\frac{1}{2}, \frac{1}{2}]^d \to \R$, and we consider
the following discretization along each axis, also studied in
\cite{jordan2005fast,gilyen2019optimizing}:
\begin{equation*}
  G_q := \left\{\frac{j}{2^q} - \frac{1}{2} + \frac{1}{2^{q+1}} : j =
  0,\dots,2^q-1 \right\}.
\end{equation*}
Thus, $G_q^d = G_q \times \dots \times G_q$ denotes a $d$-dimensional
grid where each axis has $2^q$ grid points.
\begin{example}
  With $q=3$, $G_3 = \{-\frac{7}{16}, -\frac{5}{16},
  -\frac{3}{16}, -\frac{1}{16}, \frac{1}{16}, \frac{3}{16},
  \frac{5}{16}, \frac{7}{16}\}$: one can see that the discretization
  is symmetric around the origin, and the points are equally spaced
  with distance $\frac{1}{2^q}$.
\end{example}

\noindent There is an obvious bijection between $G_q$ and the
discretization $\{\frac{j}{2^q} : j=0,\dots,2^q-1\}$ that we used for
the case of (exactly) linear functions. For $\vj \in \{0,1\}^q$, we
write $\widehat{\jmath}$ to denote the point $\frac{j}{2^q} -
\frac{1}{2} + \frac{1}{2^{q+1}}$ (a more formal treatment should
define a function mapping the integer $j$ to the corresponding point
in the shifted grid; $\widehat{\cdot}$ is a shorthand notation that is
sufficient for our exposition). We define the QFT on the transformed
domain $G_q$ as the unitary $Q_{G_q}$ that implements the following
map:\index{quantum!Fourier transform|(}\index{Fourier transform|(}
\begin{align*}
  Q_{G_q} \ket{\vk} &= \frac{1}{\sqrt{2^q}} \sum_{\vj \in \{0,1\}^q} e^{2\pi i 2^q \widehat{\jmath}\, \widehat{k}} \ket{\vj} \\
  &= \frac{1}{\sqrt{2^q}} \sum_{\vj \in \{0,1\}^q} \exp\left(2\pi i \left(\frac{jk}{2^q} - (j + k) \left(\frac{1}{2} + \frac{1}{2^{q+1}}\right) + \left(2^{q-2} - \frac{1}{2} + \frac{1}{2^{q+2}}\right)\right) \right) \ket{\vj}.
\end{align*}
Using the definition, it is easy to verify that $Q_{G_q} = U_q Q_q
U_q$, where $Q_q$ is the standard QFT as defined in
Def.~\ref{def:qft}, and $U_q$ is the $q$-qubit unitary:
\begin{equation*}
  U_q \ket{\vj} = \exp\left(2\pi i -j\left(\frac{1}{2} + \frac{1}{2^{q+1}}\right) + \frac{1}{2}\left(2^{q-2} - \frac{1}{2} + \frac{1}{2^{q+2}}\right)\right) \ket{\vj} \qquad \forall \vj \in \{0,1\}^q.
\end{equation*}
A circuit for $U_q$ can be obtained with $q$ phase shift gates
(Def.~\ref{def:phaseshiftgate}), therefore we can implement $Q_{G_q}$
with the same asymptotic gate complexity, in $\bigO{\cdot}$ notation,
as the standard QFT. The quantum gradient algorithm for approximately
linear functions then works in the same way as the one for exactly
linear functions described earlier, with the difference that we
evaluate $f$ at $(\widehat{x_1},\widehat{x_2},\dots,\widehat{x_d})$
instead of $(x_1,x_2,\dots,x_d)$, and apply the inverse
$Q_{G_q}^{\dag}$ of the transformed QFT instead of the standard
QFT. We give the pseudocode of the algorithm in
Alg.~\ref{alg:gradient}. For simplicity, we assume that we have access
to a quantum circuit that maps $\ket{\vx} \to e^{2\pi i 2^q
  f(\widehat{x})} \ket{\vx}$: this type of oracle is formally defined
in Def.~\ref{def:phaseoracle} in Sect.~\ref{sec:jordanpoly}. Note that
if we have a circuit to evaluate $f$ in binary (i.e., $U_f
\ket{\vx}\ket{\vy} = \ket{\vx} \ket{\vy \boxplus \vv{f(x)}}$), then
the desired circuit can be constructed with phase kickback, as in the
circuit in Fig.~\ref{fig:gradient}, although one has to be careful due
to the factor $2^q$ in the exponent, see Rem.~\ref{rem:phaseoracleexp}
as well as \cite{jordan2005fast,gilyen2019optimizing}.
\begin{algorithm2e}[htb]
  \SetAlgoLined
  \LinesNumbered
  \KwIn{Dimension $d$, digits of precision $q$, unitary $U_f$ to evaluate $f : G_q^d \to \R$ such that:
    \begin{equation*}
      U_f \ket{\vv{x_1}}\ket{\vv{x_2}}\cdots\ket{\vv{x_d}} = e^{2 \pi i 2^q f(\widehat{x_1},\widehat{x_2},\dots,\widehat{x_d})} \ket{\vv{x_1}}\ket{\vv{x_2}}\cdots\ket{\vv{x_d}}.
    \end{equation*}
  } 
  \KwOut{Vector $\tilde{g}$ that is close to a linear approximation of $f$ at the origin.}
  \textbf{Initialize}: Prepare $d$ $q$-qubit registers in the state $\ket{\v{0}}$, and apply $H^{\otimes q}$ to each of them.\\
  Apply the $(dq)$-qubit unitary $U_f$ to all registers.\\
  Apply $Q_{G_q}^{\dag}$ to each of the $d$ registers. \\
  Apply a measurement to all qubits. Calling $\vv{k_1}, \vv{k_2}, \dots, \vv{k_d} \in \{0,1\}^q$ the measurement outcomes from the $d$ registers, set $\tilde{g} \leftarrow (\widehat{k_1}, \widehat{k_2}, \dots, \widehat{k_d}) \in \R^d$. \\
  \Return the vector $\tilde{g}$.
  \caption{Quantum gradient algorithm for approximately linear functions.}
  \label{alg:gradient}
\end{algorithm2e}
\begin{proposition}[Based on Lem.~5.1 in \cite{gilyen2019optimizing}]
  \label{prop:jordanlinearapx}
  Let $q \in \N, q > 0$, $b \in \R, g \in \R^d$ with $\nrm{g}_{\infty}
  \le \frac{1}{3}$. Let $f : [-\frac{1}{2}, \frac{1}{2}]^d \to
  \R$. If:
  \begin{equation*}
    \abs{ f(x) - \dotp{g}{x} -b } \le \frac{1}{20 \pi 2^q}
  \end{equation*}
  for at least 99.9\% of the points $x \in G_q^d$, then the output of
  the gradient algorithm (Alg.~\ref{alg:gradient}) satisfies:
  \begin{equation*}
    \Pr( \abs{\tilde{g}_j - g_j} \le \frac{1}{2^q} ) > \frac{3}{5} \qquad \forall j=1,\dots,d.
  \end{equation*}
\end{proposition}
\begin{proof}
  Similarly to our discussion for the exact linear case, we consider
  the ``ideal'' quantum state:
  \begin{equation*}
    \ket{\psi} = \frac{1}{\sqrt{2^{dq}}} \sum_{\v{x} \in \{0,1\}^{dq}} e^{2 \pi i 2^q(\dotp{g}{\widehat{x}} + b)} \ket{\vv{x_1}}\ket{\vv{x_2}}\dots\ket{\vv{x_d}},
  \end{equation*}
  which is a product state:
  \begin{align*}
    \ket{\psi} &= e^{2 \pi i 2^q b} \left(\frac{1}{\sqrt{2^{q}}} \sum_{\vv{x_1} \in \{0,1\}^q} e^{2 \pi i 2^q g_1 \widehat{x_1}} \ket{\vv{x_1}}\right) \otimes \left(\frac{1}{\sqrt{2^{q}}} \sum_{\vv{x_2} \in \{0,1\}^q} e^{2 \pi i 2^q g_2 \widehat{x_2}} \ket{\vv{x_2}}\right) \otimes \dots\\
    &\dots \otimes \left(\frac{1}{\sqrt{2^{q}}} \sum_{\vv{x_d} \in \{0,1\}^q} e^{2 \pi i 2^q g_d \widehat{x_d}} \ket{\vv{x_d}}\right).
  \end{align*}
  The global phase can be ignored. Recalling the relationship between
  $Q_{G_q}$ and $Q_q$, by Thm.~\ref{thm:qpesimple}, if we apply
  $Q_{G_q}^{\dag}$ to each of the $d$ registers, measure all qubits,
  and interpret the resulting binary strings as points in the grid $G_q$, we obtain numbers
  $\widehat{k_1} = \tilde{g}_1,\dots, \widehat{k_d} = \tilde{g}_d \in [-\frac{1}{2},\frac{1}{2}]$ such that:
  \begin{equation*}
    \Pr( \abs{\tilde{g}_j - g_j} \le \frac{1}{2^q} ) \ge \frac{8}{\pi^2} > \frac{4}{5} \qquad \forall j=1,\dots,d.
  \end{equation*}
  (The proof of Thm.~\ref{thm:qpesimple} also applies to the
  transformed inverse QFT $Q_{G_q}^{\dag}$, after accounting for the
  unitaries $U_q$ in its definition; see the proof of
  \cite[Lem.~5.1]{gilyen2019optimizing}.) \index{quantum!Fourier transform|)}\index{Fourier transform|)}Alg.~\ref{alg:gradient}
  constructs:
  \begin{equation*}
    \ket{\phi} = \frac{1}{\sqrt{2^{dq}}} \sum_{\v{x} \in \{0,1\}^{dq}} e^{2 \pi i 2^qf(\widehat{x})} \ket{\vv{x_1}}\ket{\vv{x_2}}\dots\ket{\vv{x_d}},
  \end{equation*}
  so if we can show that $\nrm{\ket{\psi} - \ket{\phi}} \le
  \frac{1}{5}$, by Prop.~\ref{prop:euclideantotvd} the probabilities
  of the measurement outcomes cannot change by more than
  $\frac{1}{5}$, ensuring that each component of $\tilde{g}$ is
  sufficiently accurate with probability at least $\frac{4}{5} -
  \frac{1}{5} = \frac{3}{5}$. By assumption, for at least 99.9\% of
  the points $x \in G_q^d$ we have that $\abs{ f(x) - \dotp{g}{x} -b }
  \le \frac{1}{20 \pi 2^q}$; let $S \subseteq G_q^d$ be the set of
  points that satisfy the inequality. Then we have: \begingroup
  \allowdisplaybreaks
  \begin{align*}
    \nrm{\ket{\psi} - \ket{\phi}}^2 &= \frac{1}{2^{dq}} \sum_{x \in G_q^d} \abs{e^{2 \pi i 2^q f(x)} - e^{2 \pi i 2^q (\dotp{g}{x} + b)}}^2 \\
    &= \frac{1}{2^{dq}} \sum_{x \in S} \abs{e^{2 \pi i 2^q f(x)} - e^{2 \pi i 2^q (\dotp{g}{x} + b)}}^2 + \frac{1}{2^{dq}} \sum_{x \in G_q^d \setminus S} \abs{e^{2 \pi i 2^q f(x)} - e^{2 \pi i 2^q (\dotp{g}{x} + b)}}^2 \\
    &\le \frac{1}{2^{dq}} \sum_{x \in S} \abs{2 \pi 2^q f(x) - 2 \pi 2^q (\dotp{g}{x} + b)}^2 + \frac{1}{2^{dq}} \sum_{x \in G_q^d \setminus S} 4 \\
    &= \frac{1}{2^{dq}} \sum_{x \in S} (2 \pi 2^q)^2 \abs{ f(x) - (\dotp{g}{x} + b)}^2 + \frac{1}{2^{dq}} \sum_{x \in G_q^d \setminus S} 4 \\
    &\le \frac{1}{2^{dq}} \left( \sum_{x \in S} \left(\frac{1}{10}\right)^2 + \sum_{x \in G_q^d \setminus S} 4\right) \le \left(\frac{1}{10}\right)^2 + \frac{4}{1000} \le \left(\frac{1}{5}\right)^2.
  \end{align*}
  \endgroup
  In the above chain, for the first inequality we used $\abs{e^{iy} - e^{iz}} \le \abs{y - z}$, then we used the error bound for the points in $S$ and the assumption on the cardinality of $G_q^d \setminus S$.
\end{proof}

\noindent Prop.~\ref{prop:jordanlinearapx} shows that the output of
Alg.~\ref{alg:gradient} is a vector such that each component is close
to a linear approximation $g$ of $f$ with probability at least
$3/5$. By repeating the algorithm $\bigO{\log \frac{d}{\delta}}$ times
and taking the median component-wise, we boost the probability of
success so that all components are close to $g$ with probability at
least $1-\delta$ (for a formal proof of this result: suppose that we
perform $2m + 1$ repetitions of the algorithm, and we require that for
each component $j$, at least $m+1$ of the values $\tilde{g}_j$ are
within distance $\frac{1}{2^q}$ from $g_j$; if this is the case, the
median of the values satisfies the desired bound. Then we apply the
Chernoff-Hoeffding theorem, choosing $m$ to ensure that the
probability that each coordinate does not satisfy the desired
condition is at most $\delta/d$; this gives $m = \bigO{\log
  \frac{d}{\delta}}$). In this way, we obtain a more robust version of
the gradient algorithm, because we allow a small deviation from
linearity of the function $f$.
\begin{remark}
  The choice ``99.9\% of the points'' in
  Prop.~\ref{prop:jordanlinearapx} is of course arbitrary, and only
  affects the final probability of success. The main benefit of
  allowing $f(x)$ to be far from $\dotp{g}{x} + b$ for some $x$ is
  that we do not need $f$ to be computed exactly at all points. For
  example, if $f$ is computed via an algorithm that has a small
  probability of failure (as is often the case with quantum circuits),
  the gradient algorithm still works because, with high probability,
  the computation of $f$ fails only at a small fraction of the points.
\end{remark}

\subsection{Polynomial functions with other types of oracles}
\label{sec:jordanpoly}
In the first part of Sect.~\ref{sec:jordanlinear} we worked with a
linear function, and we had access to some form of a binary
oracle\index{binary!oracle}\index{oracle!function} for its value,
i.e., a unitary that outputs a binary description of the function
value. This allowed us to perform phase kickback using the property
that the Fourier state corresponding to the all-ones binary string is
an eigenvector of addition modulo the largest integer in the
register. In the second part of Sect.~\ref{sec:jordanlinear} we
directly assumed access to an oracle that applies the necessary phase,
essentially bypassing phase kickback for simplicity: the purpose of
the binary oracle in the gradient algorithm is to construct the
necessary phases via kickback. In general we may not always have a
linear or approximately linear function, and furthermore, we may not
have access to a binary oracle for the function value; note that the
binary oracle is usually available in situations where we know a
(classical) Boolean circuit to compute the function value, but may be
difficult to obtain otherwise. We now discuss some other possible
computational models for the function $f$, and also relax the
linearity assumption. The different computational models for the
function discussed in this section cover the situations that are most
commonly found when implementing quantum circuits to carry out a
computation. Results in this section are based on
\cite{gilyen2019optimizing}.

Let us first discuss how to deal with nonlinearity. If we could
somehow construct an approximation of the function $\dotp{\nabla
  f(\zeroes)}{x}$, starting from an evaluation oracle for the
nonlinear function $f$, then we could apply Alg.~\ref{alg:gradient} to
this new function, which is linear in $x$ and such that its gradient
is precisely $\nabla f(\zeroes)$. We can turn to ideas from calculus
and numerical analysis to attain this goal, relying on a central
difference approximation of the function $f$. A central difference
approximation is a higher-order version of the simple central
difference formula $\frac{\partial f}{\partial x_j}(\zeroes) \approx
\frac{f(\eta e_j) - f(-\eta e_j)}{2\eta}$. We define it below, where
we also state --- without proof --- that its value approximates
$\dotp{\nabla f(\zeroes)}{x}$. For details on the error bound obtained
with this formula, see \cite{gilyen2019optimizing}.
\begin{definition}[Central difference approximation]
  \label{def:centraldiff}
  The \emph{degree-$2m$ central difference approximation} of a
  function $f : \R^d \to \R$ is the function defined as:
  \begin{equation*}
    f^{(2m)}(x) := \sum_{k=-m}^{m} a^{(2m)}_k f(k x) \approx \nabla f(\zeroes)^{\top} x,
  \end{equation*}
  where the coefficients $a^{(2m)}_k$ are defined as:
  \begin{equation*}
    a^{(2m)}_k := \frac{(-1)^{k-1} \binom{m}{|k|}}{k \binom{m+|k|}{|k|}}
  \end{equation*}
  for $k \neq 0$, and $a^{(2m)}_0 := 0$.
\end{definition}
By computing $f^{(2m)}$ we obtain an approximation of $\dotp{\nabla
  f(\zeroes)}{x}$: the value of $m$ that is necessary for a good
approximation depends on the desired error tolerance, and on the
degree of nonlinearity of $f$. Note that evaluating $f^{(2m)}$ at one
point requires evaluating $f$ at $2m$ points, thus the value for $m$
determines the cost (query complexity) of implementing an oracle for
$f^{(2m)}$. We can then apply the gradient algorithm
(Alg.~\ref{alg:gradient}) to $f^{(2m)}$, which is linear in $x$. An
example of such a result is given below in
Thm.~\ref{thm:gradientpoly}.

\begin{figure}
  \centering
  \ifcompilefigs
  \begin{tikzpicture}[
      ell/.style 2 args={
        ellipse,
        minimum width=4cm,
        minimum height=2cm,
        draw,
        label={[name=#1]center:#2}
      },
      connection/.style={thick,densely dotted, latex-latex}
    ]
    \node [ell={A}{Binary oracle},   fill=green!20] (a) at (4,4) {};
    \node [ell={B}{Probability oracle},fill=red!20]   (b) at (8,0) {};
    \node [ell={C}{Phase oracle},   fill=blue!20]  (c) at (0,0) {};

    \draw [-latex,dashed] (b.120) to[bend right] (a.330);
    \draw [-latex] (a.300) to[bend right] (b.150);

    \draw [-latex] (b.165) to[bend right] (c.15);
    \draw [-latex] (c.345) to[bend right] (b.195);

    \draw [-latex,dashed] (c.30) to[bend right] (a.240);
    \draw [-latex] (a.210) to[bend right] (c.60);

    \node [align=left] at (8.2,2.2) {amplitude\\estimation};
    \node [align=left] at (3.1,2) {phase\\estimation};

    \node [align=left] at (0,2.2) {phase\\kickback};
    \node [align=left] at (5.4,2) {controlled\\rotations};

    \node [align=left] at (4,0.7) {Thm.~\ref{thm:probtophase}};
    \node [align=left] at (4,-0.7) {Thm.~\ref{thm:phasetoprob}};
  \end{tikzpicture}
  \else
  \includegraphics{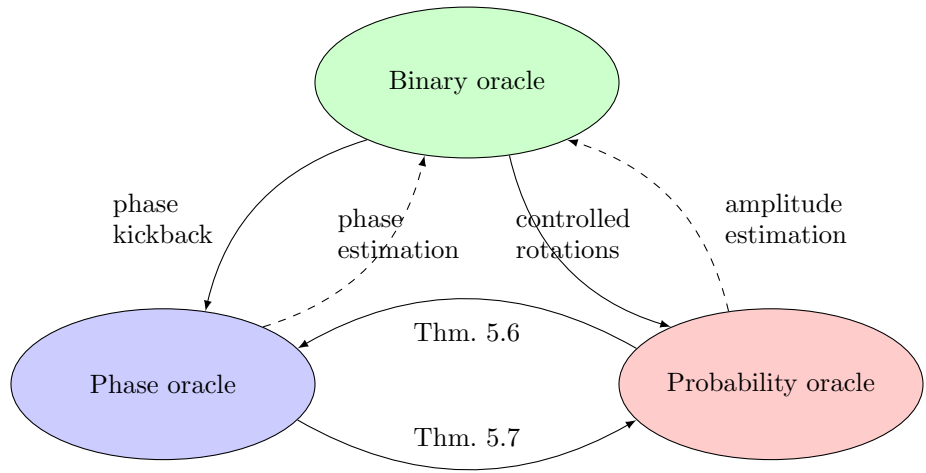}
  \fi
  \caption{Oracle conversion. Solid lines indicate efficient
    conversions (polylogarithmic cost in $1/\epsilon$), dashed lines indicate inefficient conversions (polynomial cost in $1/\epsilon$).}
  \label{fig:oracleconv}
\end{figure}

We now move to discussing different input models for the function
$f$. Besides binary oracles, there are two natural way to encode
functions that have been used in one way or another in the literature
on quantum algorithms. These are probability oracles and phase
oracles. We have already used phase oracles without a proper
definition, in Alg.~\ref{alg:gradient}. Below, $x$ is a vector and
$\vx$ is a binary encoding the vector $x$, e.g., by listing its
components in binary in fixed precision.
\begin{definition}[Probability oracle]
  \label{def:proboracle}
  A probability oracle for a function $f:
  [0, 1] \to [0, 1]$ is a unitary $U_f$ mapping $\ket{\v{0}}\ket{\vx}
  \to \sqrt{f(x)} \ket{1}\ket{\psi_x^{(1)}} + \sqrt{1-f(x)}
  \ket{0}\ket{\psi_x^{(0)}}$ for all $x$, where $\ket{\psi_x^{(0)}},
  \ket{\psi_x^{(1)}}$ are some arbitrary quantum states.
\end{definition}
Note that according to the definition of probability oracle, the
probability of observing $\ket{1}$ in the first qubit is precisely
$f(x)$.
\begin{definition}[Phase oracle]
  \label{def:phaseoracle}
  A phase oracle for a function $f: [0, 1] \to [-1, 1]$ is a unitary
  $U_f$ mapping $\ket{\v{0}}\ket{\vx} \to e^{if(x)}\ket{\v{0}}\ket{\vx}$
  for all $x$.
\end{definition}
Conversion between these oracles is possible, with variable
cost. Converting from a probability oracle to a binary oracle can be
expensive (it can be done with amplitude estimation). We can
efficiently convert a binary oracle to a phase oracle using phase
kickback, as shown in Sect.~\ref{sec:jordanlinear}. Converting from a
probability oracle to a phase oracle is also efficient: we give a
conversion result below.\index{oracle!conversion}
\begin{theorem}[Converting probability oracles into phase oracles; Cor.~4.1 in \cite{gilyen2019optimizing}]
  \label{thm:probtophase}
  Let $f: [0, 1] \to [0, 1]$ and suppose we have access to a
  probability oracle $U_f$ for $f$. Then we can implement an
  $\epsilon$-approximate phase oracle for $f$ using $\bigO{\log
    \frac{1}{\epsilon}}$ applications of $U_f$ and $U_f^{\dag}$, i.e.,
  a unitary whose output on any valid input state is at most
  $\epsilon$-away (in Euclidean norm) from the output of an exact
  phase oracle.
\end{theorem}
There are likely many constructive proofs for
Thm.~\ref{thm:probtophase}. The approach used in
\cite{gilyen2019optimizing} approximates the exponential function
$e^{i f(x)}$ with a Taylor series, then constructs each term in the
series (which is a sinusoidal function) relying on an analogy with
Grover's algorithm. Indeed, a probability oracle constructs a
superposition of a ``good'' and a ``bad'' state, marked by the first
qubit, and we can rotate in the plan spanned by these two. The terms
are then combined using a linear combination of unitaries, see
Sect.~\ref{sec:lcu}. To complement Thm.~\ref{thm:probtophase}, we note
that conversion from a phase to a probability oracle is also
efficient, provided the probability is bounded away from $0$ and $1$
by a constant.
\begin{theorem}[Converting phase oracles into probability oracles; Lem.~16 in the arXiv version of \cite{gilyen2019optimizing}]
  \label{thm:phasetoprob}
  Let $f: [0, 1] \to [\delta, 1-\delta]$ and suppose we have access to
  a phase oracle $U_f$ for $f$. Then we can implement an
  $\epsilon$-approximate probability oracle for $f$ using
  $\bigO{\frac{1}{\delta}\log \frac{1}{\epsilon}}$ applications of
  $U_f$ and $U_f^{\dag}$.
\end{theorem}
These conversion results are summarized in
Fig.~\ref{fig:oracleconv}. As indicated in the picture, conversions
between oracle types are generally efficient in the inverse precision
(i.e., they run in time polylogarithmic in $1/\epsilon$) except when
trying to convert a probability or a phase oracle to a binary oracle:
such an ``analog to digital'' transformation can be resource-intensive
(i.e., it runs in time polynomial in $1/\epsilon$). 

Putting everything together, the cost of gradient computation for
nonlinear functions $f$ with an extension of the gradient algorithm of
Sect.~\ref{sec:jordanlinear} is no longer as simple as for the linear
case, and it depends on how the function $f$ is specified, i.e., its
input model. A detailed analysis is given in
\cite{gilyen2019optimizing}; we report a version of their results
below.
\begin{theorem}[Gradient estimation for polynomial functions; based on Thm.~5.2 in \cite{gilyen2019optimizing}]
  \label{thm:gradientpoly}
  Let $f : [-1, 1]^d \to \R$ be a multivariate polynomial of degree
  $k$, and suppose we have access to a phase oracle for $f$. Then with
  $\bigOt{\frac{k}{\epsilon}}$ calls to the phase oracle, and
  $\bigOt{\frac{dk}{\epsilon}}$ additional gates, we can compute an
  $\epsilon$-approximation (in $\ell^\infty$-norm) of $\nabla
  f(\zeroes)$ with probability at least $1-\delta$.
\end{theorem}
\begin{remark}
  \label{rem:phaseoracleexp}
  For the case $k=1$, i.e., linear functions, there is a notable
  difference between Thm.~\ref{thm:gradientpoly} and the approach that
  we discussed in Sect.~\ref{sec:jordanlinear} after
  Prop.~\ref{prop:jordanlinearapx} (i.e., applying
  Alg.~\ref{alg:gradient}, taking $\bigO{\log \frac{d}{\delta}}$
  repetitions, and returning the median): Thm.~\ref{thm:gradientpoly}
  has $\frac{1}{\epsilon}$ dependence, whereas $\frac{1}{\epsilon}$
  does not show up in Prop.~\ref{prop:jordanlinearapx}. This apparent
  discrepancy is easily explained by the difference in input model: in
  Alg.~\ref{alg:gradient} we assume access to a phase oracle for $2^q
  f(x)$, whereas in Thm.~\ref{thm:gradientpoly} we assume access to a
  phase oracle for $f(x)$. Converting a phase oracle for $f(x)$ into
  one for $2^q f(x)$ leads to $\bigO{2^q} = \bigO{\frac{1}{\epsilon}}$
  applications of the weaker phase oracle, because we need to --- in
  some sense --- ``amplify the signal.''  (Recall that if we want
  error at most $\epsilon$, we need $q = \ceil{\log
    \frac{1}{\epsilon}}$ digits of precision.) On the other hand,
  suppose we start with a binary oracle for $f$, under the same
  assumption of Def.~\ref{def:phaseoracle} that the codomain of $f$ is
  $[-1, 1]$. Suppose further that the binary oracle outputs a number
  with $h$ digits of precision, i.e., it writes the value of $f(x)$ as
  a fraction $0.\v{\ell}$ with $\v{\ell} \in \{0,1\}^h$, where $h$ can
  be decided. Then, the string $\v{\ell}$ represents the integer $\ell
  = 2^h f(x)$, so we can implement the operation $U_f
  \ket{\vx}\ket{\vy} = \ket{\vx} \ket{\vy \boxplus \vv{2^hf(x)}}$: the
  value of $f(x)$ is initially fractional, we treat it as an integer
  by shifting the binary point to the right, i.e., multiplying by
  $2^h$, and perform modular addition with the last register over an
  appropriate number of digits. Thus, given a sufficiently accurate
  binary oracle for $f$, we can choose $h$ implement a phase oracle
  for $2^q f(x)$ with phase kickback and the construction of
  Fig.~\ref{fig:gradient}, using a single call to the binary
  oracle. In the first part of Sect.~\ref{sec:jordanlinear} we ignored
  this detail for simplicity.
\end{remark}
Note that for $k>1$ the performance of the algorithm degrades, due to
the cost of computing the central difference approximation. Still, as
long as the degree $k$ is small, the dimension $d$ does not appear in
the query complexity, except in polylogarithmic factors (but it
appears in the gate complexity); this is because the central
difference approximation gives a precise gradient approximation for
multivariate polynomials. The result of Thm.~\ref{thm:gradientpoly}
can be extended to more general, nonpolynomial analytic functions by
using their Taylor series approximation, and relying on error bounds
for the Taylor series. However, we need the higher-order derivatives
to be bounded (they are always bounded for polynomial functions over a
compact domain), because otherwise the error terms of the Taylor
series may make it difficult to accurately determine by how much we
are deviating from the ``ideal'' linear case. We do not pursue the
analysis of this more general case.

We also highlight that the quantum gradient algorithm can be made to
return an unbiased estimate of the gradient, by modifying the phase
estimation part. The reason why bias may occur is the following. In an
ideal world, we are able to prepare exactly the state $\ket{\psi}$ in
Eq.~\eqref{eq:jordanstate}, then we apply phase estimation with a
sufficient number of digits of precision to store the exact value of
the phases. When this happens, the phase estimation algorithm returns
a finite-precision representation of the gradient with probability
1. However, errors can occur in either step of the computation: we may
not be able to prepare $\ket{\psi}$ exactly (this is especially common
whenever the function $f$ is nonlinear, and we rely on the central
difference approximations), or we may not have enough digits of
precision to store the phase. In this case the output of phase
estimation is a random variable, whose expected value is not
necessarily the gradient --- even if we know that it is close enough
to the gradient. By making phase estimation unbiased, for a
``sufficiently linear'' (in the sense of
Prop.~\ref{prop:jordanlinearapx}) function we can obtain a gradient
algorithm whose output is an unbiased estimate of the gradient. For
details, see \cite[Sect.~6.5]{nannitomography}.\index{gradient computation|)}\index{algorithm!gradient|)}

\subsection{The gradient algorithm for quantum state tomography}
\label{sec:gradienttomo}
A perhaps surprising application of the gradient algorithm is to obtain
a classical description of an unknown quantum state that can be
prepared with a given unitary; the process of obtaining a description
of a quantum state is called \emph{quantum state tomography}.\index{state!tomography|(} In
general we can only obtain information on a quantum state via
measurements, as we discussed in Ch.~\ref{ch:intro}, and measuring a
$q$-qubit quantum state only yields $q$ bits of information. But there
are algorithms to recover a full description of the quantum state, up
to some specified level of precision: obviously, multiple measurements
or measurements of a larger number of qubits become necessary. We
describe such an algorithm that uses the gradient algorithm as a
subroutine. This algorithm, a version of which is described in
\cite{nannitomography}, is optimal in some settings, but not in all
settings: it is targeted at the case in which we have a quantum
circuit to prepare the state of interest. In other situations, other
algorithms are more efficient. Still, we find the idea to be
pedagogical, and the result can be useful as a subroutine in several
optimization algorithms.
\begin{remark}
  Quantum state tomography is another possible approach to output a
  vector with a quantum algorithm. If we have a quantum algorithm that
  encodes its solution in the amplitudes of a quantum state, and we
  want a classical description of the solution, we can obtain it with
  a tomography algorithm. For example, quantum linear systems
  algorithms encode the solution of a linear system in the amplitudes
  of a quantum state, see Ch.~\ref{ch:blockenc} and the notes
  therein. Note that the cost of a quantum state tomography algorithm
  to obtain a classical description of the desired state may be large,
  and therefore it is usually a good idea to try to avoid performing
  an expensive tomography step, see the discussion in
  Sect.~\ref{sec:hhlir}.
\end{remark}

The idea for the algorithm is the following. In
Sect.~\ref{sec:jordanlinear} we have seen that the gradient algorithm
(Alg.~\ref{alg:gradient}) is efficient for some form of gradient
computation: it outputs the gradient of a linear function with a
single application of (some implementation of) that function. If we
can construct a unitary implementing a function such that its gradient
is a description of the quantum state, we can use the gradient
algorithm to obtain that description. Let $U\ket{\v{0}} = \ket{\psi} =
\sum_{\vj \in \{0,1\}^n} \alpha_{j} \ket{\vj}$ be an $n$-qubit quantum
state constructed by the unitary $U$, and let $d = 2^n$. The
function $f(x) := \sum_{\vj \in \{0,1\}^n} \alpha_j^{\dag} x_j =
\bra{\psi} \left( \sum_{\vj \in \{0,1\}^n} x_j \ket{\vj} \right)$ is a
linear function such that its gradient is precisely the vector
$\alpha^{\dag}$, i.e., a classical description of $\ket{\psi}$. If we
can construct a phase oracle for $f(x)$, then we can apply the
gradient algorithm to it. (In fact, the algorithm described below
obtains a description of the real part of the entries of
$\alpha^{\dag}$, so the complex conjugate is irrelevant.)

To construct the phase oracle we use a construction for the inner
product of two quantum states, given access to unitaries that prepare
them. This can be done in many ways, including via a probability
oracle inspired by a construction usually known as \emph{Hadamard
test}; we give below such a construction. Readers familiar with the
Hadamard test may recognize its basic structure. (We use $\Re$ for the
real part and $\Im$ for the imaginary part of a complex number or
vector, component-wise.)
\begin{figure}[ht!]
    \leavevmode
    \centering
    \ifcompilefigs
    \Qcircuit @C=1em @R=0.7em {
      \lstick{\ket{0}}     & \qw & \gate{H} & \qw & \gate{X} & \ctrl{1} & \gate{X} & \ctrl{1} & \qw & \gate{H} & \gate{X} & \qw \\
      \lstick{\ket{\v{0}}} & \qw & \qw      & \qw & \qw      & \gate{U} & \qw      & \multigate{1}{V} & \qw & \qw      & \qw      & \qw \\
      \lstick{\ket{\v{x}}} & \qw & \qw      & \qw & \qw      & \qw      & \qw      & \ghost{V}        & \qw & \qw      & \qw      & \rstick{\ket{\v{x}}} \qw\\
    }
    \else
    \includegraphics{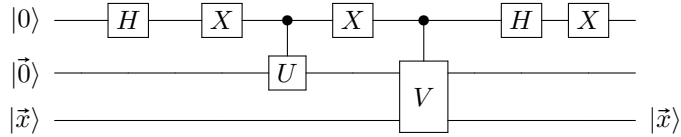}
    \fi
    \caption{Probability oracle for a function encoding a quantity proportional to the inner product of two quantum states.}
    \label{fig:probinnerprod}
\end{figure}
\begin{proposition}
  \label{prop:probinnerprod}
  Suppose we have two unitaries $U, V$ such that $U \ket{\v{0}} =
  \ket{\psi}$ and $V \ket{\v{0}} \ket{\v{x}} = \ket{\phi_x}
  \ket{\v{x}}$, and controlled version of them. Then the circuit in
  Fig.~\ref{fig:probinnerprod} is a probability oracle for the
  function $f(x) := \frac{1}{2}(1 + \Re \braket{\psi}{\phi_x})$, and
  it uses a single application of controlled-$U$ and controlled-$V$,
  plus a constant number of single-qubit gates.
\end{proposition}
\begin{proof}
  Let us analyze the circuit in Fig.~\ref{fig:probinnerprod}. After
  the first Hadamard gate, the state of the system is
  $\frac{1}{\sqrt{2}}(\ket{0} + \ket{1})\ket{\v{0}}\ket{\vx}$.
  Controlled-$U$ acts on the bottom qubit lines when the top qubit is
  $\ket{0}$, because it is sandwiched between $X$ gates, whereas
  controlled-$V$ acts when the top qubit is $\ket{1}$. Thus, after
  controlled-$V$, we are in the state
  $\frac{1}{\sqrt{2}}(\ket{0}\ket{\psi}\ket{\vx} +
  \ket{1}\ket{\phi_x}\ket{\vx})$. The final Hadamard gate produces:
  \begin{align*}
    \frac{1}{2}\left(\ket{0}(\ket{\psi}\ket{\vx} + \ket{\phi_x}\ket{\vx}) +
    \ket{1}(\ket{\psi}\ket{\vx} - \ket{\phi_x}\ket{\vx})\right).
  \end{align*}
  The probability of observing $\ket{0}$ when measuring the first
  qubit is therefore:
  \begin{align*}
    \nrm{\frac{1}{2}\left(\ket{\psi}\ket{\vx} +
      \ket{\phi_x}\ket{\vx}\right)}^2 = \frac{1}{4}\left(\bra{\psi}\bra{\vx}
    + \bra{\phi_x}\bra{\vx}\right)\left(\ket{\psi}\ket{\vx} +
    \ket{\phi_x}\ket{\vx}\right) &= \frac{1}{4}(2 + \braket{\psi}{\phi_x} +
    \braket{\phi_x}{\psi}) \\ &= \frac{1}{2}(1 + \Re
    \braket{\psi}{\phi_x}).
  \end{align*}
  The final $X$ gate bit-flips the first qubit, ensuring that the
  probability of observing $\ket{1}$ is the above expression. 
  This concludes the proof.
\end{proof}

\noindent Prop.~\ref{prop:probinnerprod} shows how to implement a
probability oracle for a function that involves the inner product
$\braket{\psi}{\phi_x}$; using Thm.~\ref{thm:probtophase}, we can then
convert it to a phase oracle, which in turn can be employed directly
as an input for Alg.~\ref{alg:gradient}. However, for
Prop.~\ref{prop:probinnerprod} to implement exactly the gadget that we
need, we still need to specify the precise form of $\ket{\phi_x}$.

Recall that our idea is to implement a tomography algorithm by
constructing a phase oracle for the function:
\begin{equation*}
  f(x) = \bra{\psi}\left(\sum_{\vj \in \{0,1\}^n} x_{j}
  \ket{\vj}\right),  
\end{equation*}
i.e., the inner product of $x$ with $\ket{\psi}$: if we can do this,
the gradient algorithm recovers $\nabla_x f(\zeroes)$, which is
precisely $\ket{\psi}$. In the context of
Prop.~\ref{prop:probinnerprod}, this means that we need $\ket{\phi_x}$
to be the state $\sum_{\vj \in \{0,1\}^n} x_{j} \ket{\vj}$. Hence, we
need a unitary that maps $\ket{\vx} \in \{0,1\}^{dq}$ to an $n$-qubit
state with amplitudes $x_0, x_1, \dots, x_{d-1}$, where the string
$\vx$ is interpreted as a vector with each component encoded on $q$
bits (we use zero-based indices for $x$, because we index its
components with binary strings). We call this operation
\emph{amplitude encoding}, as defined in Def.~\ref{def:ampenc}: we
postpone a proper definition to Sect.~\ref{sec:ampencalg} because here
we need a slightly different normalization, so to avoid confusion, we
do not introduce our shorthand notation for the amplitude encoding
until a bit later. The important part, however, is that such a unitary
is not difficult to construct, using controlled single-qubit
rotations: we give a full description of a circuit to implement it in
Sect.~\ref{sec:ampencalg}.

An important detail that must be considered is the normalization
factor for the state $\sum_{\vj \in \{0,1\}^n} x_{j}
\ket{\vj}$. Unitaries can only construct proper quantum states, i.e.,
unit vectors, thus we must ensure that the mapping $\ket{\vx} \to
\sum_{\vj \in \{0,1\}^n} x_{j} \ket{\vj}$ is unitary. In
Prop.~\ref{prop:jordanlinearapx} the function $f$ is assumed to have
domain $[-\frac{1}{2}, \frac{1}{2}]^d$, and in
Thm.~\ref{thm:gradientpoly} it has domain $[-1,1]^d$; indeed,
Alg.~\ref{alg:gradient} starts by constructing a superposition of grid
points symmetric about the origin. The argument of $f(x)$ therefore
lives in $[-1,1]^d$; the maximum Euclidean norm of such a vector is
$\sqrt{d}$. (The smaller box $[-\frac{1}{2}, \frac{1}{2}]^d$ would be
sufficient, but this only makes a constant-factor difference in the
norm, so we keep $[-1,1]^d$ to simplify calculations.) Each vector $x
\in [-1,1]^d$ may have a different norm, so we cannot hope to
construct the ``quantum state'' $\frac{1}{\sqrt{d}} \sum_{\vj \in
  \{0,1\}^n} x_{j} \ket{\vj}$ for all $x \in [-1,1]^d$: it is not
always a properly normalized quantum state. On the other hand, we want
to keep the same normalization factor regardless of the value of $x$,
otherwise $f(x)$ would not be linear. We resolve this issue by relying
on a construction for $\frac{1}{\sqrt{d}} \sum_{\vj \in \{0,1\}^n}
x_{j} \ket{\vj}$ that is not always successful. Specifically, to
ensure that some version of the mapping $\ket{\vx} \to
\frac{1}{\sqrt{d}} \sum_{\vj \in \{0,1\}^n} x_{j} \ket{\vj}$ is
well-defined (and unitary) for all input values, we define a unitary
that acts as follows:
\begin{equation*}
  V_{\text{amp}} \ket{0}\ket{\v{0}}_n\ket{\vx}_{dq} =
  \left(\ket{0}\left(\frac{1}{\sqrt{d}} \sum_{\vj \in \{0,1\}^n} x_{j}
  \ket{\vj}\right) + \ket{1}\left(\frac{1}{\sqrt{d}} \sum_{\vj \in \{0,1\}^n} \sqrt{1-x_{j}^2}\ket{\vj}\right)\right)\ket{\vx}.
\end{equation*}
In this way, the output is a normalized quantum state, and the first
register acts as a flag register: when its value is $\ket{0}$, we have
produced the desired state $\frac{1}{\sqrt{d}} \sum_{\vj \in
  \{0,1\}^n} x_{j} \ket{\vj}$, and when its value is $\ket{1}$, the
circuit has failed to produce the desired state. This unitary can be
constructed with $\bigOt{dq}$ gates, where --- as before --- $q$ is
the number of bits for each component of $x$, i.e., $\vx$ is a
$dq$-digit bitstring. It can be rigorously proven that it is
sufficient to pick $q$ polynomial in the size of the input unitary
$U$, because this already yields the necessary precision for each
component of $x$ (intuitively: the precision of each number increases
exponentially with the number of binary digits, so with $\bigO{n}
= \bigO{\log d} = \bigOt{1}$ digits we already achieve
exponentially-high precision $\bigO{2^{-d}}$). This simplifies the
expression for the number of gates to $\bigOt{d}$.

Putting everything together, we do the following:
\begin{itemize}
\item We apply the circuit described in
  Prop.~\ref{prop:probinnerprod}, choosing $U$ to be the unitary that
  prepares the state $\ket{\psi}$ of which we want a description, and
  choosing $V$ to be the unitary $V_{\text{amp}}$ that produces the
  vector $\frac{1}{\sqrt{d}} \sum_{\vj \in \{0,1\}^n} x_{j} \ket{\vj}$
  with a flag register. This yields a probability oracle for
  \begin{equation}
    \label{eq:tomofunc}
    f(x) = \bra{\psi}\left(\frac{1}{\sqrt{d}} \sum_{\vj \in \{0,1\}^n}
    x_{j} \ket{\vj}\right).
  \end{equation}
\item We convert it to a phase oracle using
  Thm.~\ref{thm:probtophase}.
\item We apply the gradient algorithm, Alg.~\ref{alg:gradient}.
  Because the function $f(x)$ is linear, we can set $k = 1$ in
  Thm.~\ref{thm:gradientpoly}. To recover the (real
  part of the) amplitudes $\alpha_{j}$ of $\ket{\psi}$ to precision
  $\epsilon$ we need to set the precision in
  Thm.~\ref{thm:gradientpoly} to $\epsilon/\sqrt{d}$, because in the
  definition of the function of Eq.~\eqref{eq:tomofunc} each component
  is scaled down by $\sqrt{d}$, requiring us to increase
  precision. This recovers the real part of $\alpha_{j}$ up to
  precision $\epsilon$ for all $j$.
\item We repeat the same algorithm adding a phase gate in the
  controlled operation to multiply $\ket{\psi}$ by the imaginary unit
  $i$, and recover the imaginary part of $\alpha_{j}$.
\end{itemize}
In this description we have performed a few simplifications. Notably,
the fact that $V_{\text{amp}}$ has a flag register introduces some
difficulty, because repeating the calculations of
Prop.~\ref{prop:probinnerprod} with the added flag register yields an
undesirable extra term in the probability oracle. This should be
expected: in some sense, the mapping $V_{\text{amp}}$ is not always
successful, because it produces $\frac{1}{\sqrt{d}} \sum_{j} x_{j}
\ket{\vj}$ only with some probability, i.e., when the flag register is
$\ket{0}$.  Nonetheless, the undesirable extra term can be eliminated
when converting from probability to phase oracle: we leave these
details as an exercise. We obtain the following.
\begin{theorem}[Quantum state tomography with element-wise error; Thm.~5.1 in \cite{nannitomography}]
  \label{thm:tomography}
  Let $U\ket{\v{0}} = \ket{\psi} = \sum_{\vj \in \{0,1\}^n} \alpha_{j}
  \ket{\vj}$ be a quantum state. Let $d = 2^n$. Using the quantum
  gradient algorithm in the way described in this section, with
  probability at least $1 - \delta$ we output $\tilde{\alpha} \in
  \R^{d}$ such that $\abs{\Re(\alpha_{j}) - \tilde{\alpha}_{j}} \le
  \epsilon$ for all $j$, using $\bigOt{\sqrt{d}/\epsilon}$
  applications of $U$ and $U^{\dag}$, and $\bigOt{d^{1.5}/\epsilon}$
  additional gates. A small modification of the same algorithm outputs
  $\tilde{\alpha} \in \R^{d}$ such that $\abs{\Im(\alpha_{j}) -
    \tilde{\alpha}_{j}} \le \epsilon$ for all $j$ with the same
  running time.
\end{theorem}
The gate count can be determined by noticing that we use precision
$\sqrt{d}/\epsilon$ in Thm.~\ref{thm:gradientpoly}, and each call to
$V_{\text{amp}}$ takes $\bigOt{d}$ gates. 

Due to the relationship between Euclidean distances between quantum
states and total variation distance stated in
Prop.~\ref{prop:euclideantotvd}, as well as the pervasiveness of the
Euclidean distance in many contexts, often one is interested in
obtaining a classical description of a quantum state with a bound on
the maximum error in Euclidean distance. It is sufficient to set the
error in Thm.~\ref{thm:tomography} to $\epsilon/\sqrt{d}$: if each
amplitude is estimated with that precision, the resulting vector has
Euclidean distance at most $\epsilon$ from the true vector. Formally,
we have the following corollary.
\begin{corollary}
  \label{cor:l2tomography}
  Let $U\ket{\v{0}} = \ket{\psi} = \sum_{\vj \in \{0,1\}^n} \alpha_{j}
  \ket{\vj}$ be a quantum state. Let $d = 2^n$. Using the quantum
  gradient algorithm in the way described in this section, with probability at least $1 - \delta$ we output
  $\tilde{\alpha} \in \R^{d}$ such that $\nrm{\Re(\alpha) -
    \tilde{\alpha}} \le \epsilon$ using $\bigOt{d/\epsilon}$
  applications of $U$ and $U^{\dag}$, and $\bigOt{d^{2}/\epsilon}$
  additional gates. A small modification of the same algorithm outputs
  $\tilde{\alpha} \in \R^{d}$ such that $\nrm{ \Im(\alpha) -
    \tilde{\alpha}} \le \epsilon$ with the same running time.
\end{corollary}
Additional discussion on quantum state tomography can be found
in the notes in Sect.~\ref{sec:gradientnotes}.\index{state!tomography|)}

\subsection{Separation from membership oracles}
\label{sec:gradientsepa}
The seminal work of Gr\"otschel, Lov\'asz and Schrijver
\cite{grotschel1988geometric} highlighted a powerful connection
between membership, separation\index{separation|(}, optimization,
violation, and validity in the context of convex optimization. Each of
these five concepts admits a proper mathematical definition, and can
be turned into a question such as ``does the point $x$ belong to a
given set $K$'' (membership) and ``is there a hyperplane separating
$x$ from $K$'' (separation). As \cite{grotschel1988geometric} shows,
all the resulting questions are equivalent under polynomial-time
transformations. In this section we show that a crucial transformation
(separation using a membership oracle) can be made more efficient if
one has access to a membership oracle implemented on a quantum
computer, by means of the gradient algorithm. We focus our discussion
on three of the five oracles: membership, separation, and
optimization. For the relationship between the two remaining oracles
(violation and validity) and the aforementioned three, we refer to
\cite{grotschel1988geometric}.

Formally, consider the following type of optimization problem:
\begin{equation}
  \label{eq:convexopt}
  \min_{x \in K} \dotp{c}{x},
\end{equation}
where $K \subset \R^d$ is a convex set, and $c \in \R^d$. It is well
known that algorithms that can solve problem \eqref{eq:convexopt}
can also solve the more general convex optimization
problem $\min_{x \in K} f(x)$, with $f : \R^d \to \R$ convex, via a
simple lifting to the $(n+1)$-dimensional problem:
\begin{equation*}
  \min_{(x,y) \in K'} y, \qquad \text{ with } K' := \left\{(x, y) \in K \times \R : f(x) \le y \le F\right\},
\end{equation*}
where $F \in \R$ is some given upper bound on $f(x)$. Thus, problem
\eqref{eq:convexopt} encompasses a large class of optimization
problems. Under fairly general assumptions, if we can separate a point
$x \not\in K$ from $K$, then we can optimize $\dotp{c}{x}$ over $K$;
and conversely, if we can optimize a linear function over $K$, then we
can separate. Similarly, being able to separate from $K$ and being
able to determine membership of a point in $K$ are two aspects of the
same question, and we can transform between the two. For a fixed set
$K$, the efficiency of the transformation between oracles that
determine membership, separation and optimization can vary depending
on the algorithm used, and, as discussed in this section, on the model
of computation. An application of the quantum gradient algorithm
provides an exponential query complexity advantage in constructing a
separating hyperplane using a membership oracle for the set $K$,
compared to classical algorithm: the classical separation algorithm
uses $\bigOt{n}$ calls to a membership oracle, whereas the quantum
algorithm uses $\bigOt{1}$ (but the membership oracle is implemented
by a unitary that can be called in superposition, of course). The
relationship between the three oracles, defined formally below, is
summarized in Fig.~\ref{fig:oraclerel}.
\begin{figure}[tb]
  \centering
  \ifcompilefigs
  \begin{tikzpicture}[node distance = 5cm]
    \tikzstyle{block} = [rectangle, draw, text width=6em, text centered, rounded corners, minimum height=3em]
    \node [block] (mem) {\small $\mathrm{Membership}$};
    \node [block, right of=mem] (sep) {\small $\mathrm{Separation}$};
    \node [block, right of=sep] (opt) {\small $\mathrm{Optimization}$};
		
    \path [draw,transform canvas={yshift=-0.35cm},->] (mem) to node[below] {Classical: $\bigOt{d}$} (sep);
    \path [draw,->,bend right=45] (mem.south) to node[below] {Quantum: $\bigOt{1}$} (sep.south);
    \path [draw,transform canvas={yshift=0.35cm},->] (sep) to node[above] {$\bigOt{1}$} (mem);
    \path [draw,transform canvas={yshift=-0.35cm},->] (sep) to node[below] {$\bigOt{d}$}  (opt);
    \path [draw,transform canvas={yshift=0.35cm},->] (opt) to node[above] {$\bigOt{d}$}  (sep);
  \end{tikzpicture}
  \else
  \includegraphics{figures/oraclerel.pdf}
  \fi
  \caption{Relationship between membership, separation and optimization oracles for a fixed $d$-dimensional convex set. An arrow from oracle $a$ to oracle $b$ indicates that one can simulate oracle $b$ with a certain number of calls to oracle $a$; the number of calls is indicated next to the arrow. These results are based on \cite{grotschel1988geometric,lee2018efficient,apeldoorn2020convex,chakrabarti2020quantum}.}
  \label{fig:oraclerel}
\end{figure}

In this section we give an overview of the main idea that leads to
this quantum advantage. We sketch some of the proofs, while keeping
the discussion at a high-enough level to leave out cumbersome details:
although the framework is elegant and most of the ideas admit an
intuitive explanation, all rigorous treatments of the subject that we
are aware of rely --- to some extent --- on heavy technical tools, and
lead to unwieldy constants that would detract from the pedagogical
purpose of this section. A detailed development can be found in our
main references for this section:
\cite{grotschel1988geometric,lee2018efficient} for the classical
algorithm, \cite{apeldoorn2020convex,chakrabarti2020quantum} for the
quantum algorithm.
\begin{remark}
  An algorithm to solve the convex optimization problem
  \eqref{eq:convexopt} using only a membership oracle for the set $K$
  is extremely general: typically, even a ``basic working
  description'' of a set lets us determine if a given point belongs to
  it or not. However, it is not the most efficient approach for most
  situations: with some knowledge of the structure of $K$, it is
  usually possible to rely on, or construct, more efficient
  algorithms. Convex optimization with a cutting plane method, i.e.,
  via separation (not necessarily through a membership oracle), can
  lead to algorithms that are theoretically very efficient, e.g.,
  \cite{lee2015faster} (although to the best of our knowledge, the
  algorithm of \cite{lee2015faster} has not led to practical
  implementations of algorithms for the numerical solution of convex
  optimization problems).
\end{remark}

\paragraph{Preliminaries.} We need to introduce several concepts to formally characterize our assumptions for problem \eqref{eq:convexopt}, and to define the oracles that constitute the main topic for the section. First, we define the concept of a ball, with distance measured according to the $\ell^p$-norm. Then, we define the neighborhood of a set (i.e., the set containing all small balls centered at points in the set) and the interior of a set (i.e., the set of all points such that the ball centered at them is fully contained in the set). Finally, we define the three fundamental oracles: membership, to determine if a point is inside or outside $K$; separation, to return a hyperplane separating a point outside $K$ from $K$; and optimization, to return a maximizer of a linear function over $K$ (equivalently, we could have defined this oracle to return a minimizer).
\begin{definition}[Ball in $\ell^p$-norm]
  \label{def:ball}
  Given $r \in \R, r > 0$, and $x \in \R^d$, we define the \emph{ball
  of radius $r$ in the $\ell^p$-norm} centered at $x$ as:
  \begin{equation*}
    B_p(x, r) := \{y \in \R^d : \nrm{x-y}_p \le r \}. 
  \end{equation*}
\end{definition}
\begin{definition}[Neighborhood and interior of a convex set]
  \label{def:kball}
  Given $K \subset \R^d$, $K$ convex, and $r \in \R, r > 0$, we define
  the \emph{$r$-neighborhood of $K$} as:
  \begin{equation*}
    B_p(K, r) := \{y \in \R^d : \nrm{x-y}_p \le r \text{ for some } x \in K \}. 
  \end{equation*}
  We also define the \emph{$r$-interior of $K$} as:
  \begin{equation*}
    B_p(K, -r) := \{y \in \R^d : B_p(y, r) \subseteq K\}. 
  \end{equation*}
\end{definition}
\begin{definition}[Membership, separation and optimization oracles]
  \label{def:3oracles}
  Fix a convex set $K \subset \R^d$. Let $\epsilon, \delta > 0$ be given. The return values of the oracles defined below are all assumed to be correct with probability at least $1 - \delta$.

  A \emph{membership oracle}, denoted
  $\text{MEM}_{\epsilon,\delta}(x)$, given $x \in \R^d$ asserts that
  $x \in B_2(K, \epsilon)$ or that $x \not\in B_2(K, -\epsilon)$.

  A \emph{separation oracle}, denoted
  $\text{SEP}_{\epsilon,\delta}(x)$, given $x \in \R^d$ asserts that
  $x \in B_2(K, \epsilon)$ or it returns a vector $g \in \R^d$ such
  that $\dotp{g}{y} \le \dotp{g}{x} + \epsilon$ for all $y \in B_2(K,
  -\epsilon)$.

  An \emph{optimization oracle}, denoted
  $\text{OPT}_{\epsilon,\delta}(c)$, given $c \in \R^d$ asserts that
  $B_2(K, -\epsilon)$ is empty, or it returns a vector $x \in B_2(K,
  \epsilon)$ such that $\dotp{c}{y} \le \dotp{c}{x} + \epsilon$ for
  all $y \in B_2(K, -\epsilon)$.
\end{definition}
It is clear that an optimization oracle directly solves problem
\eqref{eq:convexopt} up to some desired precision $\epsilon$. It is
also known that a separation oracle can be used to construct an
optimization oracle with a polynomial number of
calls. \cite{grotschel1988geometric} already showed how to convert
between the three types of oracles in Def.~\ref{def:3oracles}.  The
following result characterizes the most efficient known constructive
algorithm (at the time of writing this) to build an optimization
oracle from a separation oracle.
\begin{theorem}[From separation to optimization; Thm.~42 in \cite{lee2015faster}, Thm.~15 in \cite{lee2018efficient}]
  \label{thm:sepaopt}
  Let $K \subset \R^d$ be a convex set satisfying
  $B_2(\zeroes, r) \subset K \subset B_2(\zeroes, R)$ for given $r, R \in \R$, $0
  < r < R$. Let $c \in \R^d$. For any
  $\epsilon, \delta \in \R$, $0 < \epsilon < 1$, $0 < \delta < 1$,
  with probability at least $1 - \delta$, we can compute $x \in B_2(K,
  \epsilon)$ such that:
  \begin{equation*}
    \dotp{c}{x} \le \min_{y \in K} \dotp{c}{y} + \epsilon \nrm{c},
  \end{equation*}
  with $\bigOt{d}$ calls to $\text{SEP}_{\delta',\epsilon'}(K)$ and
  $\bigOt{d^3}$ additional arithmetic operations, in expectation,
  where $\delta', \epsilon'$ are polynomially large in the instance
  parameters.
\end{theorem}
Thus, construction of a separation oracle is sufficient to perform
optimization; this explains the arrow between separation and
optimization in Fig.~\ref{fig:oraclerel}. Note that in this setting,
we know the radius of a ball contained in $K$, and of a ball
containing $K$, which avoids some degenerate cases and ensures the
origin is contained in $K$. If we only have access to a membership
oracle for $K$, which is weaker than a separation oracle, to perform
optimization we can attempt to construct a separation oracle, and then
invoke Thm.~\ref{thm:sepaopt}. As we discuss in the remainder of this
section, one possibility to construct a separation oracle from a
membership oracle is to use a subgradient of the so-called height
function, measuring the distance of a point from the boundary of $K$
along a given direction. The subgradient can be computed via finite
differences (with a classical computer), or with Jordan's gradient
algorithm (with a quantum computer). The concepts of subgradient and
approximate subgradient, used repeatedly in this section and in
Ch.~\ref{ch:mmwu}, are formally defined below. The notation $V^{\ast}$
refers to the \emph{dual vector
space}\index{notation!dualvspace@\ensuremath{V^{\ast}}} of the origin
vector space $V$ (i.e., the space of all linear forms on $V$).
\begin{definition}[Subgradient and $\epsilon$-subgradient]
  \label{def:subgradient}
  Let $V$ be a vector space. Given $f : V \to \R$ convex, $g \in
  V^{\ast}$ is called a \emph{subgradient} of $f$ at $\bar{x}$ if, for
  every $x \in V$, we have:
  \begin{equation*}
    f(x) \ge f(\bar{x}) + \dotp{g}{x - \bar{x}}.
  \end{equation*}
  $g \in V^{\ast}$ is called an \emph{$\epsilon$-subgradient} of $f$
  at $\bar{x}$ if, for every $x \in V$, we have:
  \begin{equation*}
    f(x) \ge f(\bar{x}) + \dotp{g}{x - \bar{x}} - \epsilon.
  \end{equation*}
  The set of all subgradients at $\bar{x}$, called the
  \emph{subdifferential}, is denoted $\partial f(\bar{x})$. The set of
  all $\epsilon$-subgradients at $\bar{x}$, called the
  \emph{$\epsilon$-subdifferential}, is denoted $\partial_{\epsilon}
  f(\bar{x})$.
\end{definition}

\paragraph{Classical separation oracle.} We give a rough idea of the approach delineated in \cite{lee2018efficient}, with some simplifications from \cite{apeldoorn2020convex}, to construct a separation oracle from a membership oracle. We construct a function, called the \emph{height function}, that measures a notion of distance between a given point $z$ and the boundary of $K$. Suppose we are trying to separate a point that lies on the ``vertical'' axis (see below for why this is w.l.o.g.). An $\bar{\epsilon}$-subgradient of the height function at the origin gives the separating hyperplane, for an appropriate value of $\bar{\epsilon}$. Approximating the subgradient can be difficult because the height function is not necessarily continuous, but it is Lipschitz continuous. Fortunately, it can be shown that the expected value of the subgradient at points randomly sampled from a small ball centered at the origin is a good approximation of the sought subgradient. 

We assume that we are trying to separate the point $x = -\tau e_d$
from $K$ with $\tau \in \R, \tau > 0$, i.e., a point where the first
$d-1$ coordinates are zero and the last coordinate is negative. This
is w.l.o.g., because we can always rotate the coordinate system to
verify this condition; the rotation can be determined without much
effort, and it would not affect the query complexity of the membership
oracle anyway. (Imagining the situation in three dimensions gives a
good intuition of how to compute a rotation that zeroes the first
$d-1$ coordinates, and makes the last one negative, using a sequence
of two-dimensional rotations.) The height function $h : \R^{d-1} \to
\R \cup \{\infty\}$ is defined as:
\begin{equation*}
  h(y) := \inf_{(y, t) \in K} t,
\end{equation*}
i.e., $h(y)$ is the smallest value such that $(y, h(y)) \in K$ if such
a value exists, and it is $\infty$ if no such value exists. This is
the ``height'' of the $(d-1)$-dimensional hyperplane where the last
coordinate is zero with respect to the boundary of the $d$-dimensional
set $K$, see Fig.~\ref{fig:heightfun}.
\begin{figure}[tb]
  \centering
  \ifcompilefigs
  \begin{tikzpicture}

    \draw[-, gray] (-4, 0) -- (4, 0);
    \draw[-, gray] (0, -3.5) -- (0, 1.5);
    
    \draw[thick] (0,0) circle (1);
    \draw[dashed,rotate=135] (0,0) -- (1,0);
    \node at (-0.5, 0.2) {$r$};
    \node at (-3, -2) {\large $K$};

    \filldraw[black] (0,0) circle (2pt) node[below right] {$(0, 0)$};
    
    \draw[thick] (-4,-1) arc (0:180:-4cm and -2cm); 

    \filldraw[black] (2,0) circle (2pt) node[above] {$(y, 0)$};
    \filldraw[black] (2,-2.73) circle (2pt) node[below right] {$(y, h(y))$};
    
    \draw[dashed] (2,0) -- (2,-2.73);        
  \end{tikzpicture}
  \else
  \includegraphics{figures/heightfun.pdf}
  \fi
  \caption{Visual representation of the height function $h(y)$ in two dimensions, i.e., with $K \subset \R^2$ and $h : \R \to \R$.}
  \label{fig:heightfun}
\end{figure}
This function has several crucial properties,
which we state without proof.
\begin{proposition}[Based on Lem.~19 and Lem.~21 in \cite{apeldoorn2020convex}]
  \label{prop:heightfun}
  Let $K \subset \R^d$ be a convex set satisfying $B_2(\zeroes, r) \subset K
  \subset B_2(\zeroes, R)$ for given $r, R \in \R$, $0 < r < R$. Let $x =
  -\tau e_d \not\in B(K, -\epsilon)$. The function $h : \R^{d-1} \to
  \R \cup \{\infty\}$, $h(y) := \inf_{(y, t) \in K} t$ satisfies the following:
  \begin{enumerate}[(i)]
  \item it is convex.
  \item it is Lipschitz continuous with Lipschitz constant $\frac{2R}{r}$
    over $B_2(\zeroes, r/2)$;
  \item if we take $g \in \partial_{\bar{\epsilon}}h(\zeroes)$ for an
    appropriately chosen value of $\bar{\epsilon}$, and let $a = (g,
    1) \in \R^d$, the half-space $\dotp{a}{z} \ge \dotp{a}{x} -
    \bar{b}$ separates $x$ and $K$ for an appropriately chosen
    $\bar{b} \in \R$, i.e., $\dotp{a}{z} \ge \dotp{a}{x} - \bar{b}$
    for all $z \in K$.
  \end{enumerate}
\end{proposition}
We omit a discussion on the values of $\bar{\epsilon}, \bar{b}$ for
ease of exposition. It is not difficult to see that the height
function can be approximated using binary search exploiting the
membership oracle for $K$: given $y \in \R^{d-1}$, we consider points
in $\R^d$ of the form $(y, t)$, where $t \in \R$, and we perform
binary search on $t$ using the membership oracle to determine the
smallest value of $t$ such that $(y, t) \in K$. Then we can estimate a
subgradient of $h$ at the origin with a finite-difference
approximation, and by Prop.~\ref{prop:heightfun} (in particular
property (iii)), this yields the sought separating hyperplane.
\begin{definition}[Finite-difference gradient approximation]
  \label{def:finitediff}
  Let $f : \R^d \to \R$, and $\eta \in \R$, $\eta > 0$. The \emph{finite-difference gradient approximation} of $f$ at $x \in \R^d$ with step $\eta$ is the vector:
  \begin{equation*}
    \nabla_{(\eta)} f (x) := \begin{pmatrix}
      \frac{f(x + \eta e_1) - f(x - \eta e_1)}{2\eta} \\
      \frac{f(x + \eta e_2) - f(x - \eta e_2)}{2\eta} \\
      \vdots \\
      \frac{f(x + \eta e_d) - f(x - \eta e_d)}{2\eta}
    \end{pmatrix}.
  \end{equation*}
\end{definition}
Def.~\ref{def:finitediff} gives a symmetric formula for a gradient
approximation. The error of $\nabla_{(\eta)} f (x)$ can be upper bounded
in terms of an approximation of the trace of the Hessian matrix.
\begin{definition}[Finite-difference Laplace approximation]
  \label{def:laplaceapx}
  Let $f : \R^d \to \R$, and $\eta \in \R$, $\eta > 0$. The \emph{finite-difference Laplace approximation} of $f$ at $x \in \R^d$ with step $\eta$ is the scalar:
  \begin{equation*}
    \Delta_{(\eta)} f (x) := \sum_{j=1}^{d} \frac{f(x + \eta e_j) - 2f(x) + f(x - \eta e_j)}{\eta^2}.
  \end{equation*}
\end{definition}
\begin{lemma}[Lem.~10 in \cite{apeldoorn2020convex}]
  \label{lem:finitedifflaplace}
  Let $\eta \in \R, \eta > 0$, $x \in \R^d$, $f : B_1(x, \eta) \to \R$
  convex. Then:
  \begin{equation*}
    \sup_{g \in \partial f(x)} \nrm{g - \nabla_{(\eta)} f(x)}_1 \le \frac{1}{2} \eta \Delta_{(\eta)} f(x).
  \end{equation*}
\end{lemma}
Thus, if $\Delta_{(\eta)} f(x)$ is small, $\nabla_{(\eta)} f(x)$ is a good
approximation of a subgradient. We want to apply this result to
obtain a subgradient of the height function $h(y)$ at the origin by
computing $\nabla_{(\eta)} h(\zeroes)$, but $h(y)$ does not have enough
structure to guarantee a straightforward application of
Lem.~\ref{lem:finitedifflaplace}: we only know that $h(x)$ is convex
and Lipschitz continuous, and with just these two properties,
$\Delta_{(\eta)} f(x)$ could be unbounded as $\eta \to 0$.
\begin{example}
  Consider $f: \R \to \R, f(x) = \abs{x}$, which is convex and
  $1$-Lipschitz. We have:
  \begin{equation*}
    \nabla_{(\eta)} f(0) = \frac{f(\eta) - f(-\eta)}{2\eta} = 0,
  \end{equation*}
  but:
  \begin{equation*}
    \Delta_{(\eta)} f (0) = \frac{f(\eta) - 2f(0) + f(-\eta)}{\eta^2} = \frac{2}{\eta},
  \end{equation*}
  so $\Delta_{(\eta)} f (0) \to 0$ as $\eta \to 0$.
\end{example}

\noindent Thus, convexity and Lipschitz continuity alone may not
suffice to show that $\nabla_{(\eta)} f(x)$ is a good approximation of
a subgradient at $x$. As a consequence, our task of constructing a
sufficiently accurate approximation of a subgradient of the height
function at the origin is not complete yet. Fortunately, it can be
shown that picking a uniformly random point in a ball around the
origin yields small values of the Laplace approximation, in
expectation.
\begin{lemma}[Lem.~11 in \cite{apeldoorn2020convex}]
  \label{lem:laplacerandom}
  Let $\eta, r \in R$, $0 < \eta < r$, and let $f : B_{\infty}(x, r +
  \eta) \to \R$ be convex and $L$-Lipschitz. Then:
  \begin{equation*}
    \mathbb{E}_{z \in B_{\infty}(x, r)} \Delta_{(\eta)} f(z) \le \frac{dL}{r}.
  \end{equation*}
\end{lemma}
This leads to the following strategy: pick a uniform random point in a
small ball around the origin, and compute the finite-difference
approximation of the subgradient of the height function at that
point. In expectation, the Laplace approximation is small, so the
finite-difference approximation is an accurate estimate of a
subgradient at that point. Because the ball is small, it is also an
accurate (but slighty worse) estimate of a subgradient at the
origin, i.e., it belongs to $\partial_{\epsilon} h(\zeroes)$ for a
larger $\epsilon$ (recall Def.~\ref{def:subgradient}). Additionally,
it is not necessary that the finite-difference approximation uses the
exact value of $h$: we can use the estimate obtained with the binary
search procedure discussed after Prop.~\ref{prop:heightfun}, taking
advantage of the fact that the complexity of binary search scales
polylogarithmically with the reciprocal of the error. We obtain the
following result.
\begin{proposition}[Lem.~12 in \cite{apeldoorn2020convex}]
  \label{prop:randomsubgradisgood}
  Let $r, L, \epsilon, \delta \in \R$, $r > 0, L > 0$, $0 < \delta <
  1/3$, $0 < \epsilon < rL\sqrt{d}/\delta$. Let $f : \R^d \to \R$ be a
  convex function that is $L$-Lipschitz on $B_{\infty}(\zeroes, 2r)$,
  and let $\tilde{f} : \R^d \to \R$ be such that $\abs{f(x) -
    \tilde{f}(x)} \le \epsilon$ for all $x \in B_{\infty}(\zeroes,
  2r)$. Then, for a uniformly random $z \in B_{\infty}(\zeroes, r)$, we
  can compute parameters $0 < \eta < r, \epsilon' > \epsilon$ such
  that, with probability at least $1-\delta$, we have $\nabla_{(\eta)}
  f(z) \in \partial_{\epsilon'} f(\zeroes)$.
\end{proposition}
\begin{proof}
  The proof is not difficult, but it would require us to give
  cumbersome expressions for $\eta$ and $\epsilon'$. We give a
  high-level sketch: we use Lem.s~\ref{lem:finitedifflaplace} and
  \ref{lem:laplacerandom} to show that the finite-difference gradient
  approximation at a random point $z$ approximates $\partial f(z)$
  with error $\frac{\eta d L}{2r}$; since $z$ and $\zeroes$ are close,
  and $f$ is $L$-Lipschitz, it is then straightforward to compute an
  $\tilde{\epsilon}$ so that $\nabla_{(\eta)} f(z) \in
  \partial_{\tilde{\epsilon}} f(\zeroes)$ in expectation. Using
  Markov's inequality, with probability $1-\delta$ a sample from the
  random variable (i.e., a $z$ picked at random) achieves a gradient
  error bounded by $\frac{\tilde{\epsilon}}{\delta}$, which gives the
  final error parameter $\epsilon'$.
\end{proof}

With Prop.~\ref{prop:randomsubgradisgood}, the construction of the
separation oracle is essentially finished: we compute the
finite-difference gradient approximation at a random point in a small
ball around the origin, and return the hyperplane specified in
Prop.~\ref{prop:heightfun}. By Def.~\ref{def:finitediff}, the
computation of the finite-difference gradient approximation involves
$\bigOt{d}$ calls to the membership oracle (because we estimate the
height function at $\bigOt{d}$ points, and each estimation runs in
$\bigOt{1}$ iterations of binary search). This motivates the
``classical'' arrow between membership and separation in
Fig.~\ref{fig:oraclerel}.

\paragraph{Quantum separation oracle.} We can construct a separation oracle using a quantum computer in a similar manner. Prop.~\ref{prop:randomsubgradisgood} states that in expectation, the finite-difference gradient approximation of $h$ at points picked randomly in a ball around the origin is close to $\partial_{\epsilon} h(\zeroes)$. However, the expression for the finite-difference gradient approximation contains $h$ evaluated at $d$ different points, leading to the $\bigOt{d}$ query complexity (i.e., number of calls to the membership oracle) of the classical algorithm. We rely on Jordan's gradient algorithm to take advantage of a quantum computer and reduce the query complexity.

To be able to apply Jordan's gradient algorithm to estimate
$\partial_{\epsilon} h(\zeroes)$, we must first show that the function
$h$ is close to linear over an appropriate
domain. Lem.~\ref{lem:finitedifflaplace} does not suffice: it bounds
the distance of $\nabla_{(\eta)} h(z)$ from to $g \in \partial h(z)$
in terms of $\Delta_{(\eta)} h(z)$, but for Jordan's algorithm to work
we need that $h$ is close to linear almost everywhere in a small
$\ell^{\infty}$-ball, not just at a specific point
$z$. Lem.~\ref{lem:laplacerandom} still helps: because
$\Delta_{(\eta)} h(z)$ is small in expectation, it means that the rate
of change of the subgradients is also small (recall that
$\Delta_{(\eta)} h(z)$ is a finite-difference approximation of the
magnitude of the second-order derivatives). More formally, it can be
shown that $\Delta_{(\eta)} h(z)$ bounds by how much $h$ deviates from
linearity around $z$.
\begin{lemma}[Lem.~17 in \cite{apeldoorn2020convex}]
  \label{lem:gradientmaxerror}
  Let $\eta \in \R$, $\eta > 0$, let $z \in \R^d$, and let $f :
  B_1(z,\eta) \to \R$ convex. Then:
  \begin{equation*}
    \sup_{y \in B_1(\zeroes, \eta)} \abs{f(z + y) - f(z) - \dotp{\nabla_{(\eta)} f(z)}{y} } \le \frac{1}{2} \eta^2 \Delta_{(\eta)} f(z).
  \end{equation*}
\end{lemma}
\begin{proof}
  Define $g(y) := f(z + y) - f(z) - \dotp{\nabla_{(\eta)} f(z)}{y}$, i.e., $g(y)$ is the difference between $f(z+y)$ and the value at $z+y$ of the linear approximation of $f$ constructed at $z$. Note that $g(\zeroes) = 0$, and for every $j=1,\dots,d$, using the definition of $\nabla_{(\eta)} f(z)$ we have:
  \begin{align*}
    g(\pm \eta e_j) &= \frac{f(z + \eta e_j) - 2f(z) + f(z - \eta e_j)}{2}  \\
    &= \frac{\eta^2}{2} (\Delta_{(\eta)} f(z))_j \le \frac{\eta^2}{2} \Delta_{(\eta)} f(z).
  \end{align*}
  The points $\pm \eta e_j$ for $j=1,\dots,d$ are the extreme points of the set $B_1(\zeroes, \eta)$ over which we are taking the supremum. Using convexity of $g(y)$ and of $B_1(\zeroes, \eta)$, it is straightforward to show $\sup_{y \in B_1(\zeroes, \eta)} \abs{g(y)} \le \frac{1}{2} \eta^2 \Delta_{(\eta)} f(z)$.
\end{proof}

\noindent We use Lem.~\ref{lem:gradientmaxerror} to ensure that we
satisfy the conditions for Prop.~\ref{prop:jordanlinearapx}, when
applying the gradient algorithm to the height function. In particular,
Prop.~\ref{prop:jordanlinearapx} only requires that a function admits
some accurate linear approximation, and if that is the case, it
returns such an approximation: it does not require the function to be
differentiable. Thus, we can apply it to the height function. There
are still a few more obstacles to overcome; in particular, in
Alg.~\ref{alg:gradient} and consequently
Prop.~\ref{prop:jordanlinearapx}, it is assumed that $f$ is computed
exactly, but here we apply Alg.~\ref{alg:gradient} to the function
$h$, which is evaluated up to some error. The remaining issues can be
addressed using standard tools, leading to the following result.
\begin{theorem}[Separation oracle via quantum gradient; Thm.~23 in \cite{apeldoorn2020convex}]
  \label{thm:quantumsepa}
  Let $K \subset \R^d$ be a convex set satisfying $B_2(\zeroes, r)
  \subset K \subset B_2(\zeroes, R)$ for given $r, R \in \R$, $0 < r <
  R$, and let $x \not\in B(K, -\epsilon)$. For any $\epsilon, \delta
  \in \R$, $0 < \epsilon < R$, $0 < \delta < \frac{1}{3}$, we can
  implement $\text{SEP}_{\epsilon, \delta}(x)$ with $\bigOt{1}$ calls
  to a quantum oracle implementing $\text{MEM}_{0,\epsilon'}(K)$, for
  an appropriately chosen $\epsilon'$.
\end{theorem}
\begin{proof}
  We sketch the proof. By Prop.~\ref{prop:heightfun}, the height
  function $h$ is $\frac{2R}{r}$-Lipschitz over $B_2(\zeroes,
  r/2)$. By Lem.s~\ref{lem:laplacerandom} and
  \ref{lem:gradientmaxerror}, if we pick a random point in a small
  ball around the origin, the function $h$ is close to linear, so we
  can apply Prop.~\ref{prop:jordanlinearapx} to compute an approximate
  (sub)gradient. Prop.~\ref{prop:jordanlinearapx} evaluates $h$ only
  once, and each evaluation of $h$ uses $\bigOt{1}$ evaluations of the
  membership oracle.
\end{proof}

\noindent Thm.~\ref{thm:quantumsepa} motivates the ``quantum'' arrow
between membership and separation in Fig.~\ref{fig:oraclerel}, and
shows a quantum speedup in terms of query
complexity.\index{separation|)}

\section{Encoding an arbitrary vector in a quantum state}
\label{sec:ampencalg}
Given a $d$-dimensional vector $x \in \C^{d}$, in many
optimization-related contexts we may need access to its encoding as a
quantum state, i.e., as a quantum state with amplitudes corresponding
to the components of $x$.\index{amplitude!encoding|(} Already in
Sect.~\ref{sec:gradienttomo} we needed a way to map a binary
description of $x \in \R^{d}$ to $\frac{1}{\sqrt{d}} \sum_{\vj} x_j
\ket{\vj}$, in the context of state tomography (with a slightly
different normalization than the canonical $\frac{1}{\nrm{x}}$ to get
a unit vector, but this has little impact on the discussion presented
in this section); a similar construction is useful, for example, in
our discussion of quantum linear systems algorithms in
Ch.~\ref{ch:blockenc}. We introduce a shorthand notation for this type
of encoding of a vector, because it is used multiple times in the rest
of this \book{}.
\begin{definition}[Amplitude encoding]
  \label{def:ampenc}
  Given a vector $x \in \C^d$, we denote its {\em amplitude
    encoding} by
  \begin{equation*}
    \ket{\amp{x}} := \sum_{\vj \in \{0,1\}^n}
    \frac{x_{j}}{\nrm{x}} \ket{\vj},
  \end{equation*}
  where $n = \ceil{\log d}$.
\end{definition}
In this section we describe a classical procedure that, starting from
a classical description of the vector $x$, produces the description of
a quantum circuit that maps $\ket{\v{0}} \to \ket{\amp{x}}$.
\begin{remark}
  \label{rem:quantumimplstateprep}
  We describe a classical procedure for the amplitude encoding that
  works as follows: the procedure takes as input a binary description
  of $x$, and outputs a quantum circuit that produces $\ket{\amp{x}}$.
  Because of this, we can also write a quantum circuit for the same
  task, i.e., a quantum circuit that starts from a basis state
  encoding $x$ as a binary string, and outputs (in a different
  register) the state $\ket{\amp{x}}$ (that is, a quantum state with
  coefficients given by $x$). To do so, we run a quantum circuit
  that performs the same steps as the classical procedure, and rather
  than simply outputting the description of the quantum circuit to
  construct $\ket{\amp{x}}$, we apply the corresponding operations
  with controlled gates onto a fresh register initialized as
  $\ket{\v{0}}$. Similar considerations apply to any classical
  procedure that outputs the description of a quantum circuit to
  perform a given task.
\end{remark}

The construction given in this section is essentially a specialized
version of the scheme for creating the quantum encoding of efficiently
integrable distributions described in
\cite{grover2002creating}. Assume $d = 2^n$ for simplicity (we can
always pad $x$ with zero entries if its dimension is not a power of
two), and assume $\nrm{x} = 1$ because we can only amplitude-encode
unit vectors; with this assumption we do not have to keep track of the
factor $\frac{1}{\nrm{x}}$ in Def.~\ref{def:ampenc}, and of course an
algorithm for $x: \nrm{x} = 1$ can be applied to a general $x$ by
normalizing it in a preprocessing step. The construction performs a
classical preprocessing and then produces a quantum circuit that maps
$\ket{\v{0}} \to \ket{\amp{x}}$ for real $x$, see
Rem.~\ref{rem:quantumimplstateprep}. We discuss the case for complex
$x$ subsequently.

\begin{figure}[tb]
  \centering
  \ifcompilefigs
  \begin{tikzpicture}[
      ell/.style 2 args={
        ellipse,
        minimum width=2.25cm,
        minimum height=1.4cm,
        draw,
        label={[name=#1]center:#2},
      },
      connection/.style={thick,densely dotted, latex-latex},
      scale=0.9
    ]
    \node [ell={la}{$(x_0^2, \text{sign}(x_0))$},   fill=blue!20] (a) at (0,0) {};
    \node [ell={lb}{$(x_1^2, \text{sign}(x_1))$},   fill=blue!20] (b) at (4,0) {};
    \node [draw=none] (na) at (6,0) {\dots\dots};
    \node [ell={lc}{$\begin{array}{c}(x_{d-2}^2,\\ \text{sign}(x_{d-2}))\end{array}$},   fill=blue!20] (c) at (8,0) {};
    \node [ell={ld}{$\begin{array}{c}(x_{d-1}^2,\\ \text{sign}(x_{d-1}))\end{array}$},   fill=blue!20] (d) at (12,0) {};

    \node [ell={le}{$x_0^2 + x_1^2$},   fill=yellow!20] (e) at (2,2) {};
    \node [draw=none] (nb) at (6,2) {\dots\dots};
    \node [ell={lf}{$x_{d-2}^2 + x_{d-1}^2$},   fill=yellow!20] (f) at (10,2) {};

    \node [draw=none] (nc) at (2,3.5) {\vdots};
    \node [draw=none] (nd) at (10,3.5) {\vdots};

    \node [ell={lg}{$\displaystyle \sum_{j=0}^{d/4-1} x_j^2$},   fill=yellow!20] (g) at (2,5) {};
    \node [draw=none] (ne) at (6,5) {\dots\dots};
    \node [ell={lh}{$\displaystyle \sum_{j=3d/4}^{d-1} x_j^2$},   fill=yellow!20] (h) at (10,5) {};

    \node [ell={li}{$\displaystyle \sum_{j=0}^{d/2-1} x_j^2$},   fill=yellow!20] (i) at (4,7) {};
    \node [ell={ll}{$\displaystyle \sum_{j=d/2}^{d-1} x_j^2$},   fill=yellow!20] (l) at (8,7) {};

    \node [ell={lm}{$\displaystyle \sum_{j=0}^{d-1} x_j^2$},   fill=yellow!20] (m) at (6,9) {};

    \node [draw=none] (la) at (-2.5,0) {Level $n$};
    \node [draw=none] (lb) at (-2.5,2) {Level $n-1$};
    \node [draw=none] (lc) at (-2.5,5) {Level $2$};
    \node [draw=none] (ld) at (-2.5,7) {Level $1$};
    \node [draw=none] (le) at (-2.5,9) {Level $0$};
    \node [draw=none] (lf) at (-2.5,3.5) {\vdots};

    \node [draw=none] (lg) at (0,-1.1) {Element $0$};
    \node [draw=none] (lh) at (4,-1.1) {Element $1$};
    \node [draw=none] (li) at (8,-1.1) {Element $d-2$};
    \node [draw=none] (lj) at (12,-1.1) {Element $d-1$};

    \draw[-] (a) -- (e);
    \draw[-] (b) -- (e);
    \draw[-] (c) -- (f);
    \draw[-] (d) -- (f);

    \draw[-] (g) -- (i);
    \draw[-] (ne) -- (i);
    \draw[-] (h) -- (l);
    \draw[-] (ne) -- (l);

    \draw[-] (i) -- (m);
    \draw[-] (l) -- (m);

  \end{tikzpicture}
  \else
  \includegraphics{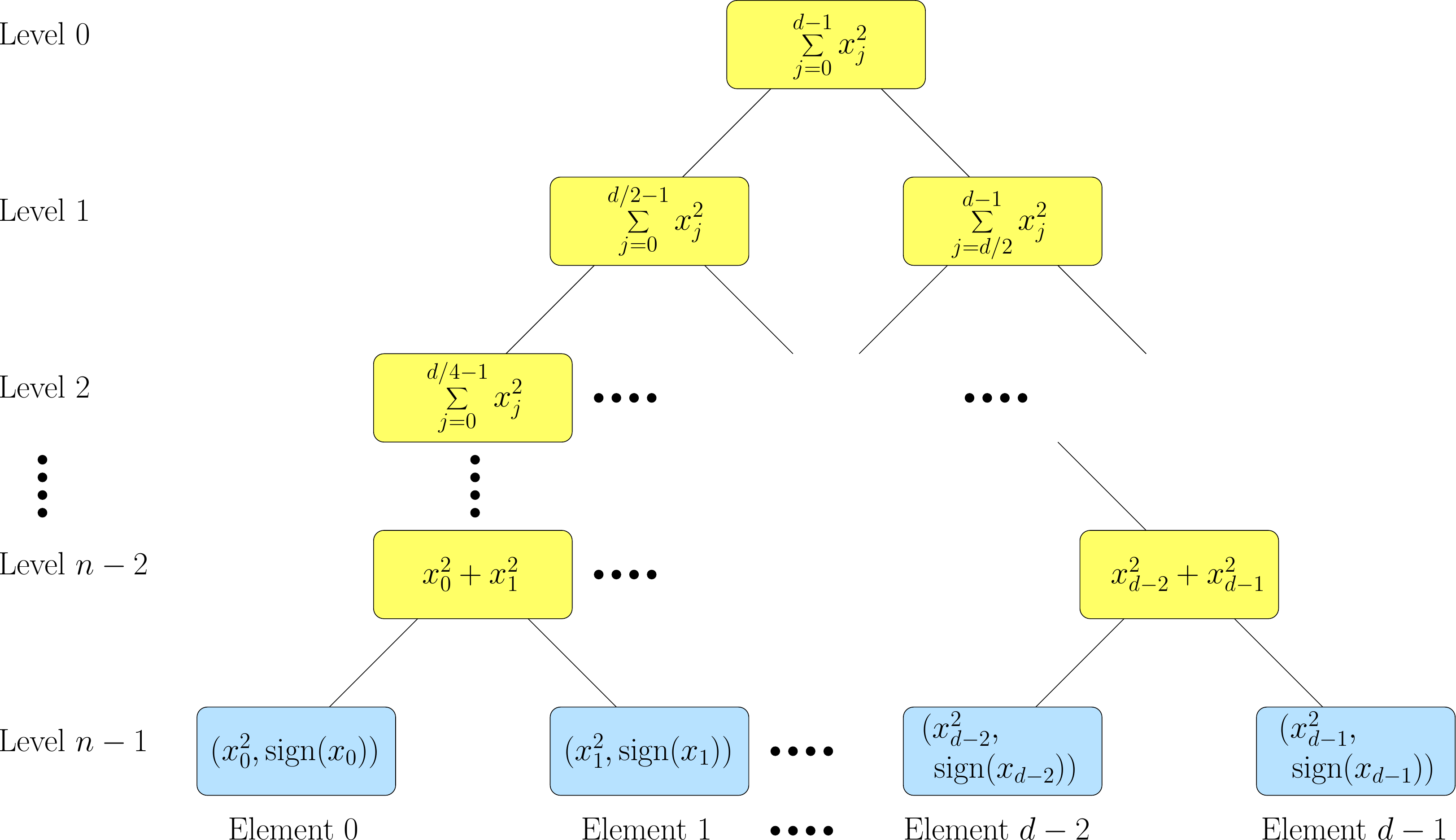}
  \fi
  \caption{Binary tree to prepare the state $\sum_{\vj \in \{0,1\}^n}
    x_j \ket{\vj}$ (assuming $\nrm{x} = 1$).}
  \label{fig:qram_tree}
\end{figure}

Starting from a classical description of $x \in \R^d$ in finite
precision, we begin by creating a binary tree, illustrated in
Fig.~\ref{fig:qram_tree}, that is used to determine the angles of some
rotations. The tree has $n+1$ levels, labeled $0$ to $n$ from top to
bottom. At the bottom level there are $d = 2^n$ nodes, each node
containing a value and its sign; the value contained in the leaf nodes
is the square of the corresponding entry of $x$. For every level
$k=n-1,\dots,0$, there are $2^k$ nodes, with each node containing the
sum of the values of the nodes below it. Note that the tree has
$2^{n+1}-1$ nodes in total. We index each node with its level and its
position in the level, labeled from left to right; for example, node
$(0, 0)$ is the root, nodes $(1, 0)$ and $(1, 1)$ are the left and
right child of the root respectively, and so on. The value contained
in each node is denoted $N(k, j)$, where $(k, j)$ is the index of the
node as described above.

\begin{algorithm2e}[htb]
  \SetAlgoLined
  \LinesNumbered
\KwIn{Vector $x \in \R^d$ with $\nrm{x} = 1$, described in finite precision.} 
\KwOut{Quantum circuit on $n$ qubits mapping $\ket{\v{0}} \to
\ket{\amp{x}}$.}
\textbf{Initialize}: Compute the binary tree illustrated in Fig.~\ref{fig:qram_tree}.\\
Set
$\displaystyle \ket{\psi_1} \leftarrow \sqrt{\frac{N(1,0)}{N(1,0) + N(1,1)}} \ket{0} + \sqrt{\frac{N(1,1)}{N(1,0) + N(1,1)}}\ket{1}.$ \\
\For{$k=1,\dots,n-1$}{
  Prepare a fresh qubit in the state $\ket{0}$, and append it to $\ket{\psi_k}$.\\
  Set: \label{step:ampenciterstep}
  \begin{equation*}
    \ket{\psi_{k+1}} \leftarrow \left(\sum_{\vj \in \{0,1\}^k} \ketbra{\vj}{\vj} \otimes R_Y\left(
    2\arccos\left(\sqrt{\frac{N(k+1, 2j)}{\substack{(N(k+1, 2j) +\\ N(k+1, 2j+1))}}}\right)
    \right) \right) \ket{\psi_k}\ket{0}.
  \end{equation*}
}
Set \label{step:ampencfinal}
$\displaystyle
  \ket{\psi_n} \leftarrow \left(\sum_{\vj \in \{0,1\}^{n}} \text{sign}(x_j) \ketbra{\vj}{\vj}\right) \ket{\psi_n}.$\\
\Return the circuit that produces the state $\ket{\psi_n}$.
\caption{Algorithm to construct the amplitude encoding of a given vector $x \in \R^d$.}
\label{alg:ampenc}
\end{algorithm2e}

The full algorithm to produce a circuit constructing $\ket{\amp{x}} =
\sum_{\vj \in \{0,1\}^n} x_j \ket{\vj}$ is described in
Alg.~\ref{alg:ampenc}; the first step of the algorithm is the
construction of the binary tree, as the algorithm references the node
values throughout. In the ``for'' loop $k=1,\dots,n-1$, we always
add a new qubit, and on line \ref{step:ampenciterstep} of
Alg.~\ref{alg:ampenc}, we apply a controlled operation where the first
$k$ qubits --- containing $\ket{\psi_k}$ --- are the control (where
$\ketbra{\vj}{\vj}$ acts), and the last qubit --- the fresh qubit ---
is the target, onto which the rotation $R_Y$ acts. (Recall the
definition of $R_Y$ in Def.~\ref{def:yrot}.) Note that the last
operation of the algorithm, on line \ref{step:ampencfinal}, can be
merged with the operation on line \ref{step:ampenciterstep} into a
single unitary, as the last operation simply adjusts the sign of the
elements of the rotation matrix $R_Y$; we write it separately for ease
of exposition. We show that this scheme leads to the desired amplitude
encoding.
\begin{proposition}
  \label{prop:ampenccorrect}
  Alg.~\ref{alg:ampenc} returns a quantum circuit on $n$ qubits
  mapping $\ket{\v{0}}$ to $\ket{\amp{x}}$.
\end{proposition}
\begin{proof}
  We first prove by induction that each state $\ket{\psi_k}$ produced
  by Alg.~\ref{alg:ampenc} up until we exit the ``for'' loop (i.e.,
  before we execute line \ref{step:ampencfinal}) satisfies the
  following:
  \begin{equation*}
    \ket{\psi_k} = \sum_{\vj \in \{0,1\}^k} \sqrt{N(k,
      j)} \ket{\vj}.
  \end{equation*}
  The base step $k=1$ follows directly from the
  initialization step of the algorithm, remembering that $N(1, 0) +
  N(1, 1) = \sum_{j=0}^{2^n-1} x_j^2 = 1$.

  For the induction step, let $\ket{\psi_{k+1}} = \sum_{\vh \in
    \{0,1\}^{k+1}} \alpha_{h} \ket{\vh}$, and consider a specific
  coefficient $\alpha_{h}$. Write $\ket{\vh}_{k+1} =
  \ket{\vj}_{k}\ket{z}_1$, i.e., we isolate the last digit $z$ of the
  binary string $\vh$. By construction, $\alpha_{h}$ is equal to the
  product of $\braket{\vj}{\psi_{k}}$ (i.e., the coefficient of
  $\ket{\vj}$ in $\ket{\psi_{k}}$) and the coefficient produced by the
  rotation inside the ``for'' loop on line
  \ref{step:ampenciterstep}. Using the induction hypothesis and the
  definition of $R_Y$, this is equal to:
  \begin{align*}
    \sqrt{N(k, j)} \left(\sqrt{\frac{N(k+1, 2j)}{N(k+1, 2j) + N(k+1, 2j+1)}}\right)    
     \qquad \text{ if } z = 0\phantom{.} \\
    \sqrt{N(k, j)} \left(\sqrt{\frac{N(k+1, 2j+1)}{N(k+1, 2j) + N(k+1, 2j+1)}}\right)    
    \qquad \text{ if } z = 1.
  \end{align*}
  Furthermore, $N(k, j) = N(k+1, 2j) + N(k+1, 2j+1)$ by construction
  of the binary tree, and $h = 2j$ if $z = 0$, $h = 2j+1$ if
  $z=1$. Thus, we can combine and simplify the above expressions,
  obtaining:
  \begin{equation*}
    \alpha_{h} = \sqrt{N(k+1, h)}.
  \end{equation*}
  This concludes the proof of the claim $\ket{\psi_k} = \sum_{\vj \in
    \{0,1\}^k} \sqrt{N(k, j)} \ket{\vj}$ for all $k$.
  
  The final operation, applied to $\ket{\psi_n}$ when we exit the
  ``for'' loop, i.e., the almost-final $n$-qubit state produced by the
  algorithm, adjusts the signs of the coefficients of the quantum
  state to match those of the vector $x$, concluding the proof.
\end{proof}

\noindent Because this construction applies one controlled operation
for every inner node of the binary tree, and the binary tree has
$\bigO{2^n} = \bigO{d}$ nodes, it can be implemented with $\bigO{d}$
multiply-controlled rotations. If we decompose them into basic gates,
the total gate complexity becomes $\bigOt{d}$. Note that there are two
sources of error in the implementation of the circuit to take into
account. The first source is the discretization error, because we
assume $x \in \R^d$ and so the values $N(k, j)$ may not be
representable exactly in finite precision. However, we assumed that we
start from a classical description of $x$ in finite precision, so this
error is already accounted for: our goal is to construct the amplitude
encoding of the finite-precision representation of $x$. The second
source of error is the implementation of the controlled rotations to
the required precision, starting from a binary description of the
angles. As usual after our discussion in Thm.~\ref{thm:sk} and
Sect.~\ref{sec:vardistance}, we simply assume that all operations are
performed with high (and sufficient) accuracy, as the cost is merely
polylogarithmic in the precision: this gets absorbed in the
$\bigOt{\cdot}$ notation.

We described the procedure for $x \in \R^d$ because in this \book{} we
only use it for real input, and this simplifies the
notation, but it is straightforward to amend the construction for
complex input: quantities of the form $x_j^2$ (i.e., the inner nodes
of the binary tree data structure in Fig.~\ref{fig:qram_tree}) should
be replaced by $\abs{x_j}^2$, and instead of adjusting for
$\text{sign}(x_j)$, we also adjust for its complex phase. We obtain
the following result.
\begin{corollary}
  \label{cor:ampenc}
  Given a classical description of $x \in \C^{d}$ in finite precision,
  there is a circuit that implements the mapping $\ket{\v{0}} \to
  \ket{\amp{x}}$ with error at most $\epsilon$ with gate complexity
  $\bigOt{d}$.
\end{corollary}
\begin{proof}
  We follow the construction whose correctness is proven in
  Prop.~\ref{prop:ampenccorrect}, with the only difference that in the
  last step, rather than adjusting only $\text{sign}(x_j)$, we apply a
  general single-qubit unitary to adjust the phase of each
  amplitude. As there are $\bigO{d}$ such operations, this does not
  affect the running time of $\bigOt{d}$. We emphasize that the
  parameter $\epsilon$ does not appear in the expression for the gate
  complexity because the dependence is polylogarithmic in
  $1/\epsilon$.
\end{proof}

\begin{example}
  Let $x = (0.4, 0.4, 0.8, 0.2) \in \R^4$, and note that $\nrm{x} =
  1$. We want to construct $\ket{\amp{x}}$, i.e., the state:
  \begin{equation*}
    \ket{\psi} = 0.4 \ket{00} + 0.4 \ket{01} + 0.8 \ket{10} + 0.2 \ket{11}.
  \end{equation*}
  \begin{figure}[h!t]
    \centering
    \ifcompilefigs
    \begin{tikzpicture}[
      ell/.style 2 args={
        ellipse,
        minimum width=2.1cm,
        minimum height=1.3cm,
        draw,
        label={[name=#1]center:#2},
      },
      connection/.style={thick,densely dotted, latex-latex},
      scale=0.9
    ]
    \node [ell={la}{$(0.16, \text{``+''})$},   fill=blue!20] (a) at (0,0) {};
    \node [ell={lb}{$(0.16, \text{``+''})$},   fill=blue!20] (b) at (4,0) {};
    \node [ell={lc}{$(0.64, \text{``+''})$},   fill=blue!20] (c) at (8,0) {};
    \node [ell={ld}{$(0.4, \text{``+''})$},   fill=blue!20] (d) at (12,0) {};

    \node [ell={le}{$0.32$},   fill=yellow!20] (e) at (2,2) {};
    \node [ell={lf}{$0.68$},   fill=yellow!20] (f) at (10,2) {};

    \node [ell={lg}{$1.00$},   fill=yellow!20] (g) at (6,4) {};

    \draw[-] (a) -- (e);
    \draw[-] (b) -- (e);
    \draw[-] (c) -- (f);
    \draw[-] (d) -- (f);

    \draw[-] (e) -- (g);
    \draw[-] (f) -- (g);
   
    \end{tikzpicture}
    \else
    \includegraphics{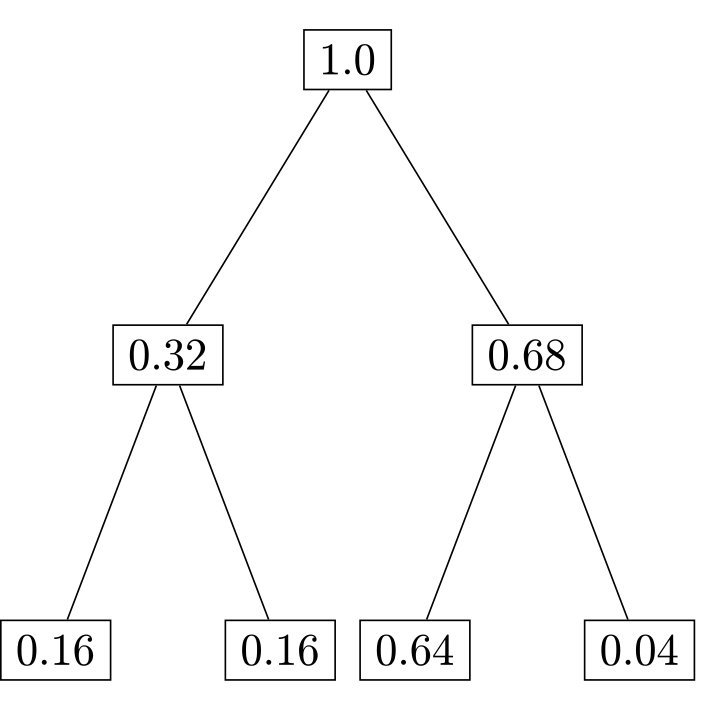}
    \fi
    \caption{Binary tree to prepare the state $\ket{\psi} = 0.4 \ket{00} + 0.4 \ket{01} + 0.8 \ket{10} + 0.2 \ket{11}$.}
    \label{fig:ex_tree}
  \end{figure}
  The binary tree corresponding to this vector is illustrated in
  Fig.~\ref{fig:ex_tree}. Because all coefficients of $x$ are
  nonnegative, we ignore the sign information at the leaf nodes, i.e.,
  we do not have to implement the corresponding sign-adjusting
  operations.  We can construct the amplitude encoding of $x$ by
  performing the following mappings, each of which requires a
  controlled rotation (one per line in the equation below):
  \begin{align*}
    \ket{0} \otimes \ket{0} &\to (\sqrt{0.32} \ket{0} + \sqrt{0.68} \ket{1}) \otimes \ket{0} \\
    (\sqrt{0.32} \ket{0} + \sqrt{0.68} \ket{1}) \otimes \ket{0} &\to  \sqrt{0.16} \ket{00} + \sqrt{0.16} \ket{01} + \sqrt{0.68} \ket{10} \\
    \sqrt{0.16} \ket{00} + \sqrt{0.16} \ket{01} + \sqrt{0.68} \ket{10} &\to \sqrt{0.16} \ket{00} + \sqrt{0.16} \ket{01} + \sqrt{0.64} \ket{10} + \sqrt{0.04} \ket{11}.
  \end{align*}
  In total, this takes three controlled rotations --- one for each
  inner node in the tree.
\end{example}

A modified version of this procedure, described in
\cite{grover2002creating}, assumes that the vector $x$ contains the
square root of the probabilities of a probability distribution over
$\{0,\dots,d-1\}$, and the distribution is efficiently
integrable. Indeed, if one takes this view, the values corresponding
to inner nodes of the binary tree of Fig.~\ref{fig:qram_tree} are just
integrals of the density function with certain lower and upper limits:
at level 1 we take the integral between $0$ and $d/2-1$, and between
$d/2$ and $d-1$; at level 2 we halve these two intervals, and so
on. It is not difficult to see that we can then avoid constructing the
binary tree a priori, and rather, construct it on the fly invoking an
oracle that computes the integrals corresponding to each inner node:
details can be found in \cite{grover2002creating}.

\section{Quantum RAM and faster amplitude encoding}
\label{sec:qram}
The input model of quantum algorithms usually requires the ability to
access data in superposition.\index{quantum!RAM|see{QRAM}}\index{QRAM} This can create slowdowns, because
theoretically any piece of the input data can be queried, requiring
availability of all the input data in the circuit that implements a
query to the input data. For instance, consider the case in which data
is accessed via a table containing $d$ pieces of data: a circuit
implementing this table has $\bigOt{d}$ gates in general, because it
must contain the entire table. The situation is similar to a classical
singly-linked list of size $d$: accessing an element takes $\bigO{d}$
time in the worst case. Classically, this issue is solved by using
data structures with constant-time access, such as arrays stored in
random-access memory (RAM). The fundamental principle of a RAM is that
any piece of data contained in it can be accessed in near-constant
time. This concept can be translated to the quantum world. We do so in
the next subsection, and then discuss a direct application for faster
amplitude encoding of vectors.

\subsection{Definition}
\label{sec:qramdefs}
Quantum RAM (QRAM) is the quantum equivalent of a classical RAM: it
allows constant-time (or in any case, very fast) access to classical
data stored in memory. In the literature one can find several possible
definitions for QRAM, as there are subtle details that matter
depending on the implementation. Here we adopt a straightforward
definition. In this section we assume $d = 2^n$.
\begin{definition}[QRAM]
  \label{def:qram}
  Let $\vv{M_1},\dots,\vv{M_d} \in \{0,1\}^q$. A
  \emph{classical-write, quantum-read quantum RAM (QRAM)} of size $dq$
  is a device that implements the unitary operation $U_{\text{QRAM}}$
  satisfying the following property:
  \begin{equation*}
    U_{\text{QRAM}} \ket{\vj} \ket{\vy}
    = \ket{\vj} \ket{\vy \oplus \vv{M_j}} \qquad \forall \vj \in \{0,1\}^n, \vy \in \{0,1\}^q.
  \end{equation*}
\end{definition}
\begin{remark}
  We call Def.~\ref{def:qram} a \emph{classical-write, quantum-read}
  QRAM bacause the only operations performed in superposition on the
  storage device is to read its data: we do not allow writing content
  in superposition. In fact, all data $\vv{M_1},\dots,\vv{M_d}$, which
  is assumed to be classically available in some form, is already
  stored in the QRAM when we perform the operation
  $U_{\text{QRAM}}$. In principle we could also define a quantum-read,
  quantum-write device that allows changing the data
  $\vv{M_1},\dots,\vv{M_d}$ stored in the device in superposition, but
  this would be an even more powerful device than the classical-write,
  quantum-read QRAM. Generally, in the literature the term QRAM stands
  for ``classical-write, quantum-read QRAM,'' unless otherwise
  specified.
\end{remark}
\begin{remark}
  In principle the operation $U_{\text{QRAM}}$ should be parameterized
  by the data $\vv{M_1},\dots,\vv{M_d}$, but we do not write this
  dependence explicitly for the sake of simplicity. In other words, a
  QRAM of a given size is not just a single unitary, but rather, a
  family of unitaries, and the specific unitary chosen from the family
  is determined by the value of the data. This subtlety does not
  impact our subsequent discussion.
\end{remark}
Given a QRAM containing some data of interest, such as the input data
of an algorithm, the running time of an algorithm that needs access to
the data can then be given in terms of the number of times that the
QRAM is accessed. This is another form of \emph{query} (or oracle) \emph{complexity},\index{complexity!query}
where the query corresponds to accessing the input data. Many (in
fact, at the time of this writing, the vast majority of) quantum
optimization algorithms assume QRAM, and their running time is usually
given in terms of number of QRAM accesses. If the QRAM oracle can be
implemented with time complexity $\bigO{1}$, similar to the time
complexity of classical RAM, then the number of QRAM calls immediately
translates into a bound on the time complexity. This is not the same
as the time complexity in the standard circuit model, where time
corresponds to the number of gates (gate complexity).

We emphasize that the operation $U_{\text{QRAM}}$ is not possible in
$\bigO{1}$ time in the standard circuit model: something more powerful
than the standard quantum gates is necessary. Indeed, in general to
construct $U_{\text{QRAM}}$ with standard gates we need to implement a
lookup table: a circuit that, conditioned on the content $\vj$ of the
first register, ``writes'' (with binary XOR) the corresponding datum
$\vv{M_j}$ in the second register. Such a lookup table needs, at the
very least, one C$X$ gate for every bit with value $1$ among the data
$\vv{M_1},\dots,\vv{M_d}$, i.e., $\Omega(d)$ gates. It is not
difficult to give an explicit circuit construction for the lookup
table with $\bigO{dq}$ two-qubit gates. Thus, requiring that
$U_{\text{QRAM}}$ runs in time $\bigO{1}$ is a strong assumption,
yielding a more powerful input model for quantum algorithms: a quantum
algorithm with access to QRAM could run certain operations faster than
any quantum algorithm in the standard circuit model; an example of
this is discussed in Sect.~\ref{sec:qramampenc}.

\begin{remark}
  In the quantum algorithms literature that employs QRAM, it is
  standard to assume that one access to the QRAM (i.e., an application
  of $U_{\text{QRAM}}$) takes $\bigO{1}$ time. We make the same
  assumption in this \book{}, however we additionally report the
  number of QRAM accesses to allow translation into the standard
  circuit model, see Rem.~\ref{rem:qramcomplexityreport}.
\end{remark}

One may wonder why such a powerful input model was initially
conceived, and why its use is widespread in the literature: after all,
we should be interested in realistic input models only. The simple
reason is that, due to the existence of classical RAM, i.e., storage
devices that allow access to any piece of data in $\bigO{1}$ time,
QRAM also becomes plausible. For example, we could consider using the
same construction of classical RAM, replacing the classical
electronics, i.e., gates, with quantum gates. This may be a very poor
way of implementing a QRAM, and we are not suggesting that it is
practical: we are merely providing an argument as to why the enticing
possibility of constructing a QRAM is not easily refuted. Several
possible constructions for QRAM are discussed in the literature, and
some have even seen some attempts at physical construction and
experimental evaluation, but significant skepticism remains on the
prospects of successfully implementing a QRAM with good fidelity
\cite{jaques2023qram}; see the notes at the end of this chapter
(Sect.~\ref{sec:gradientnotes}) for a list of references and
historical notes on QRAM.

\subsection{QRAM for amplitude encoding}
\label{sec:qramampenc}
Let us go back to the goal of encoding a given classical vector in the
amplitudes of a quantum state; i.e., implementing the mapping
$\ket{\v{0}} \to \ket{\amp{x}}$ given a description of the classical
vector $x$ with $\nrm{x} = 1$. Suppose the data contained in the nodes
of the tree in Fig.~\ref{fig:qram_tree} is stored in QRAM as an
ordered list of the nodes $N(0,0), N(1,0), N(1,1),\allowbreak
N(2,0),\dots$, where we use the same notation as in
Sect.~\ref{sec:ampencalg}. Note that this requires a QRAM of size
$\bigOt{d}$. Given the indices that uniquely identify a node, it is
easy to determine the position of the node itself in the ordered list:
$N(i, j)$ is in position $2^i + j - 1$. Then, on line
\ref{step:ampenciterstep} of Alg.~\ref{alg:ampenc}, we replace the
operation:
\begin{equation*}
  \sum_{\vj \in \{0,1\}^k} \ketbra{\vj}{\vj} \otimes R_Y\left(
  2\arccos\left(\sqrt{\frac{N(k+1, 2j)}{N(k+1, 2j) + N(k+1, 2j+1)}}\right)
  \right),
\end{equation*}
which is the most expensive operation in the construction, by using
QRAM. Denote $\ket{\psi_k} = \sum_{\vj \in \{0,1\}^k} \alpha_j
\ket{\vj}$ the state on $k$ qubits at the beginning of the $k$-th
iteration of the ``for'' loop, as defined in Alg.~\ref{alg:ampenc},
and recall that $\ket{\psi_k} = \sum_{\vj \in \{0,1\}^k} \sqrt{N(k,
  j)} \ket{\vj}$ following the proof of
Prop.~\ref{prop:ampenccorrect}. Add two fresh registers of size equal
to the number of bits necessary to store the data in the nodes of
Fig.~\ref{fig:qram_tree}, initialized with the all-zero string and
positioned between $\ket{\psi_k}$ and the $\ket{0}$ qubit on which the
rotation is applied. We can then implement the following mapping:
\begingroup \allowdisplaybreaks
  \begin{align*}
  \sum_{\vj \in \{0,1\}^k} &\alpha_j \ket{\vj} \ket{\v{0}} \ket{\v{0}} \ket{0} \\
  &\to
  \sum_{\vj \in \{0,1\}^k} \alpha_j \ket{\vj} \ket{\vv{N(k+1, 2j)}} \ket{\vv{N(k+1, 2j+1)}}\ket{0} & \text{(QRAM queries)} \\
  &\to \sum_{\vj \in \{0,1\}^k} \alpha_j \ket{\vj} \ket{\vv{N(k+1, 2j)}} \ket{\vv{N(k+1, 2j+1)}} & \text{(rotation)} \\
  &\phantom{\to} \left(\sqrt{\frac{N(k+1, 2j)}{N(k+1, 2j) + N(k+1, 2j+1)}} \ket{0}\right.\\
  &\phantom{\to} \left. + \sqrt{\frac{N(k+1, 2j+1)}{N(k+1, 2j) + N(k+1, 2j+1)}} \ket{1} \right) \\
  &\to \sum_{\vj \in \{0,1\}^k} \alpha_j \ket{\vj} \ket{\v{0}} \ket{\v{0}}
  \left(\sqrt{\frac{N(k+1, 2j)}{N(k+1, 2j) + N(k+1, 2j+1)}} \ket{0}\right.
  & \text{(uncomputation)}  \\
  &\phantom{\to} \left. + \sqrt{\frac{N(k+1, 2j+1)}{N(k+1, 2j) + N(k+1, 2j+1)}} \ket{1} \right) \\
  &= \sum_{\vj \in \{0,1\}^k} \left(\sqrt{N(k+1, 2j)} \ket{\vj}\ket{0} \right. & \text{(reorder registers)}\\
  &\phantom{to} \left. + \sqrt{N(k+1, 2j+1)} \ket{\vj}\ket{1}\right)\ket{\v{0}} \ket{\v{0}} \\
  &= \ket{\psi_{k+1}}\ket{\v{0}}\ket{\v{0}}.
  \end{align*}
\endgroup For the sake of brevity we are skipping some minor steps in
the chain of operations above, but it is easy to fill in the gaps. We
start by performing two QRAM queries to obtain the values of the nodes
of interest $N(k+1, 2j), N(k+1, 2j+1)$; to do so, we must first
compute the corresponding indices with binary arithmetics (starting
from $\vj$) in some working register that we then uncompute, so we did
not show this working register in the sequence of operations
above. With these two values we can compute the desired rotation angle
$2\arccos\left(\sqrt{\frac{N(k+1, 2j)}{N(k+1, 2j) + N(k+1,
    2j+1)}}\right)$ in a new register, and apply the rotation
conditioned on the value of this register, achieving the same effect
as line \ref{step:ampenciterstep} of Alg.~\ref{alg:ampenc}. Then we
simply uncompute the working registers. The crucial difference with
Alg.~\ref{alg:ampenc} is that, thanks to the QRAM queries, the values
$N(k+1, 2j), N(k+1, 2j+1)$ are entangled with the register containing
$\ket{\vj}$, therefore we only need a single $R_Y$ rotation to
correctly rotate the last qubit in superposition for all $\vj$;
without QRAM, for each value of $\ket{\vj}$ we had to apply a
different rotation controlled on $\ket{\vj}$, greatly increasing the
number of gates.
\begin{remark}
  The rotation angle for the operation $R_Y$ is described in binary in
  a register. It is not difficult to implement the desired operation
  that, based on the (binary) content of a register, rotates another
  qubit by a corresponding amount: this can be done with
  controlled-$R_Y$ gates, one for each qubit in the register that
  contains the binary description of the angle, see
  Rem.~\ref{rem:ryfrombinary}.
\end{remark}
As a result, in the QRAM model the construction of
$\ket{\amp{x}}$ from $\ket{\v{0}}$ can be done with a constant number
of QRAM queries for each level of the tree in
Fig.~\ref{fig:qram_tree}, and $\bigO{q}$ gates per level, where $q$ is
the number of bits used to store each of the values $N(k, j)$; we
assume $q = \bigO{\log d}$, as this already gives precision
exponential in $d$. In total, this gives a complexity of $\bigO{n}$
QRAM queries, and $\bigO{nq}$ additional gates --- an exponential
improvement over the $\bigOt{2^n} = \bigOt{d}$ gates for the same
construction without QRAM. One should not forget that there is also an
initial preparation time of $\bigOt{d}$ to read the classical
description of the vector $x$, prepare the tree data structure, and
store it in QRAM: this cost dominates all the other ones, but it only
needs to be paid once, after which we can reuse the already-prepared
QRAM to construct $\ket{\amp{x}}$ with the stated complexity as many
times as we want.
\begin{remark}
  \label{rem:qramcomplexityreport}
  For quantum algorithms in the QRAM model, the recommended (and most
  accurate) way to describe the running time is to report the number
  of accesses to the QRAM, the number of additional (two-qubit) gates,
  and the number of classical operations that need to be performed by
  the algorithm, e.g., to read and prepare some QRAM data
  structure. In this way, the cost of each component can be properly
  assessed. This also immediately translates to an upper bound to the
  complexity in the standard circuit model without QRAM, because, as we
  have seen, a QRAM of size $dq$ can be implemented with $\bigO{dq}$
  gates.
\end{remark}
\begin{corollary}
  \label{cor:qramampenc}
  Given a classical description of $x \in \C^{d}$ in finite precision
  and a QRAM of size $\bigOt{d}$, there is a circuit that implements
  the mapping $\ket{\v{0}} \to \ket{\amp{x}}$ with error at most
  $\epsilon$ using $\bigO{\log d}$ QRAM queries, $\bigO{\log^2 d}$
  additional gates, and $\bigOt{d}$ classical arithmetic operations to
  initialize the QRAM data structure.
\end{corollary}
\begin{proof}
  Follows from Prop.~\ref{prop:ampenccorrect} and the discussion
  preceding the corollary statement.
\end{proof}

As we discuss in subsequent chapters, QRAM can accelerate many
quantum algorithms that require access to data, not just amplitude
encoding.\index{amplitude!encoding|)} It is however important to remember that the QRAM input
model is stronger than the standard circuit model, and that there are
significant hurdles to the physical construction of QRAM, hence this
stronger input model may be an even bigger ask than ``just'' a
fault-tolerant quantum computer.

\section{Notes and further reading}
\label{sec:gradientnotes}
The two main references for the development of the quantum gradient
algorithm presented in this chapter are
\cite{jordan2005fast,gilyen2019optimizing}. Some direct applications
in optimization, besides the ones already discussed in this chapter,
can be found in \cite{gilyen2019optimizing}. We also remark that our
discussion on convex optimization using quantum membership oracles,
akin to the framework of \cite{grotschel1988geometric}, is mostly
based on the approach in \cite{apeldoorn2020convex}, but a different
and equally powerful approach is developed in
\cite{chakrabarti2020quantum}. The technical tools used in
\cite{chakrabarti2020quantum} are different: the analysis therein
relies on the technique of mollifier functions, which leads to an
elegant treatment, but requires additional technical tools that we
avoided for the sake of simplicity. We encourage the reader to compare
the two approaches for a more comprehensive treatment of the subject.

From a practical point of view, depending on the type of oracle that
computes the function $f$, the quantum gradient algorithm may be less
efficient than techniques based on automatic differentiation
\cite{stamatopoulos2022towards}, which can be applied to quantum
circuits that perform arithmetic computation similarly to how they are
applied in classical computing. Note, however, that the quantum
gradient algorithm can also be applied to functions that are not easily
computable on a classical computer, e.g., the quantum state inner
product function of Sect.~\ref{sec:gradienttomo}, for which automatic
differentiation is not applicable. For a discussion on central
difference approximations of functions, an entry point is
\cite{li2005general}.

Quantum state tomography\index{state!tomography} is a fundamental
topic in quantum information theory. In the context of optimization,
quantum state tomography is useful to recover a classical description
of a solution that is encoded in a pure or mixed quantum
state. Examples of quantum optimization algorithms that rely on some
form of state tomography are
\cite{augustino2023quantum,kerenidis2020quantum,wu2023inexact}; the
mirror descent framework, or matrix multiplicative weights update
framework (see Ch.~\ref{ch:mmwu}), would also rely on tomography if a
full description of the optimal solution is needed --- as opposed to
requiring only the objective function value. Quantum state tomography
using the gradient algorithm is studied in \cite{nannitomography}; the
resulting algorithms are essentially optimal for the case where we
have a (controlled, reversible) unitary that prepares the quantum
state of interest, which is usually the case in the context of states
produced by an algorithm. Thus, Thm.~\ref{thm:tomography} gives the
best possible complexity (in terms of number of calls to a unitary
preparing the state of interest) for obtaining a classical description
of a pure quantum state in the stated context; the gate count can be
improved with some form of QRAM. The gradient algorithm can also be
applied to recover the classical description of a mixed state, but the
corresponding derivation is more involved and the complexity
increases, see \cite{nannitomography}. Optimal algorithms for mixed
states when we do not necessarily have access to a unitary preparing
the state are discussed in \cite{haah2017sample,o2016efficient}. A
simpler algorithm, using compressed sensing, is discussed in
\cite{gross2010quantum}.

Historically, the bucket brigade model of
\cite{giovannetti2008quantum} was impactful for popularizing the
possibility of constructing QRAM. \cite{giovannetti2008quantum}
proposes an implementation with a tree of gates of depth $\bigO{\log
  nd}$, resulting in a $\bigO{\log nd}$ QRAM access (wall-clock)
time. This is a slowdown compared to the ideal $\bigO{1}$, but it is
still exponentially faster than the standard circuit model and its
$\bigO{nd}$ running time. \cite{blencowe2010quantum} points to several
papers that attempt to make progress on experimental realization of
some form of quantum-accessible memory. However, so far all existing
proposals have faced significant hurdles in achieving a successful
implementation and experimental demonstration. For example, the bucket
brigade model has been labeled impractical due to considerations on
how to suppress errors in a device that may have to activate lots of
quantum gates at the same time \cite{arunachalam2015robustness}. More
specifically, \cite{arunachalam2015robustness} shows that an
application of Grover's algorithm for unstructured search using QRAM
queries to identify the marked element (which is precisely how QRAM is
used in the quantum dynamic programming scheme of
\cite{ambainis2019quantum}, see Sect.~\ref{sec:ampampnotes}) would
require exponentially-small gate errors inside the QRAM. The paper
also argues that error correction could negate several of the
purported advantages of the bucket brigade QRAM. A detailed review of
several QRAM models, as well as a discussion of their main drawbacks
and limitations that point to difficulties in experimental
realizations, can be found in \cite{jaques2023qram}. Another model,
that exponentially reduces the amount of fault-tolerant quantum
resources required to implement a QRAM, is presented in
\cite{dalzell2025distillation}; however, even the model of
\cite{dalzell2025distillation} does not answer the question of whether
it is possible to create a truly efficient QRAM, because the proposed
protocol relies on expensive (i.e., linear in the size of the data)
classical update operations in every iteration.

It is worth emphasizing that despite the notorious difficulty of
constructing QRAM, the QRAM model is widely used in the literature on
quantum algorithms, and it is particularly widespread in quantum
optimization and quantum machine learning. Many quantum optimization
algorithms are developed in the framework of a query model, where the
problem data can be accessed by querying an appropriate oracle. For
example, the quantum algorithms for semidefinite optimization
discussed Ch.~\ref{ch:mmwu} are designed in such a
model. Unfortunately, the query model typically loses most of its
advantage without QRAM: for a discussion on the impact of QRAM on
optimization, or more generally, data-driven problems, see
Sect.s~\ref{sec:sparseblockenc} and \ref{sec:qramblockenc}.

\chapter{Hamiltonian simulation}
\label{ch:hamsim}
\thispagestyle{fancy}
Hamiltonian simulation is the problem for which quantum computers were
initially proposed by Richard Feynman \cite{feynman1982simulating},
and it is a crucial problem for applications in physics: it
corresponds to the simulation of the evolution of a quantum-mechanical
system over time. Although this \book{} aims to be physics-free, it is
useful to give at least the mathematical foundations of this problem,
because Hamiltonian simulation is a central component of many quantum
algorithms. We do not discuss the precise physical meaning of these
concepts, and aim to provide just enough vocabulary to read the
quantum computing literature without too much confusion.
\begin{definition}[Hamiltonian]
  \label{def:hamiltonian}
  The \emph{Hamiltonian} of a quantum-mechanical system with $n$
  qubits is a Hermitian operator $\ham \in \C^{2^n \times 2^n}$
  representing the total energy of the system. The corresponding
  \emph{expected energy} of the $n$-qubit system described by the
  state $\ket{\psi}$ is the value $\bra{\psi} \ham \ket{\psi}$.
\end{definition}
\begin{remark}
  It is an unfortunate fact that in the quantum computing literature,
  both Hamiltonians and Hadamard gates are usually indicated with the
  letter $H$. Usually it is clear from the context which of these
  mathematical objects is being referred to. To further avoid
  ambiguities, we use calligraphic $\ham$ for Hamiltonians and $H$ for
  the Hadamard gate, but this is not always the convention in the
  literature.
\end{remark}
The evolution of a physical system follows the
\emph{Schr\"odinger equation}\index{Schr\"odinger equation}:
\begin{equation}
  \label{eq:schrodinger}
  i \frac{\di \ket{\psi(t)}}{\di t} = \ham \ket{\psi(t)},
\end{equation}
where $\ham$ is the Hamiltonian of the system, $\ket{\psi(t)}$ is the
state of the system at time $t$, and the initial condition
$\ket{\psi(0)}$ is given; it is implicitly assumed that time starts
evolving from $t=0$. In quantum mechanics Eq.~\eqref{eq:schrodinger}
is usually stated with the Planck constant $\hslash$ multiplying the
l.h.s., but for our purposes the constant is unnecessary: we can think
of it as being absorbed into the Hamiltonian, yielding the
mathematically-equivalent expression in
Eq.~\eqref{eq:schrodinger}. The solution to the differential equation
in Eq.~\eqref{eq:schrodinger} is:
\begin{equation}
  \label{eq:schrodingersol}
  \ket{\psi(t)} = e^{-i\ham t} \ket{\psi(0)},
\end{equation}
hence if the initial state $\ket{\psi(0)}$ is given (as a quantum
state), to determine the state of the system after time $t$ we need to
implement the operator $e^{-i\ham t}$; this is called \emph{time
evolution} of a Hamiltonian, or \emph{Hamiltonian simulation}, see
Def.~\ref{def:hamsim}. In this chapter we limit ourselves to a
time-independent Hamiltonian $\ham$; the situation is considerably
more involved when $\ham$ is time-dependent, a case that we only
address for the very specific case of adiabatic evolution in
Ch.~\ref{ch:adiabatic}.

\section{Problem definition and preliminaries}
Before we properly introduce the Hamiltonian simulation problem, it
may be helpful to recall some basic facts about the matrix
exponential.
\begin{definition}[Matrix exponential]
  \label{def:matexp}
  The matrix exponential\index{matrix!exponential} of a square matrix $A$ is defined as:
  \begin{equation*}
    e^{A} := \sum_{k=0}^{\infty} \frac{A^k}{k!}.
  \end{equation*}
\end{definition}
If $A$ is diagonalizable $A = U D U^{-1}$, then it is easy
to prove that $\exp(A) = U \exp(D) U^{-1}$, where $\exp(D)$ is a
diagonal matrix with elements $(\exp(D))_{jj} = \exp(D_{jj})$. Thus,
we are simply applying the exponential function to the eigenvalues
of $A$. Now we can define Hamiltonian simulation.
\begin{definition}[Hamiltonian simulation]
  \label{def:hamsim}
  Given a Hermitian matrix $\ham$, a duration $t$, and a precision
  parameter $\epsilon$, the problem {\em Hamiltonian simulation}\index{simulation!Hamiltonian|(}\index{algorithm!Hamiltonian simulation|(} is
  that of implementing (i.e., providing a quantum circuit for) a
  unitary $U$ such that $\nrm{U - e^{-i\ham t}} \le \epsilon$.
\end{definition}
\begin{remark}
  $e^{-i\ham t}$ is a unitary operation: $\ham$ is Hermitian, hence it is
  diagonalizable and $e^{-i\ham t}$ transforms the eigenvalues $\lambda_j$
  of $\ham$ into $e^{-i\lambda_j t}$. Thus, all eigenvalues of $e^{-i\ham t}$
  are complex numbers with unit modulus.
\end{remark}
\begin{remark}
  Because the only restriction imposed on $\ham$ is that it is Hermitian,
  we can equivalently consider the problem of simulating $e^{i\ham t}$
  rather than $e^{-i\ham t}$: the negative sign can be absorbed into
  $\ham$. In the following, we often neglect the minus sign in the
  exponent for brevity.
\end{remark}

\subsection{The class BQP and Hamiltonian simulation}
The complexity class BQP is the class of decision problems that can be
solved efficiently by a quantum computer; it is the quantum analog of
the class BPP (bounded-error probabilistic polynomial time), that
contains problems that can be solved efficiently by a classical
computer.
\begin{definition}[Bounded-error Quantum Polynomial time (BQP) class]
  \label{def:bqp}
  \emph{BQP}\index{complexity!class BQP} (Bounded-error Quantum Polynomial time) is the class of
  decision problems that can be solved by a polynomial-time quantum
  algorithm with probability at least $\frac{2}{3}$.
\end{definition}
It turns out that Hamiltonian simulation is truly a fundamental
problem, as it is BQP-complete; this implies that any problem that can
be solved efficiently by a quantum computer can be (efficiently)
reduced to a Hamiltonian simulation.
\begin{remark}
  To show that Hamiltonian simulation is in BQP, it suffices to
  provide a polynomial-time quantum algorithm for the problem. Several
  such algorithms have been known since the early days of the field,
  see, e.g., \cite{lloyd1996universal}, or some of the references
  discussed subsequently in this chapter. To show that Hamiltonian
  simulation is BQP-complete, we additionally need to prove that any
  efficient quantum computation can be performed via Hamiltonian
  simulation.
\end{remark}

Suppose we have a quantum circuit that applies unitaries
$U_1,\dots,U_{N-1} $ onto the $q$-qubit state $\ket{\v{0}}$, and we
want to show that we can simulate the effect of this circuit via
Hamiltonian simulation. To do so, we construct a specific Hamiltonian
such that $e^{-i\ham t} \ket{\v{0}} = U_{N-1} U_{N-2}\cdots U_1
\ket{\v{0}}$ for some choice of $t$. If this construction can be done
for any choice of unitaries $U_1,\dots,U_{N-1}$, and the size of the
Hamiltonian simulation instance is at most polynomially larger than
the size of a description of the quantum circuit (i.e.,
$U_1,\dots,U_{N-1}$), this would prove that Hamiltonian simulation is
BQP-complete: any problem instance that can be solved by a
polynomial-size quantum circuit can also be solved as a polynomial-size
Hamiltonian simulation problem instance. It is therefore natural to
study Hamiltonian simulation, and many algorithmic advances in quantum
computing originated from the study of this problem.

We do not give a full proof of BQP-completeness, but we show most of
it, in particular to showcase a possible approach to turn a general
circuit into a Hamiltonian simulation instance; the ideas discussed
here date back to Feynman's initial vision for quantum computers
\cite{feynman1982simulating,feynman2018lectures}. Recall that we are
given $N$ unitaries that we want to apply, and for which we assume
that we have an efficient description. Let us introduce $N$ auxiliary
qubits; for $j=1,\dots,N$, define binary strings $\v{u}^{(j)} \in
\{0,1\}^{N}, \v{u}^{(j)}_h = 1$ if $j=h$, $0$ otherwise. Each of these
$N$-digit strings encodes an integer from 1 to $N$ by having a $1$ in
the corresponding position, and $0$ elsewhere. We use them as the
content of a ``clock register''\index{register!clock} to remember the position of the
simulation algorithm in the sequence of unitaries, so that each
unitary is applied exactly once and in the right order. Construct the
following Hamiltonian:
\begin{equation}
  \label{eq:statetransferh}
  \ham := \frac{1}{2} \sum_{j=1}^{N-1} \sqrt{j(N-j)}\left(\ketbra{\v{u}^{(j+1)}}{\v{u}^{(j)}} \otimes U_j + \ketbra{\v{u}^{(j)}}{\v{u}^{(j+1)}} \otimes U^{\dag}_j\right).
\end{equation}
We can check that this is Hermitian by construction, as it is a sum of
Hermitian terms. Note that this Hamiltonian acts on two registers: the
clock register containing the strings $\v{u}^{(j)}$, and a second
register initialized in the state $\ket{\v{0}}$, onto which we apply
the unitaries $U_j$. Define a set of states:
\begin{equation*}
  \ket{\psi_k} := \ket{\v{u}^{(k)}} \otimes \left(U_{k-1} U_{k-2} \cdots
  U_1 \ket{\v{0}}\right).
\end{equation*}
If we compute $\ket{\psi_N}$ then we have obtained, in the second
register, $U_{N-1}\cdots U_1 \ket{\v{0}}$, precisely the effect of the
circuit that we wish to simulate. First, we show by induction that:
\begin{equation*}
 \ham \ket{\psi_k} = \frac{1}{2} \sqrt{(k-1)(N+1-k)} \ket{\psi_{k-1}}
 + \frac{1}{2} \sqrt{k(N-k)} \ket{\psi_{k+1}} \text{ for }  k=1,\dots,N-1,
\end{equation*}
where we define $\ket{\psi_{0}} := 0$ for convenience (i.e., this is
not a quantum state, but rather the scalar 0 which simply disappears
from the expression). Indeed, for the base step:
\begin{align*}
  \ham \ket{\psi_1} &= \frac{1}{2} \sum_{j=1}^{N-1} \sqrt{j(N-j)}\left(\ket{\v{u}^{(j+1)}} \bra{\v{u}^{(j)}} \otimes U_j + \ket{\v{u}^{(j)}}\bra{\v{u}^{(j+1)}} \otimes U^{\dag}_j\right) \ket{\v{u}^{(1)}} \otimes \ket{\v{0}} \\
    &= \frac{1}{2} \sqrt{(N-1)} \ket{\v{u}^{(2)}} \otimes (U_1 \ket{\v{0}}) = \frac{1}{2} \sqrt{(N-1)} \ket{\psi_2},
\end{align*}
and for the induction step:
\begin{align*}
  \ham \ket{\psi_k} &= \frac{1}{2} \sum_{j=1}^{N-1} \sqrt{j(N-j)}\left(\ket{\v{u}^{(j+1)}} \bra{\v{u}^{(j)}} \otimes U_j + \ket{\v{u}^{(j)}}\bra{\v{u}^{(j+1)}} \otimes U^{\dag}_j\right) \ket{\v{u}^{(k)}} \otimes \left(U_{k-1} \cdots
  U_1 \ket{\v{0}}\right) \\
  &= \frac{1}{2} \sqrt{k(N-j)}\ket{\v{u}^{(k+1)}}\otimes (U_k U_{k-1} \cdots U_1 \ket{\v{0}}) \, +\\ &\phantom{=} \;\; \frac{1}{2} \sqrt{(k-1)(N+1-k)} \ket{\v{u}^{(k-1)}} \otimes U_{k-1}^{\dag} (U_{k-1} \cdots U_1 \ket{\v{0}}) \\
  &= \frac{1}{2} \sqrt{k(N-j)} \ket{\psi_{k+1}} +
  \frac{1}{2} \sqrt{(k-1)(N+1-k)} \ket{\psi_{k-1}}.
\end{align*}
This means that $\ham$ acts on the subspace spanned by the states
$\ket{\psi_k}$, and the effect of $e^{-i\ham t}$ can be understood in
this subspace. We claim that $e^{-i\ham t} \ket{\psi_1} =
\ket{\psi_N}$ if we choose $t = \pi$. A full proof of this result is
can be found in \cite{kay2010perfect}. A high-level sketch of the
proof is the following. The effect of $\ham$ on the first $N$ qubits
(first register), when expressed in the basis $\ket{\v{u}^{(j)}}$, can
be written in this form:
\begin{equation*}
  \ham_B = 
  \frac{1}{2}
  \begin{pmatrix}
    0 & \sqrt{N-1} & 0 & \dots & 0 & 0\\
    \sqrt{N-1} & 0 & \sqrt{2(N-2)} & \dots & 0 & 0\\
    0 & \sqrt{2(N-2)} & 0 & \dots & 0 & 0\\
    \vdots & \vdots & \vdots & \ddots & \vdots & \vdots\\
    0 & 0 & 0 & \dots & 0 & \sqrt{N-1} \\
    0 & 0 & 0 & \dots & \sqrt{N-1} & 0
  \end{pmatrix},
\end{equation*}
which is a symmetric tridiagonal matrix with zeroes on the
diagonal. We denote it by $\ham_B$ because this is $\ham$ expressed in
the basis $\ket{\v{u}^{(j)}}$, and we perform all the analysis in this
basis. The eigenvalues of the matrix $\ham_B$ are $-\frac{1}{2}(N-1),
-(\frac{1}{2}(N-1)-1),\dots,\frac{1}{2}(N-1)-1,\frac{1}{2}(N-1)$,
i.e., they range from $-\frac{1}{2}(N-1)$ to $\frac{1}{2}(N-1)$ spaced
by 1. Note that $\ham_B$ commutes with the matrix $M_s =
\sum_{j=1}^{N} \ket{\v{u}^{(j)}}\bra{\v{u}^{(N+1-j)}}$, which sends
the $j$-th element to the $(N+1-j)$-th and viceversa: this is easy to
check. As a consequence of this property, the eigenvectors of $\ham_B$
can be divided into symmetric and antisymmetric, based on how $M_s$
acts on them (suppose $\ket{\psi}$ is an eigenvector with eigenvalue
$\lambda$; then $\lambda M_s \ket{\psi} = M_s \lambda \ket{\psi} = M_s
\ham_B \ket{\psi} = \ham_B M_s \ket{\psi}$, so $M_s \ket{\psi}$ must
be equal to $\ket{\psi}$ or to $-\ket{\psi}$. We call the first type
of eigenvalue symmetric, the second antisymmetric). We can then state
the following.
\begin{proposition}
  \label{prop:statetransfereig}
  Let $S \subseteq \{1,\dots,N\}$ be the set of indices of symmetric
  eigenvalues of $\ham_B$. Suppose there exists some time $t$ and angle
  $\phi$ such that, for every eigenpair $(\lambda_j, \ket{\psi_j})$ such
  that $\braket{\psi_j}{\v{u}^{(1)}} \neq 0$, we have $e^{-i \lambda_j t}
  = e^{i\phi}$ if $j \in S$, and $e^{-i \lambda_j t} = -e^{i\phi}$ if
  $j \not\in S$. Then $e^{-i\ham_B t} \ket{\v{u}^{(1)}} = e^{i\phi}
  \ket{\v{u}^{(N)}}$.
\end{proposition}
\begin{proof}
  We consider the decomposition of $\ket{\v{u}^{(1)}}$ in terms of the
  eigenvectors $\ket{\psi_j}$ that have nonzero overlap with it. Let
  $S' := \{j \in S: \braket{\psi_j}{\v{u}^{(1)}} \neq 0\}$ be the set
  of symmetric eigenvalues with nonzero overlap, $A' := \{j \in
  \{1,\dots,N\} \setminus S : \braket{\psi_j}{\v{u}^{(1)}} \neq 0\}$
  the set of antisymmetric eigenvalues with nonzero overlap. Let
  $\ket{\v{u}^{(1)}} = \sum_{j \in S' \cup A'} \alpha_j \ket{\psi_j}$
  for some coefficients $\alpha_j$, which clearly exist because the
  eigenvectors $\ket{\psi_j}$ form a basis and we are considering all
  eigenvectors that are not orthogonal to $\ket{\v{u}^{(1)}}$. The
  time evolution according to the Hamiltonian is then:
  \begin{align*}
    e^{-i\ham_B t} \ket{\v{u}^{(1)}} &= e^{-i\ham_B t} (\sum_{j \in S'} \alpha_j \ket{\psi_j} + \sum_{j \in A'} \alpha_j \ket{\psi_j}) = \sum_{j \in S'} e^{-i \lambda_j t} \alpha_j \ket{\psi_j} + \sum_{j \in A'} e^{-i \lambda_j t} \alpha_j \ket{\psi_j} \\
    &= e^{i \phi} (\sum_{j \in S'} \alpha_j \ket{\psi_j} - \sum_{j \in A'} \alpha_j \ket{\psi_j}) = e^{i \phi} M_s (\sum_{j \in S'} \alpha_j \ket{\psi_j} + \sum_{j \in A'} \alpha_j \ket{\psi_j}) \\
    &= e^{i \phi} M_s \ket{\v{u}^{(1)}} = e^{i \phi} \ket{\v{u}^{(N)}}. 
  \end{align*}
\end{proof}

\noindent Prop.~\ref{prop:statetransfereig} shows that if a certain
property is satisfied, then evolving the state $\ket{\v{u}^{(1)}}$
with the Hamiltonian $\ham$ for a specific time $t$ yields the state
$\ket{\v{u}^{(N)}}$ (the global phase factor $e^{i \phi}$ is
unimportant, as usual). The property is the following: the length $t$
of the time evolution is such that it yields the same phase for all
symmetric eigenvalues, and there is only a sign difference between the
symmetric and antisymmetric eigenspaces. As it turns out, for a matrix
of the form $\ham_B$ (symmetric tridiagonal with positive off-diagonal
elements) the symmetry of the eigenvectors alternates, if we examine
them in increasing order of the eigenvalues. With this fact in mind,
we can see that the following property is sufficient to satisfy the
conditions of Prop.~\ref{prop:statetransfereig}:
\begin{equation}
  \label{eq:statetransferspacing}
  \lambda_j - \lambda_{j-1} = (2k + 1)\pi/t \qquad \forall j =
  2,\dots,N,
\end{equation}
where $k \in \N$. Indeed, with this property we have:
\begin{equation*}
  e^{-i \lambda_j t} = e^{-i (\lambda_1 + (j-1)(2k+1)\pi/t)t} = \begin{cases}
    e^{-i \lambda_1 t}  & \text{if } j \text{ odd} \\
    -e^{-i \lambda_1 t}  & \text{if } j \text{ even,}
  \end{cases}
\end{equation*}
because the phase $(j-1)(2k+1)\pi$ yields a multiplicative
factor $1$ or $-1$ depending on the parity of $j$. Now recall that by
construction, the eigenvalues of $\ham_B$ (which alternate between
the symmetric and antisymmetric eigenspaces) are spaced by exactly
$1$. Then, choosing $t = \pi$ satisfies
Eq.~\eqref{eq:statetransferspacing} with $k=0$, thereby showing that
evolving the Hamiltonian $\ham_B$ for time $t = \pi$ starting from
$\ket{\v{u}^{(1)}}$ yields $\ket{\v{u}^{(N)}}$. Going back to the
original Hamiltonian $\ham$ defined in Eq.~\eqref{eq:statetransferh},
this means that we are evolving $\ket{\psi_1}$ into $\ket{\psi_N}$,
which contains the state $U_{N-1} U_{N-2} \cdots U_1 \ket{\v{0}}$ in
its second register, and therefore the output of the circuit that we
wanted to simulate. Thus, we can reduce any problem that can by solved
by an efficient (polynomial-time) quantum circuit into an instance of
the Hamiltonian simulation problem, showing that Hamiltonian
simulation is BQP-complete.

\subsection{Basic remarks on Hamiltonian simulation}
We state here some basic observations on Hamiltonian simulation,
helping to set the stage for the simulation algorithms discussed in
the remainder of this chapter. First, intuitively it should be clear
that the difficulty of simulating a Hamiltonian may depend not only on
$\ham$ itself, but also on the evolution time $t$: simulating a
quantum-mechanical system for a longer duration cannot be easier than
simulating it for a shorter time.

Let $\ham$ act on $n$ qubits. For a simulation algorithm to be
efficient, we require that the running time of the algorithm is
polynomial in $n$, $t$, and $\frac{1}{\epsilon}$. In fact, some of the
algorithms discussed in the remainder of this chapter even depend
polylogarithmically on $\frac{1}{\epsilon}$. Note that $\ham$ is a
$2^n \times 2^n$ matrix, and the running time of an efficient
Hamiltonian simulation algorithm is polynomial in $n$, which is
exponentially faster than classically computing the matrix exponential
via its definition. The following remarks follow directly from the
requirement, stated above, that we impose on efficient Hamiltonian
simulation algorithms.
\begin{remark}
  \label{rem:hamsimnormub}
  Hamiltonian simulation algorithms generally assume some upper bound
  on $\nrm{\ham}$, and the reason for this is easily
  explained. Suppose we want to compute $e^{i\ham t}$: if we define a
  new Hamiltonian $\ham' = \ham t$, then $e^{i\ham t} = e^{i\ham'}$,
  i.e., the time parameter $t$ can now be set to $1$, but note that
  $\nrm{\ham'} = t\nrm{\ham}$. Thus, we can decrease the time
  parameter $t$ if we increase the norm of the Hamiltonian. Because the
  running time of a simulation algorithm generally depends on $t$, we
  need to impose some normalization so that different algorithms are
  comparable under similar conditions. The convention in the
  literature is to upper bound the norm of $\ham$ ($\nrm{\ham} \le
  1$), and analyze the dependence of the running time of a Hamiltonian
  simulation algorithm on the parameter $t$.
\end{remark}
\begin{remark}
  If we have an efficient algorithm to simulate $e^{i\ham t}$, we can
  also efficiently simulate $e^{ic\ham t}$ for any constant $c$ which is
  polynomial in $n$, simply by absorbing it into $t$.
\end{remark}

\section{Overview of simulation algorithms}
In this section we discuss several methods for Hamiltonian simulation,
based on properties of the Hamiltonian or on the input model. In fact,
the way in which the Hamiltonian is specified often has an impact on
what techniques are suitable. Initially we assume that the
Hamiltonian is diagonalizable or expressible as a sum of ``local''
terms (i.e., tensor products of a few single-qubit operators), and
subsequently generalize to Hamiltonians that may not have that
structure. For additional resources on Hamiltonian simulation, we
refer the reader to the excellent lecture notes
\cite{childs2017lecture,dewolf2019quantum}, which inspired parts of
our presentation.

\subsection{Diagonalizable Hamiltonians}
\label{sec:hamsimdiagonal}
If we know how to diagonalize $\ham$, then we can simulate $e^{i\ham t}$, as
we show next. We need the following basic fact about matrix
exponentials.
\begin{proposition}
  For any unitary $U$ and Hamiltonian $\ham$, $e^{iU\ham U^{\dag}t} = U
  e^{i\ham t} U^{\dag}$.
\end{proposition}
\begin{proof}
  We have:
  \begin{equation*}
  e^{iU\ham U^{\dag}t} = \sum_{k=0}^{\infty} \frac{(iU\ham U^{\dag}t)^k}{k!}
  = U \left(\sum_{k=0}^{\infty} \frac{(i\ham t)^k}{k!}\right) U^{\dag} = U
  e^{i\ham t} U^{\dag},
  \end{equation*}
  by definition of the matrix exponential and because $(U\ham U^{\dag})^k
  = U \ham^k U^{\dag}$, because in the expansion of the product we can
  simplify $U U^{\dag} = U^{\dag} U = I$.
\end{proof}

\noindent Then, if we know how to efficiently construct a unitary $U$
that diagonalizes $\ham$, i.e., $\ham = U D U^{\dag}$ where $D$ is
diagonal, and we can compute a binary description of the diagonal
elements $D_{j} = \bra{\vj}U^{\dag} \ham U \ket{\vj}$, we can
implement a unitary that performs the following operations:
\begin{align*}
  \ket{\vj} \ket{\v{0}} &\to \ket{\vj}  \ket{\vv{D_{j}}} & \text{(because we know how to compute the diagonal elements)}\\
  &\to e^{i D_{j} t} \ket{\vj}  \ket{\vv{D_{j}}} & \text{(using controlled phase gates and the bitstring $\vv{D_{j}}$)}\\
  &\to e^{i D_{j} t} \ket{\vj}  \ket{\v{0}} & \text{(uncomputing the second register)} \\
  &= e^{i U^{\dag} \ham U t} \ket{\vj}  \ket{\v{0}} & \text{(because $U^{\dag} \ham U$ acts on $\ket{\vj}$ as $D_{j}$)} \\
  &= U^{\dag} e^{i \ham t} U \ket{\vj}  \ket{\v{0}}.
\end{align*}
The unitary constructed above applies $U^{\dag} e^{i \ham t} U$ to any
basis state $\ket{\vj}$. Thus, by linearity, this operation simulates
$U^{\dag} e^{i\ham t} U$ on an arbitrary state.  To accomplish our goal
of applying $e^{i\ham t}$ to a given initial state $\ket{\psi}$, we simply
need to compute $U (U^{\dag} e^{i\ham t} U) U^{\dag} \ket{\psi}$, which is
easy given our assumption that we can construct $U$ and we have shown
above how to obtain $U^{\dag} e^{i\ham t} U$. Overall, we need one
application of $U, U^{\dag}$ and $(U^{\dag} e^{i\ham t} U)$.
\begin{remark}
  This simplified Hamiltonian simulation procedure only works for a
  diagonal Hamiltonian, and, by the preceding discussion, for
  Hamiltonians that we know how to diagonalize. Otherwise, it is not
  obvious how to identify the eigenvectors and act as $e^{i \ham t}$
  on them.
\end{remark}

\subsection{Product formulas: Lie-Suzuki-Trotter decomposition}
\label{sec:trotter}
One of the most common ways of expressing a Hamiltonian is as a sum of
``simple'' terms. Here, ``simple'' can mean several different things;
examples of simple terms are sparse matrices (i.e., with at most a
given number of nonzero elements per row), or tensor products of Pauli
matrices. In particular, for Hamiltonians arising from physical
models, it is often the case that the Hamiltonian is described as a
summation of several ``local'' terms, i.e., terms that act only on a
small number of qubits (corresponding to particles) and are therefore
described by small matrices tensored with identity.  For this reason,
Hamiltonian simulation of a sum of simple terms is particularly well
studied. In other applications the Hamiltonian may be a general
matrix, but there is usually an assumption of (some type of) sparsity,
because Hamiltonian simulation algorithms rely on decomposing the
Hamiltonian in simpler terms, and simulating these terms individually.

Let us study the case where $\ham = \ham_1 + \ham_2$, and the discussion can
naturally be extended to a sum of multiple terms. If the Hamiltonians
$\ham_1$ and $\ham_2$ can be efficiently simulated, we can try to devise
approaches to simulate $\ham_1 + \ham_2$ using simulation for $\ham_1$ and
$\ham_2$ individually.  Suppose $\ham_1$ and $\ham_2$ commute; then
$e^{\ham_1+\ham_2} = e^{\ham_1} e^{\ham_2}$, as can be seen from the definition of
matrix exponential (Def.~\ref{def:matexp}), therefore the simulation is
trivial. Suppose now that $\ham_1$ and $\ham_2$ do not commute, which is the
more general and difficult case. We can rely on the Lie product
formula:\index{decomposition!Lie-Suzuki-Trotter}\index{Trotter formula}
\begin{equation*}
  e^{i(\ham_1+\ham_2)t} = \lim_{h \to \infty} \left(e^{i\ham_1t/h} e^{i\ham_2t/h}\right)^h.
\end{equation*}
While this is an infinite formula, it can be truncated by picking a
finite $h$, thereby introducing some error. The error depends on the
choice of $h$, because the error of approximating $e^{\ham_1+\ham_2}$
with $e^{\ham_1}e^{\ham_2}$ is $\bigO{\nrm{\ham_1} \nrm{\ham_2}}$;
this result is a consequence of the Campbell-Baker-Hausdorff theorem,
see \cite{bhatia2013matrix}. We do not prove it rigorously, but an
intuitive explanation can be given by taking the first-order Taylor
series approximation of the matrix exponential:
\begin{equation*}
  e^{\ham_1}e^{\ham_2} - e^{\ham_1+\ham_2} \approx (I+\ham_1) (I+\ham_2) - (I + \ham_1 +
  \ham_2) = \ham_1\ham_2,
\end{equation*}
so if $\nrm{\ham_1} \nrm{\ham_2}$ is small, $e^{\ham_1}e^{\ham_2}$ is close to
$e^{\ham_1+\ham_2}$. Then for any chosen integer $h$ we have:
\begin{equation}
  \label{eq:lstexact}
  e^{i\ham t} = (e^{i(\ham_1/h+\ham_2/h)t})^h = (e^{i\ham_1t/h}e^{i\ham_2t/h} + E)^h,
\end{equation}
where $E$ is the error matrix, with $\nrm{E} = \bigO{\nrm{i\ham_1t/h}
  \nrm{i\ham_2t/h}} = \bigO{\nrm{\ham_1} \nrm{\ham_2}t^2/h^2}$. Note that
Eq.~\eqref{eq:lstexact} holds as equality, because we are explicitly
incorporating the error term $E$. Consider what happens if we replace
the expression on the r.h.s.\ of Eq.~\eqref{eq:lstexact} with
$(e^{i\ham_1t/h}e^{i\ham_2t/h})^h$, which is simply the repeated application
of the unitaries $e^{i\ham_1t/h},e^{i\ham_2t/h}$ a total of $h$ times each,
in an interleaved way: this is known as the Lie-Suzuki-Trotter
first-order approach \cite{lloyd1996universal}. By
Prop.~\ref{prop:unitaryerror}, the error of approximating each unitary
in a sequence is at most the sum of the errors of the individual
approximations (i.e., errors in a sequence of unitaries increase at
most additively, rather than multiplicatively). By neglecting the
error term $E$ a total of $h$ times, we incur error at most:
\begin{equation}
  \label{eq:trottererror}
  h \nrm{E} = \bigO{h\nrm{\ham_1} \nrm{\ham_2}t^2/h^2} = \bigO{\nrm{\ham_1} \nrm{\ham_2}t^2/h}.
\end{equation}
Thus, for any given $\epsilon > 0$, choosing $h =
\bigO{\nrm{\ham_1}\nrm{\ham_2} \frac{t^2}{\epsilon}}$ guarantees:
\begin{equation*}
  \nrm{ e^{i(\ham_1+\ham_2)t} - \left(e^{i\ham_1t/h} e^{i\ham_2t/h}\right)^h } \le
  \epsilon.
\end{equation*}
\begin{remark}
  It is possible to obtain an $\epsilon$-approximation with fewer
  terms (i.e., fewer unitaries depending on $\ham_1$ or $\ham_2$ only)
  in the product, e.g., by using higher-order terms in the Taylor
  series of the matrix exponential, which leads to a more accurate
  approximation for the same number of product terms. The first-order
  approach suffices for us to get the main idea and to state the
  approximation results used in Ch.~\ref{ch:adiabatic}.
\end{remark}
Finally, we note that the product formula can be extended to a
summation of several terms, using the fact that $e^{i(\sum_j \ham_j)t}
= \lim_{h \to \infty} \left(\prod_j e^{i\ham_jt/h}\right)^h$. If the
Hamiltonian is a summation of $m$ terms, then to obtain error
$\epsilon$ it is sufficient to choose $h = \bigO{m^2
  t^2/\epsilon}$. The gate complexity of a Hamiltonian simulation
circuit obtained with the Lie-Suzuki-Trotter first-order approach is
$h$ times the gate complexity of each individual simulation piece
$\prod_j e^{i\ham_jt/h}$, so in the best case where each of the $m$
terms $e^{i\ham_jt/h}$ can be simulated with a constant number of
gates, we end up with gate complexity $\bigO{m^3 t^2/\epsilon}$; and
if each term is more expensive to simulate, the total cost increases
accordingly.
\begin{remark}
  The gate complexity of implementing a circuit for $e^{i\ham_jt/h}$
  depends on $\ham_j$, but it is not unrealistic to think that it
  might be constant. For example, for many Hamiltonians arising from
  physical models, each term $\ham_j$ acts nontrivially on a constant
  number of qubits, and acts as the identity on the remaining ones. In
  that case, using properties of the tensor product, $e^{i\ham_jt/h}$
  is a constant-size matrix tensored with identity matrices, and as
  such, it can be implemented in a constant number of gates: we only
  need to act on the qubits on which $\ham_j$ acts nontrivially.
\end{remark}
\begin{example}
  Let us look at the case in which $\ham$ is a summation of terms, of the
  following form: $\ham = \sum_{(j,k) \in E} C_{jk}$, where $E$ is the
  edge set of some graph $G = (V, E)$ and $C_{jk}$ acts on two qubits
  only:
  \begin{equation*}
    C_{jk} := I \otimes \dots \otimes I \otimes \underbrace{Z}_{\text{pos. $j$}} \otimes I \otimes \dots \otimes I \otimes \underbrace{Z}_{\text{pos. $k$}} \otimes I \otimes \dots \otimes I.
  \end{equation*}
  In other words, $C_{jk}$ acts as the identity on all qubits, except
  on qubits $j$ and $k$, where it acts with the Pauli $Z$ matrix. In
  the literature, such a Hamiltonian would typically be written more
  compactly as $\ham = \sum_{(j,k) \in E} \sigma^Z_j \sigma^Z_k$ (see
  Eq.~\ref{eq:sigmazdef}) or as $\ham = \sum_{(j,k) \in E} Z_j
  Z_k$. This type of Hamiltonian appears in a certain formulation of
  quadratic unconstrained binary optimization problems, and it is
  discussed thoroughly in Ch.~\ref{ch:adiabatic}; in particular, see
  Sect.~\ref{sec:combopteig}.

  Suppose we want to implement $e^{i \ham t}$ for some value of
  $t$. By our discussion above, it is sufficient to implement $e^{i
    C_{jk} t}$ for each term $C_{jk}$, because these terms commute and
  so the matrix exponential of the entire Hamiltonian is equal to the
  product of the matrix exponentials for the individual
  terms. Regarding the implementation of $e^{i C_{jk} t}$, note that
  it acts trivially (i.e., as the identity) on all qubits except $j$
  and $k$, by definition of the matrix exponential. We can therefore
  limit ourselves to understanding the effect of $e^{i C_{jk} t}$ on
  the two qubits $j$ and $k$, which amounts to computing $e^{i t Z
    \otimes Z}$. This is straightforward:
  \begin{equation*}
    Z \otimes Z = \begin{pmatrix} 1 & 0 & 0 & 0 \\ 0 & -1 & 0 & 0 \\ 0 & 0 & -1 & 0 \\ 0 & 0 & 0 & 1 \end{pmatrix}
    \qquad
    e^{i t Z \otimes Z} = \begin{pmatrix} e^{i t} & 0 & 0 & 0 \\ 0 & e^{-i t} & 0 & 0 \\ 0 & 0 & e^{-i t} & 0 \\ 0 & 0 & 0 & e^{i t} \end{pmatrix}.
  \end{equation*}
  We can decompose this unitary into basic gates like any other
  unitary, with C$X$ gates and single-qubit gates: an explicit circuit
  for this is given in Sect.~\ref{sec:qaoaimp}. The gate complexity of
  simulating the entire Hamiltonian in this way is $\bigOt{|E|}$ for
  error $\epsilon$, where the errors are due to the decomposition of
  each term $C_{jk}$ into basic gates (with
  polylogarithmic scaling in $\frac{1}{\epsilon}$).

  If instead of $Z \otimes Z$ we had a more complicated (say,
  non-diagonal) two-qubit gate in the exponent, we can still perform
  Hamiltonian simulation thanks to the Lie-Suzuki-Trotter formula. For
  example, still assuming $\ham = \sum_{(j,k) \in E} C_{jk}$, let us
  now suppose $C_{jk}$ is a general $4 \times 4$ matrix (that we know)
  when restricted to the space of qubits $j$ and $k$, and it acts as
  the identity on all other qubits. Now the terms $C_{jk}$ no longer
  commute in general. However, the matrix exponential of a single
  $C_{jk}$ term is still a $4 \times 4$ matrix on qubits $i$ and $j$
  tensored with identity on the other qubits, and any such matrix can
  be approximated to high precision (say, error $< \epsilon'$) with
  $\bigOt{1}$ gates: we can analytically compute $e^{i C_{jk} t}$,
  which is unitary, and decompose it into single- and two-qubit gates,
  see Thm.~\ref{thm:sk} and the surrounding discussion. Then we apply
  the Lie-Suzuki-Trotter formula for a summation of multiple terms,
  implementing the Hamiltonian simulation with the circuit:
  \begin{equation*}
    \left(\prod_{(j,k) \in E} e^{i C_{jk} t/h}\right)^h,
  \end{equation*}
  where $h = \bigO{|E|^2 t^2 /\epsilon}$ to guarantee that
  \begin{equation*}
    \nrm{e^{it\sum_{(j,k) \in E} C_{jk}} - \left(\prod_{(j,k) \in E}
      e^{i C_{jk} t/h}\right)^h} \le \epsilon.
  \end{equation*}
  The gate complexity of the resulting circuit is $\bigOt{|E|^3 t^2
    /\epsilon}$.
\end{example}

\subsection{Linear combination of unitaries}
\label{sec:lcu}
Throughout this section we assume $\nrm{\ham} \le 1$, see
Rem.~\ref{rem:hamsimnormub}, and we want to give a circuit
implementation for $e^{i\ham t}$.

It is easy to apply the product of two unitary matrices, for which we
have a circuit implementation, to a quantum state: we simply apply
them one after the other. But how do we apply a linear combination of
those unitaries? In this section we give a possible answer to this
question, but before we dive into that, it is useful to discuss the
connection between linear combination of unitaries and Hamiltonian
simulation. The acronym LCU is often used in the literature to refer
to ``linear combination of unitaries.''\index{unitary matrix!linear combination (LCU)}

\paragraph{Connection between Hamiltonian simulation and LCU.} Suppose we know a decomposition of $\ham$
in terms of some unitary matrices: $\ham = \sum_{j=1}^m \beta_j U_j$,
where the coefficients $\beta_j$ are real numbers, which we can assume
w.l.o.g.\ because any complex phase can be absorbed into $U_j$.
\begin{remark}
  A decomposition of $\ham$ in terms of unitaries always exist: the
  Pauli matrices, appropriately tensored, form a basis for the space of
  multi-qubit operators, and they are unitary. However, it is usually
  better (for computational efficiency) if one can find a
  \emph{simple} decomposition as a linear combination of unitaries,
  i.e., with few terms. To consider a similar situation as in
  Sect.~\ref{sec:trotter}, if $\ham$ is a summation of a few terms
  that act only on a constant number of qubits, then each of these
  terms can be written as a linear combination of a constant number of
  unitaries: although this latter constant is exponential in the
  number of qubits involved, it is still a constant. This approach
  does not scale well if the number of qubits on which each term of
  $\ham$ acts is given as a (variable) input to the simulation
  algorithm.
\end{remark}
By definition of matrix exponential we have:
\begin{equation}
  \label{eq:hamsimlcu}
  e^{i\ham t} = \sum_{k=0}^{\infty} \frac{(i\ham t)^k}{k!} = \sum_{k=0}^{\infty} \frac{(it)^k}{k!} \left(\sum_{j=1}^m \beta_j U_j\right)^k = \sum_{k=0}^{\infty} \frac{(it)^k}{k!} \sum_{j_1, j_2, \dots, j_k \in \{1,\dots,m\}} \beta_{j_1}\beta_{j_2}\cdots \beta_{j_k} U_{j_1} U_{j_2} \cdots U_{j_k}.
\end{equation}
At the r.h.s.\ of Eq.~\eqref{eq:hamsimlcu} we have an infinite series,
but consider what happens if we truncate the Taylor series at $k = c(t
+ \log \frac{1}{\epsilon}) = \bigO{t + \log \frac{1}{\epsilon}}$, for
some constant $c$. Recalling that $k! \ge (k/e)^k$, we have:
{\allowdisplaybreaks
\begin{align*}
  \nrm{ e^{i\ham t} - \sum_{k=0}^{c(t + \log \frac{1}{\epsilon}) - 1} \frac{(i\ham t)^k}{k!} } &= \nrm{\sum_{k=c(t + \log \frac{1}{\epsilon})}^{\infty} \frac{(i\ham t)^k}{k!} } \le \sum_{k=c(t + \log \frac{1}{\epsilon})}^{\infty} \nrm{\frac{(i\ham t)^k}{k!} } \\
  &\le \sum_{k=c(t + \log \frac{1}{\epsilon})}^{\infty} \nrm{\frac{t^k}{k!} } = \sum_{k=c(t + \log \frac{1}{\epsilon})}^{\infty} \frac{t^k}{k!}  \\
  &\le \sum_{k=c(t + \log \frac{1}{\epsilon})}^{\infty} \left(\frac{et}{k}\right)^k 
  \le \sum_{k=c(t + \log \frac{1}{\epsilon})}^{\infty} \left(\frac{et}{ct}\right)^k \\
  &\le \sum_{k=c(t + \log \frac{1}{\epsilon})}^{\infty} \left(\frac{e}{c}\right)^k \le \frac{(e/c)^{c(t + \log (1/\epsilon))}}{1-e/c} = \frac{(e/c)^{ct}(e/c)^{c\log (1/\epsilon)}}{1-e/c}.
\end{align*}
}
Simple calculations show that $c=2e$ suffices to ensure that the above
expression is $\le \epsilon$, and note that this choice is independent
of $t$ or $\epsilon$. Thus, if we can implement the first $\bigO{t +
  \log \frac{1}{\epsilon}}$ terms of the expression at the r.h.s.\ of
Eq.~\eqref{eq:hamsimlcu}, we obtain an $\epsilon$-approximation of
$e^{i\ham t}$. In summary, one way to solve the Hamiltonian simulation
problem is to implement:
\begin{equation}
  \label{eq:hamsimlcutruncated}
  \sum_{k=0}^{\bigO{t + \log \frac{1}{\epsilon}}} \sum_{j_1, j_2, \dots, j_k \in \{1,\dots,m\}} \frac{(it)^k}{k!}  \beta_{j_1}\beta_{j_2}\cdots \beta_{j_k} U_{j_1} U_{j_2} \cdots U_{j_k},
\end{equation}
which is a linear combination of the unitaries $(U_{j_1} U_{j_2}
\cdots U_{j_k})$ with coefficients $\frac{(it)^k}{k!}
\beta_{j_1}\beta_{j_2}\cdots \beta_{j_k}$. Note that if we know how to
implement all the matrices $U_j$, then implementing $U_{j_1} U_{j_2}
\cdots U_{j_k}$ is straightforward, as it is just a sequence of
operations that we know how to apply. Thus, we now discuss the task of
implementing a linear combination of unitaries with given coefficients
in the linear combination.

\begin{figure}[h!]
  \leavevmode
  \centering
  \ifcompilefigs
  \Qcircuit @C=1em @R=0.7em @!R {
   \lstick{\ket{\v{0}}} & \qw      & \gate{W} & \multigate{1}{V} & \gate{W^{\dag}} & \qw \\
   \lstick{\ket{\psi}}  & \qw      & \qw      & \ghost{V}        & \qw            & \qw 
  }
  \else
  \includegraphics{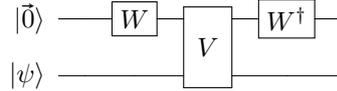}
  \fi
  \caption{Circuit for the implementation of a linear combination of
    unitaries.}
  \label{fig:lcu}
\end{figure}

\paragraph{Implementing a LCU.} To simplify the discussion, let us rename some of the quantities
involved. We can recast the problem of implementing a linear
combination of unitaries as implementing $M = \sum_{j=0}^{2^q-1}
\alpha_j V_j$ with $\alpha_j$ nonnegative and real (as before, we can
absorb everything else into $V_j$), and where the $V_j$ are $n$-qubit
unitaries. $M$ may not be unitary in general, so if we want to apply
it to some state $\ket{\psi}$, we must instead aim to implement
$M\ket{\psi}/\nrm{M\ket{\psi}}$, where the normalization ensures that
we obtain a proper quantum state. Because $\alpha_j$ are nonnegative
and real, we have $\nrm{\alpha}_1 = \sum_{j=0}^{2^q-1} \alpha_j$ for
the $\ell^1$-norm of the vector of coefficients $\alpha$. Because the
coefficients $\alpha_j$ are known, we can construct --- for example
using the amplitude encoding algorithm discussed in
Sect.~\ref{sec:ampencalg} --- a $q$-qubit unitary $W$ that implements
the following map:
\begin{equation*}
  W \ket{\v{0}}_q = \frac{1}{\sqrt{\nrm{\alpha}_1}} \sum_{\vj \in
    \{0,1\}^q} \sqrt{\alpha_j} \ket{\vj}.
\end{equation*}
Note that the state on the r.h.s.\ is a proper quantum state due to
the normalization chosen.  Suppose we have access to a unitary $V$
that can implement all the $V_j$ in a controlled manner: $V :=
\sum_{\vj \in \{0,1\}^q} \ketbra{\vj}{\vj} \otimes V_j$. The
effect of $V$ is precisely that of applying $V_j$ onto the second
register if the first register contains $\ket{\vj}$:
\begin{equation*}
  V \ket{\vj} \ket{\psi} = \ket{\vj} V_j \ket{\psi}.
\end{equation*}
Now consider the effect of the circuit in Fig.~\ref{fig:lcu}.
After $W$, we are in the state $\frac{1}{\sqrt{\nrm{\alpha}_1}}
\sum_{\vj \in \{0,1\}^q} \sqrt{\alpha_j} \ket{\vj} \ket{\psi}$. After
$V$, we are in the state $\frac{1}{\sqrt{\nrm{\alpha}_1}} \sum_{\vj
  \in \{0,1\}^q} \sqrt{\alpha_j} \ket{\vj} V_j \ket{\psi}$. The effect
of the final application of $W^{\dag}$ on the quantum state is more
difficult to write down analytically, but it yields some state
$\ket{\varphi}$. Let us find an expression for the part of
$\ket{\varphi}$ that contains $\ket{\v{0}}$ in the first register;
formally, this can be expressed as:
\begin{align*}
  (\bra{\v{0}} \otimes I^{\otimes n}) \ket{\varphi} &= (\bra{\v{0}} \otimes I^{\otimes n})
  (W^{\dag} \otimes I^{\otimes n}) V (W \otimes I^{\otimes n}) \ket{\v{0}}\ket{\psi} =
  (\bra{\v{0}}W^{\dag} \otimes I^{\otimes n}) V (W \otimes I^{\otimes n}) \ket{\v{0}}\ket{\psi} \\
  &=\left(\frac{1}{\sqrt{\nrm{\alpha}_1}} \sum_{\vj \in \{0,1\}^q} \sqrt{\alpha_j}
  \bra{\vj} \otimes I^{\otimes n} \right) \left(\frac{1}{\sqrt{\nrm{\alpha}_1}}
  \sum_{\vj \in \{0,1\}^q} \sqrt{\alpha_j} \ket{\vj} V_j \ket{\psi}\right) \\
  &= \frac{1}{\nrm{\alpha}_1} \sum_{\vj \in \{0,1\}^q} \alpha_j V_j \ket{\psi}.
\end{align*}
Thus, the final state $\ket{\varphi}$ can be written as:
\begin{equation*}
  \frac{1}{\nrm{\alpha}_1} \ket{\v{0}} M \ket{\psi} + \sqrt{1 - \frac{\nrm{M\ket{\psi}}^2}{\nrm{\alpha}_1^2}} \ket{\phi},
\end{equation*}
where $\ket{\phi}$ is some quantum state that has no support on
$\ketbra{\v{0}}{\v{0}} \otimes I^{\otimes n}$ and that we do not care
about. (The coefficient under the square root is computed by noting
that the entire state must have unit norm, hence the amplitude of the
second part in the expression for the state must be the square root of
$1 - (\text{norm of the first part})$.) In other words, we have
implemented $M \ket{\psi}$, but only if a measurement of the first
register yields $\ket{\v{0}}$: this happens with probability
$\frac{\nrm{M\ket{\psi}}^2}{\nrm{\alpha}_1^2}$. We can amplify this
probability of success to some value very close to 1 with
$\bigO{\nrm{\alpha}_1/\nrm{M\ket{\psi}}}$ rounds of oblivious
amplitude amplification (i.e., $\bigO{1/\sqrt{p}}$ rounds for an
algorithm with probability of success $p$: the usual quadratic
advantage of amplitude amplification), after which we can be almost
certain that we produced the state $M\ket{\psi}/\nrm{M\ket{\psi}}$, as
desired. Note that we need oblivious amplitude amplification because
$\ket{\psi}$ is given and we do not assume that we have access to a
unitary to prepare it, so vanilla amplitude amplification cannot be
applied directly, but fortunately we satisfy the more forgiving
assumptions of oblivious amplitude amplification, see
Sect.~\ref{sec:obliviousampamp}.

\paragraph{Resource analysis.} We can now get back to Hamiltonian simulation and conclude our
discussion on the linear combination of unitaries. We want to
implement Eq.~\eqref{eq:hamsimlcutruncated} using the algorithm
outlined above, which takes $\bigO{\nrm{\alpha}_1/\nrm{M\ket{\psi}}}$
applications of $W$ and $V$. We can ignore $\nrm{M\ket{\psi}}$,
because the $M$ that we want to implement is (an approximation of)
$e^{i\ham t}$, thus $\nrm{M\ket{\psi}} \approx 1$. Regarding
$\nrm{\alpha}_1$, because the components of $\alpha$ are the
coefficients of the linear combination in
Eq.~\ref{eq:hamsimlcutruncated}, we have:
\begin{align*}
  \nrm{\alpha}_1 &= \sum_{k=0}^{\bigO{t + \log \frac{1}{\epsilon}}} \sum_{j_1, j_2, \dots, j_k \in \{1,\dots,m\}} \frac{t^k}{k!}
  \beta_{j_1}\beta_{j_2}\cdots \beta_{j_k} \le \sum_{k=0}^{\infty} \sum_{j_1, j_2, \dots, j_k \in \{1,\dots,m\}} \frac{t^k}{k!}
  \beta_{j_1}\beta_{j_2} \cdots \beta_{j_k}\\
  &= \sum_{k=0}^{\infty} \frac{(t\nrm{\beta}_1)^k}{k!} \le e^{t\nrm{\beta}_1},
\end{align*}
which is exponential in $t$ and $\nrm{\beta}_1$. This would lead to
$e^{t\nrm{\beta}_1}$ applications of $V$ and $W$, but we can reduce
this cost by taking advantage of the logarithmic error dependence
($\log \frac{1}{\epsilon}$) of the algorithm. Indeed, note that if $t$
is small then the complexity of the algorithm is also small. Thus, we
can divide the Hamiltonian simulation into $t\nrm{\beta}_1$ blocks
where each block evolves the Hamiltonian for time $\tau =
1/\nrm{\beta}_1$ and uses precision $\epsilon'$. The end result is the
same because $(e^{i\ham \tau})^{t\nrm{\beta}_1} = e^{i\ham t}$, but
the complexity of each block (in terms of the number of applications
of $V$ and $W$) is now $\bigO{e^{\tau\nrm{\beta}_1}} = \bigO{1}$. Also
note that the gate complexity of $W$ inside each block is small,
because the vector of coefficients of the linear combination includes
$\bigO{\tau + \log \frac{1}{\epsilon'}} = \bigO{\log
  \frac{1}{\epsilon'}}$ nonzero terms, and similarly, $V$ inside each
block applies $\bigO{\tau + \log \frac{1}{\epsilon'}} = \bigO{\log
  \frac{1}{\epsilon'}}$ input unitaries $U_j$ (this is a consequence
of the truncation for Eq.~\eqref{eq:hamsimlcu}). Finally, we note that
it is sufficient to choose error $\epsilon' =
\epsilon/(t\nrm{\beta}_1)$ in each block to achieve total error at
most $\epsilon$, and the total cost is $t\nrm{\beta}_1$ times the cost
of each block. This yields an algorithm with complexity
$\bigO{t\nrm{\beta}_1 \log \frac{t\nrm{\beta}_1}{\epsilon}}$, in terms
of the number of applications of the unitaries $U_j$ and additional
elementary gates.

\subsection{Hamiltonian simulation for sparse matrices with oracle access}
\label{sec:hamsimsparse}
Our goal in this section is to introduce a natural input model for
matrices, and give an intuition for Hamiltonian simulation algorithms
based on this model. We do not go into the details because in the rest
of this \book{} we only need the input model from this section, and
not the algorithms.

In the preceding sections we assumed that the Hamiltonian can be
expressed as a sum of local terms, i.e., operators that act only on a
small number of qubits. That model is suitable for several
applications originating from the study of physical systems and
general abstract models, but it is not necessarily suitable for
data-driven applications where the data may have little structure. In
classical (i.e., non-quantum) scientific computing, data is often
represented in a compact form by listing only the nonzero
elements. The analog of that representation in the quantum world is
the {\em sparse-oracle}\index{oracle!sparse} input model. In this model, we assume that we
have access to the following quantum circuits, generally called
oracles in this context, that give a description of the Hamiltonian
$\ham$.
\begin{itemize}
\item The first oracle, mapping $\ket{\vj} \ket{\v{\ell}} \to \ket{\vj}
  \ket{\vv{c_{j\ell}}}$, provides the index $c_{j\ell}$
  of the $\ell$-th nonzero element of column $j$,
\item The second oracle, mapping $\ket{\vj} \ket{\vk} \ket{\v{z}} \to
  \ket{\vj} \ket{\vk} \ket{\v{z} \oplus \vv{\ham_{jk}}}$, provides the
  value of the element in position $j, k$ of the Hamiltonian. 
\end{itemize}
In other words, one map provides the indices of the nonzero elements
in each column, and one map provides the values of arbitrary
elements.
\begin{remark}
  The definition of the oracle providing the value for the nonzero
  entries of $\ham$ implicitly assumes that such entries are
  integer-valued; this is not a restrictive assumption, because as
  long as the entries are rational (as in any finite-precision
  representation), we can rescale them to integer and adjust the value
  of the simulation time $t$ to compensate for any scaling.
\end{remark}
The sparse-oracle input model is the main subject of
Sect.~\ref{sec:sparseblockenc}, where we discuss a circuit to
construct a unitary that (in some sense, to be specified later)
encodes a given matrix; that type of encoding, called block-encoding,
also leads to Hamiltonian simulation algorithms, see
Sect.~\ref{sec:signalproc}. However, there are algorithms for
Hamiltonian simulation tailored to the sparse-oracle input model and
that do not rely on the block-encoding framework: we give an example
of such an algorithm next.

Consider the following approach: we first decompose the Hamiltonian
into a sum of roughly $s$ terms, where $s$ is the maximum number of
nonzero elements per row/column. That is, we write $\ham =
\bar{\ham}_{\text{diag}} + \sum_{j} \bar{\ham}_j$, where
$\bar{\ham}_{\text{diag}}$ is the diagonal part of the Hamiltonian,
and $\bar{\ham}_j$ are matrices containing only off-diagonal terms. To
determine these matrices, we interpret the rows and columns of $\ham$
as nodes in a graph, with an edge between two nodes if and only if
there is a nonzero element in $\ham$ in the corresponding position;
i.e., if $\ham \in \C^{2^n \times 2^n}$, we construct a graph $G =
(V,E)$ with $V = \{0,1\}^n$, $E = \{(\v{u}, \v{v}) : \v{u}, \v{v} \in
\{0,1\}^n, \ham_{uv} \neq 0, u \neq v\}$. Then, an edge coloring of
$G$ gives a decomposition of $\ham$ into a sum of element-wise
disjoint matrices, where each matrix contains all elements
corresponding to a certain color, and by definition of edge coloring
(no two edges connected to the same vertex have the same color), each
matrix has at most one nonzero element per row or column. Because of
this, the matrices in the decomposition are easy to diagonalize. We
then simulate each matrix in the decomposition independently, e.g.,
using the methodology of Sect.~\ref{sec:hamsimdiagonal}, and use some
approach (e.g., product formula) to compose the simulation of
$\bar{\ham}_{\text{diag}}$ and $\bar{\ham}_j \, \forall j$ into a
simulation for the original Hamiltonian. Note that edge colorings can
be computed in polynomial time, and there always exists an edge
coloring of a graph of size at most $\Delta + 1$, where $\Delta$ is
the maximum degree of the graph --- which, by construction, is the
number $s$ of nonzero elements per row/column of the Hamiltonian. For
details of this approach, see \cite[Chapter 2]{childs2004thesis}.

Although the above sketch is only a brief, high-level overview of a
Hamiltonian simulation approach tailored to the sparse-oracle input
model, it gives a concrete example where it is clear why we obtain an
algorithm with a complexity that scales with the maximum number of
nonzero elements in each column/row. In fact, this is very common for
this input model, explaining why this particular model leads to
algorithms with competitive performance only if the Hamiltonian is
sparse (i.e., the number of nonzero terms per row or column is at most
polylogarithmic in the size of Hamiltonian). Further references on
Hamiltonian simulation with this input model can be found in
Sect.~\ref{sec:hamsimnotes}.

\subsection{Signal processing and the block-encoding framework}
\label{sec:signalproc}
Depending on the specific situation at hand (i.e., the characteristics
of the Hamiltonian and how it is specified), the fastest quantum
algorithms for Hamiltonian simulation generally rely on the signal
processing framework, proposed in \cite{low2019hamiltonian}. The
signal processing approach is best understood within the
block-encoding framework. The input Hamiltonian can be specified in
many possible ways in this framework. For example, it can be described
in the same sparse-oracle input model as in
Sect.~\ref{sec:hamsimsparse}, but this is not the only possibility:
the only requirement is that we can construct a circuit that acts as
the desired Hamiltonian on a certain subspace.

We do not have the necessary background to describe Hamiltonian
simulation algorithms that rely on this input model yet, but we
revisit this topic in Sect.~\ref{sec:blockencop}, after introducing
the block-encoding framework. At a very high level, it is based on the
idea of implementing a polynomial transformation of the singular
values of the Hamiltonian, approximating the exponential function. The
polynomial approximation is constructed by acting on a quantum circuit
that implements a (possibly scaled down) version of the Hamiltonian
itself, so that we can implement a matrix function of it. We refer the
reader to Sect.~\ref{sec:blockencop} and the notes in
Sect.~\ref{sec:blockencnotes} for a more detailed discussion of the
block-encoding framework, and Hamiltonian simulation within that
framework; as a byproduct, the discussion also provides the main
intuition of the signal processing technique.\index{algorithm!Hamiltonian simulation|)}\index{simulation!Hamiltonian|)}

\section{Notes and further reading}
\label{sec:hamsimnotes}
Feynman's groundbreaking proposal and discussion of a quantum computer
to simulate the evolution of a quantum-mechanical system can be found
in \cite{feynman1982simulating,feynman2018lectures}. His work also set
the foundations for showing that Hamiltonian simulation is BQP-hard,
i.e., that every problem that can be efficiently solved by a quantum
computer can be cast as a Hamiltonian simulation problem. Throughout
this chapter we gave a few references to efficient (i.e.,
polynomial-time) quantum algorithms for Hamiltonian simulation, and
several more are given below: each of these algorithms already shows
that Hamiltonian simulation is in BQP, hence we conclude that it is
BQP-complete. To read about additional BQP problems,
\cite{wocjan2006several} is a good starting point. As it turns out,
even the problem of inverting a Hermitian matrix, specified in an
appropriate manner, is BQP-complete: \cite{harrow2009quantum} shows
that the problem of simulating an arbitrary quantum circuit can be
cast as the problem of applying the inverse of a certain matrix. The
matrix inversion algorithm of \cite{harrow2009quantum} is discussed in
Ch.~\ref{ch:blockenc}, in the context of quantum algorithms for linear
systems.

In the context of optimization, Hamiltonian simulation is used as a
building block for several useful subroutines: see
Ch.~\ref{ch:blockenc} where it is at the heart of matrix manipulation
algorithms, Ch.~\ref{ch:mmwu} where matrix manipulation is used to
build algorithms for semidefinite optimization problems, and
Ch.~\ref{ch:adiabatic}, where it is used to solve minimum or maximum
eigenvalue problems. In a recent line of work, initiated in
\cite{leng2023quantum}, the solution of the Schr\"odinger equation is
used \emph{directly} to solve continuous optimization problems: this
is done by defining a Hamiltonian whose evolution follows a descent
direction for the optimization problem. The inspiration for this work
is \cite{wibisono2016variational}, describing a dynamical system that
follows the natural steepest descent direction. \cite{leng2023quantum}
proposes a quantum Hamiltonian that closely mimics such a dynamical
system, and shows that simulation of the Schr\"odinger equation with
such a Hamiltonian converges to the global minimum for both convex and
nonconvex problems --- although for nonconvex problems the simulation
time may need to be exponentially large, as expected (we do not expect
quantum computers to solve nonconvex optimization problems in
polynomial time). This algorithm for nonconvex optimization is further
investigated in \cite{leng2023quantumsepa}, whereas a software package
for its utilization and a numerical evaluation can be found in
\cite{kushnir2025qhdopt}.  \cite{augustino2023central} extends this
work to simulate the central path of interior point methods for linear
optimization problems, yielding a provably convergent quantum
algorithm for linear programs with a favorable running time compared
to several classical algorithms with respect to the dimension of the
problem (but $1/\epsilon$ dependence on the final optimality
gap). Notably,
\cite{leng2023quantum,leng2023quantumsepa,augustino2023central} do not
assume access to QRAM.

The development of an approach for Hamiltonian simulation based on a
linear combination of unitaries is due to
\cite{childs2012hamiltonian}, and the LCU technique has found numerous
applications in a variety of quantum algorithms; for further
discussion, see also \cite{childs2017quantum}.

Several examples of efficient Hamiltonian simulation algorithms for
the sparse-oracle input model are described in
\cite{berry2007efficient,berry2014exponential,berry2015hamiltonian,low2019hamiltoniansparse}.

\chapter{Matrix manipulation with quantum algorithms}
\label{ch:blockenc}
\thispagestyle{fancy}
Operations on matrices are at the heart of a vast number of
optimization algorithms. Quantum computers can only apply unitary
matrices directly, but there exist a vast toolbox of quantum algorithms
to perform complex operations on non-unitary matrices as well. In
this chapter we discuss two aspects of non-unitary matrix manipulation
that have been featured prominently in existing quantum algorithms for
optimization: algorithms for linear systems (i.e., matrix inversion),
and the block-encoding framework. Matrix inversion can also be
performed within the block-encoding framework, and it is actually more
efficient in that framework than in the phase-estimation-based
approach that we initially present. Because the phase estimation
approach to matrix inversion is historically important, elegant, and
showcases a number of powerful ideas for the design of quantum
algorithm, we discuss it anyway, and in fact we begin our discussion
in this chapter with it.

\section{Quantum linear system solvers}
\label{sec:linear_system}
The subject of this section is the solution of linear
systems\index{algorithm!linear system|(}\index{linear system!algorithm|(}\index{algorithm!HHL|see{linear system}} of
equations, one of the most ubiquitous problems in engineering. We
describe and analyze an algorithm for this task introduced by
\cite{harrow2009quantum}. Our description is mostly based on the
original version of \cite{harrow2009quantum} --- typically
referred to as the HHL algorithm, from the last name of the authors,
Harrow, Hassidim and Lloyd --- but there have been many refinements of
that scheme over the years. We discuss notable improvements to the HHL
algorithm in Sect.~\ref{sec:hhlimprovements}, yielding significantly
better gate complexity bounds (even exponentially better, in some of
the input parameters); see also the notes at the end of this chapter
(Sect.~\ref{sec:blockencnotes}).

We use the acronym QLSA to refer to a quantum linear system
algorithm. (The abbreviation QLSA is relatively common in the
literature.) The problem solved by a QLSA can be stated as follows:
given a Hermitian invertible matrix $A \in \C^{2^n \times 2^n}$, a
vector $b \in \C^{2^n}$, a precision parameter $\epsilon > 0$, compute
an $n$-qubit state $\ket{\psi}$ such that $\|\ket{\psi} -
\ket{\amp{A^{-1} b}}\| \le \epsilon$ (recall
Def.~\ref{def:ampenc}). We assume that the input data is described by
suitable quantum oracles $P_A, P_b$; we discuss the exact nature of
these oracles in Sect.~\ref{sec:hhldetails}. Furthermore, we assume
that $\|A\| \le 1$ and the condition number $\kappa$ of $A$, or an
upper bound on it, is known, see Sect.~\ref{sec:hhlkappa}. Note that
the assumption $\|A\| \le 1$ may in general require normalizing the
linear system, which in turn requires adjusting the precision. In
\cite{harrow2009quantum} there is also a somewhat hidden assumption
that $A$ is positive semidefinite, in which case all eigenvalues lie
in $[\frac{1}{\kappa}, 1]$; this assumption shows up subtly in a few
crucial points of the analysis, for example because the filter
functions for the eigenvalues (described later) are defined for
positive values only. This is handled carefully in several subsequent
works, such as
\cite{childs2017quantum,chakraborty2019power,gilyen2019quantum}, where
the assumption is that the spectrum of the matrix is in $[-1,
  -\frac{1}{\kappa}] \cup [\frac{1}{\kappa}, 1]$. For now we keep the
more restrictive assumption.

\subsection{Algorithm description: simplified exposition}
\label{sec:hhlsimple}
We give a simplified exposition of the algorithm that is not entirely
accurate, but it conveys the main ideas without getting bogged down in
details. Most of the inaccuracies of our exposition are eventually
discussed in subsequent sections, in particular
Sect.s~\ref{sec:hhldetails} and \ref{sec:hhlkappa}.

\begin{figure}[t!]
  \leavevmode
  \centering
  \ifcompilefigs
  \Qcircuit @C=1em @R=0.7em {
    \lstick{\ket{\v{0}}}       & {/^m} \qw  & \gate{H^{\otimes m}} & \ctrl{1}            & \gate{Q_m^\dag} & \qw \\
    \lstick{\ket{\amp{b}}}     & {/^n} \qw  & \qw                  & \gate{U_\text{evo}} & \qw                 & \qw \\
  }
  \else
  \includegraphics{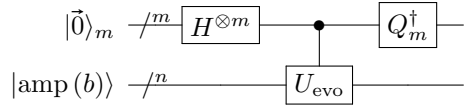}
  \fi
  \caption{Determining an eigendecomposition of $\ket{\amp{b}}$.}
  \label{fig:hhlqpe}
\end{figure}

The HHL algorithm for the solution of a linear system relies on
quantum phase estimation and Hamiltonian simulation. The first step of
this QLSA is to decompose $\ket{\amp{b}}$ into an eigenbasis of
$A$. For this, we use phase estimation. Let $m$ be a number of qubits,
to be determined later, to store the phases in quantum phase
estimation; the first $m$ qubits form the first register. We also use
an $n$-qubit register (second register) to store the eigenstate, and
an additional qubit for a certain rotation that we ignore for now and
introduce only when necessary. We apply phase estimation using the
circuit given in Fig.~\ref{fig:hhlqpe}, where the controlled
$U_\text{evo}$ block (with the first register as control) is the
unitary defined as:
\begin{equation*}
  \text{C}U_\text{evo} := \sum_{\vk \in \{0,1\}^m} \ketbra{\vk}{\vk} \otimes e^{2\pi iA k}.
\end{equation*}
This unitary applies Hamiltonian simulation for a variable amount of
time, and in particular, the length of the simulation is determined by
the content of the first register, i.e., the value of the index
$\ket{\vk}$ in the above expression.

\begin{figure}[h!]
\leavevmode
\centering
\ifcompilefigs
\Qcircuit @C=1em @R=0.7em {
  \lstick{\ket{\v{0}}} & {/^m} \qw  & \gate{H^{\otimes m}} & \ctrl{1}            & \gate{Q_m^\dag} & \ctrl{2}              & \qw  & \gate{Q_m} & \ctrl{1}                   & \gate{H^{\otimes m}} & \qw & \rstick{\ket{\v{0}}} \\
  \lstick{\ket{\amp{b}}}     & {/^n} \qw  & \qw                  & \gate{U_\text{evo}} & \qw                 & \qw                   & \qw  & \qw            & \gate{U_\text{evo}^{\dag}} & \qw                  & \qw & \rstick{C' \sum_{h=0}^{2^n-1} \frac{\beta_h}{\lambda_h} \ket{\psi_h}}\\
  \lstick{\ket{0}}     & \qw        & \qw                  & \qw                 & \qw                 & \gate{U_{\text{rot}}} & \qw  & \qw            & \qw                        & \qw                  & \qw & \rstick{\text{postselect } \ket{1}}\\
}
\else
\includegraphics{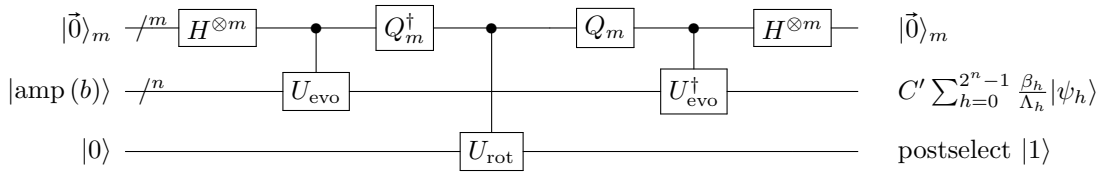}
\fi
\caption{HHL algorithm for linear systems.}
\label{fig:hhlfull}
\end{figure}

Let $\ket{\psi_h}, h=0,\dots,2^n-1$ be an orthonormal eigenbasis of
$A$, with corresponding real eigenvalues $\lambda_h, h=0,\dots,2^n-1$;
the eigenbasis exists because $A$ is Hermitian. We make the
simplifying assumption that all $\lambda_h$ can be expressed exactly
on $m$ binary digits, i.e., there exists a binary string $\vv{\ell_h}$
such that $0.\vv{\ell_h}$ is an exact representation of $\lambda_h$
(thus, $\lambda_h = \ell_h/2^m$).
\begin{remark}
  Because $\nrm{A} \le 1$ and $A$ is Hermitian, all eigenvalues are real
  numbers $\le 1$. Although in general these eigenvalues may not be
  representable in finite precision, here we are (for now) assuming
  that they can be written as $\ell_h/2^m$, with $\ell_h \in
  \{0,\dots,2^m-1\}$, for some appropriate value of $m$ and suitable
  choice of $\ell_h$. Note that some values cannot ever be represented
  in this way, such as irrational numbers or even simply the number
  $1$ (because $\ell_h/2^m < 1$ for any $m$), but the approximation
  error is exponentially small in $m$.
\end{remark}
Now suppose, for the sake of the analysis, that the input state for
the bottom part of the circuit in Fig.~\ref{fig:hhlqpe} is one of the
eigenstates of $A$, say $\ket{\psi_h}$; in other words, let us analyze
what happens if we execute the circuit setting $\ket{\amp{b}} =
\ket{\psi_h}$. The effect of the circuit is the following. After the
Hadamard gates, the state of the entire system is:
\begin{equation*}
  \frac{1}{\sqrt{2^m}} \sum_{\vk \in \{0,1\}^m} \ket{\v{k}} \ket{\psi_h}
\end{equation*}
Applying C$U_\text{evo}$ on this state yields:
\begin{align*}
  \text{C}U_\text{evo} \left(\frac{1}{\sqrt{2^m}} \sum_{\vk \in \{0,1\}^m} \ket{\v{k}} \ket{\psi_h}\right) &= \frac{1}{\sqrt{2^m}} \sum_{\vk \in \{0,1\}^m} \left(\ket{\v{k}} e^{2\pi iA k} \ket{\psi_h}\right) \\
  &= \frac{1}{\sqrt{2^m}} \sum_{\vk \in \{0,1\}^m} \left(\ket{\v{k}} e^{2\pi i\ell_h k/2^m} \ket{\psi_h}\right) \\
  &= \frac{1}{\sqrt{2^m}} \sum_{\vk \in \{0,1\}^m} \left(e^{2\pi i \ell_h k/2^m} \ket{\v{k}} \right)  \ket{\psi_h} \\
  &= Q_m \ket{\vv{\ell_h}} \ket{\psi_h}.
\end{align*}
Thus, the first register contains the quantum Fourier transform of
$\ket{\vv{\ell_h}}$. It follows that applying the inverse QFT yields:
\begin{equation*}
  Q_m^{\dag} \otimes I^{\otimes n} \left(Q_m \ket{\vv{\ell_h}} \ket{\psi_h} \right) = \ket{\vv{\ell_h}} \ket{\psi_h}.
\end{equation*}
Because this is true for each one of the eigenstates, it also applies to
a superposition of the eigenstates. Let $\ket{\amp{b}} =
\sum_{h=0}^{2^n-1} \beta_h \ket{\psi_h}$ be the decomposition of
$\ket{\amp{b}}$ in terms of the eigenbasis of $A$ (we do not need to
know the coefficients $\beta_h$, but we know that they exist). The
effect of the circuit in Fig.~\ref{fig:hhlqpe} is therefore:
\begin{align*}
  (Q_m^{\dag} \otimes I^{\otimes n}) \text{C}U_\text{evo} \left(\frac{1}{\sqrt{2^m}} \sum_{\vk \in \{0,1\}^m} \ket{\v{k}} \ket{\amp{b}}\right) &= (Q_m^{\dag} \otimes I^{\otimes n}) \text{C}U_\text{evo} \left(\frac{1}{\sqrt{2^m}} \sum_{\vk \in \{0,1\}^m} \ket{\v{k}} \sum_{h=0}^{2^n-1} \beta_h \ket{\psi_h}\right) \\
  &= \sum_{h=0}^{2^n-1} \beta_h (Q_m^{\dag} \otimes I^{\otimes n}) \text{C}U_\text{evo} \left( \frac{1}{\sqrt{2^m}} \sum_{\vk \in \{0,1\}^m}  \ket{\v{k}} \ket{\psi_h}\right) \\
  &= \sum_{h=0}^{2^n-1} \beta_h \left( \ket{\vv{\ell_h}} \ket{\psi_h}\right).
\end{align*}
Next, we introduce an auxiliary qubit in state $\ket{0}$, say at the
bottom of the circuit so that it becomes the last qubit, and perform
the mapping:
\begin{equation}
  \label{eq:hhlfinalmap}
  \sum_{h=0}^{2^n-1} \beta_h \left( \ket{\vv{\ell_h}} \ket{\psi_h} \ket{0} \right) \xrightarrow{U_{\text{rot}}} \sum_{h=0}^{2^n-1} \beta_h \left( \ket{\vv{\ell_h}}  \ket{\psi_h} \left(\sqrt{1 - \frac{C^2}{\lambda_h^2}} \ket{0} + \frac{C}{\lambda_h} \ket{1}\right)\right),
\end{equation}
where $C$ is a constant of normalization, to be discussed later. This
mapping is composed of two parts: first, we map $\ket{\vv{\ell_h}}
\to \ket{\vv{\frac{1}{\pi}\sin^{-1}\frac{C}{\lambda_h}}}$, defining
$\vv{\frac{1}{\pi}\sin^{-1}\frac{C}{\lambda_h}}$ to be a binary
representation of $\frac{1}{\pi}\sin^{-1}\frac{C}{\lambda_h}$ on $m'$
qubits. Second, we make use of a unitary $U_Y$ that implements the following operation, where
$0.\v{\theta}$ is some number in $[0,1]$:
\begin{align*}
  U_Y (\ket{\v{\theta}}_{m'} \ket{0}) := \ket{\v{\theta}} (\cos (2\pi 0.\v{\theta}) \ket{0} + \sin (2\pi 0.\v{\theta}) \ket{1}).
\end{align*}
\begin{remark}
  \label{rem:ryfrombinary}
  The operation $U_Y$ can be implemented using one controlled-$R_Y$ gate
  (Def.~\ref{def:yrot}) for each bit of $\v{\theta}$. Specifically,
  defining binary strings $\v{u}^{(j)} \in \{0,1\}^{m'}, \v{u}^{(j)}_h =
  1$ if $j=h$, $0$ otherwise, we can write $U_Y$ as:
  \begin{equation*}
    U_Y (\ket{\v{\theta}}_{m'} \ket{0}) = \prod_{j=1}^{m'} \left(\ketbra{\v{u}^{(j)}}{\v{u}^{(j)}} \otimes R_Y(2^{2-j}\pi)\right),
  \end{equation*}
  which is a sequence of controlled rotations acting on the last
  qubit, where we successively halve the angle of rotation and we
  condition on one digit of the binary representation of
  $\v{\theta}$. This expression can be easily verified by writing
  $0.\v{\theta}$ as a summation of terms that depend on a single
  binary digit each, and recalling that the angles in consecutive
  rotations get summed to each other.
\end{remark}
After taking care of normalization factors, and applying some
necessary linear transformations of the domain, we are able to rotate
the last qubit by $\sin \sin^{-1} \frac{C}{\lambda_h} =
\frac{C}{\lambda_h}$. Here we skipped a few details for the sake of
exposition, but the remaining obstacles (e.g., the value of
$\frac{1}{\pi}\sin^{-1}\frac{C}{\lambda_h}$ is in
$[-\frac{1}{2},\frac{1}{2}]$ rather than $[0,1]$) can be easily
resolved using quantum circuits for binary arithmetics, similarly to
what is done in classical digital computers. Putting everything
together, we can see that the map in Eq.~\eqref{eq:hhlfinalmap} can be
implemented efficiently with the building blocks that we just
described. After applying the map in Eq.~\eqref{eq:hhlfinalmap}, we
have the following state:
\begin{equation*}
  \sum_{h=0}^{2^n-1} \beta_h \left( \ket{\vv{\ell_h}}  \ket{\psi_h} \left(\sqrt{1 - \frac{C^2}{\lambda_h^2}} \ket{0} + \frac{C}{\lambda_h} \ket{1}\right)\right).
\end{equation*}
\begin{remark}
  The normalization constant $C$ is necessary to ensure that what we
  obtain is a valid quantum state. In particular, because we need
  $\frac{C}{\lambda_h} \in [-1,1]$ and we know that $\frac{1}{\kappa}
  \le \abs{\lambda_h} \le 1$, we choose $C = \bigO{1/\kappa}$.
\end{remark}
We then uncompute the register containing the outcome of the phase
estimation, so that we leave the working register in the initial state
(and unentangled with the rest), and apply a measurement on the
register corresponding to the auxiliary qubit for the
rotation. This is depicted in Fig.~\ref{fig:hhlfull}.
We claim that when the outcome of the measurement on the last qubit is
$\ket{1}$, the register initially containing $\ket{\amp{b}}$ now
contains the solution $\ket{\amp{A_B^{-1} b}}$. Indeed, after
uncomputing the register containing the eigenvalues
$\ket{\vv{\ell_h}}$, we obtain the following state:
\begin{equation}
  \label{eq:hhlfinalstate}
  \sum_{h=0}^{2^n-1} \beta_h \left( \ket{\v{0}}  \ket{\psi_h} \left(\sqrt{1 - \frac{C^2}{\lambda_h^2}} \ket{0} + \frac{C}{\lambda_h} \ket{1}\right)\right),
\end{equation}
and if the outcome of the measurement in the last register is
$\ket{1}$, we obtain $C' \sum_{h=0}^{2^n-1}
\frac{\beta_h}{\lambda_h} \ket{\psi_h}$ in the register that initially
contained $\ket{\amp{b}}$, where the constant of normalization $C'$
changes from $C$ because we need to renormalize the state after the
measurement --- in particular, the state after measurement has unit
norm, so $C' = 1/\nrm{\sum_{h=0}^{2^n-1} \frac{\beta_h}{\lambda_h}
  \ket{\psi_h}}$. The only part left now is to note that
$C' \sum_{h=0}^{2^n-1} \frac{\beta_h}{\lambda_h} \ket{\psi_h}$ is exactly
the solution $\ket{\amp{A^{-1} b}}$. Indeed:
\begin{equation*}
  A = \sum_{h=0}^{2^n-1} \lambda_h \ketbra{\psi_h}{\psi_h}, \qquad
  A^{-1} = \sum_{h=0}^{2^n-1} \frac{1}{\lambda_h}
  \ketbra{\psi_h}{\psi_h},
\end{equation*}
so that:
\begin{equation*}
  A^{-1} b = \sum_{h=0}^{2^n-1} \frac{1}{\lambda_h} \ketbra{\psi_h}{\psi_h} \sum_{h=0}^{2^n-1} \beta_h \ket{\psi_h} = \sum_{h=0}^{2^n-1} \frac{\beta_h}{\lambda_h} \ket{\psi_h}.
\end{equation*}
(This also simplifies the expression for $C'$ to $C' = 1/\|A^{-1}
b\|$.) Thus, the circuit in Fig.~\ref{fig:hhlfull} constructs the
state $\ket{\amp{A^{-1} b}}$.

\subsection{Complexity analysis}
\label{sec:hhldetails}
Let us analyze the complexity\index{linear system!complexity} of the HHL algorithm for linear
systems. We start with a discussion of the simplified version of the
algorithm presented in Sect.~\ref{sec:hhlsimple}. Recall that after
the controlled-$U_{\text{evo}}$ operation, we are in the state:
\begin{align*}
   \frac{1}{\sqrt{2^m}} \sum_{h=0}^{2^n-1} \beta_h \sum_{\vk \in \{0,1\}^m} \left(e^{2\pi i \ell_h k/2^m} \ket{\v{k}} \ket{\psi_h} \right).
\end{align*}
In Sect.~\ref{sec:hhlsimple} we assumed that $\lambda_h = \ell_h/2^m$
for some integer $\ell_h$ on $m$ bits, i.e., $\lambda_h$ is
representable on $m$ bits; now we drop this assumption. We need to
choose the number of qubits $m$ to represent the eigenvalues, which
directly determines the time duration of Hamiltonian simulation, to
ensure that we obtain a sufficiently accurate representation of the
eigenvalues. First, recall from phase estimation that if we want to
compute a phase to some precision $\epsilon'$ we pick $m = \bigO{\log
  \frac{1}{\epsilon'}}$, see Thm.~\ref{thm:qpe} and
Rem.~\ref{rem:qpeuexp}. The smallest eigenvalue of $A$ is at least
$1/\kappa$. If we want such a small eigenvalue to have error
$\epsilon$, i.e., $\log \frac{1}{\epsilon}$ digits of precision, we
choose $m = \bigO{\log \frac{\kappa}{\epsilon}}$: this implies that
the longest time duration $t$ for Hamiltonian simulation is $2^m =
\bigO{\frac{\kappa}{\epsilon}}$. (Intuitively: in the Fourier state
$\frac{1}{\sqrt{2^m}}\sum_{\vk \in \{0,1\}^m} e^{2\pi i \ell_h k/2^m}
\ket{\v{k}}$ we need $k \approx \frac{\kappa}{\epsilon}$ to ensure
that at least $\ceil{\log \frac{1}{\epsilon}}$ digits of $\ell_h$ are
``picked up'' in the state, if $\ell_h$ is of order $1/\kappa$; so we
choose $m = \bigO{\log \frac{\kappa}{\epsilon}}$ as the size of the
register containing $k$, which is a sufficient number of bits to
contain integers of the desired size.)

For the sake of accuracy, we mention that, rather than preparing the
first register in the state $\frac{1}{\sqrt{2^m}} \sum_{\vk \in
  \{0,1\}^m} \ket{\v{k}}$ using $H^{\otimes m}$, as indicated in
Sect.~\ref{sec:hhlsimple}, \cite{harrow2009quantum} instead suggests
preparing the state: $$\ket{\xi} := \sqrt{\frac{2}{2^m}} \sum_{\v{r}
  \in \{0,1\}^m} \sin \frac{\pi(r + \frac{1}{2})}{2^m} \ket{\v{r}},$$
where $m$ has to be chosen appropriately according to the discussion
in this section. Although this complicated expression might appear
unnecessary, it is chosen because it leads to a clean analysis of the
errors in the Fourier states that we want to construct, with small
errors. After applying the conditional Hamiltonian evolution
C$U_{\text{evo}}$ onto $\ket{\xi} \ket{\amp{b}}$, we then obtain:
\begin{align*}
  \text{C}U_\text{evo} \left(\sqrt{\frac{2}{2^m}} \sum_{h=0}^{2^n-1} \beta_h
  \sum_{\v{r} \in \{0,1\}^m} \sin \frac{\pi(r + \frac{1}{2})}{2^m} \ket{\v{r}}
  \ket{\psi_h}\right) =\\ \sqrt{\frac{2}{2^m}} \sum_{h=0}^{2^n-1}
  \beta_h \sum_{\v{r} \in \{0,1\}^m} e^{i \lambda_h rc/2^m} \sin \frac{\pi(r +
    \frac{1}{2})}{2^m} \ket{\v{r}} \ket{\psi_h},
\end{align*}
where $c$ is a suitably chosen parameter representing the time
duration parameter of the Hamiltonian evolution. Note that $c$
determines the resolution of the estimation of the eigenvalues,
because the ``step'' in the exponent (i.e., the difference between two
exponents when $t$ increases by one) is of size $c/2^m$. We choose $c
= \bigO{\kappa/\epsilon}$ because this leads to error $\le \epsilon$
in the final state, see \cite{harrow2009quantum} as well as the
intuitive explanation above regarding the choice of $m$ for the
simplified case. From the last equation, applying the inverse
QFT yields:
\begin{equation*}
  \sqrt{\frac{2}{2^m}} \sum_{h=0}^{2^n-1} \beta_h \sum_{\vk \in
    \{0,1\}^m} \sum_{\v{r} \in \{0,1\}^m} e^{\frac{ir}{2^m} (\lambda_h
    c - 2\pi k)} \sin \frac{\pi(r + \frac{1}{2})}{2^m} \ket{\v{k}}
  \ket{\psi_h}.
\end{equation*}
At this point we need to show that the coefficients $e^{\frac{ir}{2^m}
  (\lambda_h c - 2\pi k)}$ are only large when $k \approx
\frac{\lambda_h c}{2\pi}$, i.e., that these coefficients are bounded
above and small whenever $|k - \frac{\lambda_h c}{2\pi}| \ge 1$.  The
proof is long and technical, and is not discussed here, see
\cite{harrow2009quantum} for details; some of the calculations are
similar to those in the proof of Lem.~\ref{lem:ampestfinalstate} to
handle the geometric sums, see also
\cite[Sect.~5.2.1]{nielsen02quantum}. The outcome is as desired, and
allows us to focus on the basis states $\ket{\vk}$ when $k \approx
\frac{\lambda_h c}{2\pi}$. This lets us extract (a multiple of) the
eigenvalues $\lambda_h$ of $A$. Fortunately, the preparation of this
more complicated initial state does not increase the gate complexity
by a significant amount, because the state can be prepared efficiently
with the procedure of \cite{grover2002creating} (this is the same
algorithm that we described at the end of Sect.~\ref{sec:ampencalg},
whereby instead of the precomputed binary tree of
Cor.~\ref{cor:ampenc}, we compute the rotation angles for each inner
node of the tree on-the-fly, with a circuit exploiting the
analytically-known form of the coefficients).

The complexity of the algorithm can be analyzed as follows. The input
data of the algorithm is given by two oracles $P_A, P_b$: one that
describes the entries of $A$, and one that prepares the state
$\ket{\amp{b}}$ corresponding to the r.h.s.\ vector $b$. The
complexity is given in terms of the number of calls to these oracles,
plus the number of additional gates. Let us assume that $P_b$ runs in
time $T_b$, and $P_A$ runs in time $T_A$ (i.e., those are their
respective gate complexities). The only operation that we need to
perform on the matrix $A$ is Hamiltonian simulation (i.e., controlled
$U_{\text{evo}}$), and the gate complexity of the algorithm depends on
the complexity of performing the desired Hamiltonian simulation. Under
the sparse-oracle input model (see Sect.~\ref{sec:hamsimsparse}), the
Hamiltonian simulation algorithm employed in \cite{harrow2009quantum} uses
$\bigOt{t s^2 T_A}$ gates, where $s$ is the maximum number of
nonzeroes in one row of $A$; using better Hamiltonian simulation
techniques, such as Hamiltonian simulation in the block-encoding
framework starting from a sparse-access oracle (see
Thm.~\ref{thm:blockenchamsim} and Sect.~\ref{sec:sparseblockenc}), the
complexity can be reduced to $\bigOt{t s T_A}$. Because $t$ is chosen as
$\bigO{\kappa/\epsilon}$, this gives total complexity $\bigOt{T_b +
  \frac{\kappa}{\epsilon} s T_A}$ for one execution of the circuit in
Fig.~\ref{fig:hhlfull}: we prepare $\ket{\amp{b}}$ once, and the most
expensive operation in Fig.~\ref{fig:hhlfull} is Hamiltonian
simulation --- everything else takes polylogarithmic number of gates
and is therefore hidden by the $\bigOt{\cdot}$ notation. We must,
however, also consider what the probability of success is. Recall that
we obtain the solution to the matrix inversion only if we obtain the state
$\ket{1}$ in the register used for the final rotation (or $\ket{11}$
when we use filter functions, see Sect.~\ref{sec:hhlkappa}). As it
turns out, a careful analysis shows that the probability to obtain
$\ket{1}$ in such register is at least
$\Omega\left(\frac{1}{\kappa^2}\right)$: this is a lower bound on the
norm of the component of Eq.~\eqref{eq:hhlfinalstate} with $\ket{1}$
in the last qubit, because we know that $C = \bigO{\frac{1}{\kappa}},
\lambda_h \le 1 \; \forall h$. Thus, to obtain the solution with some
constant probability close to 1 we apply $\bigO{\kappa}$ (i.e.,
$1/\sqrt{p}$ where $p = 1/\kappa^2$ is the probability of success)
iterations of amplitude amplification. Summarizing, the algorithm uses
$\bigOt{\kappa(T_b + \frac{\kappa}{\epsilon} s T_A)}$ gates. As
discussed at the beginning of this section, there have been several
papers that followed up on the work of \cite{harrow2009quantum} and
improved its running time, sometimes significantly: we discuss some of
these in Sect.~\ref{sec:hhlimprovements}.

\subsection{Non-Hermitian matrices}
\label{sec:nonhermitian}
The problem definition and the discussion in preceding sections assume
that $A$ is Hermitian, but we would like to be able to apply the QLSA
also to non-Hermitian matrices. In this section we discuss how to lift
this assumption, by showing a modification of the algorithm that works
for non-Hermitian matrices. Suppose then that $A \in \C^{2^m \times
  2^n}$, with $m \le n$, not necessarily Hermitian. Let the singular
value decomposition of $A$ be:
\begin{equation*}
  A = \sum_{h=0}^{2^m-1} \sigma_h \ketbra{\psi_h}{\phi_h},
\end{equation*}
where $\ket{\psi_h}$ are the left singular vectors, $\ket{\phi_h}$ the right
singular vectors. Consider the matrix:
\begin{equation*}
  A' = \begin{pmatrix} 0 & A \\ A^{\dag} & 0 \end{pmatrix}.
\end{equation*}
This matrix $A' \in \C^{(2^m+2^n) \times (2^m+2^n)}$ is Hermitian with
$2^{m+1}$ nonzero eigenvalues $\pm \sigma_h$ (i.e., for each singular
value, we take it with both a plus and a minus sign) and corresponding
eigenvectors: $$\ket{w^{\pm}_h} = \frac{1}{\sqrt{2}} \left(\ket{0}
\ket{\psi_h} \pm \ket{1} \ket{\phi_h}\right).$$ (One can check that
these are eigenvectors with the stated eigenvalue by performing the
matrix-eigenvector multiplication.) $A'$ also has $2^m + 2^n - 2^{m+1}
= 2^n - 2^m$ zero eigenvalues. .

We then apply the QLSA using matrix $A'$ and with the r.h.s.\ vector
(i.e., the value of the ``eigenstate'' register for phase estimation)
set to:
\begin{equation*}
  \ket{0} \ket{\amp{b}} = \ket{0} \otimes \left(\sum_{h=0}^{2^m-1}
  \beta_h \ket{\psi_h} \right) = \sum_{h=0}^{2^m-1} \beta_h
  \frac{1}{\sqrt{2}} (\ket{w^+_h} + \ket{w^-_h}).
\end{equation*}
Due to the structure of the matrix $A'$, the QLSA produces a state
proportional to:
\begin{equation*}
  \sum_{h=0}^{2^m-1} \frac{\beta_h}{\sigma_h} \frac{1}{\sqrt{2}}(\ket{w^+_h} - \ket{w^-_h}) = \sum_{h=0}^{2^m-1} \frac{\beta_h}{\sigma_h} \ket{1}\ket{\phi_h},
\end{equation*}
where the sign $-$ in front of $\ket{w^-_h}$ comes from the fact that
$\ket{w^-_h}$ has negative eigenvalues. This state is a solution to
the linear system after dropping the first $\ket{1}$, because:
\begin{equation*}
  \underbrace{\left(\sum_{h=0}^{2^m-1} \sigma_h \ketbra{\psi_h}{\phi_h}\right)}_{A} \underbrace{\left(\sum_{h=0}^{2^m-1} \frac{\beta_h}{\sigma_h} \ket{\phi_h}\right)}_{\text{output state}} = \sum_{h=0}^{2^m-1} \beta_h \ket{\psi_h} = \ket{\amp{b}}.
\end{equation*}
Thus, the QLSA applied to the modified problem $A'$ recovers a
solution to the original problem $Ax=b$ for non-Hermitian $A$. Note
that the number of columns/rows of $A'$ is at most double the number
of columns/rows of $A$, so the representation of $A'$ and its
eigenvectors only requires one additional qubit compared to $A$. Thus,
applying the QLSA to $A'$ is not much more resource-intensive than
applying it to $A$ (of course, because $A$ did not satisfy the
Hermiticity assumption, we could not have applied the algorithm
directly to $A$).

\subsection{Unknown condition number}
\label{sec:hhlkappa}
In Sect.s~\ref{sec:hhlsimple} and \ref{sec:hhldetails} we saw that
many of the quantities involved in the QLSA depend on $\kappa$, and we need to
know their value to define the quantum circuit that executes the
algorithm. Thus, knowledge of $\kappa$, or an upper bound on $\kappa$,
is necessary.
\begin{remark}
  An upper bound $\kappa' \ge \kappa$ is sufficient for the following
  reason: the assumption that eigenvalues are in the interval
  $[\frac{1}{\kappa}, 1]$ is clearly satisfied if we use $\kappa'$
  instead of $\kappa$; $\kappa$ is also used as a normalization factor
  to ensure that certain operations are well defined (and unitary),
  e.g., Eq.~\eqref{eq:hhlfinalmap}, and substituting $\kappa'$ gives
  subnormalized, but valid, operations --- the subnormalization can be
  taken care of with amplitude amplification (whose complexity would
  also depend on $\kappa'$). From a computational complexity point of
  view, we want to ensure that $\kappa' = \bigO{\kappa}$, so that,
  overall, the QLSA is slowed down by no more than a constant factor.
\end{remark}
Under the assumption that eigenvalues are in $[\frac{1}{\kappa}, 1]$,
and $\kappa$ is known, we may not need additional work compared to
Sect.~\ref{sec:hhlsimple}, as we can choose the number of qubits
appropriately to represent all eigenvalues up to a small
error. Suppose, however, that we do not know precisely the condition
number $\kappa$; or suppose that we only want to perform inversion of
eigenvalues between certain values, because we know that the
corresponding eigenspaces already span the r.h.s.\ vector $b$. In such
cases we need a different approach to ensure that we are inverting the
matrix correctly.  There are several natural approaches to deal with
an unknown condition number; we discuss them here.

\paragraph{Filter functions.} To alleviate the issue of an unknown condition number,
\cite{harrow2009quantum} proposes the use of {\em filter functions}, a
concept that is also used in numerical analysis. We remark that these
are not needed with the right assumption on the spectrum of $A$,
however it is very instructive to discuss the idea of filter functions
because they show some of the techniques that can be applied to
implement matrix functions on quantum computers; the block-encoding
framework presented in Sect.~\ref{sec:blockencoding} provides
additional ways to do so.

The main complication is the fact that if some eigenvalue $\lambda_h$
is very small, say $\frac{1}{\kappa}$, then a small relative error in
the computation of the eigenvalue in phase estimation may lead to a
very large error in the computation of the inverse $\sum_{h}
\frac{1}{\lambda_h} \ketbra{\psi_h}{\psi_h}$ (i.e., the reciprocal of
some eigenvalue is of the order of $\kappa$, which may be
affected by large absolute error). In other words, the computation of
the inverse matrix is not numerically stable for small eigenvalues,
and small errors in the estimation of eigenvalues may yield large
error in the solution of the linear system. So if we do not use the
correct value for the condition number, we risk using the wrong number
of bits for phase estimation, and making large errors in the matrix
inversion.

The idea for filter functions is to have a pair of functions $f, g$
that identify the eigenspaces where the inverse is well conditioned,
and the eigenspaces where it is not. Intuitively, we pick a threshold
value $\approx \kappa$ and essentially decide that we are performing
inversion of the matrix only in the eigenspaces corresponding to
eigenvalues that are approximately larger than $1/\kappa$, ignoring
any smaller ones because they might be affected by too large an
error. The advantages are that we no longer need to assume that the
spectrum of $A$ is contained in $[\frac{1}{\kappa}, 1]$ (the algorithm
is well-defined regardless of this assumption), and that errors in the
estimation of small eigenvalues are handled gracefully. The drawback
is that the result of the algorithm may not be a solution to the
linear system if the eigenvalues do not lie in that interval, as we
discuss below --- but it is relatively close to a solution if the
eigenvalues violate the assumption only by a small amount.

The filter function $f$ that identifies the well-conditioned
subspace must satisfy these criteria:
\begin{itemize}
\item The value of the filter function must be proportional to
  $\frac{1}{\lambda}$ for $\lambda \ge \frac{1}{\kappa}$.
\item The value of the filter function must be zero for $\lambda \le
  \frac{1}{\kappa'}$, where $\kappa'$ is some appropriately chosen
  value; say, $\kappa ' = 2\kappa$.
\item In the interval $[\frac{1}{\kappa'}, \frac{1}{\kappa}]$, the
  filter function should interpolate between the two cases above.
\end{itemize}
The other filter function $g$, identifying the ill-conditioned
subspace, must instead satisfy:
\begin{itemize}
\item The value of the filter function must be 0 for $\lambda \ge
  \frac{1}{\kappa}$.
\item The value of the filter function must be a constant for $\lambda
  \le \frac{1}{\kappa'}$.
\item In the interval $[\frac{1}{\kappa'}, \frac{1}{\kappa}]$, the
  filter function should interpolate between the two cases above.
\end{itemize}
One can notice that $f$ and $g$ are in some sense complementary. We
additionally require that $(f(\lambda))^2 + (g(\lambda))^2 \le 1$ for
all $\lambda$, for normalization. An example of filter functions with
these characteristics is:
\begin{equation*}
  f(\lambda) = \begin{cases}
    \frac{1}{2\kappa \lambda} & \lambda \ge \frac{1}{\kappa} \\
    \frac{1}{2} \sin\left(\frac{\pi}{2} \frac{\lambda - \frac{1}{\kappa'}}{\frac{1}{\kappa} - \frac{1}{\kappa'}} \right) & \frac{1}{\kappa} > \lambda > \frac{1}{\kappa'} \\
    0 & \lambda < \frac{1}{\kappa'}.
  \end{cases}
  \hspace{2em}
  g(\lambda) = \begin{cases}
    0 & \lambda \ge \frac{1}{\kappa} \\
    \frac{1}{2} \cos\left(\frac{\pi}{2} \frac{\lambda - \frac{1}{\kappa'}}{\frac{1}{\kappa} - \frac{1}{\kappa'}} \right) & \frac{1}{\kappa} > \lambda > \frac{1}{\kappa'} \\
    \frac{1}{2} & \lambda < \frac{1}{\kappa'}.
    \end{cases}
\end{equation*}
We can apply these functions to compute the reciprocal of the
eigenvalues: rather than using a single bit for the final rotation
$U_{\text{rot}}$, we instead use a two-bit flag register, and produce
the following quantum state after rotation and uncomputation:
\begin{equation*}
    \sum_{h=0}^{2^n-1} \beta_h \left(\ket{\psi_h} \left(\sqrt{1 - (f(\lambda_h))^2 - (g(\lambda_h))^2} \ket{00} + g(\lambda_h) \ket{01} + f(\lambda_h)\ket{11} \right)\right),
\end{equation*}
where the final two-bit register identifies the following:
\begin{itemize}
\item If the last register is in the state $\ket{00}$, inversion did
  not take place.
\item If the last register is in the state $\ket{01}$, we inverted the
  matrix but we are in the ill-conditioned subspace, where some of the
  eigenvalues are very small (smaller than $\frac{1}{\kappa}$).
\item If the last register is in the state $\ket{11}$, we inverted the
  matrix and we are in the well-conditioned subspace: this identifies
  the part of the quantum state where we can find the solution to the
  linear system, and that gets amplified by amplitude amplification.
\end{itemize}
\begin{remark}
  The use of filter functions implicitly introduces a dependence on
  $\kappa$, because we need to choose a threshold for the eigenvalue
  filters. Thus, even if we do not make the assumption that the
  spectrum of $A$ is contained in $[\frac{1}{\kappa}, 1]$, we still
  need to choose a value of $\kappa$ before we run the algorithm, and
  inversion of $A$ only takes place for eigenvalues $\ge
  \frac{1}{\kappa}$.
\end{remark}

\paragraph{Solution verification.} This approach is applicable when we have a
computationally efficient procedure to verify if the correct solution
to the linear system has been found. Suppose we have such a procedure,
that takes as input the quantum state encoding the (potential)
solution of the linear system, and outputs ``yes'' or ``no'' to
indicate if the linear system has been satisfactorily solved. Then, if
we do not know $\kappa$, we can start with an estimate $\tilde{\kappa}
= 2$, and repeatedly execute a loop where we apply the QLSA setting
$\kappa = \tilde{\kappa}$, run the verification procedure, and if the
verification procedure outputs ``no'', we double the current estimate
$\tilde{\kappa}$. Because $\tilde{\kappa}$ increases exponentially
fast, it reaches a value that is at most twice the true value of
$\kappa$ in a logarithmic number of iterations, and this is guaranteed
to yield the correct solution.

\paragraph{Estimation of $\kappa$.} The last approach is more involved
and requires estimating $\kappa$ first. This can be done using
amplitude estimation (Sect.~\ref{sec:ampest}), thanks to the fact that
the probability of success of the QLSA before amplitude amplification
is proportional to $1/\kappa^2$: by estimating this
probability with amplitude estimation, we estimate $\kappa$. Note that
this requires a flag register that indicates what is the subspace
whose probability (after measurement) must be estimated, and such
register is available in this case: in the exposition of
Sect.~\ref{sec:hhlsimple}, it is the last qubit, containing $\ket{1}$
after rotation if the rotation has been successful. The complexity of
executing the estimation procedure for $\kappa$ is a factor
$\bigOt{1/\epsilon}$ larger than the complexity of running the QLSA,
where $\epsilon$ is the precision of the estimation: intuitively, this
is consistent with the fact that amplitude estimation has
$\bigO{1/\epsilon}$ scaling. Such a result also holds for the improved
QLSA algorithms discussed in Sect.~\ref{sec:hhlimprovements}. For
details and a precise statement of the time complexity, see
\cite{chakraborty2019power}.

\subsection{Improvements to the running time}
\label{sec:hhlimprovements}
Two important improvements over the seminal work of
\cite{harrow2009quantum} have led to much faster quantum algorithms
for linear systems. In this section we give an overview of the
corresponding improvements. We also drop the ``hidden'' assumption
that $A$ is positive semidefinite, and allow $A$ to have spectrum in
$[-1, -\frac{1}{\kappa}] \cup [\frac{1}{\kappa}, 1]$.

The first improvement concerns the dependence on the precision
parameter $\epsilon$. The HHL algorithm relies on phase estimation to
perform an eigendecomposition of $A$, but this has the inherent
$\bigO{1/\epsilon}$ dependence on precision of phase estimation, and
it becomes a major bottleneck for the algorithm because the factor
$1/\epsilon$ then shows up directly in the running time. 
\begin{remark}
  In the HHL algorithm, phase estimation is used to obtain a
  description of the eigenvalues, which is then used to apply the
  matrix function $f(x) = 1/x$ to $A$ (i.e., apply the transformation
  $A = \sum_{h=0}^{2^n-1} \lambda_h \ket{\psi_h} \bra{\psi_h} \to
  A^{-1} = \sum_{h=0}^{2^n-1} \frac{1}{\lambda_h} \ket{\psi_h}
  \bra{\psi_h}$).
\end{remark}
In principle, if we could apply the matrix function $f(x) = 1/x$
without phase estimation (and therefore without an ``explicit''
eigendecomposition), then we could get rid of the $\bigO{1/\epsilon}$
scaling in the running time, because no other step of the algorithm
has $\bigO{1/\epsilon}$ scaling. This idea is explored in
\cite{childs2017quantum}, and subsequently in
\cite{chakraborty2019power,gilyen2019quantum}. There are several
algorithmic approaches to implement this idea, although all of them
share the goal of approximating the matrix function $f(x) = 1/x$ in
some way. One of the approaches proposed in \cite{childs2017quantum},
and further developed in subsequent work, is to compute a polynomial
approximation of the function $1/x$, showing that a sufficiently
accurate approximation can be constructed with a low-degree
polynomial. Then we can directly implement the matrix function
corresponding to the polynomial in place of the function $1/x$.
\begin{remark}
  To construct a sufficiently accurate polynomial approximation of
  $1/x$, so that it can be applied as a transformation of the
  eigenvalues, it is important to know the domain of $x$, which in
  this case represents eigenvalues and therefore the domain is the
  spectrum of $A$. We rely on the assumptions that $\|A\| \le 1$ and
  an upper bound $\kappa$ on the condition number is known: this
  implies that we need to compute a polynomial approximation of $1/x$
  only over $[-1, -\frac{1}{\kappa}] \cup [\frac{1}{\kappa}, 1]$,
  because all eigenvalues lie in this set.
\end{remark}
Crucially, the complexity of implementing the desired polynomial
(which directly depends on the degree of such polynomial) scales only
polylogarithmically in $1/\epsilon$, improving over the $1/\epsilon$
dependence of the HHL algorithm described in the preceding sections. A
more general approach is taken in the singular value transformation
framework \cite{gilyen2019quantum}, directly connected to
block-encodings (Sect.~\ref{sec:blockencoding}). With singular value
transformation we can implement polynomial functions to the singular
values of an appropriately block-encoded matrix, and this includes a
sufficiently accurate approximation of the function $f(x) = 1/x$.

The second improvement is a reduction of the dependence on $\kappa$,
from quadratic to linear. To achieve this reduction we note that the
$\kappa^2$ dependence in the complexity of the HHL algorithm comes
from two separate sources: a factor of $\kappa$ in the cost is
incurred because of Hamiltonian simulation with $t =
\bigO{\kappa/\epsilon}$, which appears necessary because we want to
estimate the eigenvalues with error $\approx \epsilon/\kappa$ (i.e.,
$\lambda_h$ should be affected by error at most $\epsilon \lambda_h$,
and the smallest eigenvalue is of order $\approx 1/\kappa$, see
Sect.~\ref{sec:hhldetails}); an additional factor $\kappa$ is due to
the rotation to apply the reciprocal of the eigenvalues, because of
the constant $C$ in Eq.~\eqref{eq:hhlfinalmap}. Indeed, before
amplitude amplification the rotation is successful only with
probability $\kappa^2$: we want the last qubit on the r.h.s.\ of
Eq.~\eqref{eq:hhlfinalmap} to be $\ket{1}$ after measurement, and to
boost the probability of this event to almost 1 we need
$\bigO{\kappa}$ rounds of amplitude amplification. However, the worst
case for these two parts of the HHL algorithm occurs in opposite
situations. Suppose all the eigenvalues are small, $\lambda_h \approx
1/\kappa$: in this case estimating the eigenvalues is difficult and
forces us to perform Hamiltonian simulation with $t =
\bigO{\kappa/\epsilon}$, but amplitude amplification is easy, because
the factor $\frac{C}{\lambda_h}$ in Eq.~\eqref{eq:hhlfinalmap} is a
constant (recall $C = \bigO{1/\kappa}$) so the rotation is successful
with large probability. On the other hand, suppose all the eigenvalues
are large, $\lambda_h \approx 1$: to estimate the eigenvalues to
precision $\epsilon \lambda_h \approx \epsilon$, it would be enough to
perform Hamiltonian simulation with $t = \bigO{1/\epsilon}$, but
amplitude amplification requires more iterations because the factor
$\frac{C}{\lambda_h}$ in Eq.~\eqref{eq:hhlfinalmap} is
$\bigO{1/\kappa}$. Thus, depending on $\lambda_h$, one of the steps,
among eigenvalue estimation and amplification of the success
probability of the rotation, is not time-consuming. The pessimistic
factor $\kappa^2$ factor in the running time of HHL takes the worst
case for both steps, but these worst case scenarios do not
simultaneously apply to the same eigenvalue. Unfortunately, standard
amplitude amplification is applied onto the entire eigenvalue
estimation circuit, therefore we have to pay the cost for both
accurate eigenvalue estimation, and for amplification of the large
eigenvalues.

We can improve the running time with a technique known as
\emph{variable-time amplitude amplification}. We divide the eigenvalue
computation into several steps. In each step we use a different value
of the length $t$ of the time horizon for the corresponding
Hamiltonian simulation. We start with a short constant time, and
attempt to estimate the eigenvalues. Based on $t$, some eigenvalues
may be estimated correctly, while others may not. We use a subroutine
to try to assess which ones are correct: those are no longer
modified. After that we double the length of the time horizon, and
repeat the procedure. The amplitude estimation part is applied
differently to the different values of $t$. The details of this
procedure are quite involved, and we do not treat them in
detail. Normally, if an algorithm has success probability $p$,
amplitude amplification would perform $1/\sqrt{p}$ iterations; thus,
normally we would execute the algorithm $1/\sqrt{p}$ times for its
maximum value of $t$. Instead, with variable-time amplitude
amplification we can reduce this to $1/\sqrt{p}$ multiplied by the
\emph{average} value of $t$. For a detailed description of
variable-time amplitude amplification we refer to
\cite{ambainis2010variable}, see also
\cite{childs2017quantum,chakraborty2019power} for further discussion
in the context of QLSAs.\index{linear system!algorithm|)}

\subsection{Extracting the solution and iterative refinement}
\label{sec:hhlir}
The running time $\bigOt{\kappa(T_b + \frac{\kappa}{\epsilon} s T_A)}$
indicated in Sect.~\ref{sec:hhldetails} refers to a QLSA in its native
form: the algorithm outputs a quantum state that is close to the
amplitude encoding of the solution of the linear system. (As noted in
Sect.s~\ref{sec:hhlimprovements} and \ref{sec:blockencnotes}, much
faster versions of the algorithm have been developed; at the time of
this writing, the fastest known QLSA runs in $\bigO{\max\{T_A,T_b\}
  \kappa \log 1/\epsilon}$ time.) In some applications, one may want
to get a classical description of the solution to the linear
system. Such a description can be obtained with a straightforward
application of a quantum state tomography algorithm. For the general
setting considered here, the most computationally-efficient tomography
algorithm (in terms of number of calls to a unitary preparing the
state of interest) is precisely the one described in
Thm.~\ref{thm:tomography}; more efficient algorithms might be possible
if we know some properties of the solution to the linear
system.\index{state!tomography} Unfortunately Thm.~\ref{thm:tomography} introduces linear
scaling in $1/\epsilon$: this means that obtaining an accurate
classical description of the solution vector (e.g., $\epsilon =
10^{-10}$) would be prohibitively expensive. Thus, even if
state-of-the-art QLSAs have polylogarithmic scaling in $1/\epsilon$,
extracting the solution with quantum state tomography loses such
favorable scaling. This fundamental issue can sometimes be
circumvented with a technique known in the classical literature as
\emph{iterative refinement} \cite{wilkinson1963rounding}\index{linear system!iterative refinement}\index{iterative refinement}. This
technique has proven to be useful for the design of classical and
quantum optimization algorithms, see the notes in
Sect.~\ref{sec:blockencnotes}. We describe here the main idea.

Suppose we have a linear system:
\begin{equation}
  \label{eq:irls}
  Ax = b.
\end{equation}
Let us use $x$ to denote the vector of unknowns, and $\hat{x}$ to
denote a candidate solution with precision $\delta$, i.e., $\nrm{b -
  A\hat{x}} \le \delta$. Consider the linear system obtained by
subtracting $A\hat{x}$ from both sides of the equation: we obtain $A(x
- \hat{x}) = b - A\hat{x}$ (the vector $b - A\hat{x}$ is often called
the vector of \emph{residuals}). Define the linear system:
\begin{equation}
  \label{eq:irls2}
  Ay = \frac{1}{\delta} \left(b - A\hat{x}\right).
\end{equation}
Suppose we solve Eq.~\eqref{eq:irls2} to precision $\delta$, i.e., we
obtain $\hat{y}$ such that $\nrm{\frac{1}{\delta} \left(b -
  A\hat{x}\right) - A\hat{y}} \le \delta$. Now consider $\hat{x} +
\delta \hat{y}$. This vector satisfies the following:
\begin{align*}
  \nrm{b - A\left(\hat{x} + \delta \hat{y}\right)} =
  \delta \nrm{\frac{1}{\delta} (b - A \hat{x}) - A\hat{y}} \le \delta^2.
\end{align*}
Therefore, $\hat{x} + \delta \hat{y}$ is a $\delta^2$-precise solution
to the linear system in Eq.~\eqref{eq:irls}, and we obtained it by
solving two linear systems each with precision $\delta$. The idea can
be iterated, so that with $k$ solves we obtain a solution with
precision $\delta^k$. We can pick any constant $\delta$, and if we aim
to obtain a solution with final precision $\epsilon$, we can achieve
this goal with $k = \bigO{\log \frac{1}{\epsilon}}$ iterations, see
Prop.~\ref{prop:qlsair}.
\begin{example}
  Suppose we have the linear system:
  \begin{align*}
    x_1 + x_2 &= 3 \\
    3x_1 + x_2 &= 6,
  \end{align*}
  and we use a solver for this problem that guarantees precision
  $\delta = 1/\sqrt{2}$. In particular, we obtain the solution
  $\hat{x}_1 = 1.5, \hat{x}_2 = 2$, which has residuals $(-0.5, -0.5)$. To
  find an adjustment to the current solution, we solve the problem:
  \begin{align*}
    y_1 + y_2 &= -\frac{1}{\sqrt{2}} \\
    3y_1 + y_2 &= -\frac{1}{\sqrt{2}},
  \end{align*}
  again with precision $\delta = 1/\sqrt{2}$. For example, we obtain
  the solution $\hat{y}_1 = 0, \hat{y}_2 = -\frac{5}{4\sqrt{2}}$, with
  residuals $(\frac{1}{4\sqrt{2}}, \frac{1}{4\sqrt{2}})$. The vector:
  \begin{align*}
    \hat{x} + \delta \hat{y} = \begin{pmatrix}
      1.5 \\ 2
    \end{pmatrix}
    + \frac{1}{\sqrt{2}}
    \begin{pmatrix}
      0 \\ -\frac{5}{4\sqrt{2}}
    \end{pmatrix}
    = \begin{pmatrix}
      1.5 \\ 1.375
    \end{pmatrix}
  \end{align*}
  is now a solution with precision $\frac{1}{8} \le \delta^2$,
  obtained with the solution of two linear systems each with precision
  $\delta = 1/\sqrt{2}$ (in general we can only guarantee precision
  $\delta^2$ after two iterations).
\end{example}

We can apply the same scheme using a QLSA to solve the linear system
at each iteration, followed by state tomography to extract a classical
description of the solution. We use the classical description to keep
track of the updated candidate solution $x^{(k)}$ at iteration $k$,
and to compute its residual vector $b - Ax^{(k)}$ exactly (in fact, we
could get away with computing the residual with some error, but it
complicates the analysis slightly, so we do not discuss this
case). The scheme of the algorithm is described in
Alg.~\ref{alg:qlsair}.
\begin{algorithm2e}[htb]
  \SetAlgoLined
  \LinesNumbered
\KwIn{Matrix $A$, r.h.s.\ $b$ with $\nrm{b}=1$, target precision $\epsilon$, per-iteration precision $\delta$ with $0 < \epsilon < \delta < 1$.} 
\KwOut{Vector $x^*$ satisfying $\nrm{A x^* - b} \le \epsilon$.}
\textbf{Initialize}: Set $x^{(0)} \leftarrow \zeroes, r^{(0)} \leftarrow  b, k \leftarrow 1$. \\
\While{$\nrm{r^{(k-1)}} > \epsilon$}{
  Solve the system $A x = \frac{1}{\nrm{r^{(k-1)}}} r^{(k-1)}$ and obtain solution $\hat{x}^{(k)}$ satisfying: $$\nrm{A \hat{x}^{(k)} - \frac{1}{\nrm{r^{(k-1)}}} r^{(k-1)}} \le \delta.$$ \label{alg:qlsairsolve}\\
  Update candidate solution: $x^{(k)} \leftarrow x^{(k-1)} + \nrm{r^{(k-1)}} \hat{x}^{(k)}$. \label{alg:qlsairsolupdate} \\
  Update residual vector: $r^{(k)} \leftarrow b - A x^{(k)}$.\\
  Let $k \leftarrow k+1$.
}
\Return $x^{(k-1)}$.
\caption{Iterative refinement for linear systems.}
\label{alg:qlsair}
\end{algorithm2e}
We show below that the number of iterations performed by the algorithm
is polylogarithmically small.
\begin{proposition}
  \label{prop:qlsair}
  The iterative refinement scheme for linear systems
  (Alg.~\ref{alg:qlsair}), using constant per-iteration precision
  parameter $\delta$, returns a solution $x^*$ satisfying $\nrm{A x^*
    - b} \le \epsilon$ in $\bigO{\log \frac{1}{\epsilon}}$
  iterations.
\end{proposition}
\begin{proof}
  We show by induction that at iteration $k$ with $k \ge 1$, we have
  $\nrm{r^{(k)}} \le \delta^k$ (which implies $\nrm{A x^{(k)} - b} \le
  \delta^k$). For the base step, we know that $\nrm{r^{(1)}} = \nrm{b
    - A x^{(1)}} \le \delta$ because this is the first solution that
  we obtain on line \ref{alg:qlsairsolve} (recall that $\nrm{b} = 1$
  by assumption, so on line \ref{alg:qlsairsolupdate} we set $x^{(1)}
  \leftarrow \hat{x}^{(1)}$). Then we have:
  \begin{align*}
    \nrm{r^{(k+1)}} &= \nrm{A x^{(k+1)} - b} =
    \nrm{A (x^{(k)} + \|r^{(k)}\| \hat{x}^{(k+1)}) - b} =
    \nrm{r^{(k)}}\nrm{\frac{1}{\nrm{r^{(k)}}}(A x^{(k)} -b)  +  A\hat{x}^{(k+1)}} \\
    &= \nrm{r^{(k)}}\nrm{A\hat{x}^{(k+1)} - \frac{1}{\nrm{r^{(k)}}}r^{(k)}} \le \delta^k \cdot \delta,
  \end{align*}
  where for the final inequality, we used $\nrm{r^{(k)}} \le \delta^k$
  (induction hypothesis) and $\nrm{A\hat{x}^{(k+1)} -
    \frac{1}{\nrm{r^{(k)}}}r^{(k)}} \le \delta$ (by definition of
  $\hat{x}^{(k+1)}$ on line \ref{alg:qlsairsolve} of
  Alg.~\ref{alg:qlsair}). This concludes the proof of the inductive
  claim. Then, to obtain final precision $\epsilon$, i.e.,
  $\nrm{r^{(k)}} \le \epsilon$, it is sufficient to run the algorithm
  for $k = \ceil{\log_{\delta} \epsilon} = \bigO{\log
    \frac{1}{\epsilon}}$ iterations.
\end{proof}

\noindent Prop.~\ref{prop:qlsair} only discusses the number of
iterations, not the total running time of Alg.~\ref{alg:qlsair}. To
carefully analyze the running time we would have to determine the
exact cost of line \ref{alg:qlsairsolve}; this involves committing to
a specific QLSA, to a specific tomography algorithm, and determining
their complexity when performed with sufficient precision. Note that
the complexity of state tomography (using, e.g.,
Cor.~\ref{cor:l2tomography}) scales linearly with the required inverse
precision, but a precision of $\delta$ alone may not be enough to guarantee that the condition on line \ref{alg:qlsairsolve} of Alg.~\ref{alg:qlsair} is satisfied: other
numerical parameters (such as the norm of possible solutions) may also
intervene in the final expression, depending on the normalization of
the problem and the guarantee provided by the QLSA. Because we do not
use Alg.~\ref{alg:qlsair} again in the rest of this \book{}, we do not
pursue a detailed analysis. The interested reader can find one in
\cite{mohammadisiahroudi2024thesis}. \index{algorithm!linear system|)}

\section{Block-encodings}
\label{sec:blockencoding}
We now provide an introduction to the \emph{block-encoding}\index{block-encoding!framework|(} framework,
highlighting several results concerning basic operations on
block-encodings, and the corresponding computational
complexity. Throughout, although we skip some proofs, we try to
provide intuition on why these results hold. The framework of
block-encoded operators encompasses several other input models for
matrices, and it is generally efficient for many forms of matrix
manipulation, including Hamiltonian simulation. Results in this
section are adapted from
\cite{chakraborty2019power,gilyen2019thesis,van2020sdp,van2020quantum}.

Let us formally define a block-encoding.
\begin{definition}[Block-encoding]
  \label{def:blockenc}
  Let $A \in \C^{2^q \times 2^q}$ be a $q$-qubit operator.  Then, a
  $(q + a)$-qubit unitary $U$ is an $(\alpha, a, \xi)$-block-encoding of
  $A$ if $U = \begin{pmatrix} \widetilde{A} & \cdot \\ \cdot & \cdot
  \end{pmatrix}$, where $\cdot$ represent arbitrary entries of the matrix,
  with the property that
  $$\| \alpha \widetilde{A} - A \| \leq \xi.$$
\end{definition}
By definition, $U$ is a block-encoding of $A$ if $U$ acts as a
scaled-down version of $A$ on some part of the vector space in which
quantum states live.  In particular, we can rephrase the main property
in the definition as follows.
\begin{definition}[Block-encoding (alternative-definition)]
  \label{def:blockencalt}
  Let $A \in \C^{2^q \times 2^q}$ be a $q$-qubit operator.  Then, a
  $(q + a)$-qubit unitary $U$ is an $(\alpha, a, \xi)$-block-encoding
  of $A$ if $$\| \alpha (\bra{\v{0}}_a \otimes I^{\otimes q}) U (\ket{\v{0}}_a
  \otimes I^{\otimes q}) - A\| \le \xi.$$
\end{definition}
Each block-encoding has three parameters: the subnormalization factor
$\alpha$, the number of auxiliary qubits $a$, and the error of the
block-encoding $\xi$. It is important to keep track of these
parameters when manipulating block-encodings. However, one can often
simplify the exposition by reporting only the subnormalization factor,
as long as the number of auxiliary qubits and error $\xi$ scale at
most polylogarithmically with all the relevant parameters of a problem
instance. We try to be formal as much as possible, and keep track of
the three parameters of the block-encodings; however, sometimes
writing the exact expression for some of the parameters is cumbersome,
and in those cases we resort to $\bigO{\cdot}$ notation to indicate
only the relevant asymptotic scaling.

Def.s~\ref{def:blockenc} and \ref{def:blockencalt} imply that if we
limit ourselves to the subspace where the first $a$ qubits are
$\ket{0}$, $U$ implements the desired operation $A$, which may not be
unitary and in fact does not even need to be square, because we could
pad some rows or columns of $A$ with zeroes. The price to pay for this
flexibility is that we may need to scale down $A$, because clearly not
every matrix can be embedded into a unitary without scaling, and we
may need to do some work to ``select'' the part of the space of
quantum states on which the block-encoding acts as $A$.
\begin{remark}
  It is important to note that the structure of a block-encoding $U$
  of $A$ does not guarantee that it takes states of the form
  $\ket{\v{0}}\ket{\psi}$ to
  $\ket{\v{0}}\frac{A}{\alpha}\ket{\psi}$. In fact,
  $U\ket{\v{0}}\ket{\psi}$ (i.e., the image of $\ket{\v{0}}\ket{\psi}$
  via the block-encoding) generally has nonzero support on states that
  do not start with $\ket{\v{0}}$. Thus, when we say that $U$ acts as
  $A$ ``if we limit ourselves to the subspace where the first $a$
  qubits are $\ket{0}$,'' not only do we need to start in some state
  $\ket{\v{0}}\ket{\psi}$, but also we have to postselect or amplify
  the ``correct'' output subspace, to ensure that the first qubits are
  $\ket{\v{0}}$.
\end{remark}
\begin{example}
  Let $A = \begin{pmatrix} 2 & 0 \\ 0 & 0 \end{pmatrix}$. This matrix
  is clearly not unitary, hence there is no circuit that acts as
  $A$. The following unitary matrix $U$ is a $(2, 1,
  0)$-block-encoding of $A$:
  \begin{equation*}
    \begin{pmatrix} 1 & 0 & 0 & 0 \\ 0 & 0 & 1 & 0 \\ 0 & 1 & 0 & 0 \\
    0 & 0 & 0 & 1 \end{pmatrix},
  \end{equation*}
  because:
  \begin{equation*}
    \alpha (\bra{0} \otimes I) U (\ket{0}
  \otimes I) = 2 \left(\begin{pmatrix} 1 & 0 \end{pmatrix} \otimes \begin{pmatrix} 1 & 0 \\ 0 & 1 \end{pmatrix} \right) \begin{pmatrix} 1 & 0 & 0 & 0 \\ 0 & 0 & 1 & 0 \\ 0 & 1 & 0 & 0 \\
    0 & 0 & 0 & 1 \end{pmatrix} \left(\begin{pmatrix} 1 \\ 0 \end{pmatrix}
  \otimes \begin{pmatrix} 1 & 0 \\ 0 & 1 \end{pmatrix}\right) = A.
  \end{equation*}
  Note that $A$ is a single-qubit operator, whereas $U$ is a two-qubit
  operator, and we have to scale $A$ down by a factor 2. Readers
  may correctly recognize that $U$ is a SWAP gate.

  Suppose we want to obtain a representation of $A \ket{\psi}$, where
  $\ket{\psi} = \beta \ket{0} + \gamma \ket{1}$; note that $A
  \ket{\psi}$ may not even be a valid quantum state. In the
  block-encoding framework we can obtain (some representation of) $A
  \ket{\psi}$ by applying $U$ onto the state $\ket{0}\ket{\psi}$: the
  first qubit must be $\ket{0}$ to be in the right subspace. We have:
  \begin{equation*}
    U \ket{0}\ket{\psi} = \begin{pmatrix} \beta \\ 0 \\ \gamma
      \\ 0 \end{pmatrix},
  \end{equation*}
  and if we look at the restriction of this state onto the basis
  states beginning with $\ket{0}$ (i.e., we apply $\bra{0} \otimes
  I$), we obtain $\beta \ket{0}$, which is precisely $\frac{A}{2}
  \ket{\psi}$: note once again the scaling factor $2$ picked up when
  applying the block-encoding.

  Intuitively, because we chose $U$ to be a SWAP gate, we can see how
  this approach works: we apply a SWAP gate to a state of the form
  $\ket{0}\ket{\psi}$, so we are ``zeroing out'' the amplitude of
  $\ket{\psi}$ that corresponds to $\ket{1}$ ($\ket{01}$ in the
  two-qubit state). Of course this amplitude does not disappear: it
  simply gets ``moved'' to the part of the two-qubit state that has
  $\ket{1}$ in its first digit, i.e., to $\ket{10}$.
\end{example}

\subsection{Operations on block-encodings}
\label{sec:blockencop}
We now discuss several useful operations on block-encodings, starting
with the product of two block-encoded matrices, which is trivial to
construct.
\begin{proposition}
  \label{prop:blockencprod}
  If $U_A$ is an $(\alpha, a, \xi_A)$-block-encoding of a
  $q$-qubit operator $A$, and $U_B$ is an $(\beta, b,
  \xi_B)$-block-encoding of an $q$-qubit operator $B$, then $(I_b
  \otimes U_A)(I_a \otimes U_B)$ is an $(\alpha \beta, a +
  b, \alpha \xi_B + \beta \xi_A)$-block-encoding of $AB$.
\end{proposition}
The proof of this result follows easily from the definition of
block-encoding.
\begin{remark}
  In the above proposition and in subsequent results regarding
  block-encodings, we abuse the tensor product notation to avoid
  overly complicated expressions. The expression $(I_b \otimes U_A)$
  should be interpreted as ``identity on the $b$ auxiliary qubits for
  $U_B$, and $U_A$ on all the qubits affected by $U_A$,'' and
  similarly, $(I_a \otimes U_B)$ means ``identity on the $a$ auxiliary
  qubits for $U_A$, and $U_B$ on all the qubits affected by $U_B$.''
  Writing this accurately in tensor product notation is cumbersome:
  e.g., if we have three registers $\ket{\cdot}_a \ket{\cdot}_b
  \ket{\cdot}_q$, how do we express the matrix acting as $(I_b \otimes
  U_A)$ described above?  It would have to act as the matrix $U_A$ on
  the first and third register with $a$ and $q$ qubits respectively,
  but we have not defined a matrix that does so while tensored with
  identity on the second register. Hence, we use an imprecise, but
  considerably simpler notation, whose meaning is usually clear from
  the context; see also Fig.~\ref{fig:blockencab} and the surrounding
  discussion. Note that $I_b$ here is a $2^b \times 2^b$ identity
  matrix, which we usually denote $I^{\otimes b}$: we use the
  subscript precisely because the meaning is different with this
  ``improper'' --- but considerably simpler --- tensor product
  notation, where the subscript also indicates to which register the
  matrix should be applied. For block-encodings, sometimes we use this
  improper notation even when our usual notation would be sufficient.
\end{remark}
\begin{figure}[h!t]
  \leavevmode
  \centering
  \ifcompilefigs
  \Qcircuit @C=1em @R=.7em {
    \lstick{\ket{\v{0}}_a} & \qw  & \qw                & \multigate{2}{U_A}  & \qw & \meter & \rstick{\text{postselect } \ket{\v{0}}_a} \\
    \lstick{\ket{\v{0}}_b} & \qw  & \multigate{1}{U_B} & *!{\hdashrule[-0.3pt][x]{2.3em}{0.3pt}{0.5pt 1.5pt}} \qw & \qw & \meter & \rstick{\text{postselect } \ket{\v{0}}_b} \\
                           & \qw  & \ghost{U_B}        & \ghost{U_A}         & \qw & \qw &  \\
  }
  \else
  \includegraphics{figures/blockencab.pdf}
  \fi
  \caption{Block-encoding for the product of two matrices $AB$, given
    block-encodings of $A$ and $B$. Note the dashed line through
    $U_A$: $U_A$ does not get applied to the second register.}
  \label{fig:blockencab}
\end{figure}
In circuit form, the block-encoding of $AB$ can be implemented as
depicted in Fig.~\ref{fig:blockencab}.  The dashed line through $U_A$
is used to indicate that the second register (of dimension $b$) does
not interact with $U_A$, which only acts on the first and third
register. ``Postselect $\ket{\v{0}}$'' indicates that the product $AB$
is successfully applied onto the third register if the first two
registers are measured and we observe $\ket{\v{0}}$; one can of course
applied amplitude amplification to amplify the probability that
$\ket{\v{0}}$ is observed in the two registers.

In addition to the product of block-encodings, linear
combinations\index{unitary matrix!linear combination (LCU)} of
block-encodings can also be constructed in a relatively simple manner,
and the cost for doing so is merely logarithmic in the dimension: the
construction is very similar to the one presented in
Sect.~\ref{sec:lcu}. We first define a \emph{state-preparation pair},
which encodes the coefficients to be used in the linear combination of
block-encodings, then give the complexity of constructing the linear
combination. A state-preparation pair is simply a pair of unitaries
$P_L, P_R$ such that the inner product of $P_L \ket{\v{0}}$ and of
$P_R \ket{\v{0}}$ yields a vector with the desired coefficients (or
something close to it). This specific form is useful for the
construction of linear combinations, as shown after a formal
definition and an example. In the definition below, for convenience we
use zero-based indices for the vector $y$.
\begin{definition}[State-preparation pair]
  \label{def:statepreppair}
  Let $y \in \C^m$ and $\| y \|_1 \leq \beta$. The pair of unitaries
  $(P_L, P_R)$ is called a $(\beta, q, \xi)$-state-preparation-pair
  for $y$ if $P_L \ket{\v{0}}_q = \sum_{\vj \in \{0,1\}^q} c_{j}
  \ket{\vj}$ and $P_R \ket{\v{0}}_q = \sum_{\vj \in \{0,1\}^q} d_{j}
  \ket{\vj}$, where the coefficients $c_j, d_j$ satisfy the following
  properties: $\sum_{j=0}^{m-1} | \beta (c_j^\dag d_j) - y_j | \leq
  \xi$, and $c_j^\dag d_j = 0$ for all $j \in m, \dots, 2^q -1$.
\end{definition}
\begin{example}
  Just as in Sect.~\ref{sec:lcu}, if we aim to construct a linear
  combination with nonnegative real coefficients
  $\alpha_0,\dots,\alpha_{m-1}$, we can define a $\ceil{\log m}$-qubit
  unitary $W$ such that
  \begin{equation*}
      W \ket{\v{0}} = \frac{1}{\|\alpha\|_1} \sum_{\vj \in
        \{0,1\}^{\ceil{\log m}}} \sqrt{\alpha_j} \ket{\vj}.
  \end{equation*}
  Then, setting $P_L = W, P_R = W$, we see that this is a
  $(\|\alpha_1\|, \ceil{\log m}, 0)$-state-preparation-pair for the
  vector of coefficients $\alpha$. If $\alpha \not\ge 0$, we can
  adjust the signs by modifying at least one of the unitaries $P_L,
  P_R$.
\end{example}

\noindent Using a state-preparation pair, we can construct a linear
combination of unitaries with coefficients prescribed by the
state-preparation pair. To do so, we also need a controlled unitary
that prepares the (unweighted) unitaries appearing in the linear combination, i.e.,
if we want to construct a combination of $m$ block-encodings, we need
a circuit that, given the binary string $\vj$, implements the $j$-th
block-encoding of the linear combination. This is once again similar
to the approach discussed in Sect.~\ref{sec:lcu}. The circuit that
applies the $j$-th block-encoding, controlled on some register
containing $\ket{\vj}$, is labeled $V$ in the next proposition; its
formal definition is somewhat convoluted because we need to specify
that its action is the identity if the control register contains an
index $\ge m$, but its behavior should be intuitively clear.
\begin{proposition}
  \label{prop:lincombblock}
  Let $A = \sum_{j=0}^{m-1} y_j A^{(j)}$ be a $q$-qubit operator
  expressed as a linear combination of matrices $A^{(j)}$ with
  coefficients $y_j$. Suppose we are given a $(\beta, p,
  \xi_1)$-state-preparation pair $P_L, P_R$ for $y$, and a $(q + a +
  p)$-qubit unitary:
  \begin{equation*}
    V := \sum_{\substack{\vj \in \{0,1\}^p\\ j \le m-1}}
    \ketbra{\vj}{\vj}\otimes U^{(j)} + \left(I_p - \sum_{\substack{\vj \in
        \{0,1\}^p\\ j \le m-1}} \ketbra{\vj}{\vj}\right) \otimes I_a \otimes
    I_q,
  \end{equation*}
  with the property that $U^{(j)}$ is an $(\alpha, a,
  \xi_2)$-block-encoding of $A^{(j)}$. Then we can implement an
  $(\alpha\beta, a+p, \alpha \xi_1 + \alpha \beta
  \xi_2)$-block-encoding of $A$ with a single use of $V, P_R$ and
  $P_L^{\dag}$.
\end{proposition}
\begin{proof}
  The desired block-encoding is given by $(P_L^{\dag} \otimes I_{q+a})
  V (P_R \otimes I_{q+a})$. To verify this, using the same notation as
  in Def.~\ref{def:statepreppair}, we apply the definition
  of block-encoding and compute:
  \begin{align*}
    &(\bra{\v{0}}_{a+p} \otimes I_q)  (P_L^{\dag} \otimes I_{q+a}) V (P_R
    \otimes I_{q+a}) (\ket{\v{0}}_{a+p} \otimes I_q) \\
    &=\left(\sum_{\vj \in \{0,1\}^p} c_j^{\dag}
    \bra{\vj} \otimes \bra{\v{0}}_a \otimes I_q \right) V \left(
    \sum_{\vj \in \{0,1\}^p} d_j \ket{\vj} \otimes \ket{\v{0}}_a \otimes
    I_q  \right) \\
    &= \left(\sum_{\vj \in \{0,1\}^p} c_j^{\dag}
    \bra{\vj} \otimes \bra{\v{0}}_a \otimes I_q \right) \left(
    \sum_{\substack{\vj \in \{0,1\}^p\\j \le m-1}} d_j \ket{\vj} \otimes U^{(j)}
    \left(\ket{\v{0}}_a \otimes I_q \right) +
    \sum_{\substack{\vj \in \{0,1\}^p\\j \ge m}} d_j \ket{\vj} \otimes
    \ket{\v{0}}_a \otimes I_q \right) \\
    &= \sum_{j=0}^{m-1} c_j^{\dag}d_j \tilde{A}^{(j)},
  \end{align*}
  where $\tilde{A}^{(j)}$ is such that $\|\alpha \tilde{A}^{(j)} -
  A^{(j)}\| \le \xi_2$. For the last equality, we used the fact that
  the terms in the summation with $j \ge m$ disappear because of the
  definition of state-preparation pair (Def.~\ref{def:statepreppair}),
  and the fact that $U^{(j)}$ is an $(\alpha, a,
  \xi_2)$-block-encoding of $A^{(j)}$. Because we also have
  $\|\beta(c_j^{\dag}d_j) - y_j\| \le \xi_1$ by
  Def.~\ref{def:statepreppair}, simple calculations show that this is
  indeed a $(\alpha\beta, a+p, \alpha \xi_1 + \alpha \beta
  \xi_2)$-block-encoding of $A = \sum_{j=0}^{m-1} y_j A^{(j)}$.
\end{proof}

At this point, we have shown how to perform products of
block-encodings and linear combinations of block-encodings: this means
that we can implement polynomial functions of block-encoded
matrices. We can therefore implement polynomial approximations of
matrix functions, which allows us to manipulate the block-encoded
input matrix in a plethora of ways. For example, we can perform
Hamiltonian simulation, as discussed in Sect.~\ref{sec:signalproc},
and the corresponding complexity essentially matches the fastest known
Hamiltonian simulation algorithms --- for certain input models better
results might be obtained without going through block-encodings, but
the difference is typically small. A more precise statement of the
complexity is the following.
\begin{theorem}[Hamiltonian simulation via block-encodings]
  \label{thm:blockenchamsim}
  Suppose that $U$ is an $(\alpha, a, \xi/|2t|)$-block-encoding of the
  Hamiltonian $\ham$. Then, we can implement a $\xi$-precise Hamiltonian
  simulation unitary $V$, i.e., an $(1, a + 2, \xi)$-block-encoding of
  $e^{it \ham}$, with $\bigO{ \alpha|t| + \log (1/\xi)}$ uses of
  controlled-$U$ and its inverse, and $\bigO{ a(\alpha|t| + \log
    (1/\xi))}$ two-qubit gates.
\end{theorem}
To perform Hamiltonian simulation we apply the function $e^{ix}$ to
the eigenvalues of $\ham$, by adding together (via linear combination of
block-encodings) an approximation of $\cos x$ and an approximation of
$i \sin x$. A full proof of this result can be found in
\cite{gilyen2019quantum}, together with a discussion of the complexity
of other matrix functions, such as the inverse (i.e., the computation
of a block-encoding of $A^{-1}$); the complexity essentially depends
on the degree of the polynomial that is necessary to obtain a
sufficiently accurate approximation, and the polynomial is implemented
as discussed above, using products and linear combinations of
block-encodings.

We can also construct the block-encoding of a diagonal matrix that
contains inner products of different quantum states on the
diagonal. We show this construction because it can serve as an
inspiration for other, similar building blocks. The construction
involves unitaries that prepare the quantum states of which we want
the inner products; in the statement of the lemma, we call these
state-preparation unitaries to emphasize their role.
\begin{lemma}
  \label{lem:blockencinnerprod}
  Let the matrices $U^{(j)}, V^{(j)}$ be $(a+1)$-qubit
  state-preparation unitaries for some $a$-qubit quantum states
  $\ket{\psi_{j}}, \ket{\phi_{j}}, \ket{\tilde{\psi}_{j}},
  \ket{\tilde{\phi}_{j}}$, i.e., matrices that implement the following
  maps:
  \begin{align*}
    U^{(j)}
    \ket{0}\ket{\v{0}}_a =
    \ket{0}\ket{\psi_{j}}+\ket{1}\ket{\tilde{\psi}_{j}} \\
    V^{(j)}
    \ket{0}\ket{\v{0}}_a =
    \ket{0}\ket{\phi_{j}}+\ket{1}\ket{\tilde{\phi}_{j}},
  \end{align*}
  where the states $\ket{\tilde{\psi}_{j}}, \ket{\tilde{\phi}_{j}}$
  are garbage states that we are not interested in.  Let $U:=\sum_{\vj
    \in \{0,1\}^p} U^{(j)} \otimes \ketbra{\vj}{\vj}$ and
  $V:=\sum_{\vj \in \{0,1\}^p} V^{(j)} \otimes \ketbra{\vj}{\vj}$ be
  $(a+p+1)$-qubit unitaries controlled by the second register. Then
  the unitary:
  \begin{equation*}
    (I \otimes V^\dagger) (\text{SWAP} \otimes I_{a+p}) (I
    \otimes U)
  \end{equation*}
  is an $(a+2)$-block-encoding of the diagonal matrix
  $\text{diag}(\{\braket{\phi_{j}}{\psi_{j}}\}_{j=0,\dots,2^p-1})$,
  where the single-qubit identity matrix $I$ acts on the first qubit,
  the $\text{SWAP}$ gate acts on the first two qubits, and the
  $(a+p)$-qubit identity $I_{a+p}$ acts on the last $a+p$ qubits.
\end{lemma}
\begin{proof}
  We apply the definition of block-encoding.
  \begin{align*}
    &\bra{\v{0}}_{a+2}\bra{\vj}_p(I \otimes V^\dagger) (\text{SWAP} \otimes I_{a+p}) (I \otimes U)\ket{\v{0}}_{a+2}\ket{\vk}_p \\
    &=\bra{0}\left(\bra{0}\bra{\phi_{j}}+\bra{1}\bra{\tilde{\phi}_{j}}\right)\bra{\vj} \left(\text{SWAP} \otimes I_{a+p}\right) \ket{0}\left(\ket{0}\ket{\psi_{k}}+\ket{1}\ket{\tilde{\psi}_{k}}\right)\ket{\vk} \\
    &=\left(\bra{00}\bra{\phi_{j}}+\bra{01}\bra{\tilde{\phi}_{j}}\right)\bra{\vj} \left(\text{SWAP} \otimes I_{a+p}\right) \left(\ket{00}\ket{\psi_{k}}+\ket{01}\ket{\tilde{\psi}_{k}}\right)\ket{\vk} \\
    &=\left(\bra{00}\bra{\phi_{j}}+\bra{01}\bra{\tilde{\phi}_{j}}\right)\bra{\vj}  \left(\ket{00}\ket{\psi_{k}}+\ket{10}\ket{\tilde{\psi}_{k}}\right)\ket{\vk}.
  \end{align*}
  This last expression is equal to $\braket{\phi_{j}}{\psi_{k}}$ if
  $\vj = \vk$, and it is 0 otherwise. This concludes the proof.
\end{proof}

\noindent For example, we can recast the gradient-based quantum state
tomography algorithm of Ch.~\ref{ch:quantumgradient} in the
block-encoding framework: recall that we want to obtain a phase oracle
for the function $f(x)$ defined in Eq.~\eqref{eq:tomofunc}, which is
an inner product of quantum states. Given the structure of $f(x)$, we
can apply Lem.~\ref{lem:blockencinnerprod} to obtain a block-encoding
of a diagonal matrix containing $f(x)$ on the diagonal for all $x$,
then use Hamiltonian simulation of the block-encoded matrix to
transform them into phase factors of the form
$e^{if(x)}$, thereby obtaining an implementation of the phase oracle.\index{block-encoding!framework|)}

\subsection{Block-encoding from sparse matrices}
\label{sec:sparseblockenc}
The sparse-oracle access\index{oracle!sparse}\index{block-encoding!construction|(} model discussed in
Sect.~\ref{sec:hamsimsparse} is a natural model (even in the classical
world) to describe arbitrary sparse matrices. We now discuss how to
implement a block-encoding of a matrix that is described in this
model; in some cases, this is the first step of a quantum algorithm
that uses the block-encoding framework to work on matrices, allowing
us to convert a classical description of the matrix into a suitable
representation on the quantum computer. The block-encoding framework
therefore encompasses the sparse-oracle access model, in the sense
that sparse-access oracles can be turned into a block-encoding. This
conversion is computationally quite efficient, although there can be
cases where working with the sparse model directly leads to
computational savings, as is often the case when simulating an access
model with a different one: this must be analyzed on a case-by-case
basis.

A precise presentation of how to construct a block-encoding from a
sparse-access oracle requires care about several details. The data of
the matrix to be block-encoded is described in the usual way: we have
quantum oracles (i.e., quantum circuits, so that they can be queried
in superposition) that list the position of the nonzero elements of
the matrix in each row/column, and an oracle that provides the
corresponding values. To create certain superpositions that are
crucial for the construction we need to know upper bounds on the
number of nonzero elements in each row and column: we denote these
upper bounds by $s_r, s_c$ respectively. We also need a way of
indicating if some rows/columns have fewer nonzero entries that the
maximum allowed number $s_r, s_c$, because we want to construct a
superposition over the indices of all the nonzero entries: such a
superposition always has $s_r$ or $s_c$ terms, and we need a way
to ``mark'' terms that do not actually correspond to a nonzero entry.
We do so by using some indices larger than the size of the matrix,
which eventually result in some inner product being zero (see the
construction in the proof), thereby correctly producing a zero in the
corresponding position.

In this and subsequent sections we ignore issues related to the
precision of the representation of each entry of the matrix: we simply
assume that the number of bits of each register is large enough to
perform sufficiently accurate calculations. This is no different from
how similar operations are performed on a classical computer, and the
number of (qu)bits that is sufficient to obtain a certain level of
precision scales polylogarithmically with the reciprocal of the
allowed error. We also assume that all entries are complex numbers
with modulus at most 1, because otherwise a certain rotation in the
construction is not well defined; of course, to block-encode a matrix
that has entries with larger modulus, one can always rescale the
entries and adjust the subnormalization factor of the final
block-encoding.
\begin{proposition}\label{prop:sparseblockenc}
  Let $A \in \mathbb{C}^{2^q \times 2^q}$ be a matrix that has at most
  $s_r$ nonzero elements per row, and at most $s_c$ nonzero elements
  per column, with each element $a_{ij}$ satisfying $\abs{a_{ij}} \le
  1$. Suppose that we have access to the following sparse-access
  oracles acting on two $(q+1)$-qubit registers:
  \begin{align*}
    O_{r} : \ket{\vi} \ket{\vk}  &\to \ket{\vi} \ket{\vv{r_{ik}}}\quad \forall i=0,\dots,2^q -1, k=0,\dots,s_r-1, \\
    O_{c} : \ket{\v{\ell}} \ket{\vj}  &\to \ket{\vv{c_{\ell j}}} \ket{\vj}\quad \forall \ell=0,\dots,s_c-1, j=0,\dots,2^q-1,
  \end{align*}
  where $r_{ik}$ is the index of the $k$-th nonzero entry of the
  $i$-th row of $A$, or if there are less than $k$ nonzero entries,
  then it is $k + 2^q$, and similarly, $c_{\ell j}$ is the index of
  the $\ell$-th nonzero entry of the $j$-th column of $A$, or if there
  are less than $\ell$ nonzero entries, then it is $\ell +
  2^q$. Additionally, assume that we have access to an oracle $O_{A}$
  that returns a binary description of the entries of $A$:
  \begin{equation*}
    O_{A} : \ket{\vi} \ket{\vj} \ket{\v{0}}_p \to \ket{\vi}
    \ket{\vj} \ket{\vv{a_{ij}}},\quad \forall i,j=0,\dots,2^q-1,
  \end{equation*}
  where $\vv{a_{ij}}$ is a $p$-digit binary description of the matrix
  element $(i,j)$ of $A$. Then, we can implement a $(\sqrt{s_r s_c},
  q+3, \xi)$-block-encoding of $A$ with a single use of $O_r$, $O_c$,
  $O_{A}$, $O_{A}^{\dag}$, additionally using $\bigOt{q}$ one and
  two-qubit gates, and $\bigOt{p}$ auxiliary qubits.
\end{proposition}
\begin{proof}
  The proof is constructive and adapted from
  \cite{gilyen2019quantum}. See Rem.~\ref{rem:blockencinplace}
  for remarks about the construction of the oracles $O_r, O_c$.

  We work with three registers (plus some auxiliary space, introduced
  when necessary): a single-qubit register, and two $(q+1)$-qubit
  registers. All indices are zero-based, as usual. Define
  $(q+1)$-qubit operators $W_{r}, W_{c}$ satisfying the following:
  \begin{align*}
    W_{r} \ket{\v{0}} = \frac{1}{\sqrt{s_r}} \sum_{\substack{\vk \in \{0,1\}^{q+1}\\k \le s_r-1}} \ket{\vk}, \qquad
    W_{c} \ket{\v{0}} = \frac{1}{\sqrt{s_c}} \sum_{\substack{\v{\ell} \in \{0,1\}^{q+1}\\\ell \le s_c-1}} \ket{\v{\ell}}.
  \end{align*}
  These operators construct uniform superpositions and can be
  implemented with $\bigO{q}$ gates, because we can always assume that
  $s_r, s_c$ are powers of $2$ (padding the list of nonzero entries
  with zeroes, if necessary) and use an appropriate number of Hadamard
  gates. Let $\text{SWAP}_{q+1}$ be the operator that swaps the first
  $(q+1)$-qubit register with the second $(q+1)$-qubit register (i.e.,
  by applying a SWAP gate to $q+1$ pairs of qubits). Define the
  following $2(q+1)$-qubit operators:
  \begin{equation*}
    V_L := O_{r}(I_{q+1} \otimes W_r) \text{SWAP}_{q+1}, \qquad
    V_R := O_{c}(W_c \otimes I_{q+1}),
  \end{equation*}
  and note that their action is the following:
  \begin{align*}
    V_L \ket{\v{0}}_{q+2} \ket{\vi}_{q} &= \sum_{k=0}^{s_r-1} \frac{1}{\sqrt{s_r}} \ket{\vi}_{q+1}\ket{\vv{r_{ik}}}_{q+1} \qquad \forall i \in \{0,1\}^q \\
    V_R \ket{\v{0}}_{q+2} \ket{\vj}_{q} &= \sum_{\ell=0}^{s_c-1} \frac{1}{\sqrt{s_c}} \ket{\vv{c_{\ell j}}}_{q+1}\ket{\vj}_{q+1} \qquad \forall j \in \{0,1\}^q.
  \end{align*}
  (Note that $\vi, \vj$ are $q$-digit binary strings, but on the
  r.h.s., we store them in registers of length $q+1$: this is used
  below because we want to take the inner product of those registers
  with other registers of size $q+1$, and we need their size to match;
  thus, the registers are padded with a zero in front, although we do
  not indicate this explicitly.) Then $V_L^{\dag}V_R$ block-encodes a
  matrix that is nonzero only in those positions where $A$ is also
  nonzero; formally:
  \begin{align}
    \bra{\v{0}}_{q+2}\bra{\vi} V_L^{\dag} V_R \ket{\v{0}}_{q+2} \ket{\vj} &=
    \left(\sum_{k=0}^{s_r-1} \frac{1}{\sqrt{s_r}} \bra{\vi}\bra{\vv{r_{ik}}}\right)
    \left(\sum_{\ell=0}^{s_c-1} \frac{1}{\sqrt{s_c}} \ket{\vv{c_{\ell j}}}\ket{\vj}\right) \label{eq:blockencvlvr} \\
    &=\frac{1}{\sqrt{s_r s_c}} \text{ if } a_{ij} \neq 0, 0 \text{ otherwise.} \notag
  \end{align}
  To see the last equality, note that the terms in parentheses are
  simply superpositions over all the nonzero elements in a row or
  column, and the resulting inner product is nonzero precisely if
  there exists an index $r_{ik} = j$ and an index $c_{\ell j} = i$,
  which implies that $a_{ij} \neq 0$ (when $r_{ik} \ge 2^q$ because
  there are not enough nonzero entries in a row, we have
  $\braket{\vv{r_{ik}}}{\vj} = 0$, and similarly
  $\braket{\vi}{\vv{c_{\ell j}}}$ if $c_{\ell j} \ge 2^q$). Now we
  construct two more unitaries on $2q+3$ qubits: $U_L := I \otimes
  V_L$, and the unitary $U_R$ that satisfies the following property:
  \begin{equation}
    \label{eq:sparsetoblockur}
    U_R \ket{\v{0}}_{q+3}\ket{\vj}_q = \frac{1}{\sqrt{s_c}} \sum_{\ell=0}^{s_c-1} \left(a_{c_{\ell j} j} \ket{0} + \sqrt{1 - |a_{c_{\ell j} j}|^2} \ket{1}\right)\ket{\vv{c_{\ell j}}}_{q+1} \ket{\vj}_{q+1}.
  \end{equation}
  To construct $U_R$, we first apply $I \otimes V_R$ to the three
  registers indicated at the beginning of the proof; then we apply
  $O_A$ to write a binary description of $\ket{\vv{a_{c_{\ell j} j}}}$
  in a $p$-bit auxiliary register, and with this binary description we
  perform a rotation on the first qubit as indicated in
  Eq.~\ref{eq:sparsetoblockur} (see, e.g., \cite{berry2015hamiltonian}
  for details on how to do this, or the discussion in
  Rem.~\ref{rem:ryfrombinary} regarding the final rotation of the HHL
  algorithm, where we describe a construction that is sufficient
  whenever $a_{c_{\ell j} j}$ is a real number). Finally we uncompute
  the auxiliary register, which takes one use of $O_A^{\dag}$. The
  desired block-encoding is given by $U_L^{\dag} U_R$:
  \begin{align*}
    \bra{\v{0}}_{q+3}\bra{\vi} U_L^{\dag} U_R \ket{\v{0}}_{q+3}
    \ket{\vj} &= \bra{0}\left(\sum_{k=0}^{s_r-1} \frac{1}{\sqrt{s_r}} \bra{\vi}\bra{\vv{r_{ik}}}\right)
    \left(\sum_{\ell=0}^{s_c-1} \left(a_{c_{\ell j} j} \ket{0} + \sqrt{1 - |a_{c_{\ell j} j}|^2} \ket{1}\right)\ket{\vv{c_{\ell j}}} \ket{\vj}\right) \\
    &= \frac{a_{ij}}{\sqrt{s_r s_c}}. 
  \end{align*}
\end{proof}
\begin{remark}
  \label{rem:sparseblockencerror}
  The error $\xi$ in Prop.~\ref{prop:sparseblockenc} comes from the
  final rotation, which is preceded by the computation of the
  appropriate angles to rotate $\ket{0}$ to $a_{c_{\ell j} j} \ket{0}
  + \sqrt{1 - |a_{c_{\ell j} j}|^2} \ket{1}$. Similarly to classical
  computers, this operation may not be doable exactly in finite
  precision (we may have to deal with complex numbers), but the error
  in the computation is exponentially small in the number of qubits
  (i.e., binary digits) of precision.
\end{remark}

\begin{example}
  \label{ex:blockencmat}
  Suppose we want to block-encode the following matrix:
  \begin{equation*}
    \begin{pmatrix}
      0 & -1 & 1  & 0 \\
      1 & 1  & -1 & 0 \\
      0 & 0  & 0  & 1 \\
      -1& 0  & 0  &-1
    \end{pmatrix}.
  \end{equation*}
  Then $s_r = 3, s_c = 2$, and $q = 2$ because this is a $2^q \times
  2^q$ matrix with at most $s_r$ nonzeroes per row and $s_c$ nonzeroes
  per column. The oracles $O_r, O_c$ are defined over two 3-qubit
  registers: we use integers $\ge 4$ to ``mark'' invalid inputs
  because for a $4 \times 4$ matrix, the only valid indices are the
  numbers $\{0, 1, 2, 3\}$.  For ease of exposition, throughout this
  example we use integer numbers rather than the corresponding binary
  representation on three digits; so, for example, we write $\ket{3}$
  instead of $\ket{011}$. $O_r$ acts as follows (writing only the output for nonzero element indices $\le s_r-1$):
  \begin{align*}
    O_r \ket{0}\ket{0} &= \ket{0}\ket{1} &     O_r \ket{0}\ket{1} &= \ket{0}\ket{2}  &    O_r \ket{0}\ket{2} &= \ket{0}\ket{6}  \\
    O_r \ket{1}\ket{0} &= \ket{1}\ket{0} &     O_r \ket{1}\ket{1} &= \ket{1}\ket{1}  &    O_r \ket{1}\ket{2} &= \ket{1}\ket{2}  \\
    O_r \ket{2}\ket{0} &= \ket{2}\ket{3} &     O_r \ket{2}\ket{1} &= \ket{2}\ket{5}  &    O_r \ket{2}\ket{2} &= \ket{2}\ket{6}  \\
    O_r \ket{3}\ket{0} &= \ket{3}\ket{0} &     O_r \ket{3}\ket{1} &= \ket{3}\ket{3}  &    O_r \ket{3}\ket{2} &= \ket{3}\ket{6},
  \end{align*}
  and similarly, $O_c$ acts as follows (writing only the output for nonzero element indices $\le s_c-1$):
  \begin{align*}
    O_c \ket{0}\ket{0} &= \ket{1}\ket{0} &     O_c \ket{1}\ket{0} &= \ket{3}\ket{0}  \\
    O_c \ket{0}\ket{1} &= \ket{0}\ket{1} &     O_c \ket{1}\ket{1} &= \ket{1}\ket{1}  \\
    O_c \ket{0}\ket{2} &= \ket{0}\ket{2} &     O_c \ket{1}\ket{2} &= \ket{1}\ket{2}  \\
    O_c \ket{0}\ket{3} &= \ket{2}\ket{3} &     O_c \ket{1}\ket{3} &= \ket{3}\ket{3}.
  \end{align*}
  Thus, Eq.~\eqref{eq:blockencvlvr} for position $(0, 2)$ in the
  matrix (i.e., the third element of the first row) reads:
  \begin{align*}
    \bra{0}\bra{0} V_L^{\dag} V_R \ket{0}\ket{2} &=
    \frac{1}{\sqrt{3}} \left( \bra{0} \bra{1} + \bra{0} \bra{2} + \bra{0} \bra{6} \right)
    \frac{1}{\sqrt{2}} \left( \ket{0} \ket{2} + \ket{1} \ket{2} \right) = \frac{1}{\sqrt{6}} = \frac{1}{\sqrt{s_r s_c}},
  \end{align*}
  whereas for position $(0, 3)$ in the matrix
  (i.e., the last element of the first row, which is empty) it reads:
  \begin{align*}
    \bra{0}\bra{0} V_L^{\dag} V_R \ket{0}\ket{3} &=
    \frac{1}{\sqrt{3}} \left( \bra{0} \bra{1} + \bra{0} \bra{2} + \bra{0} \bra{6} \right)
    \frac{1}{\sqrt{2}} \left( \ket{2} \ket{3} + \ket{3} \ket{3} \right) = 0
  \end{align*}
  Finally, the oracle $O_A$, when queried at a given position $(i,
  j)$, outputs a binary description of the value of the corresponding
  element of $A$.
\end{example}
\begin{remark}
  \label{rem:blockencinplace}
  When constructing binary oracles, in all our previous discussions we
  almost always acted with $\oplus$ (binary XOR) on the register onto
  which we want to write, i.e., with operations of the form
  $\ket{\vx}\ket{\vy} \to \ket{\vx}\ket{\vy \oplus f(\vx)}$. In
  Prop.~\ref{prop:sparseblockenc} and Ex.~\ref{ex:blockencmat},
  however, the oracles $O_r, O_c$ modify the value of the output
  register ``in place''. To convince ourselves that this is
  implementable with a quantum circuit, observe that the action of
  $O_r: \ket{\vi} \ket{\vk} \to \ket{\vi} \ket{\vv{r_{ik}}}$ is simply
  a permutation of the possible basis states, as by definition it maps
  binary strings one-to-one (note that to have a one-to-one mapping,
  it is important that, when there are less than $k$ nonzero entries
  in row $i$, we set $r_{ik}$ to a value that depends on $k$ --- in
  this case, $k + 2^q$ --- rather than a fixed, $k$-independent
  value). Every permutation matrix is unitary and can be implemented
  with Boolean logic, so a quantum circuit with only $X$, C$X$ and
  CC$X$ suffices. The number of gates is polynomial in the number of
  bits. (To see that a polynomial upper bound is possible: we can
  express the value of each bit of the output of the permutation as a
  Boolean formula of the input bits. Implementing the formula for each
  bit separately already gives a naive polynomial upper bound. See
  \cite[Sect.~4.5.2]{nielsen02quantum} for another construction.)
\end{remark}

Prop.~\ref{prop:sparseblockenc} shows that a block-encoding of a
matrix in the sparse-access model can be constructed with a constant
number of calls to the oracles describing the position and values of
the nonzero matrix elements. However, one should carefully consider
the cost of implementing those oracles. There are two natural
situations that arise when evaluating the cost of the sparse-access
oracles (a third situation can occur in the QRAM input model, see
Sect.~\ref{sec:qramblockenc}).
\begin{itemize}
\item Efficient algorithmic description of the matrix: for certain
  structured matrices, it is possible to determine the value of a
  matrix element with an efficient algorithm given the corresponding
  row and column index. In other words, given $(i,j)$ we can determine
  if the element in position $(i,j)$ is nonzero and if so, its
  value. In this case, the circuits for $O_r, O_c, O_A$ implement a
  version of the algorithm to compute the positions and values of the
  nonzero elements given the input indices. Such an implementation is
  often extremely efficient and runs in time polynomial in the number
  of bits of the indices, although the details depend on the structure
  of the matrix that we aim to block-encode; see
  Ex.~\ref{ex:blockencoracle}. This is the ideal scenario, because then
  all oracles of Prop.~\ref{prop:sparseblockenc} have low or even
  negligible cost (such as $\bigOt{1}$).
\item Data-driven representation: if an efficient algorithmic
  description starting from the input indices is not available, we can
  in general assume that the matrix is given as an unstructured list
  of position/value pairs for the nonzero elements. From such a list
  we can implement $O_r, O_c, O_A$ using a lookup table. The drawback
  of this approach is that the lookup table is rather
  inefficient. Because the oracles can be queried in superposition,
  the lookup table must contain information about all the nonzero
  elements. This implies that the cost of each of these oracles is
  roughly linear in the number $\text{nnz}(A)$ of nonzero elements of
  the matrix $A$, i.e., linear in the number of entries in the lookup
  table, so that the gate complexity of $O_r, O_c, O_A$ is
  $\bigOt{\text{nnz}(A)}$.
\end{itemize}
\begin{example}
  \label{ex:blockencoracle}
  Let us look at the cost for implementing sparse-access oracles for
  the constraint matrix of the assignment problem:
  \begin{equation*}
    \left.
    \begin{array}{rrcl}
      \min & \sum_{i=1}^n \sum_{j=1}^n c_{ij} x_{ij} & & \\
      \text{s.t.:} \quad \forall i=1,\dots,n & \sum_{j=1}^n x_{ij} &=& 1 \\
      \forall j=1,\dots,n & \sum_{i=1}^n x_{ij} &=& 1 \\
      \forall i,j=1,\dots,n & x_{ij} &\ge& 0.
    \end{array}
    \right\}
  \end{equation*}
  The constraint matrix of this problem has an efficient algorithmic
  description.  Let us assume the columns (corresponding to decision
  variables) are ordered in the ``natural'' way: $x_{11}, x_{12},
  \dots, \allowbreak x_{1n}, x_{21}, x_{22}, \dots$. It is convenient to write an
  example for small $n$ to understand the structure of the constraint
  matrix; for $n=3$, the matrix $A$ is as follows:
  \begin{equation*}
    \begin{pmatrix}
      1 & 1 & 1 & 0 & 0 & 0 & 0 & 0 & 0 \\
      0 & 0 & 0 & 1 & 1 & 1 & 0 & 0 & 0 \\
      0 & 0 & 0 & 0 & 0 & 0 & 1 & 1 & 1 \\
      1 & 0 & 0 & 1 & 0 & 0 & 1 & 0 & 0 \\
      0 & 1 & 0 & 0 & 1 & 0 & 0 & 1 & 0 \\
      0 & 0 & 1 & 0 & 0 & 1 & 0 & 0 & 1 
    \end{pmatrix}.
  \end{equation*}
  Then, if we use zero-based indices for the decision variables and
  the constraint matrix, the matrix $A$ has the following elements:
  \begin{equation*}
    a_{ij} = \begin{cases}
      1 & \text{if } i \le n-1 \text{ and } ni \le j \le n(i+1)-1 \\
      1 & \text{if } i \ge n \text{ and } j \equiv (i-n) \mod n \\
      0 & \text{otherwise.}
      \end{cases}
  \end{equation*}
  We can therefore easily construct a circuit that computes the
  indices of the nonzero elements in a row, given the row index; and
  similarly for the columns. These circuits have cost $\bigOt{1}$:
  polynomial in the number of bits used to represent the indices. If
  the coefficients of the matrix were not all $1$, however, and they
  did not follow some pattern that is easy to encode in a computer
  program, we would have to store them in a lookup table, increasing
  the cost of the sparse-access oracle that provides the element
  values to approximately linear in the number of nonzero elements.
\end{example}
\begin{example}
  Suppose we want to block-encode the adjacency matrix of a hypercube
  graph (i.e., the graph formed by the edges and vertices of an
  $n$-dimensional hypercube). The matrix has an efficient algorithmic
  description: we label the nodes with binary strings, with each bit
  indicating whether the corresponding coordinate is $0$ or $1$ (e.g.,
  for $n = 3$, $000$ is the origin, $001$ is the point where the third
  coordinate is $1$, and so on). Then, given indices $(i, j)$,
  corresponding to two vertices, based on their binary representations
  $\vi, \vj$ we can determine if they are neighbors in the hypercube
  graph, by scanning the bits of the bitstrings, and checking whether
  they differ in exactly one position. Thus, we can efficiently output
  the indices of the neighbors of each node. In this case the cost of
  the sparse-access oracles is $\bigOt{n}$, which is logarithmic in
  the number of vertices and edges in the graph.
\end{example}
\begin{remark}
  An intuitive way to recognize if a certain matrix --- or, more in
  general, a piece of data --- admits an efficient algorithmic
  description is to determine if it is possible to write a short
  computer code that, when queried with a certain position, outputs
  the corresponding element. There is nothing inherently ``quantum''
  about this property: the computer code might as well be written in
  any classical language, e.g., Python. For the code to be
  efficient, the length of the code must be short: typically,
  polylogarithmic in the size of the problem whose data the code is
  intended to describe.
\end{remark}

\subsection{Block-encoding with QRAM access}
\label{sec:qramblockenc}
We can accelerate the construction of a block-encoding of a
classically-available matrix if we have access to a QRAM\index{QRAM} of an
appropriate size. The sparse-access oracle model can of course be
directly accelerated: the oracles $O_r, O_c, O_A$ can be implemented
with a single query to a QRAM containing the corresponding data, so
the QRAM query complexity of Prop.~\ref{prop:sparseblockenc} is
exactly as stated in the proposition --- similarly to the ``efficient
algorithmic description'' case. An even more efficient strategy is
described next, and it is based on the idea of constructing quantum
states encoding each row of $A$ with the QRAM-based amplitude
encoding technique of Sect.~\ref{sec:qramampenc} for vectors. Below,
we denote by $A_j$ the $j$-th column of $A$, and by $p$ the number of
bits used to represent each entry of the data structure of
Sect.~\ref{sec:qramampenc} (which, as usual, is assumed to be large
enough that we can perform all calculations with negligible error ---
the required precision is only polylogarithmically large anyway). We
recall the definition of Frobenius norm as it appears in the next
proposition.
\begin{definition}[Frobenius norm]
  \label{def:frobenius}
  Given a matrix $A \in \C^{m \times n}$ with entries $a_{ij}$, its
  \emph{Frobenius norm}\index{norm!matrix}\index{norm!Frobenius} is the quantity $\|A\|_F = \sqrt{\sum_{i=1}^m
    \sum_{j=1}^n \abs{a_{ij}}^2}$.
\end{definition}
\begin{proposition}[Adapted from \cite{chakraborty2019power,kerenidis2016quantumofficial}.]
  \label{prop:qramblockenc}
  Let $d = 2^q$ and $A \in \C^{d \times d}$. Suppose we have a
  classical finite-precision description of the entries of $A$ with
  $p$ bits each, and assume $p = \bigOt{1}$. Then, given a QRAM of
  size $\bigO{d^2p}$, we can implement a $(\|A\|_F, \bigOt{q},
  \xi)$-block-encoding of $A$ using $\bigOt{1}$ accesses to the QRAM
  and additional gates, and $\bigOt{d^2}$ classical arithmetic
  operations to initialize the QRAM data structures.
\end{proposition}
\begin{proof}
  We use two unitaries $V_L, V_R$ satisfying the following properties:
  \begin{align*}
    V_L\ket{\v{0}}_q \ket{\vi}_q &= \ket{\vi} \sum_{\v{\ell} \in \{0,1\}^q} \frac{\|A_{\ell}\|}{\|A\|_F} \ket{\v{\ell}} \qquad \forall i \in \{0,1\}^q\\
    V_R\ket{\v{0}}_q \ket{\vj}_q &= \sum_{\vk \in \{0,1\}^q} \frac{a_{k j}}{\|A_{j}\|} \ket{\vk} \ket{\vj} \qquad \forall j \in \{0,1\}^q.
  \end{align*}
  The first unitary $V_L$ can be constructed using SWAP gates and
  Cor.~\ref{cor:qramampenc} for the vector with entries
  $\|A_{\ell}\|$, yielding the quantum state $\sum_{\v{\ell} \in
    \{0,1\}^q} \frac{\|A_{\ell}\|}{\|A\|_F} \ket{\vk}$ (because the
  sum of the squares of the column norms is the Frobenius norm
  squared). The second unitary can be constructed using
  Cor.~\ref{cor:qramampenc} for the vectors with entries $a_{k j}$,
  conditioned on the value of the second register $\ket{\vj}$ to
  address the correct binary tree data structure in QRAM (i.e., the
  one for column $j$), yielding the vectors $\sum_{\vk \in \{0,1\}^q}
  \frac{a_{k j}}{\|A_{j}\|} \ket{\vk}$. For this, auxiliary registers
  of size $p$ are necessary to temporarily hold the data queried from
  the QRAM, before being used for some controlled operations and
  eventually uncomputed, as in Cor.~\ref{cor:qramampenc}. Note that
  the normalization factors once again work out, as $\sum_{k}
  |a_{kj}|^2 = \|A_{j}\|^2$. We now show that $V_L^{\dag} V_R$ is the
  desired block-encoding. We have:
  \begin{align*}
    \bra{\v{0}}\bra{\vi} V_L^{\dag} V_R \ket{\v{0}}\ket{\vj} &=
    \left(\bra{\vi} \sum_{\v{\ell} \in \{0,1\}^q} \frac{\|A_{\ell}\|}{\|A\|_F} \bra{\v{\ell}}\right) \left(\sum_{\vk \in \{0,1\}^q} \frac{a_{k j}}{\|A_{j}\|} \ket{\vk} \ket{\vj}\right) \\
    &= \frac{a_{ij}}{\|A_{j}\|}\braket{\vi}{\vi} \frac{\|A_j\|}{\|A\|_F} \braket{\vj}{\vj} = \frac{a_{ij}}{\|A\|_F}.
  \end{align*}
  Regarding the complexity, we are applying Cor.~\ref{cor:qramampenc}
  to construct the amplitude encoding of $d$-dimensional vectors,
  which can be done with $\bigO{\log d}$ QRAM calls, $\bigO{\log^2 d}$
  additional gates, and $\bigOt{d}$ classical arithmetic operations to
  initialize the QRAM data structure for each vector. The quantum
  operations can be performed in $\bigOt{1}$ time (we apply
  $V_L^{\dag}, V_R$ once, each of them calls the amplitude-encoding
  subroutine), and the classical initialization procedure is repeated
  $d+1$ times (for the $d$ columns plus the vector of column norms
  used in $V_L$), giving the stated complexity. Note that the target
  maximum error $\xi$ determines the error $\epsilon$ with which we
  implement the amplitude encoding circuits in
  Cor.~\ref{cor:qramampenc}; the dependence on $\epsilon$ is
  polylogarithmic, so the impact of $\xi$ on the final complexity is
  negligible (in $\bigOt{\cdot}$ notation).
\end{proof}
\begin{example}
  Let us have a closer look at the matrices $V_L, V_R$ of
  Prop.~\ref{prop:qramblockenc} to construct a block-encoding of the
  following matrix:
  \begin{equation*}
    A = \begin{pmatrix} 1 & -2 \\ 0 & -2 \end{pmatrix} = \begin{pmatrix} a_{00} & a_{01} \\ a_{10} & a_{11} \end{pmatrix}.
  \end{equation*}
  By definition, $V_R$ must have the following action:
  \begin{align}
    \label{eq:qramblockencvr}
    \begin{split}
    V_R \ket{00} &= \left(\frac{a_{00}}{\|A_0\|}\ket{0} + \frac{a_{10}}{\|A_0\|} \ket{1}\right) \ket{0} = \ket{00} \\
    V_R \ket{01} &= \left(\frac{a_{01}}{\|A_1\|}\ket{0} + \frac{a_{11}}{\|A_1\|} \ket{1}\right) \ket{1} = -\frac{1}{\sqrt{2}} \ket{01} -\frac{1}{\sqrt{2}} \ket{11}.
    \end{split}
  \end{align}
  This defines two columns of the matrix $V_R$. A unitary that
  implements the operation described in Eq.~\eqref{eq:qramblockencvr}
  is:
  \begin{equation*}
    V_R = \begin{pmatrix} 1 & 0 & 0 & 0 \\ 0 & -\frac{1}{\sqrt{2}} & 0 & -\frac{1}{\sqrt{2}} \\ 0 & 0 & 1 & 0 \\ 0 & -\frac{1}{\sqrt{2}} & 0 & \frac{1}{\sqrt{2}} \end{pmatrix}.
  \end{equation*}
  By definition, $V_L$ must have the following action:
  \begin{align}
    \label{eq:qramblockencvl}
    \begin{split}
    V_L \ket{00} &= \ket{0} \left(\frac{\|A_0\|}{\|A\|_F} \ket{0} + \frac{\|A_1\|}{\|A\|_F} \ket{1}\right) = \frac{1}{3} \ket{00} + \frac{2\sqrt{2}}{3} \ket{01} \\
    V_L \ket{01} &= \ket{1} \left(\frac{\|A_0\|}{\|A\|_F} \ket{0} + \frac{\|A_1\|}{\|A\|_F} \ket{1}\right) = \frac{1}{3} \ket{10} + \frac{2\sqrt{2}}{3} \ket{11}.
    \end{split}
  \end{align}
  Once again, this defines two columns of $V_L$, and a unitary that
  implements the operation is:
  \begin{equation*}
    V_L = \begin{pmatrix} \frac{1}{3} & 0 & -\frac{2\sqrt{2}}{3} & 0 \\ \frac{2\sqrt{2}}{3} & 0 & \frac{1}{3} & 0 \\ 0 & \frac{1}{3} & 0 & -\frac{2\sqrt{2}}{3} \\ 0 & \frac{2\sqrt{2}}{3} & 0 & \frac{1}{3} \end{pmatrix}.
  \end{equation*}
  Finally, we can verify that these unitaries allow us to compute a $(\|A\|_F, \bigOt{q}, \xi)=(3, 1, 0)$-block-encoding of $A$ as $V_L^{\dag} V_R$:
  \begin{equation*}
    V_L^{\dag} V_R = \begin{pmatrix} \frac{1}{3} & -\frac{2}{3} & 0 & -\frac{2}{3}  \\
      0 & -\frac{2}{3} & \frac{1}{3} & \frac{2}{3} \\
      -\frac{2\sqrt{2}}{3} & -\frac{1}{3\sqrt{2}} & 0 & -\frac{1}{3\sqrt{2}} \\
      0 & -\frac{1}{3\sqrt{2}} & -\frac{2\sqrt{2}}{3} & \frac{1}{3\sqrt{2}}
    \end{pmatrix}.
  \end{equation*}
  Notice the scaled-down copy of $A$ in the top-left block of the
  above matrix. For this specific case the final error $\xi$ of the
  block-encoding is 0, because the given matrices $V_L, V_R$ implement
  the corresponding maps exactly. In general this may not be possible
  in actual implementations, because the coefficients in
  Eq.s~\eqref{eq:qramblockencvr}-\eqref{eq:qramblockencvl} are
  constructed from finite-precision representations of $a_{ij},
  \|A_j\|, \|A\|_F$, and therefore so are the corresponding unitaries.
  These finite-precision representations lead to errors that are
  exponentially small (but not necessarily zero) in the number of
  qubits used to represent each number. Furthermore, the operations
  used in the procedure of Cor.~\ref{cor:qramampenc} may not be
  implemented exactly, as discussed in
  Sect.~\ref{sec:ampencalg}.\index{block-encoding!construction|)}
\end{example}

\subsection{Sampling from Gibbs distributions and trace estimation}
\label{sec:gibbssampling}
Before discussing the Gibbs distribution, we introduce an additional
type of matrix norm that is needed in the following.
\begin{definition}[Trace norm]
  \label{def:tracenorm}
  For a given matrix $A$, we denote its \emph{trace norm} by
  $\nrm{A}_{\Tr} :=
  \trace{\sqrt{A^{\dag}A}}$.\index{norm!matrix}\index{norm!trace} This
  is the same as the Schatten $1$-norm, i.e., the sum of the singular
  values of $A$.
\end{definition}
\begin{remark}
  The \emph{trace distance}\index{variation distance, total}\index{trace!distance} between
  two density matrices $\rho, \rho'$, i.e., the distance measured with
  the trace norm $\nrm{\rho - \rho'}_{\Tr}$, is
  generally used to measure the distance between mixed quantum
  states. One of the reasons for this choice is the fact that the
  trace distance is a generalization of the total variation distance
  for pure states, see Def.~\ref{def:tvd} and
  Prop.~\ref{prop:euclideantotvd}, because:
  \begin{equation*}
    \nrm{\ketbra{\psi}{\psi} - \ketbra{\phi}{\phi}}_{\Tr} = 2\sqrt{1 - \abs{\braket{\psi}{\phi}}^2} \le 2 \nrm{\ket{\psi} - \ket{\phi}},
  \end{equation*}
  for any pure states $\ket{\psi}, \ket{\phi}$ (the inequality above
  is due to Pythagoras' theorem). Often, the trace distance is defined
  with a factor $\frac{1}{2}$ in front, i.e., $\text{dist}(\rho,
  \rho') = \frac{1}{2} \nrm{\rho - \rho'}_{\Tr}$, so that the
  correspondance with the total variation distance becomes direct when
  $\rho, \rho'$ are diagonal (i.e., they encode classical probability
  distributions).
\end{remark}

The Gibbs distribution plays an important role in certain classes of
optimization algorithms, for example in the optimization framework
discussed in Ch.~\ref{ch:mmwu}.\index{Gibbs!state|(}
\begin{definition}[Gibbs distribution]
  \label{def:gibbsdist}
  Given a finite set $G$ (called \emph{ground set}), a function
  $f : G \to \R$, and a parameter $\beta > 0$ (called
  \emph{inverse temperature}), the corresponding \emph{Gibbs
  distribution}\index{Gibbs!distribution} is the discrete probability
  distribution over $G$ defined by:
  \begin{equation*}
    \Pr(x) = \frac{1}{\sum_{x \in G} e^{-\beta f(x)}} e^{-\beta f(x)}
    \qquad \forall x \in G.
  \end{equation*}  
\end{definition}
The function $f$ in the above definition is often called a Hamiltonian
in quantum physics: Hamiltonians encode energy levels of a system (see
Def.~\ref{def:hamiltonian}), and $f$ associates a value with $x \in
G$ (in physics, this would be an energy level).  There is a
natural and compact notation for a Gibbs distribution as a density
matrix (i.e., a mixed quantum state). Suppose $G = \{0,1\}^q$,
and let $\ham$ be a diagonal matrix such that $\bra{\vj} \ham
\ket{\vj} = -\beta f(\vj)$; we use $\ham$ to denote it due to its now
obvious connection to Def.~\ref{def:hamiltonian}. Then $\exp(\ham)$ is
a matrix that has diagonal elements equal to $e^{-\beta f(\vj)}$, and
$\exp(\ham)/\trace{\exp(\ham)}$ has, on its diagonal, exactly the
probability values of Def.~\ref{def:gibbsdist}. Note that
$\exp(\ham)/\trace{\exp(\ham)}$ is also a density matrix, because it
is positive definite matrix with unit trace, therefore it describes a
mixed quantum state. This is called a Gibbs state; we can in fact
relax the requirements that $\ham$ is diagonal, and obtain the
following definition.
\begin{definition}[Gibbs state]
  \label{def:gibbsstate}
  Given a Hermitian matrix $\ham \in \C^{2^q \times 2^q}$, usually called
  Hamiltonian, the \emph{Gibbs state} corresponding to $\ham$ is the
  (possibly mixed) quantum state $\rho$ with density matrix:
  $$ \rho  = \frac{\exp(\ham)}{\trace{\exp (\ham)}}.$$
\end{definition}
Note that performing a measurement of all qubits from a quantum
register in the state $\rho$ yields a sample from the Gibbs
distribution encoded by $\ham$, therefore constructing a Gibbs state
effectively allows us to sample from a Gibbs distribution.  In this
section we discuss how to construct the Gibbs state $\rho$ given the Hamiltonian
$\ham$, how to block-encode a (subnormalized version of a) Gibbs state,
and also how to perform certain operations on $\rho$. The discussion
is based on \cite{gilyen2019thesis,van2020quantum}: all proofs not
given here can be found in one of these two references. We first need
to define the concept of subnormalized density matrix, which is useful
because in some situations we may not want to (or cannot) work with a
density matrix, but we can work with a scaled-down version of it.
\begin{definition}[Subnormalized density matrix]
  \label{def:subnormdo}
  A \emph{subnormalized density matrix}\index{density matrix!subnormalized} $\rho$ is a positive
  semidefinite matrix of trace at most 1. A \emph{purification}\index{density matrix!purification}
  $\varrho$ of a subnormalized density matrix $\rho \in \C^{2^q \times
    2^q}$ is a pure state $\ket{\psi}$ over three registers $A, B, C$
  of size, respectively, $q$, $1$ and $p \le q$ such that
  \begin{equation*}
    (I^{\otimes q} \otimes \bra{0}) \trace[C]{\ketbra{\psi}{\psi}} (I^{\otimes q} \otimes \ket{0}) = \rho,
  \end{equation*}
  i.e., tracing out the third register (of size $p$) and projecting
  onto the subspace where the second register (of size 1) is $\ket{0}$
  yields $\rho$.
\end{definition}
\begin{remark}
  The difference between a density matrix and a subnormalized density
  matrix is subtle, and is worth pointing out. We have seen in
  Sect.~\ref{sec:mixedstate}, in particular
  Thm.~\ref{thm:purification}, that it is possible to express every
  density matrix as the partial trace of a pure state, called
  purification (Def.~\ref{def:purification}). In
  Thm.~\ref{thm:purification} we only need two registers, in
  particular we do not need the single-qubit register used in
  Def.~\ref{def:subnormdo}. In a subnormalized density matrix the
  trace does not need to be 1, whereas in a density matrix it is
  always 1. The second register serves the purpose of allowing a
  smaller trace, by defining a subspace of the entire density matrix
  that contains the part of interest; i.e., we have a ``larger''
  density matrix of the form $\rho^{(00)} \otimes \ketbra{0}{0} +
  \rho^{(01)} \otimes \ketbra{0}{1} + \rho^{(10)} \otimes
  \ketbra{1}{0} + \rho^{(11)} \otimes \ketbra{1}{1}$, with
  $\trace{\rho^{(00)} + \rho^{(11)}} = 1$, and the subnormalized
  density matrix of interest is $\rho^{(00)}$. The order and size of
  the register that we project on are just a convention: the concept
  can easily be generalized, but the generalization is not necessary
  for the discussion in this \book{}.
\end{remark}


Next, we give an algorithmic result that characterizes the complexity
of constructing a Gibbs state from the block-encoding of some
Hermitian matrix; however, for the sake of completeness we first need
to define the degree of a certain polynomial that approximates the
exponential function sufficiently well.
\begin{lemma}[Lem.~4.14 in \cite{van2020quantum}]
  \label{lem:exppolydeg}
  Let $\xi \in (0, 1/6]$ and $\beta \geq 1$. There exists a polynomial
    $P(x)$ such that:
    \begin{itemize}
    \item For all $x \in [-1,0]$, we have $\left| P(x) - \exp(2 \beta x)/4 \right| \leq \xi$.
    \item For all $x \in [-1,1]$, we have $\left| P(x)  \right| \leq 1/2$.
    \item $\text{deg}(P) = \bigOt{\beta}$.
    \end{itemize}
\end{lemma}
The polynomial of Lem.~\ref{lem:exppolydeg} appears in the error
requirement of the input block-encoding, because the matrix
exponential (rather, the polynomial approximation of the matrix
exponential) could amplify errors
significantly. Lem.~\ref{lem:exppolydeg} simply states that we can
construct the desired polynomial approximation in the eigenvalue
interval $[-1, 0]$, including an amplification factor $\beta$ for the
block-encoding, which is used to cancel out the subnormalization
factor of the input block-encoding. We are now in a position to state
the complexity of constructing a Gibbs state
$\exp(\ham)/\trace{(\exp(\ham))}$.
\begin{proposition}[Lem.~4.15 in \cite{van2020quantum}]
  \label{prop:gibbsstateconstr}
  Let $\ham \in \R^{n \times n}$ be a Hermitian matrix. Let $\theta
  \in (0, 1/3]$, $\beta > 1$, and let $d$ be the degree of the
    polynomial from Lemma \ref{lem:exppolydeg} when we let $\xi =
    \frac{\theta}{128n}$. Let $U$ be a $(\beta, a, \frac{\theta^2
      \beta}{1024^2 d^2 n^2})$-block-encoding of $\ham$. Then we can
    create a purification of a state $\tilde{\rho}$ such that
  $$ \left\| \tilde{\rho} - \frac{\exp(\ham)}{\trace{(\exp(\ham))}}
  \right\|_{\Tr} \leq \theta$$ using $\bigOt{\sqrt{n} \beta}$
  applications of $U$ and $\bigOt{\sqrt{n} \beta a}$ elementary gates.
\end{proposition}
The construction of Prop.~\ref{prop:gibbsstateconstr} is based on
polynomial approximations of the exponential function, which can be
obtained using quantum singular value transformation techniques
introduced in \cite{gilyen2019thesis,gilyen2019quantum}. The
construction requires a very high precision for the block-encoding $U$
for $\ham$, but this should not be a deterrent: we already described
situations where the running time of a block-encoding construction
scales polylogarithmically in the desired inverse precision, e.g.,
Prop.~\ref{prop:sparseblockenc}, therefore obtaining a high-precision
block-encoding is not necessarily a bottleneck. The result described
in Prop.~\ref{prop:gibbsstateconstr} is based on the following idea:
assume $n = 2^q$ for simplicity, so that $q$ is the number of qubits
of the operator $\ham$; we start by constructing the state $\sum_{\vj
  \in \{0,1\}^q} \ket{\vj}\ket{\vj}$, which requires only Hadamards
and CNOTs, and is often called \emph{maximally-mixed} state\index{state!maximally mixed} in the
literature. If we trace out the second register, we find that the
density matrix describing the first register is the rescaled identity
$I^{\otimes q}/2^q$. Then if we construct and apply a block-encoding of
$e^{\ham/2}$ to the first register (keeping the auxiliary register for
the block-encoding separate, as usual), the state evolves to
$e^{\ham/2} I^{\otimes q} e^{\ham/2} = e^{\ham}$ in the ``correct''
subspace, modulo normalization: this follows from the rules of
evolution of a mixed state in Sect.~\ref{sec:mixedstate}, together
with the fact that $\ham^{\dag} = \ham$. A somewhat similar idea is
described in the proof of Prop.~\ref{prop:traceest}. The construction
of the block-encoding of $e^{\ham/2}$ additionally requires shifting
the spectrum of $\ham$ before applying the (polynomial approximation
of the) exponential function, to ensure $\nrm{\exp(\ham)} \le 1$. We
achieve this goal by shifting the spectrum of $\ham$ to be negative,
which is why in Lem.~\ref{lem:exppolydeg} we are only concerned about
approximating the exponential for $x \in [-1, 0]$. The spectrum shift
is not an issue, because $\exp(\ham + \lambda I)/\trace{\exp(\ham +
  \lambda I)} = \exp(\ham)/\trace{\exp(\ham)}$ for every $\lambda \in
\R$. After selecting the correct subspace via amplitude amplification,
the resulting state is a density matrix proportional to $e^{\ham}$,
hence it is $\exp(\ham)/\trace{(\exp(\ham))}$.

We can also construct a block-encoding of a density matrix $\rho$ from
a unitary that prepares a purification of it.
\begin{lemma}[Block-encoding of a (subnormalized) density operator; Lem.~6.4.4 in \cite{gilyen2019thesis}]
  \label{lem:blockencpurif}
  Let $U$ be a $(q + a)$-qubit unitary that, given the input state
  $\ket{\v{0}}_q \ket{\v{0}}_a$, prepares a purification
  $\ket{\varrho}$ of the (possibly subnormalized) $q$-qubit density
  matrix $\rho$. Then we can implement a $(1, q + a,
  0)$-block-encoding of $\rho$ with a single use of $U$ and its
  inverse, and $q+1$ two-qubit gates.
\end{lemma}
Finally, we can construct a trace estimator to compute quantities of
the form $\trace{A\rho}$ using the block-encoding model. This result
is extremely useful in the quantum algorithms for semidefinite
optimization discussed in Ch.~\ref{ch:mmwu}, because the constraints
and objective function of the problems discussed in that chapter
involve expressions of the form $\trace{A\rho}$: this gives us a way
of estimating their value without having explicit knowledge of $\rho$,
as long as we can construct a state with density matrix $\rho$. To
understand some of the ideas for the construction we start with a
simpler example that does not produce what we need, but it allows us
to see the power of block-encoded matrices, and some of the operations
that can be performed with them.
\begin{example}
  \label{ex:blockenctrace}
  In this example we construct a block-encoding of the scalar
  $\trace{A\rho}$, i.e., of a $1 \times 1$ matrix, using access to a
  purification for $\rho$ and a block-encoding of $A$. While we do not
  directly make use of this, it is a relatively simple construction
  that is instructive.
  
  Let $U$ be a $(q + m)$-qubit unitary that, given the input state
  $\ket{\v{0}}_m \ket{\v{0}}_q$, prepares a purification
  $\ket{\varrho}$ of the $q$-qubit density matrix $\rho$; the
  $m$-qubit register is the purifying register. Let $V$ be a $(\alpha,
  a, \xi)$-block-encoding of a $2^q \times 2^q$ matrix $A$ (where $A$
  acts on the same $q$-qubit register defined above for $U$).

  Consider the circuit $(I_{a} \otimes U^{\dag}) (V \otimes I_{m})
  (I_{a} \otimes U)$; in this expression we are using the same
  convention whereby the subscript of the identity matrices indicates
  not only their size, but also which register they are applied
  to. This circuit first uses the unitary $U$ to create the
  purification of $\rho$, then applies the block-encoding of $A$,
  finally applies the inverse unitary $U^{\dag}$. The reason for the
  inverse unitary $U^{\dag}$ is that it allows us to ``sandwich'' the
  block encoding with $\bra{\varrho}$ on the left and $\ket{\varrho}$
  on the right, which results in block-encoding
  $\trace{A\rho/\alpha}$. To see this, let us apply the definition of
  block-encoding, using the $(a+m)$ auxiliary qubits as the ones that
  are ``hit'' by the all-zero basis state.  We have:
  \begin{align*}
    \bra{\v{0}}_{a} \bra{\v{0}}_{m} \bra{\v{0}}_{q} (I_{a} \otimes U^{\dag}) (V \otimes I_{m}) (I_{a} \otimes U) \ket{\v{0}}_{a} \ket{\v{0}}_{m} \ket{\v{0}}_{q} = \bra{\v{0}}_{a} \bra{\varrho} (V \otimes I_{m}) \ket{\v{0}}_{a} \ket{\varrho} \\
    = \bra{\varrho} (A/\alpha \otimes I_{m}) \ket{\varrho} = \trace{\bra{\varrho} (A/\alpha \otimes I_{m}) \ket{\varrho}} = \trace{(A/\alpha \otimes I_{m}) \ketbra{\varrho}{\varrho}} = \trace{A/\alpha \rho}.
  \end{align*}
  For the above chain of equalities, we used the fact that by
  definition of $\ket{\varrho}$, if we trace out the $m$-qubit
  auxiliary (purifying) register we obtain $\rho$, and it is easy to
  see that $\trace{(A/\alpha \otimes I_{m})
    \ketbra{\varrho}{\varrho}}$ corresponds to tracing out the
  $m$-qubit register (because $I_{m} = \sum_{\vj \in \{0,1\}^m}
  \ketbra{\vj}{\vj})$. Thus, we obtained the desired block-encoding,
  with subnormalization factor $\alpha$.
\end{example}

\noindent Unfortunately, it is not obvious if one can obtain an
estimate of $\trace{A\rho/\alpha}$ from the block-encoding in
Ex.~\ref{ex:blockenctrace}: the block-encoding in itself ensures that
$\trace{A\rho/\alpha}$ is the coefficient of $\ket{\v{0}}$ after
applying the circuit to the state $\ket{\v{0}}$, but from there we can
only recover $|\trace{A\rho/\alpha}|^2$, which is equal to the
probability of observing $\ket{\v{0}}$ when measuring; for example, we
can recover it by taking repeated measurements or with amplitude
estimation (Sect.~\ref{sec:ampest}). Because of the absolute value and
the square, there is a sign problem and amplitude estimation would not
be able to tell if $\trace{A\rho}$ is positive or negative. There is a
more astute construction that leads to a much better algorithm for the
estimation of $\trace{A\rho}$. Because in Ex.~\ref{ex:blockenctrace}
we end up obtaining the square of the quantity of interest, we use the
matrix square root of $A$. We start with the mixed state described by
the density matrix $\rho$, then apply the matrix square root of an
appropriately shifted version of $A$, obtaining a state corresponding
to the density matrix $\sqrt{A} \rho \sqrt{A}^{\dag}$ (after selecting
the correct part of the space, i.e., the one with support on
$\ket{\v{0}}$ in certain registers); if the construction is applied to
a Hermitian matrix, the $\dag$ disappears. After applying the circuit
based on this idea, we show that the probability of observing
$\ket{\v{0}}$ is approximately a shifted version of
$\trace{\sqrt{A}\rho\sqrt{A}}$. A more precise description is given in
the proof of the next result.
\begin{proposition}[Cor.~6.4.5 in \cite{gilyen2019thesis}]
  \label{prop:traceest}
  Let $\rho$ be the density matrix\index{trace!estimation} representing a given $q$-qubit
  quantum state, and $U$ an $(\alpha, a, \theta/2)$-block-encoding of
  a Hermitian matrix $A \in \R^{2^q \times 2^q}$ with $\|A \| \leq
  1$. We can construct a quantum circuit that, after measurement,
  outputs a (binary-encoded) sample from a random variable $Y$ with
  the property that the expected value of $Y$ is at most $\theta/4$
  away from $\trace{A\rho}$, and its standard deviation is $\sigma =
  \bigO {1}$. The quantum circuit uses $\bigOt{\alpha}$ applications
  of $U$ and $U^{\dagger}$, $\bigOt{\alpha}$ two-qubit gates, and one
  copy of $\rho$.
\end{proposition}
\begin{proof}
  We provide a sketch of the proof, referring to
  \cite{gilyen2019thesis} for details.

  As a first step, we amplify the block-encoding $U$ to obtain a
  block-encoding of $A/2$ (rather than $A/\alpha$) using
  $\bigOt{\alpha}$ applications of $U$. We do not provide a precise
  statement of this result, but this can be done by applying the
  matrix function $f(x) = \alpha x /2$ to the block-encoding (again,
  via a polynomial approximation). This is analogous to oblivious
  amplitude amplification to amplify the subspace where $U$ acts as
  $A/\alpha$, but amplitude amplification would only work for unitary
  matrices. Then, by linear combination of block-encodings
  (Prop.~\ref{prop:lincombblock}) with uniform weights, we transform
  it into a block-encoding of $A/4 + I/2$: this requires a single use
  of the block encoding of $A/2$ and of $I$ (which is trivial). Because
  $\|A \| \leq 1$ the smallest eigenvalue of $A/4$ is $\ge -1/4$, so
  $A/4 + I/2 \succeq 0$. At this point, using a polynomial
  approximation of the square root function, we construct a
  block-encoding $W$ of $\sqrt{A/4 + I/2}/2$: this requires
  $\bigOt{1}$ applications of the block-encoding for $A/4 +
  I/2$. Throughout these constructions we accumulate some error
  $\bigO{\theta}$, but as the complexity of the operations depends
  polylogarithmically on $\theta$, we use $\bigOt{\cdot}$ notation and
  neglect it for simplicity. In summary, $W$ satisfies the following
  property:
  \begin{equation*}
    (\bra{\v{0}} \otimes I_q) W(\ket{\v{0}} \otimes
    I_q) \approx \frac{1}{2}\sqrt{\frac{A}{4} + \frac{I}{2}}.
  \end{equation*}

  Let $w$ be the number of auxiliary qubits of the block-encoding $W$
  ($w$ is of the same order of magnitude as $a$). Now we apply $W$ to
  $\rho$, initially setting the auxiliary qubits to $\ket{\v{0}}_w$,
  as usual. The state of the system is $W(\ketbra{\v{0}}{\v{0}}
  \otimes \rho)W^{\dag}$, and the probability of observing $\ket{\v{0}}_w$
  when performing a measurement in the auxiliary register is:
  \begin{align*}
    \trace{(\ketbra{\v{0}}{\v{0}} \otimes I_q)
      W(\ketbra{\v{0}}{\v{0}} \otimes \rho)W^{\dag}} &=
    \trace{(\bra{\v{0}} \otimes I_q)
      W(\ketbra{\v{0}}{\v{0}} \otimes \rho)W^{\dag} (\ket{\v{0}} \otimes I_q)} \\
    &= \Tr \Bigl(\underbrace{(\bra{\v{0}} \otimes I_q) W^\dag (\ket{\v{0}} \otimes I_q)}_{\approx \sqrt{A/4+I/2}/2} \underbrace{(\bra{\v{0}} \otimes I_q)
      W (\ket{\v{0}} \otimes I_q)}_{\approx \sqrt{A/4+I/2}/2}  \rho \Bigr)\\
    &= \frac{1}{8} + \frac{\trace{A\rho}}{16} + \bigO{\theta},
  \end{align*}
  where the term $\bigO{\theta}$ comes from the errors accumulated
  throughout the construction. (In the chain of equations, we used the
  cyclic property of the trace to bring $W^{\dag}$ to the left of
  $\rho$.) Define a random variable $Y$ that takes value $14$ if the
  auxiliary register contains $\ket{\v{0}}_w$ after measurement, and
  $-2$ otherwise. We can easily construct a circuit that looks at the
  first $w$ qubits and outputs $14$ or $-2$ depending on the value
  contained in the register: this is the sample from $Y$ defined in
  the statement of the proposition. The expected value of $Y$
  satisfies:
  \begin{align*}
    \mathbb{E}[Y] = \frac{14}{8} + \frac{14\trace{A\rho}}{16} -2(1 -
    \frac{1}{8} - \frac{\trace{A\rho}}{16}) + \bigO{\theta} =
    \trace{A\rho} + \bigO{\theta},
  \end{align*}
  and with similar calculations, ensuring the constant in
  $\bigO{\theta}$ is chosen sufficiently small, we can also guarantee
  that the variance is $\bigO{1}$.
\end{proof}
\begin{remark}
  \label{rem:meanestimation}
  The construction in Prop.~\ref{prop:traceest} prepares a random
  variable via a quantum circuit. The general structure of an
  algorithm that prepares a random variable via a quantum circuit is
  the following: the algorithm, starting from the state $\ket{\v{0}}$,
  produces a quantum state and concludes with a single
  measurement. From this single measurement, a (classical) computation
  outputs the value of the random variable. This allows us to obtain a
  sample.

  If we want to estimate properties of the random variable, e.g., its
  expected value $\mathbb{E}[Y]$ (to estimate $\trace{A\rho}$), in
  general we need multiple copies of the state $\rho$: each sample of
  the random variable requires a measurement, and each measurement
  ``consumes'' a copy of the state $\rho$, which is entangled with the
  measured qubits. For example, if we use amplitude estimation to
  estimate $\mathbb{E}[Y]$ --- which is possible under some conditions
  --- then we need multiple applications of a circuit that produces
  $\rho$, and the inverse circuit; see Rem.~\ref{rem:ampestunbiased}
  and Sect.~\ref{sec:ampampnotes} regarding controlling the bias with
  amplitude estimation. However, sometimes it is possible to get away
  with fewer copies of $\rho$. An example of this is discussed in
  Sect.~\ref{sec:qmwuimpr}, but the technique is rather involved and
  relies on the specific setting discussed in that section. Different
  approaches for mean estimation are possible, and depending on the
  properties of the specific problem at hand, one might be able to get
  away with more efficient algorithms. For a discussion of quantum
  algorithms to estimate the mean of a random variable depending on
  its properties, we refer to \cite{montanaro2015montecarlo}.\index{Gibbs!state|)}
\end{remark}

\section{Notes and further reading}
\label{sec:blockencnotes}
Our discussion of quantum algorithms for linear systems is mainly
based on \cite{harrow2009quantum}, but we incorporate some subsequent
developments to relax some of the strict assumptions. In addition to
the massive improvements introduced in
\cite{chakraborty2019power,childs2017quantum,gilyen2019quantum}, and
already discussed in Sect.~\ref{sec:hhlimprovements}, recent work has
tightened existing bounds \cite{costa2022optimal} and improved the
constants \cite{dalzell2024shortcut}. The query complexity\index{linear system!complexity} of these
last two papers is $\bigO{\kappa \log \frac{1}{\epsilon}}$, without hidden
polylogarithmic factors, where each query is a call to an oracle
block-encoding the matrix $A$, or preparing the state $\ket{\amp{b}}$. Note
that this complexity is optimal, due to matching lower bounds (a lower
bound of $\Omega(\kappa)$ oracle calls is proven in
\cite{somma2021complexity}, and \cite{costa2022optimal} claims that a
lower bound of $\Omega\left(\kappa \log \frac{1}{\epsilon}\right)$ oracle calls can also
be shown, although to the best of our knowledge, such a proof has not
appeared in the open literature yet).

The block-encoding framework is established in
\cite{gilyen2019quantum}, building on previous work on
\emph{qubitization} \cite{low2017hamiltonian,low2019hamiltonian} and
\emph{quantum signal processing}
\cite{low2016methodology,low2017optimal}. All the results on matrix
manipulation discussed in this chapter can be found in
\cite{gilyen2019quantum}, or derived directly from that
framework. However, the exposition in \cite{gilyen2019quantum} is very
technical and building intuition on how the main results work may
prove to be a difficult task. Different expositions of the fundamental
concepts, which may turn out to be more accessible to some readers,
are given in \cite{martyn2021grand,tang2024cs}. The Ph.D.~thesis
\cite{gilyen2019thesis} may also be a suitable starting point, as well
as the excellent lecture notes \cite{lin2022lecture}. A direct
application of the block-encoding framework is discussed in
Ch.~\ref{ch:mmwu}.

Quantum linear systems algorithms and matrix manipulation via
block-encoding are the main source of quantum speedup for quantum
interior point methods, an algorithmic scheme that attempts to
accelerate classical interior point iterations using a quantum
computer for linear algebra. This is motivated by the fact that in
classical interior point methods, the most expensive step is the
solution of the (large and typically dense) Newton linear system in
every iteration
\cite{roos2005interior,terlaky2013interior,wright1997primal}.  The
idea of using quantum computers for linear algebra in the context of
interior point methods for semidefinite optimization was pioneered in
\cite{kerenidis2020quantum} and extended in
\cite{kerenidis2021quantum} for second-order cone programs, although
issues remained in showing convergence to a feasible solution in the
usual sense. Convergence was addressed in \cite{augustino2023quantum}
for semidefinite programs, using a reformulation of the Newton linear
system into orthogonal bases for the primal and dual space to ensure
reduction of the complementarity violation. Two major sources of
slowdown of this methodology are: the reliance on quantum state
tomography to extract the solution of the linear system, which incurs
$1/\epsilon$ scaling in the precision; and the linear dependence of
the QLSA on the condition number $\kappa$, which grows as we approach
optimality. The first source of slowdown is addressed using iterative
refinement, the second source is addressed with preconditioning or
additional iterative refinement for certain classes of problems, see
\cite{mohammadisiahroudi2025quantum,mohammadisiahroudi2024efficient,mohammadisiahroudi2025improvements,wu2023inexact}
as well as Sect.~\ref{sec:hhlir}.\index{iterative refinement}

For a discussion on iterative refinement in the context of classical
linear algebra, see \cite{saad2003iterative,wilkinson1963rounding}. In
classical optimization, iterative refinement is used in, e.g.,
\cite{gleixner2016iterative,weber2019solving}: both papers also offer
accessible introductions to the topic. On the quantum side, besides
the references on interior point methods cited above, a strategy based
on iterative refinement is used in \cite{chen2024quantum} within an
algorithm to approximate the top eigenvalues of a block-encoded
matrix.

\chapter{Quantum algorithms for SDP using mirror descent}
\label{ch:mmwu}
\thispagestyle{fancy}
Mirror descent is a powerful framework to solve nonlinear optimization
problems by taking advantage of the geometry of the space in which the
problem lives \cite{nemirovski1983problem}. Mirror descent relies on
the same concepts as projected (sub)gradient descent, with the
significant variation that the descent takes place in a ``mirror''
space, which is the vector space dual to the original space. When
dealing with optimization problems over the cone of positive
semidefinite matrices, it is possible to apply mirror descent using
the {\em quantum relative entropy} as the mirror map, resulting in an
iterative optimization scheme for semidefinite programming that
simultaneously operates in two spaces: the space of primal iterates,
which are positive semidefinite matrices expressed as matrix
exponentials, and the space of (vector) dual iterates (i.e., the
mirror space), which are the matrix logarithms of the primal
iterates. This choice of the mirror map is natural in the context of
quantum algorithms that want to take advantage of the ability of
quantum computers to efficiently compute Gibbs states, i.e., matrix
exponentials of a certain kind, see Sect.~\ref{sec:gibbssampling} and
in particular Prop.~\ref{prop:gibbsstateconstr}. Since the mirror
descent framework has proven very fruitful for the development of
quantum optimization algorithms, in this chapter we discuss its main
components, with an emphasis on those that (almost) directly lead to
quantum algorithms with an expectation of quantum advantage of some
type.

In the following we analyze two quantum algorithms for semidefinite
optimization that use the mirror descent framework. The first
algorithm, to which the majority of the chapter is devoted, is in fact
also an instantiation of the Multiplicative Weights Update (MWU)
algorithm, a meta-algorithm that has found many applications in
optimization \cite{arora2012multiplicative}; for some references on
the MWU, mostly of theoretical nature, see the notes at the end of
this chapter, Sect.~\ref{sec:mmwunotes}. The MWU algorithm is a method
to find an optimal strategy in a certain game against an adversary,
and can be thought of as an update rule for a probability
distribution. A specific version of the MWU algorithm can solve
semidefinite optimization problems (SDPs)
\cite{arora2005fast,arora2016combinatorial}, and that version is
equivalent to mirror descent with a specific choice of the mirror map.
The approach can be turned into a quantum algorithm for SDPs that
obtains a different running time tradeoff as compared to the classical
MWU algorithm. The second algorithm, based on \cite{brandao2022faster}
and originally described as an instantiation of the
matrix-exponentiated gradient updates approach of
\cite{tsuda2005matrix}, applies to the SDP relaxation of MaxCut and
other quadratic unconstrained binary optimization problems (as is
customary, we use the acronym ``SDP'' to refer to both a
``semidefinite optimization problem,'' and ``semidefinite
programming,'' i.e., semidefinite optimization: the context should
clarify any ambiguity). This second algorithm is also a mirror descent
approach with the same mirror map as the first algorithm.

Because this chapter is devoted to SDP\index{semidefinite programming}, we formally introduce the class
of optimization problems that we aim to solve. (The second algorithm
discussed in this chapter solves a restricted class of SDP: we
describe it in Sect.~\ref{sec:qubosdp}.) Given Hermitian matrices $C,
A^{(1)},\dots,A^{(m)} \in \C^{n \times n}$ and reals $b_1,\dots,b_m
\in \R$, we define the primal SDP problem as:
\begin{equation}
  \tag{P-SDP}
  \label{eq:psdp}
  \left.
  \begin{array}{rrcl}
    \max & \trace{CX} && \\
    \text{s.t.:} \quad \forall j=1,\dots,m & \trace{A^{(j)} X} &\le& b_j \\
    & X &\succeq& 0.
  \end{array}
  \right\}
\end{equation}
The corresponding dual is:
\begin{equation}
  \tag{D-SDP}
  \label{eq:dsdp}
  \left.
  \begin{array}{rrcl}
    \min & b^{\top} y && \\
    \text{s.t.:} & \sum_{j=1}^{m} y_j A^{(j)} - C&\succeq& 0 \\
    & y &\ge& \zeroes.
  \end{array}
  \right\}
\end{equation}
If strong duality holds, the optimal values of \eqref{eq:psdp} and
\eqref{eq:dsdp} are the same. Strong duality may not hold in general
but it holds under mild conditions, for example if Slater's condition
holds (the primal and dual have strictly feasible solutions)
\cite{boyd2}. For the general-purpose SDP solver described in
Sect.~\ref{sec:classicalmmwu}, we modify the problem in a suitable way
so that strictly feasible solutions exist, and this ensures that
strong duality holds; this is discussed in
Sect.~\ref{sec:mmwusdp}. Note that SDP generalizes linear programming:
a linear optimization problem is simply an SDP where all the matrices
$A^{(j)}, C$ are diagonal.

\section{The mirror descent framework}
\label{sec:mirror}
\emph{Mirror descent},\index{algorithm!mirror descent|(}\index{mirror descent!base algorithm|(} first introduced in
\cite{nemirovski1983problem}, is an algorithm akin to steepest descent
and its variants to solve continuous optimization problems. Here we
discuss mirror descent starting from projected subgradient descent;
one can think of projected subgradient descent as a natural variation
of steepest descent for constrained optimization (as opposed to
unconstrained optimization) of not-necessarily-differentiable
functions.  A reader not familiar with projected subgradient descent
should still be able to follow the discussion by relying on intuition
from the well-known steepest descent algorithm. The main difference
between mirror descent and steepest descent is that mirror descent
uses a \emph{mirror map} with the goal of adapting steepest descent to
the geometry of the space. If the mirror map is chosen appropriately,
mirror descent can yield an advantage over vanilla steepest descent,
for example by handling some difficult constraints in a simple way. In
the next section we introduce the most important concepts of the
mirror descent framework, but we do not give a fully detailed
description, and we skip many of the proofs; the notes in
Sect.~\ref{sec:mmwunotes} contain references for readers interested in
the details.

\subsection{Mirror descent as a generalization of steepest descent}
\label{sec:mirrorgen}
We start with a discussion of the mirror descent framework as a
generalization of projected subgradient descent. In this chapter we
repeatedly use the concepts of subgradients and approximate
subgradients, formally defined in Def.~\ref{def:subgradient}. Let us
consider the problem of minimizing a convex function $f(x)$ using an
iterative algorithm, with access to subgradients of the function. For
now we consider the unconstrained case; at the end of this section, we
discuss the case in which we want to minimize over a convex set
$K$. Let $x^{(t)}$ be the current iterate and $g^{(t)} \in \partial
f(x^{(t)})$ a subgradient of $f$ at $x^{(t)}$. It is well known that
we can construct an iterative algorithm to minimize $f$ by using the
following update rule:
\begin{equation}
  \label{eq:subgradupdateexplicit}
  x^{(t+1)} = x^{(t)} - \eta g^{(t)},
\end{equation}
i.e., we add to the current iterate a negative multiple of the
subgradient. If the function $f$ is differentiable, the subgradient
coincides with the gradient, and the above algorithm is usually called
\emph{steepest descent}. For a comprehensive discussion of
gradient-based methods, see \cite{bertsekasnlp}.

It is also well known that the explicit update rule in
Eq.~\eqref{eq:subgradupdateexplicit} is equivalent to the following
implicit rule, where the next iterate is expressed as the solution of
an optimization problem:
\begin{equation}
  \label{eq:subgradupdate}
  x^{(t+1)} = \arg \min_{x} \left\{\frac{1}{2} \nrm{x - x^{(t)}}^2 + \eta
    \dotp{g^{(t)}}{x-x^{(t)}}\right\}.
\end{equation}
The equivalence can be verified by taking the gradient with respect to
$x$ for the expression inside the $\min$ (the constant term $\eta
\dotp{g^{(t)}}{x^{(t)}}$ disappears), and setting it equal to zero
component-wise: this yields precisely
Eq.~\eqref{eq:subgradupdateexplicit}. Readers familiar with the
\emph{proximal operator}, defined as:
\begin{equation}
  \label{eq:prox}
  \text{prox}_f(x^{(t)}) := \arg \min_{x} \left\{\frac{1}{2} \nrm{x -
    x^{(t)}}^2 + \eta f(x)\right\},
\end{equation}
might recognize Eq.~\eqref{eq:subgradupdate} as a linearization of
$\text{prox}_f(x^{(t)})$. Comparing Eq.~\eqref{eq:prox} to
Eq.~\eqref{eq:subgradupdate}, the only difference is that the
objective function term $f(x)$ in $\text{prox}_f(x^{(t)})$ is replaced
with its linearization $f(x^{(t)}) + \dotp{g^{(t)}}{x-x^{(t)}}$ using
$g^{(t)} \in \partial f(x^{(t)})$; the constant term $f(x^{(t)})$ does
not affect the value of the minimizer of the expression, so the $\arg
\min$ is the same whether or not we add $\eta f(x^{(t)})$. Replacing
the squared Euclidean distance term in Eq.~\eqref{eq:prox} with more
general distance functions yields generalizations of the proximal
operator \cite{teboulle1992entropic}. The main motivation for doing so
is that one can sometimes fully eliminate some of the constraints of
the problem simply by appropriately choosing the distance
function. This statement may seem abstract at this point, so we give a
quick preview of how this fact is used in the remainder of this
chapter: by using a distance function called quantum relative entropy,
we will automatically ensure that all the iterates are positive
semidefinite matrices. This allows us to construct algorithms for
optimizing over the set of density matrices without having to
explicitly handle positive semidefiniteness and trace
normalization. (The trace normalization is taken care of by projection
onto a feasible set, still using the same distance function.)

Let us now get back to the derivation of mirror descent from projected
subgradient descent. To summarize the above discussion: the implicit
update in Eq.~\eqref{eq:subgradupdate} corresponds to minimizing a
weighted combination of two terms. The first term is the distance from
the current iterate $x^{(t)}$. The second term is a linear
approximation of the objective function obtained using a subgradient
at $x^{(t)}$, $g^{(t)} \in \partial f(x^{(t)})$:
\begin{equation*}
  f(x^{(t)}) + \dotp{g^{(t)}}{x - x^{(t)}}.
\end{equation*}
If we were to minimize the linear approximation term only, we would
move indefinitely in the direction opposite to a subgradient at
$x^{(t)}$, because this is an unconstrained problem. This, however, is
not a good idea: the linear approximation at $x^{(t)}$ is unlikely to
be accurate once we move far away from $x^{(t)}$. For a
non-differentiable function, the subgradient may not give a descent
direction at all: the objective function can sometimes increase
if we add a negative multiple of the subgradient to $x^{(t)}$, see
Ex.~\ref{ex:poorsubg}. 
\begin{example}
  \label{ex:poorsubg}
  Consider the univariate function $f(x) = |x|$ and let $x^{(1)} = 0$
  be our current iterate. The scalar $1$ is a subgradient of $f(x)$ at
  $x^{(1)}$, but adding \emph{any} multiple of the subgradient to
  $x^{(1)}$ increases the objective function value.
\end{example}

\noindent For a differentiable function, although the gradient gives a
descent direction, we know from Taylor's theorem that the error of the
linear approximation depends on the distance from $x^{(t)}$, hence the
approximation may lead us astray when we are far from $x^{(t)}$.
Adding a penalty for increasing the distance from the current iterate
$x^{(t)}$ ensures that we stay relatively close to it, which is a
desirable feature of the descent scheme for the reason discussed
above: by including the penalty term, hopefully we do not go too far
along poor directions, as might occur if the linear approximation is
inaccurate.

Now let us generalize Eq.~\ref{eq:subgradupdate}. Rather than use the
Euclidean distance function, we measure proximity with the
\emph{Bregman divergence}, which estimates the error between the value
of a certain \emph{mirror map}, and a linearization of the mirror map
at a given point. Because the iterate is chosen by optimizing a
problem involving a linear approximation of the objective function, we
intuitively expect the quality of the iterates to depend on how good
such an approximation is; thus, it is reasonable to use a distance
function that estimates the quality of the linear approximation of a
function --- in this case, the mirror map. Potentially, the
Bregman divergence might help us determine how far we can go from the
current iterate before our model for the objective function (the
linear approximation) becomes too inaccurate to be useful. The mirror
map is usually chosen in such a way that the Bregman divergence is
easy to compute, with the goal of achieving or maintaining
computational tractability.
\begin{definition}[Bregman divergence]
  \label{def:bregmandiv}
  Given a continuously differentiable and strictly convex function $h
  : \R^d \to \R$, called \emph{mirror map}, and two points $x, y \in
  \R^d$, the \emph{Bregman divergence}\index{Bregman divergence} from
  $x$ to $y$ is defined as:
  \begin{equation*}
    D_h(y \| x) := h(y) - h(x) - \dotp{\nabla h(x)}{y-x}.
  \end{equation*}
\end{definition}
Equipped with this notion, a natural generalization of
Eq.~\eqref{eq:subgradupdate} is:
\begin{equation}
  \label{eq:subgradupdatebreg}
  x^{(t+1)} = \arg \min_{x} \left\{D_h(x \| x^{(t)}) + \eta
    \dotp{g^{(t)}}{x-x^{(t)}}\right\},
\end{equation}
where we are using $D_h(x \| x^{(t)})$ instead of the squared Euclidean
distance between $x^{(t)}$ and $x$. By taking the gradient of the
expression inside the $\min$ and setting it equal to zero, we find:
\begin{equation*}
  \eta g^{(t)} + \nabla h(x) - \nabla h(x^{(t)}) = \zeroes,
\end{equation*}
so if $\nabla h$ is invertible, the solution to
Eq.~\ref{eq:subgradupdatebreg} leads to the following update rule:
\begin{equation}
  \label{eq:mirrorstep}
  x^{(t+1)} = (\nabla h)^{-1}\left(\nabla h(x^{(t)}) -\eta g^{(t)}\right).
\end{equation}
This expression can be interpreted as follows. To find the new iterate
$x^{(t+1)}$, we map the current iterate $x^{(t)}$ to dual space using
$\nabla h$, and we take a descent step using the subgradient
$g^{(t)}$. Then, we map back to primal space using $(\nabla h)^{-1}$:
this gives us the new point. We say that $x^{(t)}$ is mapped to a dual
space using $\nabla h$ because the gradient of a function lives in the
dual vector space. Note that the update rule in
Eq.~\ref{eq:mirrorstep} is derived by solving the optimization problem
in Eq.~\ref{eq:subgradupdatebreg} over the entire space: if we are
interested in minimizing $f(x)$ over some convex set $K$, once we map
back to primal space we must also project onto $K$ using the same
divergence $D_h$ as the distance function. This is similarly derived
starting from the projected (sub)gradient descent update rule:
\begin{equation}
  \label{eq:subgradupdateconstr}
  x^{(t+1)} = \arg \min_{x \in K} \left\{\frac{1}{2} \nrm{x - x^{(t)}}^2 + \eta
  \dotp{g^{(t)}}{x-x^{(t)}}\right\} = \arg \min_{x \in K} \left\{ \frac{1}{2} \nrm{ x - (x^{(t)} - \eta g^{(t)}) }^2 \right\}
\end{equation}
(the expressions inside the two $\arg \min$ are equivalent up to terms
that do not depend on $x$, so their $\arg \min$ is the same), which
generalizes to:
\begin{equation}
  \label{eq:subgradupdatebregconstr}
  x^{(t+1)} = \arg \min_{x \in K} \left\{D_h(x \| x^{(t)}) + \eta
    \dotp{g^{(t)}}{x-x^{(t)}}\right\} = \arg \min_{x \in K} \left\{D_h(x \| (\nabla h)^{-1}(\nabla h(x^{(t)}) -\eta g^{(t)}))\right\}
\end{equation}
when replacing the squared Euclidean distance with the Bregman
divergence. \index{mirror descent!base algorithm|)}

\subsection{Online mirror descent and the entropy mirror map}
\label{sec:mirroronline}
The mirror descent\index{mirror descent!online|(} algorithm can easily be turned into an online
algorithm, where the objective function is a summation of terms that
are discovered one at a time. Suppose we have $T$ convex functions
$f_1,\dots,f_T$, and let $x^\ast \in \arg \min \sum_{t=1}^T f_t(x)$. We
want to choose a sequence $x^{(1)},\dots,x^{(T)}$ that minimizes the
\emph{regret} with respect to the best (single) solution in hindsight:
\begin{equation*}
  \min_{x^{(1)},\dots,x^{(T)}} \sum_{t=1}^T f_t(x^{(t)}) - \sum_{t=1}^T f_t(x^\ast),
\end{equation*}
with the additional caveat that $f_t$ is revealed at iteration $t$
only after $x^{(t)}$ is determined, and therefore the choice $x^{(t)}$
can only depend on the already-seen terms $f_1,\dots,f_{t-1}$ as well
as previous iterates. Let $g^{(t)} \in \partial f_t(x^{(t)})$. We can
directly use Eq.~\ref{eq:subgradupdateconstr}, leading to the update
rule:
\begin{equation*}
  x^{(t+1)} = \arg \min_{x \in K} \left\{ \frac{1}{2} \nrm{ x - (x^{(t)} - \eta g^{(t)}) }^2 \right\} = \text{Proj}_{K} \left(x^{(t)} - \eta g^{(t)}\right),
\end{equation*}
and replacing the Euclidean distance with the Bregman divergence
yields Eq.~\ref{eq:subgradupdatebregconstr}, which we restate below:
\begin{equation}
  \label{eq:mirrorstepconst}
  x^{(t+1)} = \arg \min_{x \in K} \left\{D_h(x \| (\nabla h)^{-1}(\nabla h(x^{(t)}) -\eta g^{(t)}))\right\} = \text{Proj}_{K}^{D_h}\left((\nabla h)^{-1}(\nabla h(x^{(t)}) -\eta g^{(t)})\right).
\end{equation}
In the two above equations, we use $\text{Proj}_K$ to emphasize that
$x^{(t+1)}$ is simply the projection onto $K$ of the unconstrained
iterate: using the Euclidean norm for the standard projected
subgradient descent approach, and using the Bregman divergence $D_h$,
denoted $\text{Proj}_{K}^{D_h}$, for the more general mirror descent
approach. It is immediate to verify that the value of $x^{(t+1)}$ is
determined using the subgradient of $f_t$ but not the subgradient of
$f_{t+1}$, so this update rule can be applied to the online setting:
at time $t$, the iterate $x^{(t)}$ is chosen by combining $x^{(t-1)}$
and a subgradient of $f_{t-1}$, so we are not using the
yet-to-be-revealed term $f_t$.
\begin{remark}
  An algorithm for the online setting can be applied to the offline
  setting by letting $f_t = f$ for all $t$, and considering the
  average of the iterates $\frac{1}{T} \sum_{t=1}^{T} x^{(t)}$. If the
  regret is asymptotically smaller than $T$, the regret bound can be
  directly applied to get convergence to a desired error tolerance:
  this can be verified with straightforward algebraic manipulations.
\end{remark}
The convergence of mirror descent in the online setting is well
known. To formally state a convergence result, we need the concept of
dual norm, because the (sub)gradients of $f_t$ live in the dual vector
space.
\begin{definition}[Dual norm]
  \label{def:dualnorm}
  Given a vector space $V$ with inner product $\dotp{\cdot}{\cdot}$
  and norm $\nrm{\cdot}$, the \emph{dual norm}\index{norm!dual} on $V^\ast$ is defined
  as:
  \begin{equation*}
    \nrm{y}_{\ast} := \max_{x : \nrm{x} = 1} \dotp{y}{x},
  \end{equation*}
  for any $y \in V^\ast$.
\end{definition}
\begin{theorem}[Convergence of mirror descent; based on Thm.~4.2 in \cite{bansal2019potential} and Thm.~4.1 in \cite{beck2003mirror}]
  \label{thm:mirrorconv}
  Let $h$ be $\alpha$-strongly convex. The mirror descent algorithm
  with step size $\eta$, starting at the point $x^{(1)}$, produces a
  sequence $x_2,\dots,x_T$ with subgradients $g^{(t)} \in \partial
  f_t(x^{(t)})$ such that:
  \begin{equation*}
    \sum_{t=1}^T f_t(x^{(t)}) - \sum_{t=1}^T f_t(x^\ast) \le \frac{1}{\eta} D_h(x^\ast \| x^{(1)}) + \frac{\eta}{2\alpha} \sum_{t=1}^T \nrm{g^{(t)}}_{\ast}^2.
  \end{equation*}
\end{theorem}

\paragraph{Mirror descent over the set of density matrices.}
We now describe a specific instantiation of the online mirror descent
framework, that we use twice in the rest of this chapter and that
serves as the basis for obtaining quantum algorithms. Let ${\cal
  S}^n_{+,1}$ be the set of density matrices, i.e., positive
semidefinite matrices with unit trace (see
Sect.~\ref{sec:mixedstate}), of size $n \times n$. We apply the
framework to problems of this form:
\begin{equation*}
  \min_{\rho^{(1)},\dots,\rho^{(T)} \in {\cal S}^n_{+,1}} \sum_{t=1}^T
  f_t\left(\rho^{(t)}\right),
\end{equation*}
where the feasible set is $K = {\cal S}^n_{+,1}$. Recall that $f_1$ is
revealed only after $x^{(1)}$ is chosen; thus, $x^{(1)}$ is chosen
independently of the objective function, and can be taken to be a
given initial point; we set $\rho^{(1)} = I/n$. (Throughout this
chapter, to shorten notation we denote by $I$ the $n \times n$ density
matrix, i.e., $I = I_{n \times n}$.) We assume we are able to compute
$G^{(t)} \in \partial f_t(\rho^{(t)})$; we denote it with a capital
letter because now, given that $\rho^{(t)}$ is a matrix, the subgradient
is a matrix as well. We use the update rule in
Eq.~\ref{eq:mirrorstep}, where our iterate $x^{(t)}$ should now be
interpreted as the vectorization of the matrix $\rho^{(t)}$, and the
inner product is the trace inner product $\dotp{A}{B} :=
\trace{AB^{\dag}}$.
\begin{remark}
  \label{rem:traceherm}
  Considering how we defined the inner product in Ch.~\ref{ch:intro},
  it would be more consistent to define $\dotp{A}{B} =
  \trace{A^{\dag}B}$; typically, in engineering it is common to apply
  the $\dag$ to the first argument of the inner product. However, in
  this chapter the definition $\dotp{A}{B} := \trace{AB^{\dag}}$ makes
  the notation considerably less cumbersome, because for the cases of
  interest here, the matrix $B$ is always Hermitian, so $\dotp{A}{B} =
  \trace{AB}$. In the literature, depending on the field, both
  definitions are used. It is possible to use the definition
  $\dotp{A}{B} = \trace{A^{\dag}B}$ throughout this chapter and obtain
  the same results. (In fact, for the cases where we use the inner
  product, the matrix $A$ is also Hermitian --- even though it is not
  required to be so --- and changing the definition would not impact
  the conclusions.)
\end{remark}
We need an appropriate mirror map. The von Neumann negative
entropy\index{Bregman divergence}:
\begin{equation*}
  h(\rho) = \trace{\rho \log \rho - \rho}
\end{equation*}
is strictly convex and satisfies the conditions of
Def.~\ref{def:bregmandiv}, leading to a divergence $D_h$ that is known
as the \emph{quantum relative entropy}\index{quantum!relative entropy} \cite{nielsen02quantum}:
\begin{equation*}
 D_h(\rho \| \sigma) = \trace{\rho \log \rho - \rho \log \sigma - \rho
   + \sigma},
\end{equation*}
which simplifies to $\trace{\rho \log \rho - \rho \log \sigma}$ if
$\rho, \sigma$ are density matrices.  Plugging $h$ and $D_h$ in
Eq.~\ref{eq:mirrorstepconst}, using the fact that $\nabla h (\rho) =
\log \rho$ and $(\nabla h)^{-1}(M) = \exp(M)$, the mirror descent
algorithm follows these steps at iteration $t$: given the current
iterate $\rho^{(t)}$,
\begin{itemize}
\item we use $\nabla h$ to map $\rho^{(t)}$ to dual space, i.e., we
  compute its matrix logarithm (which yields $-\eta
  \sum_{\tau=1}^{t-1} G^{(\tau)}$ up to normalization, see below for a
  more detailed analysis);
\item we add add a multiple of the subgradient, specifically $-\eta
  G^{(t)}$, obtaining the new iterate in dual space;
\item we use the inverse $(\nabla h)^{-1}$ to map back to primal
  space, i.e., we apply the matrix exponential;
\item we project the candidate solution onto the set of density
  matrices, by normalizing its trace so that it has unit trace.
\end{itemize}
The steps of the algorithm are depicted in Fig.~\ref{fig:mdsteps}.
\begin{figure}[htb]
  \center
  \ifcompilefigs
  \begin{tikzpicture}
    \draw[very thick] (-8, -1) arc(130:-130:2.5cm and 1.6cm);
    \draw[very thick, rotate = -45] (4,0) arc(65:270:3cm and 2cm);
    \node[circle,
      draw, thick,
      text = blue,
      fill = green!20, 
      minimum size = 2.5cm] (p) at (0,-2.25) {}; 
    \draw (.8,-2.35) node {${\cal S}_{+,1}^n$};

    \draw (-.2,-1.5) node {$\bullet$};
    \draw (-.2,-1.5) coordinate (xt) node[right] {$\rho^{(t)}$};

    \draw (-.5,-2.85) node {$\bullet$};
    \draw (-.5,-2.85) coordinate (xt1) node[above] {$\rho^{(t+1)}$};
    
    \draw (-4.5,-1.6) node {$\bullet$};
    \draw (-4.5,-1.6) coordinate (Pxt) node[left] {$\log \rho^{(t)}$};

    \draw (-5,-3) node {$\bullet$};
    \draw (-5,-3) coordinate (yt) node[left] {$\log \rho^{(t)} - \eta G^{(t)}$};

    \draw (-0.75,-4.2) node {$\bullet$};
    \draw (-0.75,-4.2) coordinate (txt) node[right] {$$};

    \draw[->,  thick, shorten >=3pt](xt) to [bend right=45] (Pxt) node[above, midway, xshift=-2.45cm, yshift=-.7cm]{$\nabla h = \log$};
    
    \draw[->, thick, shorten >=3pt](yt) to [bend right=45] (txt) node[below, xshift=-1.5cm, yshift=-.5cm]{$\left(\nabla h \right)^{-1} = \exp$};

    \draw[->, thick, shorten >=3pt](Pxt) to (yt) node[left, yshift=.75cm]{subgradient step};
    
    \draw[->, dashed, thick, shorten >=3pt](txt) to [bend right=15] (xt1) node[right, yshift=-1.cm]{projection};
    
    \draw (1,-4.5) coordinate (txtq) node[right] {PSD matrices};
    \draw (-8,-4.2) coordinate (txtq) node[right] {Hermitian matrices};

  \end{tikzpicture}
  \else
  \includegraphics{figures/mdsteps.pdf}
  \fi
  \caption{Steps of the mirror descent algorithm using the von Neumann negative entropy as the mirror map.}
  \label{fig:mdsteps}  
\end{figure}

We can simplify the algorithm further by keeping track of the iterate
in the dual space, i.e., in its matrix logarithm form, and noting
that the iterates $\rho^{(t)}$ are Gibbs states
(Def.~\ref{def:gibbsstate})\index{Gibbs!state}. Below, $\allzeroes^{n \times n}$ denotes
the all-zero matrix of size $n \times n$.
\begin{proposition}
  \label{prop:mirrorhamiltonian}
  For a sequence of Hermitian matrices $\ham^{(1)},\ham^{(2)},\dots$, define:
  \begin{equation*}
    \rho^{(t)} := \exp(\ham^{(t)})/\trace{\exp(\ham^{(t)})},
  \end{equation*}
  i.e., $\rho^{(t)}$ is the Gibbs state corresponding to the
  Hamiltonian $\ham^{(t)}$. If we let:
  \begin{equation*}
    \begin{split}
      \ham^{(1)} &= \allzeroes^{n \times n}, \\
      \ham^{(t+1)} &= \ham^{(t)} - \eta G^{(t)},
    \end{split}
  \end{equation*}
  then the sequence $\rho^{(1)},\rho^{(2)},\dots$ coincides with the
  iterates of the mirror descent algorithm using the von Neumann
  negative entropy as the mirror map, and $I/n$ as the initial point.
\end{proposition}
\begin{proof}
  By induction. For $t=1$, $\rho^{(1)} = \exp(\allzeroes^{n \times
    n})/\trace{ \exp(\allzeroes^{n \times n})} = I/n$, which is the
  starting point of the mirror descent algorithm. Now suppose the
  inductive hypothesis holds for some $t$; then we have:
  \begin{equation*}
    \log \rho^{(t)} = \log\left(\exp(\ham^{(t)}) / \trace{\exp(\ham^{(t)})}
    \right) = \ham^{(t)} - \log \trace{\exp(\ham^{(t)})} I.
  \end{equation*}
  Exploiting the fact that $\nabla h$ is well-defined over ${\cal
    S}^n_{+,1}$, and its inverse is also well-defined for all
  Hermitian matrices, Eq.~\eqref{eq:mirrorstepconst} has the following
  explicit solution, see \cite{tsuda2005matrix}:
  \begin{equation}
    \label{eq:mdupdateexp}
    \rho^{(t+1)} = \frac{\exp(\log \rho^{(t)} - \eta G^{(t)})}{\trace{\exp(\log \rho^{(t)} - \eta G^{(t)})}}  \text{ where } G^{(t)} \in \partial f_t(\rho^{(t)}).
  \end{equation}
  Intuitively, this can be verified by following the step-by-step
  description of the algorithm just before the proposition
  statement. Then:
  \begin{align*}
    \rho^{(t+1)} &= \frac{\exp(\log \rho^{(t)} - \eta G^{(t)})}{\trace{\exp(\log \rho^{(t)} - \eta G^{(t)})}} \\
    &= \frac{\exp(\ham^{(t)} - \log \trace{\exp(\ham^{(t)})} I - \eta G^{(t)})}{\trace{\exp(\ham^{(t)} - \log \trace{\exp(\ham^{(t)})} I - \eta G^{(t)})}} \\
    &= \frac{\exp(\ham^{(t)} - \eta G^{(t)})}{\trace{\exp(\ham^{(t)} - \eta G^{(t)})}} = \frac{\exp(\ham^{(t+1)})}{\trace{\exp(\ham^{(t+1)})}},
  \end{align*}
  where the third equality is due to the fact that $\exp(\ham+\lambda
  I)/\trace{\exp(\ham+\lambda I)} = \exp(\ham)/\trace{\exp(\ham)}$ for all
  $\lambda \in \mathbb{R}$, so we can eliminate the term $- \log
  \trace{\exp(\ham^{(t)})} I$ appearing at the numerator and denominator.
\end{proof}

\noindent Using the expression for the matrices $\ham^{(t)}$ in
Prop.~\ref{prop:mirrorhamiltonian}, in Alg.~\ref{alg:mmwu} we give the
pseudocode for the mirror descent algorithm with the von Neumann
negative entropy as the mirror map.
\begin{algorithm2e}[htb]
  \SetAlgoLined
  \LinesNumbered
\KwIn{Parameter $\eta \le 1$, number of rounds $T$, dimension $n$.} 
\KwOut{Sequence of density matrices $\rho^{(1)},\dots,\rho^{(t)} \in {\cal S}^{n}_{+,1}$.}
\textbf{Initialize}: $\ham^{(1)} \leftarrow \allzeroes^{n \times n}$.\\
\For{$t=1,\dots,T-1$}{
  Compute $\rho^{(t)} = \exp(\ham^{(t)})/\trace{\exp(\ham^{(t)})}$.\\
  Obtain subgradient $G^{(t)} \in \partial f_t(\rho^{(t)})$.\\
  Compute $\ham^{(t+1)} = \ham^{(t)} - \eta G^{(t)}$, leading to the explicit update rule:
  \begin{equation*}
    \rho^{(t+1)} = \exp\left(- \eta \sum_{\tau=1}^{t} G^{(\tau)}\right) / \trace{\exp\left(- \eta \sum_{\tau=1}^{t} G^{(\tau)}\right)}. 
  \end{equation*}

}
\Return $\rho^{(1)},\dots,\rho^{(T)}$.
\caption{Online mirror descent with the von Neumann negative entropy as the mirror map. Also known as Matrix Multiplicative Weights Update (MMWU) algorithm.}
\label{alg:mmwu}
\end{algorithm2e}
The convergence of Alg.~\ref{alg:mmwu} follows directly from
Thm.~\ref{thm:mirrorconv}. This gives a regret bound of:
\begin{equation*}
  \frac{\log n}{\eta} + \frac{\eta}{2} \sum_{t=1}^T \nrm{G^{(t)}}_{\ast}^2,
\end{equation*}
because $D_h(\rho \| I/n) = \log n - \sum_{j} \lambda_j(\rho) \log
\frac{1}{\lambda_j(\rho)} \le \log n$ for every $\rho$, where
$\lambda_j(\rho)$ are the eigenvalues of $\rho$ (see, e.g.,
\cite{tsuda2005matrix}), and the strong convexity parameter $\alpha$
is $1$ \cite{yu2013strong}. Alg.~\ref{alg:mmwu} is also known as the
Matrix Multiplicative Weights Update (MMWU) algorithm: we discuss this
interpretation in Sect.~\ref{sec:mmwumirror}, where we additionally
prove the convergence of the algorithm (obtaining a result akin to
Thm.~\ref{thm:mirrorconv}) from first principles.

\section{Classical MMWU algorithm for SDP}
\label{sec:classicalmmwu}
As mentioned at the beginning of this chapter, the Multiplicative
Weights Update (MWU)\index{matrix!multiplicative weights update|(}\index{algorithm!multiplicative weights update|(}\index{algorithm!MWU|see{multiplicative weights update}} algorithm is a meta-algorithm; here, we are
interested in its application to semidefinite optimization. Rather
than describing the traditional MWU algorithm, we directly proceed
with a description of the \emph{matrix} MWU (MMWU) algorithm, which is
the relevant framework for the quantum algorithms that constitute the
central topic of this chapter.

\subsection{From mirror descent to the MMWU algorithm}
\label{sec:mmwumirror}
The MMWU algorithm can be derived from the mirror descent framework in the
context of a certain two-player game; the connection to the solution
of \eqref{eq:psdp} is not obvious from the definition of the game, but
we make it explicit in Sect.~\ref{sec:mmwusdp}. The game is defined as
follows. It is a two-player game; in each round $t$ we choose a
density matrix $\rho^{(t)}$, and an adversary chooses a 
matrix $M^{(t)}$ satisfying $\nrm{M^{(t)}} \le 1$. The matrix
$\rho^{(t)}$ is allowed to depend on the adversary's choices in
previous rounds $M^{(1)},\dots,M^{(t-1)}$, but not in the current
round; i.e., we choose $\rho^{(t)}$ before $M^{(t)}$ is revealed. At
the end of each round we pay $\trace{M^{(t)} \rho^{(t)}}$, and our
objective is to pay as little as possible over $T$ rounds, i.e.,
minimize the total loss accumulated over $T$ rounds.

Formally, the problem of determining an optimal strategy for this game
can be formulated as follows. We want to choose
$\rho^{(1)},\dots,\rho^{(T)} \in {\cal S}^n_{+,1}$ so that:
\begin{equation}
  \label{eq:mirrorobjfun}
  \sum_{t=1}^T f_t(\rho^{(t)}) := \sum_{t=1}^T \trace{M^{(t)}
    \rho^{(t)}}
\end{equation}
is minimized, with the restriction that the density matrices
$\rho^{(t)}$ must be chosen sequentially and each matrix $M^{(t)}$,
and therefore the function $f_t$, is revealed after we choose the
corresponding $\rho^{(t)}$, so $\rho^{(t)}$ can only depend on
$M^{(1)},\dots,M^{(t-1)}$ and on previous iterates. This is exactly
the setting for online mirror descent using the von Neumann negative
entropy as the mirror map, discussed in Sect.~\ref{sec:mirroronline}.
We therefore have an algorithm to attain low regret, i.e., a strategy
that performs relatively well compared to the best single solution in
hindsight (called $x^\ast$ in Sect.~\ref{sec:mirroronline}). The best
single density matrix (i.e., round-independent choice) in hindsight is
the rank-1 matrix corresponding to the unit eigenvector $v$ of
$\sum_{t=1}^{T} M^{(t)}$ with the smallest eigenvalue: indeed,
\begin{equation*}
  \trace{\sum_{t=1}^{T} M^{(t)} v v^{\top}} =
  \lambda_{\min}\left(\sum_{t=1}^{T} M^{(t)}\right),
\end{equation*}
where $\lambda_{\min}\left(\sum_{t=1}^{T} M^{(t)}\right)$ denotes the
smallest eigenvalue, and it is immediate to observe that the objective
function value of any given density matrix --- if the same matrix is
chosen in every round --- is at least as large as
$\lambda_{\min}\left(\sum_{t=1}^{T} M^{(t)}\right)$.  Of course, in
general we cannot hope to construct a solution that attains value
$\lambda_{\min}\left(\sum_{t=1}^{T} M^{(t)}\right)$, because to
determine $v v^{\top}$ we would need to know all the matrices
$M^{(t)}$ in advance. Thus, we attempt to minimize (or at least bound)
the regret, which for this problem is:
\begin{equation*}
  \sum_{t=1}^T \trace{M^{(t)} \rho^{(t)}} -
  \lambda_{\min}\left(\sum_{t=1}^{T} M^{(t)}\right).
\end{equation*}
This ensures that the performance attained at the task at hand is always
satisfactory in some sense.
\begin{remark}
  In the game described above we are allowed to play a different
  $\rho^{(t)}$ in every round, so if we knew what the opponent is
  about to play, it would be optimal to choose $\rho^{(t)}$ such that
  $\trace{M^{(t)} \rho^{(t)}} = \lambda_{\min}(M^{(t)})$ in every
  round: this minimizes the objective function contribution $f_t$ for
  each $t$. However, as usual in the context of two-player games, we
  assume that we do not know the opponent's play in advance, and we
  want a strategy that performs well even if the opponent plays
  optimally against us.
\end{remark}

We apply Alg.~\ref{alg:mmwu}, as described in
Sect.~\ref{sec:mirroronline}: the feasible set is $K = {\cal
  S}^n_{+,1}$, and for the initial point we take $\rho^{(1)} =
I/n$. At the end of each iteration $t$ we are presented with the term
$M^{(t)}$ that specifies $f_t$ in the objective function, and note
that $M^{(t)} \in \partial f_t(\rho^{(t)})$ because $f_t(\rho^{(t)}) =
\trace{M^{(t)} \rho^{(t)}}$ is linear in $\rho^{(t)}$ with respect to
the trace inner product.
\begin{remark}
  \label{rem:tracelinearity}
  It is helpful to think of the iterates $\rho^{(t)}$ as vectors,
  obtained by vectorizing the corresponding matrices, i.e., taking the
  columns of the matrix and stacking them on top of each other to
  obtain a vector. With this transformation in mind, it is easy to see
  why $f_t(\rho^{(t)}) = \trace{M^{(t)} \rho^{(t)}} =
  \dotp{M^{(t)}}{(\rho^{(t)})^{\dag}}$ is linear: the trace inner
  product is equal to the standard inner product on complex spaces
  (i.e., $\dotp{a}{b} = b^{\dag} a = \sum_{j} a_j \bar{b}_j$) between
  the vectorizations of $M^{(t)}$ and $\rho^{(t)}$ --- recall
  Rem.~\ref{rem:traceherm}. Because the second argument of the inner
  product is a density matrix and therefore Hermitian, we can drop the
  $\dag$: $\trace{M^{(t)} \rho^{(t)}} = \dotp{M^{(t)}}{\rho^{(t)}}$.
\end{remark}
We use the update rule in Eq.~\ref{eq:mdupdateexp}. By following
Alg.~\ref{alg:mmwu}, at every iteration $t$ we choose a properly
normalized version of $\exp\left(- \eta \sum_{\tau=1}^{t-1}
M^{(\tau)}\right)$ for some parameter $0 < \eta \le 1$; the
normalization ensures that we output matrices with unit trace, which
is a requirement of the game because we are only allowed to play density
matrices. Alg.~\ref{alg:mmwu} is usually called ``MMWU algorithm'' in
this context. Although we can derive a regret bound using
Thm.~\ref{thm:mirrorconv}, see the discussion in
Sect.~\ref{sec:mirroronline} as well as the proof in
\cite{tsuda2005matrix} using the quantum relative entropy as a
potential function, in Thm.~\ref{thm:mmwuregret} we provide a tailored
and self-contained proof. The main purpose of the proof is to showcase
some helpful inequalities and techniques for handling similar cases.
\begin{theorem}[Regret bound of MMWU algorithm; Thm.~3.1 in \cite{arora2016combinatorial}]
  \label{thm:mmwuregret}
  For any sequence of loss matrices $M^{(1)},\dots,M^{(T)}$ with
  $\nrm{M^{(t)}} \le 1$, suppose we run Alg.~\ref{alg:mmwu} setting
  $f_t(\rho) = \trace{M^{(t)} \rho}$ with subgradients $M^{(t)} \in
  \partial f_t(\rho^{(t)})$. Then, the algorithm generates density
  matrices $\rho^{(1)},\dots,\rho^{(T)}$ such that:
  \begin{equation}
    \label{eq:mmwuregret}
    \trace{\sum_{t=1}^{T} M^{(t)} \rho^{(t)}} \le
    \lambda_{\min}\left(\sum_{t=1}^{T} M^{(t)}\right) + \eta \trace{\sum_{t=1}^{T} (M^{(t)})^2 \rho^{(t)}} + \frac{\ln n}{\eta}.
  \end{equation}
\end{theorem}
\begin{proof}
  For the main chain of inequalities in the proof, we need the
  Golden-Thompson inequality:
  \begin{equation}
    \label{eq:goldenthompson}
    \trace{\exp(A + B)} \le \trace{\exp(A) \exp(B)},
  \end{equation}
  as well as the following inequality:
  \begin{equation}
    \label{eq:expquadub}
    \exp(-A) \preceq (I - A + A^2),
  \end{equation}
  which holds for every Hermitian matrix $A$ with $\nrm{A} \le 1$
  because the corresponding scalar inequality $\exp(-a) \le 1 -a +a^2$
  also holds for every $|a| \le 1$, and diagonalizing the matrix $A$
  shows that Eq.~\eqref{eq:expquadub} holds.
  
  To simplify the calculations, it is convenient to define
  \begin{equation*}
    W^{(t)} := \exp\left(- \eta \sum_{\tau=1}^{t-1} M^{(\tau)}\right),
  \end{equation*}
  and note that:
  \begin{equation}
    \label{eq:wtdenormalized}
    W^{(t)} = \trace{W^{(t)}} \rho^{(t)}.
  \end{equation}
  Thus, $W^{(t)}$ is just a de-normalized version of the density
  matrices constructed by Alg.~\ref{alg:mmwu}, where we only keep the
  numerator and therefore these matrices do not necessarily have unit
  trace.

  We now proceed to the core of the proof. To prove the desired
  result, we define a potential function $\Phi(W^{(t))})$, and derive
  upper and lower bounds to its value as $t$ increases. Combining the
  bounds for $t = T+1$ yields the expression that we aim to obtain. We
  define the potential function as: $$\Phi(W^{(t))}) :=
  \trace{W^{(t)}}.$$ We have: \begingroup \allowdisplaybreaks
  \begin{align*}
    \Phi(W^{(t+1)}) &= \trace{\exp\left(- \eta \sum_{\tau=1}^{t} M^{(\tau)}\right)} \\
    &\le \trace{\exp\left(- \eta \sum_{\tau=1}^{t-1} M^{(\tau)}\right) \exp(-\eta M^{(t)})} && \text{(by \eqref{eq:goldenthompson})}\\
    &= \trace{W^{(t)} \exp(-\eta M^{(t)})} \\
    &\le \trace{W^{(t)} (I - \eta M^{(t)} + \eta^2 (M^{(t)})^2} && \text{(by \eqref{eq:expquadub})}\\
    &= \trace{W^{(t)}} (1 - \eta \trace{M^{(t)} \rho^{(t)}} + \eta^2 \trace{(M^{(t)})^2 \rho^{(t)}}) && \text{(by \eqref{eq:wtdenormalized})}\\
    &\le \Phi(W^{(t)}) \exp\left(-\eta \trace{M^{(t)} \rho^{(t)}} + \eta^2 \trace{(M^{(t)})^2 \rho^{(t)}} \right). && \text{(using } e^x \ge 1+x \text{)}
  \end{align*}
  \endgroup
  We can now recursively apply
  \begin{equation*}
    \Phi(W^{(t+1)}) \le \Phi(W^{(t)}) \exp\left(-\eta \trace{M^{(t)} \rho^{(t)}} + \eta^2 \trace{(M^{(t)})^2 \rho^{(t)}} \right),
  \end{equation*}
  expanding the r.h.s.\ down to $t=1$, and use $\Phi(W^{(1)}) =
  \Phi(I) = n$ to obtain:
  \begin{equation*}
    \Phi(W^{(T+1)}) \le n \exp\left(-\eta \sum_{t=1}^{T} \trace{M^{(t)} \rho^{(t)}} + \sum_{t=1}^{T} \eta^2 \trace{(M^{(t)})^2 \rho^{(t)}} \right).
  \end{equation*}
  We also have:
  \begin{equation*}
   \Phi(W^{(T+1)}) =  \trace{\exp\left(- \eta \sum_{\tau=1}^{T} M^{(\tau)}\right)} \ge \exp\left( \lambda_{\min}\left(- \eta \sum_{\tau=1}^{T} M^{(\tau)}\right) \right),
  \end{equation*}
  because the trace is the sum of the eigenvalues. Combining the lower bound and upper bound for $\Phi(W^{(T+1)})$, we obtain:
  \begin{align*}
    \exp\left( \lambda_{\min}\left(- \eta \sum_{\tau=1}^{T} M^{(\tau)}\right) \right)
    \le n \exp\left(-\eta \sum_{t=1}^{T} \trace{M^{(t)} \rho^{(t)}} + \sum_{t=1}^{T} \eta^2 \trace{(M^{(t)})^2 \rho^{(t)}} \right).
  \end{align*}
  Taking the natural logarithm on both sides, using linearity of the trace, and rearranging the terms yields:
  \begin{equation*}
    \trace{\sum_{t=1}^{T} M^{(t)} \rho^{(t)}} \le
    \lambda_{\min}\left(\sum_{t=1}^{T} M^{(t)}\right) + \eta \trace{\sum_{t=1}^{T} (M^{(t)})^2 \rho^{(t)}} + \frac{\ln n}{\eta}. 
  \end{equation*}
\end{proof}

\noindent Note that Thm.~\ref{thm:mmwuregret} requires
$\nrm{M^{(t)}} \le 1$; in Thm.~\ref{thm:mirrorconv}, there is no bound on
the norm of the subgradients, but the final regret bound depends on
such norms. A tradeoff of this kind is generally inescapable in the
framework of online mirror descent or MMWU: large subgradients (i.e.,
subgradients with large norm) typically lead to worse theoretical
performance of the algorithm. Indeed, the maximum subgradient norm
directly appears in the running time of the algorithm for SDP
developed in Sect.~\ref{sec:mmwusdp}.\index{algorithm!mirror descent|)}\index{mirror descent!online|)}

\subsection{Turning the MMWU algorithm into an SDP solver}
\label{sec:mmwusdp}
We apply the MMWU algorithm to the primal-dual SDP\index{semidefinite programming|(} pair
\eqref{eq:psdp}-\eqref{eq:dsdp}, by developing a game such that a good
strategy leads to (approximate) primal and dual solutions for the
SDP. We make the following assumptions:
\begin{enumerate}[(a)]
\item $A^{(1)} = I$ and $b_1 = R > 0$;
\item $\nrm{C} \le 1$.
\end{enumerate}
Assumption (a) ensures that any solution $X$ satisfies $\trace{X} \le
R$. It also ensures that the dual has a strictly feasible solution,
i.e., there exists a vector $y$ satisfying $\sum_{j=1}^{m} y_j A^{(j)}
- C \succ 0$, and this suffices to guarantee that strong duality
holds. Thus, we can solve the primal or the dual and both yield the
same optimal objective function value. Note that with these
assumptions, the value of $R$ is known in advance, as it is part of the
input; the running time of the algorithms that we obtain depends
on it.

We reduce the optimization question (``what is the optimal objective
function value of \ref{eq:psdp}'') into a sequence of feasibility
questions: for some given scalar $\gamma$, does there exist a primal
feasible solution with value at least $\gamma$? It is well known that
we can approximately answer the optimization question by using binary
search on $\gamma$. In particular, for the objective function the
following inequality holds (recall Def.~\ref{def:tracenorm}):
\begin{equation*}
  \abs{\trace{CX}} \le \nrm{C} \nrm{X}_{\Tr} \le R,
\end{equation*}
where the first inequality is known as a matrix H\"older inequality,
and for the second inequality we used $\nrm{X}_{\Tr} = \trace{X}
\le R$ (this requires $X \succeq 0$), and $\nrm{C} \le 1$.  Thus, we
know that the optimal objective function value for the primal-dual
pair lies in $[-R, R]$, and we can perform binary search\index{binary!search} to determine
the optimal value:
\begin{itemize}
\item Set $\ell \leftarrow - R, u \leftarrow R$.
\item Repeat until $u - \ell \le \epsilon$:
  \begin{itemize}
  \item Set $\gamma \leftarrow (\ell + u)/2$.
  \item Solve a feasibility problem to determine if there exists a
    solution to \eqref{eq:psdp}-\eqref{eq:dsdp} with value at least
    $\gamma$.
  \item If ``yes'', set $\ell \leftarrow (\ell + u)/2$. If ``no'', set
    $u \leftarrow (\ell + u)/2$.
  \end{itemize}
\end{itemize}
This algorithm halves the search interval $[\ell, u]$ for the optimal
value $\gamma$ at every iteration, and therefore takes $\bigO{\log
  (R/\epsilon)}$ iterations to determine an interval of size
$\epsilon$ that contains the optimal objective function value.

We have just seen that we can solve problem \eqref{eq:psdp} if we can
answer the question ``does there exist a primal feasible solution with
value at least $\gamma$?'' for arbitrary $\gamma \in [-R, R]$. To
answer the question using the MMWU algorithm (i.e., the mirror descent
scheme of Alg.~\ref{alg:mmwu}), a sketch of the idea is as follows. We
start from a candidate primal solution $\rho^{(0)} \succeq 0$. At each
step $t$, we generate a vector $y^{(t)}$ of dual variables such that
$\trace{(\sum_{j=1}^m y^{(t)}_j A^{(j)} - C) \rho^{(t)}} \ge 0$, and
such that $b^{\top} y \le \gamma$. We use $M^{(t)} = \sum_{j=1}^m
y^{(t)}_j A^{(j)} - C$ as the adversary response. Then
Eq.~\ref{eq:mmwuregret} from Thm.~\ref{thm:mmwuregret} states that:
\begin{equation}
  \label{eq:mwumineig}
  \lambda_{\min}\left(\sum_{t=1}^{T} M^{(t)}\right) \ge \trace{\sum_{t=1}^{T} M^{(t)} \rho^{(t)}} - \eta \trace{\sum_{t=1}^{T} (M^{(t)})^2 \rho^{(t)}} - \frac{\ln n}{\eta} \ge -\eta T - \frac{\ln n}{\eta},
\end{equation}
because $\trace{\sum_{t=1}^{T} M^{(t)} \rho^{(t)}} \ge 0$ by
construction, and $\trace{(M^{(t)})^2 \rho^{(t)}} \le 1$ by the matrix
H\"older inequality. Dividing by $T$ on both sides of
Eq.~\ref{eq:mwumineig}, we find that $\frac{1}{T} \sum_{t=1}^{T}
M^{(t)}$ is almost positive semidefinite: its smallest eigenvalue is
only slightly negative, $\ge -\eta - \frac{\ln n}{\eta T}$. Recalling
the definition of $M^{(t)}$, this implies that $\frac{1}{T}
\sum_{t=1}^T \sum_{j=1}^m y^{(t)}_j A^{(j)} - C$ is almost positive
semidefinite. Thus, if we define $y = \frac{1}{T} \sum_{t=1}^T
y^{(t)}$ (the average of the dual vectors $y^{(t)}$ generated), we
obtained an almost feasible solution $y$ for the dual \eqref{eq:dsdp},
and such that $b^{\top} y \le \gamma$. Using the fact that $A^{(1)}$
is assumed to be the identity matrix, we can make the dual solution
$y$ feasible by increasing $y_1$ until $\sum_{j=1}^m y_j A^{(j)} - C
\succeq 0$: the necessary shift is at most $\eta + \frac{\ln n}{\eta
  T}$, and with the right choice of parameters, we can ensure that the
objective function deteriorates by at most $\epsilon$, so that
$b^{\top} y \le \gamma + \epsilon$. This dual feasible solution $y$
therefore certifies that there can be no primal feasible solution with
value $> \gamma + \epsilon$, approximately answering the question that
we posed at the beginning, and allowing us to continue in the binary
search for the optimal objective function value. Otherwise, i.e., if
we cannot find a vector $y^{(t)}$ with the desired properties, the
primal solution $\rho^{(t)}$ can be shown to be primal feasible (after
rescaling), allowing us to continue in the binary search.

Let us formalize the idea sketched in the previous paragraph. We
define an oracle that is used to construct a helpful adversary matrix
$M^{(t)}$ at iteration $t$, based on the information available up to
iteration $t$. We call this {\sc PIC-Oracle}, for
``Primal-Infeasiblity-Certificate Oracle''. The purpose of the oracle
is to either prove that the primal solution $X^{(t)}$ is feasible, or
give some dual information regarding its infeasibility. (We use $\rho$
for density matrices, $X$ for general --- not necessarily
trace-normalized --- matrices.) The dual information, in the form of a
vector $y^{(t)}$, is used to construct $M^{(t)}$. The oracle makes use
of a certain polytope, defined next.
\begin{definition}[Primal-infeasibility-certificate polytope]
  \label{def:picpolytope}
  We define $P_{\epsilon}(X)$ as the following polytope, parameterized
  by a scalar $\epsilon > 0$ and matrix $X \succeq 0$:
  \begin{equation}
    \label{eq:picpolytope}
    P_{\epsilon}(X) := 
    \left\{ y \in \R^m :
    \begin{array}{rcl}
      b^{\top} y &\le& \gamma \\
      \trace{(\sum_{j=1}^m y_j A^{(j)} - C)X } &\ge& -\epsilon \\
      y&\ge& \zeroes
    \end{array}
    \right\}.
  \end{equation}
\end{definition}
The definition of the polytope should depend on $\gamma$, but in every
iteration of the algorithm (using the reduction from optimality to
feasibility) $\gamma$ is fixed, so to ease notation we neglect this
detail: $\gamma$ can be considered part of the problem data (within a
single iteration) just as $A^{(j)}$ and $C$. The oracle
$\textsc{PIC-Oracle}_{\epsilon}(X)$ is the function defined as follows:
\begin{equation*}
  \textsc{PIC-Oracle}_{\epsilon}(X) := \begin{cases}
    y \in P_{\epsilon}(X) & \text{if } P_{\epsilon}(X) \neq \emptyset \\
    \text{``failure''} & \text{otherwise.}
  \end{cases}
\end{equation*}
Two properties of $P_{\epsilon}(X)$ are important to understand why
$\textsc{PIC-Oracle}_{\epsilon}(X)$ gives information about the
feasibility or infeasibility of a given primal solution $X$.
\begin{lemma}[Lem.~4.2 in \cite{arora2016combinatorial}]
  \label{lem:polytopeprop}
  Let $X \succeq 0$. Suppose $P_{\epsilon}(X)$
  (Eq.~\ref{eq:picpolytope}) is empty. Then, up to rescaling, the
  matrix $X$ is feasible for \eqref{eq:psdp} with objective function
  value at least $\gamma$. On the contrary, suppose $P_{\epsilon}(X)$
  is nonempty. Then $X$ is either not feasible for \eqref{eq:psdp}, or
  it has objective function value at most $\gamma + \epsilon$.
\end{lemma}
\begin{proof}
  Suppose $P_{\epsilon}(X)$ is empty. Consider the following LP:
  \begin{equation}
    \label{eq:oraclelpprim}
    \left.
    \begin{array}{rrcl}
      \min & b^{\top} y & & \\
      & \trace{\sum_{j=1}^m y_j A^{(j)} X } &\ge& \trace{CX} -\epsilon \\
      & y &\ge& \zeroes,
    \end{array}
    \right\}
  \end{equation}
  and its dual:
  \begin{equation}
    \label{eq:oraclelpdual}
    \left.
    \begin{array}{rrcl}
      \max & (\trace{CX} - \epsilon) z & & \\
      \forall j=1\,\dots,m & \trace{A^{(j)}X } z&\le& b_j \\
      & z &\ge& 0.
    \end{array}
    \right\}
  \end{equation}
  Note that solving problem \eqref{eq:oraclelpprim} is equivalent to
  determining if $P_{\epsilon}(X)$ is empty: indeed, if
  $P_{\epsilon}(X)$ is empty, it must be the case that the optimal
  value of problem \eqref{eq:oraclelpprim} is greater than $\gamma$
  (at least one feasible solution for problem \eqref{eq:oraclelpprim}
  exists, because $A^{(1)} = I$). So the optimum $z^*$ of the dual,
  problem \eqref{eq:oraclelpdual}, has value $\ge \gamma$. Then $z^*
  X$ is a feasible solution to the primal (this is directly implied by
  the constraints in problem \eqref{eq:oraclelpdual}), and:
  \begin{equation*}
    \trace{CXz^*} \ge \gamma + \epsilon z^* \ge \gamma,
  \end{equation*}
  showing the first half of the result.

  Suppose now $P_{\epsilon}(X)$ is nonempty and $y \in
  P_{\epsilon}(X)$. Assume $X$ is primal feasible --- if not, the
  statement of the lemma already holds. Then from the constraints of
  problems \eqref{eq:picpolytope} and \eqref{eq:psdp} we have:
  \begin{equation*}
    \trace{CX} \le \trace{(\sum_{j=1}^m y_j A^{(j)})X} + \epsilon \le
    \sum_{j=1}^m y_j b_j + \epsilon = b^{\top} y + \epsilon \le \gamma
    + \epsilon,
  \end{equation*}
  showing that the objective function value of $X$ is at most $\gamma
  + \epsilon$ and concluding the proof.
\end{proof}

\noindent Lem.~\ref{lem:polytopeprop} shows that
$\textsc{PIC-Oracle}_{\epsilon}(X)$ can provide very useful
information: it either shows that we already have a primal feasible
solution with value at least $\gamma$, or it gives an infeasibility
certificate: the dual vector $y$ that it returns, which we call
$y^{(t)}$ at iteration $t$, is the infeasibility certificate. We can
use the infeasibility certificate $y^{(t)}$ to construct the matrix
$M^{(t)}$ for the current iteration of Alg.~\ref{alg:mmwu}, which,
from our earlier discussion, is chosen as:
\begin{equation}
  \label{eq:adversarymatrixunscaled}
  M^{(t)} = \sum_{j=1}^m y^{(t)}_j A^{(j)} - C.
\end{equation}
There remains a technical issue to resolve: in
Thm.~\ref{thm:mmwuregret} and the surrounding discussion, we assumed
$\nrm{M^{(t)}} \le 1$, but there is no guarantee that the
choice in Eq.~\ref{eq:adversarymatrixunscaled} satisfies this
bound. Thus, we may have to rescale $M^{(t)}$. It turns out that the
magnitude of the scaling factor affects the convergence speed of the
algorithm: if we must scale aggressively, the algorithm makes less
progress toward feasibility, and as a result it converges more
slowly. Formally, the magnitude of the scaling parameter is called the
\emph{width} of the primal-infeasibility-certificate oracle.
\begin{definition}[Width of {\sc PIC-Oracle}]
  \label{def:widthoracle}
  The \emph{width} of $\textsc{PIC-Oracle}_{\epsilon}(X)$ is the
  smallest $w^*$ such that $\nrm{\sum_{j=1}^m y_j A^{(j)} - C} \le
  w^*$ for every $y$ returned by $\textsc{PIC-Oracle}_{\epsilon}(X)$,
  when $\textsc{PIC-Oracle}_{\epsilon}(X)$ is called with $\epsilon >
  0$, $X \succeq 0$ and $\gamma \in [-R, R]$.
\end{definition}
The unscaled choice of $M^{(t)}$ in
Eq.~\ref{eq:adversarymatrixunscaled} may not satisfy $\nrm{M^{(t)}}
\le 1$, but it clearly does if we divide the r.h.s.\ by the width
$w^*$ defined in Def.~\ref{def:widthoracle}, i.e., if we choose:
\begin{equation}
  \label{eq:adversarymatrix}
  M^{(t)} = \frac{1}{w^*} \left(\sum_{j=1}^m y^{(t)}_j A^{(j)} - C\right).
\end{equation}
This is the matrix $M^{(t)}$ that defines the objective function in
iteration $t$ for Alg.~\ref{alg:mmwu}.
\begin{remark}
  The choice of $M^{(t)}$ immediately suggests that any upper bound
  for $w^*$ suffices to guarantee the desired property $\nrm{M^{(t)}}
  \le 1$. Because the convergence speed of the algorithm depends on
  the magnitude of the scaling, we should still aim to find a tight
  bound on $w^*$.
\end{remark}
At this point we have all the necessary components to give the
pseudocode of the MMWU algorithm for SDP, see Alg.~\ref{alg:mmwusdp},
and show its convergence. The algorithm closely follows the informal
exposition given earlier in this section.
\begin{algorithm2e}[htb]
  \SetAlgoLined
  \LinesNumbered
\KwIn{Description of \eqref{eq:psdp}, trace bound $R > 0$, objective function guess $\gamma$, tolerance $\epsilon > 0$, width bound $w > 0$, oracle $\textsc{PIC-Oracle}_{\epsilon}(X)$ with width at most $w$.} 
\KwOut{Either a feasible solution for \eqref{eq:psdp} with objective function value at least $\gamma$, or a feasible solution for \eqref{eq:dsdp} with objective function value at most $\gamma + \epsilon$.}
\textbf{Initialize}: $\rho^{(1)} \leftarrow I/n$, $\eta \leftarrow \sqrt{\frac{\ln n}{T}}$, $T \leftarrow \ceil{\frac{9 w^2 R^2 \ln n}{\epsilon^2}}$.\\
\For{$t=1,\dots,T$}{
  \If{$\textsc{PIC-Oracle}_{\epsilon/3}(R \rho^{(t)})$ outputs ``failure''}{
    \Return $R \rho^{(t)}$ after rescaling as described in Lem.~\ref{lem:polytopeprop}.
  }
  Otherwise, let $y^{(t)}$ be the vector generated by $\textsc{PIC-Oracle}_{\epsilon/3}(R \rho^{(t)})$.\\
  Compute $M^{(t)} = \frac{1}{w} \left(\sum_{j=1}^m y^{(t)}_j A^{(j)} - C\right)$.\\
  Compute $\rho^{(t+1)} = \exp\left(- \eta \sum_{\tau=1}^{t} M^{(\tau)}\right) / \trace{\exp\left(- \eta \sum_{\tau=1}^{t} M^{(\tau)}\right)}$.
}
\Return $\frac{1}{T} \sum_{t=1}^T y^{(t)} + \frac{\epsilon}{R} e_1$, where $e_1 = (1, 0, 0, \dots) \in \R^m$.
\caption{Matrix Multiplicative Weights Update (MMWU) algorithm for SDP.}
\label{alg:mmwusdp}
\end{algorithm2e}
\begin{theorem}[Convergence of MMWU algorithm for SDP; Thm.~4.4 in \cite{arora2016combinatorial}, Thm.~5 in \cite{van2020quantum}]
  \label{thm:mmwusdpconv}
  Alg.~\ref{alg:mmwusdp} returns either a feasible solution for
  \eqref{eq:psdp} with objective function value at least $\gamma$, or
  a feasible solution for \eqref{eq:dsdp} with objective function
  value at most $\gamma + \epsilon$.
\end{theorem}
\begin{proof}
  There are two possible exit points of the algorithm: either
  $\textsc{PIC-Oracle}_{\epsilon/3}(R \rho^{(t)})$ outputs ``failure'' at
  some iteration, in which case we return a primal solution, or
  $\textsc{PIC-Oracle}_{\epsilon/3}(R \rho^{(t)})$ never outputs
  ``failure'', and we return a dual vector in the last line of
  Alg.~\ref{alg:mmwusdp}.

  In case $\textsc{PIC-Oracle}_{\epsilon/3}(R \rho^{(t)})$ outputs
  ``failure'' at some iteration, Lem.~\ref{lem:polytopeprop} shows that a
  properly scaled version of $R \rho^{(t)}$ is primal feasible and
  satisfies the desired conditions.

  In the other case, in every iteration we return $M^{(t)}$ according
  to Eq.~\ref{eq:adversarymatrix}, so the following inequality holds
  by definition of $\textsc{PIC-Oracle}_{\epsilon/3}(R \rho^{(t)})$
  (recall Def.~\ref{def:picpolytope}):
  \begin{equation*}
    \trace{\left(\sum_{j=1}^m y^{(t)}_j A^{(j)} - C\right)R\rho^{(t)} } \ge -\frac{\epsilon}{3},
  \end{equation*}
  so by rearranging we find:
  \begin{equation}
    \label{eq:mrhoinprodlb}
    \trace{M^{(t)} \rho^{(t)}} = \trace{\frac{1}{w}\left(\sum_{j=1}^m y^{(t)}_j A^{(j)} - C\right)\rho^{(t)} } \ge -\frac{\epsilon}{3wR}.
  \end{equation}
  Following the same argument that we used to obtain Eq.~\ref{eq:mwumineig}, we use Eq.~\ref{eq:mmwuregret} from Thm.~\ref{thm:mmwuregret}, divided by $T$ on both sides:
  \begin{equation*}
    \lambda_{\min}\left(\frac{1}{T}\sum_{t=1}^{T} M^{(t)}\right) \ge \trace{\frac{1}{T} \sum_{t=1}^{T} M^{(t)} \rho^{(t)}} - \eta \trace{\frac{1}{T} \sum_{t=1}^{T} (M^{(t)})^2 \rho^{(t)}} - \frac{\ln n}{\eta T} \ge -\frac{\epsilon}{3wR} -\eta - \frac{\ln n}{\eta T},
  \end{equation*}
  where we used Eq.~\ref{eq:mrhoinprodlb}, and $\trace{(M^{(t)})^2
    \rho^{(t)}} \le 1$. Plugging in the values for $\eta$ and $T$, as
  defined in Alg.~\ref{alg:mmwusdp}, we finally obtain:
  \begin{equation*}
    \frac{1}{w} \lambda_{\min}\left(\frac{1}{T}\sum_{t=1}^{T} \left(\sum_{j=1}^m y^{(t)}_j A^{(j)} - C\right) \right) \ge -\frac{\epsilon}{3wR} - \frac{\epsilon}{3wR} - \frac{\epsilon}{3wR} \ge -\frac{\epsilon}{wR},
  \end{equation*}
  hence:
  \begin{equation}
    \label{eq:mineigdualsdp}
    \lambda_{\min}\left(\frac{1}{T} \sum_{t=1}^{T}\left(\sum_{j=1}^m y^{(t)}_j A^{(j)} - C\right) \right) \ge -\frac{\epsilon}{R}.
  \end{equation}
  Using $A^{(1)} = I$ and the fact that Alg.~\ref{alg:mmwusdp} returns
  $\bar{y} = \frac{1}{T} \sum_{t=1}^T y^{(t)} + \frac{\epsilon}{R} e_1$, we have:
  \begin{equation*}
    \lambda_{\min}\left(\sum_{j=1}^m \bar{y}^{(t)}_j A^{(j)} - C\right) = \lambda_{\min}\left(\frac{1}{T} \sum_{t=1}^{T} \left(\sum_{j=1}^m y^{(t)}_j A^{(j)} - C\right) + \frac{\epsilon}{R} I \right) \ge -\frac{\epsilon}{R} + \frac{\epsilon}{R} \ge 0,
  \end{equation*}
  where the first inequality is due to
  Eq.~\ref{eq:mineigdualsdp}. This implies that the solution returned
  by Alg.~\ref{alg:mmwusdp} is dual feasible, i.e.,
  \begin{equation*}
    \sum_{j=1}^m \bar{y}^{(t)}_j A^{(j)} - C \succeq 0.
  \end{equation*}
  Using the definition of $\textsc{PIC-Oracle}_{\epsilon/3}(R \rho^{(t)})$ and $b_1 = R$, the objective function value of this solution satisfies:
  \begin{equation*}
    b^{\top} \bar{y} = b^{\top}\left(\frac{1}{T} \sum_{t=1}^T y^{(t)} + \frac{\epsilon}{R} e_1\right) \le \frac{1}{T} \sum_{t=1}^T \gamma + \epsilon = \gamma + \epsilon. 
  \end{equation*}
\end{proof}

\noindent Thm.~\ref{thm:mmwusdpconv} shows that the MMWU algorithm
correctly determines the feasibility of \eqref{eq:psdp} in $T =
\ceil{\frac{9 w^2 R^2 \ln n}{\epsilon^2}}$ iterations. In each
iteration the most expensive operation is the computation of the
matrix exponential, that can be carried out in
$\bigOt{n^{\omega}/\epsilon}$ time, where $\omega$ is the matrix
multiplication exponent, by truncating the corresponding Taylor series
after $\bigO{1/\epsilon}$ terms.
\begin{remark}
  \label{rem:matrixexp}
  The best current estimate for $\omega$ is $\approx 2.37$
  \cite{williams2024new}. In practice, however, the computational
  complexity is $\bigO{n^3}$, i.e., the same as the ``usual''
  complexity for LU factorization. In other words, although in theory
  we can multiply two $n \times n$ matrices in time
  $\bigO{n^{2.37\dots}}$, the algorithms used in practice run in time
  $\bigO{n^3}$, so the running time $\bigOt{n^{\omega}/\epsilon}$ for
  computing the matrix exponential is only aspirational.
\end{remark}
The binary search scheme calls Alg.~\ref{alg:mmwusdp} as a subroutine
$\bigO{\log(\nrm{C}R/\epsilon)}$ times, and by storing the primal or
dual solutions returned by Alg.~\ref{alg:mmwusdp} in each iteration of
binary search, we can eventually return the primal feasible solution
with the largest objective function value, and the dual solution with
the tightest bound.

\section{Quantum MMWU algorithm for SDP}
\label{sec:qmwu}
Looking at the classical algorithm for semidefinite optimization
described in Alg.~\ref{alg:mmwusdp}, one step appears as a natural
candidate for quantization: the preparation of the Gibbs states
$\rho^{(t)}$, that can be prepared and sampled from with a complexity
that scales as $\sqrt{n}$ on a quantum computer, depending on the
input model --- this is discussed in
Sect.~\ref{sec:gibbssampling}. This seems a clear advantage over
classical algorithms: the Gibbs states are $n \times n$ matrices, and
the fastest known general algorithm to construct them relies on the
computation of the matrix exponential, which runs in
$\bigO{n^{\omega}}$ time, see Rem.~\ref{rem:matrixexp}. But exploiting
this advantage is not straightforward. In each iteration of the
algorithm we must be able to output a dual vector $y^{(t)}$, computed
via $\textsc{PIC-Oracle}$: the construction of a Gibbs state by itself
may not be helpful. Finding a quantum algorithm that can execute all
the necessary steps requires some additional effort.

For a proper discussion of the quantum algorithm we must specify the
input model. We assume the following:
\begin{itemize}
\item For each of the matrices $A^{(j)}$, we have access to a
  controlled unitary that implements a block-encoding of the
  matrix. That is: we have a circuit with a control register such that
  if the control register contains $\ket{\vj}$, we apply a
  block-encoding of $A^{(j)}$ on a register of appropriate size.
\item For the matrix $C$, we have access to a unitary that implements
  a block-encoding of it.
\end{itemize}
For simplicity, we assume that each of these block-encodings has the
same subnormalization factor $\alpha$, uses $p$ auxiliary qubits, and
has negligible error $\xi_a \ll \epsilon$. We do not discuss the error
of the block-encodings in too much detail because we have seen in
Prop.s~\ref{prop:sparseblockenc} and \ref{prop:qramblockenc} that when
we construct a block-encoding from classical data, using the
sparse-oracle access model or the QRAM model, we can reduce the error
of a block-encoding at merely polylogarithmic cost. Thus, although in
principle we have to pay attention to the error parameter, to keep our
exposition simple we just assume that the error is chosen small
enough, and this affects the running time only polylogarithmically. In
the following, we label as ``negligible'' errors that have
polylogarithmic scaling under these assumptions, and do not keep track
of them (as usual in $\bigOt{\cdot}$ notation).

\subsection{Dealing with inexact trace values}
\label{sec:qmwutraceerror}
Careful examination of Alg.~\ref{alg:mmwusdp} reveals that full
knowledge of the iterates $\rho^{(t)}$ is not strictly necessary: in
every iteration we simply need to be able to compute an infeasibility
certificate $y^{(t)}$ from the dual problem. The Gibbs state
$\rho^{(t)}$ has an effect on the algorithm only insofar as it defines
the feasible dual vectors $y^{(t)}$, returned by
$\textsc{PIC-Oracle}$. In the language of mirror descent, knowledge of
the primal solution does not matter as long as we can compute a
subgradient. Therefore we can focus on constructing
$\textsc{PIC-Oracle}$, taking advantage of a quantum computer.

The scheme that we would like to use is to exploit
Prop.~\ref{prop:gibbsstateconstr} to construct each Gibbs state
$\rho^{(t)}$, then apply the trace estimation procedure of
Prop.~\ref{prop:traceest} to compute all the trace inner products
$\trace{A^{(j)}\rho^{(t)}}, \allowbreak \trace{C\rho^{(t)}}$ involved
in $\textsc{PIC-Oracle}$. This immediately raises an issue: the trace
estimation\index{trace!estimation|(} incurs some error, therefore we must analyze the stability
of the algorithm to errors in the definition of the polytope
$P_{\epsilon}(X)$ describing $\textsc{PIC-Oracle}$.
\begin{remark}
  Some specialized variants of the classical algorithm of
  Sect.~\ref{sec:mmwusdp} (i.e., Alg.~\ref{alg:mmwusdp}) described in
  \cite{kale2007efficient} also rely on perturbed versions of the
  polytope describing $\textsc{PIC-Oracle}$: this is not a uniquely
  ``quantum'' feature. The advantage of using a perturbed polytope
  lies in the fact that we can get away with imprecise trace
  estimation procedures, which can lead to faster classical algorithms
  as well.
\end{remark}

We need a further assumption, on top of assumptions (a) and (b) given
at the beginning of Sect.~\ref{sec:mmwusdp}.
\begin{enumerate}[(a)]
  \setcounter{enumi}{2}
\item There exists an optimal solution to \eqref{eq:dsdp} satisfying
  $\nrm{y}_1 \le r$, and the parameter $r \ge 1$ is part of the input to
  the algorithm.
\end{enumerate}
Assumption (c) ensures that the set of feasible dual vectors is
bounded. Using this assumption we define a hierarchy of relaxations of
the feasible region of \eqref{eq:dsdp}. For this, we need a more
general version of the polytope $P_{\epsilon}(X)$, where the entries
of the defining constraints are not fixed: subsequently, we apply the
general version of $P_{\epsilon}(X)$ using some estimates of
$\trace{A^{(j)}X}, \trace{CX}$ (possibly affected by error) as the
entries of the defining constraints.
\begin{definition}[Generalized primal-infeasibility-certificate polytope]
  \label{def:genpicpolytope}
  We define $\hat{P}(a, c)$ as the following polytope, parameterized by
  a vector $a \in \R^m$ and a scalar $c \in \R$:
  \begin{equation}
    \label{eq:genpicpolytope}
    \hat{P}(a, c) := 
    \left\{ y \in \R^m :
    \begin{array}{rcl}
      b^{\top} y &\le& \gamma \\
      \sum_{j=1}^m y_j &\le& r \\
      \sum_{j=1}^m a_j y_j &\ge& c \\
      y&\ge& \zeroes
    \end{array}
    \right\}.
  \end{equation}
\end{definition}
Note that $\hat{P}(\trace{A^{(1)}X},\dots,\trace{A^{(m)}X},\trace{CX})
= P_0(X) \cap \{y : \nrm{y}_1 \le r\}$. The hierarchy of relaxations is
described in the following proposition.
\begin{proposition}[\cite{van2020quantum}]
  \label{prop:polytopeinc}
  Let $X \succeq 0$ and $\rho = X/\trace{X}$, with $\trace{X} \le
  R$. Let $\theta \ge 0$. Let $\tilde{a} \in \R^m, \tilde{c} \in \R$
  satisfy:
  \begin{equation*}
    \abs{\trace{C\rho} - \tilde{c}} \le \theta, \qquad \abs{\trace{A^{(j)}\rho} - \tilde{a}_j} \le \theta \quad \forall j=1,\dots,m.
  \end{equation*}
  The following chain of inclusions holds:
  \begin{equation*}
    \left(P_0(X) \cap \{y : \nrm{y}_1 \le r\}\right) \subseteq \hat{P}(\tilde{a}, \tilde{c}-(r+1)\theta) \subseteq \left(P_{4Rr\theta}(X)\cap \{y : \nrm{y}_1 \le r\}\right).
  \end{equation*}
\end{proposition}
\begin{proof}
  Let $\bar{y} \in P_0(X) \cap \{y : \nrm{y}_1 \le r\}.$ We have:
  \begin{align*}
    \sum_{j=1}^m \tilde{a}_j \bar{y}_j \ge \sum_{j=1}^m \left(\trace{A^{(j)}\rho}-\theta\right) \bar{y}_j \ge \trace{C\rho} - \theta \nrm{y}_1 \ge \tilde{c} - (r+1) \theta,
  \end{align*}
  where the first inequality used the definition of $\tilde{a}$ with
  nonnegativity of $\bar{y}$, the second inequality used
  $\trace{\left(\sum_{j=1}^m y_j A^{(j)} - C\right)X } \ge 0$ (from
  the constraints of Eq.~\ref{eq:picpolytope}) and $\bar{y} \ge \zeroes$,
  the last inequality used $\nrm{y}_1 \le r$. This shows the first
  inclusion.

  For the second inclusion, let $\bar{y} \in \hat{P}(\tilde{a},
  \tilde{c}-(r+1)\theta)$. Then:
  \begin{align*}
    \trace{\left(\sum_{j=1}^m \bar{y}_j A^{(j)} - C\right)\rho } &\ge \sum_{j=1}^m (\tilde{a}_j - \theta) \bar{y}_j - \tilde{c} - \theta \ge \sum_{j=1}^m \tilde{a}_j  \bar{y}_j - \tilde{c} - \theta(\nrm{y}_1+1) \\
    &\ge -2(r+1)\theta \ge -4r\theta,
  \end{align*}
  where we used the definition of $\tilde{a}, \tilde{c}$ in the first
  inequality, the definition of $\hat{P}(\tilde{a},
  \tilde{c}-(r+1)\theta)$ in the third inequality, $r \ge 1$ in the
  last inequality, and $\bar{y} \ge \zeroes$. Multiplying this
  inequality by $\trace{X}$, using $\rho = X/\trace{X}$ and
  $\abs{\trace{X}} \le R$, yields:
  \begin{equation*}
    \trace{\left(\sum_{j=1}^m \bar{y}_j A^{(j)} - C\right)X } \ge -4 \trace{X} r \theta \ge -4Rr\theta. 
  \end{equation*}
\end{proof}

\noindent Prop.~\ref{prop:polytopeinc} gives a precise way to deal
with errors in the trace estimation: suppose we choose $\theta =
\epsilon/(12Rr)$ as the maximum allowed tolerance in trace
estimation. Then $\hat{P}(\tilde{a}, \tilde{c}-(r+1)\theta) \subseteq
P_{\epsilon/3}(X)$, so if we can give an algorithm to compute a point
in $\hat{P}(\tilde{a}, \tilde{c}-(r+1)\theta)$, we can directly
implement Alg.~\ref{alg:mmwusdp}. From now on, we use the following
shorthand:
\begin{equation}
  \label{eq:qmwutheta}
  \theta = \frac{\epsilon}{12Rr}.
\end{equation}

\subsection{Computing the dual vector}
\label{sec:qmwudual}
Following the discussion in Sect.~\ref{sec:qmwutraceerror}, one way to
implement a quantum SDP solver is to give a quantum algorithm to
compute a point in $\hat{P}(\tilde{a}, \tilde{c}-(r+1)\theta)$, after
which we have all the necessary components for the MMWU framework. The
simplest approach is to estimate the trace values on a quantum
computer and then use a classical algorithm to find the required dual
vector.

Let us do a back-of-the-envelope calculation of the running time of
such an approach. Alg.~\ref{alg:mmwusdp} runs for $T = \ceil{\frac{9
    w^2 R^2 \ln n}{\epsilon^2}}$ iterations: recalling that the oracle
width parameter can be chosen as $w = r$ (due to the constraint
$\nrm{y}_1 \le r$ in the generalized PIC polytope), we write this as
$\bigOt{(Rr/\epsilon)^2}$. In each iteration the algorithm uses
$M^{(t)} = \frac{1}{r} \left(\sum_{j=1}^m y^{(t)}_j A^{(j)} -
C\right)$, and the Hamiltonian of the Gibbs state $\rho^{(t)}$ used
at iteration $t$ is a linear combination of the matrices
$M^{(\tau)}, \tau=1,\dots,t-1$; thus, at every iteration, by taking the
appropriate linear combination we can compute a vector $\hat{y}^{(t)}
\in \R^{m+1}$ such that:
\begin{equation*}
  \rho^{(t)} = \exp\left(\sum_{j=1}^m \hat{y}^{(t)}_j A^{(j)} - \hat{y}^{(t)}_{m+1} C\right)/\trace{\exp\left(\sum_{j=1}^m \hat{y}^{(t)}_j A^{(j)} - \hat{y}^{(t)}_{m+1} C\right)}.  
\end{equation*}
We can construct a state-preparation pair
(Def.~\ref{def:statepreppair}) for $\hat{y}^{(t)}$ with $\bigO{m}$
gates. Then, using linear combination of block-encodings
(Prop.~\ref{prop:lincombblock}) followed by Gibbs state preparation
(Prop.~\ref{prop:gibbsstateconstr}), we construct a purification of
$\rho^{(t)}$, which we use to estimate the trace values with
Prop.~\ref{prop:traceest}. We can upper bound the cost of these
operations as follows:
\begin{itemize}
\item The state-preparation pair for $\hat{y}$ can be constructed with
  negligible error, but its subnormalization factor (which then
  affects the subnormalization factor of the linear combination of
  block-encodings) is $\nrm{\hat{y}}_1$. We can bound the
  subnormalization factor as:
  \begin{equation*}
    \nrm{\hat{y}}_1 \le \sum_{t=1}^{T} \frac{\eta}{r} \nrm{y^{(t)}}_1 \le \sqrt{T} \ln n = \bigOt{Rr/\epsilon},
  \end{equation*}
  where the first inequality is due to the scaling of the vectors
  $y^{(t)}$ in Alg.~\ref{alg:mmwusdp}, the second inequality uses
  $\nrm{y}_1 \le r$ and $\eta = \sqrt{\ln n/T}$, and the final
  equality is by definition of $T$.
\item The linear combination of block-encodings then has
  subnormalization factor $\bigOt{\alpha Rr/\epsilon}$, because we
  need to multiply the subnormalization factor $\alpha$ of the input
  matrices $A^{(j)}, C$ with the subnormalization factor of the
  state-preparation pair; the additional resource consumption of this
  step (gates, auxiliary qubits) is negligible, and so is the error.
\item We construct the purification of $\rho^{(t)}$ starting from
  the linear combination of block-encodings at the previous step; this
  gives a running time of $\bigOt{\alpha Rr \sqrt{n}/\epsilon}$, where
  the running time is in terms of number of accesses to the circuit
  implementing controlled block-encodings of the input matrices, and a
  similar number of additional gates.
\item Finally, we prepare a random variable with expected value very
  close to $\trace{A^{(j)} \rho^{(t)}}$ using $\bigOt{\alpha}$
  applications of the block encoding of $A^{(j)}$, and similarly for
  $\trace{C \rho^{(t)}}$; crucially, we can obtain one sample from one
  of these random variables with only a single copy of $\rho^{(t)}$,
  i.e., we do not need to repeat the construction of $\rho^{(t)}$ for
  this step. Thus, the cost for the circuit to obtain one sample is
  dominated by the cost for preparing $\rho^{(t)}$, which is
  $\bigOt{\alpha Rr \sqrt{n}/\epsilon}$; the cost for preparing the
  random variable is negligible, compared to this. Note that the
  random variable obtained with Prop.~\ref{prop:traceest} has a small
  bias, but the running time dependence on the bias is
  polylogarithmic, so we can make the bias very small, and the total
  error accumulated by the algorithm is dominated by the error in the
  next step.
\item Each of the trace values must be estimated to error $\theta =
  \epsilon/(12Rr)$: we use mean estimation (see
  Rem.~\ref{rem:meanestimation}, in particular we can use the
  algorithm of \cite{montanaro2015montecarlo}) to compute the expected
  values. The mean estimation algorithm applies the circuit to obtain
  a sample from the random variable a total of $\bigO{1/\theta}$
  times. Overall, this implies that the estimation of a single value
  among $\trace{A^{(j)} \rho^{(t)}}$ or $\trace{C \rho^{(t)}}$ takes
  $\bigOt{\alpha \sqrt{n} (Rr /\epsilon)^2}$ applications of the input
  block-encodings. Estimating all of these values, because there are
  $m+1$ of them, increases the total complexity to $\bigOt{\alpha m
    \sqrt{n} (Rr /\epsilon)^2}$.
\end{itemize}
Multiplying the above cost by the number of iterations $T$, we obtain
the running time:
\begin{equation*}
  \bigOt{\alpha m \sqrt{n} (Rr /\epsilon)^4}.
\end{equation*}
\begin{remark}
  For such an algorithm to work we have to assume that all the
  subroutines are successful: for example, all the trace values must
  be estimated within the required precision, otherwise the algorithm
  may not find the correct solution. This is generally not an issue as
  long as the complexity of all subroutines scales polylogarithmically
  with the inverse of the maximum failure probability: if the
  algorithm executes $K$ subroutines in total, we set the maximum
  failure probability of each subroutine to $\delta/K$. By the union
  bound, the probability that any subroutine fails is then at most
  $\delta$. As long as the term $\polylog{\delta/K}$ is acceptable in
  the running time expressions, this allows us to assume that all
  subroutines are successful when analyzing the algorithm. We used a
  similar approach in Sect.~\ref{sec:qpeiterative}.
\end{remark}

The simple algorithm described above has linear scaling in $m$,
because it estimates all the values $\trace{A^{(j)}\rho^{(t)}}$ for
$j=1,\dots,m$. We can reduce this to $\sqrt{m}$ based on the
observation that if the generalized PIC polytope $\hat{P}(\tilde{a},
\tilde{c}-(r+1)\theta)$ is nonempty, then it contains a sparse
vector. This can be seen by reformulating the question of emptiness of
$\hat{P}(\tilde{a}, \tilde{c}-(r+1)\theta)$ as a linear program, using
the definition in Eq.~\ref{eq:genpicpolytope}:
\begin{equation}
  \label{eq:mmwusparselp}
  \left.
  \begin{array}{rrcl}
    \min & \sum_{j=1}^m y_j & & \\
    & b^{\top} y &\le& \gamma \\
    & \tilde{a}^{\top} y &\ge& \tilde{c}-(r+1)\theta \\
    & y &\ge& \zeroes.
  \end{array}
  \right\}
\end{equation}
If this problem has a solution with value $\le r$, then
$\hat{P}(\tilde{a}, \tilde{c}-(r+1)\theta)$ is nonempty and contains
the desired vector $y^{(t)}$. It is known that any feasible linear
program with two constraints (besides nonnegativity) has a solution
with at most two nonzero elements --- a so-called \emph{basic
solution}. Thus, we can choose to solve problem~\ref{eq:mmwusparselp}
in each iteration of Alg.~\ref{alg:mmwusdp} to obtain the vector
$y^{(t)}$, and we take advantage of the existence of a sparse solution
to improve the performance of the algorithm.

Finding a sparse solution requires a rather sophisticated approach,
and we do not discuss it in detail here. The important features of
this approach that need to be highlighted are that it its main idea is
entirely classical (i.e., it is based on the the geometry of
problem~\ref{eq:mmwusparselp}, and it gives a classical algorithm as
well), and it reduces the solution of problem \ref{eq:mmwusparselp} to
the problem of searching over the points $(b_j, \tilde{a}_j)$. Indeed,
as is shown in \cite{van2020sdp}, calling $\nrm{y} = L$ and using a
change of variables $z = y/L$, we can reformulate $\hat{P}(\tilde{a},
\tilde{c}-(r+1)\theta)$, and therefore problem
\eqref{eq:mmwusparselp}, as the following problem:
\begin{equation*}
  \left.
  \begin{array}{rrcl}
    & b^{\top} z &\le& \gamma/L \\
    & \tilde{a}^{\top} z &\ge& (\tilde{c}-(r+1)\theta)/L \\
    & \nrm{z} &=& 1 \\
    & z &\ge& \zeroes \\
    & 0 <L &\le& r.
  \end{array}
  \right\}  
\end{equation*}
Note that $z$ defines a convex combination of the points $(b_j,
\tilde{a}_j)$, and to satisfy the constraints, we want such a
combination that lies to the upper left of the point $(\gamma/L,
(\tilde{c}-(r+1)\theta)/L)$. If such a combination exist, there is one
that has nonzero coefficients for only two points (a basic solution
for the linear program). We can find these two points with a procedure
that we summarize as follows:
\begin{itemize}
\item Check if $\gamma \ge 0$ and $\tilde{c}-(r+1)\theta \le 0$. If
  so, return $z = \zeroes$ as a feasible solution.
\item Scan the points $(b_j, \tilde{a}_j)$ to see if any of them is a
  solution. If so, return $z = e_j$ as a feasible solution.
\item Find two points $(b_j, \tilde{a}_j)$, $(b_k, \tilde{a}_k)$ such
  that the line segment connecting them intersects the feasible
  region. If so, return the corresponding convex combination as a
  feasible solution. Crucially, these two points can be found
  independently of each other, i.e., we do not need to search over all
  pairs, but rather search over the list of points at most twice: this
  exploits a geometric idea described in \cite{van2020sdp}.
\item If a feasible solution was not found in the steps above, return
  ``failure''.
\end{itemize}
With this procedure we obtain the following statement.
\begin{proposition}[Informal; see Lem.~16 in \cite{van2020sdp} for a precise statement]
  \label{prop:qmwufastsearch}
  Assume that we have access to a quantum circuit $U$ such that:
  \begin{equation*}
    U \ket{\vj}\ket{\v{0}}\ket{\v{0}} = \ket{\vj}\ket{\vv{\tilde{a}_j}}\ket{\psi_j},
  \end{equation*}
  where $\tilde{a}_j$ is such that $|\trace{A^{(j)}\rho} -
  \tilde{a}_j| \le \theta$. There is a quantum algorithm that uses
  $\bigOt{\sqrt{m}}$ calls to $U$, and a number of gates of the same
  order, and with high probability returns a vector in
  $P_{4Rr\theta}(X)\cap \{y : \nrm{y}_1 \le r\}$ if $P_{0}(X)\cap \{y :
  \nrm{y}_1 \le r\}$ is nonempty, and returns ``failure'' if $P_{0}(X)\cap
  \{y : \nrm{y}_1 \le r\}$ is empty.
\end{proposition}
In fact, the above proposition also works if $U$ outputs a
superposition of possible trace values, as long as the probability of
obtaining a wrong estimate is exponentially small --- which is easy to
achieve with Prop.~\ref{prop:traceest}. Thus, we construct $U$ with
Prop.~\ref{prop:traceest} and a mean estimation algorithm, such as the
one in \cite{montanaro2015montecarlo}, that has scaling
$\bigOt{1/\theta}$ for error $\theta$.

Summarizing, we can reduce the problem of finding a point in the
generalized PIC polytope $\hat{P}(\tilde{a}, \tilde{c}-(r+1)\theta)$
to the problem of performing $\bigOt{\sqrt{m}}$ trace estimations in
quantum superposition. With this approach, the total complexity of the
algorithm is:
\begin{equation*}
  \bigOt{\alpha \sqrt{mn} (Rr /\epsilon)^4}
\end{equation*}
calls to the block-encoding of the input matrices, and a similar
number of additional gates.
\begin{remark}
  Although the running time reported above is attractive in its
  dependence on $m$ and $n$, the poor scaling on $Rr /\epsilon$ is an
  issue. As discussed in \cite{van2020sdp}, for the majority of known
  problems formulated as SDPs at least one of the parameters $R, r$
  scales linearly in the dimensions $m, n$. The fastest classical
  optimization methods for SDPs depend polylogarithmically on the size
  of primal/dual solutions, and the precision parameter
  $\epsilon$. For example, the interior point method of
  \cite{jiang2020faster} has a running time of $\bigOt{\sqrt{n}(mn^2 +
    m^{\omega} + n^{\omega})}$, where $\omega$ is as discussed in
  Rem.~\ref{rem:matrixexp}. (More practical interior point methods for
  semidefinite optimization have running time $\bigOt{n^{6.5}}$, see,
  e.g., \cite{monteiro1998polynomial,nesterov1998primal}.) Thus, the
  quantum MMWU algorithm for SDP discussed in this section yields an
  end-to-end quantum speedup only if $Rr /\epsilon$ is very small,
  which is not the case for most problems.
\end{remark}

\subsection{Further improvements}
\label{sec:qmwuimpr}
We can reduce the complexity of the quantum MMWU algorithm even
further (at least in some parameters), using techniques that we
overview here because they could be useful in the design of other
quantum optimization algorithms. The discussion in this section is
meant to convey intuition and provide the right references, rather
than giving a detailed and mathematically precise description of the
corresponding ideas; thus, we do not give formal statements or proofs.

The major improvement that we aim to obtain in this section is to
reduce the dependence on $m$ and $n$ from $\bigOt{\sqrt{mn}}$ down to
$\bigOt{\sqrt{m} + \sqrt{n}}$. The scheme presented in
Sect.~\ref{sec:qmwudual} needs to estimate $m$ trace values of the
form $\trace{A^{(j)} \rho}$, and because the construction of the Gibbs
state $\rho$ runs in time $\bigOt{\alpha \sqrt{n}}$, already improving
over $\bigOt{m\sqrt{n}}$ requires significant ingenuity, as we
discussed. If we want to use Prop.~\ref{prop:gibbsstateconstr} for the
construction, an algorithm that scales as $\bigOt{\sqrt{m} +
  \sqrt{n}}$ is only allowed to produce, at every iteration, a number
of Gibbs states that does not scale with $m$.

To make progress on such a construction we separate the Gibbs state
preparation from the trace estimation procedure. In the setting of the
algorithm, at each iteration we construct the \emph{same} state
$\rho^{(t)}$, and we essentially want to perform a search over
multiple $\trace{A^{(j)} \rho^{(t)}}$: recall that the procedure of
Prop.~\ref{prop:qmwufastsearch} reduces to two searches over these
values. In fact, the search can be translated into a minimum finding
problem: even in the most expensive part of
Prop.~\ref{prop:qmwufastsearch}, where we look for two points $(b_j,
\tilde{a}_j), (b_k, \tilde{a}_k)$ defining a line segment with a
certain property, \cite{van2020sdp} shows that this can be
accomplished with a search over angles, and the angles can be computed
easily from the trace values. For ease of exposition, we consider the
problem of determining the maximum of $\trace{A^{(j)} \rho^{(t)}}$
over all $j$; a search over any value that can be computed easily from
$\trace{A^{(j)} \rho^{(t)}}$ has the same computational complexity,
and this includes minimum finding (minimum finding and maximum finding
are equivalent in this setting, because we can flip the sign of each
value).

Classically, estimating $\trace{A^{(j)} \rho^{(t)}}$ for all $j$ only
requires us to compute $\rho^{(t)}$ once: after explicitly computing
the entries of the Gibbs state, we can ``reuse it'' in as many
calculations as we want, for example estimating all $\trace{A^{(j)}
  \rho^{(t)}}$ while paying the cost for the construction of
$\rho^{(t)}$ only once. In the quantum setting this is not obvious,
because a measurement after constructing the Gibbs state would
generally cause the quantum state to collapse to a basis
state. However, we can do something similar to ``reusing'' the state
under certain conditions, exploiting an idea described in the
\emph{gentle search lemma} of \cite{aaronson2018shadow}. (To be more
precise: we do not actually reuse the state; rather, we use it in a
way that allows us to estimate whether $m$ trace values are larger
than a given threshold with a number of samples that does not depend
on $m$.)  Suppose we have a unitary that outputs a sample from a
random variable that estimates $\trace{A^{(j)} \rho^{(t)}}$ starting
from a copy of $\rho^{(t)}$ (as in Prop.~\ref{prop:traceest}), and ask
the question: does there exist some $j = 1,\dots,m$ such that
$\trace{A^{(j)} \rho^{(t)}} \ge \mu$ for some given value $\mu$?  If
we can answer this existence question, we can perform binary search on
the set $\{1,\dots,m\}$ to find the index of such a $j$: every time we
split the current interval (initially, $\{1,\dots,m\}$) in two sets,
check existence of $j$ with $\trace{A^{(j)} \rho^{(t)}} \ge \mu$ in
each of the two sets, and apply the search procedure recursively on
a set that gives the positive answer. This eventually yields an
index with the desired property. We are back to the existence
question: does there exist $j = 1,\dots,m$ such that $\trace{A^{(j)}
  \rho^{(t)}} \ge \mu$? Note again that $\rho^{(t)}$ is fixed and only
the $A^{(j)}$ are changing.

Recall the connection between trace values and measurement
probabilities discussed in Sect.~\ref{sec:mixedstate}, see
Rem.~\ref{rem:densitymatmeas} and the surrounding discussion. The
gentle search lemma of \cite{aaronson2018shadow} states, informally,
that if we have several two-outcome measurements $M^{(j)}$ (e.g., testing
whether a qubit is $\ket{0}$ or $\ket{1}$) with the property that
either $\trace{M^{(j)} \rho} \ge \beta$ for some $j$, or
$\trace{M^{(j)} \rho} \le \beta - \delta$ for all $j$, we can detect
which of the two cases holds with $\bigOt{1/\delta^2}$ samples, and
this lets us find $j$ such that $\trace{M^{(j)} \rho} \ge \beta$
with a similar number of samples.
\begin{remark}
  Under the stated property, for fixed $j$, the probability of
  observing the first measurement outcome (``accept'') is at least
  $\beta$ in the first case, and is at most $\beta - \delta$ in the
  second case. Thus, if $\trace{M^{(j)} \rho} \ge \beta$, by the
  Chernoff bound with very high probability at least a fraction
  $\approx (\beta - \delta/2)$ of the $\bigOt{1/\delta^2}$ samples
  outputs ``accept'', and we can detect this by counting the number of
  ``accept''. We now use a result from \cite{harrow2017sequential}:
  given multiple measurements such that either (i) at least one of
  them is very likely to output $\ket{1}$, or (ii) all of them are
  very likely to output $\ket{0}$, distinguishing the two cases (i.e.,
  determining whether there exists a measurement that outputs
  $\ket{1}$) is easy. Combining this with the previous construction
  that counts the number of ``accept'' in the $\bigOt{1/\delta^2}$
  samples, and amplitude amplification, we obtain the stated result.
\end{remark}

The above idea gives us a blueprint to separate Gibbs state\index{Gibbs!state}
preparation and trace estimation: we prepare $\bigOt{1/\theta^2}$
copies (where $\theta$ is as in Eq.~\ref{eq:qmwutheta}) in
parallel. We apply the trace estimation procedure of
Prop.~\ref{prop:traceest} to each copy, controlled on the index $j$ of
the matrix $A^{(j)}$ for which we want to compute the estimate, and
take the sample average of the $\bigOt{1/\theta^2}$ samples. By
Chernoff bound, with high probability this is a trace estimate with
precision $\bigOt{\theta}$, because the standard deviation is constant
and so the sample average is unlikely to deviate much from the
expected value (recall that by Prop.~\ref{prop:traceest}, the expected
value is at most $\theta/4$ away from $\trace{A^{(j)}
  \rho^{(t)}}$). Now construct a circuit that ``accepts'' if the
sample average is larger than a given threshold value $\mu$. If there
exists $j$ such that $\trace{A^{(j)} \rho^{(t)}} \ge \mu$, the circuit
``accepts'' with large probability, say at least $2/3$ --- we can
adjust the number of samples to ensure that the probability of
acceptance is at least this much. If, on the other hand, no such $j$
exists, the circuit ``rejects'' with the same large probability, and
so it ``accepts'' with probability at most $1/3$. Using the gentle
search lemma, with $\beta = 2/3, \delta = 1/3$ we can distinguish
these two cases with $\bigOt{1/\delta^2} = \bigOt{1}$ samples (i.e.,
measurements) from these circuits: this allows us to implement a
quantum search (with the usual quadratic speedup) over the values
$\trace{A^{(j)} \rho^{(t)}}$. The quantum search does not require the
construction of additional copies of $\rho^{(t)}$, because it can be
implemented following similar logic to oblivious amplitude
amplification (Sect.~\ref{sec:obliviousampamp}), where we amplify the
effect of some algorithm applied onto a given state, without
constructing the given state from scratch every time.  The details of
this procedure are described in \cite{van2018improvements}, with one
key result adapted from \cite{brandao2019quantum}. Overall, this gives
the following complexity of every iteration:
\begin{itemize}
\item We prepare $\bigOt{(Rr/\epsilon)^2}$ copies of $\rho^{(t)}$:
  because each purification of $\rho^{(t)}$ uses $\bigOt{\alpha
    Rr\sqrt{n} /\epsilon}$ accesses to the block-encodings describing
  the input, this brings the cost to $\bigOt{\alpha \sqrt{n}
    (Rr/\epsilon)^3}$.
\item We run the search procedure of
  Prop.~\ref{prop:qmwufastsearch}. This still takes time
  $\bigOt{\alpha \sqrt{m} (Rr/\epsilon)^2}$, as in
  Sect.~\ref{sec:qmwudual}, because of the required precision and the
  subnormalization of the block-encodings. Crucially, as we discussed
  above, this cost is now additive (rather than multiplicative) with
  the cost in the previous bullet, because we use the same
  $\bigOt{(Rr/\epsilon)^2}$ copies of $\rho^{(t)}$.\index{trace!estimation|)}
\end{itemize}
The number of iterations is still $\bigOt{(Rr/\epsilon)^2}$, giving a
total complexity of:
\begin{equation*}
  \bigOt{(\sqrt{m} + \sqrt{n} Rr/\epsilon) \alpha (Rr /\epsilon)^4}
\end{equation*}
calls to the block-encodings of the input matrices, and a similar
number of additional gates, see \cite{van2018improvements} for a
formal statement and detailed proofs.\index{matrix!multiplicative weights update|)}\index{algorithm!multiplicative weights update|)}

\section{Quantum algorithm for the SDP relaxation of MaxCut}
\label{sec:qubosdp}
Consider the following quadratic unconstrained optimization problem
with $\pm 1$ decision variables:
\begin{equation}
  \label{eq:pm1qp}
  \tag{$\pm 1$-QP}
  \left.
  \begin{array}{rrcl}
    \max & z^{\top} C z & & \\
    \text{s.t.:} & z &\in& \{-1,1\}^n,
  \end{array}
  \right\}
\end{equation}
where $C \in \R^{n \times n}$ is a symmetric matrix. In this section
we describe a quantum algorithm for a relaxation of this
problem. Problem \eqref{eq:pm1qp} finds direct application in a few
areas, see the notes in Sect.~\ref{sec:mmwunotes}. Although the
unconstrained formulation limits the modeling power of this problem,
one could potentially incorporate constraints by penalizing their
violation in the objective function.
\begin{remark}
  Exact penalization of the constraint violations in the objective
  function may require large penalty coefficients, severely hampering
  the ability to solve the problem in practice. Thus, we do not
  advocate constraint penalization in the objective function as a
  general methodology that is helpful for practical purposes. Rather,
  we point out that \emph{in theory} it is possible, and focus on the
  development of a quantum algorithm for a relaxation of the problem,
  because it lets us discuss several interesting ideas even if the
  method may not be widely applicable.
\end{remark}

Problem \eqref{eq:pm1qp} is equivalent to the combinatorial
optimization problem MaxCut\index{MaxCut!definition}, whose
description is as follows: given a weighted undirected graph $G =
(V,E)$, partition its nodes into two sets $V_1, V_2$ such that the sum
of the weights of edges that have one endpoint in $V_1$ and the other
in $V_2$ is maximized. MaxCut is known to be NP-hard
\cite{gareyjohnson}. Transforming an instance of MaxCut into an
instance of \eqref{eq:pm1qp} is easy: suppose $G$ has vertex set
$\{1,\dots,n\}$, and let $w_{ij}$ be the weight of edge $(i,j) \in E$
with $i < j$ (the weight is $0$ if the edge is not present); set the
elements of the objective function matrix $C_{ij} = C_{ji} =
-\frac{1}{2}w_{ij} $. This yields a symmetric matrix $C$, and
according to the objective function, for every $(i,j) \in E$ we either
gain $w_{ij}$ if $z_i \neq z_j$ (because in this case, $z_i z_j =
-1$), or we have to pay $w_{ij}$ if $z_i = z_j$ (because in this case,
$z_i z_j = 1$). Then, $\sum_{(i,j) \in E} w_{ij} + \max_{z \in
  \{-1,1\}^n} z^{\top} C z$ equals twice the value of the MaxCut,
because each edge contribution $w_{ij}$ disappears from this
expression if $z_i = z_j$, and is counted twice if $z_i \neq z_j$. The
reverse equivalence (from \eqref{eq:pm1qp} to MaxCut) follows from the
same construction.
\begin{remark}
  The equivalence between \eqref{eq:pm1qp} and MaxCut is in terms of
  the optimal solution vectors, not in terms of the corresponding
  objective function values: to translate the objective function
  values of the two problems we need to shift and scale, as discussed
  above.
\end{remark}
Problem \eqref{eq:pm1qp} can also be reformulated into a problem with
$\{0,1\}$ binary variables, as opposed to $\{-1,+1\}$; this yields a
quadratic unconstrained binary optimization problem (QUBO), see
Def.~\ref{def:qubo} and the discussion in Sect.~\ref{sec:combopteig}.

\subsection{Obtaining the normalized SDP relaxation}
\label{sec:qubosdprelax}
Because \eqref{eq:pm1qp} is NP-hard, its solution can be difficult, and
we can consider instead a convex relaxation of the problem to obtain a
bound on its optimal value. Convex relaxations of difficult discrete
optimization problems are also at the heart of the branch-and-bound
algorithm, so an efficient algorithm to solve a relaxation can lead to
a more effective branch-and-bound.  One way to obtain a relaxation of
\eqref{eq:pm1qp} is to define a new decision variable $X = z
z^{\top}$. Imposing $\text{rank}(X) = 1$, $X \succeq 0$ and $X_{jj} =
1$ for all $j=1,\dots,n$ suffices to ensure that $X$ is of the form $z
z^{\top}$ for some vector $z \in \{-1,+1\}^n$. Note that if $X = z
z^{\top}$ then $z^{\top} C z = \trace{CX}$. Thus, \eqref{eq:pm1qp} is
equivalent to:
\begin{equation*}  
  \left.
  \begin{array}{rrcl}
    \max & \trace{CX} & & \\
    \text{s.t.:}  & \forall j \quad X_{jj} &=& 1 \\
    & X &\succeq& 0 \\
    & \text{rank}(X) &=& 1.
  \end{array}
  \right\}
\end{equation*}
This is a nonconvex optimization problem because of the constraint
$\text{Rank}(X) = 1$; dropping this constraints yields an SDP with
special structure:\index{MaxCut!semidefinite relaxation}
\begin{equation}
  \label{eq:qubosdporig}
  \left.
  \begin{array}{r}
    \max \quad \trace{CX}  \\
    \text{s.t.:}  \quad \text{diag}(X) = \allones \\
    X \succeq 0.
    \end{array}
  \right\}
  \tag{MaxCutSDP-orig}
\end{equation}
We aim to solve this problem up to some precision $\epsilon$. In fact,
we work with a normalized version where solutions are constrained to
having unit trace.

Let $\hat{C} := C/\nrm{C}_F$ (recall Def.~\ref{def:frobenius}), and
change the objective function matrix from $C$ to $\hat{C}$, which
can be achieved without loss of generality by rescaling $C$. In
addition, define a new decision variable $\rho = X/n$, so that the
diagonal constraints become $\text{diag}(\rho) = \frac{1}{n}
\allones$.
\begin{remark}
  \label{rem:qubosdpscale}
  These two scaling operations (of the objective function and of the
  decision variables) are w.l.o.g., but we should be careful about the
  final precision because a solution that is $\epsilon$ away from
  optimality in the rescaled problem might be $(n \nrm{C}_F \epsilon)$
  away from optimality in the original problem. From now on we work
  with the rescaled problem, and our time and gate complexity
  evaluation also concerns the rescaled problem. Only at the end of
  our analysis, in Rem.~\ref{rem:qubosdpcostorig}, we discuss an
  $\epsilon$-optimal solution to the original problem
  \eqref{eq:qubosdporig}.
\end{remark}
We use $\rho$ for the new decision variable because the (rescaled)
constraints of \eqref{eq:qubosdporig} impose that it is a positive
semidefinite matrix with unit trace; thus, it is a density matrix. We
therefore obtain the following problem:
\begin{equation}
  \label{eq:qubosdp}
  \left.
  \begin{array}{r}
    \max \quad \trace{\hat{C} \rho}  \\
    \text{s.t.:}  \quad \text{diag}(\rho) = \frac{1}{n} \allones \\
    \rho \succeq 0.
    \end{array}
  \right\}
  \tag{MaxCutSDP}
\end{equation}
The optimal objective function of the rescaled problem
\eqref{eq:qubosdp} lies in the interval $[-1, 1]$, because, by the
matrix H\"older inequality (see Sect.~\ref{sec:mmwusdp}), we have:
\begin{equation*}
  \abs{\trace{\hat{C}\rho}} \le \nrm{\hat{C}}_F \nrm{\rho}_{\Tr}
  \le 1.
\end{equation*}
Similarly to our approach in Sect.~\ref{sec:mmwusdp}, we reduce the
solution of the optimization problem \eqref{eq:qubosdp} to a sequence
of feasibility problems, each of which is solved up to precision
$\epsilon$. We perform binary search on the optimal objective function
value $\gamma$ in the interval $[-1, 1]$, and solve feasibility problems to determine if a
feasible solution with value at least as large as the current guess
$\gamma$ exists. Our goal is then to solve this problem:
\begin{equation}
  \left.
  \begin{array}{rrcl}
    \min& 0 & & \\
    &\trace{\hat{C} \rho} &\ge& \gamma - \epsilon \\
    &\displaystyle \left\| \text{diag}(\rho) - \frac{1}{n} \allones \right\|_1 &\le& \epsilon  \\
    &\trace{\rho} &=& 1 \\
    &\rho& \succeq& 0.
  \end{array}
  \right\}
  \tag{MaxCutSDP-F}
  \label{eq:qubosdprenorm}
\end{equation}
Note that the tolerance $\epsilon$ determines by how much we relax two
of the constraints: the constraint that the objective function value
is at least $\gamma$, and the constraint that the diagonal elements
are $1/n$. We do not relax the constraint $\rho \succeq 0$ and the
(implied) constraint $\trace{\rho} = 1$. As a consequence, although
solutions of problem \ref{eq:qubosdprenorm} satisfy the constraints of
problem \eqref{eq:qubosdp} only approximately, they are still density
matrices; in other words, because we keep the constraints
$\trace{\rho} = 1, \rho \succeq 0$ intact, we have that $\rho \in
{\cal S}^n_{+,1}$. We can determine the optimum of problem
\eqref{eq:qubosdp} with precision $\epsilon$ by solving $\bigO{\log
  \frac{1}{\epsilon}}$ problems of the form \eqref{eq:qubosdprenorm}
for different values of $\gamma$.

\subsection{Solving the relaxation using inexact mirror descent}
\label{sec:qubosdpsolve}
Define:\index{mirror descent!inexact|(}\index{algorithm!mirror descent|(}
\begin{equation}
  \label{eq:fgammadef}
  f_{\gamma}(\rho) := \max\left\{\gamma - \trace{\hat{C} \rho}, \sum_{j=1}^{n} \abs{\rho_{jj}-\frac{1}{n}} \right\},
\end{equation}
and consider the optimization problem:
\begin{equation*}
  \left.
  \begin{array}{rr}
    \min & f_{\gamma}(\rho) \\
    \text{s.t.:} & \rho \in {\cal S}^n_{+,1}.
  \end{array}
  \right\}
\end{equation*}
It is immediate to observe that if we find $\rho$ such that
$f_{\gamma}(\rho) \le \epsilon$, $\rho$ is a solution to
\eqref{eq:qubosdprenorm}. Hence, our goal in this section is to
minimize $f_{\gamma}$ for a given value of $\gamma$, which is
iteratively modified by the binary search scheme for the optimal
objective function value.

We apply the online mirror descent scheme described in
Alg.~\ref{alg:mmwu}, see the discussion in
Sect.~\ref{sec:mirroronline}, but in this case, the objective function
is simply $f_{\gamma}$ rather than the sum of $T$ different terms
$f_1,\dots,f_T$. Thus, we pick $f_t = f_{\gamma}$ for all $t$; this
still leads to convergence to some specified precision $\epsilon$, see
Thm.~\ref{thm:qubosdpconvavg}. To take advantage of a quantum
computer, we employ a scheme whereby the iterate $\rho^{(t)}$, which
is a density matrix, is represented by a Gibbs state constructed on
the quantum computer. Classically, we keep track of the matrices
$\ham^{(t)}$ that define $\rho^{(t)}$ via matrix
exponentiation. Because the (sub)gradient updates are applied to
$\ham^{(t)}$ directly, as long as we are able to compute a classical
description of the subgradients $G^{(t)} \in \partial
f_{\gamma}(\rho^{(t)})$ we can, in principle, follow
Alg.~\ref{alg:mmwu} by updating the Hamiltonians $\ham^{(t)}$ even
without explicit classical knowledge of $\rho^{(t)}$. However, given
the definition of $f_{\gamma}$ in Eq.~\eqref{eq:fgammadef}, it is
immediate to observe that the computation of $G^{(t)} \in \partial
f_{\gamma}(\rho^{(t)})$ requires knowledge of the terms
$\trace{\hat{C}\rho^{(t)}}, \rho^{(t)}_{jj}$ appearing in the
objective function. Thus, some information about $\rho^{(t)}$ is
necessary to proceed with Alg.~\ref{alg:mmwu}. We show below that we
can use a quantum computer, together with classical knowledge of
$\ham^{(t)}$, to determine a subgradient $G^{(t)}$.
\begin{remark}
  \label{rem:qubosdphybrid}
  The crucial observation for this scheme is that we do not construct
  a full classical representation of $\rho^{(t)}$, and we do not need
  such a representation to be able to optimize: we want to avoid
  explicit classical computation of
  $\exp(\ham^{(t)})/\trace{\exp(\ham^{(t)})}$, and rely on the quantum
  computer for all calculations involving the Gibbs state. In this
  way, we do not have to classically compute the matrix exponential in
  the Gibbs state: we only compute the subgradient $G^{(t)}$.
\end{remark}
Because we aim to devise an algorithm that may not have access to an
explicit classical description of $\rho^{(t)}$, we forego the idea of
computing an exact subgradient $G^{(t)} \in \partial
f_{\gamma}(\rho^{(t)})$. Instead, we compute an \emph{inexact}
subgradient $G^{(t)} \in \partial_{\epsilon}
f_{\gamma}(\rho^{(t)})$ with the following algorithm:
\begin{itemize}
\item Estimate the following quantities:
  \begin{equation*}
    \trace{\hat{C}\rho^{(t)}}, \rho^{(t)}_{11}, \rho^{(t)}_{22},
    \dots, \rho^{(t)}_{nn},
  \end{equation*}
  with sufficient precision to guarantee that:
  \begin{equation}
    \label{eq:rhoesterorr}
    \begin{split}
    \abs{\text{est}\left(\trace{\hat{C}\rho^{(t)}}\right) - \trace{\hat{C}\rho^{(t)}}} &\le \frac{\epsilon}{4} \\
    \sum_{j=1}^n \abs{\text{est}(\rho^{(t)}_{jj})-\rho^{(t)}_{jj}} &\le \frac{\epsilon}{4},
    \end{split}
  \end{equation}
  where $\text{est}(x)$ is the computed estimate for a given quantity
  $x$.
\item If $\max\left\{\gamma -
  \text{est}\left(\trace{\hat{C}\rho^{(t)}}\right), \sum_{j=1}^n
  \abs{\text{est}(\rho^{(t)}_{jj})-\frac{1}{n}} \right\} \le
  \frac{3\epsilon}{4}$, return $G^{(t)} = \allzeroes^{n \times n}$,
  i.e., the all-zero matrix of size $n \times n$.
\item Otherwise, if $\gamma -
  \text{est}\left(\trace{\hat{C}\rho^{(t)}}\right)$ attains the
  maximum, return $G^{(t)} = -\hat{C}$.
\item Otherwise, return $G^{(t)} = \allzeroes^{n \times n} +
  \sum_{j=1}^n \left( \mathbb{I}\left( \text{est}(\rho^{(t)}_{jj}) >
  \frac{1}{n} \right) - \mathbb{I}\left( \frac{1}{n} >
  \text{est}(\rho^{(t)}_{jj})\right) \right) E_{jj}$, where $E_{jj}$
  is the $n \times n$ matrix with $1$ in position $jj$, and 0
  everywhere else (i.e., the outer product of the
  $j$-th basis vector with itself, $e_j e_j^{\top}$).
\end{itemize}
\begin{proposition}
  \label{prop:quboinexactsg}
  Let $G^{(t)}$ be computed according to the algorithm above. Then
  $G^{(t)} \in \partial_{\epsilon/2} f_{\gamma}(\rho^{(t)})$.
\end{proposition}
We prove Prop.~\ref{prop:quboinexactsg} below, but some comments are
in order first. Note that we are only interested in a subgradient
whenever $f_{\gamma}(\rho^{(t)}) > \epsilon$: if
$f_{\gamma}(\rho^{(t)}) \le \epsilon$, then $\rho^{(t)}$ is a solution
to \eqref{eq:qubosdprenorm}, so the optimization algorithm can
stop. Furthermore, if:
\begin{equation}
  \label{eq:qubohavesol}
  \max\left\{\gamma -
  \text{est}\left(\trace{\hat{C}\rho^{(t)}}\right), \sum_{j=1}^n
  \abs{\text{est}(\rho_{jj})-\frac{1}{n}} \right\} \le \frac{3\epsilon}{4},
\end{equation}
it must be the case that $f_{\gamma}(\rho^{(t)}) \le \epsilon$,
because the l.h.s.\ is an estimate of $f_{\gamma}(\rho^{(t)})$
with precision $\frac{\epsilon}{4}$ (this is easily proven with
Eq.~\ref{eq:rhoesterorr} and triangle inequalities). Thus, when
Eq.~\eqref{eq:qubohavesol} holds we can safely return $G^{(t)} =
\allzeroes^{n \times n}$, indicating that we have a solution with the
desired precision. Our proof of Prop.~\ref{prop:quboinexactsg} relies
on the following lemma.
\begin{lemma}
  \label{lem:maxsg}
  Let $h_i(x)$ be $1$-Lipschitz convex functions for $i=1,\dots,m$,
  and $f(x) = \max_{i=1,\dots,m} h_i(x)$. For given points $\bar{x},
  \hat{x}$ such that $\nrm{\bar{x} - \hat{x}} \le \frac{\epsilon}{4}$,
  let $j \in \arg \max_{i=1,\dots,m} h_i(\hat{x})$. Then, for $g \in
  \partial h_j(\hat{x})$, we have $g \in \partial_{\hat{\epsilon}}
  f(\bar{x})$, where $\hat{\epsilon} =
  \frac{\epsilon}{4}(\nrm{g}_{\ast} + 1)$.
\end{lemma}
\begin{proof}
  Using the fact that $f$ is $1$-Lipschitz because it is the maximum
  of $1$-Lipschitz functions (so $f(\bar{x}) - f(\hat{x}) \le
  \nrm{\bar{x} - \hat{x}}$), and the fact that $f(\hat{x}) \le
  h_j(\hat{x})$ because index $j$ attains the maximum in the
  expression $\max_{i=1,\dots,m} h_i(\hat{x})$, we have:
  \begin{align*}
    f(\bar{x}) + \dotp{g}{x - \bar{x}} &\le
    f(\hat{x}) + \underbrace{\nrm{\bar{x} - \hat{x}}}_{\le \frac{\epsilon}{4}} + \dotp{g}{x - \hat{x}} - \underbrace{\dotp{g}{\bar{x} - \hat{x}}}_{\le \frac{\epsilon \nrm{g}_{\ast}}{4} \text{ in abs. val.}} \\
    &\le f(\hat{x}) + \dotp{g}{x - \hat{x}} + \frac{\epsilon}{4}(\nrm{g}_{\ast} + 1) \\
    &\le \underbrace{h_j(\hat{x})  + \dotp{g}{x - \hat{x}}}_{\le h_j(x) \text{ because } g \in \partial h_j(\hat{x})} + \frac{\epsilon}{4}(\nrm{g}_{\ast} + 1)  \le h_j(x) + \frac{\epsilon}{4}(\nrm{g}_{\ast} + 1)\\
    &\le f(x) + \frac{\epsilon}{4}(\nrm{g}_{\ast} + 1).
  \end{align*}
  This shows that $g \in \partial_{\hat{\epsilon}} f(\bar{x})$.
\end{proof}

\noindent We can now prove Prop.~\ref{prop:quboinexactsg}.
\begin{proof}
  If $\max\left\{\gamma -
  \text{est}\left(\trace{\hat{C}\rho^{(t)}}\right), \sum_{j=1}^n
  \abs{\text{est}(\rho^{(t)}_{jj})-\frac{1}{n}} \right\} \le
  \frac{3\epsilon}{4}$, we return $G^{(t)} = \allzeroes^{n \times
    n}$. This is trivially an $\epsilon$-subgradient:
  $f_{\gamma}(\rho^{(t)}) \le \epsilon$ because the two terms in the
  $\max$ are estimated with error at most $\frac{\epsilon}{4}$ each,
  therefore $f_{\gamma}(\rho^{(t)}) + \dotp{\allzeroes^{n \times
      n}}{\rho - \rho^{(t)}} - \epsilon \le 0 \le f_\gamma(\rho)$ for
  all $\rho$.
  
  Now assume at least one between $\gamma -
  \text{est}\left(\trace{\hat{C}\rho^{(t)}}\right)$ and $\sum_{j=1}^n
  \abs{\text{est}(\rho^{(t)}_{jj})-\frac{1}{n}}$ is $>
  \frac{3\epsilon}{4}$. We apply Lem.~\ref{lem:maxsg} with $h_1(\rho)
  = \gamma - \trace{\hat{C}\rho}, h_2(\rho) = \sum_{j=1}^n
  \abs{\rho_{jj}-\frac{1}{n}}$ and $\bar{x} = \rho^{(t)}$. It is easy
  to see that $h_1$ and $h_2$ are 1-Lipschitz with respect to the
  trace distance (using $\nrm{\hat{C}} \le 1$). Based on our estimates
  for $\rho^{(t)}$, satisfying the guarantees in
  Eq.~\eqref{eq:rhoesterorr}, we are evaluating $h_1, h_2$ at a point
  $\text{est}(\rho^{(t)})$ that has trace distance at most
  $\frac{\epsilon}{4}$ from $\rho^{(t)}$. (Note: we may not have
  explicit knowledge of the full $\text{est}(\rho^{(t)})$, because we
  only estimate the diagonal elements as well as
  $\trace{\hat{C}\rho^{(t)}}$, but this is not necessary.) Then,
  Lem.~\ref{lem:maxsg} implies that if we take the maximum of $h_1$
  and $h_2$, and return a subgradient of the corresponding function at
  $\text{est}(\rho^{(t)})$, we have obtained $G^{(t)} \in
  \partial_{\hat{\epsilon}} f_{\gamma}(\rho^{(t)})$. Thus, we just
  need to show that we are correctly returning a subgradient of $h_1$
  or $h_2$ at $\text{est}(\rho^{(t)})$, depending on which one attains
  the maximum. The function $h_1$ is linear in $\rho$ and we return
  $-\hat{C}$ (recall Rem.~\ref{rem:tracelinearity}) if $h_1$ attains
  the $\max$. The second function is a sum of absolute values
  $\abs{\rho_{jj}-\frac{1}{n}}$, and we return the sum of a
  subgradient for each term at $\text{est}(\rho^{(t)})$ ($E_{jj}$ if
  $\text{est}(\rho^{(t)}_{jj}) > \frac{1}{n}$, $-E_{jj}$ if
  $\text{est}(\rho^{(t)}_{jj}) < \frac{1}{n}$) if $h_2$ attains the
  $\max$. Finally, note that the norm of the returned subgradient $g =
  G^{(t)}$ always satisfies $\nrm{g}_{\ast} \le 1$, because
  $\nrm{\hat{C}} \le 1$ and $\max_{a_j \in \{-1,+1\}, \rho \in {\cal
      S}^n_{+,1}} \dotp{\sum_{j=1}^n a_j E_{jj}}{\rho} = 1$. Thus,
  $\hat{\epsilon} = \frac{\epsilon}{4}(\nrm{g}_{\ast} + 1) =
  \frac{\epsilon}{2}$, and $G^{(t)} \in \partial_{\epsilon/2}
  f_{\gamma}(\rho^{(t)})$.
\end{proof}

\noindent We can now state the main convergence result, adapted from
\cite{brandao2022faster}.
\begin{theorem}[Convergence of mirror descent for MaxCut SDP]
  \label{thm:qubosdpconvavg}
  Suppose there exists $\rho^{\ast} \in {\cal S}^n_{+,1}$ such that
  $f_{\gamma}(\rho^{\ast}) = 0$, i.e., problem \eqref{eq:qubosdprenorm} is
  feasible with $\epsilon = 0$. Then, the mirror descent algorithm
  Alg.~\ref{alg:mmwu} with step size $\eta = \frac{\epsilon}{16}$,
  inexact subgradients $G^{(t)} \in \partial_{\epsilon/2}
  f_{\gamma}(\rho^{(t)})$ computed as described above, and number of
  steps $T = \frac{64}{\epsilon^2} \log n$, returns density matrices
  $\rho^{(1)},\dots,\rho^{(T)}$ such that:
  \begin{equation*}
    f_{\gamma}\left(\frac{1}{T} \sum_{t=1}^T \rho^{(t)} \right) \le 
    \epsilon.
  \end{equation*}
\end{theorem}
\begin{proof}
  We apply Thm.~\ref{thm:mirrorconv}. Because in this case we are
  using $\frac{\epsilon}{2}$-subgradients rather than exact
  subgradients, we need to modify the convergence result slighly: in
  each iteration the objective function may be worse compared to the
  exact case, by an amount that is at most
  $\frac{\epsilon}{2}$. Applying this change to
  Thm.~\ref{thm:mirrorconv}, recalling that the strong convexity
  parameter is $\alpha = 1$ (see Sect.~\ref{sec:mirroronline} and
  \cite{yu2013strong}), we obtain the following result:
  \begin{equation}
    \label{eq:epsregretbound}
    \sum_{t=1}^T f_{\gamma}(\rho^{(t)}) - \sum_{t=1}^T f_{\gamma}(\rho^\ast) \le \frac{1}{\eta} D_h(\rho^\ast \| I/n) + \frac{\eta}{2} \sum_{t=1}^T \nrm{G^{(t)}}_{\ast}^2 + \frac{\epsilon}{2} T.
  \end{equation}
  (A formal proof of this inequality can be obtained by modifying the
  convergence proof for mirror descent and using the definition of
  $\frac{\epsilon}{2}$-subgradient instead of the exact subgradient
  inequality; this makes the bound worse by an additive term
  $\frac{\epsilon}{2}$ in every iteration. See, e.g., the proof in
  \cite[Thm.~4.2]{bansal2019potential}, and the analysis for inexact
  subgradients in \cite{nedic2014stochastic}.) The function
  $f_{\gamma}$ is convex, so we can use Jensen's inequality
  $f_{\gamma}(\frac{1}{T} \sum_t \rho^{(t)}) \le \frac{1}{T} \sum_t
  f_{\gamma}(\rho^{(t)})$. Combining this with
  Eq.~\eqref{eq:epsregretbound}, after dividing by $T$ on both sides,
  and remembering that $D_h(\rho^\ast \| I/n) \le \log n$ (see
  Sect.~\ref{sec:mirroronline}), we obtain:
  \begin{equation*}
    f_{\gamma}\left(\frac{1}{T}
    \sum_{t=1}^T \rho^{(t)}\right) \le \frac{1}{T} \sum_{t=1}^T
    f_{\gamma}(\rho^{(t)}) \le \frac{1}{T} \sum_{t=1}^T f_{\gamma}(\rho^\ast) + \frac{\log n}{\eta T} + \frac{\eta}{2} + \frac{\epsilon}{2} \le \frac{\epsilon}{4} + \frac{\epsilon}{32} + \frac{\epsilon}{2} \le  \epsilon,
  \end{equation*}
  where in the second inequality we used the fact that
  $f_{\gamma}(\rho^\ast) = 0$ by assumption, and we substituted the
  values for $\eta$ and $T$ given in the theorem statement.
\end{proof}
\begin{remark}
  Using the same choices for $\eta, T, G^{(t)}$ as indicated in
  Thm.~\ref{thm:qubosdpconvavg}, it is possible to show that the last
  iterate $\rho^{(T)}$ of Alg.~\ref{alg:mmwu} (as opposed to the
  average of the iterates) satisfies $f_{\gamma}(\rho^{(T)}) \le
  \epsilon$, if an exactly feasible solution for
  \eqref{eq:qubosdprenorm} exists. The proof of this result follows
  from \cite{brandao2022faster}, in particular Theorem 2.1 and Lemma
  3.1 therein, after noting that this specific instantiation of
  Alg.~\ref{alg:mmwu} follows exactly the same steps as the algorithm
  described in \cite{brandao2022faster}. The proof technique in
  \cite{brandao2022faster} uses a potential function argument, see
  also \cite{tsuda2005matrix}.

  Typically, convergence for online mirror descent is shown for the
  average of the iterates, rather than for the last iterate only, see,
  e.g., \cite{allen2017linear}. We discuss convergence for the average
  iterate so that we can directly use Thm.~\ref{thm:mirrorconv}: the
  bound on the quality of the final iterate is similar to the bound on
  the quality of the average iterate. However, the time complexity for
  computing properties of the optimal solution can be worse when
  considering the average iterate, because to construct a mixed state
  corresponding to the average iterate, we have to construct a linear
  combination of Gibbs states rather than a single Gibbs state; we can
  do that with Prop.~\ref{prop:lincombblock}, but it may come at a
  cost, depending on the input model (see the discussion in
  Sect.~\ref{sec:qubosdptime}). Overall, in theory considering the
  last iterate only is the better choice for this specific problem: we
  consider the average of the iterates for educational purposes, and
  because the computational complexity difference is not significant
  in the QRAM model.
\end{remark}

\subsection{Complexity of the quantum algorithm}
\label{sec:qubosdptime}
Thm.~\ref{thm:mirrorconv} states that Alg.~\ref{alg:mmwu} runs for $T
= \frac{64}{\epsilon^2} \log n$ iterations. In this section we analyze
the cost (i.e., gate complexity) of each iteration, and use it to
derive the overall complexity of the algorithm.

The main bottleneck of the algorithm is the computation of the
subgradient $G^{(t)}$. As remarked in Rem.~\ref{rem:qubosdphybrid}, we
aim for a scheme where the Hamiltonians $\ham^{(t)}$ in
Alg.~\ref{alg:mmwu} are stored classically, but all computations
involving the corresponding Gibbs state $\rho^{(t)} =
\exp(\ham^{(t)})/\trace{\exp(\ham^{(t)})}$ are performed on a quantum
computer. In fact, it is straightforward to verify that the only thing
we need to proceed with Alg.~\ref{alg:mmwu} is a classical description
of the subgradient $G^{(t)}$. According to the procedure described in
Sect.~\ref{sec:qubosdpsolve}, we can classically construct $G^{(t)}$
after estimating $\trace{\hat{C}\rho^{(t)}}, \rho^{(t)}_{11},
\rho^{(t)}_{22}, \dots, \rho^{(t)}_{nn}$ with sufficient precision to
satisfy Eq.~\eqref{eq:rhoesterorr}. Thus, our next task is to analyze
the complexity of estimating these quantities. We use $\bigOt{\cdot}$
notation in the analysis below, and for simplicity we neglect small
factors that would eventually be absorbed in the $\bigOt{\cdot}$
notation anyway.

It is obvious that the estimation of $\trace{\hat{C}\rho^{(t)}},
\rho^{(t)}_{11}, \rho^{(t)}_{22}, \dots, \rho^{(t)}_{nn}$ requires the
ability to construct $\rho^{(t)}$. Suppose we have access to a
block-encoding of $\ham^{(t)}$ with subnormalization factor $\alpha$
and sufficient precision (the precision is one of those parameters
that can be swept under the rug in this analysis: it does not impact
the final running time expression). Because $\rho^{(t)}$ is a Gibbs
state, we use the Gibbs state construction procedure analyzed in
Prop.~\ref{prop:gibbsstateconstr}. Its complexity is $\bigOt{\sqrt{n}
  \alpha}$ calls to a block-encoding for $\ham^{(t)}$, and a similar
number of additional gates. Then Prop.~\ref{prop:traceest} lets us
estimate $\trace{\hat{C}\rho^{(t)}}$ with one copy of $\rho^{(t)}$ and
$\bigOt{\alpha}$ applications of the block-encoding for $\hat{C}$ and
additional gates, for a high-precision estimate. The estimation of
$\rho^{(t)}_{11}, \rho^{(t)}_{22}, \dots, \rho^{(t)}_{nn}$ is also
quite simple, after figuring out the correct strategy.
\begin{remark}
  As we are attempting to estimate elements of the density matrix
  describing a mixed quantum state, this falls under the umbrella of
  quantum state tomography. However, Thm.~\ref{thm:tomography} is not
  exactly designed for the task at hand: it assumes access to a state
  preparation unitary for a pure state, whereas we are now dealing
  with a mixed state. Furthermore, Thm.~\ref{thm:tomography} estimates
  amplitudes, but the diagonal elements $\rho^{(t)}_{11},
  \rho^{(t)}_{22}, \dots, \rho^{(t)}_{nn}$ are the probabilities of
  observing the basis states $\ket{\vj}, \vj \in \{0,1\}^{\ceil{\log
      n}}$.
\end{remark}
Because our goal is to estimate the probabilities of observing
$\ket{\vj}$, for $\vj \in \{0,1\}^{\ceil{\log n}}$, when applying
measurements to the mixed quantum state represented by $\rho^{(t)}$,
there is no need to use the complicated tomography procedure (based on
the quantum gradient algorithm) described in
Thm.~\ref{thm:tomography}. It is simpler to repeatedly construct a
purification of $\rho^{(t)}$ via Prop.~\ref{prop:gibbsstateconstr},
perform a measurement of the qubits corresponding to the state
register (i.e., we ignore the purifying register), and estimate the
probabilities by counting the observations. We need to satisfy
Eq.~\eqref{eq:rhoesterorr}, hence we require $\sum_{j=1}^n
\abs{\text{est}(\rho^{(t)}_{jj})-\rho^{(t)}_{jj}} \le
\frac{\epsilon}{4}$. This is the same as estimating, by taking
samples, an $n$-dimensional vector of probabilities with $\ell^1$-norm
distance at most $\frac{\epsilon}{4}$ from the true vector of
probabilities. It is known that this can be achieved with high
probability by taking $\bigOt{\frac{n}{\epsilon^2}}$ samples, see,
e.g., \cite{canonne2020short}. This brings the overall complexity of
the subgradient estimation procedure to:
\begin{equation*}
  \bigOt{\frac{\alpha n^{1.5}}{\epsilon^2}}
\end{equation*}
calls to a block-encoding of $\ham^{(t)}$ with subnormalization factor
$\alpha$, and a similar number of additional gates; the term comes
from estimating the diagonal elements, the cost for estimating
$\trace{\hat{C}\rho^{(t)}}$ is $\bigOt{\alpha \sqrt{n}}$, so it is
asymptotically negligible in $\bigO{\cdot}$ notation.

Now we can focus on the complexity of constructing the block-encoding
of $\ham^{(t)}$, and determining its subnormalization factor $\alpha$. For
this, we need to analyze the structure of $\ham^{(t)}$. According to
Alg.~\ref{alg:mmwu}, $\ham^{(t)}$ is simply an accumulation of
subgradients $G^{(t)}$ (more precisely,
$\frac{\epsilon}{2}$-subgradients in this case). The following
properties hold.
\begin{lemma}
  \label{lem:qubohk}
  For every iteration $t$ of Alg.~\ref{alg:mmwu} applied to
  \eqref{eq:qubosdprenorm} as described in
  Sect.~\ref{sec:qubosdpsolve}, $\ham^{(t)} = y_1 \hat{C} + y_2 D$ for
  some vector $y \in \R^2$, where $D$ is a diagonal
  matrix. Furthermore, $\nrm{y}_1 \le \frac{4}{\epsilon} \log n$.
\end{lemma}
\begin{proof}
  For each $t$, $G^{(t)}$ is either $\hat{C}$ or a diagonal matrix
  with $-1, 0, +1$ on the diagonal. Then it is clear that in every
  iteration we can express $\ham^{(t)}$ in the stated form for some
  coefficients $y_1, y_2$. In every iteration we accumulate $G^{(t)}$
  with coefficient $\eta = \frac{\epsilon}{16}$, so either $y_1$ or
  $y_2$ changes by at most $\frac{\epsilon}{16}$ (we can keep $D$
  normalized so that its entries are less than 1 in absolute
  value). Because initially $y_1 = y_2 = 0$ and we perform $T =
  \frac{64}{\epsilon^2} \log n$ iterations, we have $\nrm{y}_1 \le
  \eta T = \frac{4}{\epsilon} \log n$.
\end{proof}

\noindent The above result lets us utilize
Prop.~\ref{prop:lincombblock} to construct the block-encoding of
$\ham^{(t)}$: assuming access to a block-encoding for $\hat{C}$, and a
block-encoding for the diagonal matrix $D$ of Lem.~\ref{lem:qubohk},
linear combination of block-encodings produces the desired quantum
circuit. It is now necessary to fix the input model, so that we can
analyze the cost for block-encoding $\hat{C}$ and $D$. To simplify the
analysis and --- at the same time --- obtain the fastest asymptotic
running time, we assume that we have access to QRAM: in this way, we
can utilize the sparse-oracle model of Prop.~\ref{prop:sparseblockenc}
or the QRAM model of Prop.~\ref{prop:qramblockenc}, whichever is
faster. To block-encode $\hat{C}$ we rely on
Prop.~\ref{prop:qramblockenc}: because $\|\hat{C}\|_F = 1$, this gives
us a $(1, \bigOt{1}, 0)$-block-encoding with $\bigOt{1}$ gates and
accesses to QRAM. To block-encode $D$ we rely on
Prop.~\ref{prop:sparseblockenc}, because $D$ is diagonal: this gives
us a $(1, \bigOt{1}, \xi)$-block-encoding, where $\xi$ can be made
extremely small with little extra cost and can be assumed to be zero
to avoid burdensome details, see
Rem.~\ref{rem:sparseblockencerror}. This block-encoding uses
$\bigOt{1}$ calls to oracles describing $D$ and additional gates, but
in the QRAM model, the oracles describing $D$ have constant cost: the
sparsity of $D$ is fixed and known (it is a diagonal matrix), its
elements can be stored in QRAM and queried at unit cost. So, a $(1,
\bigOt{1}, 0)$-block-encoding of $\hat{C}$ and $D$ can be constructed
with $\bigOt{1}$ accesses to QRAM and additional
gates. Prop.~\ref{prop:lincombblock} then gives us a $(\nrm{y}_1,
\bigOt{1}, 0)$-block-encoding of $\ham^{(t)}$ using a constant number
of queries to block-encodings for $\hat{C}$, $D$, and to a
state-preparation pair for $y$. Because $y$ is $2$-dimensional, the
state-preparation pair has $\bigOt{1}$ cost for any precision.

We now have all the ingredients to state the complexity of the
algorithm.
\begin{proposition}
  \label{prop:qubosdpcost}
  Given access to a QRAM of size $\bigOt{n^2}$, we can determine a
  solution to \eqref{eq:qubosdp} with optimality and feasibility
  tolerance $\epsilon$ (i.e., problem \eqref{eq:qubosdprenorm}) using
  $\bigOt{\frac{n^{1.5}}{\epsilon^5}}$ accesses to the QRAM and
  additional gates, and $\bigOt{n^2}$ classical operations.
\end{proposition}
\begin{proof}
  The complexity of the subgradient estimation is:
  \begin{equation*}
    \bigOt{\frac{\alpha n^{1.5}}{\epsilon^2}}
  \end{equation*}
  calls to a block-encoding of $\ham^{(t)}$ with subnormalization
  factor $\alpha$, and a similar number of additional gates. Because
  $\alpha = \nrm{y}_1$, and $\nrm{y}_1 \le \frac{4}{\epsilon} \log n$
  by Lem.~\ref{lem:qubohk}, substituting in the above the expression
  gives the asymptotic complexity bound
  $\bigOt{\frac{n^{1.5}}{\epsilon^3}}$ for subgradient computation in
  each iteration. The number of iterations of the algorithm is
  $\bigOt{\frac{1}{\epsilon^2}}$; multiplying the per-iteration
  complexity by the number of iterations, we obtain the stated total
  complexity for the quantum part. The classical cost is inherited
  from Prop.~\ref{prop:qramblockenc}, to prepare the QRAM data
  structures for the block-encoding of $\hat{C}$.
\end{proof}
\begin{remark}
  Without QRAM, the main difference in the running time analysis is
  that the construction of block-encodings for $\hat{C}$ and $D$ may
  not be as efficient. For example, in the sparse-oracle access model,
  block-encoding $D$ has gate complexity $\bigOt{n}$ (because there
  are $n$ diagonal elements that can be queried, potentially in
  superposition, so the oracle that returns their value implements a
  lookup table), and block-encoding $\hat{C}$ may have gate complexity
  up to $\bigOt{n^2}$ if the matrix is dense. The density of $\hat{C}$
  may also make the subnormalization factor worse due to
  Prop.~\ref{prop:sparseblockenc}, leading to a significant
  deterioration of the performance of the algorithm.
\end{remark}
\begin{remark}
  \label{rem:qubosdpcostorig}
  Prop.~\ref{prop:qubosdpcost} analyzes the complexity of obtaining a
  solution to problem \eqref{eq:qubosdp}, but this is not the same as
  the original problem \eqref{eq:qubosdporig}. In particular, because
  we scaled down the objective function by a factor $\nrm{C}_F$, as
  well as the decision variable $X$ by a factor $n$, to obtain a
  solution to problem \eqref{eq:qubosdporig} with additive error
  $\epsilon$ it is sufficient to set the error in problem
  \eqref{eq:qubosdp} to $n \nrm{C}_F \epsilon$. This yields an
  approximate solution to problem
  \eqref{eq:qubosdporig}. \cite{brandao2019quantum} shows that if
  $\epsilon$ is chosen small enough, an approximate solution to
  problem \eqref{eq:qubosdporig} can be turned into an exactly
  feasible solution (i.e., all constraints are satisfied exactly, not
  approximately) with a small deterioration of the objective
  function. Furthermore, one can use a randomized rounding procedure
  to turn a solution to problem \eqref{eq:qubosdporig} into a solution
  for the original discrete problem \eqref{eq:pm1qp}, with an
  approximation guarantee on the expected value of the objective
  function (similar in spirit to \cite{goemans1995improved}).\index{semidefinite programming|)}\index{mirror descent!inexact|)}\index{algorithm!mirror descent|)}
\end{remark}

\section{Notes and further reading}
\label{sec:mmwunotes}
The mirror descent algorithm for continuous optimization was initially
proposed in \cite{nemirovski1983problem}, and since then, it has been
used extensively. For a derivation of mirror descent starting from the
projected subgradient algorithm, as well as a detailed convergence
analysis, we refer the reader to \cite{beck2003mirror}. A clear
exposition of proof techniques for convergence rates using potential
functions (including for mirror descent) can be found in
\cite{bansal2019potential}. The relationship between MMWU and mirror
descent is addressed in an appendix in \cite{allen2014linear}.

The MWU\index{matrix!multiplicative weights update}\index{algorithm!multiplicative weights update} algorithm has its origin in the Fictitious Play algorithm from
game theory \cite{brown1951iterative}, although it has been
rediscovered multiple times under different names in several fields.
An overview of the MWU algorithm and its applications to optimization
is given in the excellent survey \cite{arora2012multiplicative}, see
also the references mentioned therein. The classical MMWU algorithm
for SDP is described in
\cite{arora2005fast,arora2016combinatorial}. The implementation and
computational evaluation of some variants of the MWU algorithm to
mixed-integer nonlinear optimization is discussed in
\cite{mencarelli2017multiplicative}.

The quantum MMWU was initially proposed in
\cite{brandao2017quantum,van2020sdp}. The framework, in its first
instantiation, presented several limitations, and did not yield an
end-to-end speedup over classical algorithms for most problems, see
\cite{van2020sdp} for a discussion. Nonetheless, all the basic ingredients
of the framework were already present in these early
works. Subsequent work has attempted to remove some of the limitations
and improve the complexity of the algorithm, see
\cite{brandao2019quantum,van2018improvements}; these more recent works
also include a primal-only algorithm, as opposed to the primal-dual
framework discussed in this chapter. Despite these improvements, the
running time dependence of these algorithms on the final optimality
gap, as well as the size of the optimal primal and dual solution,
remains poor. 

The quadratic unconstrained $\{-1,+1\}$ optimization problem
\eqref{eq:pm1qp} finds applications in areas such as image compression
\cite{o1983digital}, correlation clustering \cite{mei2017solving},
structured principal component analysis \cite{kueng2021binary}. It is
strongly related to the Ising model
\cite{barahona1982computational}. The quantum algorithm discussed in
Sect.~\ref{sec:qubosdpsolve} to solve the SDP relaxation of
\eqref{eq:pm1qp} was first presented in \cite{brandao2022faster},
where it is called ``Hamiltonian updates''; the presentation in
\cite{brandao2022faster} does not rely on mirror descent, so
convergence is proven from first principles, outside the mirror
descent framework. Such SDP relaxation has only diagonal constraints,
i.e., constraints on the diagonal elements of the matrix, and more
specifically it imposes that the diagonal elements are equal to
$1$. SDPs with only diagonal constraints admit specialized classical
algorithms as well, see, e.g., \cite{lee2020m}. Convergence of the
final iterate of stochastic mirror descent, in addition to convergence
of the average iterate, is discussed in \cite{nedic2014stochastic}.

\chapter{Optimization with the adiabatic theorem}
\label{ch:adiabatic}
\thispagestyle{fancy}
The adiabatic theorem is a powerful result concerning the evolution of
quantum-mechanical systems. Intuitively, it states the following:
suppose we want to study the evolution of a quantum-mechanical system
governed by a given Hamiltonian $\ham$ (Def.~\ref{def:hamiltonian}),
and the initial state of the system is the eigenstate of $\ham$
corresponding to its lowest eigenvalue. The laws of quantum physics
tell us that the evolution of the system over time follows the
Schr\"odinger equation, Eq.~\eqref{eq:schrodinger}. Suppose now that
the Hamiltonian is not fixed, but rather, it changes very slowly over
time, so that the evolution of the system is described by the
differential equation $i \frac{\di \ket{\psi(t)}}{\di t} = \ham(t)
\ket{\psi(t)}$, and $\ham(t)$ is slowly changing. If, for any value of
the time parameter $t$, there is always a gap between the lowest
eigenvalue of $\ham(t)$ and all other eigenvalues, then the adiabatic
theorem asserts that the state of the system always remains in an
eigenstate with the lowest eigenvalue throughout the evolution. One
specific case of interest in this chapter occurs when the
time-dependent Hamiltonian $\ham(t)$ varies from an initial
(time-independent) Hamiltonian $\ham_{\text{I}}$ to a final
Hamiltonian $\ham_{\text{F}}$. If the change happens slowly, and the
initial state of the system is an eigenstate of $\ham_{\text{I}}$ with
minimum eigenvalue, the adiabatic theorem asserts that the evolution
eventually leads to the eigenstate with minimum eigenvalue of the
final Hamiltonian $\ham_{\text{F}}$. This result can be used for
optimization: imagine that the initial Hamiltonian encodes an easy
problem for which we can construct the eigenstate with minimum
eigenvalue, and the final Hamiltonian encodes a difficult optimization
problem whose solution is given by the eigenstate with minimum
eigenvalue. By the adiabatic theorem, evolving the eigenstate of the
initial Hamiltonian while slowly changing the Hamiltonian into the
final one leads to the solution of the difficult optimization
problem. Of course we need to know how slowly $\ham_{\text{I}}$ should
be changed into $\ham_{\text{F}}$. The adiabatic theorem is at the
heart of the quantum approximate optimization algorithm (QAOA), a
framework that has been widely used to implement optimization
algorithms on existing quantum hardware, because it has low
requirements of quantum resources --- although it does not guarantee
improvements over classical algorithms in general. In this chapter we
first discuss the adiabatic theorem, and then give an overview of
QAOA.
\begin{remark}
  The term \emph{adiabatic}\index{adiabatic!process} refers to a process that occurs without
  heat or mass transfer between a thermodynamic system and its
  environment. For the purposes of this chapter (and quantum
  algorithms more in general), it should be intended as a process
  where the evolution of the system is slow enough that the system has
  time to adapt, i.e., remain in some instantaneous eigenstate
  evolving through time, typically the one corresponding to the
  smallest eigenvalue. Conversely, in a \emph{diabatic} process the
  system evolves too rapidly, and it may not remain in an
  instantaneous eigenstate of the system.
\end{remark}

\section{The adiabatic theorem}
\label{sec:adiabatic}
Before diving into a formal statement of the adiabatic theorem, it may
be helpful to make its connection with optimization apparent. In this
chapter, optimization refers to the NP-hard problem of optimizing a
quadratic function over binary variables.

\subsection{Combinatorial optimization as an eigenvalue problem}
\label{sec:combopteig}
Suppose we have the following optimization problem:
\begin{equation}
  \label{eq:01opt}
  \min f(\vx) \qquad \vx \in \{0,1\}^n.
\end{equation}
We put no restriction on $f : \{0,1\}^n \to \R$ for now, therefore we
can encode any combinatorial optimization problem in this way (e.g.,
by assigning value $\infty$ to infeasible binary strings, if there are
any, or some large finite value that effectively amounts to
$\infty$). We encode this problem in a diagonal Hamiltonian $\ham \in
\R^{2^n \times 2^n}$ defined as follows:
\begin{equation*}
  \ham := \sum_{\vj \in \{0,1\}^n} f(\vj) \ketbra{\vj}{\vj}.
\end{equation*}
Note that $\ham$ contains all the possible objective function values
on its diagonal, each one in the position associated with the
corresponding binary string ($\bra{\vj} \ham \ket{\vj} = f(\vj) \quad
\forall \vj \in \{0,1\}^n$), and zeroes everywhere else ($\bra{\vj}
\ham \ket{\vk} = 0 \quad \forall \vj \neq \vk$). Then problem
\eqref{eq:01opt} can be trivially reformulated as follows:
\begin{equation*}
  \min_{\vj \in \{0,1\}^n} \bra{\vj} \ham \ket{\vj},
\end{equation*}
and this problem has the same optimal solution as:\index{eigenstate}
\begin{equation}
  \label{eq:mineig}
  \min_{\ket{\psi}} \bra{\psi} \ham
  \ket{\psi}.
\end{equation}
To see the equivalence, simply note that $\ket{\vj}, \vj \in
\{0,1\}^n$ is an eigenbasis for $\ham$ (because $\ham$ is diagonal), so we
can express every state $\ket{\psi}$ in terms of the eigenbasis, and
the problem becomes:
\begin{equation*}
  \min_{\substack{\alpha \in \C^{2^n}\\ \nrm{\alpha} = 1}} \left( \sum_{\vj \in \{0,1\}^n} \alpha^{\dag}_j \bra{\vj}\right) \ham \left( \sum_{\vj \in \{0,1\}^n} \alpha_j \ket{\vj}\right) = \min_{\substack{\alpha \in \C^{2^n}\\ \nrm{\alpha} = 1}} \sum_{\vj \in \{0,1\}^n} |\alpha_j|^2  \ham_{jj} = \min_{\vj \in \{0,1\}^n} f(\vj).
\end{equation*}
The same argument also works to show that for every Hamiltonian $\ham$
--- including those that are not necessarily diagonal --- problem
\eqref{eq:mineig} is a minimum eigenvalue problem, i.e., it is
equivalent to determining $\lambda_{\min}(\ham)$, the minimum
eigenvalue of $\ham$. Recall that Hamiltonians are Hermitian. Let
$\ket{\psi_0},\dots,\ket{\psi_{2^n-1}}$ be a basis of orthonormal
eigenstates of $\ham$, and $V$ a matrix with those vectors as its
columns. Then $\ham = V \Lambda V^{\dag}$, and
\begingroup
\allowdisplaybreaks
\begin{align*}
  \min_{\ket{\psi} \in \left(\C^2\right)^{\otimes n}} \bra{\psi} \ham
  \ket{\psi} &= \min_{\substack{\alpha \in \C^{2^n}\\ \nrm{\alpha} = 1}} \left( \sum_{\vj \in \{0,1\}^n} \alpha^{\dag}_j \bra{\psi_j}\right) \ham \left( \sum_{\vj \in \{0,1\}^n} \alpha_j \ket{\psi_j}\right) \\
  &= \min_{\substack{\alpha \in \C^{2^n}\\ \nrm{\alpha} = 1}} \left( \sum_{\vj \in \{0,1\}^n} \alpha^{\dag}_j \bra{\psi_j}\right) V \Lambda V^{\dag} \left( \sum_{\vj \in \{0,1\}^n} \alpha_j \ket{\psi_j}\right) \\
  &= \min_{\substack{\alpha \in \C^{2^n}\\ \nrm{\alpha} = 1}} \left( \sum_{\vj \in \{0,1\}^n} \alpha^{\dag}_j \bra{\vj}\right) \Lambda \left( \sum_{\vj \in \{0,1\}^n} \alpha_j \ket{\vj}\right) \\
  &=  \min_{\substack{\alpha \in \C^{2^n}\\ \nrm{\alpha} = 1}} \sum_{\vj \in \{0,1\}^n} |\alpha_j|^2 \Lambda_{jj} = \lambda_{\min}(\ham).
\end{align*}
\endgroup
Thus, the general combinatorial optimization problem in
Eq.~\eqref{eq:01opt} can be solved as the problem of determining the
minimum eigenvalue of a certain Hamiltonian, written compactly as
problem \eqref{eq:mineig}.

One may wonder how to construct the Hamiltonian $\ham = \sum_{\vj \in
  \{0,1\}^n} f(\vj) \ketbra{\vj}{\vj}$ starting from a ``reasonable''
description of $f$: usually listing all values of $f(\vj)$ is not
efficient, because there are $2^n$ of them. Thus, we are looking for
some implicit, more compact representation, for example by means of an
analytical description of the objective function, and that
representation should allow for efficient construction of $\ham$. An
especially straightforward way to construct $\ham$ is applicable
whenever we start from a problem stated as a quadratic unconstrained
binary optimization problem (QUBO)\index{QUBO}, see Def.~\ref{def:qubo}, because
there is a simple transformation from a QUBO to the desired
Hamiltonian. Also note that, because QUBOs are NP-complete\index{complexity!class NP}
\cite{barahona1982computational}, any problem in NP can be formulated
as a QUBO with an appropriate polynomial transformation.
\begin{remark}
  \label{rem:qubowise}
  The fact that it is \emph{possible} to formulate every problem in NP
  as a QUBO does not mean that it is \emph{wise} to do so. For
  example, it is well known that polynomial transformations (intended
  here as polynomial-time reductions, or Karp reductions) map an
  instance of a problem to an instance of a different problem with
  polynomial overhead \cite{gareyjohnson}. However, the overhead for
  such a mapping can be very large, and applying a solution algorithm
  to the transformed problem could be considerably less efficient, in
  practice, than applying a solution algorithm to the original
  problem. Thus, formulating a combinatorial optimization problem as a
  QUBO, and solving it as such, may end up being much slower than
  solving the problem in a different form.
\end{remark}
\begin{definition}[Quadratic Unconstrained Binary Optimization (QUBO) problem]\label{def:qubo}
    The following problem is called a \emph{Quadratic Unconstrained Binary Optimization (QUBO)} problem with $n$ decision variables:
    \begin{equation}
      \tag{QUBO}
      \label{eq:qubo}
      \min_{\vx \in \{0,1\}^n} \vx^{\top} Q \vx,
    \end{equation}
    where $Q \in \R^{n \times n}$ is a symmetric matrix.
\end{definition}
\noindent (In certain algebraic expressions, such as the above, we use
binary strings also as ``regular'' 0-1 column vectors, allowing them to
appear in operations such as matrix-vector multiplication. The meaning
of these expressions should be clear from the context.)

The mapping from \eqref{eq:qubo} to a Hamiltonian is via Pauli $Z$
matrices. First, transform the $\{0,1\}$-variables $\vx_j$ into
$\{-1, +1\}$-variables $z_j$, using the linear transformation $\vx_j =
(1 - z_j)/2$. With this transformation, $z = 1 - 2\vx$ so $z \in
\{-1, +1\}^n$. Problem \eqref{eq:qubo} becomes:
\begin{equation}
  \label{eq:ising}
  \min_{z \in \{-1, +1\}^n} z^{\top} A z + c^{\top} z + b
\end{equation}
where $A \in \R^{n \times n}, c \in \R^{n}, b \in \R$ are easily
computed from \eqref{eq:qubo}:
\begin{equation*}
  A = \frac{Q}{4} \qquad c^{\top} = -\frac{\allones^{\top} Q}{2} \qquad b = \frac{\allones^{\top} Q \allones}{4}.
\end{equation*}
The sought Hamiltonian can be obtained from these quantities using
matrices $\sigma^Z_j$:
\begin{equation}
  \label{eq:sigmazdef}
  \sigma^Z_j := \underbrace{I \otimes \dots \otimes I \otimes \hspace*{-1.2em} \overset{\substack{\text{position } j\\\downarrow}}{Z} \hspace*{-1.2em} \otimes I \dots \otimes I}_{n \text{ times}},
\end{equation}
where $Z$ is the Pauli $Z$ matrix as given in Def.~\ref{def:pauli}.
\begin{proposition}
  \label{prop:isingtoham}
  The Hamiltonian:
  \begin{equation*}
    \ham := \sum_{j,k=1}^n A_{jk} \sigma^Z_j \sigma^Z_k + \sum_{j=1}^n c_j \sigma^Z_j
  \end{equation*}
  satisfies the properties:
  \begin{equation*}
    \bra{\vx} \ham \ket{\vx} = z^{\top} A z + c^{\top} z \quad \forall \vx \in \{0,1\}^n, \qquad
    \bra{\vj} \ham \ket{\vk} = 0 \quad \forall \vj \neq \vk,
  \end{equation*}
  where $z = 1 - 2\vx \in \{-1,+1\}^n$.
\end{proposition}
\begin{proof}
  Both properties can be verified with simple algebraic
  manipulations. It is immediate to see that the matrices $\sigma^Z_j,
  \sigma^Z_k$ commute when $j \neq k$. By definition we have:
  \begin{equation}
    \label{eq:vxhvx}
    \bra{\vx} \ham \ket{\vx} = \bra{\vx} \left(\sum_{j,k=1}^n A_{jk} \sigma^Z_j \sigma^Z_k + \sum_{j=1}^n c_j \sigma^Z_j \right)\ket{\vx}.    
  \end{equation}
  Note that for every $h=1,\dots,n$,
  \begin{equation*}
    \bra{\vj} \sigma^Z_h \ket{\vk} = \braket{\vj_1}{\vk_1} \otimes \dots \otimes \braket{\vj_{h-1}}{\vk_{h-1}} \otimes \bra{\vj_h}Z\ket{\vk_h} \otimes \braket{\vj_{h+1}}{\vk_{h+1}} \otimes \dots \otimes \braket{\vj_n}{\vk_n},
  \end{equation*}
  so each such term is zero if $\vj \neq \vk$, and it is equal to
  $\bra{\vj_h} Z \ket{\vj_h} = (-1)^{\vj_h}$ if $\vj =
  \vk$. Similarly, for every $h,\ell=1,\dots,n$:
  \begin{equation*}
    \bra{\vj} \sigma^Z_h \sigma^Z_{\ell} \ket{\vk} = \braket{\vj_1}{\vk_1} \otimes \dots \otimes  \underbrace{\bra{\vj_h}Z\ket{\vk_h}}_{\text{position } h} \otimes \dots \otimes \underbrace{\bra{\vj_{\ell}}Z\ket{\vk_{\ell}}}_{\text{position } \ell} \otimes \dots \otimes \braket{\vj_n}{\vk_n},
  \end{equation*}
  which is zero if $\vj \neq \vk$, and it is equal to $\bra{\vj_h} Z
  \ket{\vj_h} \bra{\vj_{\ell}} Z \ket{\vj_{\ell}} = (-1)^{\vj_h}
  (-1)^{\vj_{\ell}}$ if $\vj = \vk$. (The latter expression also works
  if $h=\ell$, in which case $\bra{\vj} \sigma^Z_h \sigma^Z_h
  \ket{\vj} = 1$.) So:
  \begin{equation*}
    \bra{\vx} \sigma^Z_j \ket{\vx} = (-1)^{\vx_j} = z_j, \qquad
    \bra{\vx} \sigma^Z_j \sigma^Z_k \ket{\vx} = (-1)^{\vx_j + \vx_k} = z_j z_k,
  \end{equation*}
  and using linearity in Eq.~\eqref{eq:vxhvx}, we finally obtain
  $\bra{\vx} \ham \ket{\vx} = \sum_{j,k=1}^n A_{jk} z_j z_k +
  \sum_{j=1}^n c_j z_j = z^{\top} A z + c^{\top} z$ with $z = 1 - 2\vx
  \in \{-1,+1\}^n$.
\end{proof}

\noindent Prop.~\ref{prop:isingtoham} gives an explicit construction
of a Hamiltonian that has the possible objective function values of
Eq.~\eqref{eq:ising} on the diagonal, minus the scalar shift $b$ that
is not influential for optimization anyway. After establishing that we
can cast problem \eqref{eq:qubo} as the problem of finding the minimum
eigenvalue of a certain Hamiltonian, explicitly constructed using a
linear combination of tensor products of $Z$ and identity matrices, we
now discuss the adiabatic theorem, which gives a sufficient condition
to find the minimum eigenvalue via time-dependent Hamiltonian
simulation.

\subsection{Theorem statement}
\label{sec:adistatement}
Properly introducing the context of the adiabatic theorem requires
several pieces of notation. Our exposition in the next few sections is
based on \cite{ambainis2004elementary}. Although the result proven in
\cite{ambainis2004elementary} is not as strong as it could be, it
provides the key elements, and the proof relies on linear algebra
only. Tighter and more precise bounds can be found in, e.g.,
\cite{jansen2007bounds,childs2017lecture}, see also
Sect.~\ref{sec:adigap}.

From now on, for brevity we write \emph{minimum eigenpair} to
indicate the lowest eigenvalue and its corresponding eigenvector of a
Hamiltonian, and \emph{minimum eigenvector} to indicate the
eigenvector corresponding to the lowest eigenvalue. (Although we often
write ``eigenstate'' for ``eigenvector'', in this chapter we also use
the more generic term eigenvector because we sometimes deal with
vectors that are not normalized quantum states.)

Let\index{simulation!Hamiltonian|(} $\ham(s)$ be a time-dependent Hamiltonian, with $0 \le s \le
1$. For now we use $s$ to denote time because we reserve $t$ to denote
``unnormalized'' time outside the interval $[0, 1]$: this aspect is
clarified subsequently in this section. We assume that the entries of
$\ham(s)$ are twice differentiable. We denote the first and second
derivatives of $\ham$ at time $s$ as $\ham'(s), \ham''(s)$,
respectively: these are intended to be the matrices of element-wise
derivatives.

The adiabatic theorem concerns the situation
where we know the minimum eigenstate of $\ham(0)$, and we want to
determine the minimum eigenstate of $\ham(1)$.
\begin{remark}
  \label{rem:adiforopt}
  This setup is helpful for combinatorial optimization\index{adiabatic!optimization} in the
  following sense. As discussed in Sect.~\ref{sec:combopteig}, we know
  how to construct a Hamiltonian $\ham_{\text{F}}$ that encodes a
  combinatorial optimization problem of the form \eqref{eq:qubo}, and
  this is already a form that encompasses all optimization problems in
  the complexity class NP (although it is important to keep
  Rem~\ref{rem:qubowise} in mind). In the notation $\ham_{\text{F}}$,
  we use the subscript ``$\text{F}$'' for ``final''. Pick a simple
  Hamiltonian $\ham_{\text{I}}$ (the subscript ``$\text{I}$'' stands
  for ``initial'') for which we know the minimum eigenpair; for
  example, we could define $\ket{\phi} := \frac{1}{\sqrt{2^n}}
  \sum_{\vj \in \{0,1\}^n} \ket{\vj}$ and pick:
  \begin{equation*}
    \ham_{\text{I}} = I^{\otimes n} - \ketbra{\phi}{\phi},
  \end{equation*}
  that has minimum eigenvalue $0$ achieved by the state $\ket{\phi}$
  (we do not claim that this is a good choice for the initial
  Hamiltonian, but it is a valid choice). We can then define the
  time-dependent Hamiltonian:
  \begin{equation}
    \label{eq:adilinearint}
    \ham(s) = (1-s) \ham_{\text{I}} + s \ham_{\text{F}}.
  \end{equation}
  By construction, this Hamiltonian is such that we know the minimum
  eigenstate of $\ham(0) = \ham_{\text{I}}$, and we want to determine
  the minimum eigenstate of $\ham(1) = \ham_{\text{F}}$, as that
  corresponds to the solution of \eqref{eq:qubo} and therefore problem
  \eqref{eq:01opt}. The Hamiltonian in Eq.~\ref{eq:adilinearint} is a
  linear interpolation between the initial and final Hamiltonian, but
  some of the results that we prove in this chapter hold for more
  general types of interpolation. In fact, the discussion in
  Sect.~\ref{sec:adigap} shows that other forms of interpolation may
  be advantageous in certain situations.
\end{remark}
\begin{definition}[Norm of time-dependent quantities]
  \label{def:timedepnorm}
  We denote $\nrm{\ham} := \max_{s \in [0,1]} \nrm{\ham(s)}$, i.e., the norm
  of a time-dependent quantity, when written without the time
  variable, is intended to be the maximum norm over the time
  horizon. We extend this notation to all time-dependent quantities.
\end{definition}
We also need the concept of \emph{spectral gap}\index{spectral gap} of a Hamiltonian, as
the statement of the adiabatic theorem depends on it.  The concept of
spectral gap relies on properties of the spectrum of $\ham$. It is
known that, when the entries of $\ham$ vary continuously over
time, the eigenvalues of $\ham$ are also continuous functions of
time. In other words, one can pick continuous functions $\lambda_1(s),
\lambda_2(s), \dots$ that describe the eigenvalues of $\ham$ as $s$
varies. Intuitively, this is a consequence of the fact that the
eigenvalues are the roots of the characteristic polynomial of $\ham$,
and the coefficients of the characteristic polynomial are continuous
functions of the entries of $\ham$. For a formal proof, see
\cite[Thm.~5.2]{kato1995perturbation}.
\begin{remark}
  As stated earlier, in this chapter all Hamiltonians are assumed to
  have twice-differentiable entries, therefore their eigenvalues vary
  continuously as functions of time. We do not repeat this assumption
  in the statement of subsequent results: when we mention a
  time-dependent Hamiltonian, it has twice-differentiable entries. We
  emphasize this fact because without the assumption, some of the
  statements would not make sense, so it is important to keep it in
  mind.
\end{remark}
\begin{definition}[Instantaneous spectral gap]
  \label{def:instspectralgap}
  Let $\ham(s)$ be a time-dependent Hamiltonian, and $\lambda(s)$ an
  eigenvalue of $\ham(s)$, for $s \in [0,1]$. Let $\gamma(s), s \in
  [0,1]$ be the quantity defined in the following way:
  \begin{itemize}
  \item if $\lambda(s)$ has algebraic multiplicity $>1$, $\gamma(s): =
    0$;
  \item otherwise, let $\lambda_{<}(s)$ be the largest eigenvalue
    strictly smaller than $\gamma(s)$, $\lambda_>(s)$ the smallest
    eigenvalue strictly larger than $\gamma(s)$, and $\gamma(s) :=
    \min\{\lambda(s) - \lambda_{<}(s), \lambda_{>}(s) - \lambda(s)\}$.
  \end{itemize}
  We say that $\ham(s)$ has \emph{instantaneous spectral gap
  $\gamma(s)$ around $\lambda(s)$}.
\end{definition}
In other words, the instantaneous spectral gap around an eigenvalue is
a ``buffer zone'' in the spectrum of the Hamiltonian that contains no
other eigenvalue, at a specific time.
\begin{definition}[Spectral gap]
  \label{def:spectralgap}
  Let $\ham(s)$, $\lambda(s)$ and $\gamma(s)$ be as in
  Def.~\ref{def:instspectralgap}. We say that $\ham(s)$ has a
  \emph{spectral gap $\gamma$ around $\lambda(s)$} if $\gamma = \min_s
  \gamma(s)$. The \emph{spectral gap} of $\ham(s)$ is the spectral gap
  around its minimum eigenvalue.
\end{definition}
Note that the spectral gap could be zero, if the eigenvalues are not
distinct. Note also that if $\ham(s)$ has spectral gap $\gamma$ around
$\lambda(s)$, then all other eigenvalues of $\ham(s)$ are $\le
\lambda(s) - \gamma$ or $\ge \lambda(s) + \gamma$ for all $s$. In the
literature, the term ``spectral gap'' is sometimes used to implicitly
refer to a \emph{nonzero} spectral gap (sometimes one also reads
``gapped Hamiltonian'', to refer to a Hamiltonian that has nonzero
spectral gap); we use the more general definition above, and
explicitly require $\gamma > 0$ for the main result.

In the adiabatic theorem\index{adiabatic!theorem|(} we consider a slow time evolution according
to the Hamiltonian $\ham(s)$. Recall that the time-independent
Schr\"odinger equation, Eq.~\eqref{eq:schrodinger}, is:
\begin{equation*}
  i \frac{\di \ket{\psi(t)}}{\di t} = \ham \ket{\psi(t)},
\end{equation*}
and it has solution $\ket{\psi(t)} = e^{-i\ham t}
\ket{\psi(0)}$. Thus, if $\ket{\psi(0)}$ is an eigenstate of $\ham$,
the system remains in an eigenstate as time evolves (see
Def.~\ref{def:matexp} and the surrounding discussion: the eigenvectors
of $e^{-i\ham t}$ are the same as the eigenvectors of $\ham$, only the
eigenvalues change; this can be seen by diagonalizing $\ham$ and
applying the definition of matrix exponential). This applies to
time-independent Hamiltonians. When the Hamiltonian is time-dependent
this property is not true in general, but intuitively it is
conceivable that if the Hamiltonian changes slowly enough, the
evolution of the system under a time-dependent Hamiltonian is similar
to the evolution for time-independent Hamiltonians; thus, the system
remains in the instantaneous eigenstate for each time
instant. This is the essence of the adiabatic theorem; of course, we
need to prove that such a result holds.

For some large value $T \in \R$, we consider the following slow
evolution, modified from Eq.~\eqref{eq:schrodinger} by absorbing a
multiplicative factor $-1$ into the Hamiltonian for ease of subsequent
calculations, and by introducing a time-dependent Hamiltonian that
slowly changes between $\ham(0)$ and $\ham(1)$ (note that the larger
the value of $T$, the slower the change):\index{Schr\"odinger equation}
\begin{equation}
  \label{eq:adislowschrodinger}
  \frac{\di \ket{\varphi(t)}}{\di t} = i \ham(t/T) \ket{\varphi(t)}, \qquad t \in [0, T],
\end{equation}
where time $t$ progresses from $0$ to $T$. With a change of variable
$s = t/T$, we have $T \di s = \di t$, and therefore
Eq.~\eqref{eq:adislowschrodinger} can be rewritten as:
\begin{equation*}
  \frac{\di \ket{\varphi(sT)}}{T \di s} = i \ham(s) \ket{\varphi(sT)}, \qquad s \in [0, 1].
\end{equation*}
Defining $\ket{\psi(s)} := \ket{\varphi(sT)}$, we finally obtain:
\begin{equation}
  \label{eq:adischrodinger}
  \frac{\di \ket{\psi(s)}}{\di s} = i T \ham(s) \ket{\psi(s)}, \qquad s \in [0, 1].
\end{equation}
From now on we always refer to Eq.~\eqref{eq:adischrodinger} rather
than Eq.~\eqref{eq:adislowschrodinger}, using $s$ as our time
variable; to build intuition, it is useful to remember that $s = t/T$,
and so as $T \to \infty$, time evolves very slowly.

We want to determine a value of $T$ with the following property: if
the initial state $\ket{\psi(0)}$ in Eq.~\eqref{eq:adischrodinger} is
the minimum eigenvector of $\ham(0)$, then the final state
$\ket{\psi(1)}$ is the minimum eigenvector of $\ham(1)$.
\begin{remark}
  At this point it is not clear that such a value of $T$ exists. The
  main reason to believe that such a property may hold is the
  intuition given earlier in this section: if $T \to \infty$, the
  system should behave similarly to the case of a time-independent
  Hamiltonian, thus it should remain in an (instantaneous) eigenstate
  of the Hamiltonian.
\end{remark}
The adiabatic theorem provides sufficient conditions so that the value
of $T$ with the desired property not only exists, but can also be
lower bounded, so that picking any $T$ larger than a threshold
suffices to guarantee that the final state $\ket{\psi(1)}$ is
(approximately) the minimum eigenvector of $\ham(1)$.
\begin{theorem}[Adiabatic theorem]
  \label{thm:adiabatic}
  Let $\ham(s)$ be a time-dependent Hamiltonian, let $\lambda(s)$ be
  an eigenvalue of $\ham(s)$, and let $\ket{\varphi(s)}$ be an
  eigenstate of $\ham(s)$ with eigenvalue $\lambda(s)$. Assume that
  there is a spectral gap $\gamma > 0$ around $\lambda(s)$. Assume
  further that, from the initial state $\ket{\varphi(0)}$, we apply
  the following time evolution:
  \begin{equation*}
    \frac{\di \ket{\psi(s)}}{\di s} = i T \ham(s) \ket{\psi(s)}, \qquad s
    \in [0, 1], \qquad \ket{\psi(0)} = \ket{\varphi(0)},
  \end{equation*}
  and for some $\delta >0$, $T$ satisfies:
  \begin{equation*}
    T \ge \frac{10^4}{\delta^2} \left(\frac{\nrm{\ham'}^3}{\gamma^4} +
    \frac{\nrm{\ham'} \nrm{\ham''}}{\gamma^3} \right).
  \end{equation*}
  Then the system approximately remains in the instantaneous
  eigenstate $\ket{\varphi(s)}$ with eigenvalue $\lambda(s)$ for all $s$,
  and in particular, the final state $\ket{\psi(1)}$ has Euclidean
  distance at most $\delta$ from $\ket{\varphi(1)}$, up to global phase:
  \begin{equation*}
    \nrm{ e^{i \theta} \ket{\psi(1)} - \ket{\varphi(1)} } \le \delta
    \qquad \text{ for some } \theta.
  \end{equation*}
\end{theorem}
\begin{remark}
  \label{rem:spectralgap}
  It is intuitive to see that the nonzero spectral gap condition is
  necessary even just to have the concept of an adiabatic
  theorem. Suppose there is no spectral gap, i.e., the eigenvalue
  $\lambda(s)$ corresponding to the initial eigenstate ``crosses''
  another eigenvalue $\lambda'(s)$ as $s$ goes from $0$ to $1$. In
  that case, the corresponding instantaneous eigenstate
  $\ket{\varphi(s)}$ would not be well-defined, due to the degenerate
  eigenvalue. When there are two eigenvalues $\lambda(s) =
  \lambda'(s)$ that are identical at time $s$, the eigenstate
  $\ket{\varphi(s)}$ corresponding to $\lambda(s)$ is no longer 
  unique, and we cannot properly define how the state of the system is
  supposed to approximately follow $\ket{\varphi(s)}$. With nonzero
  spectral gap these issues do not arise: the eigenvalue of interest
  is nondegenerate, and there is always a unique eigenstate
  corresponding to $\lambda(s)$.
\end{remark}
A proof of Thm.~\ref{thm:adiabatic} is given in
Sect.s~\ref{sec:adihighlevel} and \ref{sec:adiproof}. In the
first section we give a shorter, more intuitive but less precise
version of the argument of the proof. Several gaps are filled in in
the second section, although we do not give all the details, and refer
to \cite{ambainis2004elementary} for the (very few) missing pieces.

\subsection{High-level proof}
\label{sec:adihighlevel}
Simulating Eq.~\eqref{eq:adischrodinger} directly is difficult because
it is a continuous-time dynamical system where the evolution operator
(the Hamiltonian) also changes with time. The approach that we take,
as one generally takes on all digital computers, is to discretize time
into very small steps, and perform time evolution in each of those
time steps with a fixed Hamiltonian. As the size of those steps goes
down to zero, the discrete-time evolution approaches the
continuous-time evolution, so we can analyze properties of the
discrete-time evolution instead of Eq.~\eqref{eq:adischrodinger}.
More specifically, we divide the time interval $[0, 1]$ (for the
normalized time variable $s$) into $N$ equally-spaced subintervals
with breakpoints $\frac{0}{N}, \frac{1}{N}, \dots, \frac{N-1}{N},
\frac{N}{N}$. In each interval we apply the time-independent
Hamiltonian $\ham(j/N)$ for $\frac{1}{N}$ units of time. It is easy to
verify that the solution to the differential equation with
time-independent Hamiltonian:
\begin{equation*}
  \frac{\di \ket{\psi(s)}}{\di s} = i T \ham(j/N) \ket{\psi(s)} 
\end{equation*}
is:
\begin{equation*}
  \ket{\psi(s)} = e^{i T \ham(j/N) s} \ket{\psi(0)}.
\end{equation*}
Thus, applying the time-independent Hamiltonian $\ham(j/N)$ for
$\frac{1}{N}$ units of time is equivalent to applying $e^{i T/N
  \ham(j/N)}$ to the initial state (we obtain this by setting $s =
\frac{1}{N}$ in the last equation). As a consequence, defining
\begin{equation}
  \label{eq:ujdef}
  U_j := e^{i T/N \ham(j/N)},
\end{equation}
the original continuous-time evolution in
Eq.~\eqref{eq:adischrodinger} can be approximated by applying to the
initial state a sequence of $N$ unitaries:
\begin{equation*}
  \ket{\psi(1)} \approx U_{N-1} U_{N-2} \cdots U_1 U_0 \ket{\psi(0)},
  \qquad \ket{\psi(0)} = \ket{\varphi(0)}.
\end{equation*}
As $N \to \infty$, this approximation gets better and the error goes
to zero \cite{van2001powerful}. For brevity and for notational
similarity to other vectors that appear in the rest of the analysis,
it is convenient to denote by $g_j = \ket{\varphi(j/N)}$ a unit
eigenvector of $\ham(j/N)$ (and, consequently, $U_j$) corresponding to
$\lambda(j/N)$, see Rem.~\ref{rem:spectralgap}:
\begin{equation}
  \label{eq:gjdef}
  g_j = \ket{\varphi(j/N)} \in \{ x : \ham(j/N) x = \lambda(j/N) x,
  \nrm{x} = 1 \}.
\end{equation}
The vector $g_j$ is normalized, whereas other vectors for which we do
not use the bra-ket notation are typically not normalized.  It is
important to note that due to the gap assumption $\gamma > 0$ in
Thm.~\ref{thm:adiabatic}, $\lambda(j/N)$ is an eigenvalue with
algebraic multiplicity 1, therefore there is a single eigenspace
associated with it.
\begin{remark}
  \label{rem:phasechoice}
  Eigenvectors can only be defined up to a global phase: if $g_j$ is
  an eigenvector, then $e^{i \theta} g_j$ is also an eigenvector with
  the same eigenvalue and the same magnitude, i.e., $\nrm{g_j} =
  \nrm{e^{i\theta} g_j}$. These eigenvectors are equivalent for the
  purposes of defining the final state of the adiabatic evolution, but
  we need to choose the phases appropriately to be able to compute
  Euclidean distances between eigenvectors: we could have eigenvectors
  $g_j, g_{j+1}$ of $\ham(j/N), \ham((j+1)/N)$ that are close to each
  other for some choice of their global phases, but far from each
  other for different global phases. In the high-level exposition
  given in this section we simply ignore the issue of choosing the
  global phases of the eigenvectors, so the results should be
  interpreted as ``there exists a choice of the global phases such
  that these results hold.'' A more precise discussion is given in
  Sect.~\ref{sec:adiproof}.
\end{remark}
Note that in the limit $N \to \infty$, and assuming that the adiabatic
theorem (Thm.~\ref{thm:adiabatic}) holds, the state $\ket{\psi(j/N)}$
of the system is essentially the same as $g_j =
\ket{\varphi(j/N)}$. Thus, we want to show that, as $N \to \infty$:
\begin{equation*}
  g_{N} \approx U_{N-1} g_{N-1} \approx U_{N-1} U_{N-2} g_{N-2}
  \approx U_{N-1} U_{N-2} \cdots U_1 U_0 g_0.
\end{equation*}
\vspace*{-1em}
\begin{remark}
  \label{rem:ntoinfty}
  Because $N \to \infty$, in $\bigO{\cdot}$ expressions containing $N$
  we only write the terms that depend on $N$, which is always the
  leading term. This means that, for example, instead of
  $\bigO{\nrm{\ham}/N}$ we write $\bigO{1/N}$.
\end{remark}

\begin{figure}[htb]
  \center
  \ifcompilefigs
  \begin{tikzpicture}[>=Stealth,scale=1.5]
    \draw [->,thick] (0,0) -- (0,2) node (gj) [midway,left,scale=1.2] {$g_j$};
    \draw [->,thick] (0,0) -- (1.1547,1.732) node (gjp) [midway,right,scale=1.2] {$g_{j+1}$};
    \draw [->,thick] (1.1547,1.732) -- (0.25,2.35) node (pj) [above right=-3ex and 2.7ex,scale=1.2] {$p_{j+1}$};
    \draw [->,thick,loosely dashed] (1.1547,1.732) -- (0,2) node (diff) [midway,below,scale=0.85] {$g_j - g_{j+1}$};
    \draw [-,dotted] (0,2) -- (0.25,2.35) node (perp) {};
  \end{tikzpicture}
  \else
  \includegraphics{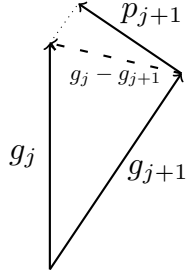}
  \fi
  \caption{Representation of the eigenvectors $g_j, g_{j+1}$, and the
    projection $p_j$ of $g_j-g_{j+1}$ onto the space orthogonal to
    $g_{j+1}$.}
  \label{fig:adivectors}
\end{figure}

For simplicity, from now on we consider the case in which $\lambda(s)
= 0$ for all $s$, i.e., we are tracking the eigenvector corresponding
to the zero eigenvalue for the entire time evolution.
\begin{remark}
  \label{rem:zeroeig}
  This is without loss of generality because, given a general
  Hamiltonian $\ham(s)$ and eigenvalue of interest $\lambda(s)$, we
  can always consider a new Hamiltonian defined as $\hat{\ham}(s) :=
  \ham(s) - \lambda(s) I$, while keeping the initial state
  fixed. These Hamiltonians define the same adiabatic evolution up to
  a time-dependent global phase, and by construction, for the new
  Hamiltonian $\hat{\ham}(s)$ the system always remains in the
  eigenvector with eigenvalue $0$. To ensure correctness of this
  argument we also need to show that the value of $T$ in
  Thm.~\ref{thm:adiabatic} applies to both $\ham(s)$ and
  $\hat{\ham}(s)$, i.e., that this transformation does not require us
  to choose a significantly larger $T$; we do this subsequently in
  Cor.~\ref{cor:zeroeig}.
\end{remark}
Consider a decomposition of $g_j$ in terms of $g_{j+1}$ and its
orthogonal complement, the subspace $G_{j+1}^{\perp}$. Define:
\begin{equation}
  \label{eq:pjdef}
  p_{j+1} := \text{Proj}_{G_{j+1}^{\perp}}(g_j - g_{j+1})
\end{equation}
as the projection of $g_j - g_{j+1}$ onto $G_{j+1}^{\perp}$, i.e., onto
the space perpendicular to $g_{j+1}$, see Fig.~\ref{fig:adivectors}
for a graphical representation of these vectors. 
Clearly $g_j = g_{j+1} + (g_j - g_{j+1})$. It turns out that $g_j -
g_{j+1}$ is almost orthogonal to $g_{j+1}$, i.e., $p_{j+1}$ almost
coincides with $g_j - g_{j+1}$:
\begin{equation}
  \label{eq:pjerror}
  \nrm{ p_{j+1} - (g_j - g_{j+1}) } = \bigO{\frac{1}{N^2}}.
\end{equation}
Thus, using the fact that $U_j g_j = g_j$, because $g_j$ is an
eigenvector with eigenvalue $e^{i T/N \lambda(j/N)} = 1$, we can
write:
\begin{equation*}
  U_j g_j - (g_{j+1} + p_{j+1}) = g_j - (g_{j+1} + p_{j+1}) =
  (g_j - g_{j+1}) - p_{j+1} = v_{j+1} \text{ with } \nrm{v_{j+1}} = \bigO{\frac{1}{N^2}}.
\end{equation*}
Rearranging the terms in the last equality yields $U_j g_j = g_{j+1} +
p_{j+1} + v_{j+1}$.  As a consequence, if we apply all the unitaries
$U_0,\dots,U_{N-1}$ in sequence, we obtain:
\begin{align*}
  U_{N-1}\cdots U_1 U_0g_0 &= U_{N-1}\cdots U_2 U_1 (g_1 + p_1 + v_1) =
  U_{N-1}\cdots U_2 U_1 g_1 + U_{N-1}\cdots U_2 U_1 (p_1 + v_1) \\
  &= U_{N-1}\cdots U_3 U_2 (g_2 + p_2 + v_2)  + U_{N-1}\cdots U_2 U_1 (p_1 + v_1) \\
  &= U_{N-1}\cdots U_3 U_2 g_2 + U_{N-1}\cdots U_3 U_2(p_2 + v_2)  + U_{N-1}\cdots U_2 U_1 (p_1 + v_1) = \dots \\
  &= g_N + \sum_{j=1}^{N} U_{N-1}\cdots U_j p_j + \underbrace{\sum_{j=1}^{N}
  U_{N-1}\cdots U_j v_j}_{\text{error term}},
\end{align*}
and the norm of the error term can be upper bounded as:
\begin{equation*}
  \nrm{\sum_{j=1}^{N} U_{N-1}\cdots U_j v_j} \le \sum_{j=1}^N \nrm{v_j} = \bigO{\frac{1}{N}}.
\end{equation*}
\begin{remark}
  \label{rem:upperlimitempty}
  The summation $\sum_{j=1}^{N} U_{N-1}\cdots U_j p_j$ is written with
  upper limit $N$, even though the sequence of matrices stops at
  $U_{N-1}$, as a shorthand for the inclusion of the term $p_N$. In
  other words, $\sum_{j=1}^{N} U_{N-1}\cdots U_j p_j = p_N +
  \sum_{j=1}^{N-1} U_{N-1}\cdots U_j p_j$. We use the same convention
  for $\sum_{j=1}^{N} U_{N-1}\cdots U_j v_j$, although the latter sum
  stops being important in the following.
\end{remark}
If we can show that $\nrm{\sum_{j=1}^{N} U_{N-1}\cdots U_j p_j}$ is
small, then $U_{N-1}\cdots U_0g_0 \approx g_N$, proving the desired
result. More precisely, to show Thm.~\ref{thm:adiabatic} we in fact
need to show that $\nrm{\sum_{j=1}^{N} U_{N-1}\cdots U_j p_j} \le
\delta$: because $N \to \infty$, the norm of the error term goes to
zero, implying that the distance between the final state produced by
the evolution and $g_N = \ket{\varphi(1)}$ is at most $\delta$, as
desired. Thus, our last task is to prove:
\begin{equation}
  \label{eq:adideltadist}
  \nrm{\sum_{j=1}^{N} U_{N-1}\cdots U_j p_j} \le \delta.
\end{equation}

To this end, we split $\sum_{j=1}^{N} U_{N-1}\cdots U_j p_j$ into
groups with $M$ terms each, where we choose an appropriate $M =
\bigO{N}$, i.e., we partition the expression into $N/M$ sums:
\begin{align}
  \label{eq:adisumpartition}
  \sum_{j=1}^{M} U_{N-1}\cdots U_j p_j, \qquad \sum_{j=M+1}^{2M} U_{N-1}\cdots U_j p_j, \qquad \sum_{j=2M+1}^{3M} U_{N-1}\cdots U_j p_j, \qquad \dots
\end{align}
The precise value of $M$ is important and is specified later, but for
now we can just take it to be some small fraction of $N$. We then show
that each group has a small norm. The reason for performing this split
is not readily apparent, and it may be useful to give some
intuition. The expression $\sum_{j=1}^{N} U_{N-1}\cdots U_j p_j$ is
not easy to analyze; however, if we could replace each $p_j$ with
$p_1$, and each $U_j$ with $U_1$, then we would obtain the much
simpler expression $\sum_{j=1}^{N-1} U_1^j p_1$, which is relatively
easy to analyze by expressing $p_1$ in terms of eigenvectors of
$U_1$. The issue is, of course, that replacing each $p_j$ with $p_1$,
and each $U_j$ with $U_1$, incurs a large total error, so this
strategy would fail. However, a more nuanced version of this strategy
can be made to work. By triangle inequality, partitioning the
``large'' sum into $M/N$ terms, we obtain:
\begin{equation*}
  \nrm{\sum_{j=1}^{N} U_{N-1}\cdots U_j p_j} \le \sum_{k=0}^{N/M-1} \nrm{\sum_{j=kM+1}^{(k+1)M} U_{N-1}\cdots U_j p_j},
\end{equation*}
so if we can show that:
\begin{equation*}
  \nrm{\sum_{j=kM+1}^{(k+1)M} U_{N-1}\cdots U_j p_j} \le \frac{\delta M}{N},
\end{equation*}
this immediately implies Eq.~\eqref{eq:adideltadist}. We look at the
case $k=0$ to understand the main idea, as the other cases are very
similar. Thus, we are looking for an upper bound to
$\nrm{\sum_{j=1}^{M} U_{N-1}\cdots U_j p_j}$. The terms $U_{N-1}\cdots
U_M$ appear in every term of the summation on the l.h.s., and because
they are unitary, they do not affect the norm and can be suppressed.
In Sect.~\ref{sec:adiproof} we show that:
\begin{equation*}
  \nrm{\sum_{j=1}^{M} U_{M-1}\cdots U_j p_j} \approx
  \nrm{\sum_{j=0}^{M-1} U_1^j p_1},
\end{equation*}
with some error that is discussed below. This lets us work with an
approximation that is much easier to analyze. To generalize this idea
for all $k=0,\dots,N/M-1$, we exploit the fact that $N \to \infty$,
so, recalling the definitions of $U_j, g_j, p_j$ in
Eq.s~\eqref{eq:ujdef}, \eqref{eq:gjdef}, \eqref{eq:pjdef}, we see that
$U_j$ and $p_j$ change relatively slowly, i.e., the differences
$\nrm{U_{j+1} - U_j}, \nrm{p_{j+1}-p_j}$ are small. Then we perform
the following substitution in Eq.~\eqref{eq:adisumpartition}:
\begin{equation}
  \label{eq:adisimplification}
  \text{for } k=0,\dots,\frac{N}{M}-1 \qquad \sum_{j=kM+1}^{(k+1)M}
U_{(k+1)M-1}\cdots U_j p_j \longrightarrow \sum_{j=0}^{M-1} U_{kM+1}^j p_{kM+1}.
\end{equation}
Note that for each $k$, we suppressed unitaries $U_{N-1}\cdots
U_{(k+1)M}$ from the l.h.s., because in the corresponding group in
Eq.~\eqref{eq:adisumpartition} those unitaries appear in all terms of
the summation and are uninfluential for the norm. The proposed
substitution incurs smaller total error (when summing over all groups
of terms, and taking the norm) than approximating the entire summation
$\sum_{j=1}^{N} U_{N-1}\cdots U_j p_j$ with $\sum_{j=0}^{N-1} U_1^j
p_1$. This is intutive, because the substitution with $N/M$ groups of
terms uses $U_{kM+1},p_{kM+1}$ for multiple values of $k$, rather than
just for $k=0$, and follows the original summation more
closely. Further, it can be shown (see Sect.~\ref{sec:adiproof}) that
the substitution indicated in Eq.~\eqref{eq:adisimplification}
introduces an error of $\frac{\delta M}{2 N}$, which is half of the
target upper bound for the norm of $\nrm{\sum_{j=1}^{M} U_{N-1}\cdots
  U_j p_j}$. The choice for the value of $T$ in
Thm.~\ref{thm:adiabatic} comes up in this step of the proof, because
we need a $T$ large enough to cancel out some other terms and
eventually upper bound the error of Eq.~\eqref{eq:adisimplification}
by $\frac{\delta M}{2 N}$. So now, restricting ourselves to the case
$k=0$ because the same argument works for all $k$, we aim to show that
$\nrm{\sum_{j=0}^{M-1} U_{kM+1}^j p_{kM+1}} = \nrm{\sum_{j=0}^{M-1}
  U_1^j p_1} \le \frac{\delta M}{2 N}$. We can express $p_1$ in an
eigenbasis of $U_1$, and because by assumption $p_1$ is orthogonal to
$g_1$, we know that $p_1$ is a combination of the eigenvectors of
$U_1$ (hence, $\ham(1/N)$) that are not $g_1$. Calling these unit
eigenvectors $v_2,\dots,v_d$ (for simplicity and without loss of
generality, we assume $v_1=g_1$) with eigenvalues
$\sigma_2,\dots,\sigma_d$, and calling $a_2,\dots,a_d$ the
coefficients such that $\sum_{h=2}^{d} a_h v_h = p_1$, we have:
\begingroup \allowdisplaybreaks
\begin{equation}
  \label{eq:adiu1p1geom}
  \begin{aligned}
  \nrm{\sum_{j=0}^{M-1} U_1^j p_1} &= \nrm{\sum_{j=0}^{M-1} U_1^j
    \sum_{h=2}^{d} a_h v_h} = \nrm{\sum_{h=2}^d a_h \sum_{j=0}^{M-1} U_1^j
    v_h} = \nrm{\sum_{h=2}^d a_h \sum_{j=0}^{M-1} e^{i T/N \sigma_h^j} v_h} \\
  &\le
  \nrm{\sum_{h=2}^d a_h v_h \abs{\sum_{j=0}^{M-1} e^{i T/N \sigma_h^j}}} \le
  \nrm{\sum_{h=2}^d a_h} \max_h \abs{\sum_{j=0}^{M-1} e^{i T/N \sigma_h^j}} \\
  &\le \nrm{p_1} \max_h \abs{\sum_{j=0}^{M-1} e^{i T/N \sigma_h^j}},
  \end{aligned}
\end{equation}
\endgroup
where the third equality uses the fact that $v_h$ is an eigenvector of
$U_1^j$ with eigenvalue $\sigma_h^j$, and the second inequality uses
the fact that the vectors $v_h$ have unit norm. It is possible to show
that
\begin{equation}
  \label{eq:p1bound}
  \nrm{p_1} \le \nrm{\ham'}/(\gamma N),
\end{equation}
see Sect.~\ref{sec:adiproof}: the quantities involved in this bound
are rather intuitive, because by Eq.~\eqref{eq:pjdef}, we expect that
the speed by which $\ham$ changes affects the difference between the
consecutive (in time) eigenvectors $g_1$ and $g_2$. Let us now focus
on the remaining term $\max_h \abs{\sum_{j=0}^{M-1} e^{i T/N
    \sigma_h^j}}$. This is simply a geometric sum, thus:
\begin{equation}
  \label{eq:geomsum1}
  \abs{\sum_{j=0}^{M-1} e^{i T/N \sigma_h^j}} = \frac{\abs{1- e^{i M T/N \sigma_h}}}{\abs{1-e^{i T/N \sigma_h}}} \le \frac{2}{\abs{1-e^{i T/N \sigma_h}}} \le \frac{4N}{T \sigma_h} \le \frac{4N}{T \gamma},
\end{equation}
where for the first equality we use the formula for the geometric sum
and the fact that $\abs{a/b} = \abs{a}/\abs{b}$ for $a,b \in \C$, for
the second inequality we use the fact that $\abs{e^{i\theta} - 1} \ge
\abs{\theta}/2$ for small $\theta$, and for the last inequality we
used the fact that $|\sigma_h| > \gamma$ for all $h$ (recall
Def.~\ref{def:spectralgap} and the assumptions of
Thm.~\ref{thm:adiabatic}: there is a spectral gap $\gamma$ around the
eigenvalue $0$ that is being ``tracked'' by the adiabatic evolution).
Using Eq.s~\eqref{eq:p1bound} and \eqref{eq:geomsum1}:
\begin{equation}
  \label{eq:adiu1p1geombound}
  \nrm{p_1} \max_h \abs{\sum_{j=0}^{M-1} e^{i T/N \sigma_h^j}} \le
  \frac{4\nrm{\ham'}}{\gamma^2 T} \le \underbrace{\frac{8 \nrm{\ham'} N}{\gamma^2 T \delta}}_{\text{this is how we choose } M} \frac{\delta}{N} \le \frac{\delta M}{N}. 
\end{equation}
In the above expression, we had to pick $M$ appropriately to ensure
that we obtain the desired upper bound $\frac{\delta M}{N}$, thereby
proving Eq.~\eqref{eq:adideltadist} and Thm.~\ref{thm:adiabatic}. (The
choice $M = \ceil{8 \nrm{\ham'} N /(\delta \gamma^2 T)} = \bigO{N}$ is
discussed in Sect.~\ref{sec:adiproof}.)

The discussion above gives the essence of a possible proof of the
adiabatic theorem. In short, starting from an eigenvector of
$\ham(j/N)$ whose eigenvalue evolves through time without crossing any
other eigenvalue, we would like to show that the eigenvector evolves
into the corresponding eigenvector in $\ham((j+1)/N)$. There is a
small component of the former eigenvector that is orthogonal to the
latter eigenvector. However, up to some small error, successive powers
of $e^{i T/N \ham((j+1)/N)}$ act on this orthogonal component, leading
to a geometric series that ends up mostly canceling out. Because the
component orthogonal to the evolution of the initial eigenvector
cancels out in these geometric sums, the only meaningful part of the
state of the system that ``survives'' through the evolution is
precisely the desired sequence of eigenvectors of $\ham(j/N)$. This
eventually brings us to the final eigenvector, i.e., the state
$\ket{\psi(1)}$ is almost equal to $\ket{\varphi(1)}$.

\subsection{Filling the gaps}
\label{sec:adiproof}
We now describe in more detail several components that were omitted
from the proof in Sect.~\ref{sec:adihighlevel}. We begin by stating a
version of Thm.~\ref{thm:adiabatic} for the case where the eigenvalue
$\lambda(s)$ is identically zero, which is the special case discussed
in Sect.~\ref{sec:adihighlevel}. For this special case we can get
somewhat tighter bounds on $T$. At the end of this section we show
that the assumption is not restrictive, and use the Hamiltonian
transformation of Rem.~\ref{rem:zeroeig} to prove the bound on $T$
given in Thm.~\ref{thm:adiabatic}.
\begin{theorem}[Special case of the adiabatic theorem]
  \label{thm:adiabaticzero}
  Let $\ham(s)$ be a time-dependent Hamiltonian, and let
  $\ket{\varphi(s)}$ be an eigenstate of $\ham(s)$ with eigenvalue
  $\lambda(s) = 0$ for all $s \in [0,1]$. Assume that there is a
  spectral gap $\gamma > 0$ around $\lambda(s) = 0$, i.e., all other
  eigenvalues are at least $\gamma$ in absolute value. Assume further
  that, from the initial state $\ket{\varphi(0)}$, we apply the
  following time evolution:
  \begin{equation*}
    \frac{\di \ket{\psi(s)}}{\di s} = i T \ham(s) \ket{\psi(s)}, \qquad s
    \in [0, 1], \qquad \ket{\psi(0)} = \ket{\varphi(0)},
  \end{equation*}
  and for some $\delta >0$, $T$ satisfies:
  \begin{equation*}
    T \ge \frac{10^3}{\delta^2} \max \left\{\frac{\nrm{\ham'}^3}{\gamma^4},
    \frac{\nrm{\ham'} \nrm{\ham''}}{\gamma^3} \right\}.
  \end{equation*}
  Then the system approximately remains in the instantaneous
  eigenstate $\ket{\varphi(s)}$ with eigenvalue 0 for all $s$,
  and in particular, the final state $\ket{\psi(1)}$ has Euclidean
  distance at most $\delta$ from $\ket{\varphi(1)}$, up to global phase:
  \begin{equation*}
    \nrm{ e^{i \theta} \ket{\psi(1)} - \ket{\varphi(1)} } \le \delta
    \qquad \text{ for some } \theta.
  \end{equation*}
\end{theorem}
To show Thm.~\ref{thm:adiabaticzero} we need the following lemma,
which we state without proof.
\begin{lemma}[Lem.~3.2 in \cite{ambainis2004elementary}]
  \label{lem:statederbound}
  In the context of Thm.~\ref{thm:adiabaticzero}, assume the phase of
  $\ket{\varphi(s)}$ is chosen so that
  $\braket{\varphi'(s)}{\varphi(s)} = 0$. Then the following holds:
  \begin{equation*}
    \nrm{\ket{\varphi'}} \le \frac{\nrm{\ham'}}{\gamma} \qquad
    \nrm{\ket{\varphi''}} \le \frac{\nrm{\ham''}}{\gamma} +
    \frac{3\nrm{\ham'}^2}{\gamma^2}.
  \end{equation*}
\end{lemma}
\begin{remark}
  $\ket{\varphi'(s)}, \ket{\varphi''(s)}$ are not normalized quantum
  states: they are the column vectors with the component-wise
  first-order and second-order derivatives of $\ket{\varphi(s)}$. We
  (exceptionally) write these vectors using the bra-ket notation
  because we frequently take their inner product with normalized
  quantum states, and this avoids the need for further notation.
\end{remark}
We can now proceed with the proof of Thm.~\ref{thm:adiabaticzero}.
\begin{proof}
  Fix $M = \ceil{8 \nrm{\ham'} N /(\delta \gamma^2 T)}$. The proof
  follows exactly the same structure given in
  Sect.~\ref{sec:adihighlevel}. There are a few intermediate results
  that are used without proof in Sect.~\ref{sec:adihighlevel}, and
  need to be proven here.
  \begin{enumerate}[(i)]
  \item The choice of the phases of the eigenvectors $g_j$ of
    Eq.~\eqref{eq:gjdef}, see Rem.~\ref{rem:phasechoice}. An
    alternative approach would be to measure distances using a metric
    that is insensitive to phase, but we stick with the familiar
    Euclidean norm.
  \item Eq.~\eqref{eq:pjerror}, stating that $g_j - g_{j+1}$ is
    approximated well by $p_{j+1}$.
  \item Eq.~\eqref{eq:adisimplification}, i.e., the approximation
    whereby we substitute:
    \begin{equation*}
      \sum_{j=kM+1}^{(k+1)M} U_{(k+1)M-1}\cdots U_j p_j
      \longrightarrow \sum_{j=0}^{M-1} U_{kM+1}^j p_{kM+1}.
    \end{equation*}
    We discuss a proof for $k=0$, as the argument works in the same
    way for all $k$.
  \item The bound on $\nrm{p_j}$ given in Eq.~\eqref{eq:p1bound}. We
    discuss the case $j=1$, the proof can be easily generalized to any
    $j$.
  \end{enumerate}

  (i). The speed by which the phase of a (generic) time-dependent,
  differentiable complex unit vector $\ket{\psi(s)}$ changes is
  measured by the quantity $\braket{\psi'(s)}{\psi(s)}$. To see this,
  note first that $\braket{\psi'(s)}{\psi(s)}$ is an imaginary number,
  therefore it is zero for real vectors. Indeed,
  $\braket{\psi(s)}{\psi(s)} = 1$, and if we take the derivative with
  respect to $s$ on both sides of this equation we obtain:
  \begin{equation*}
    \frac{\di}{\di s} \braket{\psi(s)}{\psi(s)} = \braket{\psi(s)'}{\psi(s)} + \braket{\psi(s)}{\psi'(s)} = 0,
  \end{equation*}
  so $\braket{\psi(s)'}{\psi(s)}$ has no real part. This is
  geometrically clear: a real unit vector lies on the unit sphere, and
  if it moves around the sphere, its derivative is orthogonal to the
  vector itself because it needs to remain on the sphere. Not so for
  complex vectors, where a rotation can also pick up a phase, hence
  the possibility of a nonzero (but imaginary)
  $\braket{\psi'(s)}{\psi(s)}$. We fix the evolution of the
  eigenvectors by requiring that:
  \begin{equation}
    \label{eq:phasechoice}
    \braket{\varphi'(s)}{\varphi(s)} = 0 \qquad \text{ for all } s \in
           [0,1].
  \end{equation}
  This is always possible. Consider $\ket{\phi(s)} = e^{i \beta(s)}
  \ket{\varphi(s)}$, where $\beta(s) = \int_0^s i
  \braket{\varphi'(x)}{\varphi(x)} dx$. Then $\ket{\phi(s)}$ is equal
  to $\ket{\varphi(s)}$ up to global phase, and it satisfies the
  condition in Eq.~\eqref{eq:phasechoice}:
  \begin{align*}
    \braket{\phi'(s)}{\phi(s)} &= \left(i e^{i \beta(s)} \beta'(s)\bra{\varphi(s)} + e^{i \beta(s)} \bra{\varphi'(s)}\right) e^{i \beta(s)} \ket{\varphi(s)} \\
    &= \left(- e^{i \beta(s)} \braket{\varphi'(s)}{\varphi(s)} \bra{\varphi(s)} + e^{i \beta(s)} \bra{\varphi'(s)}\right) e^{i \beta(s)} \ket{\varphi(s)} = 0.
  \end{align*}
  The evolution of $\ket{\phi(s)}$ is equivalent to the evolution of
  $\ket{\varphi(s)}$ because they are eigenvectors with the same
  eigenvalue, so we can choose to study $\ket{\phi(s)}$ instead of
  $\ket{\varphi(s)}$. This fixes our choice of phases. In the
  following, we still denote by $\ket{\varphi(s)}$ the eigenvector of
  $\ham(s)$ corresponding to $\lambda(s) = 0$, but we can now assume
  that Eq.~\eqref{eq:phasechoice} holds without loss of generality.

  (ii). We want to show
  \begin{equation*}
    \nrm{ p_{j+1} - (g_j - g_{j+1}) } = \bigO{\frac{1}{N^2}}.
  \end{equation*}
  We take the Taylor expansion of $\ket{\varphi(s)}$ as a function of
  $s$, centered at $s=(j+1)/N$, and we evaluate it at $s=j/N$:
  \begin{equation*}
    \ket{\varphi(j/N)} = \ket{\varphi((j+1)/N)} - \frac{1}{N} \ket{\varphi'((j+1)/N)} + \bigO{\nrm{\frac{j+1}{N} - \frac{j}{N}}^2}.
  \end{equation*}
  (Here and in the following, we employ the usual convention of
  indicating error terms in a Taylor series approximation as additive
  $\bigO{\cdot}$ terms, to be interpreted as: there is an error term
  whose norm is $\bigO{\cdot}$.) Remembering that $g_{j}$ is the
  eigenvector $\ket{\varphi(j/N)}$ with eigenvalue $\lambda(s) = 0$ of
  $\ham(j/N)$, we can rewrite this as:
  \begin{equation*}
    g_j - g_{j+1} = - \frac{1}{N} \ket{\varphi'((j+1)/N)} + \bigO{\frac{1}{N^2}}.
  \end{equation*}
  Project both sides of the equation onto $G_{j+1}^{\perp}$, the
  orthogonal complement of the space spanned by $g_{j+1}$, by applying the corresponding
  projector. Using the definition of $p_{j+1}$ in
  Eq.~\eqref{eq:pjdef}, we obtain:
  \begin{equation}
    \label{eq:pjpsiprime}
    \begin{aligned}
      p_{j+1} &= \text{Proj}_{G_{j+1}^{\perp}}(g_j - g_{j+1}) =
      \text{Proj}_{G_{j+1}^{\perp}}\left(- \frac{1}{N} \ket{\varphi'((j+1)/N)}\right) + \bigO{\frac{1}{N^2}} \\
      &= - \frac{1}{N} \ket{\varphi'((j+1)/N)} + \bigO{\frac{1}{N^2}},
    \end{aligned}
  \end{equation}
  because $\braket{\varphi'((j+1)/N)}{\varphi((j+1)/N)} = 0$ (recall
  Eq.~\eqref{eq:phasechoice}), implying that $\ket{\varphi'((j+1)/N)}$
  fully lies in the orthogonal complement of $\ket{\varphi((j+1)/N)} =
  g_{j+1}$. Subtracting this equation from the previous one, and
  taking the norm, yields:
  \begin{equation*}
    \nrm{ p_{j+1} - (g_j - g_{j+1}) } = \bigO{\frac{1}{N^2}}.
  \end{equation*}

  (iii). We want to show (recall that we are looking at the case $k=0$
  in Eq.~\ref{eq:adisimplification}):
  \begin{equation}
    \label{eq:adisimplification2}
    \nrm{\sum_{j=1}^{M} U_{M-1}\cdots U_j p_j - \sum_{j=0}^{M-1} U_1^j
      p_1} \le \frac{\delta M}{2 N}.
  \end{equation}
  The proof of Eq.~\eqref{eq:adisimplification2} is rather long and
  tedious, but it is important to sketch it because it is where the
  particular choice of $T$ comes into play. We first study the effect
  of replacing all $p_j$ with $p_1$. Using Eq.~\eqref{eq:pjpsiprime},
  we can write:
  \begin{equation*}
    p_{j+h} - p_{j} = - \frac{1}{N} \left(\ket{\varphi'((j+h)/N)} -
    \ket{\varphi'(j/N)} \right) + \bigO{\frac{1}{N^2}}.
  \end{equation*}
  Applying the mean value theorem to the difference within parentheses
  at the r.h.s., the following equation holds at some point
  $y \in [j/N, (j+h)/N]$:
  \begin{equation*}
    \frac{\ket{\varphi'((j+h)/N)} - \ket{\varphi'(j/N)}}{h/N} = \ket{\varphi''(y)}.
  \end{equation*}
  Therefore, taking a worst-case upper bound on the norm of
  $\ket{\varphi''(y)}$ and then using Lem.~\ref{lem:statederbound}, we
  obtain:
  \begin{equation}
    \label{eq:pjnormdiff}
    \begin{aligned}
      \nrm{p_{j+h} - p_{j}} &\le \frac{h}{N^2} \nrm{\ket{\varphi''(y)}} +
      \bigO{\frac{1}{N^2}} \le \frac{h}{N^2} \nrm{\ket{\varphi''}} +
      \bigO{\frac{1}{N^2}} \\
      &\le \frac{h}{N^2}
      \left(\frac{\nrm{\ham''}}{\gamma} + \frac{3\nrm{\ham'}^2}{\gamma^2}
      \right) + \bigO{\frac{1}{N^2}}.
    \end{aligned}
  \end{equation}
  Analyzing the very first summation appearing in
  Eq.~\eqref{eq:adisimplification2}, using the triangle inequality and
  Eq.~\eqref{eq:pjnormdiff}, we have:
  \begin{align*}
    \nrm{\sum_{j=1}^{M} U_{M-1}\cdots U_j p_j - \sum_{j=1}^{M} U_{M-1}\cdots U_j
      p_1} &\le \sum_{j=1}^{M} \nrm{U_{M-1}\cdots U_j p_j - U_{M-1}\cdots U_j p_1} = \sum_{j=1}^{M} \nrm{p_j - p_1} \\
    &\le \sum_{j=1}^{M} \left(\frac{j}{N^2}
    \left(\frac{\nrm{\ham''}}{\gamma} + \frac{3\nrm{\ham'}^2}{\gamma^2}
    \right) + \bigO{\frac{1}{N^2}} \right) \\
    &\le \frac{M^2}{N^2}
    \left(\frac{\nrm{\ham''}}{\gamma} + \frac{3\nrm{\ham'}^2}{\gamma^2}
    \right) + \bigO{\frac{1}{N}}.
  \end{align*}
  Recall our definition of $M = \ceil{8 \nrm{\ham'} N /(\delta
    \gamma^2 T)}$, and substitute into the last equation (but, for
  ease of subsequent calculations, keep one ``copy'' of $M$; i.e.,
  write $M^2 = M \ceil{8 \nrm{\ham'} N /(\delta \gamma^2 T)}$). We
  get:
  \begin{equation}
    \label{eq:adipjsubbound}
    \frac{M^2}{N^2} \left(\frac{\nrm{\ham''}}{\gamma} +
    \frac{3\nrm{\ham'}^2}{\gamma^2} \right) = \frac{8 M \nrm{\ham'}}{N
      \delta \gamma^2 T} \left(\frac{\nrm{\ham''}}{\gamma} +
    \frac{3\nrm{\ham'}^2}{\gamma^2} \right) = \frac{\delta M}{4 N }
    \left(\frac{32\nrm{\ham'}\nrm{\ham''}}{\delta^2 \gamma^3 T} +
    \frac{96\nrm{\ham'}^3}{\delta^2 \gamma^4 T} \right).
  \end{equation}
  Because, by assumption,
  \begin{equation}
    \label{eq:aditbounds}
    T \ge \frac{10^3}{\delta^2} \frac{\nrm{\ham'}^3}{\gamma^4}, \qquad T
    \ge \frac{10^3}{\delta^2} \frac{\nrm{\ham'} \nrm{\ham''}}{\gamma^3},
  \end{equation}
  the term in parentheses at the r.h.s.\ in
  Eq.~\eqref{eq:adipjsubbound} is upper bounded by $1$. Thus, thanks
  to our choice of $M$ and $T$, we obtain:
  \begin{equation}
    \label{eq:adireplacepjgoal}
    \nrm{\sum_{j=1}^{M} U_{M-1}\cdots U_j p_j - \sum_{j=1}^{M}
      U_{M-1}\cdots U_j p_1} \le \frac{\delta M}{4 N } +
    \bigO{\frac{1}{N}}.
  \end{equation}
  This bounds the approximation error incurred by substituting $p_1$
  to replace each $p_j$. Finally, we analyze the effect of replacing
  each $U_j$ with $U_1$; in particular, we want to show:
  \begin{equation}
    \label{eq:adireplaceujgoal}
    \begin{aligned}
      \nrm{\sum_{j=1}^{M} U_{M-1}\cdots U_j p_1 - \sum_{j=0}^{M-1} U_1^j
        p_1} &= \nrm{\sum_{j=1}^{M} U_{M-1}\cdots U_j p_1 - \sum_{j=1}^{M} U_1^{M-j}
        p_1} \\
      &= \nrm{\sum_{j=1}^{M-1} U_{M-1}\cdots U_j p_1 - \sum_{j=1}^{M-1} U_1^{M-j}
        p_1} \\
      &\le \frac{\delta M}{4 N} + \bigO{\frac{1}{N^2}}.
    \end{aligned}
  \end{equation}
  In Eq.~\ref{eq:adireplaceujgoal}, the second equality is due to the
  fact that one term in each summation is equal to $p_1$ (recall
  Rem.~\ref{rem:upperlimitempty}), so it cancels out.  We prove
  Eq.~\ref{eq:adireplaceujgoal} by induction. Note that in the first
  summation, i.e., $\sum_{j=1}^{M} U_{M-1}\cdots U_j p_1$, $U_1$
  appears once, $U_2$ appears twice, $U_3$ appears three times, and so
  on. In this summation we replace each unitary up to $U_h$ with
  $U_1$, and proceed by induction on $h$. The inductive statement is:
  \begin{equation}
    \label{eq:adireplaceind}
    \nrm{\sum_{j=1}^{h} U_{h} U_{h-1} \dots U_j p_1 - \sum_{j=1}^{h} U_1^{h+1-j}
      p_1} \le \frac{2 (h + 1)h \nrm{\ham'}^2}{\gamma^2 N^2} + \bigO{h/N^3}.
  \end{equation}
  We prove the statement in Lem.~\ref{lem:adireplaceind}. Applying it
  for $h = M-1$ yields:
  \begin{equation*}
    \nrm{\sum_{j=1}^{M-1} U_{M-1} U_{M-2} \dots U_j p_1 - \sum_{j=1}^{M-1} U_1^{M-j}
      p_1} \le \frac{2 M^2 \nrm{\ham'}^2}{\gamma^2 N^2} + \bigO{\frac{1}{N^2}}.
  \end{equation*}
  By definition of $M = \ceil{8 \nrm{\ham'} N /(\delta \gamma^2 T)}$,
  and by our choice of $T$ (see Eq.~\eqref{eq:aditbounds}), we finally
  upper bound the r.h.s.\ of the above equation as follows:
  \begin{align*}
    \frac{2 M^2 \nrm{\ham'}^2}{\gamma^2 N^2} = \frac{16 M \nrm{\ham'}^3}{\delta \gamma^4 N T} = \frac{\delta M}{4 N} \left(\frac{64 \nrm{\ham'}^3}{\delta^2 \gamma^4 T} \right)\le \frac{\delta M}{4 N}.
  \end{align*}
  This proves Eq.~\ref{eq:adireplaceujgoal}. Putting together
  Eq.s~\ref{eq:adireplacepjgoal} and \ref{eq:adireplaceujgoal} with a
  final triangle inequality completes the proof of (iii): up to error
  terms that go to zero as $N \to \infty$, replacing all $p_j$ with
  $p_1$ introduces an error of $\frac{\delta M}{4 N}$, and replacing
  all $U_j$ with $U_1$ introduces an error of $\frac{\delta M}{4 N}$,
  yielding \eqref{eq:adisimplification2}.

  (iv). The inequality $\nrm{p_1} \le \nrm{\ham'}/(\gamma N)$ in
  Eq.~\eqref{eq:p1bound} follows immediately by applying
  Lem.~\ref{lem:statederbound} to Eq.~\ref{eq:pjpsiprime}.
\end{proof}

\noindent The lemma below is used in the proof of
Thm.~\ref{thm:adiabaticzero}.
\begin{lemma}
  \label{lem:adireplaceind}
  Let $p_j,U_j$ for $j=1,\dots,N$ be defined as in
  Thm.~\ref{thm:adiabaticzero}. Then:
  \begin{equation*}
    \nrm{\sum_{j=1}^{k} U_{k} U_{k-1} \dots U_j p_1 - \sum_{j=1}^{k} U_1^{k+1-j}
      p_1} \le \frac{2 (k + 1)k \nrm{\ham'}^2}{\gamma^2 N^2} + \bigO{k/N^3}.    
  \end{equation*}
\end{lemma}
\begin{proof}
  We proceed by induction on $k$. For $k=1$, the statement is trivial:
  the two summations inside the norm are equal to $U_1 p_1$, hence the
  l.h.s.\ is zero. To inductively go from $k-1$ to $k$, we use the
  triangle inequality as follows:
  \begingroup \allowdisplaybreaks
  \begin{equation}
    \label{eq:adiupindstep}
    \begin{aligned}
      &\phantom{=\|} \nrm{\sum_{j=1}^{k} U_{k} U_{k-1} \dots U_j p_1 - \sum_{j=1}^{k} U_1^{k+1-j}
        p_1}\\
      &= \nrm{\sum_{j=1}^{k} U_{k} U_{k-1} \dots U_j p_1 - \sum_{j=1}^{k} U_{k} U_1^{k-j} p_1 + \sum_{j=1}^{k} U_{k} U_1^{k-j} p_1 - \sum_{j=1}^{k} U_1^{k+1-j}
        p_1} \\
      &\le \underbrace{\nrm{\sum_{j=1}^{k} U_{k} U_{k-1} \dots U_j p_1 - \sum_{j=1}^{k} U_{k} U_1^{k-j} p_1}}_{\text{term } (a)} + \underbrace{\nrm{\sum_{j=1}^{k} U_{k} U_1^{k-j} p_1 - \sum_{j=1}^{k} U_1^{k+1-j} p_1}}_{\text{term } (b)}.
    \end{aligned}
  \end{equation}
  \endgroup
  For term $(a)$, note that the terms of the two summations for $j=k$
  are both equal to $U_k p_1$, so they cancel out. Thus, we can write:
  \begin{align*}
    \nrm{\sum_{j=1}^{k} U_{k} U_{k-1} \dots U_j p_1 - \sum_{j=1}^{k} U_{k} U_1^{k-j} p_1} &= \nrm{U_{k} \left(\sum_{j=1}^{k-1} U_{k-1} \dots U_j p_1 - \sum_{j=1}^{k-1} U_1^{k-j} p_1\right)} =\\
    \nrm{\sum_{j=1}^{k-1} U_{k-1} \dots U_j p_1 - \sum_{j=1}^{k-1} U_1^{k-j} p_1}
    &\le \frac{2 k(k-1) \nrm{\ham'}^2}{\gamma^2 N^2} + \bigO{k/N^3}.
  \end{align*}
  where for the second equality we used the fact that $U_{k}$ is
  unitary, and the final inequality applies the inductive
  hypothesis. For term $(b)$ we have:
  \begin{equation}
    \label{eq:aditermb}
    \nrm{\sum_{j=1}^{k} U_{k} U_1^{k-j} p_1 - \sum_{j=1}^{k} U_1^{k+1-j}
      p_1} =  \nrm{(U_{k}- U_1)\sum_{j=1}^{k} U_1^{k-j} p_1 } \le \nrm{U_k- U_1}\nrm{\sum_{j=1}^{k} U_1^{k-j} p_1}.
  \end{equation}
  We bound $\nrm{U_{k} - U_1}$ using a telescopic sum:
  \begin{equation}
    \label{eq:adiuktelescopic}
    \nrm{U_{k} - U_1} = \nrm{ \sum_{j=1}^{k-1} U_{j+1} - U_{j}} \le \sum_{j=1}^{k-1} \nrm{U_{j+1} - U_{j}}.
  \end{equation}
  Recall that $U_{j+1} = e^{i \frac{T}{N} \ham((j+1)/N)}$
  (Eq.~\ref{eq:ujdef}). Using a Trotter formula
  (Sect.~\ref{sec:trotter}), we have:
  \begin{equation*}
    U_{j+1} = e^{i \frac{T}{N} \left(\ham((j+1)/N)-\ham(j/N)+\ham(j/N)\right)} \approx e^{i \frac{T}{N} \left(\ham((j+1)/N)-\ham(j/N)\right)} e^{i \frac{T}{N} \ham(j/N)},
  \end{equation*}
  with error:
  \begingroup
  \allowdisplaybreaks
  \begin{align*}
    \bigO{ \nrm{\frac{T}{N} \left(\ham((j+1)/N)-\ham(j/N)\right)} \nrm{\frac{T}{N} \ham(j/N)}} &= \bigO{\frac{T^2}{N^2} \nrm{\left(\ham((j+1)/N)-\ham(j/N)\right)} \nrm{\ham(j/N)}} \\
    &= \bigO{\frac{T^2}{N^2} \nrm{\frac{1}{N} \ham'(j/N)} \nrm{\ham(j/N)}} \\
    &= \bigO{\frac{T^2\nrm{\ham}\nrm{\ham'}}{N^3}} = \bigO{\frac{1}{N^3}},
  \end{align*}
  \endgroup where for the second equality we used the fact that
  $\lim_{N \to \infty} \frac{\ham((j+1)/N) - \ham(j/N)}{1/N} =
  \ham'(j/N)$, and for the final equality we neglected al terms except
  the ones that depend on $N$, see Rem.~\ref{rem:ntoinfty}. Thus:
  \begin{equation}
    \label{eq:adiujsingle}
    \begin{aligned}
      \nrm{U_{j+1} - U_{j}} &= \nrm{e^{i \frac{T}{N} \ham((j+1)/N)} - e^{i \frac{T}{N} \ham(j/N)} } \\
      &= \nrm{e^{i \frac{T}{N} \ham(j/N)}\left(e^{i \frac{T}{N} (\ham((j+1)/N) - \ham(j/N))} - I  \right)} + \bigO{\frac{1}{N^3}} \\
      &= \nrm{\frac{T}{N} \left(\ham((j+1)/N) - \ham(j/N)\right)} + \bigO{\frac{1}{N^3}}
    \le \frac{T \nrm{\ham'}}{N^2} + \bigO{\frac{1}{N^3}}.
    \end{aligned}
  \end{equation}
  In the chain above, for the third equality we used the fact that
  $e^{i \frac{T}{N} \ham(j/N)}$ is unitary plus the fact that
  $e^{i\frac{T}{N} (\ham((j+1)/N) - \ham(j/N))} - I = i\frac{T}{N}
  (\ham((j+1)/N) - \ham(j/N)) + \bigO{\frac{T^2}{N^2}
    \frac{\nrm{\ham'}^2}{N^2} }$ by definition of matrix exponential
  and because $\lim_{N \to \infty} \frac{\ham((j+1)/N) -
    \ham(j/N)}{1/N} = \ham'(j/N)$, and for the inequality we again
  used the limit defining $\ham'(j/N)$. Using Eq.~\eqref{eq:adiujsingle} in
  Eq.~\eqref{eq:adiuktelescopic}, we finally obtain:
  \begin{equation*}
    \nrm{U_{k} - U_1} \le \frac{k T \nrm{\ham'}}{N^2} + \bigO{\frac{k}{N^3}}.
  \end{equation*}
  We plug this into Eq.~\eqref{eq:aditermb}. For the remaining term in
  Eq.~\eqref{eq:aditermb}, i.e., $\nrm{\sum_{j=1}^{k} U_1^{k-j} p_1}$,
  we apply exactly the same argument used in
  Eq.s~\eqref{eq:adiu1p1geom}-\eqref{eq:adiu1p1geombound} to show
  that:
  \begin{equation*}
    \nrm{\sum_{j=1}^{k} U_1^{k-j} p_1} = \nrm{\sum_{j=0}^{k-1} U_1^{j} p_1} \le \frac{4 \nrm{\ham'}}{T \gamma^2},
  \end{equation*}
  see in particular the first inequality in
  Eq.~\eqref{eq:adiu1p1geombound}.  Eq.~\eqref{eq:aditermb} then
  yields:
  \begin{equation*}
    \nrm{\sum_{j=1}^{k} U_{k} U_1^{k-j} p_1 - \sum_{j=1}^{k} U_1^{k+1-j}
      p_1} \le \left(\frac{kT \nrm{\ham'}}{N^2} + \bigO{\frac{k}{N^3}}\right) \frac{4 \nrm{\ham'}}{T \gamma^2} = \frac{4 k\nrm{\ham'}^2}{\gamma^2 N^2} + \bigO{\frac{k}{N^3}}.
  \end{equation*}
  We can now continue the chain of inequalities in
  Eq.~\eqref{eq:adiupindstep}, using the upper bounds for terms $(a)$
  and $(b)$ that we derived:
  \begin{align*}
    \nrm{\sum_{j=1}^{k} U_{k} U_{k-1} \dots U_j p_1 - \sum_{j=1}^{k} U_1^{k+1-j}
      p_1} &\le \frac{2 k(k-1)\nrm{\ham'}^2}{\gamma^2 N^2} + \frac{4 k\nrm{\ham'}^2}{\gamma^2 N^2} + \bigO{\frac{k}{N^3}}\\
    &= \frac{2 (k+1)k \nrm{\ham'}^2}{\gamma^2 N^2} + \bigO{\frac{k}{N^3}}.
  \end{align*}
  This concludes the proof.
\end{proof}

We end the section by formally proving the general version of the
adiabatic theorem, Thm.~\ref{thm:adiabatic}, using the special case
proven in this section. Specifically, in Thm.~\ref{thm:adiabaticzero}
we assume $\lambda(s) = 0$ for all $s$, but Thm.~\ref{thm:adiabatic}
applies to any eigenpair. We show that the assumption $\lambda(s) = 0$
is not restrictive by using the Hamiltonian transformation sketched in
Rem.~\ref{rem:zeroeig}: for any given Hamiltonian $\ham$ and eigenpair
$(\ket{\varphi(s)}, \lambda(s))$ for the adiabatic evolution, we
transform $\ham$ into a new Hamiltonian $\hat{\ham}$ with eigenpair
$(\ket{\varphi(s)}, 0)$, apply Thm.~\ref{thm:adiabaticzero}, and the
value of $T$ required by Thm.~\ref{thm:adiabaticzero} for $\hat{\ham}$
corresponds to the value of $T$ stated in Thm.~\ref{thm:adiabatic}.
\begin{corollary}
  \label{cor:zeroeig}
  The lower bound for $T$ stated in Thm.~\ref{thm:adiabatic} suffices
  to ensure adiabatic evolution, i.e., a final state such that:
  \begin{equation*}
    \nrm{ e^{i \theta} \ket{\psi(1)} - \ket{\varphi(1)} } \le \delta
    \qquad \text{ for some } \theta.
  \end{equation*}
\end{corollary}
\begin{proof}
  Consider the Hamiltonian $\hat{\ham}(s) = \ham(s) - \lambda(s) I$. Because
  $\ket{\varphi(s)}$ is an eigenvector of $\ham(s)$ with eigenvalue
  $\lambda(s)$, we have, for all $s$:
  \begin{equation*}
    \hat{\ham}(s) \ket{\varphi(s)} = \ham(s) \ket{\varphi(s)} - \lambda(s) \ket{\varphi(s)} = 0,
  \end{equation*}
  thus $\ket{\varphi(s)}$ is an eigenvector of $\hat{\ham}(s)$ with
  eigenvalue 0. \cite{ambainis2004elementary}[Lem.~4.1] shows that for
  any $s$:
  \begin{equation}
    \label{eq:ldbound}
    \lambda'(s) \le \nrm{\ham'}, \qquad \lambda''(s) \le \nrm{\ham''} + 4 \nrm{\ham'}^2/\gamma.
  \end{equation}
  Using Eq.~\eqref{eq:ldbound}, we therefore have:
  \begin{equation*}
    \nrm{\hat{\ham}'} = \max_{s \in [0,1]} \nrm{ \frac{\di(\ham(s) - \lambda(s) I)}{\di s} } \le \nrm{\ham'} + \nrm{\lambda'} \le 2 \nrm{\ham'},
  \end{equation*}
  and:
  \begin{equation*}
    \nrm{\hat{\ham}''} = \max_{s \in [0,1]} \nrm{ \frac{\di^2(\ham(s) - \lambda(s) I)}{\di s^2} } \le \nrm{\ham''} + \nrm{\lambda''} \le 2 \nrm{\ham''} + 4 \nrm{\ham''}^2/\gamma.
  \end{equation*}
  Applying Thm.~\ref{thm:adiabaticzero} to the Hamiltonian
  $\hat{\ham}$, and using the above bounds for the norm of the
  derivatives of $\hat{\ham}$, we see that it is sufficient to choose
  $T \ge \frac{1000}{\delta^2}
  \max\left\{\frac{\nrm{\hat{\ham}'}^3}{\gamma^4},
  \frac{\nrm{\hat{\ham}'} \nrm{\hat{\ham}''}}{\gamma^3} \right\}$.
  We upper bound the r.h.s.\ of this inequality as follows:
  \begin{align*}
    \frac{1000}{\delta^2} \max\left\{\frac{\nrm{\hat{\ham}'}^3}{\gamma^4}, \frac{\nrm{\hat{\ham}'} \nrm{\hat{\ham}''}}{\gamma^3} \right\} &\le
    \frac{1000}{\delta^2} \max\left\{\frac{8\nrm{\ham'}^3}{\gamma^4}, \frac{2 \nrm{\ham'} (2\nrm{\ham''} + 4 \nrm{\ham'}^2/\gamma)}{\gamma^3} \right\} \\
    &= \frac{1000}{\delta^2} \left(\frac{8 \nrm{\ham'}^3}{\gamma^4} + \frac{4 \nrm{\ham'} \nrm{\ham''}}{\gamma^3} \right) \\
    &\le \frac{10^4}{\delta^2} \left(\frac{\nrm{\ham'}^3}{\gamma^4} + \frac{\nrm{\ham'} \nrm{\ham''}}{\gamma^3} \right). 
  \end{align*}
  This is the lower bound for $T$ stated in Thm.~\ref{thm:adiabatic},
  and it completes its proof.\index{simulation!Hamiltonian|)}
\end{proof}

\subsection{Spectral gap dependence and estimation}
\label{sec:adigap}
To understand if the adiabatic theorem\index{spectral gap|(} can lead to an effective
optimization algorithm in theory, we must analyze its running time
relative to the problem instance parameters. Recall how optimization
with the adiabatic theorem works: we start in the ground state of a
known, ``easy'' Hamiltonian (i.e., one for which the ground state is
known and can be easily prepared), then we transform this initial
Hamiltonian into a target Hamiltonian that encodes the desired
optimization problem; see the discussion in
Rem.~\ref{rem:adiforopt}. When applying the adiabatic theorem to solve
an optimization problem in the above manner, the value of $T$ (i.e.,
the length of the simulation of the Schr\"odinger equation) determines
the running time: as we have seen in Ch.~\ref{ch:hamsim}, the time
complexity of efficient Hamiltonian simulation algorithms is linear in
the length of the time horizon --- the parameter that we called
``$t$'' in Ch.~\ref{ch:hamsim}.
\begin{remark}
  In fact, no quantum algorithm can solve the Hamiltonian simulation
  problem with fewer than $\Omega(t)$ operations, due to existing lower bounds
  \cite{berry2007efficient,berry2015hamiltonian}. So in general we do
  not expect that it is possible to perform optimization using the
  adiabatic theorem with fewer than $T$ operations on a quantum
  computer.
\end{remark}
\noindent In turn, the spectral gap $\gamma$ is often the crucial
parameter that determines the value of $T$.

According to Thm.~\ref{thm:adiabatic}, to ensure that the slow time
evolution in Eq.~\eqref{eq:adischrodinger} always leaves the system in
an instantaneous eigenstate of the Hamiltonian (with at most a small
error), the choice of $T$ satisfies $T \ge \nrm{\ham'}^3/\gamma^4 +
\nrm{\ham'} \nrm{\ham''}/\gamma^3$. Thus, our proof of the adiabatic
theorem yields a dependence on the spectral gap parameter $\gamma$
that is in the order of $1/\gamma^4$. As stated in
Sect.~\ref{sec:adihighlevel}, this is not tight. The folklore result
is that for the main statement of Thm.~\ref{thm:adiabatic} to hold, it
is sufficient to choose:
\begin{equation*}
  T \gg \int_{0}^{1} \frac{\nrm{\ham'(s)}}{\gamma^2} \di s,
\end{equation*}
see, e.g., \cite{van2001powerful,reichardt2004quantum}. It should be
noted that in the open literature, occasionally doubt has been cast on
the sufficiency of the above condition on $T$ for the general
case. This is likely due to the paucity of rigorous proofs and the
appearance of counterexamples under specific conditions; see, e.g.,
the discussion in \cite{ambainis2004elementary}, as well as
\cite[Sect.~5]{jansen2007bounds}. A more precise characterization of a
sufficient value of $T$, taken from \cite{childs2017lecture} which is
itself based on \cite{teufel2003adiabatic}, is the following (recall
Def.s~\ref{def:instspectralgap} and \ref{def:spectralgap}):
\begin{theorem}[Adiabatic theorem with tighter spectral bound; Thm.~28.1 in \cite{childs2017lecture}]
  \label{thm:adiabatictighter}
  There exists some constant $c$ such that the statement of
  Thm.~\ref{thm:adiabatic} holds if we choose $T$
  satisfying:
  \begin{equation*}
    T \ge \frac{c}{\delta} \left(\frac{\nrm{\ham'(0)}}{\gamma(0)^2} + \frac{\nrm{\ham'(1)}}{\gamma(1)^2} + \int_{0}^{1} \left(\frac{\nrm{\ham'(s)}^2}{\gamma(s)^3} +
    \frac{\nrm{\ham''(s)}}{\gamma(s)^2}\right) \di s \right).
  \end{equation*}
\end{theorem}
An essentially identical result is also proven in
\cite{jansen2007bounds}. This result tightens the dependence on the
spectral gap parameter to the order of $1/\gamma^3$. The expression in
Thm.~\ref{thm:adiabatictighter} depends on some instantaneous
quantities, in particular the instantaneous spectral gap $\gamma(s)$
as well as $\nrm{\ham'(s)}$ evaluated at specific points. We can
simplify it in the case of linear interpolation between the initial
and final Hamiltonian: from Eq.~\eqref{eq:adilinearint} we compute
$\ham'$ and $\ham''$, and taking some pessimistic bounds to eliminate
$s$ from the expression, we obtaining the following.
\begin{corollary}[Adiabatic theorem for linear interpolation]
  \label{cor:adiabaticlinear}
  There exists some constant $c$ such that, when the Hamiltonian
  $\ham(s)$ performs linear interpolation between an initial and a
  final Hamiltonian, as in Eq.~\eqref{eq:adilinearint}, the statement
  of Thm.~\ref{thm:adiabatic} holds if we choose $T$ satisfying:
  \begin{equation*}
    T \ge \frac{c}{\delta} \left(\frac{\nrm{\ham_{\text{F}} - \ham_{\text{I}}}}{\gamma^2} + \frac{\nrm{\ham_{\text{F}} - \ham_{\text{I}}}^2}{\gamma^3} \right).
  \end{equation*}
\end{corollary}

Unfortunately estimating $\gamma$ is often very difficult. Even when
using the linear interpolation strategy of
Eq.~\eqref{eq:adilinearint}, gap estimation requires the analysis of
the spectrum of a time-dependent matrix that is the sum of two terms,
i.e., the initial and final Hamiltonian. This is notoriously
difficult, because there is no precise relationship between the
spectrum of each term and the spectrum of their sum: although several
results to bound the eigenvalues of a sum of two Hermitian matrices
are known (e.g., Weyl's inequality, see also \cite{bhatia2013matrix}),
they generally do not provide useful characterizations of the spectral
gap. Note that we are interested in \emph{lower bounds} to the
spectral gap, because $\gamma$ appears at the denominator of the
expression for the simulation running time $T$.

The spectral gap for just a few time-dependent Hamiltonians that solve
combinatorial optimization problems is known. Typically, a lot of
structure is required and ad-hoc procedures are necessary. We provide
two illustrative examples below: in one case, Ex.~\ref{ex:adihamming},
the gap is polynomially small, leading to a polynomial-time algorithm
to solve a trivial optimization problem; in the other case,
Ex.~\ref{ex:adigrover}, the gap is exponentially small, and leads to
an algorithm that is slower than Grover's algorithm for
black-box search, unless we modify the time-dependent Hamiltonian and
do something more sophisticated than the linear interpolation of
Rem.~\ref{rem:adiforopt}.
\begin{example}
  \label{ex:adihamming}
  Let us apply adiabatic optimization to the problem of minimizing the
  Hamming weight (i.e., the number of ``1''s) of a binary string. This
  problem has the all-zero binary string $\v{0}$ as the obvious
  solution, therefore it is not a difficult problem to solve --- we
  can determine the solution analytically. Nonetheless, the
  application of adiabatic optimization is instructive, and gives us
  an opportunity to showcase the choices involved in the application
  of the approach, and the type of analysis that is necessary.

  For $\vj \in \{0,1\}^n$, let $w(\vj) := \sum_{k=1}^n \vj_k$ be its
  Hamming weight. We aim to solve the following optimization problem:
  \begin{equation}
    \label{eq:adihammingobj}
    \min_{\vj \in \{0,1\}^n} w(\vj).
  \end{equation}
  We choose the final Hamiltonian, for which we want to find the
  minimum eigenvector, as:
  \begin{equation*}
    \ham_{\text{F}} = \sum_{\vj \in \{0,1\}^n} w(\vj) \ketbra{\vj}{\vj}.
  \end{equation*}
  It is straightforward to observe that $\ham_{\text{F}} \ket{\vj} = w(\vj)
  \ket{\vj}$, therefore the eigenvector with minimum eigenvalue
  encodes the global optimum of problem \eqref{eq:adihammingobj}.

  To optimize via the adiabatic algorithm we also need to choose an
  initial Hamiltonian. We choose:
  \begin{equation*}
    \ham_{\text{I}} = \sum_{\vj \in \{0,1\}^n} w(\vj) H^{\otimes n}
    \ketbra{\vj}{\vj} H^{\otimes n},
  \end{equation*}
  where $H^{\otimes n}$ denotes the tensor of $n$ Hadamard
  gates. (Recall: $\ham$ denotes a Hamiltonian, $H$ a Hadamard gate.)
  The minimum eigenpair of $\ham_{\text{I}}$ is the eigenstate
  $H^{\otimes n} \ket{\v{0}}$ with eigenvalue $w(\v{0}) = 0$.

  We want to analyze the spectrum of the time-dependent Hamiltonian:
  \begin{equation*}
    \ham(s) = (1-s) \ham_{\text{I}} + s \ham_{\text{F}},
  \end{equation*}
  so that we have an expression for the spectral gap $\gamma$ to plug
  into Cor.~\ref{cor:adiabaticlinear}. We can do so by decomposing
  $\ham_{\text{I}}$ and $\ham_{\text{F}}$ into sums of single-qubit Hamiltonians. By
  definition, $w(\vj)$ is a sum of terms that depend on a single digit
  of the string $\vj$. We can then write $\ham_{\text{I}}$ as sums of $n$
  Hamiltonians dependent on a single digit:
  \begin{align*}
    \ham_{\text{I}} &= \sum_{\vj \in \{0,1\}^n} w(\vj) H^{\otimes n}
    \ketbra{\vj}{\vj} H^{\otimes n} = \sum_{\vj \in \{0,1\}^n}
    \left(\sum_{k=1}^n \vj_k\right) H^{\otimes n} \ketbra{\vj}{\vj} H^{\otimes n} \\
    &= \sum_{k=1}^n \underbrace{I \otimes \dots \otimes I \otimes \underbrace{\left(\sum_{x \in \{0,1\}} x H \ketbra{x}{x} H\right)}_{\text{position } k} \otimes \, I \otimes \dots \otimes I}_{n \text{ terms}}, \\
    \ham_{\text{F}} &= \sum_{\vj \in \{0,1\}^n} w(\vj) \ketbra{\vj}{\vj} = \sum_{\vj \in \{0,1\}^n}
    \left(\sum_{k=1}^n \vj_k\right) \ketbra{\vj}{\vj} \\
    &= \sum_{k=1}^n \underbrace{I \otimes \dots \otimes I \otimes \underbrace{\left(\sum_{x \in \{0,1\}} x \ketbra{x}{x} \right)}_{\text{position } k} \otimes\, I \otimes \dots \otimes I}_{n \text{ terms}},
  \end{align*}
  where we used the facts that $\sum_{x \in \{0,1\}}
  \ketbra{x}{x} = I$ and $\sum_{x \in \{0,1\}} H
  \ketbra{x}{x} H = H \left(\sum_{x \in \{0,1\}}
  \ketbra{x}{x}\right) H = I$. Thus:
  \begin{equation}
    \label{eq:adihamminghs}
    \ham(s) = \sum_{k=1}^n I \otimes \dots \otimes I \otimes \underbrace{\left(\sum_{x \in \{0,1\}} x \left((1-s) H \ketbra{x}{x} H + s \ketbra{x}{x}\right) \right)}_{\text{position } k} \otimes I \otimes \dots \otimes I.
  \end{equation}
  Let us analyze a single term of the summation, i.e., a term for
  fixed $k$ --- the other Hamiltonians are similar. For each $k$ the
  only nontrivial action is on the $k$-th qubit. The corresponding
  single-qubit Hamiltonian on the $k$-th digit is:
  \begin{equation*}
    \sum_{x \in \{0,1\}} x \left((1-s) H \ketbra{x}{x} H + s \ketbra{x}{x}\right) = \frac{1}{2} \begin{pmatrix} 1-s & s-1 \\ s-1 & 1+s \end{pmatrix}.
  \end{equation*}
  The eigendecomposition of this matrix is straightforward to
  calculate, yielding eigenvalues:
  \begin{equation*}
    \lambda_0(s) = \frac{1}{2}\left(1 - \sqrt{2s^2 - 2s +1}\right)
    \qquad
    \lambda_1(s) = \frac{1}{2}\left(1 + \sqrt{2s^2 - 2s +1}\right),
  \end{equation*}
  with corresponding eigenstates that we label $\ket{\psi_0(s)},
  \ket{\psi_1(s)}$ respectively. Having established the eigenvalues of
  a single term in Eq.~\eqref{eq:adihamminghs}, we can easily
  establish the eigenvalues of $\ham(s)$. Indeed, for every $\vx \in
  \{0,1\}^n$, the state:
  \begin{equation*}
    \ket{\psi_{\vx}(s)} := \ket{\psi_{\vx_1}(s)} \otimes \ket{\psi_{\vx_2}(s)} \otimes \dots \otimes \ket{\psi_{\vx_n}(s)},
  \end{equation*}
  i.e., a tensor product of the eigenstate $\ket{\psi_0(s)}$ or
  $\ket{\psi_1(s)}$ for each qubit, is an eigenstate of $\ham(s)$ with
  eigenvalue $(n-w(\vx))\lambda_0(s) + w(\vx) \lambda_1(s)$. This
  follows from the fact that the $k$-th term in $\ham(s)$ has
  $\ket{\psi_{\vx}(s)}$ as an eigenstate, with eigenvalue $\lambda_0(s)$
  or $\lambda_1(s)$ depending on $\vx_k$. The vectors of the form
  $\ket{\psi}_{\vx}$ are $2^n$ linearly independent eigenstates,
  therefore we have characterized the full set of eigenvectors for
  $\ham(s)$. It follows that the instantaneous spectral gap is:
  \begin{equation*}
    \gamma(s) = \left((n-1)\lambda_0(s) + \lambda_1(s)\right) -
    n\lambda_0(s) = \sqrt{2s^2 - 2s +1}.
  \end{equation*}
  Solving $\min_{s \in [0,1]} \gamma(s)$ yields $\gamma = \min_{s \in
    [0,1]} \gamma(s) = 1/\sqrt{2}$, attained at $s = 1/2$: the
  spectral gap $\gamma$ is constant. Using
  Cor.~\ref{cor:adiabaticlinear}, we see that we must choose $T =
  \bigO{\nrm{\ham_{\text{F}}-\ham_{\text{I}}}^2} = \bigO{n^2}$. With
  this choice of the initial and final Hamiltonians, problem
  \eqref{eq:adihammingobj} is solved in polynomial time via adiabatic
  optimization. As remarked at the beginning of this example, the
  problem is trivial so this discussion is simply meant to illustrate
  a problem with constant spectral gap. In \cite{van2001powerful},
  which was the inspiration for this example, the reader can find a
  proof that a perturbed version of problem \eqref{eq:adihammingobj}
  requires exponential time.
\end{example}
\begin{example}
  \label{ex:adigrover}
  We now study the application of adiabatic optimization to the
  black-box search problem solved with Grover's algorithm in
  Sect.~\ref{sec:grover}. Let $\v{\ell} \in \{0,1\}^n$ be the marked
  element, which we assume to be unique for simplicity, and define:
  \begin{equation}
    \label{eq:adigroverf}
    f(\vj) := \begin{cases} 0 & \text{if } \vj = \v{\ell} \\
      1 & \text{if } \vj \neq \v{\ell}.
    \end{cases}
  \end{equation}
  Our goal is to determine $\v{\ell}$, which we can do by solving the
  following optimization problem and finding its global optimum:
  \begin{equation*}
    \min_{\vj \in \{0,1\}^n} f(\vj).
  \end{equation*}
  We want to solve this problem using the adiabatic theorem. As a
  Hamiltonian, the problem can be encoded by:
  \begin{equation*}
    \ham_{\text{F}} = I^{\otimes n} - \ketbra{\v{\ell}}{\v{\ell}}.
  \end{equation*}
  This is the projector onto states orthogonal to $\ket{\v{\ell}}$.
  The state $\ket{\v{\ell}}$ is an eigenstate of $\ham_{\text{F}}$ with
  eigenvalue $0$, whereas every state orthogonal to $\ket{\v{\ell}}$ is
  an eigenstate with eigenvalue $1$. This corresponds precisely to
  the objective function in Eq.~\eqref{eq:adigroverf}. Thus, $\ham_{\text{F}}$
  is our final Hamiltonian for which we want to determine the
  eigenstate with the smallest eigenvalue. Note that $\v{\ell}$ is
  supposed to be unknown, so we assume that we have the ability to
  apply and operate on the Hamiltonian $\ham_{\text{F}}$, but we do not know its
  analytical description --- otherwise, we would know the value of
  $\v{\ell}$.

  As in the previous example, we need to choose an initial
  Hamiltonian. Define $\ket{\phi} := \left(H \ket{0}\right)^{\otimes
    n} = \frac{1}{\sqrt{2^n}} \sum_{\vj \in \{0,1\}^n} \ket{\vj}$,
  i.e., the uniform superposition state. We choose the initial
  Hamiltonian as:
  \begin{equation*}
    \ham_{\text{I}} = I^{\otimes n} - \ketbra{\phi}{\phi}.
  \end{equation*}
  This is the projector onto states orthogonal to $\ket{\phi}$;
  the state $\ket{\phi}$ is an eigenvector of $\ham_{\text{I}}$ with
  eigenvalue $0$. Linear interpolation between $\ham_{\text{I}}$ and $\ham_{\text{F}}$ yields
  the time-dependent Hamiltonian:
  \begin{equation*}
    \ham(s) = (1-s) \ham_{\text{I}} + s \ham_{\text{F}} = I^{\otimes n} + (s-1)
    \ketbra{\phi}{\phi} - s \ketbra{\v{\ell}}{\v{\ell}}.
  \end{equation*}
  The spectrum of $\ham(s)$ is easy to analyze. $\ham_{\text{I}}$ acts
  as the identity on every state orthogonal to $\ket{\phi}$, whereas
  $\ham_{\text{F}}$ acts as the identity on every state orthogonal to
  $\ket{\v{\ell}}$. Thus, if a state is orthogonal to both
  $\ket{\phi}$ and $\ket{\v{\ell}}$, $\ham(s)$ acts as the
  identity. It follows that the only nontrivial action of $\ham(s)$
  takes place in the subspace spanned by $\ket{\phi}$ and
  $\ket{\v{\ell}}$, which is two-dimensional. An orthonormal basis for
  this two-dimensional space is given by $\{\ket{\v{\ell}},
  \ket{\v{\ell}^{\perp}}\}$, where:
  \begin{equation*}
    \ket{\v{\ell}^{\perp}} := \frac{1}{\sqrt{1 - \abs{\braket{\phi}{\v{\ell}}}^2}}\left(\ket{\phi} - \braket{\phi}{\v{\ell}} \ket{\v{\ell}}\right) =
    \frac{\sqrt{2^n}}{\sqrt{2^n - 1}}\left(\ket{\phi} - \frac{1}{\sqrt{2^n}} \ket{\v{\ell}}\right)
  \end{equation*}
  is the (normalized) projection of $\ket{\phi}$ onto the space
  orthogonal to $\ket{\v{\ell}}$. We can then express $\ham(s)$ in the
  basis $\{\ket{\v{\ell}}, \ket{\v{\ell}^{\perp}}\}$:
  \begin{equation*}
    \ham_{\text{F}} = \begin{pmatrix}
      0 & 0 \\ 0 & 1
    \end{pmatrix},
    \qquad
    \ham_{\text{I}} = \begin{pmatrix}
      1 - \frac{1}{2^n} & - \frac{\sqrt{2^n-1}}{2^n} \\ - \frac{\sqrt{2^n-1}}{2^n} & \frac{1}{2^n}
    \end{pmatrix}
    ,
  \end{equation*}
  therefore:
  \begin{equation}
    \label{eq:adigroverhs}
    \ham(s) = (1-s) \ham_{\text{I}} + s \ham_{\text{F}} =  \begin{pmatrix}
      (1-s)(1 - \frac{1}{2^n}) & - (1-s) \frac{\sqrt{2^n-1}}{2^n} \\ - (1-s) \frac{\sqrt{2^n-1}}{2^n} & (1-s) \frac{1}{2^n} + s
    \end{pmatrix}.
  \end{equation}
  The eigenvalues of this matrix can be obtained with
  straightforward calculations, yielding:
  \begin{equation*}
    \lambda_0(s) = \frac{1}{2} \left(1 - \sqrt{1-4s(1-s)\left(1-\frac{1}{2^n}\right)} \right)
    \qquad
    \lambda_1(s) = \frac{1}{2} \left(1 + \sqrt{1-4s(1-s)\left(1-\frac{1}{2^n}\right)} \right).
  \end{equation*}
  The instantaneous spectral gap is the difference between $\lambda_1(s)$
  and $\lambda_0(s)$:
  \begin{equation*}
    \gamma(s) = \lambda_1(s) - \lambda_0(s) = \sqrt{1-4s(1-s)\left(1-\frac{1}{2^n}\right)},
  \end{equation*}
  and $\gamma(s)$ is minimized at $s = 1/2$, where $\gamma(1/2) =
  \frac{1}{\sqrt{2^n}}$. Thus, the spectral gap is exponentially
  small. Applying Cor.~\ref{cor:adiabaticlinear} with $\gamma =
  \frac{1}{\sqrt{2^n}}$ gives a running time of $\bigO{2^{3n/2}}$,
  worse than Grover's algorithm and worse than evaluating $f(\vj)$ for
  all values of $\vj \in \{0,1\}^n$. To get a tighter bound, we apply
  Thm.~\ref{thm:adiabatictighter} directly (remember that
  Cor.~\ref{cor:adiabaticlinear} intentionally loosened some bounds to
  get a simpler expression). Using the fact that $\ham' =
  \ham_{\text{F}} - \ham_{\text{I}}$, $\ham'' = 0$, we obtain:
  \begin{align*}
    T &\ge \frac{c}{\delta} \left(\frac{\nrm{\ham'(0)}}{\gamma(0)^2} + \frac{\nrm{\ham'(1)}}{\gamma(1)^2} + \int_{0}^{1} \left(\frac{\nrm{\ham'(s)}^2}{\gamma(s)^3} +
    \frac{\nrm{\ham''(s)}}{\gamma(s)^2}\right) \di s \right) \\
    &= \frac{c}{\delta} \left(\nrm{\ham_{\text{F}} - \ham_{\text{I}}} + \int_{0}^{1} \frac{\nrm{\ham_{\text{F}} - \ham_{\text{I}}}^2}{\left(1-4s(1-s)\left(1-\frac{1}{2^n}\right)\right)^{3/2}} \di s \right) = \bigO{2^n}.
  \end{align*}
  (The last equality is not obvious, but the computation of the
  integral is tedious and not particularly insightful, so we skip it.)
  This is still no better than evaluating $f(\vj)$ for all values of
  $\vj \in \{0,1\}^n$, therefore giving no quantum speedup.

  We can do better, while still relying on the adiabatic theorem, but
  we need to adjust our strategy. One approach would be to change the
  initial Hamiltonian $\ham_{\text{I}}$, and try to come up with an
  $\ham_{\text{I}}$ that yields a better running time. There is
  another possibility: we can change the time-dependent Hamiltonian
  $\ham(s)$, while keeping $\ham_{\text{I}}$ and $\ham_{\text{F}}$
  fixed. Our current definition of $\ham(s)$ performs linear
  interpolation between $\ham_{\text{I}}$ and $\ham_{\text{F}}$, see
  Eq.~\eqref{eq:adilinearint}; however, the adiabatic theorem as
  stated (Thm.s~\ref{thm:adiabatic} and \ref{thm:adiabatictighter})
  allows for a general time-varying Hamiltonian, provided that it is
  twice differentiable and we can bound the norm of its
  derivatives. Thus, we are allowed to perform nonlinear
  interpolation, and that can change the running time because it can
  affect the spectral gap and the derivatives of $\ham(s)$. In
  Eq.~\eqref{eq:adigroverhs} we replace the linear interpolation terms
  $(1-s), s$ with more general functions $1-h(s), h(s)$, yielding
  $\gamma(s) =
  \sqrt{1-4h(s)(1-h(s))\left(1-\frac{1}{2^n}\right)}$. With an
  appropriate choice of $h(s)$, the value of $T$ for adiabatic
  optimization to work in this context goes down to
  $\bigO{\sqrt{2^n}}$, matching the query and gate complexity of
  Grover's algorithm: this can be achieved by choosing $h(s)$ so that
  the Hamiltonian changes more slowly when the instantaneous spectral
  gap is small, i.e., around $s = 1/2$, and to compensate for the
  slowdown, the pace of the interpolation is increased around the two
  endpoints, $s=0$ and $s=1$. (Recall: a smaller spectral gap requires
  a larger evolution time to ensure that the system remains in the
  correct eigenspace; increasing the time horizon is equivalent to
  decreasing the rate of change of the Hamiltonian.) The function
  should of course also satisfy $h(0) = 0, h(1) = 1$. For example, one
  can choose $h$ so that $h'(s) \propto \gamma(s)^{3/2}$, which
  ensures that the interpolation proceeds very slowly when the gap is
  close to zero, and faster when the gap is close to 1. More details
  on this can be found in, e.g., \cite{childs2017lecture}.
\end{example}

\noindent In both examples we could characterize the spectral gap
because we managed to reduce the time-dependent Hamiltonian $\ham(s)$
to one or more two-dimensional time-dependent Hamiltonians, allowing
us to analyze the spectrum with simple linear algebra. Many
interesting problems do not have sufficient amount of structure to
give tight bounds on the gap, and as a result, they are difficult to
study in the context of adiabatic
optimization.\index{adiabatic!theorem|)}\index{spectral gap|)}

\section{The quantum approximate optimization algorithm}
\label{sec:qaoa}
The Quantum Approximate Optimization Algorithm
(QAOA)\index{algorithm!QAOA|(}, initially proposed in
\cite{farhi2014quantum}, is designed as a low-resource approximation
of adiabatic evolution, with the goal of being implementable even on
quantum computers that can only successfully execute a relatively
small number of gates --- or in any case, fewer gates than would be
necessary for an accurate simulation of the Schr\"odinger equation
involved in adiabatic optimization. The QAOA is characterized by an
algorithmic parameter $p$ that determines the number of layers of its
circuit implementation, and essentially trades quantum resources for
quality of the approximation. However, we note that, in practice,
larger $p$, while better in theory, does not always mean that a better
solution to an optimization problem is obtained; this is discussed
later in this section, after the necessary concepts have been
introduced.

\subsection{Derivation from the adiabatic theorem}
\label{sec:qaoaadi}
QAOA is traditionally discussed in the setting of solving a binary
optimization problem in maximization form:
\begin{equation}
  \label{eq:01optmax}
  \max f(\vx) \qquad \vx \in \{0,1\}^n.
\end{equation}
\vspace*{-1em}
\begin{remark}
  All of the discussion in this section can be converted to
  minimization with the usual transformation $\max f(x) = - \min
  -f(x)$, but for reasons that come up subsequently in our
  presentation, QAOA is more naturally discussed for maximization.
\end{remark}
Problem \eqref{eq:01optmax} is encoded by the following Hamiltonian:
\begin{equation}
  \label{eq:qaoahf}
  \ham_{\text{F}} = \sum_{\vj \in \{0,1\}^n} f(\vj) \ketbra{\vj}{\vj}.
\end{equation}
The maximum eigenpair of $\ham_{\text{F}}$ is a solution of problem
\eqref{eq:01optmax}. We can determine such eigenpair by using
adiabatic optimization. It should be clear from the statement of
Thm.~\ref{thm:adiabatic} that the derivation of adiabatic optimization
in Sect.~\ref{sec:adiabatic} is perfectly symmetric with respect to
minimization or maximization: if we start in a minimum eigenstate of
the initial Hamiltonian, adiabatic evolution eventually converges to a
minimum eigenstate of the final Hamiltonian, whereas if we start in a
maximum eigenstate of the initial Hamiltonian, we find a maximum
eigenstate of the final Hamiltonian. For adiabatic optimization we
need an initial Hamiltonian $\ham_{\text{I}}$; defining matrices
$\sigma^X_j$:
\begin{equation*}
  \sigma^X_j := \underbrace{I \otimes \dots \otimes I \otimes \hspace*{-1.2em} \overset{\substack{\text{position } j\\\downarrow}}{X} \hspace*{-1.2em} \otimes I \dots \otimes I}_{n \text{ times}},
\end{equation*}
where $X$ is the Pauli $X$ matrix as given in Def.~\ref{def:pauli}, we
choose the Hamiltonian:
\begin{equation}
  \label{eq:qaoahi}
  \ham_{\text{I}} = \sum_{j=1}^n \sigma^X_j,
\end{equation}
for which an eigenstate with maximum eigenvalue is given by the
product state:
\begin{equation*}
  \ket{\varphi(0)} = \left(H \ket{0}\right)^{\otimes n}.
\end{equation*}
This is easy to prove: each term $\sigma^X_j$ has eigenvalues with
absolute value at most $1$ because it is a real unitary matrix, and
the vector that has $H\ket{0}$ on the $j$-th qubit is an eigenvector
with eigenvalue 1, which is therefore the maximum. Thus:
\begin{equation*}
  \sum_{j=1}^n \sigma^X_j \left(H \ket{0}\right)^{\otimes n} = n
  \left(H \ket{0}\right)^{\otimes n}.
\end{equation*}
Now we apply the adiabatic theorem, and obtain the following result.
\begin{proposition}
  \label{prop:qaoaadi}
  Consider the time-dependent Hamiltonian:
  \begin{equation}
    \label{eq:qaoaadihs}
    \ham(s) = (1-s) \ham_{\text{I}} + s \ham_{\text{F}},
  \end{equation}
  where $\ham_{\text{F}}, \ham_{\text{I}}$ are defined as in Eq.s~\eqref{eq:qaoahf},
  \eqref{eq:qaoahi} respectively. For $p \in \N$, $p > 0$, let $\beta,
  \theta \in [0,2\pi]^p$ be $p$-dimensional vectors, and define the
  following unitary parameterized by $p$, $\beta$ and $\theta$:
  \begin{equation}
    \label{eq:uqaoa}
    U_{\text{QAOA}}(p, \beta,\theta) := e^{-i \beta_p \ham_{\text{I}}} e^{-i \theta_p \ham_{\text{F}}}
    e^{-i \beta_{p-1} \ham_{\text{I}}} e^{-i \theta_{p-1} \ham_{\text{F}}} \cdots
    e^{-i \beta_1 \ham_{\text{I}}} e^{-i \theta_1 \ham_{\text{F}}}.
  \end{equation}
  Then, for any $\delta > 0$ and for $p \to \infty$, there exists a
  choice $\beta^*, \theta^*$ of $\beta, \theta$ such that
  $U_{\text{QAOA}}(p, \beta^*,\theta^*)$ approximates an adiabatic
  evolution of $\ham(s)$, and in particular, the following holds:
  \begin{equation*}
   \nrm{\lim_{p \to \infty} U_{\text{QAOA}}(p, \beta^*,\theta^*) \left(H
   \ket{0}\right)^{\otimes n} - \ket{\varphi(1)}} \le \delta,
  \end{equation*}
  where $\ket{\varphi(1)}$ is an eigenstate of $\ham(1) = \ham_{\text{F}}$
  with maximum eigenvalue, encoding a solution to problem
  \eqref{eq:01optmax}.
\end{proposition}
In the proof of Prop.~\ref{prop:qaoaadi} we use the following
definition; the definition can be skipped if the reader is not
interested in the details of the proof.
\begin{definition}[Irreducible matrix]
  Given a matrix $M \in \R^{n \times n}$ with nonnegative entries, the
  \emph{matrix graph} associated with $M$ is the directed graph $G_M = (V, A)$
  with $V := \{1,\dots,n\}$ and $A = \{(i,j) : M_{ij} \neq 0\}$, i.e.,
  there is an arc between $i$ and $j$ if and only if the element
  $M_{ij}$ is nonzero. The matrix $M$ is said to be
  \emph{irreducible} if its matrix graph $G_M$ is strongly connected.
\end{definition}
We can now proceed with the proof of Prop.~\ref{prop:qaoaadi}.
\begin{proof}
  Thm.~\ref{thm:adiabatic} states that if we start in the state
  $\left(H \ket{0}\right)^{\otimes n}$, which is an eigenstate of
  $\ham_{\text{I}}$ with maximum eigenvalue, and simulate the time-dependent
  Schr\"odinger equation in Eq.~\eqref{eq:adischrodinger}, we remain
  in a maximum eigenvalue of the time-dependent Hamiltonian $\ham(s)$,
  provided that there is a nonzero spectral gap and $T$ is chosen
  large enough.

  First, let us address the spectral gap $\gamma$. Here it is
  sufficient to show that the spectral gap around the maximum
  eigenvalue is nonzero: we eventually choose a very large $T$, going
  to infinity, therefore we do not have to worry about the gap
  dependence even if $\gamma$ is very small --- we only need to make
  sure that the lower bound on $T$ given in Thm.~\ref{thm:adiabatic}
  is finite for this problem. To show that $\gamma > 0$ we use the
  Perron-Frobenius theorem. One version of the Perron-Frobenius
  theorem states that if a nonnegative matrix is irreducible, then its
  largest eigenvalue is positive and simple, i.e., it has algebraic
  multiplicity $1$ \cite{horn2012matrix}. Note that if such properties
  hold, then $\gamma > 0$, because the largest eigenvalue is strictly
  larger than the second largest. We can w.l.o.g.\ assume that
  $\ham_{\text{F}}$ is nonnegative: shifting the objective function
  value by a constant does not change the optimum and only translates
  the spectrum of the Hamiltonian. $\ham_{\text{I}}$ is also
  nonnegative by construction, hence $\ham(s)$ is nonnegative. Proving
  irreducibility of $\ham(s)$ is relatively straightforward:
  $\ham_{\text{F}}$ is a diagonal matrix, and does not affect the rest
  of the analysis; $\ham_{\text{I}}$ has a special structure and its
  graph is strongly connected, as we show next. Recall that
  $\ham_{\text{I}} \in \R^{2^n \times 2^n}$ and its rows and columns
  can be indexed by $n$-digit binary strings, so the nodes of the
  matrix graph of $\ham_{\text{I}}$ also correspond to $n$-digit
  binary strings. From Eq.~\eqref{eq:qaoahi}, $\ham_{\text{I}}$ is a
  sum of terms, each of which corresponds to the unitary matrix $\sigma^X_j$ that
  applies an $X$ gate on a single qubit. It is clear that $\sigma^X_j$
  connects nodes in the matrix graph such that the corresponding
  labels differ by one bit-flip in the $j$-th bit, see
  Fig.~\ref{fig:sigmax}.
  \begin{figure}[htb]
    \center
    \ifcompilefigs
    \begin{tikzpicture}[->,auto,node distance=1.5cm,
        thick,main node/.style={circle,draw}]

      \node[main node] (1) {00};
      \node[main node] (2) [right of=1] {01};
      \node[main node] (3) [right of=2] {10};
      \node[main node] (4) [right of=3] {11};
      \node[draw=none] (10) [right of=4] {};
      \node[main node] (5) [right of=10] {00};
      \node[main node] (6) [right of=5] {01};
      \node[main node] (7) [right of=6] {10};
      \node[main node] (8) [right of=7] {11};

      \path[every node/.style={font=\sffamily\small}]
      (1) edge [bend right] node {} (2)
      (2) edge [bend right] node {} (1)
      (3) edge [bend right] node {} (4)
      (4) edge [bend right] node {} (3);

      \path[every node/.style={font=\sffamily\small}]
      (5) edge [bend right] node {} (7)
      (7) edge [bend right] node {} (5)
      (6) edge [bend right] node {} (8)
      (8) edge [bend right] node {} (6);
    \end{tikzpicture}
    \else
    \includegraphics{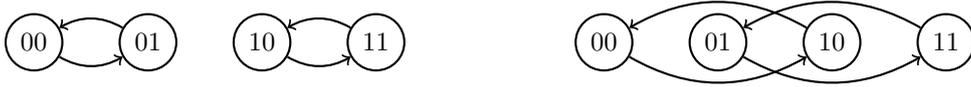}
    \fi
    \caption{Matrix graph of $\sigma^X_2$ (on the left) and $\sigma^X_1$ (on the right) over two qubits.}
    \label{fig:sigmax}
  \end{figure}    
Because $\ham_{\text{I}}$ is the sum of $\sigma^X_j$ for all $j$ and this
yields no cancellations in the matrix terms, the matrix graph of
$\ham_{\text{I}}$ is the union of the matrix graphs of all $\sigma^X_j$ (all
of these graphs have the same vertex set, we just take the union of
the sets of arcs). Then, proving that $\ham_{\text{I}}$ is strongly connected
is equivalent to proving that from any given starting bitstring, we
can reach another arbitrary bitstring using a sequence of bit-flips on
any bit, which is obviously true. Therefore $\ham(s)$ is irreducible
for every $s \in [0,1)$, and by the Perron-Frobenius theorem it has
  nonzero spectral gap $\gamma > 0$. (The spectral gap could be zero
  for $s=1$ if $\ham_{\text{F}}$ has a degenerate largest eigenvalue, i.e., if
  there are multiple optimal solutions for the optimization problem,
  but there are many ways to deal with this issue; for example, we
  can assume that the optimal solution is unique after adding a small
  random perturbation to the objective function values.) Because $\gamma > 0$,
  there exists a possibly very large but finite $T$ that satisfies the
  conditions in Thm.~\ref{thm:adiabatic} (or
  Thm.~\ref{thm:adiabatictighter}), and whose value depends only on
  the problem instance. We fix such value of $T$.
  
  We now consider the simulation of the dynamics of the Schr\"odinger
  equation (Eq.~\eqref{eq:adischrodinger}) with the given value of
  $T$, which guarantees finding the maximum eigenstate of
  $\ham_{\text{F}}$ because we start in a maximum eigenstate of
  $\ham_{\text{I}}$. Just as in Sect.~\ref{sec:adihighlevel}, we
  discretize time in an infinitely-large number of time steps $N$, and
  simulate $\ham(s)$ for an infinitesimally-small time $1/N$, where in
  each time step $\ham(s)$ is fixed. Recalling Eq.~\eqref{eq:ujdef},
  in the $j$-th time step the system evolves by applying:
  \begin{equation}
    \label{eq:qaoaadiexp}
    e^{i T/N \ham(j/N)} = \exp\left(i \frac{T}{N} \left(\frac{N-j}{N} \ham_{\text{I}} +  \frac{j}{N} \ham_{\text{F}}\right) \right).
  \end{equation}
  We now apply a product formula (Sect.~\ref{sec:trotter}) to compute
  an approximation of Eq.~\eqref{eq:qaoaadiexp}. By
  Eq.~\eqref{eq:trottererror}, the error of the approximation:
  \begin{equation}
    \label{eq:qaoatrotter}
    \exp\left(i \frac{T}{N} \left(\frac{N-j}{N} \ham_{\text{I}} +  \frac{j}{N} \ham_{\text{F}}\right) \right) \approx \left(\exp\left(i \frac{T}{hN} \frac{N-j}{N} \ham_{\text{I}}\right)\exp\left(i \frac{T}{hN} \frac{j}{N} \ham_{\text{F}}\right)\right)^h
  \end{equation}
  decreases as $N$ and $h$ grow, and in particular, with $N \to
  \infty, h \to \infty$, the error of the entire adiabatic evolution
  goes to zero. Comparing Eq.s~\eqref{eq:uqaoa} and
  \eqref{eq:qaoatrotter}, we see that the angles $\beta, \theta$ in
  Eq.~\eqref{eq:uqaoa} can be chosen to obtain
  Eq.~\eqref{eq:qaoatrotter}: pick $N$ and $h$ large enough so that
  applying Eq.~\eqref{eq:qaoatrotter} with $j=0,\dots,N-1$ guarantees
  sufficiently small error (the error is arbitrarily small), fix
  $p=Nh$, finally choose $\beta_1,\dots,\beta_p,
  \theta_1,\dots,\theta_p$ to match Eq.~\eqref{eq:qaoatrotter}.  This
  proves that there is a choice $\beta^*, \theta^*$ such that $\lim_{p
    \to \infty} U_{\text{QAOA}}(p, \beta^*,\theta^*) \left(H
  \ket{0}\right)^{\otimes n}$ follows the trajectory for adiabatic
  optimization according to Thm.~\ref{thm:adiabatic}, and therefore,
  the final state can be made to have arbitrarily small distance from
  a maximum eigenstate $\ket{\varphi(1)}$ of $\ham(1) = \ham_{\text{F}}$.
\end{proof}

\noindent The unitary $U_{\text{QAOA}}(p, \beta, \theta)$ is the main
operator of the QAOA, as we discuss in the next section.

\subsection{Algorithm description and properties}
\label{sec:qaoaalg}
Prop.~\ref{prop:qaoaadi} states that the unitary operator
$U_{\text{QAOA}}(p, \beta, \theta)$ can approximate adiabatic
evolution, for some choice of the angles $\beta, \theta$, in the limit
$p \to \infty$, i.e., when it includes an infinite sequence of matrix
exponentials of $\ham_{\text{F}}$ and $\ham_{\text{I}}$. To derive an implementable
algorithm we need to work with finite $p$. The QAOA is a general
framework that leaves many important decisions open, allowing the
details to be specified based on the properties of the optimization
problem at hand. The basic scheme followed by QAOA is: pick a finite
value of $p$; determine values of the angles $\beta, \theta \in [0,
  2\pi]^p$; construct $U_{\text{QAOA}}(p, \beta, \theta) \left(H
\ket{0}\right)^{\otimes n}$, and sample from the corresponding quantum
states a few times. We summarize this scheme in
Alg.~\ref{alg:qaoabase}.
\begin{algorithm2e}[tb]
  \SetAlgoLined
  \LinesNumbered
\KwIn{Hamiltonian $\ham_{\text{F}}$ encoding problem \eqref{eq:01optmax}, parameter $p$, number of samples $r$.} 
\KwOut{Best solution found.}
Determine angles $\beta, \theta \in [0, 2\pi]^p$.\\
Construct the state $U_{\text{QAOA}}(p, \beta, \theta) \left(H \ket{0}\right)^{\otimes n}$, and perform $r$ measurements on it. Let $M$ be the set of observed measurement outcomes.\\
\Return $\vj \in \arg \max_{\vj \in M} \{f(\vj)\}$.
\caption{Quantum approximate optimization algorithm (QAOA).}
\label{alg:qaoabase}
\end{algorithm2e}
This general description leaves open several crucial choices that have
tremendous impact on the performance of the algorithm: the choice of
$p$, the choice of the angles $\beta, \theta$, and, to a lesser
extent, the number of samples. These choices can be made based on
theoretical or practical considerations. We discuss the three
components separately. We begin with choosing the angles because the
corresponding considerations make the discussion on choosing $p$
clearer.

\paragraph{Choosing the angles $\beta, \theta$.} For this discussion assume that $p$ is fixed and given. We need to choose $\beta, \theta \in [0, 2\pi]^p$ so as to maximize the objective function value of the solution that is returned in the final step of Alg.~\ref{alg:qaoabase}, after taking $r$ samples from the quantum state constructed with $U_{\text{QAOA}}(p, \beta, \theta)$. To do so, we must fix the criterion that is used to compare different values of $\beta, \theta$: how do we measure the quality of a certain choice? A natural approach, proposed in the seminal QAOA paper \cite{farhi2014quantum}, is to compare the expected value of the measurement outcomes from the final state $U_{\text{QAOA}}(p, \beta, \theta) \left(H \ket{0}\right)^{\otimes n}$. Recall from Eq.~\eqref{eq:qaoahf} that $\ham_{\text{F}}$ is a diagonal matrix with the objective function values $f(\vj)$ on the diagonal. Given a quantum state $\ket{\psi} = \sum_{\vj \in \{0,1\}^n} \alpha_j \ket{\vj}$ (in this case $\ket{\psi}$ is the state produced by QAOA, but the property stated next holds for any state), it is easy to see that $\bra{\psi} \ham_{\text{F}} \ket{\psi}$ is precisely the expected value of the objective function $f$ with respect to the probability distribution over the measurement outcomes:
\begin{equation}
  \label{eq:expectedvalue}
  \bra{\psi} \ham_{\text{F}} \ket{\psi} = \left(\sum_{\vj \in \{0,1\}^n} \alpha^{\dag}_j \bra{\vj} \right) \ham_{\text{F}} \left(\sum_{\vj \in \{0,1\}^n} \alpha_j \ket{\vj} \right) = \sum_{\vj \in \{0,1\}^n} |\alpha_j|^2 f(\vj) = \mathbb{E}[f(X)],
\end{equation}
where $X$ is the random variable over measurement outcomes with
probability distribution $\Pr(X = \vj) = |\alpha_j|^2$. The quantity
$\bra{\psi} \ham_{\text{F}} \ket{\psi}$ can be computed in multiple
ways. For example, we can measure $\ket{\psi}$ multiple times, say $m$
times, and compute the sample average $\frac{1}{m} \sum_{k=1}^{m}
f({\vj}^{(k)})$ where ${\vj}^{(k)}$ is the $k$-th observed sample; 
if $\ham_{\text{F}}$ is a weighted summation of tensor products of $\sigma^Z_j$
terms, as in Prop.~\ref{prop:isingtoham}, we can estimate each term of
the summation by checking the corresponding bits of the measurement
outcomes across samples (i.e., computing the sample average for each single $\sigma^Z_j$ or $\sigma^Z_j \otimes \sigma^Z_k$ term, then calculating their weighted combination). The angles can then be chosen as the solution
to the following optimization problem for fixed $p$:
\begin{equation}
  \label{eq:qaoamaxangle}
  \max_{\beta,\theta} \left(\left(\bra{0} H\right)^{\otimes n} U^{\dag}_{\text{QAOA}}(p, \beta, \theta)\right)  \ham_{\text{F}} \left(U_{\text{QAOA}}(p, \beta, \theta) \left(H \ket{0}\right)^{\otimes n}\right).
\end{equation}
This is equivalent to:
\begin{equation}
  \label{eq:qaoamaxexpect}
  \max_{\beta,\theta} \mathbb{E}[f(X)],
\end{equation}
where the relationship between $\beta, \theta$ and the random variable
$X$ is via the probabilities $|\alpha_j|^2$: in other words, we aim to
maximize the expected objective function value of the binary strings
(i.e., solutions) sampled from the quantum
state. Problem~\eqref{eq:qaoamaxangle} is a continuous optimization
problem that can be solved (usually in a heuristic manner, see
Rem.~\ref{rem:qaoacontopt}) with many possible classical algorithms;
in the literature it is often solved with derivative-free optimization
techniques, using the quantum computer only to evaluate the
objective function, but derivatives can be computed, and no clearly
dominant solution strategy has emerged so far. We discuss some of the
issues related to the solution of problem~\eqref{eq:qaoamaxangle} in
Sect.~\ref{sec:qaoaimp}.
\begin{remark}
  \label{rem:qaoacontopt}
  Problem~\eqref{eq:qaoamaxangle} is continuous but \emph{nonconvex}
  in general, and can be very difficult to solve to the global
  optimum.  In very special cases, the Hamiltonian $\ham_{\text{F}}$
  may have sufficient structure that problem~\eqref{eq:qaoamaxangle}
  can be solved efficiently --- at least in practice, if not in theory
  --- but in general we can only hope for a local minimum that may be
  of poor quality. Thus, the solution of
  problem~\eqref{eq:qaoamaxangle} can be a significant obstacle. In
  fact, it may be as difficult as solving the original combinatorial
  optimization problem \eqref{eq:01optmax}: we are replacing the
  solution of a difficult problem (problem \eqref{eq:01optmax}) with the
  solution of another difficult problem (problem \eqref{eq:qaoamaxangle}), and
  may not gain any quantum advantage in the process.
\end{remark}
Although using problems
\eqref{eq:qaoamaxangle}-\eqref{eq:qaoamaxexpect} to guide the choice
of the angles is natural and well-motivated, it is not the only
possibility. In light of the fact that the ultimate goal is to obtain
a sample from $U_{\text{QAOA}}(p, \beta, \theta) \left(H
\ket{0}\right)^{\otimes n}$ with maximum objective function value in
problem~\eqref{eq:01optmax}, it may be reasonable to utilize a metric
different from the expected value. For example,
\cite{barkoutsos2020improving} proposes focusing on the lower tail of
the distribution (for a minimization problem) rather than the expected
value, using the Conditional Value-at-Risk (CVaR) of $f(X)$ at a given
level as the objective function in problem~\eqref{eq:qaoamaxexpect}.

\paragraph{Choice of $p$.} The set of quantum states that can be obtained as $U_{\text{QAOA}} (p, \beta, \theta) \left(H \ket{0}\right)^{\otimes n}$ gets larger as $p$ increases. This is easy to see: any unitary $U_{\text{QAOA}} (p, \beta, \theta)$ can also be obtained as $U_{\text{QAOA}} (p', \beta, \theta)$ for $p' > p$, simply by setting to zero $\beta_j, \theta_j : j > p$. Thus, the following relationship holds for $p' > p$:
\begin{align*}
  \max_{\beta,\theta} \left(\left(\bra{0} H\right)^{\otimes n}
  U^{\dag}_{\text{QAOA}}(p, \beta, \theta)\right) \ham_{\text{F}}
  \left(U_{\text{QAOA}}(p, \beta, \theta) \left(H
  \ket{0}\right)^{\otimes n}\right) \le\\ \max_{\beta,\theta}
  \left(\left(\bra{0} H\right)^{\otimes n} U^{\dag}_{\text{QAOA}}(p',
  \beta, \theta)\right) \ham_{\text{F}} \left(U_{\text{QAOA}}(p', \beta, \theta)
  \left(H \ket{0}\right)^{\otimes n}\right).
\end{align*}
As a consequence, at least in theory it is reasonable to
choose $p$ as large as possible. However, a large $p$ has at least two
important practical drawbacks: (i) it leads to circuits that require
more gates; (ii) it may lead to worse solutions, because heuristic
algorithms to solve the nonconvex problem \eqref{eq:qaoamaxangle} may
struggle if there are additional parameters $\beta, \theta$ to
optimize.

From a theoretical point of view, a few results are known showing
that, for small $p$, QAOA achieves an expected approximation ratio
with respect to the optimal solution. Here, approximation ratio is
meant in the usual optimization sense, as for approximation algorithms.\index{approximation!algorithm}
\begin{definition}[Approximation algorithm]
  An \emph{approximation algorithm} with \emph{approximation ratio
  $r \le 1$} is an algorithm that, given a maximization problem with optimal
  objective function value $f^*$, returns a solution with objective
  function value at least $r f^*$.
\end{definition}
There are two main approximation results that sparked the interest in
QAOA, as the first quantum algorithm with an approximation guarantee
for some optimization problem. We report them below.
\begin{theorem}[QAOA for MaxCut on 3-regular graphs; \cite{farhi2014quantum}]
  \label{thm:qaoamaxcut}
  Given a 3-regular graph $G$\index{MaxCut!approximation} (i.e., a graph
  where each node has exactly three incident edges), for $p=1$ there
  is a choice of the angles $\beta, \theta$ such that QAOA
  (Alg.~\ref{alg:qaoabase}) returns a solution to the MaxCut problem
  on $G$ achieving approximation ratio $0.6924$, in expectation.
\end{theorem}
For the second result we first define the optimization problem, called
Max E3LIN2, as it is not as well known as MaxCut. We are given a set
of linear equations modulo $2$ over $n$ binary variables. Each
equation contains exactly three variables. Thus, each equation is of
the form:
\begin{equation*}
  x_j + x_k + x_h \equiv b \mod 2,
\end{equation*}
where $b \in \{0,1\}$. Our goal is to find an assignment of the binary
decision variables that maximizes the number of satisfied equations.
\begin{theorem}[QAOA for Max E3LIN2; \cite{farhi2014bounded}]
  \label{thm:qaoamaxe3lin2}
  Given an instance of Max E3LIN2 such that
  each variable appears in no more than $D+1$ equations, for $p=1$
  there is a choice of the angles $\beta, \theta$ such that QAOA
  (Alg.~\ref{alg:qaoabase}) returns a solution achieving 
  approximation ratio $\frac{1}{2} + \frac{1}{101 \sqrt{D} \ln D}$, in
  expectation.
\end{theorem}
\begin{remark}
  The approximation ratios in Thm.s~\ref{thm:qaoamaxcut} and
  \ref{thm:qaoamaxe3lin2} are lower than (i.e., not as good as) the best
  approximation ratios that can be obtained by classical
  algorithms. Thus, QAOA with $p=1$ does not lead to provable quantum
  advantage for these problems, at least based on existing results.
\end{remark}
Although we do not give a full proof of how these approximation ratios
are obtained, it may be useful to provide a high-level overview, which
has the added benefit of illustrating the difficulties faced when
trying to extend these results to higher values of $p$ or other types
of combinatorial optimization problems. In Sect.~\ref{sec:qaoamaxcut}
we sketch the main ideas of the analysis behind
Thm.~\ref{thm:qaoamaxcut}.

\paragraph{Number of samples.} The main consideration to determine the number of samples is the variance of the random variable whose value is the objective function value of the sample. As before, let $X$ be the random variable
characterizing the measurement outcomes, and consider the distribution of
$f(X)$, i.e., the objective function values of the samples. Applying the formula $\text{Var}(X) = \mathbb{E}[X^2] - (\mathbb{E}[X])^2$, recalling problems \eqref{eq:qaoamaxangle}-\eqref{eq:qaoamaxexpect}, we see that the variance can be expressed as:
\begin{align*}
  \left(\left(\bra{0} H\right)^{\otimes n} U^{\dag}_{\text{QAOA}}(p,
  \beta, \theta)\right) \ham_{\text{F}}^2 \left(U_{\text{QAOA}}(p,
  \beta, \theta) \left(H \ket{0}\right)^{\otimes n}\right)
  -\\ \left(\left(\left(\bra{0} H\right)^{\otimes n}
  U^{\dag}_{\text{QAOA}}(p, \beta, \theta)\right) \ham_{\text{F}}
  \left(U_{\text{QAOA}}(p, \beta, \theta) \left(H
  \ket{0}\right)^{\otimes n}\right) \right)^2.
\end{align*}
Although such an expression is typically difficult to analyze, for
special cases it may have sufficient structure.
\cite{farhi2014quantum} shows that for the MaxCut problem on graphs
with bounded degree and $|E|$ edges, if $p$ is fixed, the variance is
$\bigO{|E|}$ and the standard deviation is $\sigma =
\bigO{\sqrt{|E|}}$. By central
limit theorem, the sample mean of $\bigO{|E|^{2k}}$ samples, $k > 1$,
is normally distributed with mean $\mu = \mathbb{E}[f(X)]$ and
standard deviation $\sigma/\sqrt{|E|^{2k}} =
\bigO{1/|E|^{k-1/2}}$. Applying Chebyshev's inequality (i.e.,
$\Pr(\abs{X - \mu} \ge h \sigma) \le 1/h^2$) and choosing the
constants appropriately, we then obtain:
\begin{equation*}
  \Pr\left(\abs{f(X) - \mu} \ge 1\right) \le \frac{1}{|E|^{2k-1}}.
\end{equation*}
So, for example, setting $k=1$ we find that with probability at least
$1-1/|E|$, the sample mean of $\bigO{|E|^2}$ samples estimates the
expected value with error at most $1$.
\begin{remark}
  The concentration around the mean has the benefit that we can expect
  to quickly (i.e., with few samples) obtain binary strings
  corresponding to solutions with objective function value close to
  $\mathbb{E}[f(X)]$, but it also has the drawback that we cannot
  expect to sample solutions with objective function value much better
  than $\mathbb{E}[f(X)]$.
\end{remark}

\subsection{QAOA for MaxCut with fixed $p$}
\label{sec:qaoamaxcut}
\begin{figure}[htb]
  \centering
  \ifcompilefigs
  \begin{tikzpicture}[      
    ]
    \node [circle,draw] (j1) at (0,0) {$j$};
    \node [circle,draw] (k1) at (1.5,0) {$k$};
    \node [draw=none] (a1) at (-1.5,1) {};
    \node [draw=none] (b1) at (-1.5,-1) {};
    \node [draw=none] (c1) at (3,1) {};
    \node [draw=none] (d1) at (3,-1) {};

    \draw [-] (j1) to (k1);
    \draw [-,dashed] (j1) to (a1);
    \draw [-,dashed] (j1) to (b1);
    \draw [-,dashed] (k1) to (c1);
    \draw [-,dashed] (k1) to (d1);

    \node [circle,draw] (j2) at (6,0) {$j$};
    \node [circle,draw] (k2) at (7.5,0) {$k$};
    \node [circle,draw] (a2) at (6.75,1) {$\phantom{j}$};
    \node [draw=none] (b2) at (4.5,-1) {};
    \node [draw=none] (d2) at (9,-1) {};

    \draw [-] (j2) to (k2);
    \draw [-,dashed] (j2) to (a2);
    \draw [-,dashed] (j2) to (b2);
    \draw [-,dashed] (k2) to (a2);
    \draw [-,dashed] (k2) to (d2);

    \node [circle,draw] (j3) at (12,0) {$j$};
    \node [circle,draw] (k3) at (13.5,0) {$k$};
    \node [circle,draw] (a3) at (12.75,1) {$\phantom{j}$};
    \node [circle,draw] (b3) at (12.75,-1) {$\phantom{j}$};

    \draw [-] (j3) to (k3);
    \draw [-,dashed] (j3) to (a3);
    \draw [-,dashed] (j3) to (b3);
    \draw [-,dashed] (k3) to (a3);
    \draw [-,dashed] (k3) to (b3);

  \end{tikzpicture}
  \else
  \includegraphics{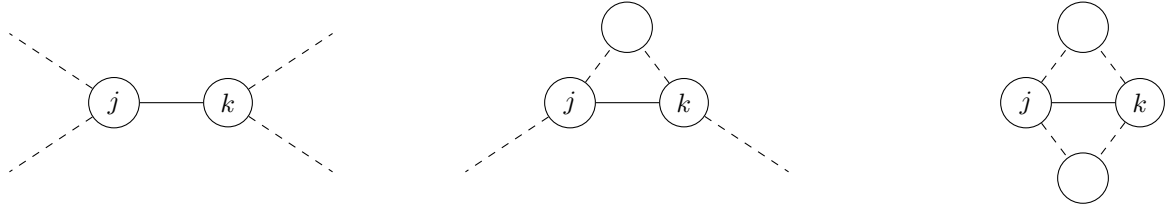}
  \fi
  \caption{Possible subgraphs at distance $1$ from $(j,k)$ in a 3-regular graph.}
  \label{fig:qaoa3regular}
\end{figure}

We present a high-level overview of the argument that leads to
Thm.~\ref{thm:qaoamaxcut}.\index{MaxCut!approximation|(} We fix $p=1$ for now. Define:
\begin{equation}
  \label{eq:qaoacijdef}
  C_{jk} := \frac{1}{2} \left(I^{\otimes n} - \sigma^Z_j \sigma^Z_k\right).
\end{equation}
By definition, $\sigma^Z_j$ acts as the identity on every qubit except
$j$, see Eq.~\eqref{eq:sigmazdef}. Restricted to the space of qubits
$j$ and $k$ (i.e., the $j$-th and $k$-th digit of the $n$-digit basis
states that we are considering), $C_{jk}$ acts as the following
matrix:
\begin{equation*}
  \begin{pmatrix}
    0 & 0 & 0 & 0 \\
    0 & 1 & 0 & 0 \\
    0 & 0 & 1 & 0 \\
    0 & 0 & 0 & 0
  \end{pmatrix}.
\end{equation*}
Thus, each basis state is an eigenstate of $C_{jk}$, with eigenvalue
$1$ if qubits $j$ and $k$ take different values, and with eigenvalue
$0$ otherwise. This is exactly the objective function contribution of
edge $(j,k)$ in the MaxCut problem: we gain $1$ if the endpoints of an
edge have different labels (e.g., $\ket{0}$ and $\ket{1}$), we gain
nothing if they have the same label. It follows that, for the
(unweighted) MaxCut problem on a graph $G = (V,E)$, where the vertex
set is $V = \{1,\dots,n\}$, the final Hamiltonian $\ham_{\text{F}}$
on $n$ qubits can be written as:
\begin{equation*}
  \ham_{\text{F}} = \sum_{(j,k) \in E} C_{jk}.
\end{equation*}
Now let us analyze the expected objective function value in
Eq.~\eqref{eq:expectedvalue} for the state produced by QAOA with
$p=1$. By definition of $\ham_{\text{F}}$, we have:
\begin{align*}
\left(\left(\bra{0} H\right)^{\otimes n} U^{\dag}_{\text{QAOA}}(p,
\beta, \theta)\right) \ham_{\text{F}} \left(U_{\text{QAOA}}(p, \beta, \theta)
\left(H \ket{0}\right)^{\otimes n}\right) =\\ \sum_{(j,k) \in E}
\left(\left(\bra{0} H\right)^{\otimes n} U^{\dag}_{\text{QAOA}}(p,
\beta, \theta)\right) C_{jk} \left(U_{\text{QAOA}}(p, \beta, \theta)
\left(H \ket{0}\right)^{\otimes n}\right).
\end{align*}
A single term $C_{jk}$ in the above expression is of the following
form --- using brackets to more easily identify the constituents:
\begin{equation*}
  \left(\bra{0} H\right)^{\otimes n} \left[e^{i \theta \sum_{(h,\ell)
        \in E} C_{h\ell}}\right] \left[e^{i \beta \sum_{h=1}^n
      \sigma^X_h} \right] C_{jk} \left[e^{-i \beta \sum_{h=1}^n
      \sigma^X_h}\right] \left[e^{-i \theta \sum_{(h,\ell) \in E}
      C_{h\ell}}\right] \left(H \ket{0}\right)^{\otimes n}.
\end{equation*}
(Recall that $p=1$ so $\beta, \theta$ are scalars, not vectors.)  By
definition of $C_{jk}$, this is equal to:
\begin{equation*}
  \left(\bra{0} H\right)^{\otimes n} \left[e^{i \theta \sum_{(h,\ell)
        \in E} C_{h\ell}}\right] \left[e^{i \beta \sum_{h=1}^n
      \sigma^X_h} \right] \frac{1}{2}\left(I^{\otimes n} - \sigma^Z_j \sigma^Z_k\right) \left[e^{-i \beta \sum_{h=1}^n
      \sigma^X_h}\right] \left[e^{-i \theta \sum_{(h,\ell) \in E}
      C_{h\ell}}\right] \left(H \ket{0}\right)^{\otimes n}.
\end{equation*}
To better understand the structure of this expression, we drop the
rescaling factor $\frac{1}{2}$, and we can drop the identity matrix
too: distributing the multiplication and looking at the term with the
identity matrix, we see that it yields a constant shift to the entire
value of the above expression (each term that multiplies the identity
from the right simplifies with a term that multiplies the identity
from the left), which does not depend on $\beta$ or $\theta$. Such a
constant term can be dropped. Thus, we are left with:
\begin{equation*}
  \left(\bra{0} H\right)^{\otimes n} \left[e^{i \theta \sum_{(h,\ell)
        \in E} C_{h\ell}}\right] \left[e^{i \beta \sum_{h=1}^n
      \sigma^X_h} \right] \left(-\sigma^Z_j \sigma^Z_k\right) \left[e^{-i \beta \sum_{h=1}^n
      \sigma^X_h}\right] \left[e^{-i \theta \sum_{(h,\ell) \in E}
      C_{h\ell}}\right] \left(H \ket{0}\right)^{\otimes n}.
\end{equation*}
The central term $-\sigma^Z_j \sigma^Z_k$ acts as the identity on
every qubit except the $j$-th and $k$-th, and the matrix exponential
$e^{-i \beta \sum_{h=1}^n \sigma^X_h}$ can be written as the product
$\prod_{h=1}^n e^{-i \beta \sigma^X_h}$ because the matrices
$\sigma^X_h$ commute; similarly for $e^{i \beta \sum_{h=1}^n
  \sigma^X_h}$. Thus, all terms $e^{-i \beta \sigma^X_h}$ except for
$h=j,k$ commute through $-\sigma^Z_j \sigma^Z_k$, and cancel out with
the respective terms in $e^{i \beta \sum_{h=1}^n \sigma^X_h}$. This
leads to the following simplification:
\begin{equation*}
  \left(\bra{0} H\right)^{\otimes n} \left[e^{i \theta \sum_{(h,\ell)
        \in E} C_{h\ell}}\right] \left[e^{i \beta (\sigma^X_j +
      \sigma^X_k)} \right] \left(-\sigma^Z_j \sigma^Z_k\right)
  \left[e^{-i \beta (\sigma^X_j + \sigma^X_k)}\right] \left[e^{-i
      \theta \sum_{(h,\ell) \in E} C_{h\ell}}\right] \left(H
  \ket{0}\right)^{\otimes n}.
\end{equation*}
Now we analyze $e^{-i \theta \sum_{(h,\ell) \in E} C_{h\ell}}$. Each
term of the summation in the exponent is diagonal, hence everything
commutes. Then, all terms in $e^{-i \theta \sum_{(h,\ell) \in E}
  C_{h\ell}}$ that do not involve qubit $j$ or $k$ commute through
$\left[e^{i \beta (\sigma^X_j + \sigma^X_k)} \right] \left(-\sigma^Z_j
\sigma^Z_k\right) \left[e^{-i \beta (\sigma^X_j +
    \sigma^X_k)}\right]$, and cancel out the corresponding term in
$e^{i \theta \sum_{(h,\ell) \in E} C_{h\ell}}$. We are left with the
following simplified expression:
\begin{multline}
  \label{eq:qaoaexpvalsimple}
  \left(\bra{0} H\right)^{\otimes n} \left[\exp\left(i \theta
    \sum_{\substack{(h,\ell) \in E\\\{h,\ell\} \cap \{j,k\} \neq
        \emptyset}} C_{h\ell}\right)\right] \left[e^{i \beta
      (\sigma^X_j + \sigma^X_k)} \right] \left(-\sigma^Z_j
  \sigma^Z_k\right)\\ \left[e^{-i \beta (\sigma^X_j +
      \sigma^X_k)}\right] \left[\exp\left(-i \theta
    \sum_{\substack{(h,\ell) \in E\\\{h,\ell\} \cap \{j,k\} \neq
        \emptyset}} C_{h\ell}\right)\right] \left(H
  \ket{0}\right)^{\otimes n}.
\end{multline}
This expression depends only on qubits $j, k$, and qubits that are
adjacent to $j$ or $k$ in the graph $G$. In other words, to compute
the value of the objective function contribution for edge $(j,k)$ we
only need to consider the subgraph containing edge $(j,k)$ and all
edges adjacent to it.
\begin{remark}
  This argument can be extended beyond $p=1$: for $p=1$, we need to
  consider the subgraph including edges at distance $1$ from node $j$
  or node $k$; if we apply exactly the same line of reasoning, we
  obtain that in general we need to consider the subgraph including
  edges at distance $p$ from node $j$ or node $k$.
\end{remark}
The above discussion helps us characterize the performance of QAOA
with $p=1$ for MaxCut on a 3-regular graph. Because the graph is
3-regular, for every term $C_{jk}$ in the expected objective function
value, and therefore for every edge $(j,k)$ in the graph, there are
only three possible subgraphs with distance $1$ from $(j,k)$,
illustrated in Fig.~\ref{fig:qaoa3regular}.
Specifically, each of the three subgraphs occurs when nodes $j$ and
$k$ have, respectively, zero, one or two neighbors in common. For
fixed $\beta, \theta$, we can compute the value of
Eq.~\eqref{eq:qaoaexpvalsimple} for the three subgraph types. Using
the fact that the graph is 3-regular and combinatorial arguments, we
can derive a formula for the number of subgraphs of each type in
Fig.~\ref{fig:qaoa3regular} that can possibly appear in the graph, in
relation to each other. Then it is a simple exercise to numerically
determine the optimal angles $\beta,\theta$ and the worst-case value
(i.e., minimum value) for the expected objective function value:
\begin{equation*}
  \min_{\substack{\text{all possible}\\\text{3-regular graphs}}} \max_{\beta,\theta} \left(\left(\bra{0} H\right)^{\otimes n} U^{\dag}_{\text{QAOA}}(1, \beta, \theta)\right)  \ham_{\text{F}} \left(U_{\text{QAOA}}(1, \beta, \theta) \left(H \ket{0}\right)^{\otimes n}\right).
\end{equation*}
This is how the approximation ratio $0.6924$ of
Thm.~\ref{thm:qaoamaxcut} is shown in \cite{farhi2014quantum}.\index{MaxCut!approximation|)}

\subsection{Implementation}
\label{sec:qaoaimp}
\begin{figure}[h!]
  \leavevmode
  \centering
  \ifcompilefigs
  \Qcircuit @C=1em @R=.7em {
    & \ctrl{1} & \qw                 & \ctrl{1} & \qw \\
    & \targ    & \gate{R_Z(2\theta)} & \targ    & \qw \\
  }
  \else
  \includegraphics{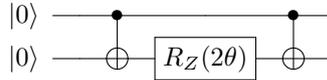}
  \fi
  \caption{Circuit implementing $e^{-i \theta \sigma^Z_1
      \sigma^Z_2}$.}
  \label{fig:qaoaexpz}
\end{figure}

The circuit $U_{\text{QAOA}}(p, \beta, \theta)$ is simple to
implement, consisting of a small number of gates relative to most of
the algorithms discussed before. Assuming a QUBO objective function
(Def.~\ref{def:qubo}), yielding the Hamiltonian of
Prop.~\ref{prop:isingtoham} as in the MaxCut discussion of
Sect.~\ref{sec:qaoamaxcut}, the unitary $e^{-i \theta
  \ham_{\text{F}}}$ can be decomposed into two-qubit blocks $e^{-i
  \theta \sigma^Z_j \sigma^Z_k}$, and single-qubit gates. The unitary
$e^{-i \theta \sigma^Z_j \sigma^Z_k}$ can be implemented with two C$X$
gates and the single qubit rotation $R_Z(2\theta) = e^{-i \theta Z}$
that implements the matrix exponential of the Pauli matrix $Z$, using
the circuit represented in Fig.~\ref{fig:qaoaexpz}.
\begin{definition}[$Z$ rotation gate]
  \label{def:rz}
  The gate $R_Z(\theta)$\index{gate!Z rotation} is defined as the matrix:
  \begin{equation*}
    R_Z(\theta) := 
    \begin{pmatrix}
      e^{-i \theta/2} & 0 \\ 0 & e^{i \theta/2}
    \end{pmatrix}.
  \end{equation*}
\end{definition}
\begin{remark}
  \label{rem:relativephasecontrol}
  The $R_Z(\theta)$ gate is equivalent, up to a global phase factor,
  to the phase shift gate $P(\theta)$ of
  Def.~\ref{def:phaseshiftgate}; in particular we can obtain $e^{-i
    \theta \sigma^Z_1 \sigma^Z_2}$ up to global phase substituting
  $P(2\theta)$ for $R_Z(2\theta)$ in Fig.~\ref{fig:qaoaexpz}.\index{phase!global}\index{phase!relative}\index{relative phase|see{phase, relative}} However,
  controlled versions of $R_Z$ are {\em not} equivalent to controlled
  versions of $P$, and vice-versa, because of relative phases:
  controlled-$R_Z(\theta)$ acts as $\ketbra{0}{0} \otimes I +
  \ketbra{1}{1} \otimes R_Z(\theta)$, and controlled-$P(\theta)$ acts
  as $\ketbra{0}{0} \otimes I + \ketbra{1}{1} \otimes P(\theta)$, so
  even if $P(\theta) = e^{i \theta/2} R_Z(\theta)$, the controlled
  unitaries are not equal up to a global phase. This remark acts as a
  reminder to be careful about equivalence up to global phase, when
  controlled operations are involved.
\end{remark}
The unitary $e^{-i \beta \ham_{\text{I}}}$, by its equivalence with
the product of single-qubit operators $\prod_{j=1}^n e^{-i \beta
  \sigma^X_j}$, can be decomposed into single-qubit gates $R_X(2\beta)
= e^{-i \beta X}$ that implement the matrix exponential of the Pauli
matrix $X$, applied to each qubit.
\begin{definition}[$X$ rotation gate]
  \label{def:rx}
  The gate $R_X(\theta)$\index{gate!X rotation} is defined as the matrix:
  \begin{equation*}
    R_X(\theta) := 
    \begin{pmatrix}
      \cos \theta/2 & -i \sin \theta/2 \\ -i \sin \theta/2 & \cos \theta/2
    \end{pmatrix}.
  \end{equation*}
\end{definition}
Thus, the number of basic gates of $U_{\text{QAOA}}(p, \beta, \theta)$
is directly proportional to $p$, the number of qubits, and the number
of terms $\sigma^Z_j \sigma^Z_k$ in $\ham_{\text{F}}$ (which depends on the
sparsity of the QUBO matrix).

For the solution of problem \eqref{eq:qaoamaxangle}, there is no known
theoretically-elegant solution, except for very few special cases
where the angles can be determined analytically. The use of classical
derivative-free algorithms is common in the literature. The
derivatives of the objective function in problem
\eqref{eq:qaoamaxangle} can be computed analytically, and evaluated
using circuits with similar building blocks as $U_{\text{QAOA}}(p,
\beta, \theta)$. The widespread use of derivative-free algorithms for
a continuous optimization problem with a differentiable objective
function is likely due to practical and numerical considerations: the
execution of quantum circuits to compute derivatives can represent a
significant time investment, from a practical point of view, and the
presence of noise because of hardware limitations may reduce the
impact of (imperfectly-estimated) partial derivatives. Because of the
nonconvexity of problem \eqref{eq:qaoamaxangle}, the problem is
typically solved without optimality guarantees. This makes it
difficult to show rigorous approximation guarantees similarly to
Thm.s~\ref{thm:qaoamaxcut} and \ref{thm:qaoamaxe3lin2}. Overall,
limited to the analysis reported in this chapter, QAOA does not yield
a provable advantage over classical algorithms, and can be considered
a heuristic with an approximation guarantee for some structured
problems.\index{algorithm!QAOA|)}

\section{Notes and further reading}
\label{sec:adiabaticnotes}
The adiabatic theorem is a foundational result dating back to the
early days of quantum mechanics \cite{born1928beweis}. Its use in the
theory of quantum computing is often associated with the
minimization of diagonal Hamiltonians for combinatorial optimization
problems (the same type of problems discussed in
Sect.~\ref{sec:qaoaadi}), but it has further and much broader
implications. Throughout this \book{} we employed the circuit model for
quantum computers. Alternatively, it is possible to model quantum
computers purely using an adiabatic evolution. Intuitively, this may
not be surprising: if a quantum computer can exist in the physical
world, its evolution must admit a description in terms of some
quantum-mechanical system, so in theory we can describe that system
and simulate its evolution using the Schr\"odinger equation. Perhaps
more surprisingly, it is possible to give an explicit construction for
an initial Hamiltonian, an initial eigenstate of the initial
Hamiltonian, and a final Hamiltonian such that the adiabatic evolution
of the system in the sense of Thm.~\ref{thm:adiabatic} simulates any
given quantum circuit, in time that is polynomial in the size of the
circuit. Because we have already shown in Sect.~\ref{sec:adiabatic} that
quantum circuits can simulate the adiabatic evolution, this implies
that the adiabatic model and the circuit model are equivalent. This
fundamental result is shown in \cite{aharonov2008adiabatic}.

There is vast literature on the subject of combinatorial optimization
with the adiabatic theorem, e.g.,
\cite{farhi2001quantum,finnila1994quantum,santoro2002theory}. An area
of particular interest for computational optimization is that of
quantum annealers \cite{johnson2011quantum}, a physical implementation
of the adiabatic model of computation that, due to hardware
restrictions, is not fully general, and cannot simulate an arbitrary
quantum circuit. It can, however, attempt to simulate the adiabatic
evolution of QUBO Hamiltonians as those in
Prop.~\ref{prop:isingtoham}. Although it does not guarantee finding
the global optimum due to hardware restrictions and evolution time, it
may heuristically find a solution with optimal or near-optimal
objective function value; see \cite{mcgeoch2020theory} for a general
introduction and \cite{crosson2021prospects} for a discussion on the
prospects of proving speedups for some type of problems. The
comparison between the computational performance of quantum annealers
and classical algorithms on meaningful combinatorial optimization
problems is the subject of many works in the past ten years, and we
refer to
\cite{albash2018demonstration,junger2021quantum,rehfeldt2023faster,tasseff2022emerging}
as entry points to survey the state of the field.

For readers interested in QAOA, in addition to the seminal articles
\cite{farhi2014quantum,farhi2014bounded}, a good starting point may be
the PhD thesis \cite{hadfield2018quantum}. The performance of QAOA is
the subject of numerous papers. One of the reasons for this interest
is the fact that QAOA circuits can produce samples from probability
distributions that are hard to construct classically; this is shown in
\cite{farhi2016quantum}, by first proving that it is \#P-hard to
compute matrix elements of a quantum circuit, and then showing that
QAOA with $p=1$ already produces distributions that are hard to sample
classically. Thus, QAOA as a framework exhibits a form of likely
quantum advantage. We emphasize that the computation of the expected
value in Eq.~\eqref{eq:expectedvalue} for the QAOA circuit (i.e., the
objective function of problem \eqref{eq:qaoamaxangle}) sometimes
admits an efficient classical algorithm; see, e.g.,
\cite{bravyi2020obstacles,ozaeta2022expectation} for analytic formulas
for $p=1$, and \cite{boulebnane2024solving} for a characterization of
the success probability of QAOA on random $k$-SAT instances. While the
expected values can sometimes be efficiently computed without a
quantum computer, sampling from the distribution created by the QAOA
circuit is believed to be difficult, as discussed above. The fact that
QAOA produces a hard-to-sample probability distribution does not
necessarily translate into good theoretical or practical performance
for combinatorial optimization. Some papers suggest that QAOA may
yield advantage over classical optimization algorithms for some
problems, e.g.,
\cite{lykov2023sampling,montanaro2024quantum,shaydulin2024evidence},
while others bring arguments in favor of the opposite conclusion,
e.g.,
\cite{hastings2019classical,bravyi2020obstacles}. \cite{bravyi2020obstacles}
additionally proposes the idea of employing the QAOA framework to
recursively identify pairs of binary variables that can be fixed to
either have the same value, or different value. (This is equivalent to
imposing constraints $x_j = x_k$ or $x_j = 1-x_k$ for a binary integer
program.) Each of these fixings reduces the size of the
problem. Assume $p$ is given and fixed. Using the MaxCut problem as an
example, and the same notation as in Sect.~\ref{sec:qaoamaxcut}, a
fixing can be identified by first choosing $\beta, \theta$ to solve
problem \eqref{eq:qaoamaxangle}, then scanning the edges and picking
the edge $(j,k)$ that maximizes the expression:
\begin{equation*}
  \left(\left(\bra{0} H\right)^{\otimes n} U^{\dag}_{\text{QAOA}}(p,
\beta, \theta)\right) \sigma^Z_j \sigma^Z_k \left(U_{\text{QAOA}}(p, \beta, \theta)
\left(H \ket{0}\right)^{\otimes n}\right),
\end{equation*}
i.e., finding the pair of variables that is maximally correlated or
anticorrelated in the QAOA solution. This idea has been shown to
outperform the traditional QAOA both numerically and theoretically on
several problems \cite{bae2024recursive,kondo2024recursive}.

\listoftheorems[ignoreall,show=definition]

\bibliographystyle{apalike}
\bibliography{quantum,phd} 


\end{document}